AIX-MARSEILLE UNIVERSITY
IUSTI UMR 7343 CNRS

A DISSERTATION
*in Partial Fulfilment of the Requirements for the Degree of*
DOCTOR OF PHILOSOPHY
*Specialty*: Energetics, Mechanics and Fluid Mechanics
*Doctoral School 353*: Engineering Sciences: Mechanics, Physics, Micro and Nanoelectronics
*presented by*

PRASHANT KUMAR

# Investigation of Kelvin-like solid foams for potential engineering applications: An attractive set of geometrical and thermo-hydraulic properties

Public defence at Aix-Marseille University on Septembre 26, 2014,
in front of a jury composed of:

| | |
|---|---|
| Prof. Dominique Baillis | Referee (INSA, Lyon, France) |
| Prof. Michel De Paepe | Referee (Universiteit Gent, Belgium) |
| Prem. Yves Bienvenu | Examiner (Mines, ParisTech, France) |
| Prof. Azita Ahmadi-Senichault | Examiner (Université de Bordeaux, France) |
| Prof. Lounès Tadrist | Examiner (Aix-Marseille Université, France) |
| Dr. Isabelle Pitault | Examiner (Université de Lyon, France) |
| Dr. Fredéric Topin | Supervisor (Aix-Marseille Université, France) |

**Acknowledgements**

I still remember, back in November 2010 when I joined IUSTI (*Institut Universitaire des Systèmes Thermiques Industriels*), CNRS laboratory of Aix-Marseille University as a research engineer and later, was offered a thesis in May 2011. And today, I want to express my deepest gratitude to the people who are involved in the successful completion of my PhD.

My first and foremost gratitude goes to my supervisor, Assistant Prof. *Frederic TOPIN*, for the trust he has in me, for providing me the opportunity to work with him. I am thankful to him for explaining my questions lucidly. His way of dealing with problems at smaller levels and then adding levels of complexity to it, to combine various effects, was really productive and I found it an efficient way of working. He always welcomed new ideas and perspectives toward research and provided feedback through experience to show alternative directions and views. I highly appreciate the freedom he gave to me to work on this research topic, in choosing my own direction to achieve the research objectives, all of which made this experience extremely efficient and productive for me. I feel spoiled due to the freedom and flexibility he has afforded to me, and I believe I have learnt to value the productive work than the mundane office hours. To say the least, I have learned a lot from him; both professionally and personally. Thank you *Fred*, for allowing me to attend many conferences, presenting my work and enjoyable vacations otherwise.

Besides my advisor, I would like to thank the jury members who made me feel honoured to accept and judge this work. I am really thankful to Prof. *Dominique BAILLIS* and Prof. *Michel DE PAEPE*, who have had read this long manuscript conscientiously and henceforth, effectuate their detailed reports. I also express my gratitude to Prof. *Azita AHMADI-SENICHAULT* to accept as jury head, and Prof. *Yves BIENVENU* and Dr. *Isabelle PITAULT* for their encouragement, insightful comments, and fruitful suggestions.

My special thanks go to Prof. *Lounes TADRIST* (head of the laboratory) and Assistant Prof. *Jerome VICENTE* for their help, useful advices and insightful discussions during this thesis.

*Jean-Michel HUGO* (founder of THEMISTh), thank you very much for sharing the knowledge of 'milleux poreux', great discussions during the travels while participating in the international conferences.

Collaborative work with other institutes as well technical assistance from their personnel was an important element towards the successful completion of the present work. In this regard, I would like to thank all the members involved in this project. The financial support from ANR is gratefully acknowledged.

My sincere gratitude is rendered to all members of IUSTI for providing a friendly and enjoyable working atmosphere. I would like thank my colleagues at IUSTI, TCM team members for the pleasant atmosphere in truly international environment, people responsible in the administration, Mrs. *Jeanne PULLINO* and Mrs. *Joyce BARTOLINI* in particular, for their care and help.

Finally, I wholeheartedly thank my family: my parents *Mahendra* and *Indo*, brother *Nishant* and sister *Sristi*, my grandparents and all my relatives. *Nishant*, thank you for the trust you have in me and supporting me to follow my dreams. I really enjoyed all the trips together; we managed to make time for, in Europe. It brings smiles to my face, whenever think about our vacations in Marseille and Enschede.

I want to thank everyone for coming from around the globe on this special occasion. There are many people, who are behind me on this day – my sincere apologies if I have forgotten to name someone, but I know you will always be there for me.

Marseille, 26 September 2014 Prashant KUMAR

**Résumé**

Les mousses à cellules ouvertes ont diverses applications industrielles, par exemple pour des échangeurs de chaleur, des réacteurs structurés, la filtration, la catalyse, récepteurs solaires volumétriques en raison de leurs propriétés uniques telles qu'une importante porosité et une surface spécifique élevée. Leurs complexes interconnexions 3-D présentent un chemin tortueux qui guide l'écoulement de fluide et amène à l'amélioration de mélange pour augmenter le transfert de chaleur et contribue à diminuer la perte de charge par rapport à milieux de type 'lit de grains sphériques'. Cependant, en dépit de leurs propriétés intéressantes qui peuvent être exploitées dans les différentes industries, les mousses à cellules ouvertes n'ont pas encore été appliquées dans les opérations commerciales à grande échelle pour remplacer les lits de grains sphériques. Cela peut être attribué à leur coût de fabrication élevé, le manque de connaissance suffisante des équations de transport dans les mousses à cellules ouvertes ainsi que le manque d'expériences.

Pour améliorer et modifier la conception de ces applications, la compréhension des caractéristiques des mousses est donc très importante. En raison de la microstructure de la matrice complexe des mousses, il est difficile de prédire leurs propriétés thermo-physiques par modélisations mathématiques. Par conséquent, des études expérimentales et numériques ont été proposées pour découvrir l'effet de la microstructure sur la perte de charge et la conductivité thermique effective des mousses à cellules ouvertes.

L'objectif principal de ce travail était d'aborder les problèmes liés à la détermination des propriétés de transport de mousses à cellules ouvertes. A cet égard, une caractérisation détaillée de mousses à cellules ouvertes par rapport à leurs paramètres morphologiques a été exécutée dans un premier temps. La matérialisation de la géométrie de mousse, conçue en CAO (conception assistée par ordinateur), est réalisée par procédé de coulé par la route de fonderie. La possibilité de prédire les mêmes propriétés géométriques de la matrice de mousse conçue en CAO, par coulées et celles mesurées en utilisant les ressources disponibles est présentée et validée.

En outre, afin de déterminer théoriquement la surface spécifique géométrique et les relations entre les paramètres géométriques de mousses à cellules ouvertes (*isotropes* métalliques et céramiques), une corrélation mathématique généralisée a été développée. A cet effet, la géométrie de la tetrakaidecahedron (efficacité de remplissage d'espace et de la géométrie de mousse largement accepté dans la communauté) a été utilisé et différentes formes de sections transversales de brins de structures en mousse ont été pris en compte de façon explicite. La corrélation dérivée pour prédire les propriétés géométriques peut facilement être étendue à des formes différentes. En outre, la connaissance de l'un des deux paramètres géométriques est suffisante pour prédire avec précision les caractéristiques géométriques des autres.

La méthodologie pour développer la nature *anisotrope* des mousses à cellules ouvertes des sections transversales de brins différentes a également été prise en compte. La corrélation mathématique dérivée pour les mousses *isotropes* a été étendue et adaptée pour prédire la surface spécifique. Les corrélations ont été validées par rapport aux données expérimentales et les données de ce travail pour n'importe quel type de mousses *isotropes* et *anisotropes* avec moins d'erreurs que les corrélations déjà présentées dans la littérature.

Des simulations numériques 3D à l'échelle des pores ont été réalisées pour étudier les caractéristiques de pertes de charges. L'écoulement du fluide à travers la mousse à cellule ouverte a été réalisé dans trois régimes différents : le régime de Darcy, le régime transitoire et le régime inertiel. La méthodologie pour extraire les caractéristiques d'écoulement, à savoir la perméabilité de Darcy et le coefficient d'inertie de Forchheimer, a été donnée. L'impact des deux approches de perméabilité à savoir la perméabilité Darcianne et de Forchheimer sur les propriétés d'écoulement est discuté. L'importance des propriétés géométriques sur les caractéristiques d'écoulement de fluide et leurs inclusions dans les corrélations proposées pour prédire la perte de charge est discutée. La question « *Les paramètres d'Ergun peuvent-ils avoir des valeurs numériques constantes ou non* ? » est aussi largement discutée.

L'applicabilité des corrélations proposées pour la perte de charge a été validée pour des mousses à cellules ouvertes de différents matériaux (métalliques et céramiques) avec une large gamme de tailles de pores, différentes formes de brins et de la porosité ouverte. A cet effet, des données numériques et expérimentales de perte de charge issues de ce travail, ainsi que de la littérature ont été utilisée. Il a été démontré que les corrélations proposées dans le présent ouvrage peuvent prédire la perte de charge dans les mousses avec plus de précision que n'importe quelles autres corrélations disponibles dans l'état de l'art, que ce soit théorique ou empirique.

Les simulations numériques 3-D pour déterminer les tensors conductivité thermique effective en équilibre thermique local sont effectuées pour des mousses *isotropes* et *anisotropes* à cellules ouvertes. L'importance de la conductivité thermique intrinsèque du solide a été démontrée pour prédire la conductivité thermique effective analytiquement.

Trois différentes corrélations étaient dérivées pour prédire la conductivité thermique effective à la fois isotrope et anisotrope des mousses à cellules ouvertes. Les paramètres géométriques de la matrice de mousse étaient introduits dans les corrélations pour prédire la conductivité thermique effective. L'un des deux modèles peut également être utilisé pour résoudre un problème en même temps comme un système de deux équations linéaires où les deux phases solides intrinsèques et la conductivité thermique effective sont inconnues. Les corrélations ont été validées par rapport à des données expérimentales, numériques et de la littérature. Elles permettent d'obtenir des résultats avec plus de précision que les corrélations présentées dans la littérature.

**Abstract**

Open cell foams have diverse industrial applications e.g. heat exchangers, structured reactors, filtration due to their unique properties such as high porosity and high specific surface area. Their complex 3-D interconnections possess a tortuous path to guide the fluid flow; enhancing the mixing to improve heat transfer and helps in lowering down the pressure drop compared to packed bed of spheres. However, despite their attractive properties which can be exploited in various industries, open cell foams have still not been applied in large-scale commercial operations to replace the conventional packed bed spheres. This can be ascribed to their high manufacturing cost, lack of sufficient knowledge of transport processes in foam like structures as well as lack of handling experience.

To improve and modify the design of such applications, understanding of the foam characteristics is of quite importance. Because of the complex microstructure of foam matrix, it is difficult to predict their thermo-physical properties by mathematical modelling analysis. Therefore, experimental and numerical studies have been suggested to discover the effect of microstructure on the structural, pressure drop and effective thermal conductivity of open-cell foams.

The main aim of the present work was to address the problems related to the determination of transport properties of open cell foams. In this regard, as a first step, a comprehensive characterization of open cell foams with respect to their morphological parameters was performed. The materialization of foam geometry designed in CAD (computer aided design) is realized by casting method through foundry route. The possibility to predict the same geometrical properties of foam matrix designed in CAD, cast ones and measured using available resources is shown and validated.

In addition, in order to theoretically determine the geometric specific surface area and relationships between geometrical parameters of *isotropic* open cell foams (metal and ceramic), a generalized mathematical correlation was developed. For this purpose the tetrakaidecahedron geometry (an efficiently space-filling and widely accepted representative geometry of foams) was used and different shapes of strut cross-sections of foam structures were taken explicitly into account. The derived correlation to predict geometrical properties can be easily extended to different strut shapes. Furthermore, knowledge of any of the two geometrical parameters is sufficient to predict accurately other set of geometrical properties.

The methodology to develop *anisotropic* nature of open cell foams of different strut cross sections was also taken into account. The mathematical correlation derived for *isotropic* foams was extended and adapted to predict the specific surface area. The correlations were validated against experimental data and data from present work for any type of *isotropic* and *anisotropic* foams with least error than correlations already presented in the literature.

3-D numerical simulations at pore scale were performed to study the pressure drop characteristics. Fluid flow through open cell foam was performed in three different regimes: Darcy regime, transition regime and inertia regime. Methodology to extract flow characteristics namely Darcian permeability and Forchheimer inertia coefficient was given. Impact of two permeability approach i.e. Darcian and Forcheimmer permeability on flow properties is discussed. Importance of geometrical properties on fluid flow characteristics and their inclusion in the proposed correlations for predicting pressure drop is discussed. "*Can Ergun parameters have constant numerical values or not*" is also extensively discussed.

The applicability of the proposed correlations for pressure drop was validated for open cell foams of different materials (metal and ceramic) on a large range of pore sizes, different strut shapes and open porosity. For this purpose, numerical data of pressure drop from the present work as well as experimental data were used. It was demonstrated that the correlations proposed in the present work can predict the pressure drop in foams with more precision than any other state of the art correlation either theoretical or empirical.

3-D numerical simulations to determine effective thermal conductivity tensors in local thermal equilibrium condition were performed for *isotropic* and *anisotropic* open cell foams. Importance of intrinsic solid phase thermal conductivity was demonstrated to predict effective thermal conductivity analytically.

Three different correlations were derived to predict the effective thermal conductivity for both, *isotropic* and *anisotropic* open cell foams. Geometrical parameters of foam matrix were introduced in the correlations to predict effective thermal conductivity. Any of the two models can also be used to solve a problem simultaneously as a system of two linear equations where both intrinsic solid phase and effective thermal conductivities are unknown. The correlations were validated against experimental, numerical and data from the literature and estimate the results with the most precision than the correlations presented in the literature.

**Table of Content**

**Appendices**

**List of Figures**

**List of Tables**

## Nomenclature

### Latin Symbols

| | | |
|---|---|---|
| $a_c$ | Specific surface area | $m^{-1}$ |
| $a_{sw}$ | Opening area of square face of foam structure | $mm^2$ |
| $a_{hw}$ | Opening area of hexagon face of foam structure | $mm^2$ |
| $a_1$ | Strut shape controlling parameter | - |
| $a_2$ | Strut size controlling parameter | $mm$ |
| $a_3$ | Strut ligament curvature controlling parameter | - |
| $d_s/d_f$ | Strut / Fiber diameter | $mm$ |
| $d_h$ | Hydraulic diameter | $mm$ |
| $d_{ps}$ | Equivalent pore diameter of square opening | $mm$ |
| $d_{ph}$ | Equivalent pore diameter of hexagon opening | $mm$ |
| $d_p$ | Pore diameter | $mm$ |
| ${D_p}^{sp}$ | Maximum totally included sphere/ball in the cell | $mm$ |
| ${d_p}^{eq}$ | Equivalent sphere diameter | $mm$ |
| $d_{cell}$ | Cell diameter | $mm$ |
| $d_w$ | Window diameter | $mm$ |
| $D_p$ | Particle diameter | $mm$ |
| $F$ | Correction factor (Eq. 5.22) | - |
| $G$ | Shape function (Equation 3.3) | - |
| $k'$ | Constant (Eqs. 3.11 and 3.12) | - |
| $k''$ | Constant (Eqs. 3.13 and 3.14) | - |
| $L_s$ | Strut length | $mm$ |
| $L$ | Node to node length | $mm$ |
| $R_{eq}$ | Equivalent circular strut radius | $mm$ |
| $S_{ligament}$ | Lateral surface area of one ligament | $mm^2$ |
| $S_{node}$ | Lateral surface area of one node | $mm^2$ |
| $S'_L$ | Total lateral surface area of ligaments (anisotropic case) | $mm^2$ |
| $S'_N$ | Total lateral surface area of nodes (anisotropic case) | $mm^2$ |
| $V_f$ | Total fluid inside the octahedron | $mm^3$ |
| $V_s$ | Total solid volume of the octahedron | $mm^3$ |
| $V_{ligament}$ | Volume of one ligament | $mm^3$ |
| $V_{node}$ | Volume of one node | $mm^3$ |
| $V_T$ | Volume of octahedron | $mm^3$ |
| $V_c$ | Volume of cubic cell | $mm^3$ |
| $K$ | Permeability of the porous medium | $m^2$ |
| $K_{For}/K_{poly}$ | Forcheimmer permeability/ Permeability obtained using polynomial curve | $m^2$ |
| $K_D$ | Darcian permeability (Eqs. 4.1 and 4.6) | $m^2$ |
| $C$ | Inertia coefficient of the porous medium | $m^{-1}$ |
| $C_{poly}$ | Inertia coefficient obtained using polynomial curve | $m^{-1}$ |
| $C_{For}$ | Forcheimmer inertia coefficient (Eqs. 4.2 and 4.8) | $m^{-1}$ |
| $V$ | Superficial fluid velocity | $m.s^{-1}$ |
| $\nabla\langle P\rangle$ | Pressure gradient | $Pa.m^{-1}$ |
| $\Delta P$ | Pressure drop | $Pa$ |
| $\Delta x$ | Length of the porous medium | $mm$ |
| $C_L$ | Cubic coefficient of flow (Equation 4.3) | - |
| $E_1$ | Ergun parameter of viscous component | - |
| $E_2$ | Ergun parameter of inertia component | - |
| $E_1]_D$ | Ergun parameter of viscous component (Dietrich et al., 2009; Eq. 4.35) | - |
| $E_2]_D$ | Ergun parameter of inertia component (Dietrich et al., 2009; Eq. 4.35) | - |
| $d_h]_D$ | Hydraulic diameter (Dietrich et al., 2009; Eq. 4.35) | $mm$ |
| $c$ | Universal inertial coefficient | - |
| $f$ | Friction factor | - |
| $T$ | Temperature | $^oC$ |
| $P$ | Power | $W$ |
| $S$ | Constant (Eq. 5.30) | - |

**Greek Symbols**

| | | |
|---|---|---|
| $\varepsilon_o$ | Fluid phase open porosity | - |
| $\varepsilon_s$ | Solid phase open porosity | - |
| $\varepsilon_{st}$ | Strut porosity | - |
| $\varepsilon_t$ | Total porosity ($\varepsilon_t = \varepsilon_o + \varepsilon_{st}$) | - |
| $\varepsilon_n$ | Nominal porosity | - |
| $\varepsilon_{sur}$ | Surface porosity (Eq. 4.9) | - |
| $\partial$ | Tortuosity | - |
| $\alpha_{eq}$ | Ratio of strut radius (or side length of strut shape) to node to node length | - |
| $\beta$ | Ratio of strut length to node to node length | - |
| $\Omega$ | Elongation factor | - |
| $\mu$ | Dynamic fluid viscosity | $Pa.s$ |
| $\nu$ | Kinematic fluid viscosity | $m^2.s^{-1}$ |
| $\rho$ | Fluid density | $kg.m^{-3}$ |
| $\rho_s$ | Solid density | $kg.m^{-3}$ |
| $\rho_b$ | Bulk density | $kg.m^{-3}$ |
| $\rho_r$ | Relative density | - |
| $\varphi$ | Golden ratio (~ 1.6180) | - |
| $\theta$ | Ratio of pore diameter to node to node length | - |
| $\theta'$ | Ratio of pore diameter to equivalent strut radius | - |
| $\Pi$ | Constant (Eq. 3.69) | - |
| $\zeta$ | Constant (Eq. 3.70) | - |
| $\kappa$ | Constant (Eq. 3.79) | - |
| $\tau$ | Dimensionless geometrical parameter (Eq. 4.36) | - |
| $\Phi$ | Heat flux | $W.m^{-2}$ |
| $\psi$ | Dimensionless geometrical parameter (Eqs. 5.27 and 5.28) | - |
| $\psi''$ | Dimensionless geometrical parameter (Eqs. 5.40 and 5.41) | - |
| $\psi'$ | Dimensionless geometrical parameter (Eqs. 5.45 and 5.46) | - |
| $\lambda_s$ | Intrinsic solid phase conductivity | $W.m^{-1}.K^{-1}$ |
| $\lambda_f$ | Fluid phase conductivity | $W.m^{-1}.K^{-1}$ |
| $\lambda_{eff}$ | Effective thermal conductivity | $W.m^{-1}.K^{-1}$ |
| $\lambda_{parallel}$ | Effective parallel thermal conductivity (Eq. 5.1) | $W.m^{-1}.K^{-1}$ |
| $\lambda_{series}$ | Effective series thermal conductivity (Eq. 5.2) | $W.m^{-1}.K^{-1}$ |

**Abbreviations**

| | |
|---|---|
| PPI | Pores per inch |
| CAD | Computer aided design |
| SEBM | Selective electron beam melting |
| CTIF | Centre technique des Industries de la fonderie |
| LTE | Local thermal equilibrium |
| μ-CT | Micro computed tomography |
| ROI | Region of interest |
| RUC | Representative unit cell |
| HW | Heywood circularity factor |
| BET | Brunauer–Emmett–Teller theory |
| MRI | Magnetic resonance imaging |
| SiC | Silicon carbide |
| RMSD | Root mean square difference |

**Dimensionless Numbers**

| | |
|---|---|
| $Re$ | Reynolds number ($Re = \frac{\rho d_h V}{\mu}$) |
| $Re_c$ | Critical Reynolds number |
| $Hg$ | Hagen number ($Hg = \frac{\Delta P}{\Delta x}\frac{{d_h}^3}{\rho V^2}$) |

**Subscript and Superscript**

| | |
|---|---|
| $eq$ | Equivalent |
| $f$ | fluid |
| $s$ | solid |
| $xx$ | X-direction |
| $yy$ | Y-direction |
| $zz$ | Z-direction |

# Chapter 1

# Introduction

## 1.1 Motivation and aims

Three-dimensional cellular materials such as open cell foams (metal or ceramic) are of potential applications to achieve substantial gains in terms of increasing heat transfer, mixtures and chemical reactions. These materials can provide simultaneously a set of excellent qualities e.g. mechanical, thermal, catalytic that lead to significant reductions in cost, weight and offer opportunities for applications in diverse areas e.g. automobiles, fuel cells, chemical engineering, etc. (Lu et al., 1998; Calmidi and Mahajan 2000; Zhu et al., 2000; Banhart 2001; Boomsma and Poulikakos, 2001, Pitault et al., 2004).

This cellular material is primarily used for its mechanical properties especially in the aerospace and transportation where lightweight and resistant structures are needed. The thermal properties of such materials are equally important in many targeted applications e.g. cooling of electronic components, thermal insulation (sandwich panels). To a lesser degree, the acoustic properties are also used in applications such as vibration damping, adjusting the sound and noise reduction. Currently, heat exchangers, reactors (including catalyst) with foam as internals are one of the most promising areas of application. The open cell foam in these exchangers can be used as a catalyst carrier.

Open cell foams (mostly metal foams) are proposed as a promoter of heat transfer for many applications such as compact heat exchangers, two-phase cooling and the spreaders (Banhart, 2001; Tadrist et al., 2004). Indeed, this material allows increasing the heat and mass transfer significantly without prohibitively increasing the flow resistance. Their field of applications have increased substantially in recent years due to the properties already highlighted and the emergence of industrial production site. For example, aluminium foams are used as mechanical reinforcement in aeronautics and the space (Sullines and Daryabeige, 2001) while the others are used as thermal protection in geothermal transactions, oil extraction (Vafai and Tien, 1982). The ceramic and metal foams are used in burners, heat pipes, the high-performance batteries and as support catalyst in fuel cell systems (Catillon et al., 2004). The control of porous material texture used in the compact and multifunctional exchangers (evaporators, spray-reformers, etc.) is a major technological challenge.

For chemical engineering applications such as reactor internals, open cell foams offer clear advantages over conventional randomly packed (irregular arrangement of individual particles) fixed-bed reactors e.g. remarkably low pressure drop due to high geometric specific surface area and enhanced heat and mass transfer due to the continuous connection of the foam structure.

Because of their novelty, particular structure, and variability of texture related to the different routes of development, open cell foams are still poorly characterized in terms of the transport properties. The accurate assessment of thermo-hydraulic properties becomes critical for various applications. The recent emergence of potential use of foams in new technologies motivates the development of tools for the characterization of foam matrix.

Studies in the field of solid mechanics do not require the data such as the actual density of the foam sample and evaluation of strut diameter in majority of the analysis. From the mechanical point of view, the approaches based on periodic patterns are very satisfactory. However, from the viewpoint of heat transfer and fluid flow, the situation is more complex. The foam structure does not allow exact analytical study to estimate geometrical and thermo-hydraulic characteristics.

The foam matrix is usually described by its morphological parameters, namely cell size, pore size, strut thickness and porosity. The most commonly used characteristic of open cell foams is the PPI (pores per inch) or so called pore count value, which can be obtained by counting the pores in a linear inch (e.g. Gibson and Ashby, 1997; Richardson et al., 2000; Mullens et al., 2006) which, actually does not define anything regarding foam properties and is generally provided by the manufacturer.

In spite of their outstanding properties which can be exploited in the numerous conventional and renewable systems, open cell foams have not still been applied in large-scale commercial operations. Although transport phenomena in porous media are studied for nearly two centuries, the work on highly porous materials are still relatively few and recent (Topin et al., 2006; Inayat et al. 2011a, b; Hugo and Topin, 2012).

In order to successfully design the advanced heat exchangers, volumetric solar receivers, reactors with foams as internals, a comprehensive knowledge of the geometric, morphological and thermo-hydraulic properties of foam structures is the primary prerequisite. Many authors (Richardson et al., 2003; Giani et al., 2005; Lacroix et al., 2007; Garrido et al., 2008; Grosse et al., 2009; Huu et al., 2009; Dietrich et al., 2009; Dietrich et al., 2010; Inayat et al., 2011 a, b) have introduced different experimental methods and proposed various geometric models and mathematical correlations in order to determine the foam parameters e.g. pore and strut diameters, specific surface area, pressure drop, and, effective thermal conductivity, which are essential parameters for a successful heat exchanger and reactor design. However, as mentioned before, despite a considerable amount of research that has been performed in this area over the past decades, no generally applicable correlations for determining the foam properties have been proposed so far. According to the recent literature, for a realistic modelling of foams and develop appropriate correlations for estimating the foam properties, it is extremely important to elect a suitable representative geometry and consider the variations in the strut cross section with/without the internal cavity.

In this regard, the first part of the present work investigates in determining/estimating the morphological and geometrical characteristics of open cell foams. The aims include:

- Design of the foam structure in CAD (computer aided design) and its materialization using cast method by foundry route in order to study the morphological properties.
- An experimental characterization of cast open cell foams with respect to their pore size, strut size, porosity, and specific surface area in order to compare and validate with the CAD model.
- Development of *isotropic* and *anisotropic* foam matrices of various strut cross sections in CAD for a wide range of low and high porosity.
- Development of a mathematical correlation in order to theoretically determine the geometrical characteristics (and relations between them) of open cell foams by taking the different strut morphologies (solid or hollow nature of foams) and different strut shapes into account.

As discussed above, the thermo-hydraulic properties are poorly characterized because of insufficient knowledge of geometrical parameters and experimental data are still in scarce. The experimental data reported in the literature are performed for a given small set of foam samples (usually very high porosity) which generally do not give close resemblances to the different authors' work. It is thus, necessary to generate a database of thermo-hydraulic properties to understand the physical mechanisms behind the transport phenomena and influence of individual geometrical parameter on them.

The second part of the present work deals with the 3-D numerical simulations at pore scale to obtain the flow properties and effective thermal conductivity. The aims include:

- Creation of a database of flow properties and effective thermal conductivity for *isotropic* and *anisotropic* foam structures.
- Development of generally applicable pressure drop correlation (different forms on a common basis according to the strut cross section) in order to predict the flow properties in a wide range of pore size and porosity.
- Development of three dependent/independent generally applicable correlations by taking the different strut morphologies (solid or hollow nature of foams) into account to predict effective thermal conductivity.
- Understanding the critical importance of intrinsic solid phase conductivity of foam material.

## 1.2 Scope and outline of the thesis

The following paragraphs present the goals set for each chapter as well as a brief overview of the experimental/numerical and theoretical approaches used.

An experimental characterization of the morphological parameters of regular and periodic open cell foams is presented in the second chapter. The cast foam sample is first designed in CAD and then materialized by casting method using foundry route. The aim is to compare and validate the foam characteristics' defined in CAD and cast sample in order to highlight the important morphological properties (necessary for geometric description and modelling) of open cell foams.

Further, virtual *isotropic* and *anisotropic* foam samples of different strut cross sections based on tetrakaidecahedron geometry are developed in CAD in order to measure their geometrical properties directly from CAD. The measured parameters of foams are further used in the following chapters for developing and validating the correlations to estimate the properties of open cell foams for their potential engineering applications.

Various geometric models as well as correlations reported in the literature are discussed in the third chapter. The applicability and validity of state of the art correlations for predicting geometrical characteristics of foam matrix are evaluated. A generic mathematical correlation for predicting the complete set of geometrical properties (porosity, strut and pore diameters, strut and node lengths, and specific surface area) of a foam matrix based on tetrakaidecahedron structure is derived. Correlation for open cell foams having hollow strut is also derived.

Nowadays, it is possible to obtain thermo-hydraulic properties of open cell foams using 3-D numerical simulations. Due to the scarcity of commercially available open cell foams of different strut shapes in a wide porosity range and experimental data, it is convenient to generate a database in order to understand the physics behind the transport phenomena.

The fourth chapter deals with the prediction/extraction of flow properties reported in the literature and to review the state of the art correlations for the prediction of pressure drop. 3-D numerical simulations at pore scale were performed and a database of extracted flow properties is generated. The applicability and validity of state of the art correlations for pressure drop prediction are evaluated using experimental and numerical data from the present work. A methodology is proposed to determine the flow properties in order to reduce dispersion in the friction factor. The purpose of this work is to develop a new generalized correlation (different forms according to the strut cross section) to predict precisely the flow characteristics that allows the theoretical determination of pressure drop with more precision for *isotropic* open cell foams of different strut shapes in the wide range of pore size and porosity. Emphasis is given to compare the numerically obtained flow properties separately i.e. permeability and inertia coefficient against correlations instead of pressure drop values for the entire velocity range. The validity of the new correlations is examined by comparing the predictions against the experimental/numerical pressure drop data of open cell foams from the present work.

In the fifth chapter, the problem of predicting the effective thermal conductivity in local thermal equilibrium condition of open cell foams is addressed. A database of effective thermal conductivity values is generated using 3-D numerical simulations at pore scale. The applicability and validity of state of the art of effective thermal conductivity correlations are examined using the data from present work. The goal is to develop a generalized correlation that allows the theoretical determination of the effective thermal conductivity precisely of different foam materials. Three different dependent or independent correlations are developed in the wide range of solid to fluid thermal conductivity ratios for *isotropic* and *anisotropic* foams. The validity and applicability of the new correlations are evaluated by comparing the predicted values against the experimental/numerical effective thermal conductivity data of open cell foams from present work as well as from the literature. Critical importance of intrinsic solid phase conductivity of foams is discussed in detail.

A general conclusion of the thesis is given in the sixth chapter, in which major findings and the concluding remarks from each chapter are presented followed by future recommendations to the present work.

# Chapter 2

# Morphological characterization of open cell foams

Parts of this work were already published or submitted to International Journal of Thermal Sciences, Journal of Porous Media and Defects and Diffusion Forum.

- **P. Kumar**, F. Topin & J. Vicente, Determination of effective thermal conductivity from geometrical properties: Application to open cell foams, *International Journal of Thermal Sciences*, 81, pp. 13-28, 2014.

- **P. Kumar** & F. Topin, About thermo-hydraulic properties of open cell foams: Pore scale numerical analysis of strut shapes, *Defects and Diffusion Forum*, 354, pp. 195-200, 2014.

- **P. Kumar**, F. Topin & L. Tadrist, Geometrical characterization of Kelvin like metal foams for different strut shapes and porosity, *Journal of Porous Media* (under review).

- **P. Kumar** & F. Topin, Impact of anisotropy on geometrical and thermal conductivity of metallic foam structures, Special issue of Heat and Mass Transfer in Porous Media, *Journal of Porous Media* (under review).

Parts of this work were already communicated in national/international conferences.

- **P. Kumar** & F. Topin, Influence of strut shape and porosity on geometrical properties and effective thermal conductivity of Kelvin-like anisotropic metal foams, *The 15$^{th}$ Heat Transfer Conference (IHTC)-2014*, Kyoto, *Japan.*

- **P. Kumar** & F. Topin, About thermo-hydraulic properties of open cell foams: Pore scale numerical analysis of strut shapes, *Diffusion in Solid and Liquids (DSL)-2013*, Madrid, *Spain.*

## 2.1 Background

The term “solid foams” is commonly used for describing 3-D cellular materials with a solid phase arranged into cells - polyhedra, which fill the 3-D space. The cells can be either open or closed. High technology foams are manufactured from polymers, ceramics, and metals and that can be used in reinforced lightweight structures, packaging, and crash-protection systems. Because of their structure, natural and synthetic cellular solid foams show unique physical properties which provide their optimal functionality. The development of mechanics of cellular foams is documented in the work of Gibson and Ashby (1997).

The topology and morphology of the foam microstructure reflect a method of its preparation which usually involves a continuous liquid phase that eventually solidifies and therefore surface tension and related interfacial effects often control the foam structure. There are two well-known elementary features of the liquid foam structure that are required to minimize surface energy. According to Bikerman (1973), three films always meet at equal angles of 120° to form a film junction called Plateau border (Gibson and Ashby (1997), where the work of J.A.F. Plateau, 1873 is cited). Four Plateau borders always join at the tetrahedral angle of 109.47°. For open cell foams, Plateau borders are identified as foam skeleton struts (Warren and Kraynik, 1997) which naturally takes the shape of convex triangle as shown in the Figure 2.1.

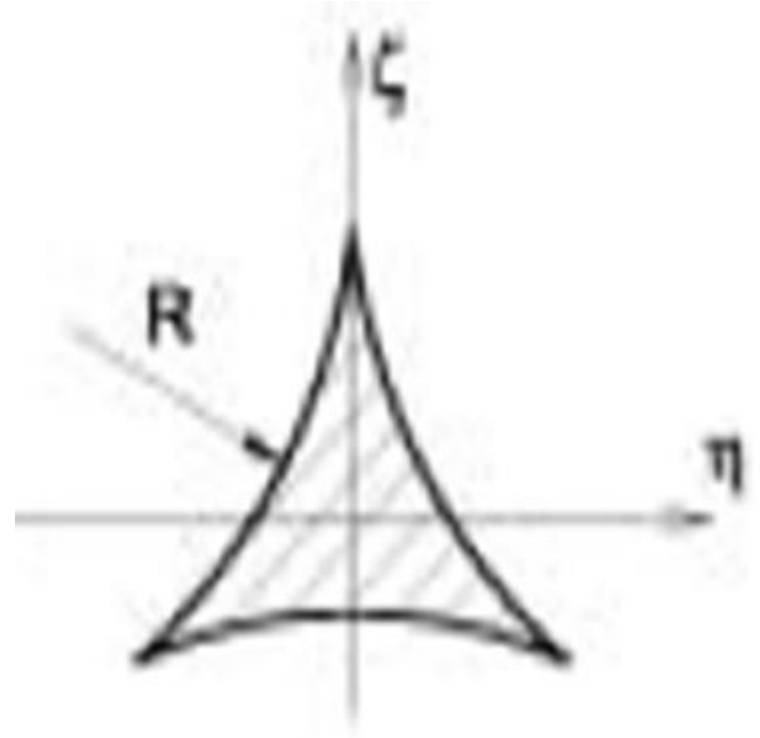

**Figure 2.1**. Strut cross section: Plateau border

Open cell foams are highly porous cellular materials with pore densities of usually 10-100 PPI and typical porosity ranges from 75% to 90% for ceramic foams and 85% to 97% for metal foams (Richardson et al., 2000; Giani et al., 2005; Twigg and Richardson, 2007). They have a complex 3-D internal architecture with widely distributed properties (Gibson and Ashby, 1997; Richardson et al., 2003; Mullens et al., 2006).

Open cell foams may exhibit considerable variations in their structure which can be caused by different factors, e.g. manufacturing route, polymer foam templates and the slurry (ceramic or metal) used. This may hold true even for foams of similar PPI (Mullens et al., 2006; Garrido, 2008). Also, the manufacturing route and the porosity range influence the foam properties to a large extent e.g. manufacturing route determines the strut morphology (solid or hollow) and porosity range determines the shape of the strut cross section (circular, convex or concave triangular). The change in the strut cross section is categorized in many works e.g. Bhattacharya et al., 2002; Scheffler and Colombo, 2005; Huu et al., 2009; Inayat et al., 2011a.

There are several manufacturing routes that exist already in the literature to fabricate replicated open cell foams that exhibit irregular cellular structures. Open cell foams are produced either by introducing voids into an initially pore-free liquid (molten) metal or by collecting subdivided material in a way that the assembly becomes highly porous (Gergely et al., 2000).

The present work is mainly focused on the methodologies that produce periodic structures and are able to produce open cell foams of desired and controlled morphological quantities. In this chapter, a brief overview of different manufacturing routes for the production of periodic open cell foams is given. The goal of this chapter is to perform a comprehensive experimental characterization of the morphological properties of open cell foams. For the complete characterization, virtual open cell foams were first designed in computer aided design (CAD) and then materialized using casting method by the foundry route. The objective was to validate the geometrical characteristics of CAD data against the measured values of cast foam samples. The second objective was to create virtual *isotropic* and *anisotropic* foam samples of desired strut shape to generate a database of geometrical characteristics that were measured directly in CAD.

## 2.2 Periodic cellular foams

Periodic cellular materials (foam-like structures) represent a class of cellular materials with defined cell/pore size, cell geometry and cell orientation. They can be produced by additive manufacturing using different techniques, e.g. selective electron beam melting (SEBM), selective laser sintering, 3-D printing etc. (Stampfl et al., 2004; Heinl et al., 2007). With these techniques, it is possible to control the morphological and geometrical parameters of foam structures with a high degree of reproducibility and vary these properties in a controlled manner. An example of periodic cellular lattices by SEBM is shown in the Figure 2.2.

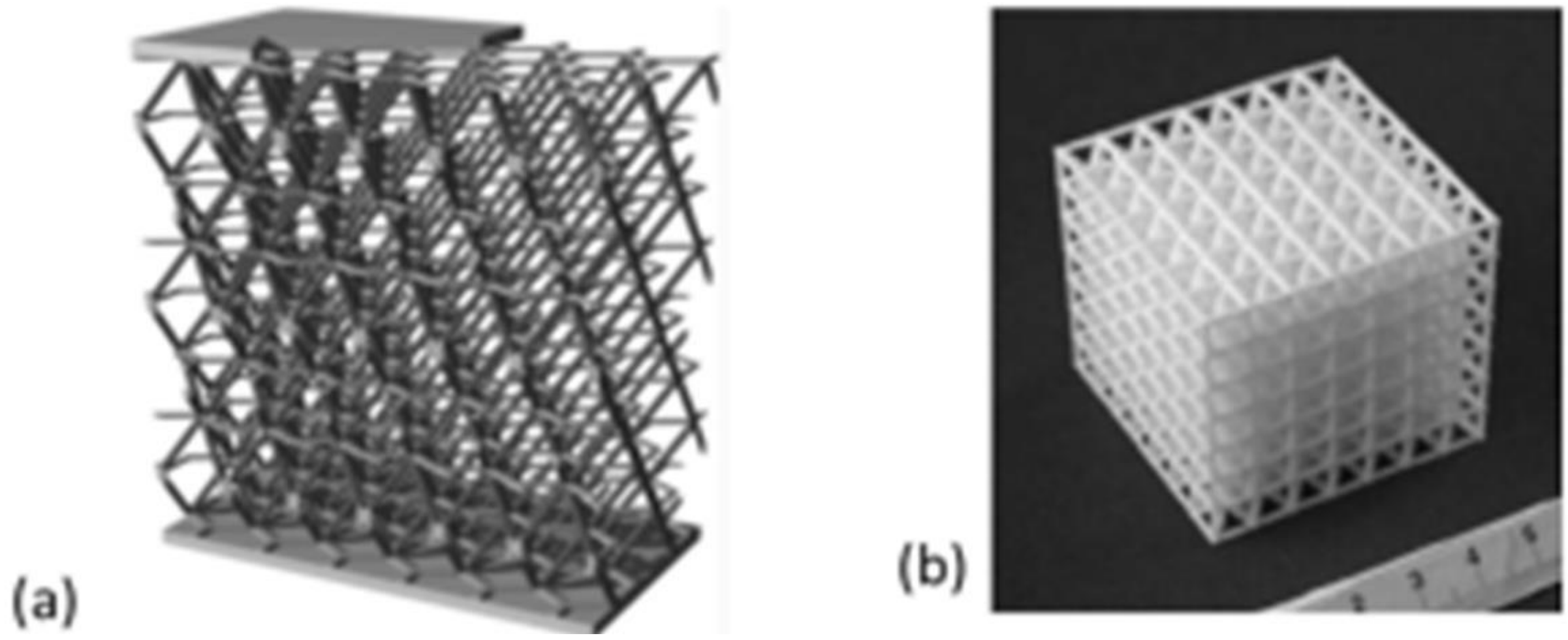

**Figure 2.2**. (a) An example of periodic cellular lattice (Wadley et al., 2003), (b) periodic cellular structure produced by using selective laser sintering (Stampfl et al., 2004).

Owing to a fine control (through manufacturing techniques) on the geometrical and morphological properties of periodic cellular structures, they can be considered as ideal systems to study the effect of morphological parameters on the mechanical, thermal and fluid transport properties of foams and foam-like structures. The knowledge obtained from such a systematic study can thus be applied in mathematical modeling as to develop appropriate correlations for the prediction of important data needed for designing a column or a reactor or a heat exchanger that uses a foam matrix as internal. The mathematical and geometrical models as well as correlations developed for ideal foam/cellular geometries can then be adapted for the non-ideal geometries as encountered in replicated open cell foams.

***Selective electron beam melting (SEBM)***

Selective electron beam melting is an additive manufacturing technique which allows the free form generation of 3-D metallic components from metal powder. In contrast to conventional machining, in additive manufacturing technique, parts are produced by successive melting of materials layers rather than removing the material (Heinl et al., 2008).

The basic requirement for applying SEBM technique is the generation of 3-D CAD component which is to be produced. In order to generate the layer information, the CAD model is sliced into layers of constant thickness. Each layer is then melted to an exact geometry as defined by the 3-D CAD model. The SEBM process is carried out in vacuum in order to avoid the exposure of the materials to atmosphere. The complete description of the procedure and the process parameters of the SEBM manufacturing technique can be found elsewhere (Heinl et al., 2007; Heinl et al., 2008). A systematic presentation of production of cellular periodic foam algorithm is presented in Figure 2.3. Using SEBM technique, samples of periodic cellular structures with ideal cubic and Kelvin-like cell geometries can be easily manufactured and an example is presented in Figure 2.4.

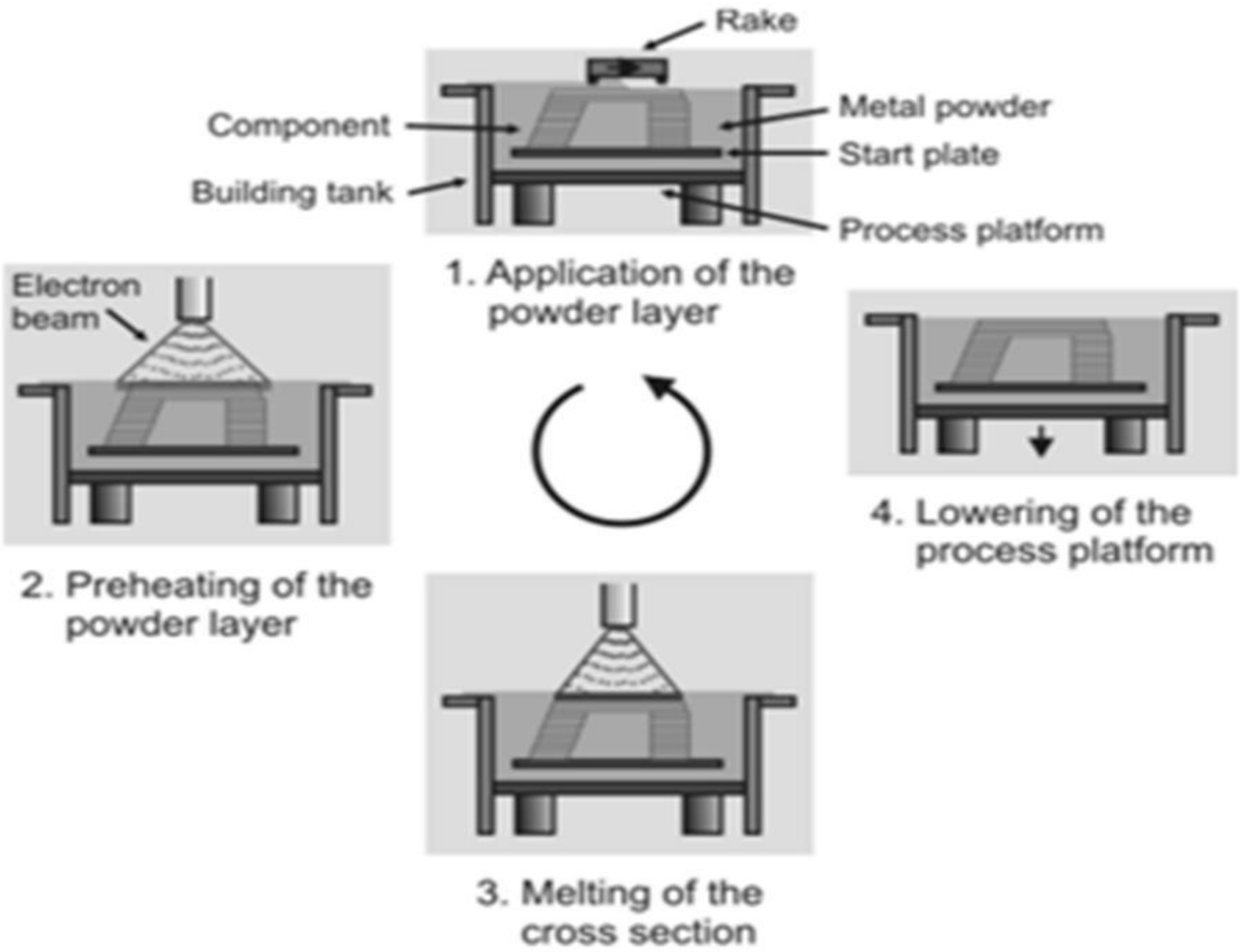

**Figure 2.3**. Component generation using the SEBM technique (Heinl et al., 2008).

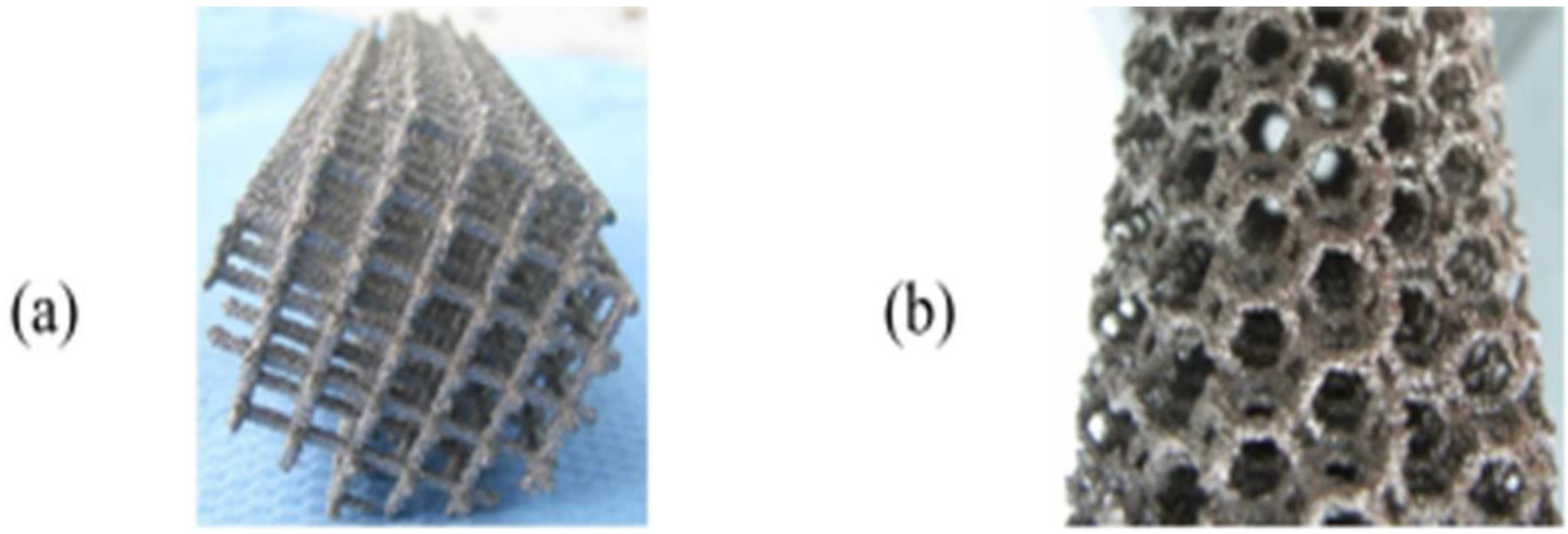

**Figure 2.4**. Periodic cellular structures with ideal (a) cubic and (b) tetrakaidecahedron packing (see Inyat et al., 2011a)

### *CTIF (Centre technique des Industries de la fonderie) casting technique*

A Kelvin cell is produced by CTIF using casting method by the foundry route (Dairon et al., 2011) that was earlier modeled in CAD. Casting processes are based on the “infiltration” technique that is similar to conventional foundry techniques, in which a preform, a sort of porous core, is infiltrated with molten metal possessing convex triangular strut shape. In the beginning, balls or aggregates, called "precursors", are placed loose into the mould. As they do not occupy all of the space, these precursors form a network of interconnected pores, i.e. the preform (Dairon et al., 2011). When precursors are spherical, this network gets a topology similar to that of foams produced by introducing a gas into a liquid. In the second case, open-pore polymer foam is used as lost pattern. The polymer foam is infiltrated by ceramic slurry, then heat-treated to solidify the slurry and burn out the polymer foam. The result is a network of pores having the same shape as the original foam, which can be then infiltrated by metal and presented in Figure 2.5.

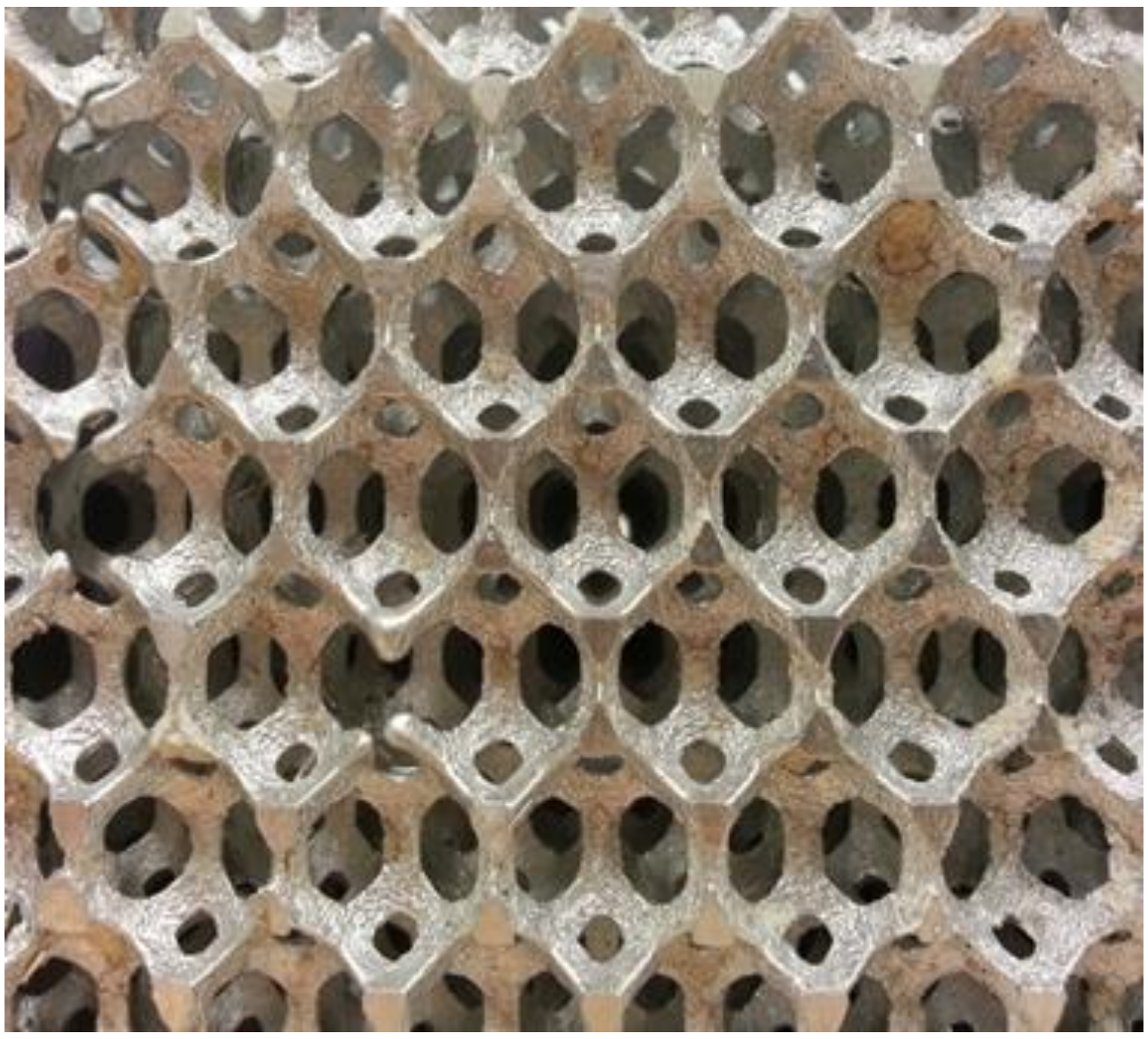

**Figure 2.5**. Image of cast regular and periodic foam fabricated by CTIF (Dairon et al., 2011).

Casting techniques developed by CTIF allow manufacturing of open cell metal foams with which morphology and metallurgy and so, properties can be tailored. The obtained structures either stochastic or regular can be included in pieces getting a solid skin and solid parts. At last, the use of casting techniques keeps the well-known possibilities of the foundry route in terms of complexity of shapes and reduced cost.

## 2.3 Matrix of open cell foams

Open cell foams either metallic or ceramic find their potential applications in many engineering sectors including liquid-metal filtration, light-weight constructions, use as packings in columns, as catalytic reactors, heat exchangers, solar receivers, or for thermal insulation (Schlegel et al., 1993; Adler and Standtke, 2003a, b; Reitzmann et al.,2006; Lacroix et al., 2007).

In the case of open cell foams, the cells are connected to each other with solid edges (struts) and cell to cell connectivity takes place via open faces (pores or windows) (e.g. Mullens et al., 2006; Sharafat et al., 2006, Vicente et al., 2006a). The basic building blocks in open cell foams are the struts, which are connected to each other by three dimensionally constituting polyhedral cells (void volumes) and generate its foam-like structure.

The strut morphology and its cross section (resulting from the manufacturing route and the porosity range respectively) affect the geometrical characteristics of foams (Mullens et al., 2006; Binner, 2006; Huu et al., 2009; Inayat et al., 2011a). There are some geometries of the open cell foams that reveal either square and hexagon or pentagon openings and that depend on the manufacturing route. A typical foam matrix of reticulated open cell foams (with square and hexagon opening in left and pentagon opening in right) is shown in Figure 2.6. The details of different strut morphologies and various strut cross sections are discussed thoroughly in the following sections.

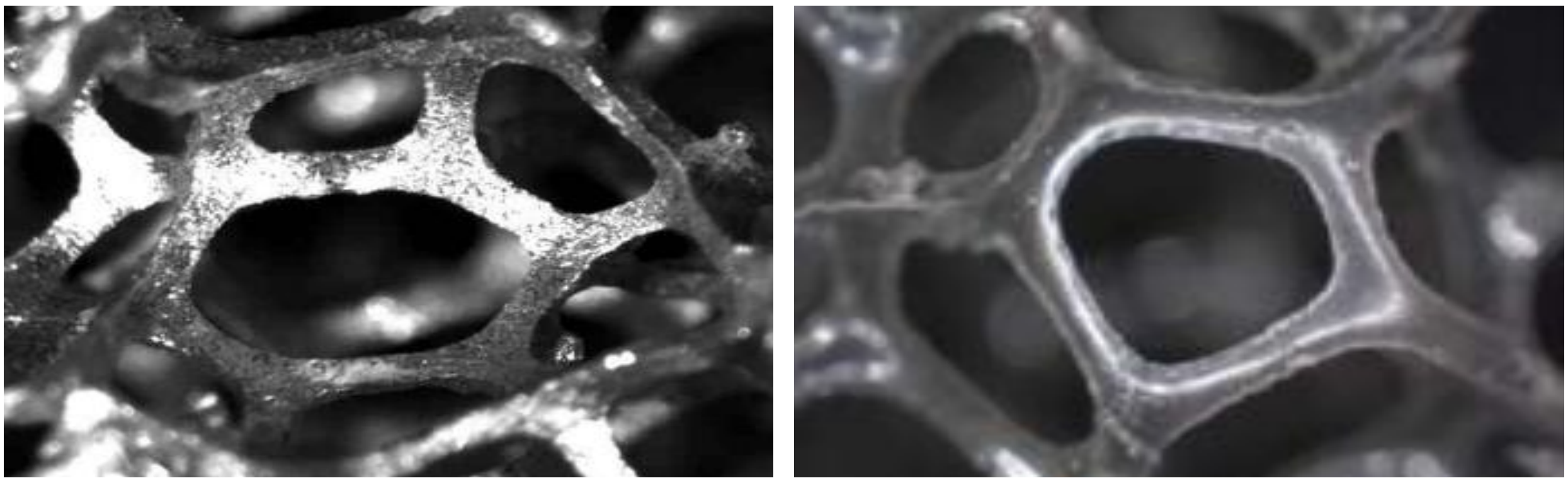

**Figure 2.6**. Presentation of square and hexagonal openings (left) and pentagonal opening (right) of cellular structures.

Manufacturing route plays a pivotal role in producing solid struts and struts with internal cavity. Open cell foams are usually manufactured by impregnating open cell polymer foams internally with ceramic slurry and then firing in a kiln, leaving only ceramic material.

Metal foams can have either solid struts (e.g. investing casting method) or struts with internal cavity (e.g. electrochemical deposition method). On the other hand, most commercial ceramic foams are prepared using a reticulation method that was first patented by Schwartzwalder and Somers (1963) in which they are obtained as positive image of the polymeric foam template (Richardson et al., 2000). This process involves infiltration of ceramic slurry into the polymer foam, removal of excess slurry, drying and burning of the polymer phase. As a result, the ceramic particles are sintered together to give a foam structure consisting of hollow struts (with internal void volume) that have a rough surface. It has been reported in the literature by many authors (e.g. Garrido et al., 2008; Grosse et al., 2009; Inayat et al., 2011 a, b; Dietrich et al., 2009, Dietrich et al., 2010) that ceramic foams possess internal void or cavity and this cavity is generally not accessible to the fluid flow.

Figure 2.7 (top) shows metal foam with solid struts while Figure 2.7 (bottom) shows ceramic foams of different PPI and porosity with internal cavity in the solid struts.

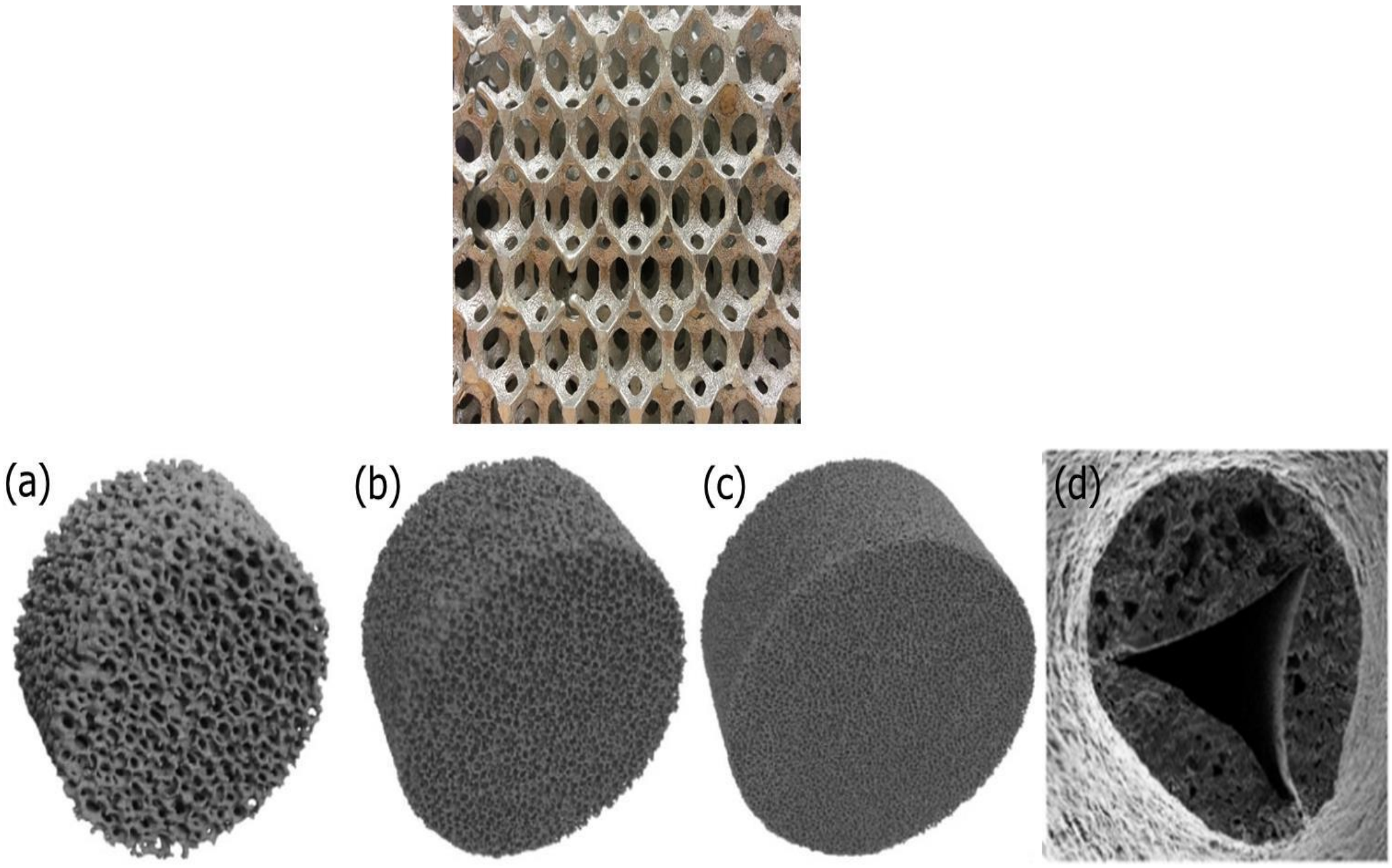

**Figure 2.7**. Top- Presentation of metal foam of $\varepsilon_o$=85% (no internal void) obtained by casting process (see Dairon et al., 2011). Bottom- Presentation of OBSiC ceramic foams of $\varepsilon_n$=80%: (a) 10 PPI, (b) 20 PPI, (c) 45 PPI and (d) hollow strut (or internal void) (see Dietrich et al., 2009).

Open cell foams are usually available in a wide range of cell/pore sizes for several cm down to µm. The cell size is one of the key parameters in designing and manufacturing of the open cell foams, as for many engineering applications, the foam performance can be directly influenced by its cell size. The conventional way of representing the cell or pore size of a foam structure is PPI (pores per linear inch), which is obtained by counting pores in a linear inch. It is also termed as pore count or pore density (Scheffler and Colombo, 2005; Twigg and Richardson, 2007; Garrido, 2008). The term PPI, pore-count or pore density can be confusing due to the unclear definition of a pore, because it could be a window or a cell.

The pore density or pore count used by the foam manufacturers does not give a precise measure but merely reflects a range of cell or pore sizes. Usually, each foam manufacturer has its own reference scale e.g. foam considered as 40 PPI by one manufacturer could be defined as 60 PPI by another. Also in practice, there can be a wide variation between the measured pore size and the pore diameter calculated from the PPI value (Richardson et al., 2000; Mullens et al., 2006).

Most of the replicated open cell foams are slightly anisotropic in nature. They may have ellipsoidal cell shape (caused by elongation) and deformed strut network. The anisotropy in foam structures may be ascribed to the manufacturing process (Garrido et al., 2008; Grosse et al., 2009; De Jaeger et al., 2011).

Open cell foams usually have high geometric specific surface area (surface to volume ratio). In a foam structure, the geometric surface area is relevant for momentum, heat and mass transfer and hence is of fundamental importance for designing various industrial systems based on foam matrices (Stemmet et al., 2007; Garrido et al., 2008; Zuercher et al., 2009; Leveque et al., 2009). Therefore, an accurate determination of the specific surface area of foams is extremely important for their successful industrial applications.

## 2.4 Morphological characterization

With the development of advanced technologies in the field of cellular materials, open cell foams with controlled morphological properties can be easily fabricated in order to study the influence of various strut shapes and their impact on thermo-hydraulic properties. Further, it is also critical to have knowledge of these designed controlled parameters and then, production and materialization of such open cell foams to identify whether they correspond to the same characteristics or not that were already modelled in CAD. This could be extremely useful to implement directly in various engineering and chemical processes like heat exchangers, filters, separation, solar receivers etc. to predict the systems' performances just by knowing the morphological properties of open cell foams.

A periodic open cell foam structure having convex triangular strut shape (see Figure 2.8) of different porosities was designed first by controlling the cell size of the foam structure in CAD. The details of geometrical parameters are presented in the Table 2.2. Using the CAD unit cell, the foam samples were then, cast by CTIF (Dairon et al., 2011).

For the morphological study in this thesis, X-ray micro computed tomography (µCT) was performed to measure the geometrical parameters of cast Kelvin cell foams using iMorph (Vicente et al., 2006a). The detailed description of the µCT principle, the measurement method and the algorithm can be found in the work of Hildebrand and Ruegsegger (1997).

To scan the totality of our sample (Length x Breadth x Height, $mm^3$), a resolution of 155µm to obtain the voxel size was fixed. The total volume in voxel size is 750 x 750 x 1178. The chosen region of interest (ROI) used for the geometrical computation is a box inside the tomography sample of voxel size 458 x 436 x 500 (see Figure 2.8-left). The ROI is centered in order to avoid edge effect due to walls and its size has been maximized in order to obtain a large number of pores and struts for geometrical statistical analysis using in-house code; iMorph (Vicente et al., 2006). The threshold choice is crucial for determining porosity ($\varepsilon_o$) and specific surface area ($a_c$) of foam samples.

The following quantities were mainly measured: porosity ($\varepsilon_o$), strut diameter ($d_s$), pore diameters (${d_p}^{eq}$ and ${D_p}^{sp}$), node to node length ($L$), spherecity, tortuosity ($\partial$) and specific surface area ($a_c$). These quantities describe well the foam properties; are inter-linked with each other and can be well related to thermo-hydraulic properties.

### 2.4.1 Characterization of porosity

In Figure 2.8 (right-top), the density distribution clearly shows two separated classes that represent the solid and pore voxels respectively. The threshold should be chosen between these two Gaussian curves. The impact of the threshold variation on the porosity value was estimated. The threshold varies between 74 and 207 and thus, porosities vary between 83% and 89% between the two Gaussian distributions. For this sample, a threshold value of 117 leads to obtain the same porosity of 85% as per weighing method of the original sample. So the uncertainty in porosity measurement using µCT is ±3% of the sample. The threshold variability concerns voxels located near the interface of the solid and pore phases. The error in porosity generally depends on voxel size and thus, impact the specific surface area that is directly linked to the number of voxels at the interface. Moreover, the uncertainty is determined from the maximum and minimum threshold values that are actually linked to addition of removal of voxels on the whole surface.

For the other foam samples, the threshold is determined using the same procedure by taking into account the experimental (weighted) porosity. Moreover, the thin or sharp edges of strut shape under voxel resolution are not visible and need to be taken into account. It is very difficult to quantify the total volume of these thin/sharp edges, taking a threshold which gives exactly the experimental porosity which may overestimate the identification of few solid voxels. This leads to a smooth strut shape. It is also well known that the tomography tends to blur data and overestimate the solid present near the thin/sharp edge of the strut. Using iMorph (Vicente et al., 2006a), the 3-D reconstruction of 85% porosity sample from computed tomography and the image of the same real cast foam sample are shown in Figure 2.8 (right-center, bottom). The same procedure for 3-D reconstruction and the measurement of geometrical parameters of other foam samples are followed (see Table 2.2).

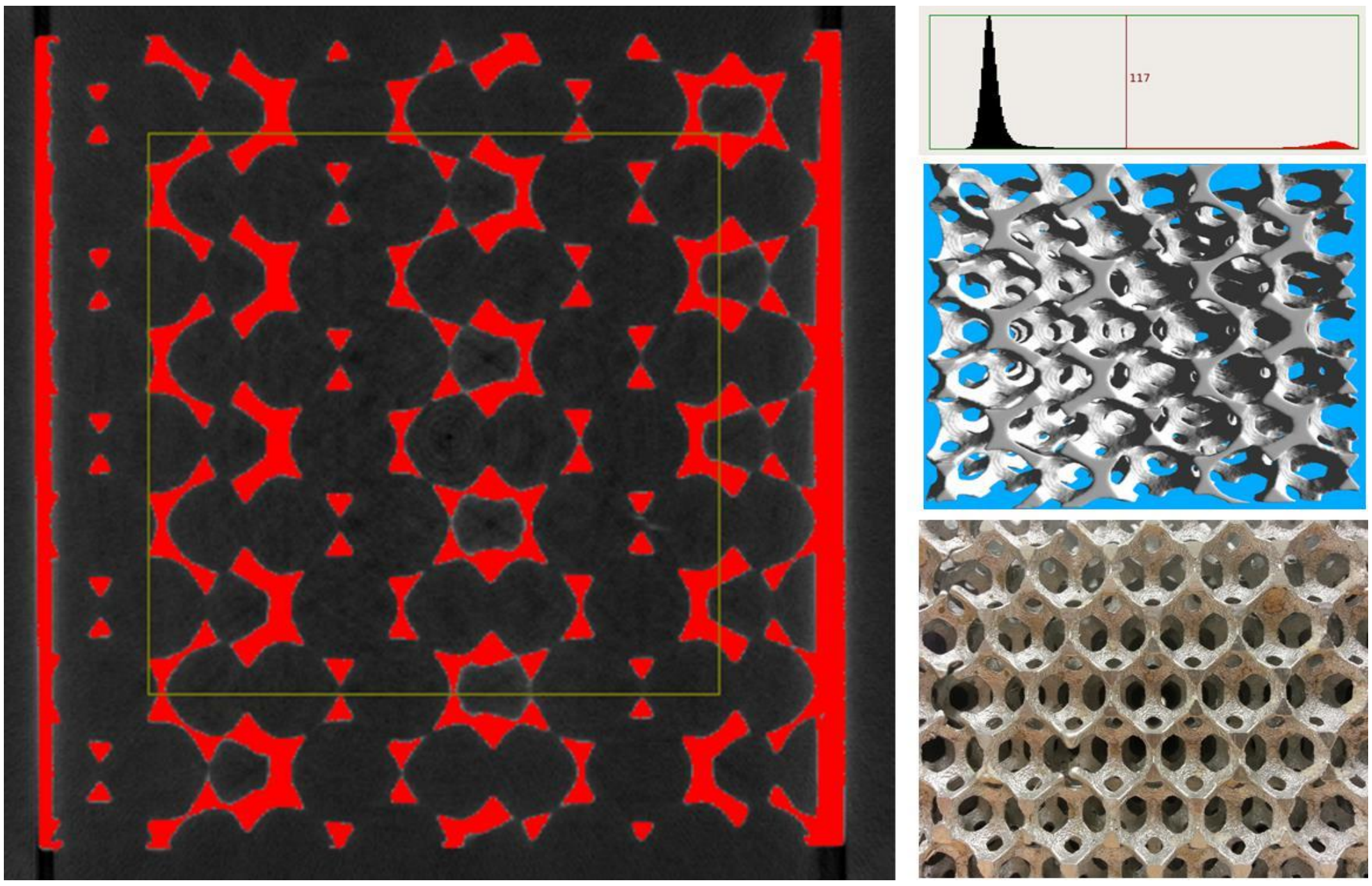

**Figure 2.8**. Left: Slice obtained from original tomography made by µCT scan. The box inside the slice represents the chosen region of interest (ROI). Right-top: Material density distribution (Grey levels) of ROI. Solid voxels with grey level upper than 117 are shown in red. Right-center: 3- D reconstruction uses the marching cubes algorithm to obtain the iso-density surface mesh of the chosen threshold used to separate the two phases. Right-bottom: Image of cast Kelvin cell foam. ROI is made on sample 4 of 85% porosity.

### 2.4.2 Characterization of strut and pore diameter

In iMorph (Vicente et al., 2006), there are two definitions to determine pore diameter: ${d_p}^{eq}$ and ${D_p}^{sp}$. ${d_p}^{eq}$ is defined as equivalent sphere diameter of same fluid volume in the cell while ${D_p}^{sp}$ is defined as maximum totally included sphere/ball in the cell. The strut diameter, $d_s$ is defined as maximum totally included sphere/ball in the solid volume of the ligament. Note that, in the analytical approach presented in section 3.4, $d_p$ (pore diameter) is used.

To estimate the strut diameter ($d_s$) and pore diameter (here ${D_p}^{sp}$), a method based on granulometry analysis is proposed. ${D_p}^{sp}$ was chosen because when the measurement tool, iMorph was developed in-house, it was supposed to obtain the parameters that can be easily measured. These types of parameters were generally reported in the literature.

For every voxel in a given phase, the diameter of the maximal ball including this voxel (see Figure 2.9-left) is calculated and called aperture diameter. The mean value of the aperture diameter distribution is used to give a first approximation of the mean phase diameter. The distribution of the maximal included balls gives information on how balls are distributed in the phase (see Figure 2.9-right). Another estimation of the mean phase diameter deduced from the distribution (number) of the full maximal included balls was then computed. In Figure 2.9, the 3-D granulometry analysis of strut on one arbitrary slice was shown.

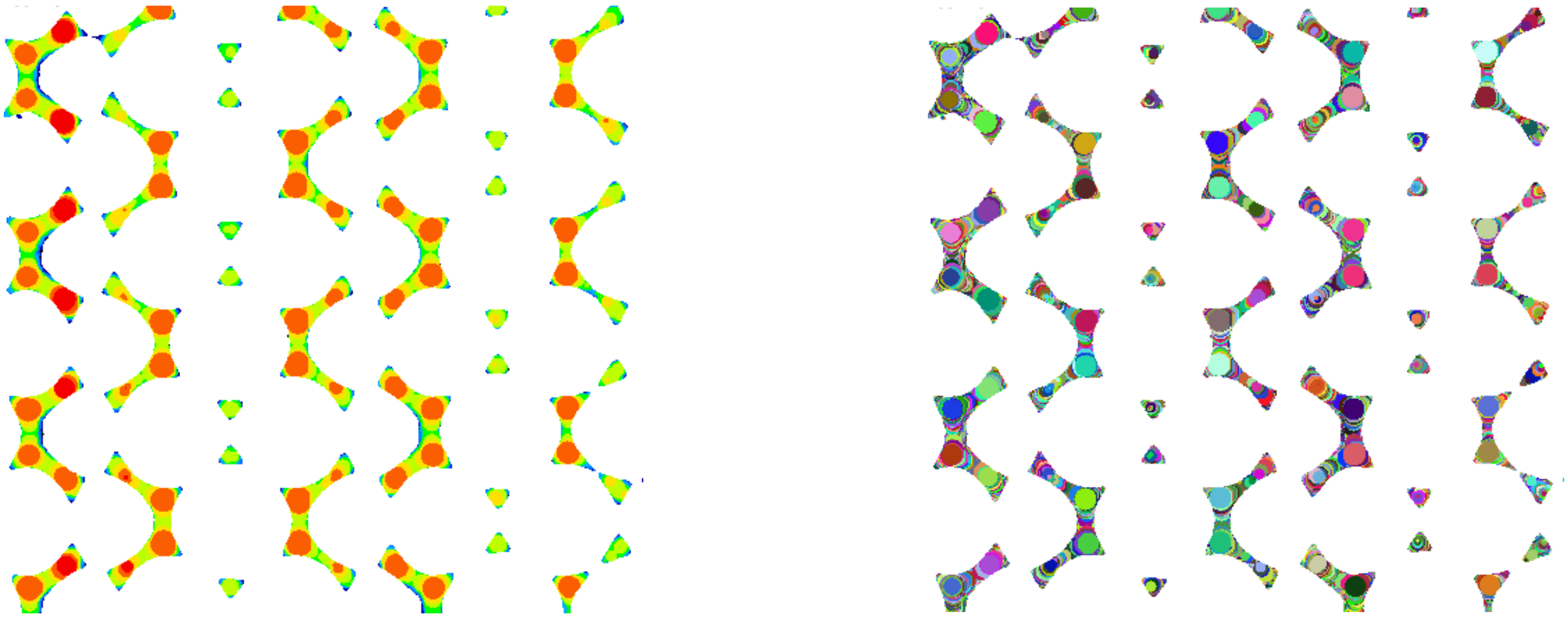

**Figure 2.9**. Left-Strut aperture Map: for every voxels of the phase, the radius of the maximal included balls that enclose the voxel is superimposed. Right- Strut maximal ball map: for every voxels of the phase, the identifier of the maximal ball is superimposed. Full balls are deduced from this map to determine strut diameter. Granulomtery analysis is shown for sample 4 of 85% porosity.

Figure 2.10 shows the different granulometry distributions of both struts and pores. Mean $d_s$ and ${D_p}^{sp}$ with the maximum balls distributions are estimated that present narrow Gaussian distribution. The aperture map distribution of the struts gives 2500μm which is size of maximum totally included ball. For the pore granulometry analysis, the distribution gives pore diameter (${D_p}^{sp}$) of 12700μm. For solid (strut) and pore phases, the standard deviation of maximal balls distributions was always less than one voxel size. Moreover, volume of the fluid phase in voxels to calculate equivalent sphere diameter, ${d_p}^{eq}$ was measured which is an included sphere of the same volume of fluid phase.

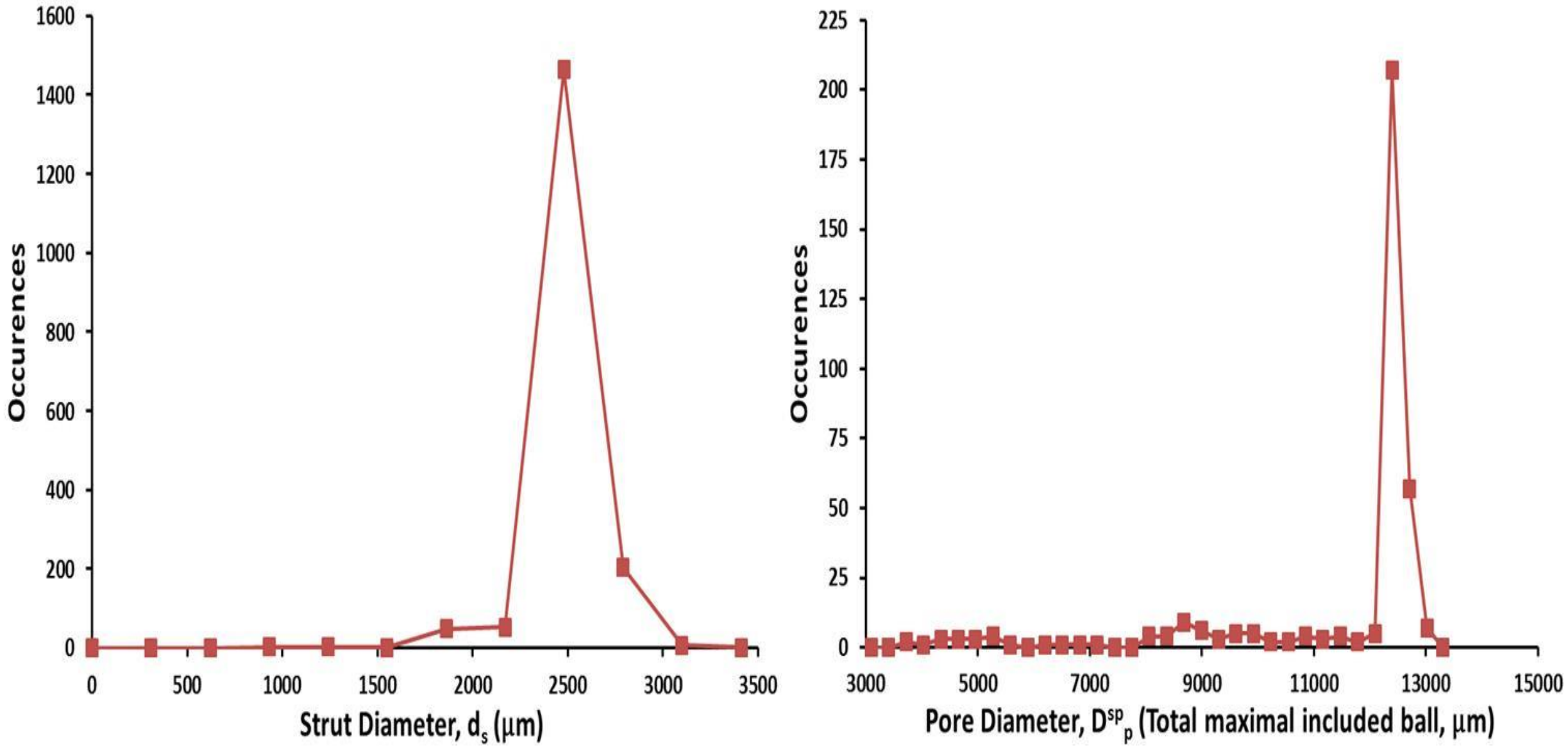

**Figure 2.10**. Mean phase diameter determination from granulometry analysis. Aperture map distribution (volume) and maximal balls distributions (numbers). In two figures, the diameter of both strut and pore are presented. Left: Solid, Right: Pores for sample 4 of 85% porosity.

### 2.4.3 Characterization of node to node length

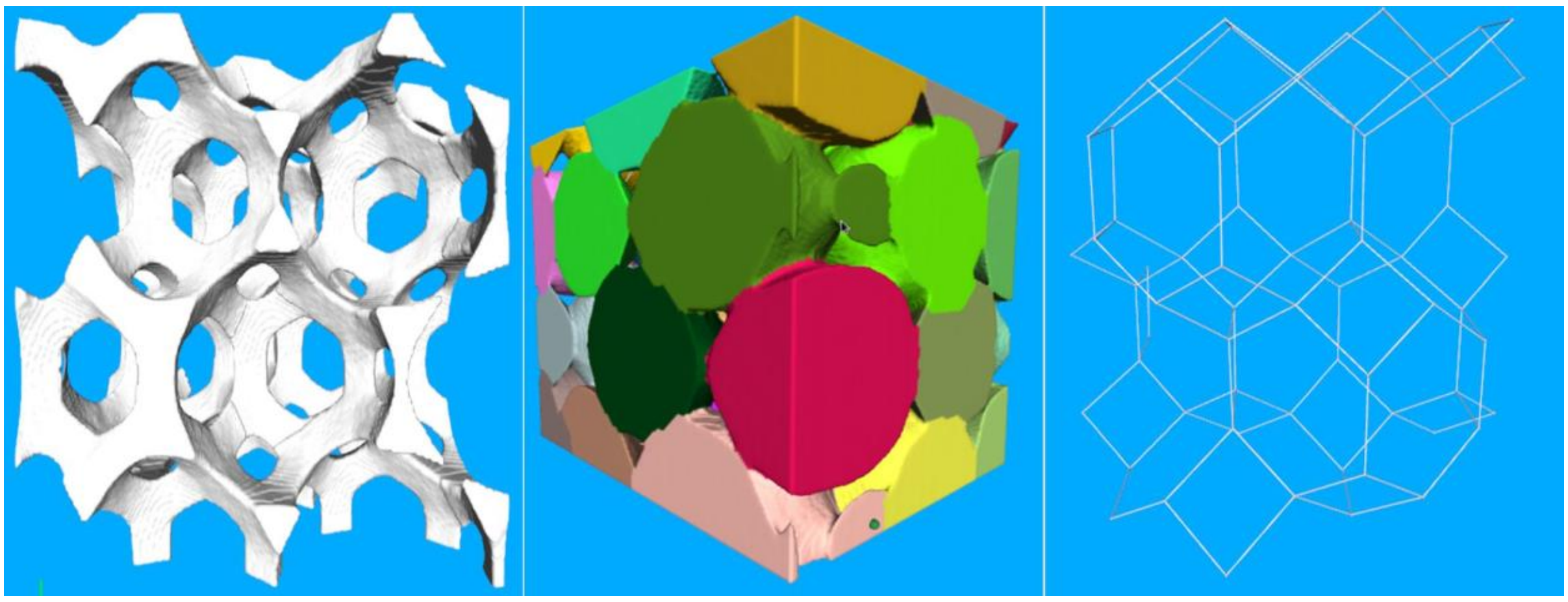

**Figure 2.11.** Ligament reconstruction and identification. Left: 3-D surface mesh, Center: Cells identification, Right: Ligament network obtained from the cells segmentation and constructed by Plateau's law for sample 4 of 85% porosity.

The ligament network is obtained from cells identification and Plateau's law as presented in the Figure 2.11. These cells are segmented using watershed method on the distance map. Watershed method needs markers that can be associated to the cell centre. These markers are obtained from maximal included balls by assuming that one cell present only one totally included maximal ball. Each labelled cell is inflated by an adapted fast-marching method and according to Plateau's law; voxels of the resulting images are divided into three categories: faces, struts and nodes of the solid skeleton.

To estimate the node to node length ($L$), the computation on 1-D strut network is proposed (see Figure 2.11-right). Based on cell segmentation and ligament construction (Figure 2.11), statistics on node to node length ($L$) is then easily calculated and presented in Figure 2.12.

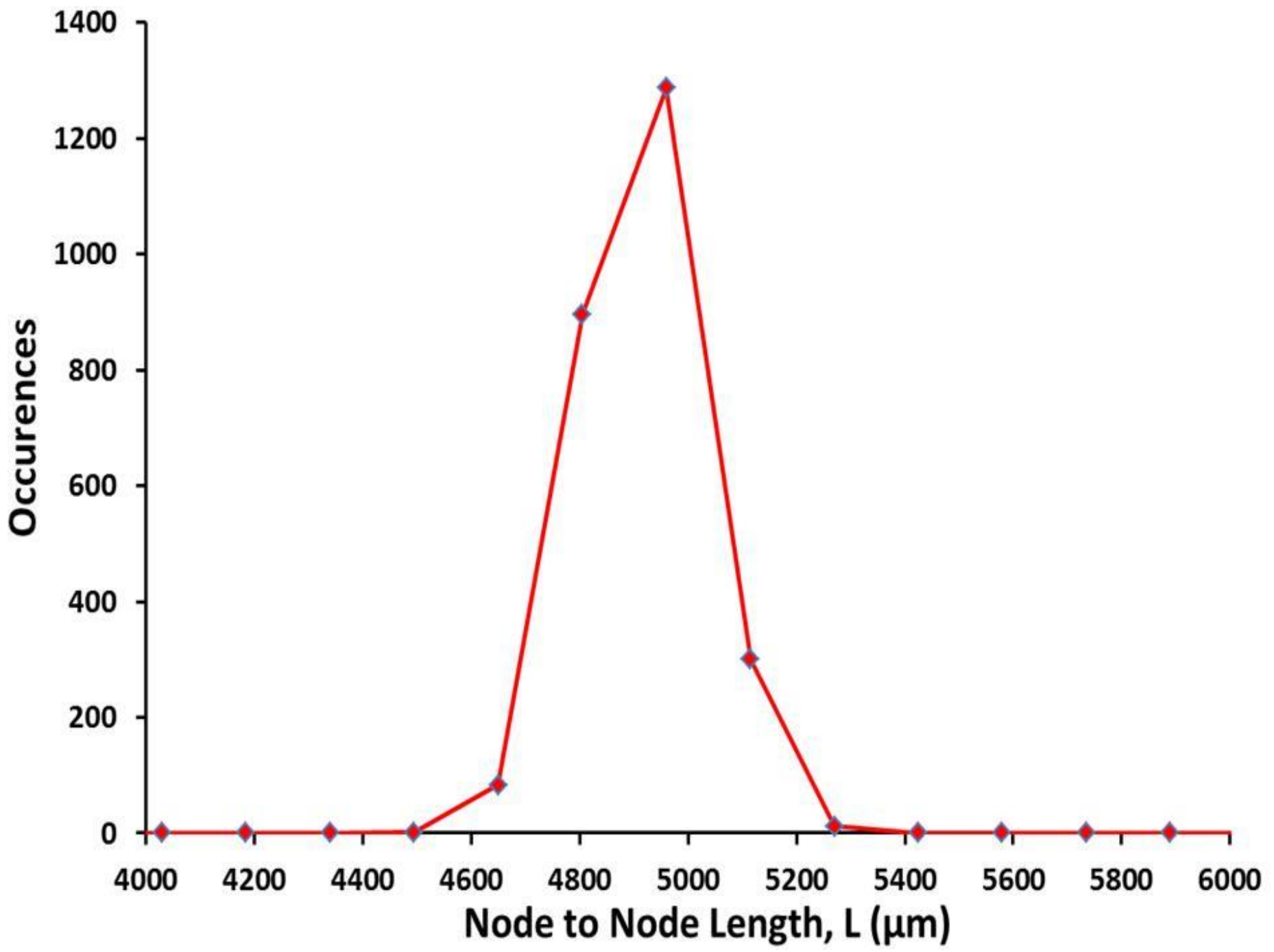

**Figure 2.12**. Node to Node length distribution of sample 4 of 85% porosity.

### 2.4.4 Characterization of Sphericity

Cells segmentation into individual object can give more morphological information. The 3-D inertia matrix is determined for each segmented cell (see Figure 2.13-left). The three Eigen values of 3-D inertia matrix are denoted respectively by $a$, $b$ and $c$ with $c < b < a$ and represent the half-length of equivalent ellipsoid in X, Y and Z directions respectively. These values can then be used as an estimator of cell-anisotropy. Half-length distribution of equivalent ellipsoid is presented in Figure 2.13-right. The mean value differences between $a$, $b$ and $c$ are 100 µm which is less than the voxel resolution (155µm) and thus, we neglect the anisotropy of the cell and consider that the pores are nearly spherical.

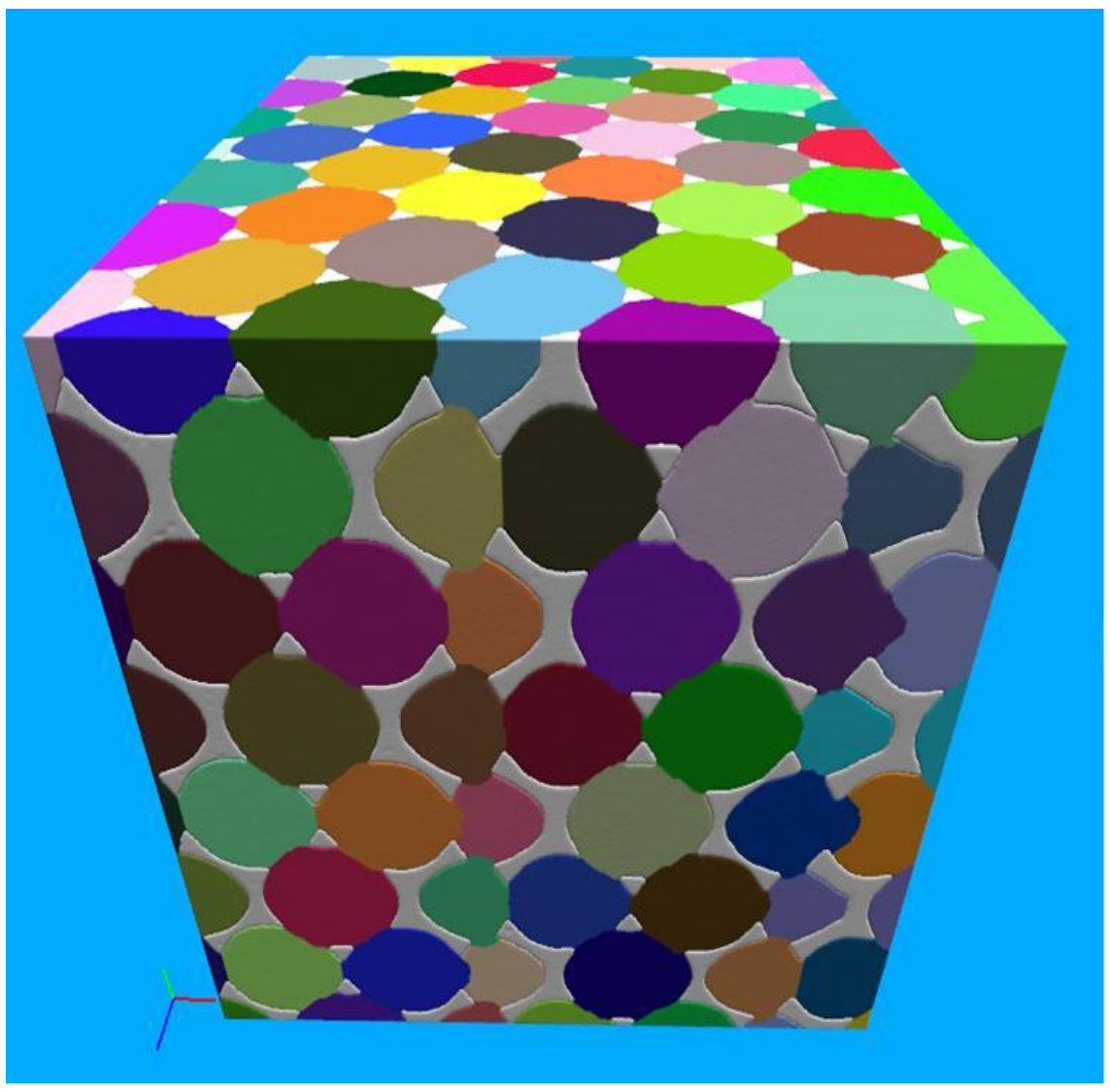

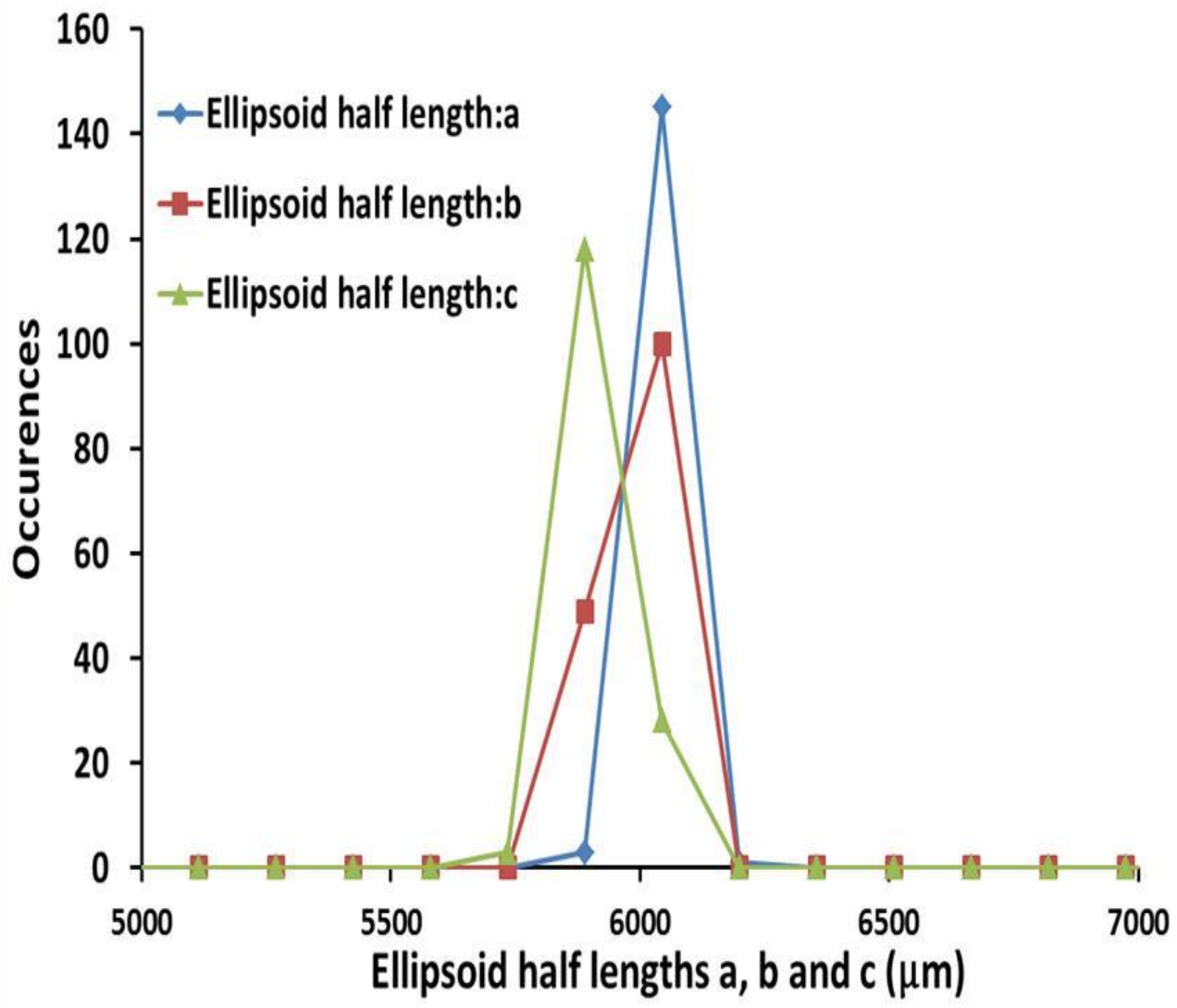

**Figure 2.13**. Left- Automatic cell extraction. Right- Equivalent ellipsoid half-length distribution for sample 4 of 85% porosity.

### 2.4.5 Characterization of tortuosity

The tortuosity ($\partial$) of both solid and fluid phases for all the samples was also determined and presented in Table 2.1. The methodology to determine solid and fluid phases tortuosity is presented in the work of Brun (2009). From Figure 2.14 (left), solid tortuosity of the Kelvin cell (cast foam samples) shows that the struts and their orientations are structured and is roughly similar to the ones obtained by Brun (2009). Skeleton tortuosity is higher than the plane tortuosity because working on skeleton impose only centered path as already fully discussed by Brun (2009).

**Table 2.1.** Presentation of plane tortuosity for solid and fluid phase of cast foam samples.

| Sample | Porosity ($\varepsilon_o$) | Fluid Tortuosity | | | Solid Tortuosity | | |
|---|---|---|---|---|---|---|---|
| | | X-Axis | Y-Axis | Z-Axis | X-Axis | Y-Axis | Z-Axis |
| 1 | 0.825 | 1.0 | 1.0 | 1.0 | 1.095 | 1.092 | 1.030 |
| 2 | 0.84 | 1.0 | 1.0 | 1.0 | 1.115 | 1.070 | 1.062 |
| 3 | 0.845 | 1.0 | 1.0 | 1.0 | 1.053 | 1.094 | 1.053 |
| 4 | 0.85 | 1.0 | 1.0 | 1.0 | 1.095 | 1.092 | 1.030 |

In Figure 2.14 (right), fluid tortuosity between different RECEMAT foams and cast Kelvin cell foam samples is compared. Fluid tortuosity is close to $1 + \delta$, where $\delta$ is of the order of $10^{-3}$. There are almost no significant differences between tested samples. The solid phase tortuosity is also low between $1 < \partial < 1.2$ according to the sample. It is thus, clear that foams are not tortious contrary to what is often reported in the literature (e.g. Bhattacharya et al., 2002; Ahmed et al., 2011). Earlier studies have assumed that foams are tortuous in order to explain the mixing phenomena to justify heat transfer characteristics but tortuosity was not measured. Recently, several authors (Brun, 2009; Haussener et al., 2010; Lawrence et al., 2010) have directly measured the tortuosity of foams using different techniques and found them non-tortuous.

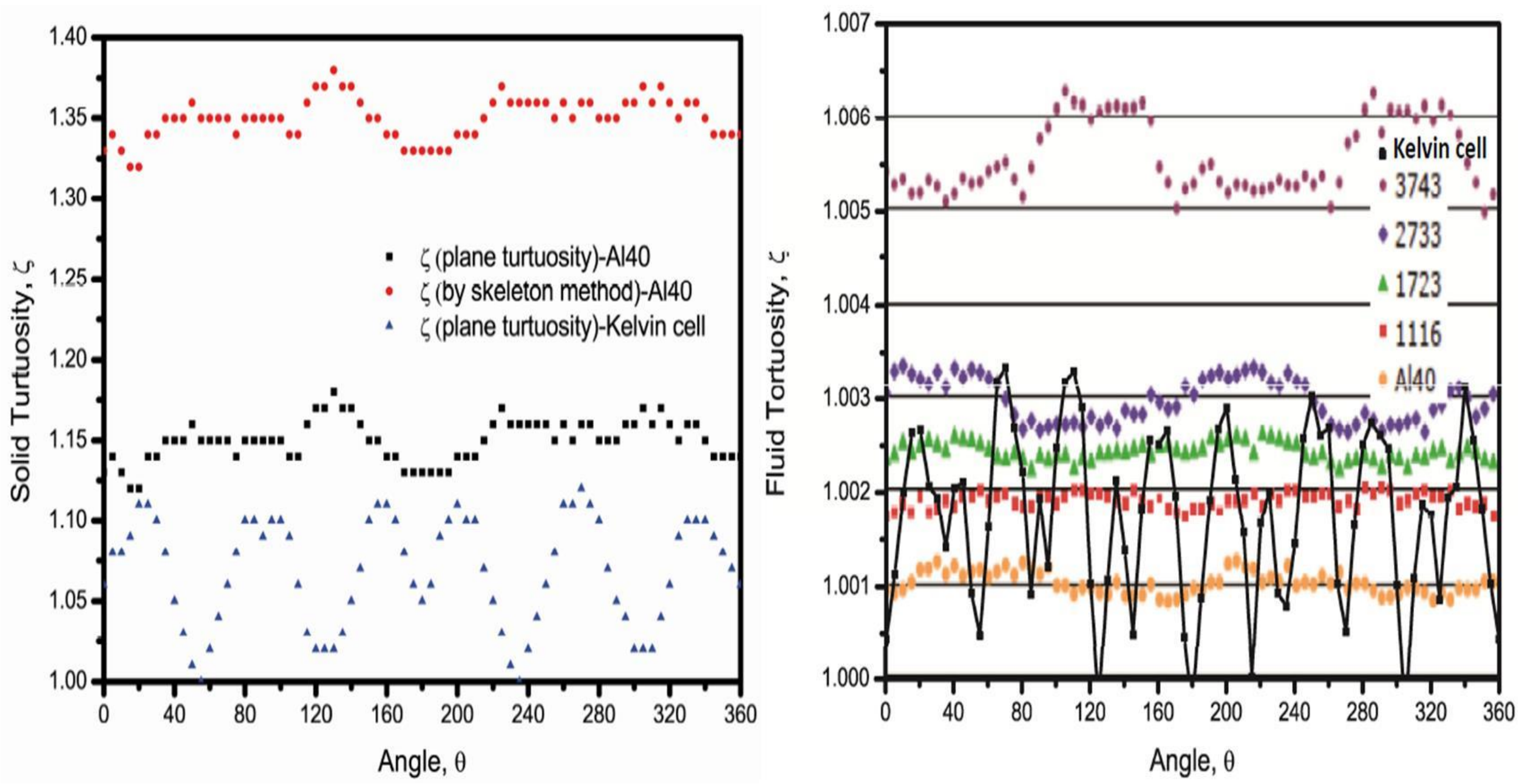

**Figure 2.14.** Left- Comparison of plane and skeleton directional tortuosity of two different foams (ERG and cast Kelvin cell foam sample) of solid phase. Right- Comparison of fluid phase plane tortuosity of different foams (RECEMAT and cast Kelvin cell foam sample). Data for figures (left and right) are taken from Brun thesis (2009).

### 2.4.6 Characterization of specific surface area

The in-house code, iMorph (Vicente et al., 2006) implements the algorithm of "*marching cubes*" to calculate solid fluid interface with sub-voxel precision. The technique created a polygonal model that approaches the iso-density surfaces to the volume for a given threshold; the latter corresponds to the value used to binarize these images. The specific surface areas of the foam samples are presented in Table 2.2.

### 2.4.7 Comparison and validation of geometrical characteristics of foam matrix measured on virtual and cast foam samples

CAD data and the data obtained using iMorph (Vicente et al., 2006) on cast foams are presented in Table 2.2. It is clearly evident that the parameters chosen to construct the foam structure in CAD resemble well with the measured properties of cast foam samples. Moreover, it was shown in the section 2.4.4 that the foams studied in the present work are spherical and thus, do not possess any anisotropy. It is insightful that virtual regular and periodic foam sample constructed in CAD could be materialized by casting method developed by CTIF even for the medium porosity range (no change in the strut cross section with porosity).

**Table 2.2**. Presentation of geometrical parameters of virtual CAD and cast foam samples.

| | | CAD Data | | | Data obtained by iMorph | | | | |
|---|---|---|---|---|---|---|---|---|---|
| Sample | $\varepsilon_o$ | $L$ (mm) | $d_p$ (mm) | $a_c$ ($m^{-1}$) | $d_s$ (mm) | $L$ (mm) | ${D_p}^{sp}$ (mm) | $d_p$ (mm) | $a_c$ ($m^{-1}$) |
| 1 | 0.825 | 3.6 | 3.72 | 373.6 | 1.9 | 3.6 | 8.7 | 3.84 | 370.9 |
| 2 | 0.84 | 3.6 | 3.88 | 360.1 | 1.9 | 3.6 | 8.7 | 3.97 | 357.4 |
| 3 | 0.845 | 5.0 | 5.51 | 255.8 | 2.5 | 5.0 | 12.4 | 5.48 | 263.9 |
| 4 | 0.85 | 5.0 | 5.53 | 251.1 | 2.5 | 5.0 | 12.7 | 5.67 | 252.2 |

### 2.5 Development of virtual *isotropic* and *anisotropic* foams

In the sections 2.3 and 2.4, it was shown that it is possible to control the morphological parameters of foam matrix using new techniques (e.g. SEBM, 3-D rapid prototyping, CTIF's casting method) and can tailor them to the desired output for numerous applications. Most of the commercially available foams possess strut shapes: convex or concave triangular (in case of metal foams) and circular (in case of ceramic foams). Using CAD, it is possible to construct foam matrix of various strut shapes for low and high porosity range ($0.60 < \varepsilon_o < 0.95$). This allows us to study the impact of strut shapes on geometrical characteristics and thermo-hydraulic properties. This way, it could also allow us to optimize strut shape and size for various engineering applications.

#### 2.5.1 Development of virtual *isotropic* foams

Strut shapes like circular, equilateral triangle, diamond (double equilateral triangle), square, hexagon and star (regular hexagram) were generated. Their cross sections are in line with the ligament axis of the truncated octahedron edge. Moreover, the rotation of square and hexagon shapes at 45 and 90 degrees respectively are also studied with respect to ligament axis where node forms a quite different shape than original ones which has subsequent effect on geometrical properties. Moreover, same strut shape with and without rotation also influences greatly the thermo-hydraulic properties. Note that, constant cross section of the ligament along its axis has been studied in this thesis.

In CAD, the modelling (or construction) of a foam structure has first started with a tetrahedral element composed of four identical half struts. This choice of a structural element is consistent with the topological feature of foam. Further, 3-D foam was modelled with this smallest repetitive element, which defines a spatially periodic structure. Micro-structural features of open cell foam are represented by a tetrahedral unit cell with a skeleton of four half nodes of length $L/2$, where dihedral angle is approximately 109.471° at edges shared by two hexagons or 125.263° at edges shared by a hexagon and a square faces (see also Figure 3.8 in the section 3.4.1.1). Figure 2.15 shows the construction of Kelvin cell which is based on a truncated octahedron. The node to node length ($L = \sqrt{2}$mm) is kept fixed for entire calculations which is based on a given cell diameter ($d_{cell}$).

For each strut shape, the same procedure of creating four half struts followed by replication to create the foam sample in the unit cell was followed. Figure 2.16 shows the different strut cross sections and their characteristics dimensions for a given porosity.

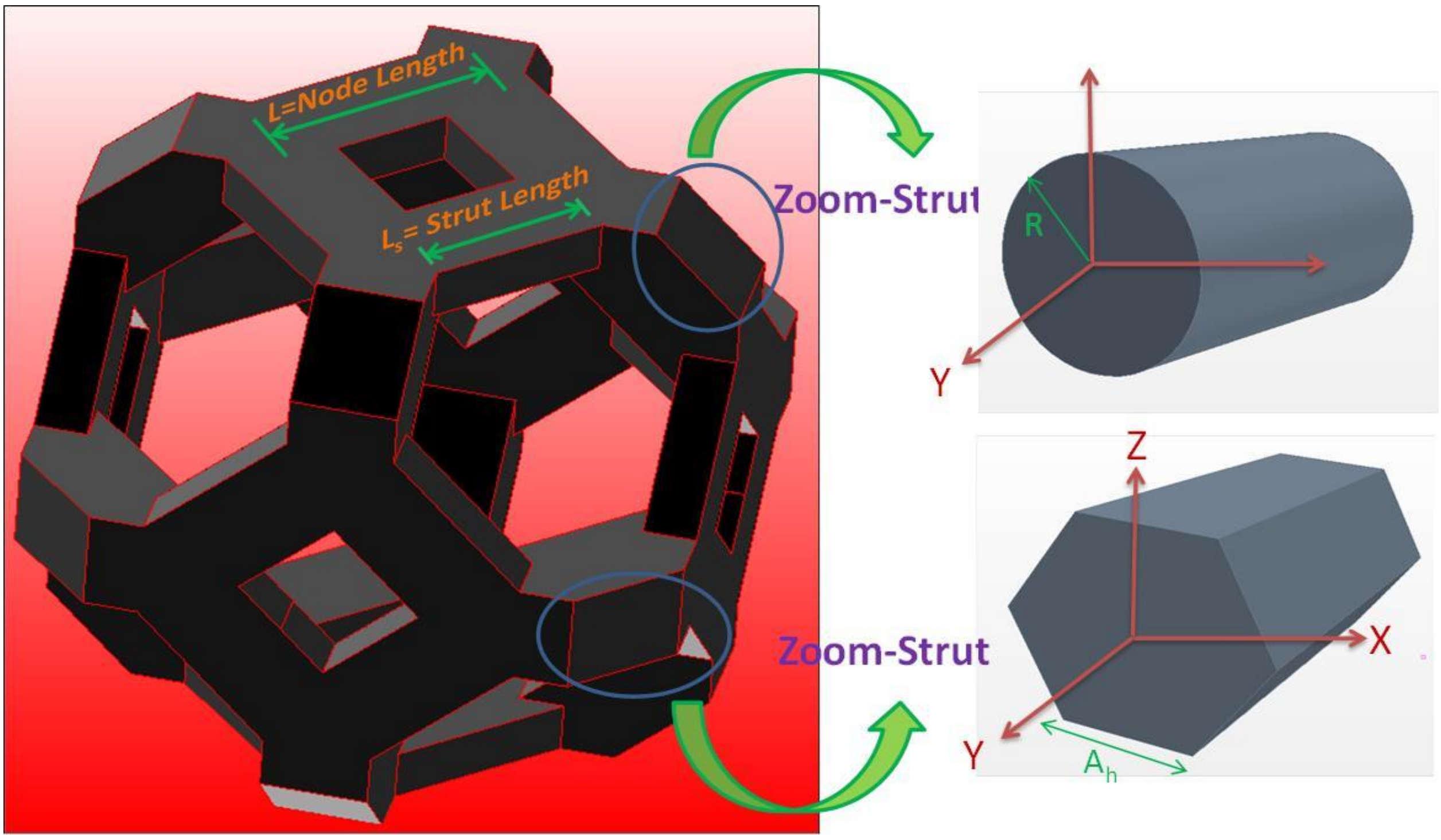

**Figure 2.15.** Presentation of tetrakaidecahedon model of Kelvin cell (left). Strut length, $L_s$ and Node Length, $L$ are clearly shown. Constant ligament cross sections with different strut shapes circular (on the top-right) and hexagon (at the bottom-right) are presented with their characteristic dimensions.

The impact of any homothetic transform on foam properties has already been studied (e.g. Bonnet al., 2008; Hugo, 2012). On the other hand, the impact of foam samples constructed or fabricated based on constant cell diameter ($d_{cell}$) on geometrical and thermo-physical properties is still unknown and is not yet reported in the literature.

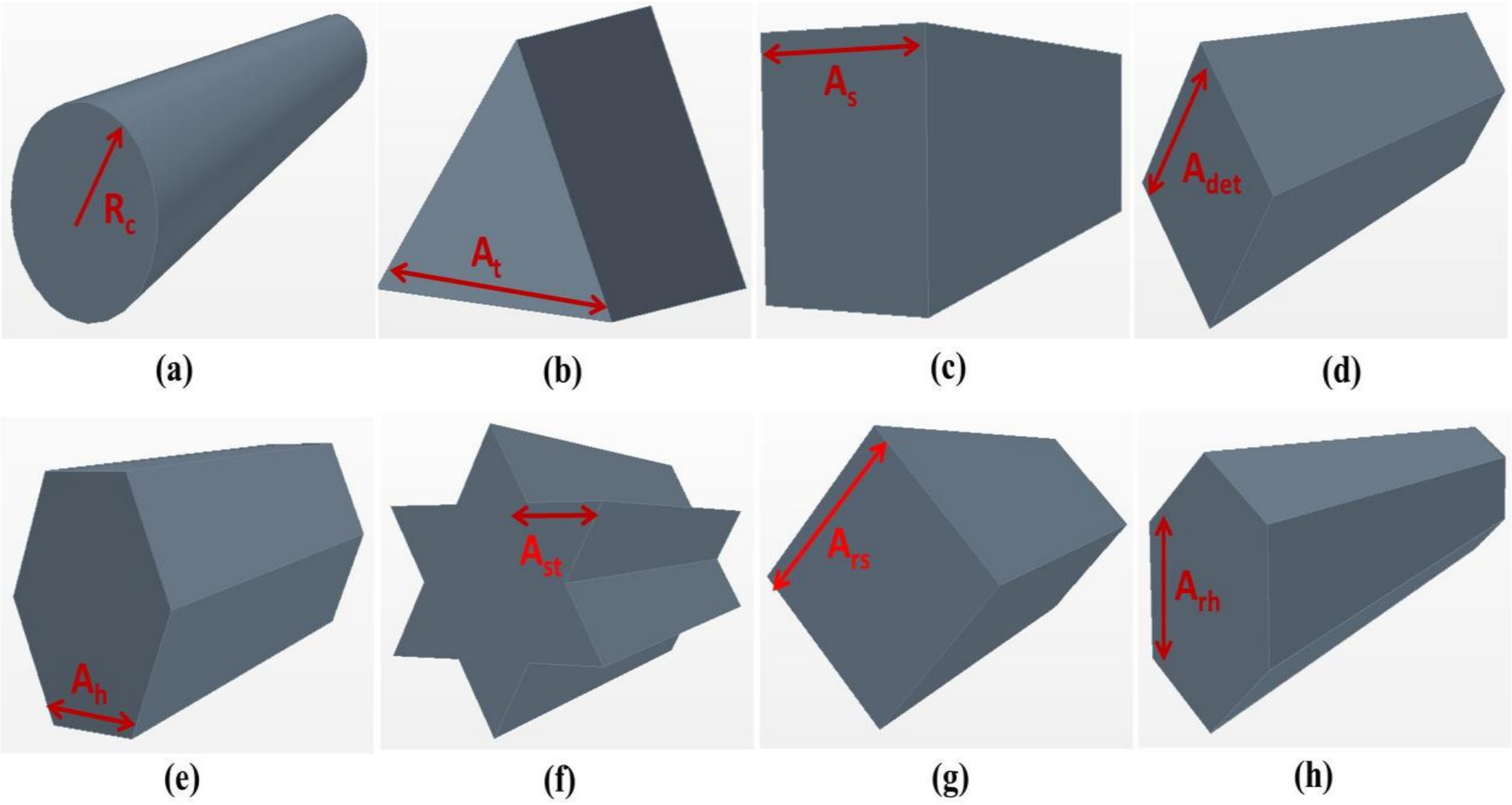

**Figure 2.16**. Representation of different 3-D strut and ligament shapes (a) Circular (b) Equilateral Triangle (c) Square (d) Diamond (Double Equilateral Triangle) (e) Hexagon (f) Star (g) Rotated Square (h) Rotated Hexagon. The characteristic dimensions of struts are also presented that are used in the section 3.4.2 and Appendix C and D for analytical solutions.

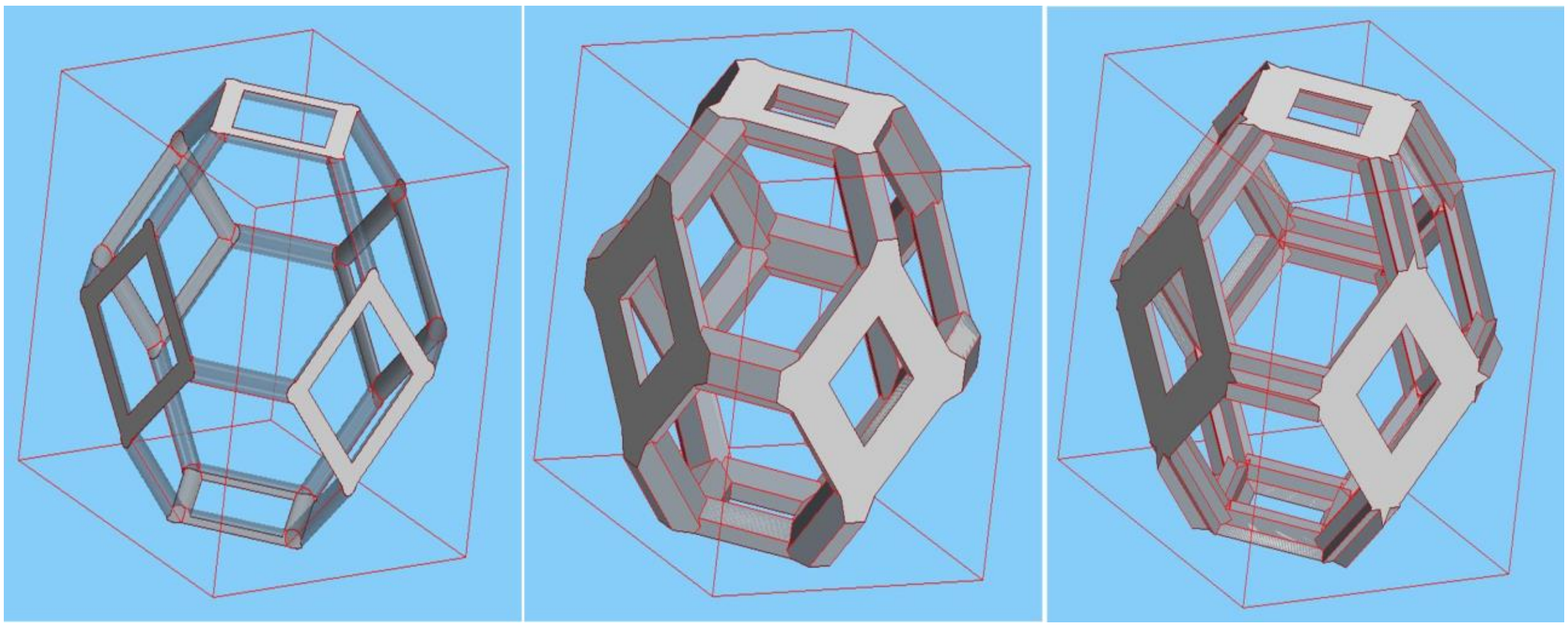

**Figure 2.17**. Presentation of 3-D circular, hexagon and star strut shape virtual Kelvin-like foams inside a periodic unit cell. Two half struts of square face, one one-fourth strut of hexagon face of all foams are joined at the node that are totally included in the cubic cell.

A construction method was defined that allows us to use strut shape and porosity as a control parameter. Thus, we use an arbitrary cell size to construct our foams. Using this construction method, foam samples of chosen porosity for any strut shape can be easily generated.

**Table 2.3**. Values of various strut shapes, porosities and their characteristic dimensions.

| | CAD Measurement | | | | | Analytical | | | | |
|---|---|---|---|---|---|---|---|---|---|---|
| Shape | $\varepsilon_o$ (%) | $d_s$ *or* $A$ (mm) | $a_{sw}$ (mm²) | $a_{hw}$ (mm²) | $a_c$ (m⁻¹) | $\alpha_{eq} = \frac{R_{eq}}{L}$ | $\beta = \frac{L_s}{L}$ | $a_{sw}$ (mm²) | $a_{hw}$ (mm²) | $a_c$ (m⁻¹) |
| Circular | 60 | 1.212 | 0.198 | 0.514 | 982 | 0.429 | 0.341 | 0.232 | 0.603 | 1037 |
| | 65 | 1.110 | 0.277 | 0.720 | 979 | 0.392 | 0.393 | 0.309 | 0.803 | 1020 |
| | 70 | 1.006 | 0.371 | 0.965 | 960 | 0.356 | 0.447 | 0.400 | 1.039 | 992 |
| | 75 | 0.900 | 0.482 | 1.252 | 926 | 0.318 | 0.504 | 0.507 | 1.318 | 949 |
| | 80 | 0.789 | 0.614 | 1.595 | 873 | 0.279 | 0.563 | 0.635 | 1.649 | 889 |
| | 85 | 0.669 | 0.773 | 2.009 | 796 | 0.236 | 0.628 | 0.790 | 2.052 | 807 |
| | 90 | 0.534 | 0.975 | 2.533 | 686 | 0.189 | 0.702 | 0.987 | 2.564 | 692 |
| | 95 | 0.367 | 1.255 | 3.261 | 515 | 0.130 | 0.794 | 1.262 | 3.279 | 518 |
| Equilateral Triangle | 80 | 4.485 | 5.457 | 14.178 | 273 | 0.735 | 0.563 | 7.935 | 20.615 | 279 |
| | 85 | 3.704 | 7.840 | 20.370 | 252 | 0.625 | 0.628 | 9.875 | 25.655 | 256 |
| | 90 | 2.882 | 10.811 | 28.088 | 221 | 0.501 | 0.702 | 12.336 | 32.050 | 223 |
| | 95 | 2.451 | 12.560 | 32.633 | 167 | 0.346 | 0.794 | 15.776 | 40.988 | 171 |
| Square | 60 | 1.075 | 0.197 | 0.511 | 1092 | 0.760 | 0.341 | 0.232 | 0.603 | 1122 |
| | 65 | 0.984 | 0.277 | 0.720 | 1091 | 0.696 | 0.393 | 0.309 | 0.803 | 1111 |
| | 70 | 0.891 | 0.372 | 0.965 | 1074 | 0.630 | 0.447 | 0.400 | 1.039 | 1085 |
| | 75 | 0.797 | 0.483 | 1.255 | 1037 | 0.563 | 0.504 | 0.507 | 1.318 | 1043 |
| | 80 | 0.698 | 0.615 | 1.598 | 979 | 0.493 | 0.563 | 0.635 | 1.649 | 982 |
| | 85 | 0.591 | 0.775 | 2.014 | 895 | 0.418 | 0.628 | 0.790 | 2.052 | 895 |
| | 90 | 0.472 | 0.977 | 2.538 | 772 | 0.334 | 0.702 | 0.987 | 2.564 | 771 |
| | 95 | 0.325 | 1.257 | 3.266 | 580 | 0.229 | 0.794 | 1.262 | 3.279 | 579 |
| Rotated Square | 80 | 0.694 | 0.632 | 1.649 | 996 | 0.491 | 0.563 | 0.635 | 1.649 | 1050 |
| | 85 | 0.589 | 0.788 | 2.052 | 906 | 0.416 | 0.628 | 0.790 | 2.052 | 944 |
| | 90 | 0.470 | 0.982 | 2.564 | 779 | 0.332 | 0.702 | 0.987 | 2.564 | 803 |
| | 95 | 0.324 | 1.261 | 3.279 | 583 | 0.229 | 0.794 | 1.262 | 3.279 | 594 |
| Diamond | 80 | 0.752 | 0.600 | 1.560 | 1070 | 0.532 | 0.563 | 0.635 | 1.649 | 1043 |
| | 85 | 0.637 | 0.762 | 1.979 | 974 | 0.450 | 0.628 | 0.790 | 2.052 | 953 |
| | 90 | 0.508 | 0.965 | 2.508 | 838 | 0.359 | 0.702 | 0.987 | 2.564 | 823 |
| | 95 | 0.349 | 1.249 | 3.244 | 627 | 0.247 | 0.794 | 1.262 | 3.279 | 620 |
| Hexagon | 60 | 0.665 | 0.199 | 0.517 | 1025 | 0.470 | 0.341 | 0.232 | 0.603 | 1070 |
| | 65 | 0.609 | 0.279 | 0.725 | 1023 | 0.431 | 0.393 | 0.309 | 0.803 | 1056 |
| | 70 | 0.552 | 0.373 | 0.970 | 1005 | 0.390 | 0.447 | 0.400 | 1.039 | 1028 |
| | 75 | 0.494 | 0.484 | 1.258 | 970 | 0.349 | 0.504 | 0.507 | 1.318 | 986 |
| | 80 | 0.432 | 0.616 | 1.602 | 915 | 0.306 | 0.563 | 0.635 | 1.649 | 925 |
| | 85 | 0.367 | 0.776 | 2.016 | 835 | 0.259 | 0.628 | 0.790 | 2.052 | 841 |
| | 90 | 0.292 | 0.977 | 2.539 | 720 | 0.207 | 0.702 | 0.987 | 2.564 | 723 |
| | 95 | 0.201 | 1.257 | 3.267 | 540 | 0.142 | 0.794 | 1.262 | 3.279 | 542 |
| Rotated Hexagon | 60 | 0.664 | 0.202 | 0.525 | 1033 | 0.469 | 0.341 | 0.232 | 0.603 | 1070 |
| | 65 | 0.608 | 0.282 | 0.733 | 1029 | 0.430 | 0.393 | 0.309 | 0.803 | 1056 |
| | 70 | 0.551 | 0.377 | 0.978 | 1009 | 0.390 | 0.447 | 0.400 | 1.039 | 1028 |
| | 75 | 0.493 | 0.488 | 1.267 | 973 | 0.348 | 0.504 | 0.507 | 1.318 | 986 |
| | 80 | 0.432 | 0.620 | 1.611 | 917 | 0.305 | 0.563 | 0.635 | 1.649 | 925 |
| | 85 | 0.366 | 0.779 | 2.025 | 837 | 0.259 | 0.628 | 0.790 | 2.052 | 841 |
| | 90 | 0.292 | 0.980 | 2.547 | 721 | 0.207 | 0.702 | 0.987 | 2.564 | 723 |
| | 95 | 0.201 | 1.260 | 3.273 | 541 | 0.142 | 0.794 | 1.262 | 3.279 | 542 |
| Star | 75 | 0.347 | 0.489 | 1.270 | 1399 | 0.246 | 0.504 | 0.507 | 1.318 | 1306 |
| | 80 | 0.305 | 0.620 | 1.611 | 1314 | 0.215 | 0.563 | 0.635 | 1.649 | 1240 |
| | 85 | 0.258 | 0.779 | 2.023 | 1195 | 0.183 | 0.628 | 0.790 | 2.052 | 1140 |
| | 90 | 0.206 | 0.979 | 2.545 | 1027 | 0.146 | 0.702 | 0.987 | 2.564 | 991 |
| | 95 | 0.142 | 1.258 | 3.269 | 769 | 0.100 | 0.794 | 1.262 | 3.279 | 751 |
| Average Deviation | | | | | | | | 6.5% | 6.5% | 0.88% |

Due to the chosen construction method, some limitations rise mainly for the complex shapes (e.g. diamond, star). This procedure creates only (in a periodic unit cell) 36 struts that are along the edge of the truncated octahedron. For certain shapes and values of porosity, mainly for low porosities, some other strut part has to be added in the unit cell which limits the construction procedure.

The geometrical parameters of 49 virtual Kelvin-like foam samples using classical CAD approach are measured (see Table 2.3). The porosities were generated for circular, square, hexagon and rotated hexagon down to 60%, for equilateral triangle, diamond and rotated square strut shapes down to 80% and down to 75% for star strut shapes. The window areas, $a_{sw}$ and $a_{hw}$ ($mm^2$) of square and hexagon openings are also presented. The equivalent window area of square and hexagon openings based on strut length are also presented (see Table 2.3) and detailed in chapter 3. Note that the analytical results of $\alpha_{eq}$, $\beta$, $a_{sw}$, $a_{hw}$ and $a_c$ in Table 2.3 are explained in the section 3.4.2.

Figure 2.17 represents Kelvin like cell foams of different strut shapes inside a cubic unit cell where the struts accumulate at the edges of the truncated octahedron and only the struts those are totally included in the truncated octahedron and cubic unit cell are kept. The length of the cubic unit cell is $2\sqrt{2}L$ (see also section 3.4.2).

### 2.5.2 Development of virtual *anisotropic* foams

Anisotropy is performed on the virtual *isotropic* foam samples that were developed in section 2.5.1. By conserving the porosity of original *isotropic* foam sample, *anisotropic* open cell foam is constructed by elongating it in X direction by a factor $\Omega$ and a factor of $1/\sqrt{\Omega}$ is applied in Y and Z directions respectively as shown in Figure 2.18. In the present work, elongation factor $\Omega$ is varied from 0.8 to 3.0. For *isotropic* open cell foams, $\Omega = 1$. A database of 555 virtual *anisotropic* foam samples was generated and specific surface areas were measured directly by classical CAD approach. Specific surface areas for all strut shapes at various elongation factors ($\Omega$=0.8-3.0) in the wide range of porosity ($0.60< \varepsilon_o <0.95$) are presented in Appendix A (see Table A.1).

The methodology proposed above to transform *isotropic* foam sample into *anisotropic* is very recently realized and materialized by CTIF for equilateral triangular strut shape. For other strut cross sections, the work by CTIF to materialize the proposed methodology is currently in progress. An image of *anisotropic* foam sample of equilateral triangular strut cross section is presented in Figure 2.19.

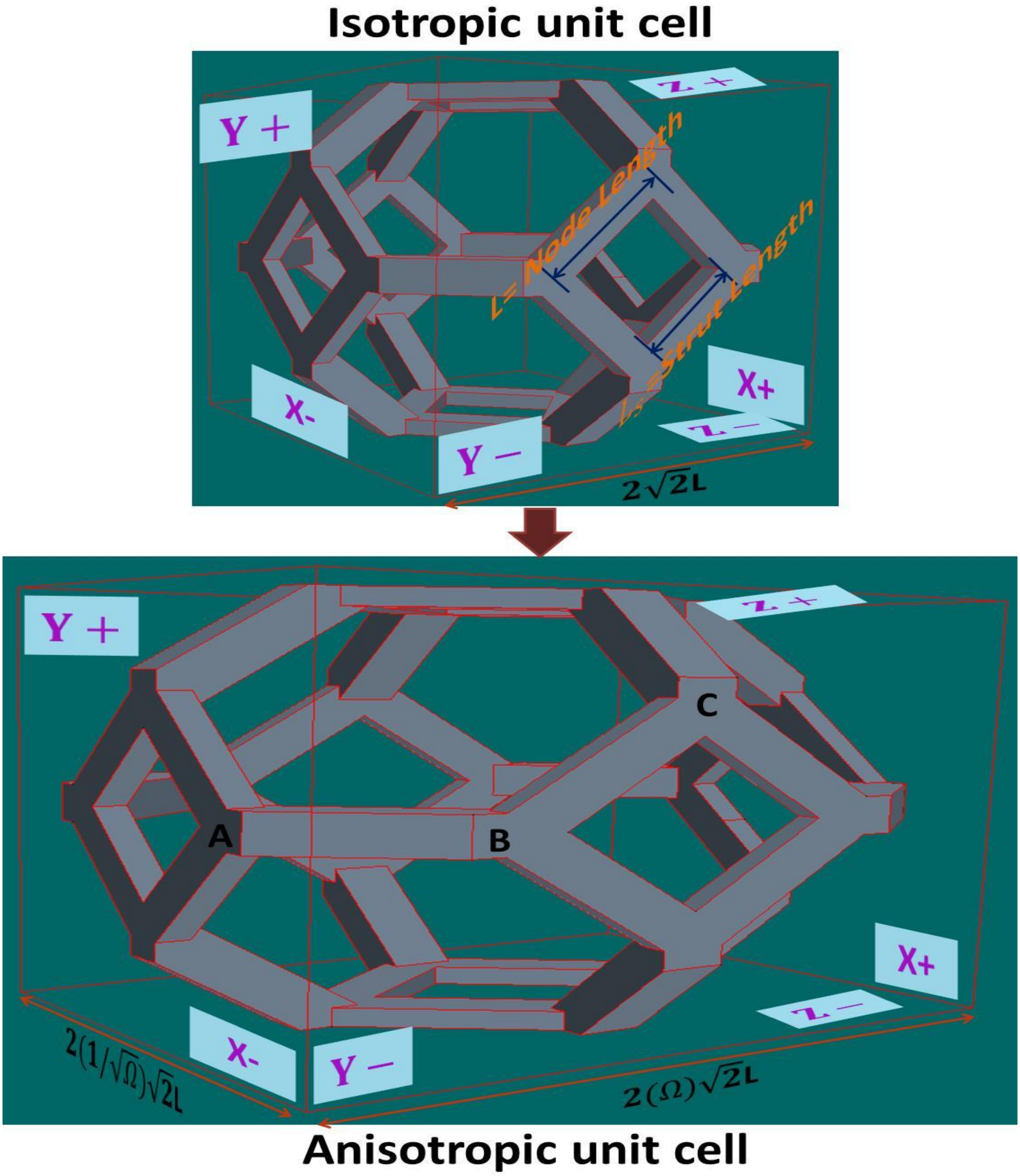

**Figure 2.18**. Top- Presentation of tetrakaidecahedon model of square strut shape Kelvin like *isotropic* open cell foam (left). Strut length ($L_s$), Node Length ($L$) are clearly shown. The sample is placed inside a cube of cubic length $2\sqrt{2}L$. Bottom- Presentation of *anisotropic* open cell foam by applying simultaneously elongation and compression to *isotropic* unit cell in order to conserve the porosity. Different sections in X, Y and Z directions are also marked for effective thermal conductivity tensors calculations (see chapter 5).

## 2.6 Summary and conclusion

Open cell foams (metal or ceramic) offer interesting properties such as large specific surface area, high mechanical strength, high porosity and low pressure drop compared to conventional packed bed spheres. Due to these properties, they are excellent candidates for many industrial applications such as gas filters, heat exchangers, volumetric solar receivers, porous burners and catalyst supports etc... A thorough description of foam geometry requires a comprehensive characterization of their structural/morphological parameters due to their complex 3-D internal architecture.

Due to the unclear definition of a pore, the conventional way of representing cell size or pore size by PPI (pores per linear inch) does not give reliable information because a pore can be a cell or an opening into the cell. Hence, PPI should be considered as merely a nominal value and is not recommended to be used in any modeling work or analytical calculations. Hence, when giving a pore size of the foam sample, it is recommended to specify it as either cell or pore size.

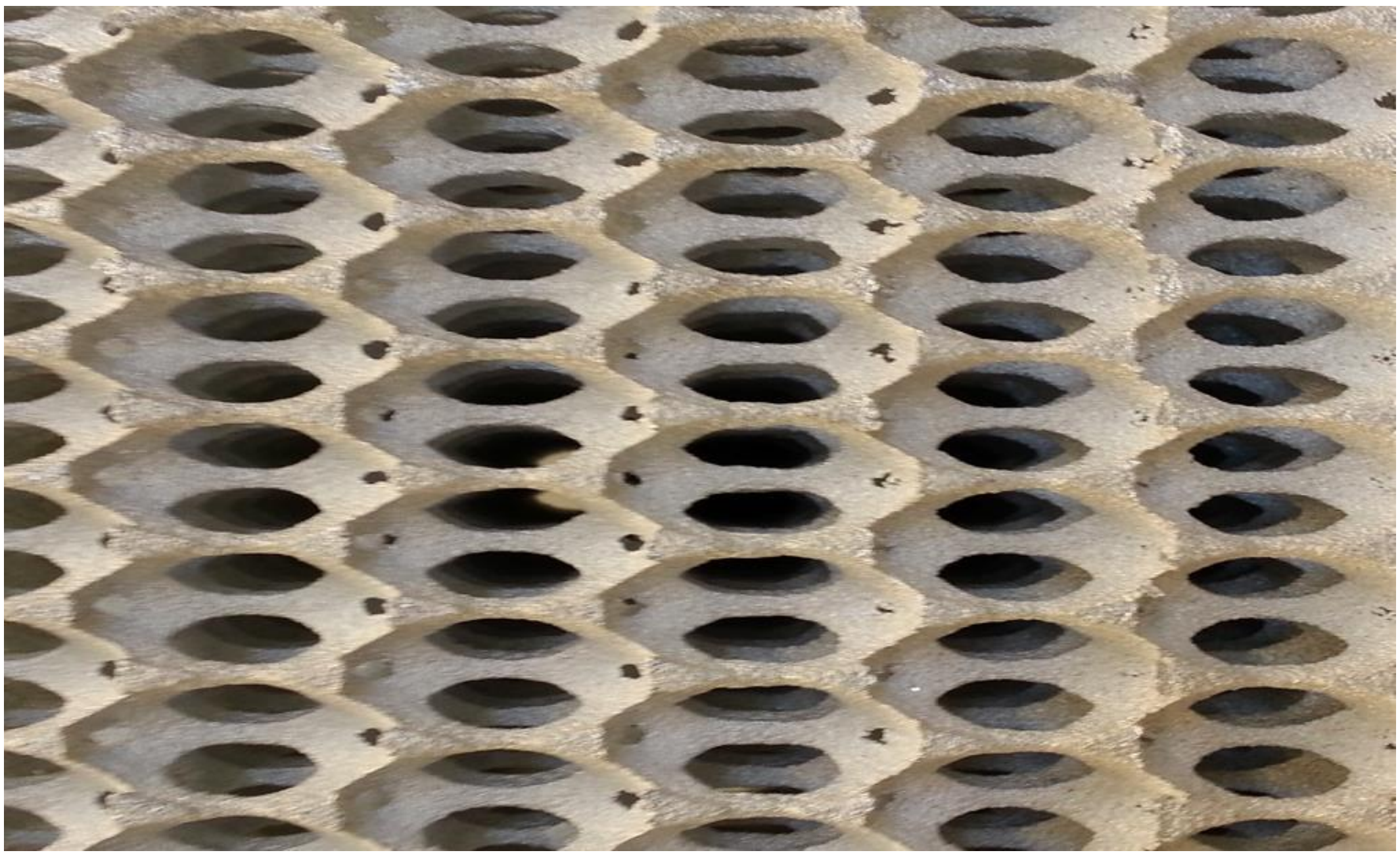

**Figure 2.19**. Image of *anisotropic* foam sample of equilateral triangular strut cross section. The cell is elongated in one direction and contracted in other two directions.

In this chapter, morphological characterization of reticulated open cell foams (cast foams) that represents periodic structure of different cell geometries, cell sizes and porosities has been presented. The studied samples have a defined cell/pore size, geometry and orientation. The foam samples were characterized with respect to their morphological parameters by using different methods including image analysis and X-ray μCT.

It has been widely quoted in the literature that foam structures exhibit different strut morphologies, namely cylindrical, equilateral triangular, convex and concave triangular in the porosity range, $0.87< \varepsilon_o <0.97$. This change in the strut shape formation was explained with porosity variation (low and high porosity) due to which the difference between the calculated and measured values of geometrical parameters is adjusted by empirical correction factors proposed by the authors. The cast foam samples in the studied work are of convex triangular strut shape, even at medium porosity range, $0.825< \varepsilon_o <0.85$. No change in the strut shape is observed and is presented in Figure 2.8 and validated in Table 2.2.

Based on equivalent characteristic dimensions between virtual and cast foam samples without any change in the strut cross section for variable porosities, virtual *isotropic* foams of different strut cross sections have been developed in CAD. It is found that different strut shapes influence strongly the geometrical properties of foams to a great extent including their geometric specific surface area at low and high porosities. Further, a methodology was proposed to transform *isotropic* foams into *anisotropic* ones by keeping the same porosity. A database of various characteristic dimensions for both *isotropic* and *anisotropic* foams is created.

# Chapter 3

# Geometrical characterization of open cell foams

Parts of this work were already published or submitted to Acta Materialia, International Journal of Thermal Sciences, Journal of Porous Media, Chemical Engineering Science and Defects and Diffusion Forum.

- **P. Kumar** & F. Topin, The geometric and thermo-hydraulic characterization of ceramic foams: An analytical approach, *Acta Materilia*, 75, pp. 273-286, 2014.

- **P. Kumar** & F. Topin, Investigation of fluid flow properties in open cell foams: Darcy and weak inertia regimes, *Chemical Engineering Science,* 116, pp. 793-805.

- **P. Kumar**, F. Topin & J. Vicente, Determination of effective thermal conductivity from geometrical properties: Application to open cell foams, *International Journal of Thermal Sciences*, 81, pp. 13-28, 2014.

- **P. Kumar** & F. Topin, About thermo-hydraulic properties of open cell foams: Pore scale numerical analysis of strut shapes, *Defects and Diffusion Forum*, 354, pp. 195-200, 2014.

- **P. Kumar**, F. Topin & L. Tadrist, Geometrical characterization of Kelvin like metal foams for different strut shapes and porosity, *Journal of Porous Media* (under review).

- **P. Kumar** & F. Topin, Impact of anisotropy on geometrical and thermal conductivity of metallic foam structures, Special issue of Heat and Mass Transfer in Porous Media, *Journal of Porous Media* (under review).

Parts of this work were already communicated in national/international conferences.

- **P. Kumar** & F. Topin, Influence of strut shape and porosity on geometrical properties and effective thermal conductivity of Kelvin-like anisotropic metal foams, *The 15$^{th}$ Heat Transfer Conference (IHTC)-2014*, Kyoto, *Japan.*

- **P. Kumar** & F. Topin, Evaluation of thermal properties of metal and ceramic foams by geometrical characterization, *27$^{th}$ European Symposium on Applied Thermodynamics (ESAT)-2014*, Eindhoven, *The Netherlands.*

## 3.1 Background

Due to the wide porosity range (0.60<$\varepsilon_o$<0.97) of open cell foams, they have low pressure drop (Dietrich et al., 2009; Inayat et al., 2011b) compared to conventional packed bed spheres. Compared to packed beds of spherical particles, open cell foams have a significantly higher thermal conductivity due to their continuous connection, which is interesting for heat exchanger applications and chemical reactors that perform highly exothermic reactions (Adler, 2003a, b). Potential technical applications of open cell foams are porous burners, solar receivers, carrier for catalysts, lightweight constructions or heat insulation (Sheppard, 1993; Reitzmann et al., 2006).

In contrast to such potential engineering applications, the morphological properties of foam structures are still difficult to determine due to their complex geometry (Haughey and Beveridge, 1969; Cybulski and Moulijn, 1994; Gibson and Ashby, 1997; Garrido et al., 2008). Due to the complexity of the strut connections at nodes, orientations, their shapes, variation in strut thickness along ligament axis plays an important role in determining geometrical characteristics of the foam structure. Small changes in the strut dimension or its characteristics greatly influence the physical properties (Hugo and Topin, 2012). A comprehensive experimental characterization of open cell foams in terms of their structural, morphological, mechanical as well as thermo-hydraulic properties can be time-consuming and expensive. Therefore, it is essential to develop geometric models and correlations that allow the accurate prediction of the foam properties which are important for the reactor and heat exchangers design etc. with foams as internals.

In this chapter, the validity and applicability of geometric models and correlations reported in the literature to characterize morphological properties of open cell foams (metal and ceramic) are examined. In addition, a generalized correlation to predict geometrical characteristics of open cell foams in case of solid struts and various relationships between them is developed by taking the different strut cross sections into account. As it is commonly discussed that open cell foam structures are slightly elongated which could be due to the production methodology employed. Considering this fact, a generalized correlation to predict specific surface area and relationships between various geometrical parameters for *anisotropic* foams is also derived. This generalized correlation is further extended to predict the geometrical properties of open cell foams for the strut cross sections with void (internal cavity) that are already known in the literature. Lastly, all the analytically calculated results are compared and validated against the experimental and measured data of both metal and ceramic foams (solid struts and struts with voids).

## 3.2 Literature review of geometrical models and correlations

A brief overview of state of the art of geometrical models of foam structure is given in this section. This section is classified into two categories: metal foams (mostly open cell foams of solid struts) and ceramic foams (mostly open cell foams of struts with voids). In this thesis, metal and ceramic foams based on their strut formations (solid or hollow) are subsequently distinguished. Various authors have ignored the impact of struts with void during the measurements of geometrical properties of foam samples (Giani et al., 2005; Lacroix et al., 2007; Huu et al., 2009).

### 3.2.1 Geometrical characterization and correlations of metal foams

In open cell foams, before the development of new techniques and methodologies to determine full set of geometrical parameters, literature review shows that metallic foams often dealt with tortuosity to describe the relation between pore or strut diameter with porosity. Authors have calculated specific surface area analytically using their empirical correlations to describe the flow properties and heat transfer coefficient. Depending on the author's choice of unit cell (polyhedral, tetrakaidecahedron and dodecahedron structure); empirical correlations have used quite significant correction factors based on the assumptions or hypotheses to describe a comparable empirical model with respect to experimental data.

There are several manufacturing routes to produce open cell foams such as electro-chemical deposition, powder technology, casting techniques, bubbling agent technology, etc. (see Banhart, 2001) which acquire a complex pore shape, ligament/strut connections at node. 2-D visualization of metal foams by means of the sizing technique (using snapshots or optical microscope) has been performed to determine some of the foam characteristics. Because of limited set of porosities and foam samples based on different manufacturing techniques led researchers to model analytical solutions to characterise other meaningful properties of the foam samples.

To account for all the geometrical characteristics, porosity is one of the parameters which is easy to measure. The solid density ($\rho_s$) was measured with a He multipycnometer (QuantaChrome Corporation), designed for measurement of the true volume of solid materials using He displacement. The bulk density ($\rho_b$) was found from the volume of the pellet calculated from the dimensions and included the volume of any closed or accessible pores within the particle. The bed porosity, $\varepsilon_o$ was found from two densities by the expression, $\varepsilon_o = 1 - \rho_b/\rho_s$. Note that, in the case of metal foams, porosity is sometimes represented as open porosity to maintain the consistency among defined parameters ($\varepsilon_o = \varepsilon_t$ in the case of metal foams).

Also, foams become more or less elongated depending on the manufacturing processes. Usually, it is usually reported that high tortuosity of pore space explains the high heat transfer performance. There is a group of authors e.g. Du Plessis et al., 1994; Calmidi, 1998; Bhattacharya et al., 2002; Fourie and Du Plessis, 2002 who presented the empirical geometrical correlations using tortuosity. The tortuosity of a porous medium for the first time was defined by Carman (1937) and given in a particular direction, like the square of the ratio of the average effective distance traversed by the fluid at the Euclidean distance between two sections. Adler (1992) defined a geometrical tortuosity for each phase (solid and pores) for a couple of points contained in the same phase according to the Equation 3.1:

$$\partial\ (p_1, p_2) = \left[\frac{L_{min}(p_1, p_2)}{\|p_1 - p_2\|}\right]^2 \tag{3.1}$$

where, $L_{min}(p_1, p_2)$ is the length of shortest path in the phase joining $p_1$ to $p_2$.

Various authors (Du Plessis et al., 1994; Calmidi, 1998; Bhattacharya et al., 2002; Fourie and Du Plessis, 2002) analytically calculated tortuosity using their representative cubic cell of different strut/fiber diameters and correlated it to the geometrical and flow properties. Du Plessis et al., (1994) characterized the microstructure of metal foams by the rectangular distribution of solid material in a representative unit cell (RUC) as a function of tortuosity, porosity, total volume and fluid–solid interface area (Table 3.1, Equation 3.2). Calmidi (1998) introduced a 3-D dodecahedron unit cell structure using cylindrical struts to establish an analytical model for the fiber diameter estimation as a function of porosity, pore diameter and shape function. The shape function (denoted by $G$) takes into account the variation of fiber cross section with porosity (Table 3.1, Equation 3.3). Further, Bhattacharya et al. (2002) extended the work of Calmidi (1998) and modified the relationship given in the Equation 3.3 by replacing pore diameter with the characteristic cell size which was evaluated by counting the number of cells in a given length of foam and repeating the procedure over two lengths to get an average value (Table 3.1, Equation 3.4). These authors established a model for tortuosity as a function of porosity and shape function, which can cover a wider range of pore densities and porosities and showed that Calmidi's model (1998) has a maximum deviation of ±7% from the measured values of pore and fiber diameters. Experimental investigation of Bhattacharya et al., (2002) indicates that the tortuosity modelled by Du Plessis et al., (1994) (Table 3.1, Equation 3.2) is accurate mainly for high pore densities. Fourie and Du Plessis (2002) enhanced the modelling procedure of Du Plessis et al., (1994) to accurately predict the hydrodynamic conditions in both, Darcy and inertia regimes, without a priori knowledge of the flow behaviour of the particular metal foams (Table 3.1, Equation 3.5).

Recent developments have explicitly expanded the concept of tortuosity. Vicente et al., (2006a) found that the tortuosity, defined for a couple of points, is not suitable to correlate transport properties to morphology. Indeed, it is the average tortuosity between two surfaces in a direction that governs the transport phenomena. These authors used the fast marching method to measure geometrical tortuosity of two phases (solid and pore) and reported that the tortuosity of pores is very low compared to solid matrix value. Their measurements show no clear influence of the pore size on tortuosity.

There is another group of researchers who have not taken tortuosity into consideration anywhere in their correlation. In this group, researchers have measured only a few geometrical properties i.e. porosity and pore diameter (or strut diameter) and subsequently, using their chosen geometrical model, they have calculated other pertinent geometrical properties. As shown in the Figure 2.6, depending upon the set of foam samples and visualization techniques, researchers have described the unit cell as a family of polyhedron namely, cubic lattice structure (Gibson and Ashby, 1997; Lu et al., 1998; Boomsma et al., 2003; Giani et al., 2005), tetrakaidecahedron structure (Gibson and Ashby, 1997; Kwon et al., 2003; Kanaun and Tkachenko, 2008; De Jaeger et al., 2011; Inayat et al., 2011 a, b) and pentagonal dodecahedron structure (Ozmat et al., 2004; Huu et al., 2009; Smorygo et al., 2011).

**Table 3.1**. A synthesis of pertinent correlations for metal foam pore and strut/fiber diameters (with tortuosity).

| Authors | Tortuosity | Correlation between geometrical parameters | Eq. No. |
|---|---|---|---|
| Du Plessis et al., (1994) | $\frac{1}{\partial} = \frac{3}{4\varepsilon_o} + \frac{\sqrt{9-8\varepsilon_o}}{2\varepsilon_o}.cos\left\{\frac{4\pi}{3} + \frac{1}{3}cos^{-1}\left(\frac{8{\varepsilon_o}^2 - 36\varepsilon_o + 27}{(9-8\varepsilon_o)^{3/2}}\right)\right\}$ | $\frac{d_p}{d} = \sqrt{\frac{\varepsilon_o}{\partial}}$ , $\frac{d_f}{d} = 1 - \sqrt{\frac{\varepsilon_o}{\partial}}$ | 3.2 |
| Calmidi (1998) | - | $\frac{d_f}{d_p} = 2\sqrt{\frac{(1-\varepsilon_o)}{3\pi}\frac{1}{1-e^{-(1-\varepsilon_o)/0.04}}}$ | 3.3 |
| Bhattacharya et al., (2002) | $\frac{1}{\partial} = \frac{\pi}{4\varepsilon_o}\left\{1 - \left(1.18\sqrt{\frac{(1-\varepsilon_o)}{3\pi}\frac{1}{1-e^{-(1-\varepsilon_o)/0.04}}}\right)^2\right\}$ | $\frac{d_f}{d_p} = 1.18\sqrt{\frac{(1-\varepsilon_o)}{3\pi}\frac{1}{1-e^{-(1-\varepsilon_o)/0.04}}}$ | 3.4 |
| Fourie and Du Plessis (2002) | $\partial = 2 + 2cos\left[\frac{4\pi}{3} + \frac{1}{3}cos^{-1}(2\varepsilon_o - 1)\right]$ | $\frac{d_p}{d} = \sqrt{\frac{\varepsilon_o}{\partial}} = \frac{3-\partial}{2}$ | 3.5 |

**Table 3.2**. A synthesis of pertinent correlations for metal foam pore/strut diameter and open porosity.

| Authors | Correlation between geometrical parameters | Correlation of geometrical parameters with porosity | Constraints/ Boundary conditions | Eq. No. |
|---|---|---|---|---|
| Lu et al., (1998) | $d_s = d_p \frac{2}{\sqrt{3\pi}}(1-\varepsilon_o)^{1/2}$ | - | - | 3.6 |
| Giani et al., (2005) | $d_s = d_p \left[\frac{4}{3\pi}(1-\varepsilon_o)\right]^{1/2}$ | - | - | 3.7 |
| Kanaun et al., (2008) | - | $\varepsilon_o = 1 - \frac{N_1 v_{l-2a_2} + N_n v_{int}}{V_o}$ | - | 3.8 |
| De Jaeger et al., (2011) | $a_2(\varepsilon_o) = \begin{cases} 34.72 \quad (a) \\ 17343\varepsilon_o{}^2 - 31809\varepsilon_o + 14622 \quad (b) \end{cases}$<br>$(a)\ for\ 0.88 \leq \varepsilon_o \leq 0.91$<br>$(b)\ for\ \varepsilon_o \geq 0.91$<br>$HW = 0.971(1-\varepsilon_o)^{-0.09}$ | $\varepsilon_o = 1 - \frac{24\int_{i_1}^{i_2} f(\xi)\,d\xi + 32\frac{\sqrt{2}}{3}R(1/2)^2}{V_o}$ | $i_1 = -\frac{1}{2} + \eta\frac{R(1/2)}{l}$<br>$i_2 = \frac{1}{2} - \eta\frac{R(1/2)}{l}$ | 3.9 |
| *Ozmat et al., (2004) | $\frac{d^2}{\sqrt{3}}.(s-1.4d).\frac{30}{3} + 0.3\frac{20}{3}d^3 = \rho_r\frac{15+7\sqrt{5}}{4}s^3$ | - | - | 3.10 |
| Huu et al., (2009) (Triangular "Slim" strut) | $k' = \frac{d_s}{l}, \frac{d_p}{l} = \frac{\varphi}{\sqrt{3-\varphi}}\left(1 - \frac{k'}{2}\sqrt{\frac{2}{3}}\right)$ | $\varepsilon_o = -1 + k'^2\frac{\sqrt{15}}{\varphi^4} - k'^3\frac{\sqrt{10}}{3\varphi^4}$ | "Slim" foam | 3.11 |
| Huu et al., (2009) (Circular "Slim" strut) | $k' = \frac{d_s}{l}, \frac{d_p}{l} = \frac{\varphi}{\sqrt{3-\varphi}}\left(1 - \frac{k'}{2}\sqrt{\frac{2}{3}}\right)$ | $\varepsilon_o = -1 + k'^2\frac{\sqrt{15}}{\varphi^4}\left(1 - \frac{k'}{2}\sqrt{\frac{2}{3}}\right) - k'^3\frac{2\sqrt{10}}{12\varphi^4}$ | | 3.12 |
| Smorygo et al., (2011) (Triangular strut) | $d_s = d_p\left[1.954\left(\frac{0.97-\varepsilon_o}{\varepsilon_o - 0.34}\right)\right]$ | $\varepsilon_o = \frac{\pi(k''^3 - 3(k''-1)^2(2k''+1))}{3\sqrt{2}}$ | $0.9 < \varepsilon_o < 0.95$<br>$k'' \approx \varepsilon_o + 0.18$ | 3.13 |
| Smorygo et al., (2011) (Circular strut) | $d_s = d_p\left[0.435\left(\frac{1-\varepsilon_o}{\varepsilon_o - 0.68}\right)\right]$ | | $0.75 < \varepsilon_o < 0.9$<br>$k'' \approx 0.57(\varepsilon_o + 1)$ | 3.14 |

*$\rho_r$ is relative density

The correlations by various researchers e.g. Lu et al., 1998; Giani et al., 2005 based on cubic lattice structure are presented in Table 3.2. Lu et al., (1998) represented a simple cubic unit cell consisting of slender cylinders to predict heat transfer parameters as a function of foam density and cell size (Table 3.2, Equation 3.6). These authors predicted the correlation according to ERG Duocel datasheets and found that the cell ligaments of the ERG foams are best described as a rod with rectangular or triangular cross section rather than a circular cylinder. Moreover, they doubted the data directly provided by ERG catalogue. For instance, these authors observed that the microstructural characterizations suggested by Duocel foam labelled by ERG as 40 PPI is actually described more closely by 30 PPI.

Giani et al., (2005) also modelled their foams using cubic lattice structure like Lu et al., (1998) using cylindrical struts. Pore diameter and porosity were measured using optical measurements and He multipycnometer respectively and correlated with strut diameter (Table 3.2, Equation 3.7). Due to unavailability of information about other set of geometrical parameters; strut and pore diameters, specific surface area etc. were determined using the analogy of particle diameter ($D_p$). Particle diameter ($D_p$) is most commonly related to equivalent packed bed of spheres but it is unclear that how spheres in packed bed are stacked and whether they are well connected or not. The relation is usually given as $D_p = 6/a_c$ but this expression does not take into account the geometrical structure/description of foam which is generally not the case in the case of open cell foams and thus, induces a lot of inconsistencies.

All the analytical correlations proposed by authors e.g. Du Plessis et al., 1994; Calmidi, 1998; Lu et al., 1998; Bhattacharya et al., 2002; Fourie and Du Plessis, 2002; Giani et al., 2005 induce a lot of discrepancies in predicting morphological characteristics because of their cubic lattice models which have no resemblance with the real foam structure. As 2-D visualization using optical microscope is of low accuracy and hence, the strut diameters (mainly cylindrical cross section) are scattered in the literature. Moreover, the errors in the strut diameters are less in very high porosity range ($\varepsilon_o$ >0.90) but it is very significant in low porosity range (0.60< $\varepsilon_o$ <0.90) for the proposed models by these authors and thus, are not applicable and cannot be generalized on different foam samples.

The above discussion suggests that the two parameters i.e. $d_p$ or $d_s$ and $\varepsilon_o$ are not sufficient to characterise the foam structure. Moreover, the representation of these parameters obtained by 2-D measurements does not provide any explicit information about the cell or strut orientations, average porosity, cell elongations (slight anisotropy), ligament variations along its axis, the windows connection openings (pentagon or hexagon), strut shape identification, average specific surface area etc. Thus, there was indeed a need of 3-D tools that can characterise completely any foam structure and reconstruct them for further analysis to determine thermo-hydraulic properties.

Recently, X-ray μCT technique has become an important tool to examine the porous media (Spowage et al., 2006; Vicente et al., 2006a) to create cross sections of a 3-D object that later can be used to recreate a virtual model without destroying the original model. The most important advantage of μCT is the possibility of 3-D visualisation, quantification of porosity and mineralogical pore-relating phases. It provides 3-D information of the scanned specimen concerning cell arrangement like orientation, gradient, homogeneity, etc. Images of contiguous planes can be stacked by mean of a reconstruction algorithm to form 3-D images of a section or if the entire part has been scanned, a full volumetric image of the specimen (see Figure 2.8). X-ray μCT has been applied by different authors (e.g. Vicente et al., 2006a; Grosse et al., 2009; Inayat et al., 2011a) to characterize the morphological parameters of foam structures.

In the case of commercially available foams (ERG, RECEMAT, ALANTUM etc.), it has been seen that the ligament exhibits a variable cross section along its axis. In the literature, different strut shapes e.g. circular, triangular, convex and concave triangular have been discussed (Bhattacharya et al., 2002; Kanaun and Tkachenko, 2008; Inayat et al., 2011a; De Jaeger et al., 2011, Dairon et al., 2011) which have strong influence on geometrical and thermo-hydraulic properties. On the other hand, variable cross section of the ligament along its axis needs to be taken in account which is generally not reported in the literature.

Few authors have derived the correlations by considering the real strut cross section in order to accommodate the real foam geometry usually obtained by μCT on tetrakaidecahedron structure. Kanaun and Tkachenko (2008) showed that for a varying cross section along the ligament axis, it is possible to accumulate more mass at the node junction that will change the specific surface area for a given porosity. These authors showed that for a given porosity, one can obtain different specific surface areas by varying ligament cross section along its axis and considered strut length as an important geometrical parameter in determining specific surface area (Table 3.2, Equation 3.8). De Jaeger et al., (2011) manufactured in-house open cell aluminum foams to characterize the geometrical parameters analytically and obtained convex triangular strut shape up for the porosity up to 88%. The variation in strut cross section was described by a fourth order polynomial curve along ligament axis where struts are thinner near the center and more material is accumulated in the vicinity of nodes. Strut cross sectional shape is known to depend on porosity (see Bhattacharya et al., 2002) and clearly is not an equilateral triangle or a circle. De Jaeger et al., (2011) introduced this fact in their analytical derivation and quantified porosity dependency via the Heywood circularity factor (HW), defined as the ratio of the strut cross section perimeter to the equivalent perimeter of a circle with the same surface area and obtained this dependency by a fitting curve on 10 different strut cross sections (Table 3.2, Equation 3.9). To determine axial shape factor, these authors obtained the best fit in relation with open porosity for porosities $\varepsilon_o \geq 0.91$ and a constant value of 34.72 in the porosity range, $0.88 \leq \varepsilon_o \leq 0.91$. All these fits were obtained to quantify the strut cross section variation and axial shape factor on in-house manufactured foams (De Jaeger et al., 2011).

Dairon et al., (2011) reported convex triangular strut shape of constant ligament cross section in the porosity range, $0.80 < \varepsilon_o < 0.95$. Using their casting technique, these authors reported neither circular nor concave triangular strut shapes.

There are some foam geometries that reveal pentagon openings (see Figure 2.6-right), and a few authors have derived the analytical solutions to accommodate such geometry using pentagonal dodecahedron structure. Ozmat et al., (2004) estimated 1.4 and 0.3 as coefficients of linear measure of nodes and nodal shape factor for three different foam samples of 10-30 PPI respectively and are empirical values (Table 3.2, Equation 3.10). Moreover, these authors compressed these foams to approximately 30% of their relative density to increase their surface area (see also Boomsma et al., 2003) but foams lose their characteristics (geometry and its networks) when compressed only in one direction. For a set of three samples and estimated coefficients using a fitting relationship, their correlation cannot be extended to apply on different foams.

To account for wide porosity range ($0.75 \leq \varepsilon_o \leq 0.95$), Huu et al., (2009) proposed to use a pentagonal dodecahedron structure as the unit cell. Their approach enables accounting for triangular or cylindrical struts and solid accumulation at their meeting points which were defined as “slim” (triangular strut) and “fat” (circular strut) foams. The “slim” and “fat” models were derived for both circular strut (0.75≤ $\varepsilon_o$ ≤0.90) and triangular strut shapes (0.90≥ $\varepsilon_o$). These models predict lower and upper bound of geometrical parameters. In other words, the "slim" model has tendency to overestimate the experimental data, while the "fat" model underestimates the experimental values (Table 3.2, Equation 3.11 and 3.12). These authors considered non-porous struts (solid struts without internal voids) in their modelling, which imply that, the total porosity ($\varepsilon_t$) is the same as the open porosity ($\varepsilon_o$). These authors compared their models with experimental data which were actually measured for ceramic foams (hollow strut nature). In their geometrical correlations, these authors did not take into account porosity due to inner void in the strut ($\varepsilon_s$: strut porosity) and thus, "slim" and "fat" models induce high errors.

An inverted open cell foam model based on hexagonal close packing symmetry was proposed by Smorygo et al., (2011) for the characterization of foams with different strut configurations based on pentagonal dodecahedron unit cell. Their model is also splitted into two strut shapes namely circular and triangular same as the "slim" and "fat" models of Huu et al., (2009). These authors used a fitting parameter, $k''$ which is the ratio of sphere diameter to cell diameter. Their circular strut configuration ensures open porosity with a relatively high $d_p/d_{cell}$ ratio at identical porosity values. The model does not allow calculation for $d_p/d_{cell}$ in the low porosity range ($\varepsilon_o$ ≤0.74). The fitting parameter, $k''$ varies from 1 to 1.155 from circular strut shape to triangular one in the porosity range (0.75≤ $\varepsilon_o$ ≤0.95). These fitting correction factors, however, are sensitive to both porosity range and foam strut configuration. The correction factors of 0.85-0.9 and 0.8-1.45 were used to avoid discrepancies in the calculated values of pore and strut diameters for their foam samples (Table 3.2, Equation 3.13 and 3.14). Due to the limit of the proposed model, the calculated pore diameter ($d_p$) is overestimated while the calculated strut diameter ($d_s$) was underestimated against the measured experimental values. These authors also ignored the strut porosity ($\varepsilon_s$) while deriving their correlations for ceramic foams.

Importantly, regular dodecahedron is not a space filling structure and thus, may not be appropriate to use as periodic unit cell as fully discussed by many authors e.g. Wells, 1991; Gibson and Ashby, 1997; Bourret et al., 1997; Steinhaus, 1999; Inayat et al., 2011 a, b. Bourret et al., (1997) showed that when the porosity of solid foam decreases, it is more difficult to obtain a perfect packing of the pentagonal dodecahedron structure as a unit cell.

Since the tetrakaidecahedron structure displays the most efficient filling space (structure) and regular network of unit cells, having the lowest geometric specific surface area (see Gibson and Ashby, 1997), it is still most probably the best idealized representation of the foam geometry. The measurements of geometrical properties are consistent with most of the foams commercially available (see Inayat et al., 2011a). Moreover, tetrakaidecahedron structure can be easily repeated and represents the closest foam geometry that exists in reality. One can carry out analytical solutions to easily describe foam shape and geometry. It is also possible to produce Kelvin-like open cell foams with this structure by changing only the strut shape.

It is thus, clear that there is a need for an analytical correlation that accounts no change in strut shape or contains a parameter which accounts for the change in strut shape if any, without any fitting parameter and should be valid for a wide range of porosity based on tetrakaidecahedron structure.

For many industrial systems, external specific surface area ($a_c$) is an important parameter which is responsible for the performance and successful design of reactors and/or heat exchangers. For solid foams, it has been widely quoted that specific surface area is typically higher than 1000 $m^{-1}$ even at very low pore densities. In case of metal foams (solid struts), the specific surface area is usually determined by conventional physisorption measurements of gases, applying the BET method (Brunauer et al., 1938). The range of specific surface area of commercially available foams is roughly 300-5000 $m^{-1}$.

Various authors have not measured specific surface area (e.g. Lu et al., 1998; Calmidi and Mahajan, 2000; Fourie and Du Plessis, 2002) and derived their analytical expressions based on tortuosity that were purely empirical (Table 3.3, Equation 3.15-3.17) using pentagonal dodecahedron or cubic lattice unit cell. However, it has been earlier discussed that tortuosity is not significant in open cell foams and thus, these correlations could not provide satisfactory results. Based on slender cylindrical struts in cubic lattice unit cell, the correlation of Lu et al., (1998) to predict specific surface area correlates well for 10 and 20 PPI foams but a difference of 25% was observed between the predicted results and measurements of 40 PPI foam (Table 3.3, Equation 3.17). These authors explained this difference may be attributed to the actual mean cell size being different from the nominal mean cell size used in the calculation.

Giani et al., (2005) adopted the analogy between solid foam structure and packed bed spherical particles with the same specific surface area and same porosity which leads to Equation 3.18 (see Table 3.3) and thus, a relationship $d_p = 1.5d_s$ was deduced.

Kanaun and Tkachenko (2008) were mainly focussed in determining porosity using actual geometrical parameters of the foam structure. These authors derived expressions that estimate well the geometrical surface area of ligament and node junction. Moreover, their model contains the variable ligament cross section function which allows us to obtain different specific surface areas for the same porosity. This is due to the extra accumulation of mass at the node and thin node at the centre.

**Table 3.3**. A synthesis of pertinent correlations of specific surface area for metal foams (with and without tortuosity).

| Authors | Specific Surface Area, $a_c$ | Remarks | Eq. No. |
|---|---|---|---|
| Calmidi and Mahajan (2000) | $a_c = \frac{3\pi d_f}{(0.59d_p)^2}\left[1 - e^{-(1-\varepsilon_o)/0.04}\right]$ | | 3.15 |
| Fourie and Du Plessis (2002) | $a_c = \frac{3}{d}(3 - \partial)(\partial - 1)$ | | 3.16 |
| *Lu et al., (1998) | $a_c = \left(\frac{2\sqrt{3\pi}}{d_p}\right){\rho_r}^{1/2}$ | | 3.17 |
| Giani et al., (2005) | $a_c = \frac{4}{d_s}(1 - \varepsilon_o) = \frac{2}{d_p}[3\pi(1 - \varepsilon_o)]^2$ | | 3.18 |
| De Jaeger et al., (2011) | $a_c = 4\frac{a_1 - 2}{\sqrt{{a_1}^2 - 2}}\sqrt{\frac{A_o}{\pi}}\int_{i_1}^{i_2}\sqrt{f(\xi)}\,d\xi$ | $A_o$- Cross sectional are between centre of nodes | 3.19 |
| Ozmat et al., (2004) | $a_c = \frac{d(p - 1.25d)30 + 0.4d^2\frac{20}{3}}{100\rho_r\frac{15 + 7\sqrt{5}}{4}s^2}$ | $p$- Hypotenuse distance between pentagonal centre to its edge | 3.20 |
| Huu et al., (2009) (Triangular strut) | $a_c = \frac{M}{\sqrt{5}\varphi^2}60k'\left(1 - \frac{1}{2}\sqrt{\frac{2}{3}}k'\right)$ | | 3.21 |
| Huu et al., (2009) (Circular foam) | $a_c = \frac{M}{\sqrt{5}\varphi^2}20\pi k'\left(1 - \frac{1}{2}\sqrt{\frac{2}{3}}k'\right)$ | $M = \frac{\left(1 - \frac{k'}{2}\sqrt{\frac{2}{3}}\right)}{\varphi\sqrt{3 - \varphi}}$ | 3.22 |
| Smorygo et al., (2011) (Triangular strut) | $a_c = d_{cell}\frac{1}{\left[0.023\left(\frac{\varepsilon_o}{1 - \varepsilon_o}\right) + 7\right]}$ | | 3.23 |
| Smorygo et al., (2011) (Circular strut) | $a_c = d_{cell}\frac{1}{\left[0.022\left(\frac{\varepsilon_o}{1 - \varepsilon_o}\right) + 14\right]}$ | $d_{cell}$- Cell diameter | 3.24 |

*$\rho_r$ is relative density

De Jaeger et al., (2011) applied Green's theorem to evaluate specific surface area that contains an elliptical function. This elliptical function can be found elsewhere (e.g. Mathematica). Their expression of specific surface area is a complex function of geometrical parameters and can be solved by numerical calculation using iterative method (Table 3.3, Equation 3.19).

Ozmat et al., (2004) derived specific surface area in relation with pore density the same way like geometrical properties. These authors introduced two correction parameters, 1.25 and 0.4 respectively that are purely empirical in order to validate their calculated results against measurements (Table 3.3, Equation 3.20).

The correlations of specific surface area proposed by Huu et al., (2009) have a general trend of overestimating the experimental data and that were observed with the lowest deviation for the case of a nominal porosity of 90% (Table 3.3, Equation 3.21 and 3.22).

Smorygo et al., (2011) proposed the analytical models to predict specific surface area same as the models of Huu et al., (2009) by ignoring the strut porosity (hollow strut nature). The models based on circular and triangular nature of strut shapes underestimated the measured data and the fitting parameter accounts for linear dependency of $a_c.d_{cell}$ with open porosity (Table 3.3, Equation 3.23 and 3.24).

For the above reasons, the direct comparison of the model-derived geometrical parameters and specific surface areas with data extracted from the literature are challenging.

### 3.2.2 Geometrical characterization and correlations of ceramic foams

Open cell foams with struts having an internal cavity (mostly ceramic foams) possess remarkable properties (e.g. high porosity, large external surface area and high mechanical strength) same as open cell foams with solid struts (mostly metal foams), and are excellent candidates for a variety of industrial applications. The first commercial application of ceramic foams reported in the literature was their use as filters for molten metals by few authors (Maiorove et al., 1984; Brockmeyer and Aubrey, 1987). In the past two decades, ceramic foams have been extensively investigated for a number of other applications including gas filters, heat exchangers, porous burners and catalyst supports (Twigg and Richardson, 1994; 2002).

Different geometric models (cubic lattice structure, tetrakaidecahedra structure, and pentagonal dodecahedron structure) to characterize ceramic foams have been reported in the literature. These models were used by authors in order to derive correlations to predict the specific surface area of foam structures using measured parameters such as pore or strut diameter and porosity. BET method (Brunauer et al., 1938), however, cannot be used for ceramic foams because they may possess internal void volumes and a rough surface, which would lead to an overestimation of the specific surface area values.

Innocentini et al., (1999) identified the relation of $D_p$ with $d_h$ (Table 3.4, Equation 3.25) and predicted specific surface area correlation based on equivalent bed spheres.

Richardson et al., (2000) measured geometrical properties of $\alpha - Al_2O_3$ foam samples. Total porosity using He multipycnometer while pore diameter using 2-D image analysis were measured. These authors had no access to measurements of specific surface area of their ceramic foams and used the relationships of Gibson and Ashby (1997) to derive specific surface area ($a_c$) based on pentagonal dodecahedron structure (Table 3.4, Equation 3.26). They used the simple assumption of hydraulic diameter method to treat the pores as uniform, parallel cylinders; each with a constant diameter equal ${d_p}^{Heq}$ (${d_p}^{Heq}$- is the diameter of a circle with an area equivalent to the hexagonal window) that leads to $a_c = 4\varepsilon_o/{d_p}^{Heq}(1-\varepsilon_o)$. This approach, however, is not appropriate to apply on foam structures.

Buciuman and Kraushaar-Czarnetzki (2003) measured geometrical properties of Alumina and Mullite foam samples. The foams properties were determined with an optical microscope. The diameters of the cell windows (or pore diameters) and thicknesses of the struts were measured by means of the sizing technique with feature-to-feature scanning while specific surface area was measured using BET. The specific surface areas were overestimated due to the presence of hollow struts. Their correlation between geometrical parameters and specific surface area was based on tetrakaidekahedron structure using the similar approach of Gibson and Ashby (1997) (Table 3.4, Equation 3.27).

Lacroix et al., (2007) used the cubic lattice model as the equivalent particle model in order to develop a direct analogy between foams and beds made of spherical particles to predict specific surface area.

**Table 3.4**. A synthesis of pertinent correlations of geometrical parameters and specific surface area for ceramic foams.

| Authors | Geometrical parameters | Specific Surface Area, $a_c$ | Eq. No. |
|---|---|---|---|
| Innocentini et al., (1999) | $D_p = 1.5\frac{1-\varepsilon_o}{\varepsilon_o}d_h$ | $a_c = \frac{6}{D_p}$ | 3.25 |
| Richardson et al., (2000) | $d_s = \frac{0.5338d_w(1-\varepsilon_o)^{0.5}}{1-0.971(1-\varepsilon_o)^{0.5}}$ | $a_c = \frac{12.979[1-0.971(1-\varepsilon_o)^{0.5}]}{d_w(1-\varepsilon_o)^{0.5}}$ | 3.26 |
| Buciuman and Kraushaar-Czarnetzki (2003) | $d_s = d_w\left[\frac{1-\varepsilon_o}{2.59}\right]^{0.5}$ | $a_c = 4.82\frac{1}{(d_w+d_s)}\sqrt{1-\varepsilon_o}$ | 3.27 |
| Lacroix et al., (2007) | $d_s = \frac{d_w[4/3\pi\,(1-\varepsilon_o)]^{0.5}}{1-[4/3\pi\,(1-\varepsilon_o)]^{0.5}}$ | $a_c = \frac{6}{D_p}(1-\varepsilon_o)$ or $a_c = \frac{4}{d_s}(1-\varepsilon_o)$ | 3.28 |
| Garrido et al., (2008) | - | $a_c = 3.84\left(\frac{d_w+d_s}{m}\right)^{-0.85}.\varepsilon_o^{-0.82}$ | 3.29 |
| Grosse et al., (2009) | - | $a_c = \frac{8.21\sqrt{1-\varepsilon_n}-1.55(1-\varepsilon_n)}{(d_w+d_s)}$ | 3.30 |
| | - | $a_c = \frac{4.84\sqrt{1-\varepsilon_o}-2.64(1-\varepsilon_o)}{(d_w+d_s)}$ | 3.31 |
| Inayat et al., (2011a) | $d_s = d_w\frac{0.6164(1-\varepsilon_o)^{0.5}}{1-0.971(1-\varepsilon_o)^{0.5}}$ | $a_c = 4.867\frac{[1-0.971(1-\varepsilon_o)^{0.5}]}{d_w(1-\varepsilon_o)^{0.5}}(1-\varepsilon_o)$ | 3.32 |

In their correlation, these authors assumed that the cellular structure is made of solid cylindrical filaments (struts) connected in the three dimensions as a regular cubic lattice that leads to $D_p = 1.5d_s$ (Table 3.4, Equation 3.28).

The empirical correlations proposed by authors e.g. Innocentini et al., 1999; Richardson et al., 2000; Lacroix et al., 2007 induce a lot of discrepancies in determining precisely specific surface area. Moreover, their correlations were based on only two geometrical parameters and do not conclude same results because of low accuracy measurements or ill-defined geometrical quantities or simplified geometric model/structure.

Recently, a few authors e.g. Garrido et al., 2008; Grosse et al., 2009; Dietrich et al., 2009 measured specific surface area of ceramic foams using magnetic resonance imaging (MRI) technique to take hollow strut into account (which was not taken care of using BET).

Garrido et al., (2008) compared the experimental results obtained for strut porosity, open porosity and specific surface area using MRI with the correlations of Buciuman and Kraushaar-Czarnetzki (2003) and Lacroix et al., (2007) and reported the overestimation of predicted results. Garrido et al., (2008) also argued that the cubic lattice model results in a stronger over-prediction of the surface area than the tetrakaidecahedron model.

Grosse et al., (2009) measured open and total porosity using mercury intrusion porosimetry and specific surface area using MRI. Their analytical model to predict specific surface area was based on a novel space-filling structure commonly known as Weaire-Phelan structure that was proposed for equilibrium foam structures by Phelan et al., (1995). Phelan et al., (1995) did not provide any correlation to calculate specific surface area. The unit cell of the Weaire-Phelan structure consists of two pentagonal dodecahedra with 12 identical pentagonal faces and six tetrakaidecahedra, each with two hexagonal faces, four pentagonal faces similar to those of the dodecahedra, and eight pentagonal faces of identical shape. On the basis of Weaire-Phelan model, Grosse et al., (2009) derived a correlation for the specific surface area following the procedure of Buciuman and Kraushaar-Czarnetzki (2003) (see Table 3.4, Equation 3.30). However, their correlation in Equation 3.30 was unable to predict their experimental results. Therefore, these authors used an empirical fitting procedure to redefine the coefficient using open porosity and obtained a semi-empirical correlation (Table 3.4, Equation 3.31) in order to account the anisotropy of foams (see Zeschky et al., 2005) which gave close agreement to their experimental data.

Inayat et al., (2011a) proposed empirical correlations (for different strut shapes namely, triangular, circular and concave triangular) using the geometrical relationships of Gibson and Ashby (1997) with open porosity ($\varepsilon_o$), window and strut diameters ($d_w$ and $d_s$) and is given by Equation 3.32 (Table 3.4). The expression of cylindrical strut shape only is presented in Equation 3.32. Their correlation predicts an excellent agreement with experimental specific surface area.

The literature survey reveals that all the correlations derived for ceramic foams are based on hypothesis considered by Gibson and Ashby (1997). Only two geometrical quantities were measured: $d_p$ or $d_w$ and $\varepsilon_o$. Other quantities were simply calculated using the geometrical relations presented by Gibson and Ashby (1997). Different correlations have different numerical values of fitting/correction factors that depend on the type of unit cell (e.g. cubic lattice model, the tetrakaidecahedra model or pentagonal dodecahedron model) to predict analytically geometrical parameters and specific surface area. Moreover, some of the correlations were derived using empirical fitting on a small set of foam samples and thus, cannot be directly applied to completely characterize the ceramic foams.

### 3.3 Performance of state of the art correlations

In this section, the validity and suitability of state of the art correlations (hence the geometric models behind them) for foams of different materials, pore size and porosity are examined. After careful review of the correlations presented in the literature, it is observed that the authors have not measured all the geometrical properties of foams simultaneously. Most of the authors have measured only two parameters: porosity ($\varepsilon_o$) and pore or strut diameter ($d_p$ or $d_s$). In Table B.1 (see Appendix B), it is evident that the geometrical properties were measured for the foam samples of very high porosity (majority of the commercially available foam samples have $\varepsilon_o$ >0.90). Thus, the authors' choice of the geometrical model used to derive correlations does not impact strongly the predicted properties. Other parameters were simply derived by using the same analytical model to complete the set of other geometrical properties. In most of the cases, the experimental data and correlations were validated within a small range of error for a given set of foams because the strut geometry (or strut cross section) does not play a significant role at high porosity ($\varepsilon_o$ >0.90). Moreover, the correlations were derived either using pore diameter or strut diameter depending upon which parameter was measured. As discussed in the literature, strut shape changes from circular to convex or concave triangular shape with porosity, the correlations have empirical numerical values which were obtained by curve fitting and these empirical values are quite different depending upon strut geometry and geometrical model used and thus, cannot be applied to determine the geometrical properties of different foam matrices.

However, the correlations developed by De Jaeger et al., (2011) and Inayat et al., (2011a) to predict geometrical properties of foam structure are satisfactory. The correlation of De Jaeger et al., (2011) is limited to 88% porosity and is only applicable for convex or concave triangular strut shape metal foams. On the other hand, correlations proposed by Inayat et al., (2011a) are applicable to both metal and ceramic foams but are limited to convex or concave triangular strut shape of the foam samples. For different nature of strut shapes, their correlations (De Jaeger et al., 2011 and Inayat et al., 2011a) cannot be extended and thus, cannot be applied to different foam structures.

### 3.3.1 Problems and investigation of correlations in case of metal foams

Various relationships between geometrical parameters and specific surface areas proposed by various authors e.g. Du Plessis et al., 1994; Calmidi, 1998; Lu et al., 1998; Fourie et Du Plessis, 2002; Bhattacharya et al., 2002; Giani et al., 2005; Huu et al., 2009; Smorygo et al., 2011; De Jaeger et al., 2011 have been investigated. It has always been difficult to characterize the metal foams due to ill-measured quantities like pore and strut diameters before the development of 3-D measurement tools. There are a very few authors (e.g. Perrot et al., 2007; Brun et al., 2009; De Jaeger et al., 2011) who have provided a complete set of experimental data of geometrical parameters; performed on tomographied images of foams using μCT and BET methods.

To validate the suitability of different correlations reported in the literature, the geometrical measurements of the sample 4 of 85% porosity studied in the present work (see Table 2.2) was chosen arbitrary (results remain true for other samples) and the calculated values obtained from various correlations are presented in Table 3.5. Table 3.5 is divided in two parts: “Known pore diameter” ($d_p$) and “Known strut/fiber diameter” ($d_s$ or $d_f$). In the literature, some correlations were derived based on a known pore diameter while the others were derived based on known strut diameter. These two approaches were based on author's choice of unit cell and available measured parameters. The calculated values based on both the approaches are presented.

In the case of "known pore diameter", it is interesting to note that the strut diameter ($d_s$) is underestimated by an approximate factor of 0.5 except the correlation proposed by Smorygo et al., (2011) where it has predicted the precise value of strut diameter. However, the calculated values of specific surface area are scattered. Most consistent values were obtained for the correlations of Fourie and Du Plessis (2002), De Jaeger et al., (2011) and Smorygo et al., (2011).

One can easily notice that the correlations derived for simple cylindrical strut cross section based on cubic lattice or dodecahedron structure overestimate the strut diameter values by 1.5-2.5 orders of magnitude. The values of specific surface areas are scattered and the correlations underestimate and overestimate the experimental data by 0.45-4.5 orders of magnitude. It is pointed out that despite the correlation of Smorygo et al., (2011) predicts accurate results but it is based on few unknowns and those were obtained by curve fitting and thus, are not appropriate for other strut cross sections. Moreover, the correlation of De Jaeger et al., (2011) predicts accurate results but was derived only for convex triangular strut shape and also cannot be applied to other strut cross sections.

In the case of "known strut/fiber diameter", the pore diameter prediction using correlations are overestimated by an approximate factor of 2 orders of magnitude and predict good results of specific surface area except for the correlations proposed by Calmidi and Mahajan (2000); Fourie and Du Plessis (2002) and Huu et al., (2009).

Main reasons of such high errors in geometrical parameters and specific surface areas are:

- Ill-measurement of geometrical parameters.
- Over-simplified strut shape and geometry of the unit cell.
- Relation between pore diameter and strut diameter is critical.

After careful evaluation, a few remarks are presented in order to derive precisely geometrical characteristics:

- New modes of extracting experimental data from 3-D reconstructed surface.
- Use of tetrakaidecahedron unit cell approach to maintain the consistency of results.

**Table 3.5**. Comparison of strut/fiber diameters and specific surface area of various correlations from the data obtained by iMorph on cast sample number 4 of 85% porosity.

| | Known pore diameter ($d_p$) | | | | | Known strut/fiber diameter ($d_s$ or $d_f$) | | | | |
|---|---|---|---|---|---|---|---|---|---|---|
| Authors | ***$\partial$(-) | *$d_s$ or $d_f$ (mm) | $d_p$ (mm) | $d_s/d_p$ (-) | $a_c$ (m$^{-1}$) | $\partial$(-) | *$d_s$ or $d_f$ (mm) | $d_p$ (mm) | $d_s/d_p$ (-) | $a_c$ (m$^{-1}$) |
| **Current work Measurements on sample 4 | 1.1 | 2.5 | 5.53 | 0.452 | 251.1 | 1.1 | 2.5 | 5.53 | 0.452 | 251.1 |
| Du Plessis et al., (1994) | 1.7 | - | 5.53 | 0.414 | - | 1.7 | 2.5 | - | 0.414 | - |
| Calmidi and Mahajan (2000) | - | 1.426 | 5.53 | 0.258 | 1233 | | 2.5 | 9.69 | 0.258 | 704 |
| Bhattacharya et al., (2002) | 1.108 | | 5.53 | 0.152 | - | 1.108 | 2.5 | | 0.152 | - |
| Fourie and Du Plessis (2002) | 1.422 | | 5.53 | 0.268 | 285 | 1.422 | 2.5 | | 0.268 | 168 |
| Lu et al. (1998) | - | 1.395 | 5.53 | 0.252 | 430 | - | 2.5 | 9.908 | 0.252 | 240 |
| Giani et al. (2005) | - | 1.395 | 5.53 | 0.252 | 430 | - | 2.5 | 1.174 | 0.252 | 240 |
| De Jaeger et al., (2011) | - | - | 5.53 | - | 282 | - | 2.5 | - | - | 282 |
| Huu et al., (2009) | - | 1.553 | 5.53 | 0.281 | 428 | - | 2.5 | 0.794 | 3.14 | 1664 |
| Smorygo et al., (2011) | - | 2.54 | 5.53 | 0.459 | 275 | - | 2.5 | 5.435 | 0.46 | 280 |

* Value of $d_s$ is taken according to the definition provided in iMoprh. For the calculations and comparison with correlations in the literature proposed by various authors was approximated as convex triangular strut shape.
**Grey block are the measurements performed on real cast sample by CTIF studied in this work. We used these values for comparison with different correlations in literature.
*** $\partial$-Tortuosity.

- Quantify the strut shape and its cross section along the ligament axis for a wide porosity range.
- Introduction of a new parameter which takes into account different strut shapes without any fitting parameter.

### 3.3.2 Problems and investigation of correlations in case of ceramic foams

Like the section 3.3.1, the applicability of the various correlations derived for ceramic foams with the experimental findings of a few authors (Garrido et al., 2008; Grosse et al., 2009) are also compared and evaluated. In Figure 3.1-left, for a "known strut diameter" ($d_s$), the correlation of Buciuman and Kraushaar-Czarnetzki (2003) overestimates the experimental results and the errors are maximum. On the other hand, very close calculated results were obtained for the correlations of Inayat et al., (2011a). In Figure 3.1-right, a complete behavioural change is observed in the calculated results in the case of "known window diameter" ($d_w$). The errors are minimal for the correlation of Buciuman and Kraushaar-Czarnetzki (2003) while they are high for the correlation of Inayat et al., (2011a).

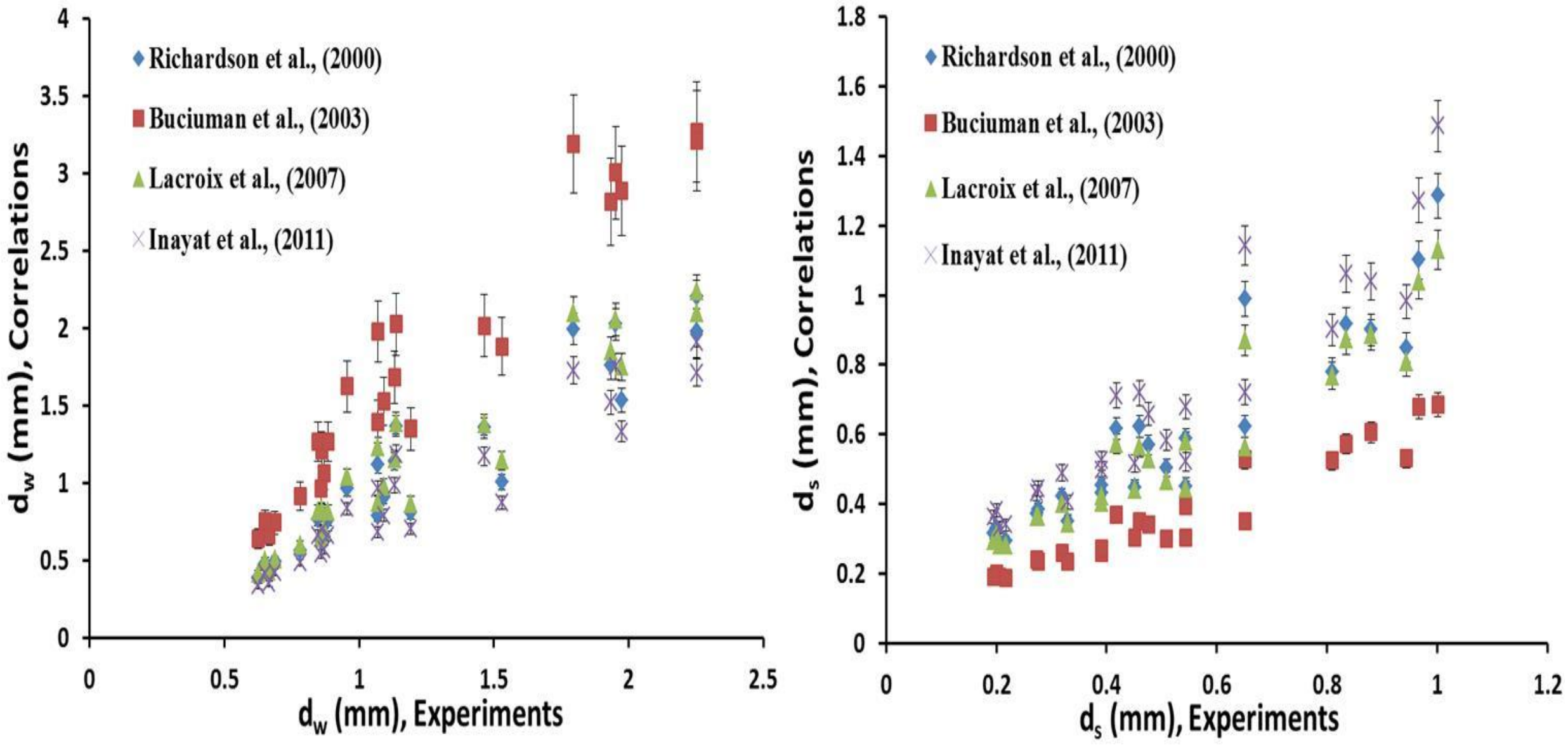

**Figure 3.1**. Left: Comparison of experimental window diameter ($d_w$) and predicted results from correlations for a "known strut diameter ($d_s$)". Right: Comparison of experimental strut diameter ($d_s$) and predicted results from correlations for a "known window diameter ($d_w$)".

Specific surface areas are also compared against the correlations derived by different authors (Richardson et al., 2000; Buciuman and Kraushaar-Czarnetzki, 2003; Lacroix et al., 2007; Grosse et al., 2009; Inayat et al., 2011a;) for both "known parameters": $d_s$ and $d_w$ in

Figure 3.2. Note that, the estimated values of specific surface area using the correlation of Richardson et al., (2000) are not presented as the calculated results have a large error range.

From the Figure 3.2, it is observed that the correlations of Grosse et al., (2009) and Inayat et al., (2011a) provide good estimates of specific surface area using Equations 3.31 and 3.32 respectively for “known window diameter” ($d_w$). Other correlations overestimate the calculated values because of their simplified geometry and geometrical models. On the other hand, it is surprising to note that all the correlations using ‘known strut diameter’ ($d_s$) overestimate the experimental values of specific surface area.

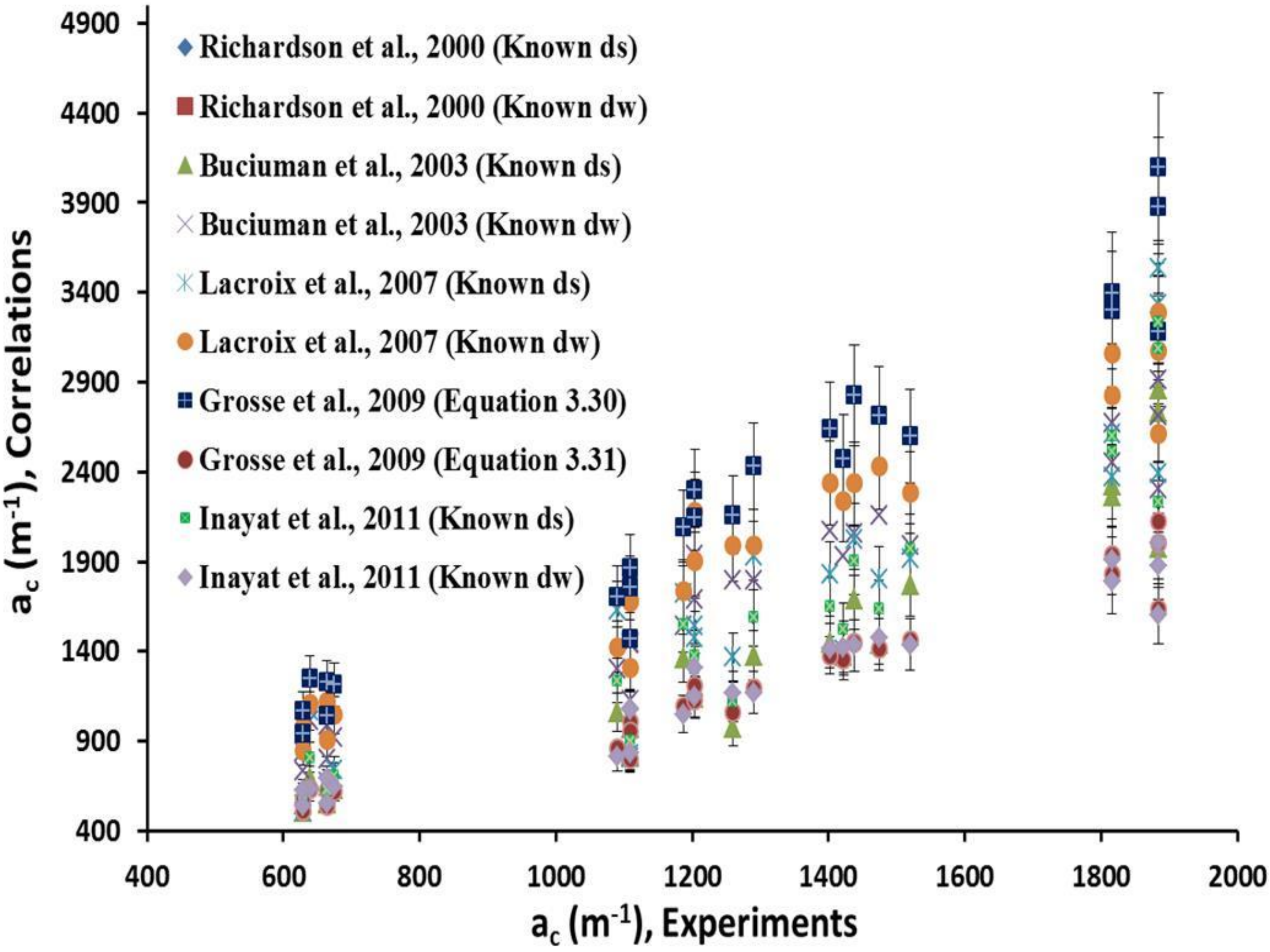

**Figure 3.2**. Comparison of experimental and calculated specific surface area ($a_c$) for “known strut diameter ($d_s$)” and “known window diameter ($d_w$)”.

The main reasons of non-adoption of the correlations are:

- Insufficient geometrical characterization of ceramic foams.
- Ill-measurements and ill-extraction of geometrical parameters.
- Hypothesis of equivalent sphere particle to relate with specific surface area.
- Use of polyhedral, cubic lattice and pentagonal dodecahedron approach.

For a better proposed model, a few remarks are suggested below:

- Better understanding of hollow strut and its measurement.
- Correct definition of open and total porosity.
- Relationship between strut and window (or pore) diameter is critical.
- Use of tetrakaidecahedron structure to maintain the consistency of results.

## 3.4 Correlation for predicting geometrical characteristics of foam matrix

The strut shape and morphology have a significant impact on the geometrical properties of foams including specific surface area. Due to the fact that foams exhibit different strut shapes at different porosities, there is no correlation proposed in the literature that could account different strut cross sections as well as low and high porosity range without any fitting parameter.

Few authors e.g. Bhattacharya et al., 2002; Inayat et al., 2011a discussed the variation in the strut shape from circular to convex or concave triangular triangle with porosity varying from low (0.70< $\varepsilon_o$ <0.90) to high porosity (0.90< $\varepsilon_o$ <0.97). The strut shape dependence with porosity actually depends on the manufacturing method employed. As discussed in the section 2.4, the foam samples studied in this work produced by casting method possess convex equilateral triangle even for a medium porosity range (0.825< $\varepsilon_o$ <0.85). It is thus, evident that porosity variation has no such impact on different strut shape formation.

This section is divided in four categories:

- Development of an analytical model for the family of circular to convex or concave triangular shape for *isotropic* metal foams.
- Development of a generalized analytical model for different strut shapes for *isotropic* metal foams.
- Development of a generalized analytical model for different strut shapes for *anisotropic* metal foams.
- Development of a generalized analytical model for *isotropic* ceramic foams.

In the first category, the family of convex or concave triangular strut shape is discussed. To accommodate the various shapes reported in the literature, a generalized correlation to take circular, convex and concave triangular strut shape into account is developed first and relationships between various geometrical properties are derived in the section 3.4.1.

In the second category, virtual *isotropic* foams of different strut shapes (modelled in CAD) are studied that give insight to tailor an optimized foam structure in order to improve heat and mass characteristics while lowering down the pressure drop. Moreover, strut shapes impact greatly on fluid flow properties which are extremely critical in the planning and designing of numerous engineering processes. With new techniques like 3-D printing or SEBM etc., open cell foams of desired strut shapes can be easily manufactured. A generalized analytical correlation has been developed in section 3.4.2 to accommodate simple and complex strut shapes to derive/predict accurate geometrical properties.

In the third category, *anisotropic* nature of foams is studied. As discussed in section 2.3, commercially available foams are slightly *anisotropic* in nature. The analytical correlation of *isotropic* foams is extended in the section 3.4.3 to predict the geometrical properties of *anisotropic* foams.

In the fourth category, an analytical correlation is developed to predict geometrical properties of ceramic foams in section 3.4.4. It is worth noting that in the present work, there is no development/fabrication of any strut shape in CAD in the case of ceramic foams. This analytical model is also an extension of analytical correlation developed in section 3.4.2 by taking hollow nature of the strut into account.

### 3.4.1 Correlation for predicting geometrical properties of *isotropic* metal foams for the family of convex or concave triangular strut shape

In this section, we investigated the impact of various geometrical parameters, their individual influences and derived various correlations to predict accurately the geometrical parameters of foam matrix.

#### 3.4.1.1 Characterization of geometrical parameters

To fully describe the strut shape and its variable cross section along the ligament axis, in fact, one needs four parameters as $a_1$, $a_2$, $a_3$ and $L$. The parameter $a_1$ controls the shape of the strut which could be either convex triangular, concave triangular or circular; parameter $a_2$ controls the size of the strut; parameter $a_3$ controls the curvature of ligament axis; and parameter $L$ is node to node length (also length of truncated octahedron in the studied model, see Figure 3.3-left) as presented by Kanaun and Tkachenko (2008) and are shown in the Figures 3.3, 3.4, 3.5 and 3.6. As the struts intersection (or node junction) is very complex to visualize and to calculate analytically, the solid volume is split into three categories namely, ligament, node and total volume of unit cell to derive a relationship between porosity and geometrical parameters. By doing so, one more parameter, $L_s$ is introduced which is length of the ligament or strut length ($L_s < L$).

The boundary of the ligament cross section is defined using Equation 3.33 (see Kanaun and Tkachenko, 2008):

$$y(\Phi, x) = R(x)\left(cos(\Phi) + \frac{cos(2\Phi)}{a_1}\right) \tag{3.33a}$$

$$z(\Phi, x) = R(x)\left(-sin(\Phi) + \frac{sin(2\Phi)}{a_1}\right) \tag{3.33b}$$

where, $\Phi$ is the angle parameter between $0 \leq \Phi < 2\pi$, (y, z) are the Cartesian coordinates in the plane of the ligament cross section and the coordinate $x$ is directed along the ligament axis.

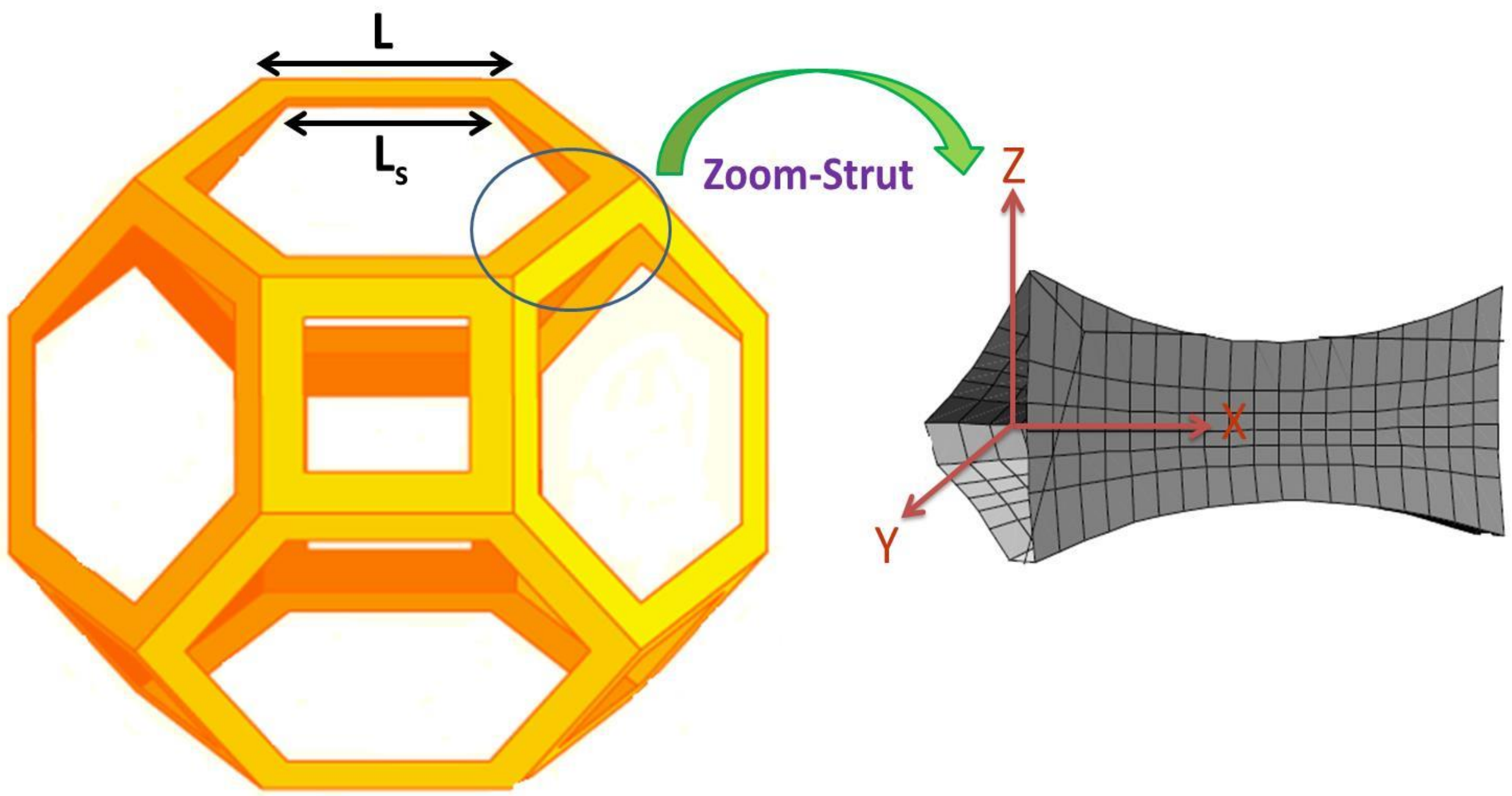

**Figure 3.3**. Truncated octahedron cell at some porosity (on left) and convex triangular shape of the strut (zoom-on right side).

The tailored strut shape designed in CAD and actually produced cast metal foam could vary in terms of strut length and sharp convex triangular shape. In this work, convex triangular strut shape was modelled in CAD and then materialized the foam geometry by casting method (see Dairon et al., 2011) as presented in Table 2.2. The analytical solution thus derived withholds convex triangular strut shape.

The function $R(x)$ that defines the ligament form along its axis (Kanaun and Tkachenko, 2008) is taken in the form of Equation 3.34:

$$R(\xi) = a_2[1 - a_3(1 - \xi^2)]\,,\, \xi = \frac{x}{L}\ ,\frac{-L}{2} < x < \frac{L}{2} \tag{3.34}$$

Equation 3.34 reflects the fact that the ligaments are thinner in the middle region than in the regions near its ends $x = \frac{-L}{2}\ ,\frac{L}{2}$ and $0 \leq a_3 < 1$.

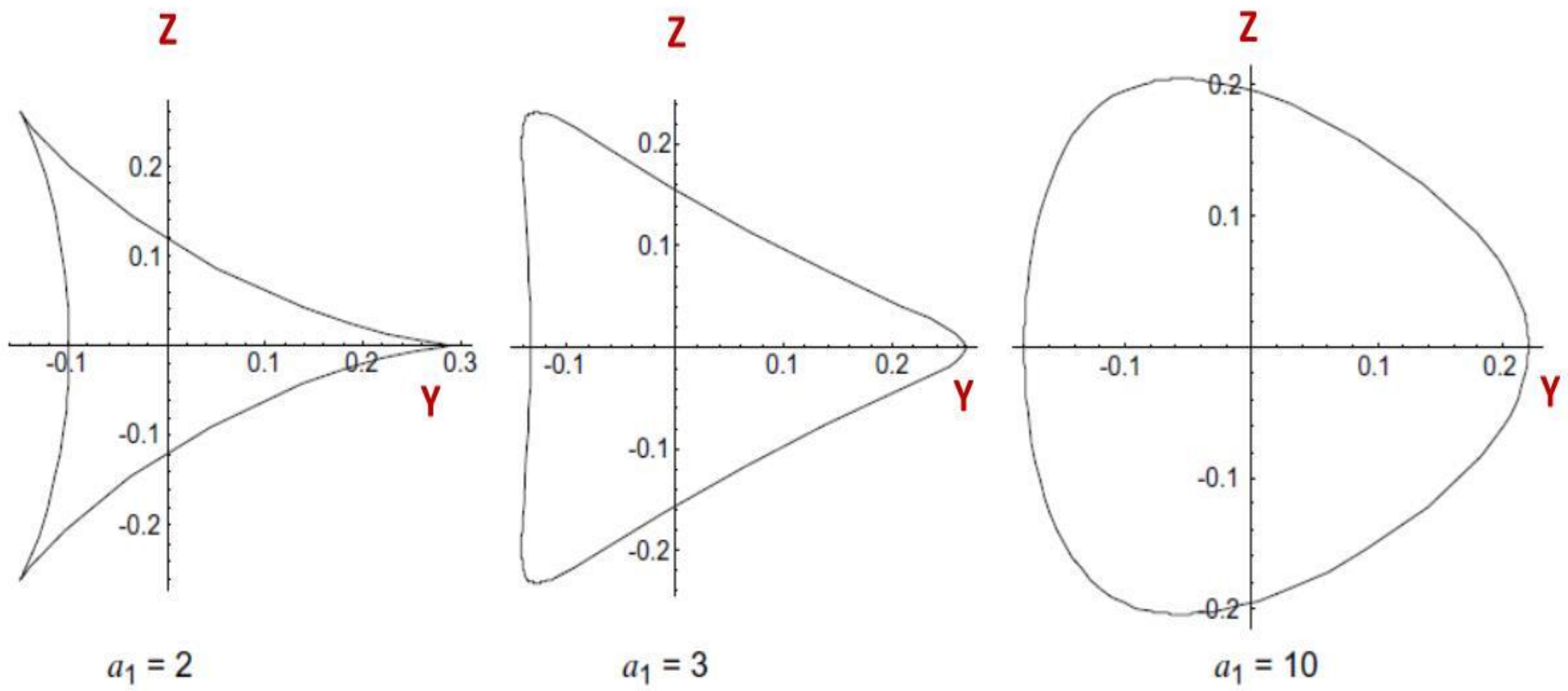

**Figure 3.4**. The area of the cross section of the ligament for different values of parameter $a_1$ in Eqs. (3.33a) and (3.33b), $a_2$=0.3 and $\xi$=0 (see Kanaun and Tkachenko, 2008). $a_1$ controls the convex or concave triangular nature of the cross section of the strut.

From Figure 3.4, it is clear that $a_1$ = 2 or 2.01 is appropriate to describe convex triangular strut shape due to the sharp edges at the vertices. In order to have convex triangular strut, one has to fix $a_1$ = 2 or 2.01 and is dimensionless. Upon increasing the value of parameter $a_1$ (2< $a_1$ <100), different forms of concave triangular strut shapes could be generated. In order to predict the parameter $a_1$ for concave triangular strut shape, it is rather difficult to quantify because of their blunt (or obtuse) edges. Inayat et al., (2011a) derived a very simple relation for a given concave triangular strut shape that was assumed to be in circumcised circle which is not the case of majority of existing foams. Moreover, for parameter $a_1$ = 100, the strut shape cross section takes circular form (see De Jaeger et al., 2011). Figure 3.5 (left) shows different convex triangular strut cross sections at different $a_2$ for $a_1$=2.01. $a_1$ <2 (or 2.01) will produce unreal convex triangular shape. Similarly, different concave triangular strut shapes at different $a_2$ for $a_1$=10 are presented in the Figure 3.5 (right).

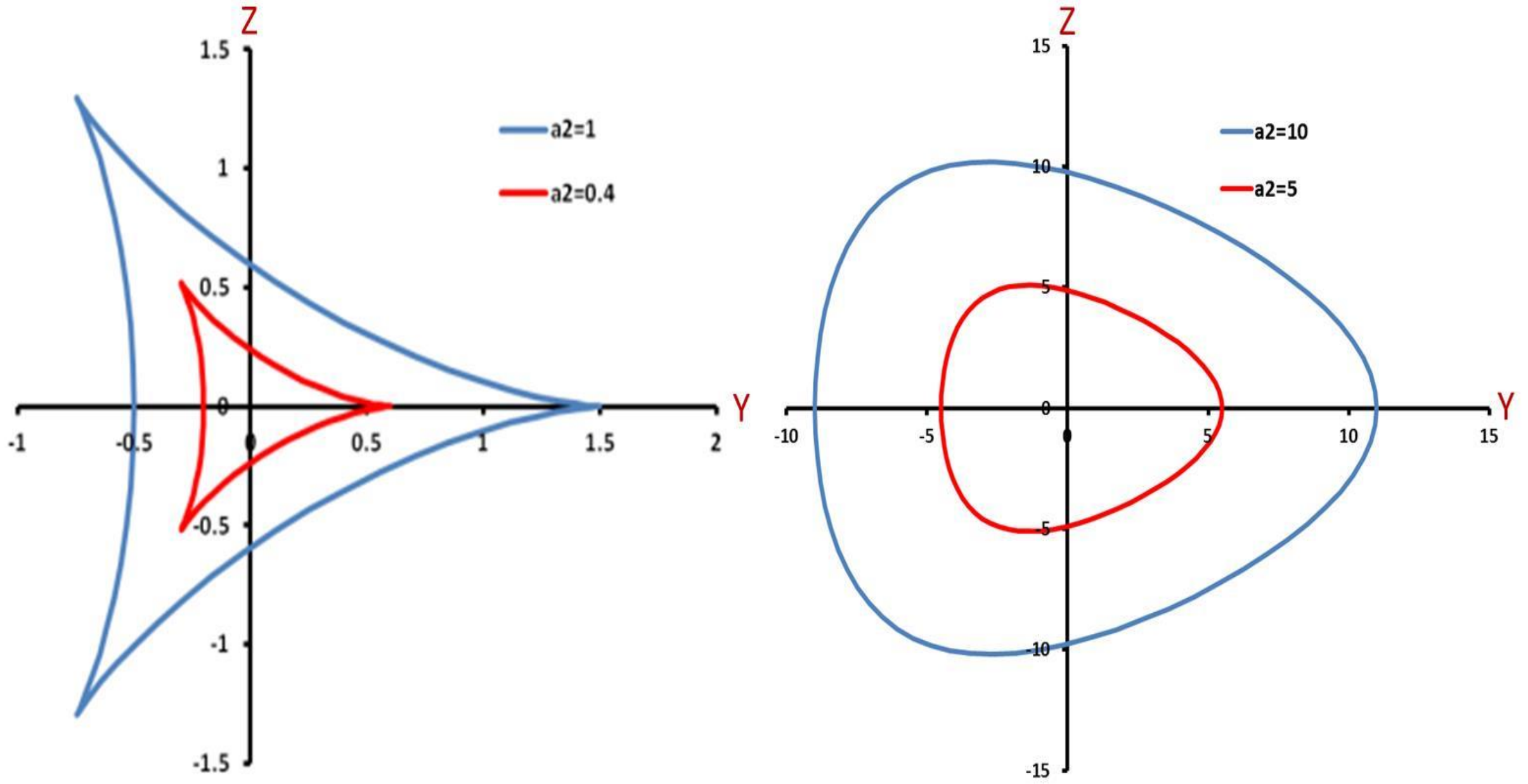

**Figure 3.5**. Left- Different convex triangular strut cross sections for $a_1$ = 2.01. Right- Different concave triangular strut cross sections for $a_1$ = 10. The parameter $a_2$ controls the size of the cross section of strut (see Kanaun and Tkachenko, 2008). Clearly, concave triangular strut shape is difficult to quantify before end to know its shape factor.

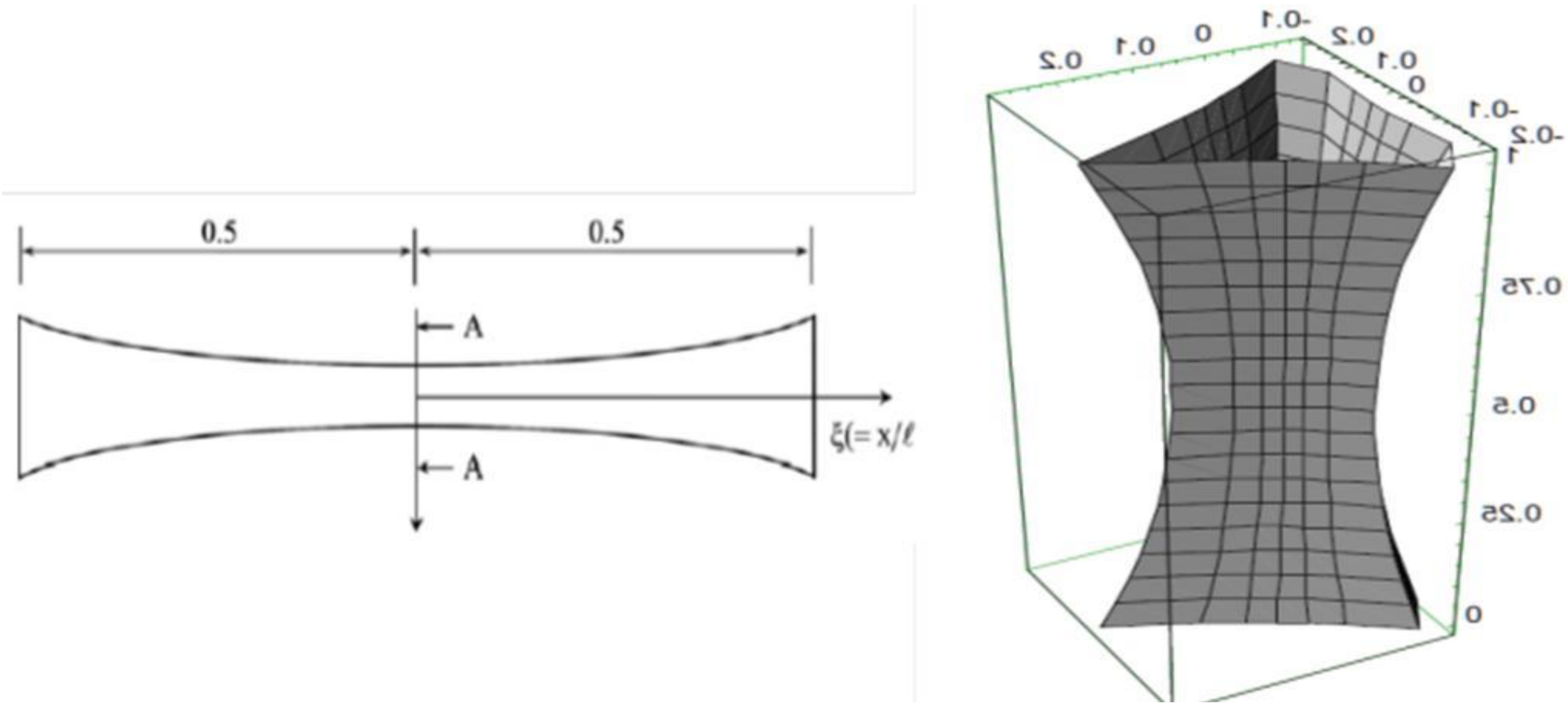

**Figure 3.6**. Left- Geometrical presentation of ligament along X-axis. Right-3-D view of ligament shape defined by Eqs. (3.33a), (3.33b) and (3.34) for $a_1 = 2.5, a_2 = 0.3$ and $a_3 = 0.4$ (see Kanaun and Tkachenko (2008). $a_3$ controls the variation of cross section size along ligament axis.

The parameter $a_3$ determines the variable cross section of the ligament having minimum surface area at the center between two nodes (see Figure 3.6) and is reported in work of Kanaun and Tkachenko (2008). These authors gave the range of $a_3$ parameter varying from 0 to 1. An apparent strut volume is calculated using Equation 3.36 and its dependence on $a_3$ is presented in Figure 3.7. It is clear that beyond, $a_3 = 0.3$, the strut surface self-intersects where the inner part of the ligament goes towards the outer side and acquires maximum volume in the center. Values of $a_3 > 0.3$ will produce unreal and unfeasible struts. For different values of $a_3$ in the range $0 \leq a_3 \leq 0.3$, one can obtain variable ligament cross sections along its axis if the porosity is known.

The proposed approximation of the strut shape and ligament nature shown in Figure 3.5 and 3.6 allows us to calculate basic geometrical characteristics of the ligament. For instance, the area $S$ of its cross section uses the mathematical formulations presented in the work of Kanaun and Tkachenko (2008):

$$S(a_1, a_2, a_3, \xi) = \pi a_2{}^2 \frac{1 + a_1{}^2}{a_1{}^2} [1 - 4a_3{}^2(1 - \xi^2)]^2 \tag{3.35}$$

and the volume of the ligament, $V_{ligament}$ is:

$$V_{ligament}(a_1, a_2, a_3) = 2\pi a_2{}^2 L_s \frac{1 + a_1{}^2}{15 a_1{}^2} (15 - 80a_3 + 128a_3{}^2) \tag{3.36}$$

The typical element of the microstructure of the idealized foam material consists of four ligaments connected at the node junction. Such an element generated using CAD model is shown in Figure 3.8.

For calculating the volume of the ligament phase in an elementary cell of the foam material, one has to take into account the volume $V_{int}$ of intersection of the ligaments in the nodes. Approximately, the volume $V_{int}$ may be evaluated as the sum of four pyramids (also referred as $V_{node}$) whose bases have the area $S_{int}$. Thus, for $S_{int}$ and $V_{int}$, we can write:

$$S_{int} = S\left(a_1, a_2, a_3, \frac{a_2}{L}\right) \tag{3.37}$$

and,

$$V_{int} = \frac{4}{3} a_2 \, S_{int} \tag{3.38}$$

The value of $S_{int}$ can be determined by using $S(a_1, a_2, a_3, \xi)$ as presented in Equation 3.35. From the above definitions of surface area, volume of ligament and volume of node at the ligaments junction, the porosity, $\varepsilon_o$ can be calculated by formula:

$$\varepsilon_o = \frac{V_f}{V_T} \ or\ \varepsilon_o = 1 - \frac{V_s}{V_T} \tag{3.39}$$

and,

$$V_T = 8\sqrt{2}\, L^3 \tag{3.40}$$

where, $V_f$ and $V_s$ are volumes of fluid and solid. $V_T$ and $L$ are total volume and length (distance between two nodes points) of the truncated octahedron respectively.

In the truncated octahedron structure (see Figure 3.3-left), there are 36 ligaments and 24 nodes but only 1/3rd of both, volume of ligament and volume of node are considered because of periodic characteristics of Kelvin cell foam.

By using Equation 3.39 and 3.40, one can relate the parameters $a_2$, $a_3$, $L_s$ and $L$ ($a_1$= 2.01, fixed as deduced from direct observation) with porosity as shown below:

$$\varepsilon_o = 1 - \frac{1}{3}\left(\frac{36V_{ligament} + 24V_{node}}{V_T}\right) \tag{3.41}$$

$$\varepsilon_o = 1 - \frac{\pi\sqrt{2}}{30}\left(\frac{1 + {a_1}^2}{{a_1}^2}\right)[3\alpha^2\beta\{15 - 80a_3 + 128{a_3}^2\} + 20\alpha^3\{1 - 4{a_3}^2(1 - \alpha^2)\}^2] \tag{3.42}$$

where, $\alpha = \frac{a_2}{L}$ and $\beta = \frac{L_s}{L}$

Equation 3.42 gives a generic relation of porosity as a function of $a_1$, $\alpha$, $\beta$ and $a_3$. On the other hand, it could be convenient to get approximate values of $\alpha$ and $\beta$ as a function of $\varepsilon_o$, $a_1$ and $a_3$. This approach can be used to determine all the geometrical properties if full set of parameters are not known. At the junction, we can approximate the node by using geometrical interpretation (see Figure 3.8):

$$a_2 + L_s = L \Rightarrow \alpha + \beta = 1 \tag{3.43}$$

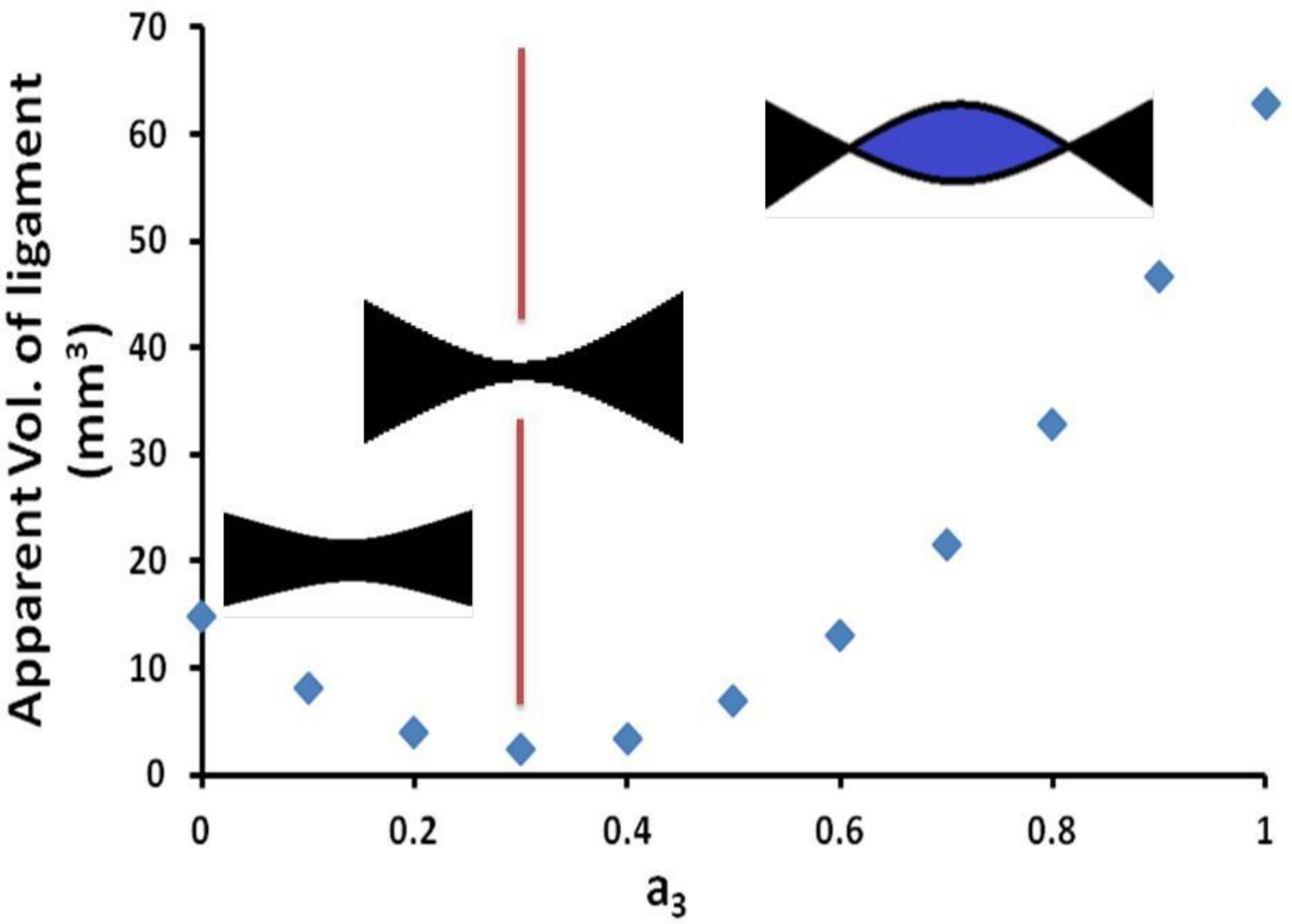

**Figure 3.7**. Apparent volume variation with $a_3$ and the strut shape formation (see Kanaun and Tkachenko, 2008).

The above Equation 3.43 is derived from the approximated geometry at the node junction of the ligaments. On solving the Equations 3.42 and 3.43 simultaneously, one can correlate $\alpha$ and $\beta$ with porosity, $\varepsilon_o$ and parameter, $a_3$. Based on that, we can rewrite the relationship between $\alpha$, $\beta$, $a_3$ and $\varepsilon_o$ ($a_1$=2 or 2.01) as presented in the Figure 3.9. In Figure 3.9 (left), the relation between $\alpha$ and $\varepsilon_o$ is clearly following a power law and can be expressed as:

$$\frac{a_2}{L} = \alpha = \sigma(1-\varepsilon_o)^{\eta} \qquad (3.44a)$$

Similarly, using Equation 3.43 and 3.44a, we can draw the correlation between $\beta$ and $\varepsilon_o$ (see Figure 3.9-right) in Equation 3.44b as:

$$\frac{L_s}{L} = \beta = 1 - \sigma(1-\varepsilon_o)^{\eta} \qquad (3.44b)$$

where, $\sigma$ and $\eta$ are the parameters which depend on $a_3$ and $\varepsilon_o$.

The parameter $a_2$ controls the size of strut cross section and depends on porosity as shown in the Figure 3.5. $a_2$ is related to strut diameter, $d_s$ (data obtained from iMorph) and is given by:

$$d_s = 4a_2 \tag{3.45}$$

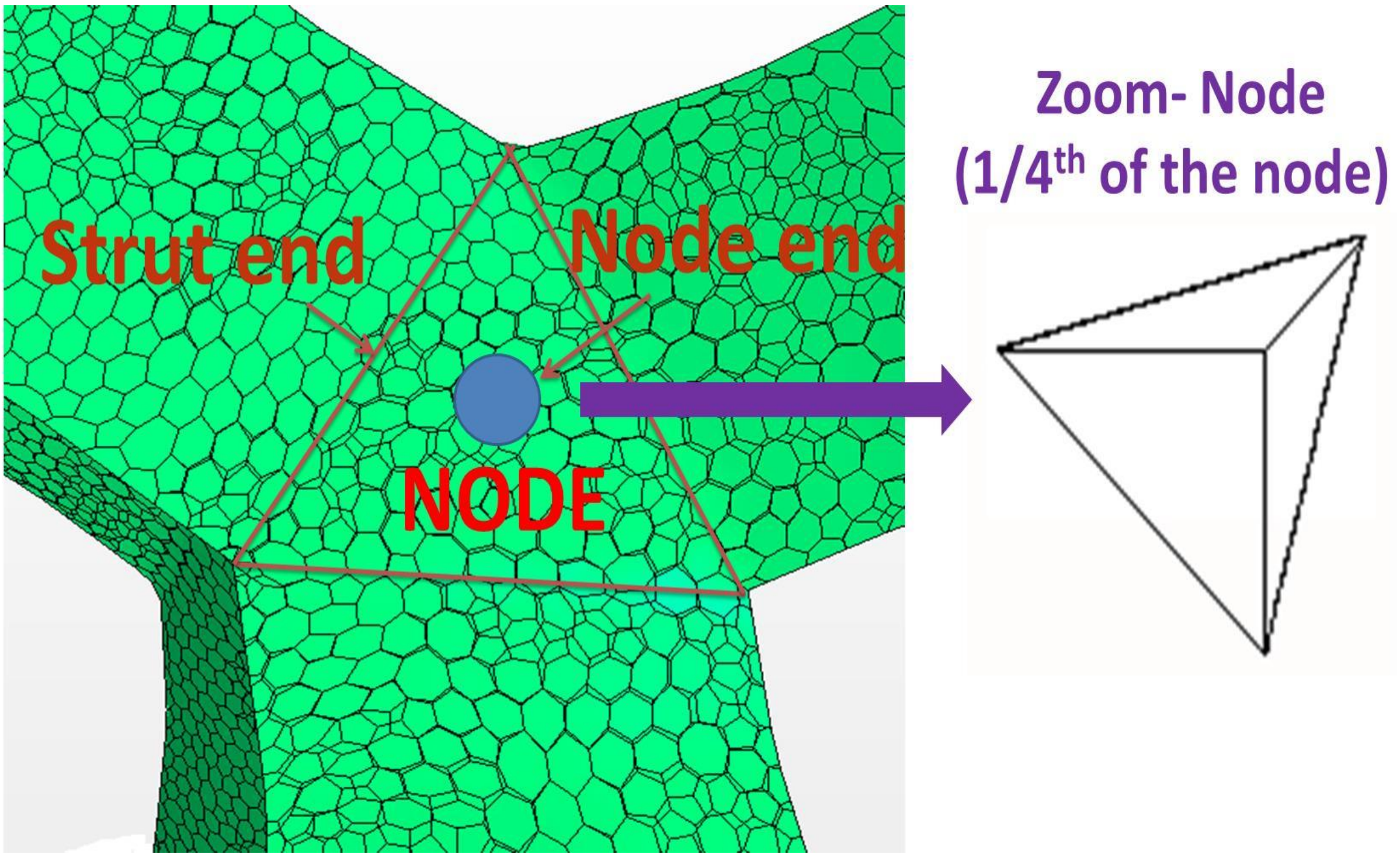

**Figure 3.8**. A typical node of the foam structure. $S_{int}$ is the surface area at the intersection of the node. $V_{int}$ corresponds to the volume of the node at intersection of four struts (Zoom-node represents one pyramid here but for calculation, four pyramids are accounted to determine a node of four struts).

Various authors (e.g. Garrido et al., 2008; Grosse et al., 2009) estimated the pore diameter as the mean diameter of square and hexagon openings of the foam structure. The opening areas of square ($a_{sw}$) and hexagon ($a_{hw}$) faces of the foam structure were measured (see Table 2.3). An approximate measured value of pore diameter ($d_p$) is estimated by the measurements of the opening window areas of six square and eight hexagon faces (see Figure 3.3- left) using Equation 3.46 as:

$$d_p = \sqrt{\frac{8{d_{ph}}^2 + 6{d_{ps}}^2}{14}} \tag{3.46}$$

where, $d_{ph}$ and $d_{ps}$ are the equivalent pore diameters of same circle area of hexagon and square faces respectively.

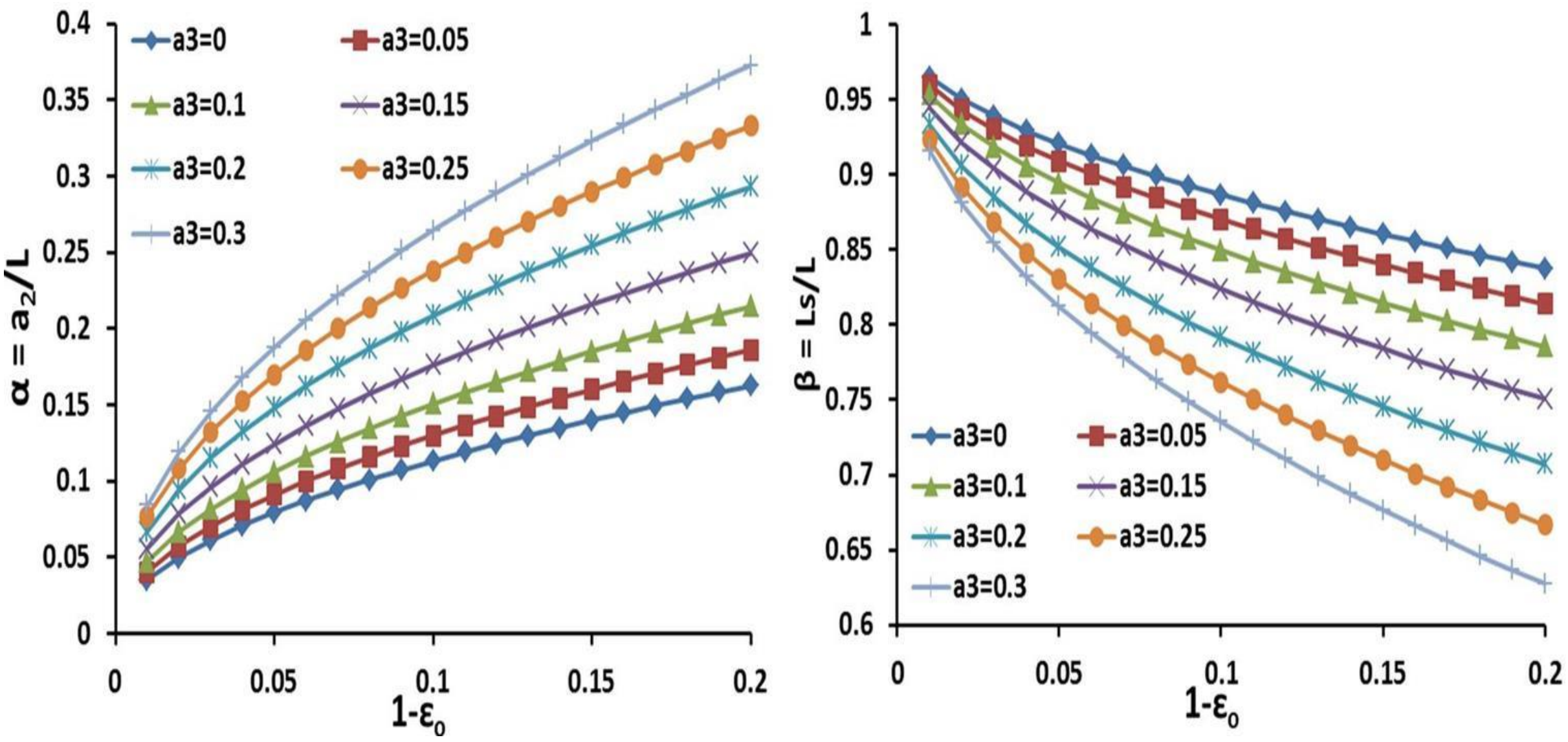

**Figure 3.9**. Left- Plot of $\alpha$ (dimensionless) vs. $1-\varepsilon_o$. Right- Plot of $\beta$ (dimensionless) vs. $1-\varepsilon_o$.

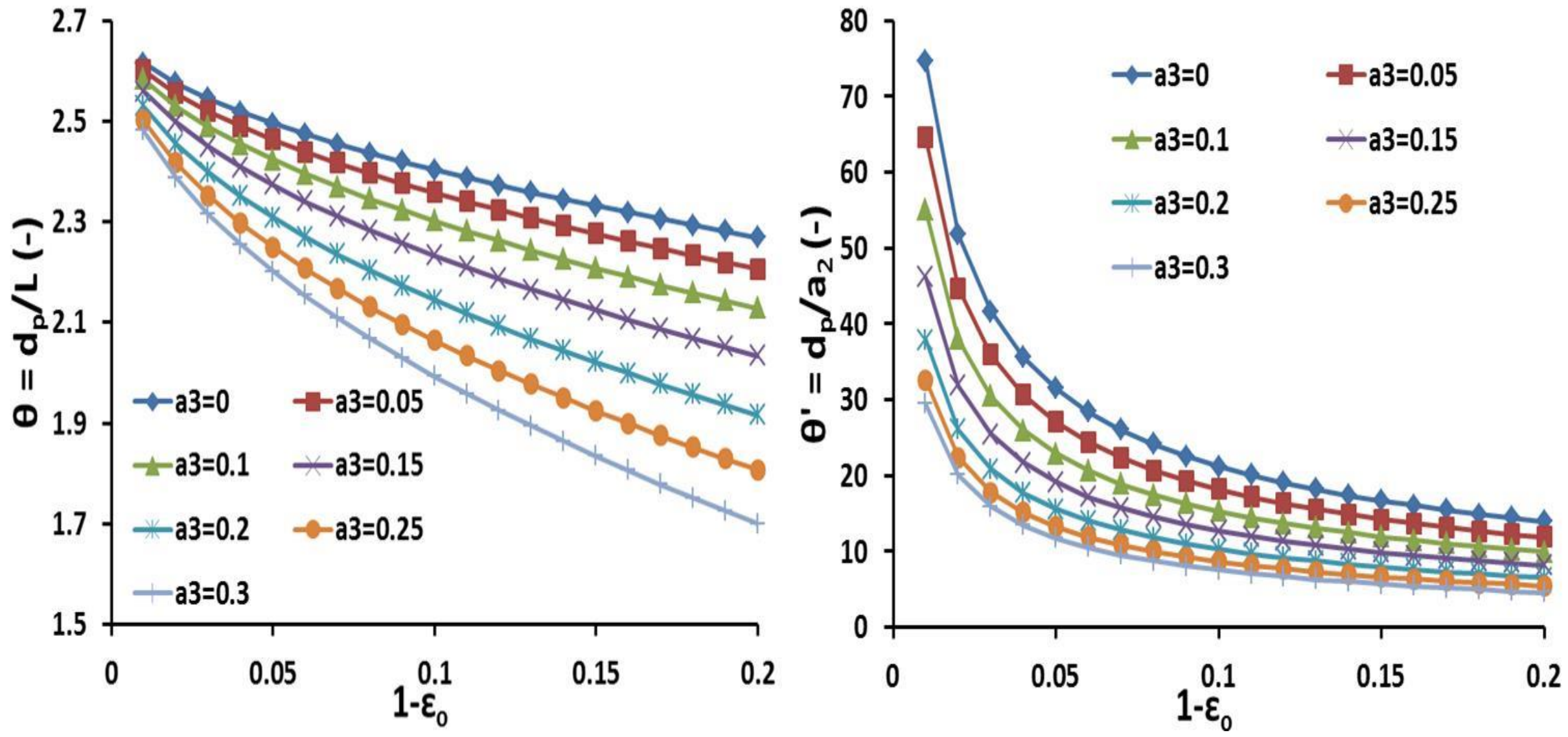

**Figure 3.10**. Left- Plot of $\theta$ (dimensionless) vs. $1-\varepsilon_o$. $\theta$ is function of $\beta$ (or $L_s$) and $a_3$. Right: Plot of $\theta'$ (dimensionless) vs. $1-\varepsilon_o$.

Upon simplification, we can rewrite the above Equation 3.46 as:

$$d_p = \sqrt{\frac{8\left(\frac{6\sqrt{3}{L_s}^2}{\pi}\right)+6\left(\frac{4{L_s}^2}{\pi}\right)}{14}} = 4L_s\sqrt{\frac{3\sqrt{3}+1.5}{14\pi}} \tag{3.47}$$

In dimensionless form, we introduce $\theta$ (=$d_p/L$) in the Equation 3.48 as:

$$\theta = \frac{d_p}{L} = 4\beta\sqrt{\frac{3\sqrt{3}+1.5}{14\pi}} \tag{3.48}$$

Similarly, $d_p$ and $a_2$ can be easily related and is given by the Equation 3.49 as:

$$\theta' = \frac{d_p}{a_2} = \frac{\theta}{\alpha} \tag{3.49}$$

The plots of $\theta$ and $\theta'$ against porosity ($\varepsilon_o$) is shown in Figure 3.10. It is clearly seen in Figure 3.10 (left) that the curve follows the same behavior as of $\beta$ (see Figure 3.9-right). It is due to the fact that $d_p$ is clearly a function of $L_s$ (Equation 3.48) and follows a decreasing trend with decrease in porosity. On the other hand, $\theta'$ is highest for $a_3$=0 and lowest for $a_3$=0.3 and increases with increase in porosity and follows an inverse function of $a_2$ in Figure 3.10 (right).

### 3.4.1.2 Specific surface area

The specific surface area is simply deduced from surface area of solid included inside the octahedron volume:

$$a_c = \frac{\mathrm{S_{ligaments}} + \mathrm{S_{nodes}}}{\text{Volume of the truncated octahedron}} \tag{3.50}$$

Perimeter of the strut cross section (see De Jaeger et al., 2011) is given by:

$$ds = \sqrt{\left(\frac{dy}{d\Phi}\right)^2 + \left(\frac{dz}{d\Phi}\right)^2}\,.d\Phi \tag{3.51}$$

Surface area of one ligament is obtained by integrating ds over the ligament length from $-L_s/2 \le \xi \le L_s/2$:

$$S_{ligament} = \int_{-L_s/2}^{L_s/2} R(x)\,dx \int_0^{2\pi} \sqrt{1 - \frac{4}{a_1} cos3\Phi + \frac{4}{{a_1}^2}}\,.d\Phi \tag{3.52}$$

On substituting the values of $R(x)$ from Equation 3.33 for $0 \le \Phi < 2\pi$, the surface area of the one ligament is calculated as:

$$S_{ligament} = L_s.a_2\left(1 - \frac{3}{4}a_3\right)H(a_1) \tag{3.53}$$

where, $H(a_1) = \frac{2((-2+a_1)EllipticE[-\frac{8a_1}{(-2+a_1)^2}]+(2+a_1)EllipticE[\frac{8a_1}{(2+a_1)^2}])}{a_1}$ is a 1st order *ellipticE* integral function.

Similarly, surface area of one node is calculated by using Equation 3.37:

$$S_{node} = \pi {a_2}^2 \frac{1 + {a_1}^2}{{a_1}^2} \left[1 - 4{a_3}^2 \left\{1 - \left(\frac{a_2}{L}\right)^2\right\}\right]^2 \tag{3.54}$$

By substituting 1/3rd surface areas of 36 ligaments and 24 nodes having four faces obtained in Equation 3.53 and 3.54 and volume of truncated octahedron (Equation 3.40) in Equation 3.50, we get:

$$a_c = \frac{1}{L\sqrt{2}} \left[\left\{1.5\alpha\beta \left(1 - \frac{3}{4}a_3\right) H(a_1)\right\} + \left\{4\pi\alpha^2 \left(\frac{1 + {a_1}^2}{{a_1}^2}\right) \{1 - 4{a_3}^2(1 - \alpha^2)\}^2\right\}\right] \tag{3.55}$$

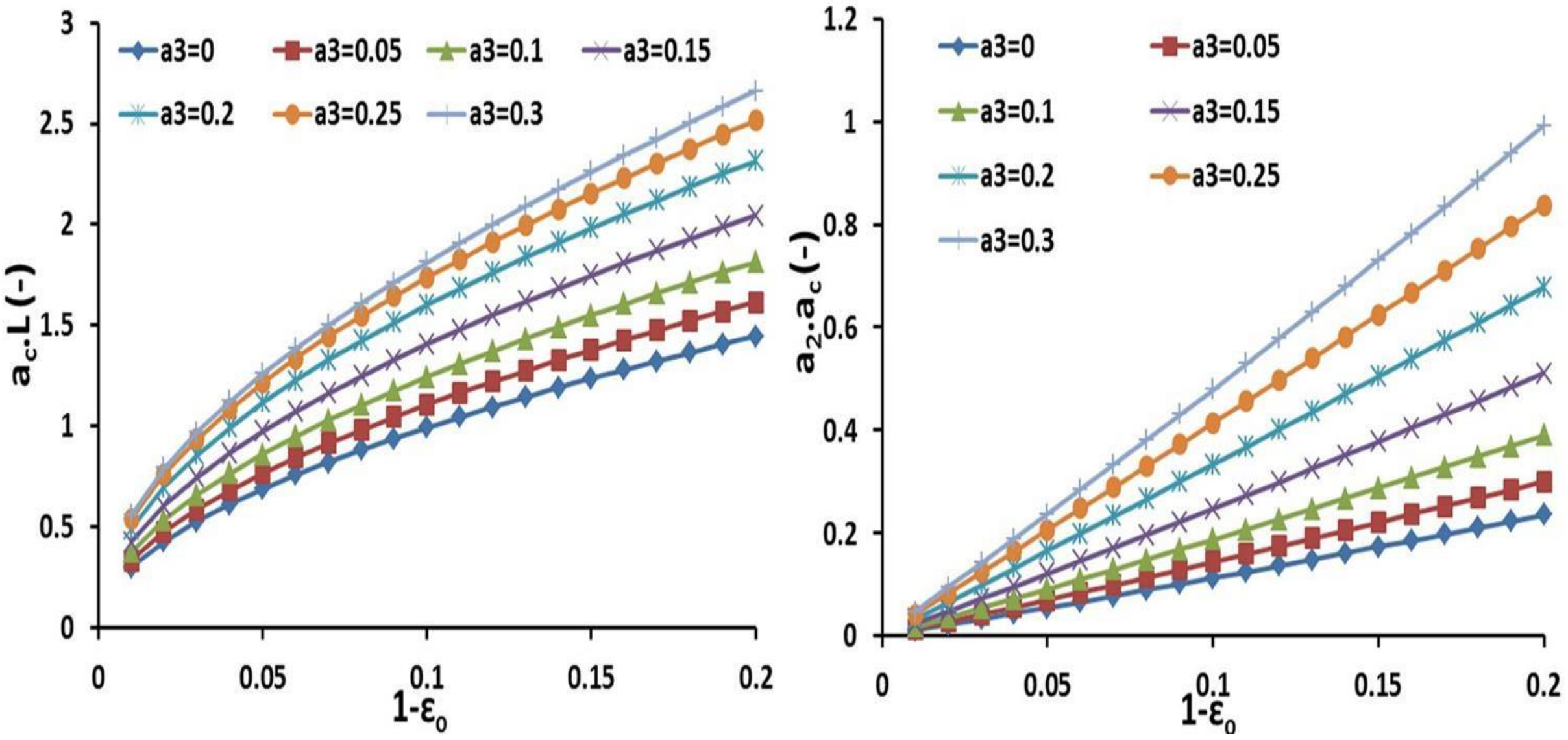

**Figure 3.11**. Left- Plot of $a_c.L$ (dimensionless) vs. $1 - \varepsilon_o$. Right- Plot of $a_2.a_c$ (dimensionless) vs. $1 - \varepsilon_o$.

For a given unit cell, node to node length is fixed. Changing $a_3$ has direct impact on $a_2$ and it will increase if $a_3$ increases which clarifies that there will be more accumulation of matter at the nodes than at the centre of the strut which in turn, responsible for increase in specific surface area by keeping the same porosity. A dimensionless curve $a_c.L$ with porosity ($\varepsilon_o$) is also presented in Figure 3.11 (left). One can determine $a_c$ or $L$ if any of these two parameters are known for a given porosity. Similarly, a dimensionless curve (see Figure 3.11-right) which relates $a_2$, $a_c$ and $\varepsilon_o$ is presented that follows a linear behaviour. One can easily determine all other geometrical parameters if any of the two geometrical quantities are known using the above established correlations.

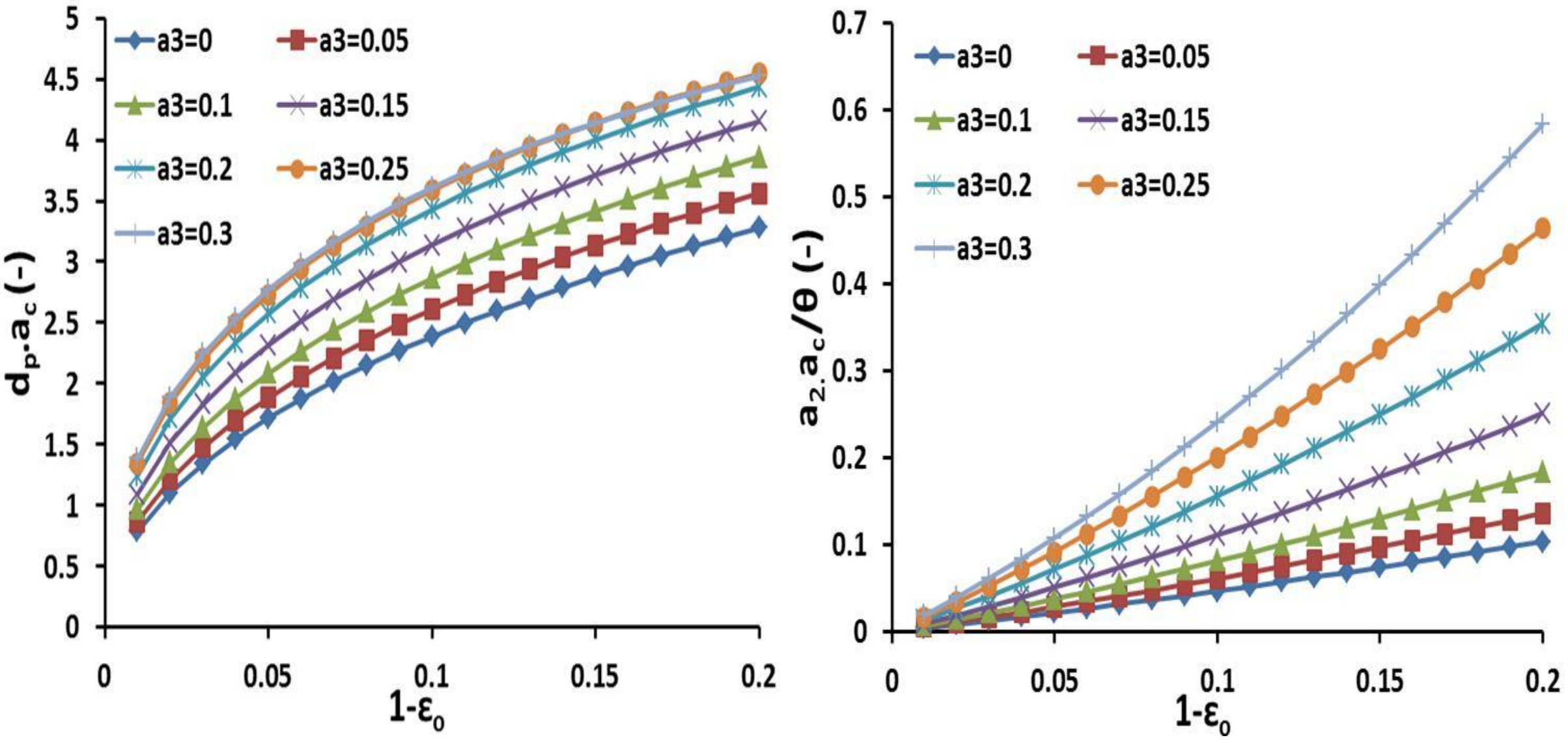

**Figure 3.12**. Left- Plot of $d_p.a_c$ (dimensionless) vs. $1-\varepsilon_o$. Right- Plot of $a_2.a_c/\theta$ (dimensionless) vs. $1-\varepsilon_o$.

Two dimensionless curves relating $d_p.a_c$ and $a_2.a_c/\theta$ with porosity ($\varepsilon_o$) are also presented in Figure 3.12. In Figure 3.12 (left), it is clear that $d_p.a_c$ decreases with increase in porosity. However, increase in $a_3$ impacts $d_p.a_c$ and follows the power law. The difference in the values between two consecutive $d_p.a_c$ starts to decrease when $a_3$ increases and it is observed that $d_p.a_c$ possess approximately the same values for $a_3$=0.25 and $a_3$=0.30. This could be due to the fact that $d_p$ increases with increase in porosity (Figure 3.10-left) while $a_c$ decreases with increase in porosity (see Figure 3.11-left) and thus, a combination of both i.e. $d_p.a_c$ acquires increasing and decreasing behavior where the values are similar for high value of $a_3$. In Figure 3.12 (right), $a_2.a_c/\theta$ increases with decrease in porosity and follows a quadratic polynomial function with porosity. This curve is critical for the cases where individual information about geometrical parameters are unknown e.g., for a given application where a coupled constraint of geometrical and hydraulic properties ($a_2.a_c$) is known, it would be easier to optimize the curvature of the ligament and specific surface area in order to increase the performance of the system. These curves are also important to choose a variable ligament along its axis as a geometrical constraint for better fluid flow and heat transfer applications.

Using these curves, one can characterize all the geometrical parameters of the foam structure for the family of triangular strut shape. The derived correlations presented above allow foams to be tailored according to different specifications. In this way, foams with desired properties can be made to meet the needs of different engineering applications.

### 3.4.2 Correlations for predicting geometrical properties of *isotropic* metal foams for different strut shapes

To the best of our knowledge, none of the studies have been carried out that took different strut cross sections like circular, equilateral triangle, diamond (double equilateral triangle), square, rotated square, hexagon, rotated hexagon and star into account. In the chapter 2, the possibility to produce foam structures with such cross sections using 3-D printing or rapid prototyping or SEBM (see also Inayat et al., 2011a) was shown. In the literature, all correlations predict pretty well the reasonable estimates of geometrical properties for a given set of foams but they fail for other type of foam samples. The reasons of their failure are because of not enough input parameters, supposed relationship between parameters and inaccessibility to wide range of foam parameters e.g. porosity of commercially materialized foams (that represent the same global structure) is nearly fixed ($\varepsilon_o$~90% ±3%).

#### 3.4.2.1 Characterisation of geometrical parameters

A generalized correlation to predict various geometrical characteristics of foam matrices is developed in this section for different strut cross sections that were modelled in CAD and already presented in section 2.5.

The node junction at different porosities of different strut shapes possesses a complex shape and is difficult to visualize. To make our analytical approach clear and user-friendly, the node shape at the junction is approximated because ligaments of different strut shapes intersect differently at the junction. Moreover, it is very difficult to derive exact calculation for each strut shape. As circular strut shape is easy to visualize at the node and do not possess complex geometry compared to other strut shapes, an equivalent radius, $R_{eq}$ is defined for each strut shape and thus, different $R_{eq}$ for different strut shapes are provided in Table 3.6. Note that, modelling of the foam structure in CAD was performed for constant ligament shape along its axis in this work.

For each porosity and strut shape, an equivalent radius ($R_{eq}$) was assumed which is the radius of the circle of same area than the strut cross section. Obviously, for a given $R_{eq}$, node volume is the same and is independent of any strut shape. It is the most important hypothesis in the analytical derivation of correlation.

In order to provide an approximate analytical solution, $L_s$ as strut length (without considering node points) and $L$ as distance between two nodes (or length of solid truncated octahedron edge) were defined as shown in Figure 2.15. For any strut shape, an equivalent circular strut shape of radius, $R_{eq}$ of the same strut cross section was considered and then its characteristic dimensional dependence was deduced (see Table 3.6).

**Table 3.6**. Representation of characteristic length, symbol and equivalent radius of various strut shapes.

| Shapes | Characteristic Length | Characteristic Symbol | Equivalent Radius |
|---|---|---|---|
| Circular | Radius | $R_c$ | $R_{eq} = R_c$ |
| Equilateral Triangle | Length of the side | $A_t$ | $R_{eq} = A_t.\sqrt{\sqrt{3}/4\pi}$ |
| Square | Length of the side | $A_s$ | $R_{eq} = A_s/\sqrt{\pi}$ |
| Rotated Square (at $45^0$) | Length of the side | $A_{rs}$ | $R_{eq} = A_{rs}/\sqrt{\pi}$ |
| Diamond | Length of the side | $A_{det}$ | $R_{eq} = A_{det}.\sqrt{\sqrt{3}/2\pi}$ |
| Hexagon | Length of the side | $A_h$ | $R_{eq} = A_h.\sqrt{3\sqrt{3}/2\pi}$ |
| Rotated Hexagon (at $90^0$) | Length of the side | $A_{rh}$ | $R_{eq} = A_{rh}.\sqrt{3\sqrt{3}/2\pi}$ |
| Star (regular Hexagram) | Length of the side | $A_{st}$ | $R_{eq} = A_{st}.\sqrt{3\sqrt{3}/\pi}$ |

The node volume calculation was chosen to base on the formulation given by Kanaun and Tkachenko (2008). Volume of node at the junction of four struts of an equivalent circular strut shape is given as (see Figure 3.13):

$$V_{node} = \frac{4}{3}\pi {R_{eq}}^3 \tag{3.56}$$

Volume of the ligament of an equivalent circular strut shape is given as:

$$V_{ligament} = \pi {R_{eq}}^2 L_s \tag{3.57}$$

At the ligaments intersection, the node can be approximated by using geometrical interpretation as (see Figure 3.13):

$$1.6R_{eq} + L_s = L \tag{3.58}$$

In dimensionless form, we can rewrite Equation 3.58 as:

$$1.6\alpha_{eq} + \beta = 1 \tag{3.59}$$

where, $\alpha_{eq} = \frac{R_{eq}}{L}$ and $\beta = \frac{L_s}{L}$

Total volume of the truncated octahedron is given as:

$$V_T = 8\sqrt{2}L^3 \tag{3.60}$$

In a truncated octahedron structure (see Figure 2.15), there are 36 ligaments and 24 nodes but only 1/3rd of both, volume of ligament and volume of node are included in the unit periodic cell.

For a periodic Kelvin like cell foam in a unit cell, solid volume $V_s$ is given as:

$$V_s = \frac{1}{3}\left(36V_{ligament} + 24.V_{node}\right) \tag{3.61}$$

Porosity of a porous medium is given as:

$$\varepsilon_o = \frac{1 - V_s}{V_T} \tag{3.62}$$

Equations 3.60, 3.61 and 3.62 represent a general methodology to evaluate geometrical properties of an equivalent circular strut shape of a foam structure. One of the studied shapes is shown here, say, for circular strut shape, $R_{eq} = R_c$ and the relation between geometrical parameters and porosity is given as:

$$\varepsilon_o = \frac{1 - \frac{1}{3}\left(36\pi {R_c}^2 L_s + 24.\frac{4}{3}\pi {R_c}^3\right)}{8\sqrt{2}L^3} \Rightarrow 12\pi{\alpha_c}^2\beta + \frac{32}{3}\pi{\alpha_c}^3 = 8\sqrt{2}(1 - \varepsilon_o) \tag{3.63}$$

where, $\alpha_c = \frac{R_c}{L}$ and $\beta = \frac{L_s}{L}$

Note that, $\alpha_c$ is ratio of strut radius to node length where subscript $c$ represents circular strut shape. Moreover, $\beta$ is found to be independent of the strut shape.

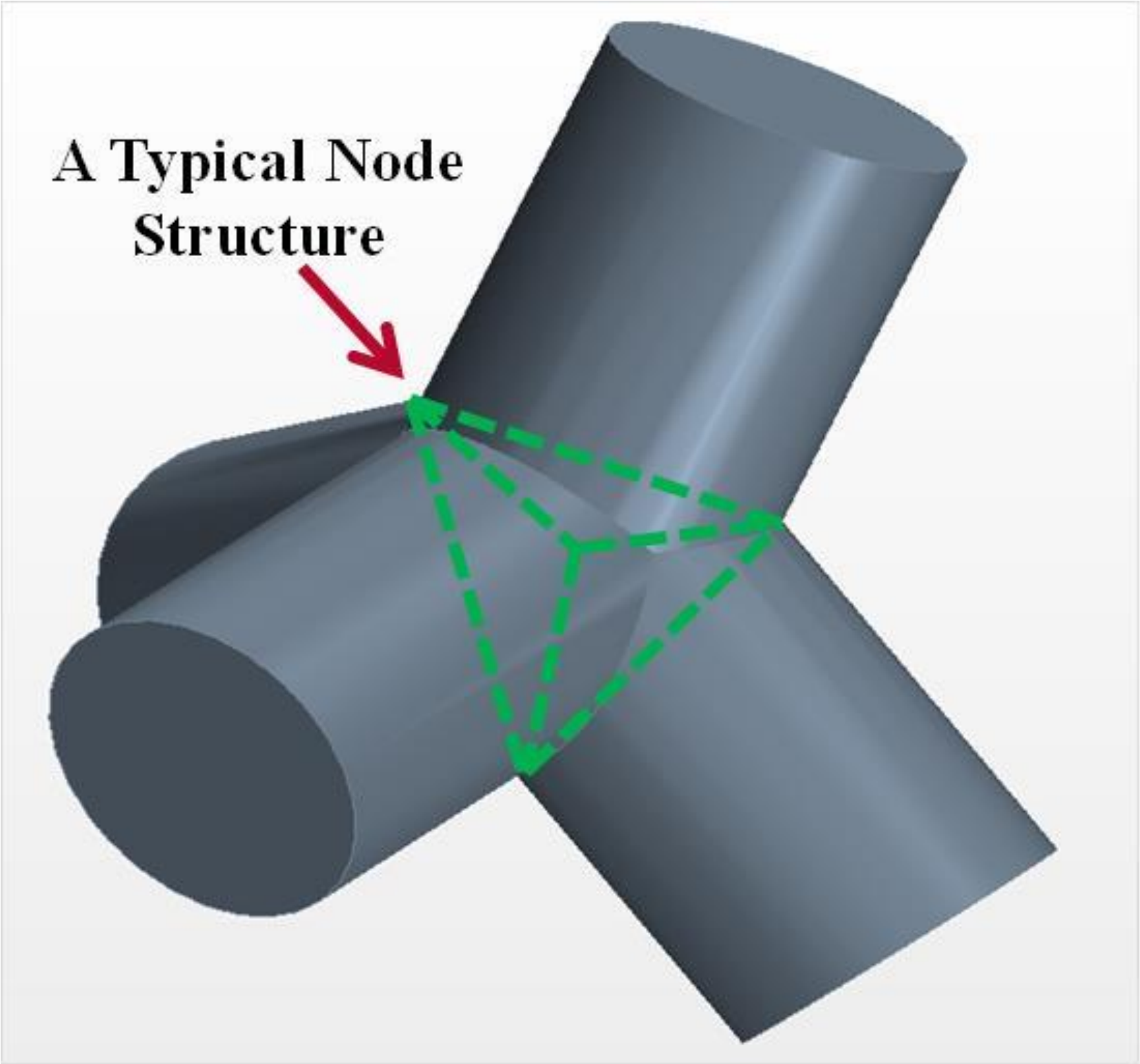

**Figure 3.13.** A typical node of foam structure. The four faces of a pyramid are shown which are taken into consideration in calculating volume of the node. The face of the nodes changes with the strut shape. Four struts of circular shape at node are shown which is approximated as triangular pyramid.

For the other strut shapes, the relation between porosity and geometrical parameters is presented in Appendix C. Equation 3.63 gives a generic relation of porosity as a function of geometrical parameters.

Equations 3.59 and 3.63 could be combined to get approximate values of $\alpha_{eq}$ and $\beta$ as a function of $\varepsilon_o$. This approach can used to determine all the geometrical properties if full set of geometrical parameters are not known. Different values of $\alpha_{eq}$ and $\beta$ for different strut shapes are plotted which clearly follow a power law as shown in Figure 3.14 (left and right) and can be expressed as:

$$\alpha_{eq} = \Lambda_1(1-\varepsilon_o)^{\Lambda_2} \quad (3.64)$$

$$\beta = 1 - 1.6\,\Lambda_1(1-\varepsilon_o)^{\Lambda_2} \quad (3.65)$$

where, $\Lambda_1$ and $\Lambda_2$ are the parameters that depend only on strut shape and also on rotation but with less impact (as the node shape change a little bit with rotation, the parameters $\Lambda_1$and $\Lambda_2$ are influenced).

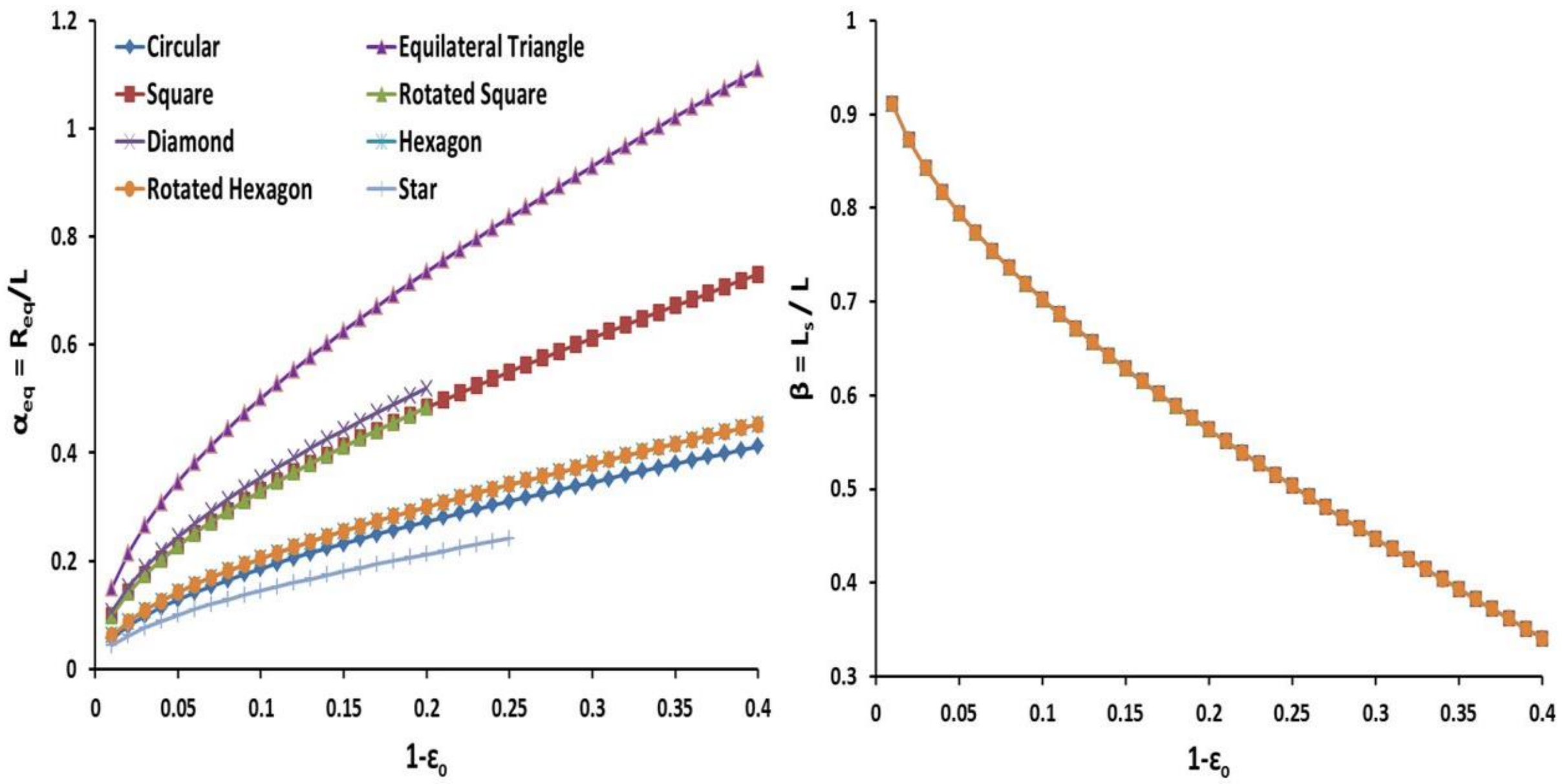

**Figure 3.14.** Left- Plot of $\alpha_{eq}$ (dimensionless) vs $1-\varepsilon_o$. $\alpha_{eq}$ for different strut shapes is limited with respect to porosity as per construction methodology. Right- Plot of $\beta$ (dimensionless) vs $1-\varepsilon_o$ (A unique curve is obtained for all the strut shapes due to the hypothesis made).

From Figure 3.14 (left), it is clearly seen that the $\alpha_{eq}$ of different strut cross sections follows the same trend of power law with porosity where the exponent of power law is always less than 1. There are nearly three groups of $\alpha_{eq}$ that mainly depend on strut shape. One group which has the highest $\alpha_{eq}$ is of equilateral triangular shape followed by moderate $\alpha_{eq}$ for diamond, square and rotated square strut shapes in the second group. The lowest $\alpha_{eq}$ is obtained for hexagon, rotated hexagon, circular and star strut shapes in the third group. In Figure 3.14 (right), a unique curve of $\beta$ is obtained for all the strut shapes. It is mainly due to the hypothesis that all the foam samples possess same node volume irrespective of the strut shape.

Pore diameter ($d_p$) is calculated in the same way as presented in section 3.4.1.1 by taking the mean diameter of six square and eight hexagon window openings.

$d_p$ and $R_{eq}$ can be easily related and is given by the Equation 3.66 as:

$$\theta' = \frac{d_p}{R_{eq}} = \frac{\theta}{\alpha_{eq}} \tag{3.66}$$

These dimensionless correlations are extremely useful in a way that any of the measured geometrical quantities leads to predict other geometrical properties simultaneously.

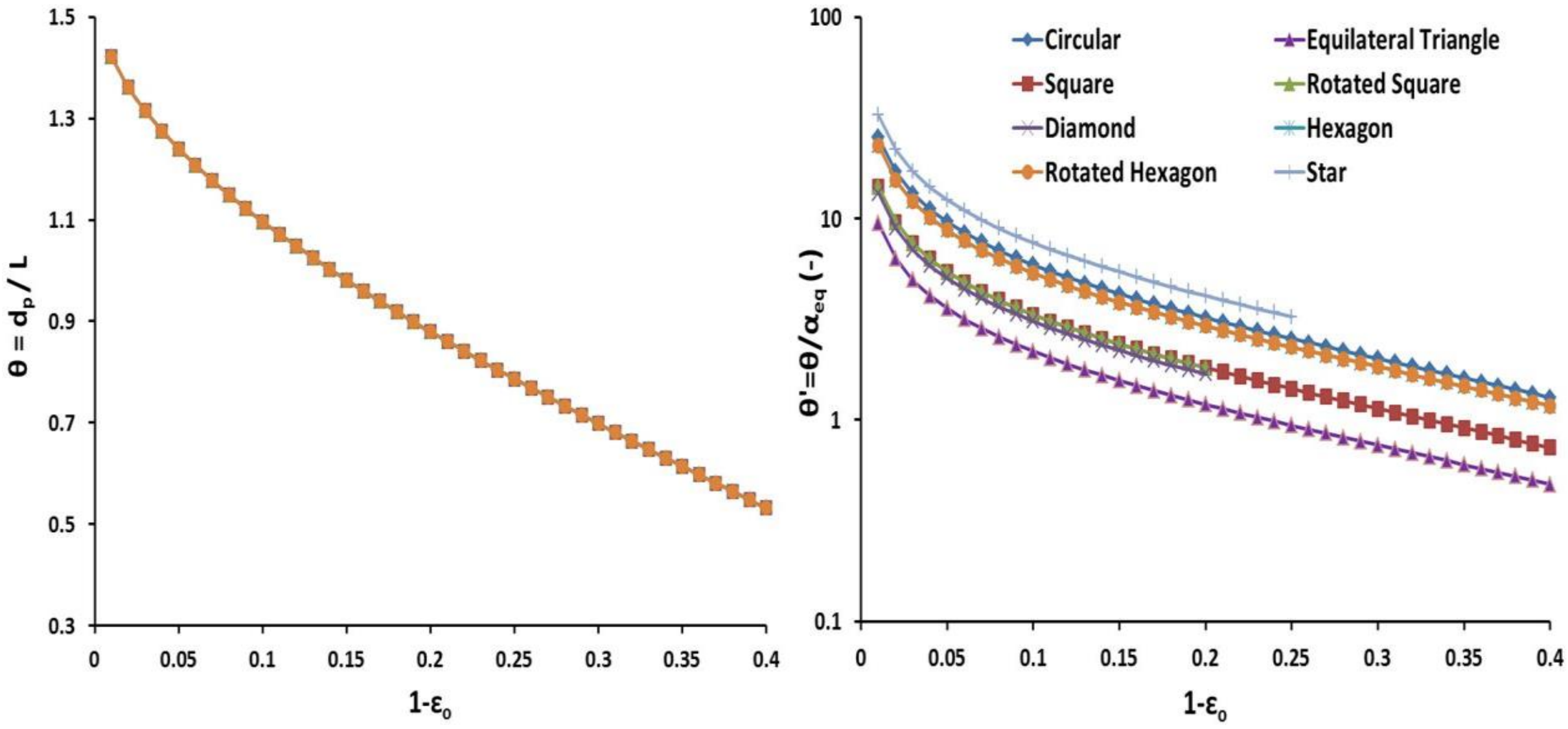

**Fig 3.15.** Left- Plot of $\theta$ (dimensionless) vs $1-\varepsilon_o$. $\theta$ is function of $\beta$ (or $L_s$) for different strut shapes and thus, a unique curve is obtained for all the strut shapes due to the hypothesis made. Right: Plot of $\theta'$ (ratio of pore diameter to equivalent strut radius) vs $1-\varepsilon_o$.

The plots of $\theta$ and $\theta'$ against porosity ($\varepsilon_o$) is shown in Figure 3.15. It is clearly seen in Figure 3.15 (left) that a unique curve is obtained for $\theta$. It is due to the fact that $d_p$ is clearly a function of $L_s$ that is based on the same node volume hypothesis, and follows a decreasing trend with decrease in porosity. On the other hand, $\theta'$ is highest for star strut shape and lowest for equilateral strut shape and increases with increase in porosity and follow an inverse function of $\alpha_{eq}$ as presented in Figure 3.15 (right).

### 3.4.2.2 Specific surface area

As it is far more convenient to calculate specific surface area using a cubic unit cell, the foam structure is considered in the cubic cell of volume $V_c$ (see Figure 2.18-top), specific surface area, $a_c$ can be written as:

$$a_c = \frac{\left(36\, S_{ligament} + 24\, S_{node}\right)}{V_c} \tag{3.67}$$

where, $S_{ligament}$ and $S_{node}$ is the surface area of one ligament and node contained in the cubic cell of volume, $V_c$ (the distance between two opposite points of the cubic cell is $2\sqrt{2}L$ and thus, $V_c\ = 2V_T$ ).

In the Figure 2.17, Kelvin-like cell foams of different strut shapes inside a cubic cell are presented. One can easily notice that there are 12 full ligaments and 24 half ligaments in a unit cell (see also Figure 2.18-top). Also, at the node junction, there are two half nodes and one one-fourth node. Specific surface area of a circular strut shape is given as:

$$a_c = \frac{\left\{48\pi R_c L_s + 24.\frac{3}{4}\left(\frac{5}{4}\pi {R_c}^2\right)\right\}}{2\left(8\sqrt{2}L^3\right)} = \frac{1}{\sqrt{2}L}\left(3\pi\alpha_c\beta + \frac{45}{32}\pi{\alpha_c}^2\right) \qquad (3.68)$$

For the other strut shapes, the relation between specific surface area ($a_c$), node to node length ($L$) and geometrical parameters is presented in Appendix D.

For a given unit cell, the node to node length is fixed and one can easily determine specific surface area of different strut shapes. A dimensionless curve $a_c.L$ against porosity ($\varepsilon_o$) is presented in Figure 3.16. From this curve, one can identify either $a_c$ or $L$ for a known porosity. While keeping the same porosity, one can increase the specific surface area by utilizing different strut shapes. The maximum increase in specific area is observed for star strut shape whilst lowest is observed for circular strut shape for a given porosity.

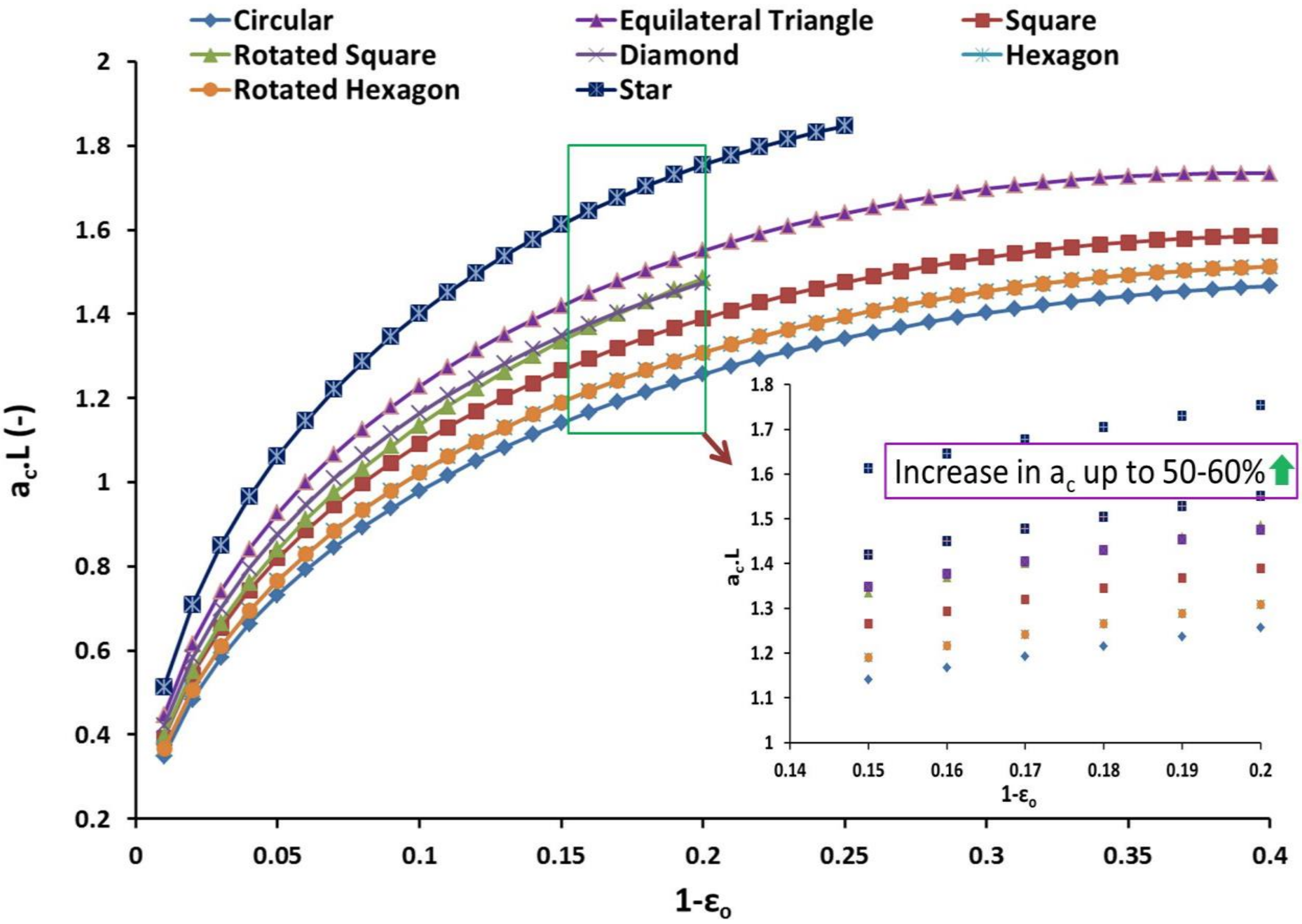

**Figure 3.16.** Plot of $a_c\,L$ (dimensionless) vs $1 - \varepsilon_o$. A sharp increase in $a_c$ is observed at lower porosity for complex shapes.

Two dimensionless curves relating $d_p.a_c$ and $\alpha_{eq}.a_c.L$ with porosity ($\varepsilon_o$) are presented in Figure 3.17. In Figure 3.17 (left), it is clear that $d_p.a_c$ has two different behaviors where it first increases and then decreases with decreasing porosity. This change in behaviour is observed at a limiting porosity of 84%. $d_p$ increases with increase in porosity (Figure 3.15-left) while $a_c$ decreases with increase in porosity (Figure 3.16) and thus, a combination of both i.e. $d_p.a_c$ acquires increasing and decreasing behavior. Due to the assumption of constant node volume irrespective of the strut shape at a given porosity allows to obtain limiting porosity close to 84% for all the trends. The limiting porosity is important for the applications where thermo-hydraulic properties (heat transfer and pressure drop) are critical. The correlations of several authors (Buciuman Kraushaar-Czarnetzki, 2003; Moreira et al., 2004; Lacroix et al., 2007; Garrido et al., 2008; Grosse et al., 2009, Huu et al., 2009) showed the continuous decreasing trend of $d_p.a_c$ with increase in porosity. These authors derived the correlations only for high porosity foam samples. This could be one of the reasons that they had observed only decreasing trend. In this work, virtual samples have been generated till low porosity and thus, both the trends have been observed. Some of the authors (e.g. Lacroix et al., 2007) used the unit cell like cubic lattice that is not an adequate unit cell and thus, could not reproduce the same trend and accurate results due to over-simplification of geometrical parameters.

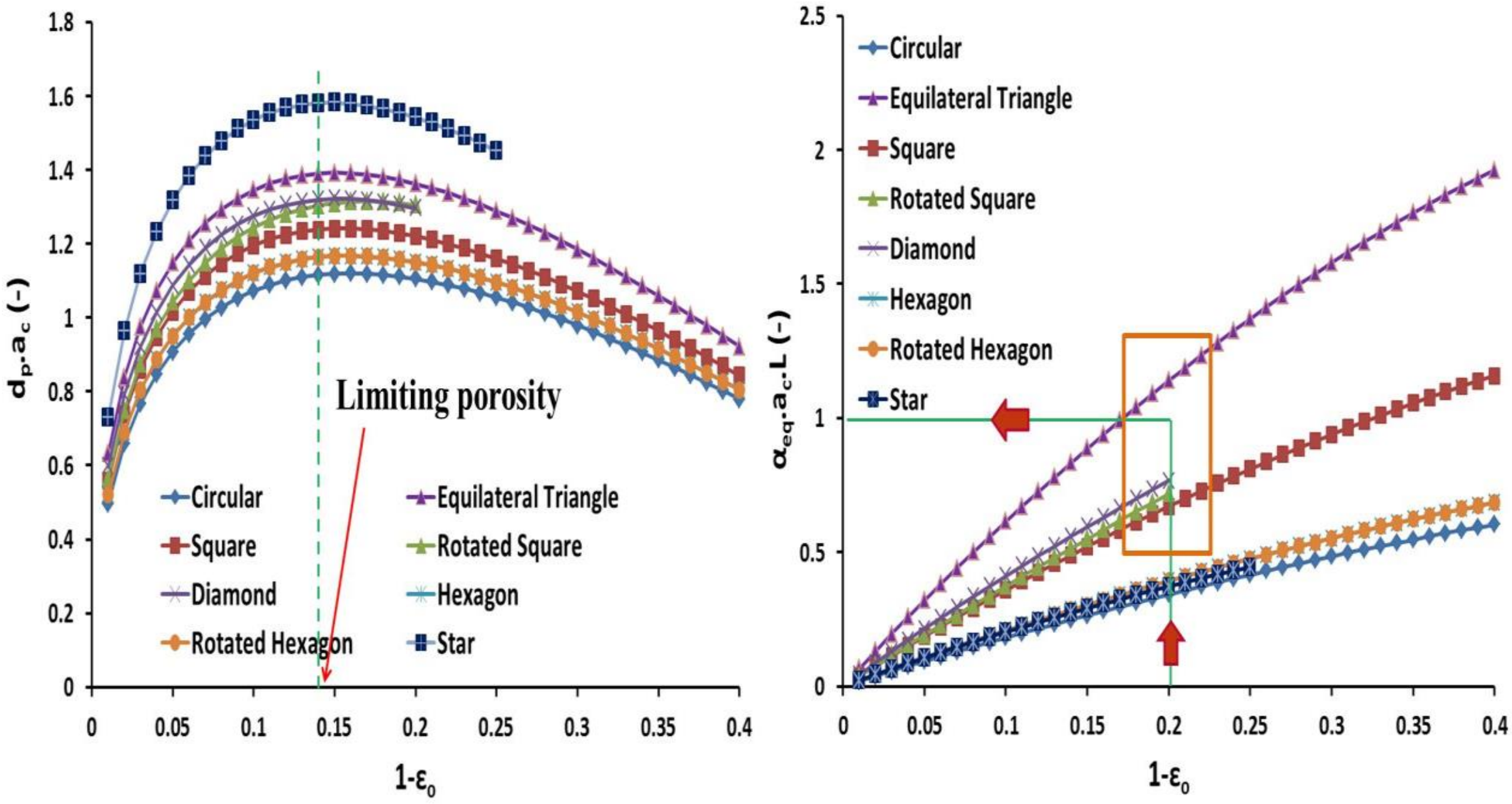

**Figure 3.17.** Left: Plot of $d_p.a_c$ (dimensionless) vs. $1-\varepsilon_o$. Two distinguished behaviors are obtained. Right: Plot of $\alpha_{eq}.a_c.L$(dimensionless) vs. $1-\varepsilon_o$. For a given application purpose of $\varepsilon_o$>0.80 (mechanical constraint) and hydraulic constraint ($\alpha_{eq}.a_c$>0.6), only square, rotated square or diamond shape can be used.

In Figure 3.17 (right), $\alpha_{eq}.a_c.L$ increases with decrease in porosity. This increasing trend with decreasing porosity is lowest for circular strut shape while it is highest for equilateral triangle strut shape (both $\alpha_{eq}$ and $a_c.L$ increase with decreasing porosity) for a known porosity. This curve is also important to choose a desired strut shape as a geometrical constraint for heat transfer applications.

Using these curves, one can characterize all the geometrical parameters of any strut shape. Any of the two quantities are known, one can easily determine all the geometrical parameters using the above established correlations and curves. Moreover, these correlations can be used to determine the shape of the strut for a given $d_p.a_c$ and $\alpha_{eq}.a_c.L$ which can insight one to tailor their own foam accordingly. In this way, one can realize any number of foams depending upon the various engineering applications needs.

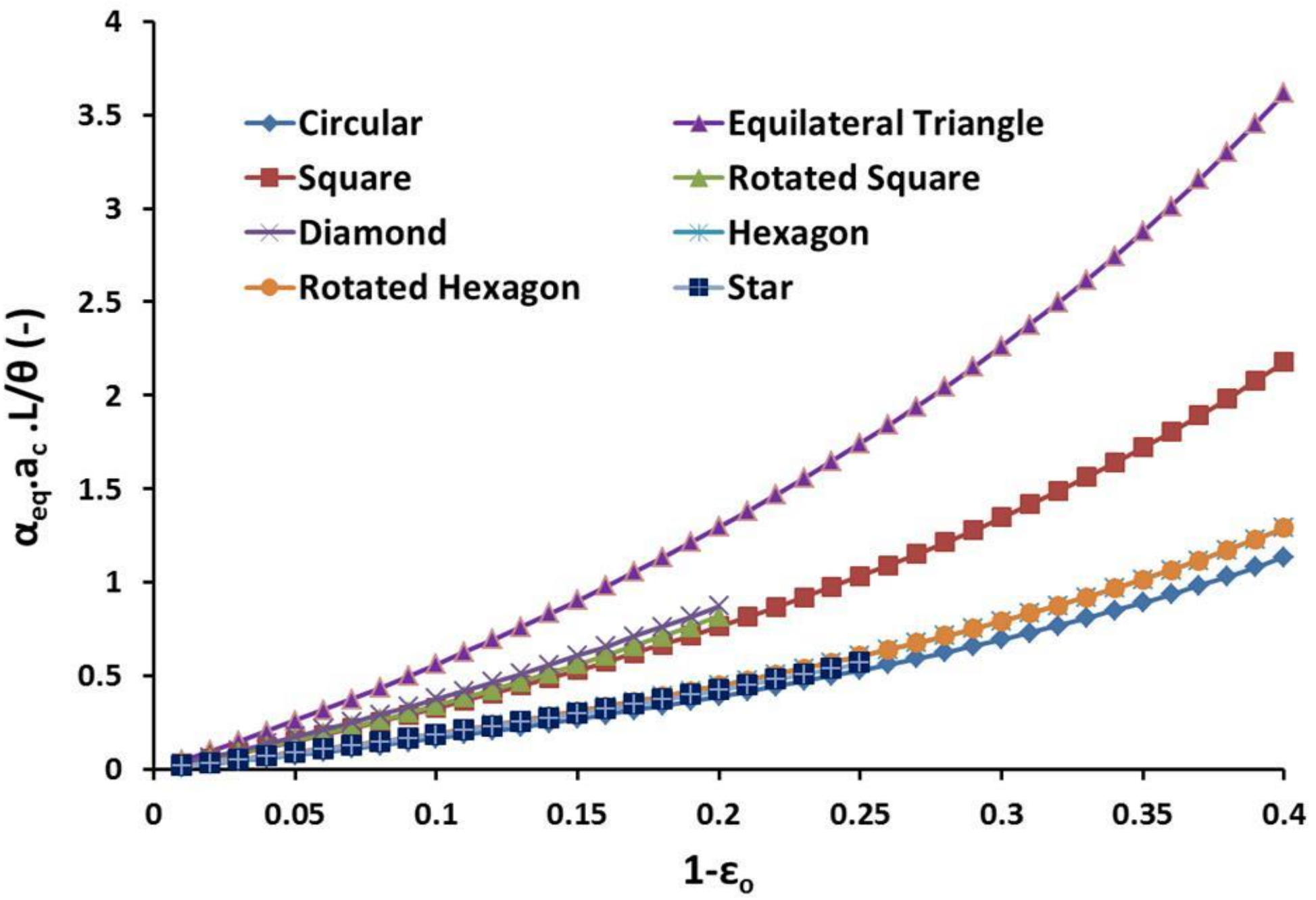

**Figure 3.18**. Plot of $\alpha_{eq}.a_c.L/\theta$ (non-dimensional) vs. $1-\varepsilon_o$

A plot of $\alpha_{eq}.a_c.L/\theta$ against porosity ($\varepsilon_o$) is shown in Figure 3.18 and this dimensionless product follows a quadratic polynomial tendancy with porosity. This curve is critical for the cases where individual information about geometrical parameters are unknown e.g., for a given application where a coupled constraint of geometrical and hydraulic properties (e.g. $\alpha_{eq}.a_c.L$) is known, it would be easier to optimize the strut shape and specific surface area in order to increase the performance of the system.

### 3.4.3 Correlations for predicting geometrical properties for different strut shapes for *anisotropic* metal foams

According to the methodology of anisotropy proposed (see section 2.5.2 and Figure 2.18-bottom), there are four different strut lengths i.e. $L_{s1}, L_{s2}, L_{s3}$ *and* $L_{s4}$. $L_{s1}$ corresponds to strut length in horizontal square faces in X-direction, $L_{s2}$ corresponds to strut length in vertical square faces in Y-direction while $L_{s3}$ and $L_{s4}$ correspond to strut lengths of hexagon faces in 3-D space. Moreover, $L_{s1} = L_{s3}$ and $L_{s2} = L_{s4}$ and are due to simultaneous elongation and compression in all the directions. Two hypotheses (H.1 and H.2) were considered to derive an accurate geometrical correlation for *anisotropic* open cell foams to predict geometrical characteristics and specific surface area as follows:

H.1 Solid ligament volume of horizontal and vertical square faces of *anisotropic* foam structure is same as the *isotropic* ones because their deformations depend only on elongation and compression.

H.2 As porosity is kept constant for any deformation of any given shape; the total solid volume of ligaments and total solid volume of nodes of *anisotropic* foams are same as of their original *isotropic* foam samples.

When elongation ($\Omega$ in X-direction) and compression factors ($1/\sqrt{\Omega}$ in Y and Z-directions) are applied simultaneously, the strut lengths ($L_{s1}$) placed in horizontal square face depend on elongation $\Omega$ in X-direction and compression $1/\sqrt{\Omega}$ in Z direction (see Figure 3.19-top). On the other hand, strut lengths ($L_{s2}$) placed in vertical square face depend only on compression $1/\sqrt{\Omega}$ in Y and Z-directions (see Figure 3.19-bottom). The dimensions of strut lengths $L_{s1}$ and $L_{s2}$ in an *anisotropic* foam structure can be determined using Figure 3.19 and are given by Equation 3.69 and 3.70:

$$L_{s1} = \frac{L_s}{\sqrt{2}}\sqrt{\Omega^2 + \frac{1}{\Omega}} = \frac{L_s}{\Pi} \tag{3.69}$$

and,

$$L_{s2} = \frac{L_s}{\sqrt{\Omega}} = \frac{L_s}{\zeta} \tag{3.70}$$

where, $\Pi = \sqrt{2}/\sqrt{\Omega^2 + {}^{1}/_{\Omega}}$ and $\zeta = \sqrt{\Omega}$

Based on hypothesis H.1 of same ligament volume in horizontal and vertical square faces as of *isotropic* case, equivalent radii namely, $R_{eq1}$ and $R_{eq2}$ of the ligaments in horizontal and vertical square faces can be easily derived by equating them to their original *isotropic* case and are given by Equation 3.71 and 3.72:

$$R_{eq1} = \sqrt{\sqrt{2}/\sqrt{\Omega^2 + {}^{1}/_{\Omega}}}\, R_{eq} = \sqrt{\Pi} R_{eq} \tag{3.71}$$

and,

$$R_{eq2} = \sqrt{\sqrt{\Omega}} R_{eq} = \sqrt{\zeta} R_{eq} \tag{3.72}$$

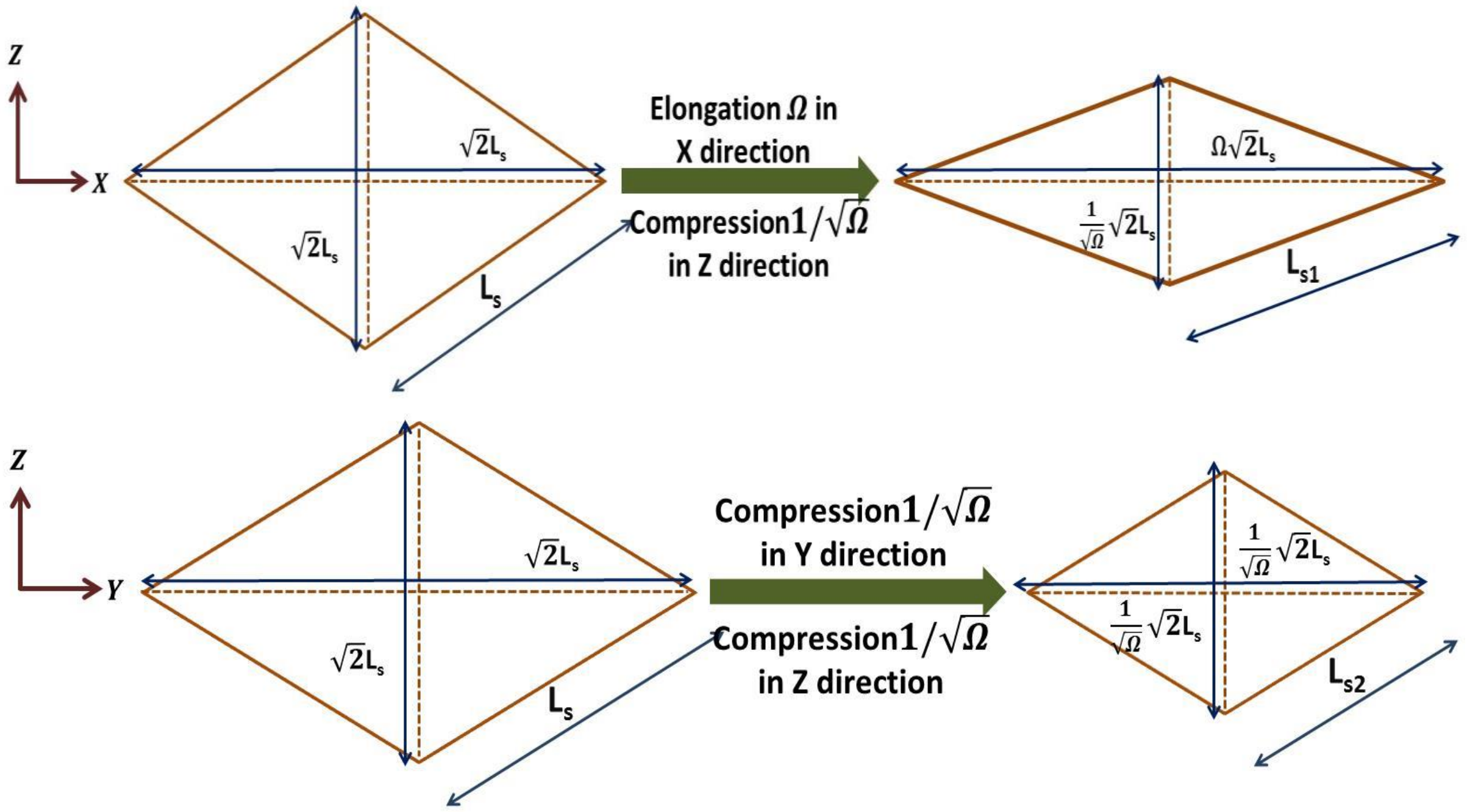

**Figure 3.19**. Representation and determination of strut lengths: $L_{s1}$ and $L_{s2}$. Top- in horizontal square face after applying elongation $\Omega$ in X-direction and compression of in $1/\sqrt{\Omega}$ in Z-direction. Bottom- in vertical square face after applying compression of in $1/\sqrt{\Omega}$ in Y and Z-direction.

Equivalent radii $R_{eq3}$ and $R_{eq4}$ can be obtained by using hypothesis H.2. Nodes ‘A’ and ‘B’ were considered first as shown in Figure 2.18 (bottom) to describe $R_{eq3}$ (assuming same node volumes at junctions ‘A and B’ as in case of *isotropic* foam structure) and is given as:

$$R_{eq3} = \sqrt[3]{\frac{4{R_{eq}}^3 - \left({R_{eq1}}^3 + {R_{eq2}}^3\right)}{2}} = \sqrt[3]{\frac{4 - (\Pi^{3/2} + \zeta^{3/2})}{2}} R_{eq} = \delta^{1/3} R_{eq} \tag{3.73}$$

where, $\delta = \frac{4-(\Pi^{3/2}+\zeta^{3/2})}{2}$

Similarly, $R_{eq4}$ (assuming same node volume at junction 'C' in case of *isotropic* foam structure) is given as:

$$R_{eq4} = \sqrt[3]{2{R_{eq}}^3 - {R_{eq1}}^3} = \sqrt[3]{2 - \Pi^{3/2}}R_{eq} = \omega^{1/3}R_{eq} \qquad (3.74)$$

where, $\omega = 2 - \Pi^{3/2}$

One can easily determine all the strut lengths and equivalent radii using $\Omega$ (or $\Pi$ and $\zeta$) given by Equations 3.69-3.74 for all different strut shapes of *anisotropic* foam structure. Importantly, all the strut lengths and equivalent radii of *anisotropic* foam structure are the functions of $R_{eq}$ and $L_s$ only for any strut shape and size of the *isotropic* foam structure.

In case of *anisotropic* foam in the unit cell (see Figure 2.18-bottom), there are 16 half strut lengths of square face in horizontal direction (length $L_{s1}$), 8 half strut lengths of square face in vertical direction (length $L_{s2}$), 8 strut lengths of hexagon face in horizontal direction (length $L_{s3}$) and 4 struts lengths of hexagon face in vertical direction (length $L_{s4}$). Also, at the node as in the case of *isotropic* foam structure, there are two-half nodes and one node of one-fourth shape.

Surface area of all the ligaments ($S'_L$) and nodes ($S'_N$) were calculated to determine specific surface of circular strut shape ($R_{eq}$=$R_c$) for any elongation and compression factors.

$S'_L$ of circular strut shape is given as:

$$S'_L = \left[\left\{\left(\frac{16}{2}\right)(2\pi R_{c1}L_{s1})\right\} + \left\{\left(\frac{8}{2}\right)(2\pi R_{c2}L_{s2})\right\} + \{8(2\pi R_{c3}L_{s3})\} + \{4(2\pi R_{c4}L_{s4})\}\right] \qquad (3.75a)$$

$$S'_L = 8\pi R_c L_s\left[\left\{\frac{2}{\sqrt{\Pi}}\right\} + \left\{\frac{2}{\sqrt{\zeta}}\right\} + \left\{\frac{2}{\Pi}(\delta)^{1/3}\right\} + \left\{\frac{1}{\zeta}(\omega)^{1/3}\right\}\right] \qquad (3.75b)$$

$S'_N$ of circular strut shape is given as:

$$S'_N = \left(\frac{3}{4}\right)\left[8\left(\pi {R_{c1}}^2 + \left(\frac{\pi {R_{c4}}^2}{4}\right)\right) + 8\left(\pi {R_{c1}}^2 + \left(\frac{\pi {R_{c3}}^2}{4}\right)\right) + 8\left(\pi {R_{c2}}^2 + \left(\frac{\pi {R_{c3}}^2}{4}\right)\right)\right] \qquad (3.76a)$$

$$S'_N = 6\pi R_c{}^2\left[2\Pi + \zeta + \frac{1}{2}(\delta)^{2/3} + \frac{(\omega)^{2/3}}{4}\right] \qquad (3.76b)$$

On substitution of Equation 3.75 b and 3.76 b in Equation 3.77, specific surface area of circular strut shape in the case of *anisotropic* foam structure is given as:

$$a_c = \frac{(S'_L + S'_N)}{V_c}$$
$$= \frac{8\pi\alpha_c\beta\left[\frac{2}{\sqrt{\Pi}} + \frac{1}{\sqrt{\zeta}} + \frac{2}{\Pi}.(\delta)^{1/3} + \frac{1}{\zeta}.(\omega)^{1/3}\right] + 6\pi\alpha_c{}^2\left[2\Pi + \zeta + \frac{1}{2}(\delta)^{2/3} + \frac{(\omega)^{2/3}}{4}\right]}{16\sqrt{2}L} \qquad (3.77)$$

Equation 3.77 presents the specific surface area of *anisotropic* foam structure of circular strut shape that is a function of geometrical parameters, $\alpha_c$ and $\beta$ of *isotropic* foam of circular strut shape (see section 3.4.2) respectively as well as elongation and compression factors, $\Omega$ and $1/\sqrt{\Omega}$. Thus, in order to characterize *anisotropic* foams, one has to know only the geometrical parameters of *isotropic* open cell foam and all other geometrical properties and their relationships for any strut shapes can be easily derived using the proposed methodology of equivalent radius, $R_{eq}$.

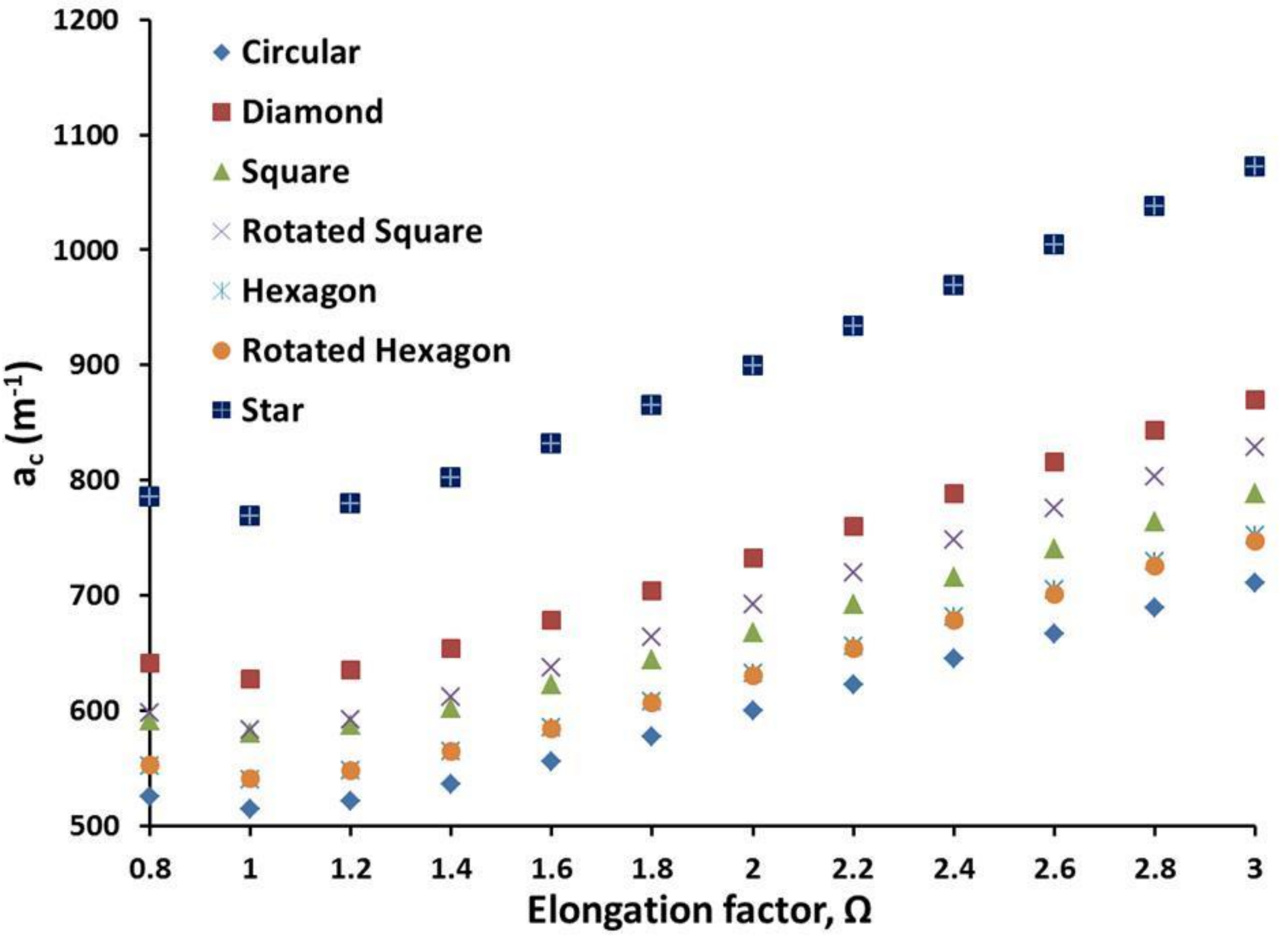

**Figure 3.20**. Variation of specific surface area ($a_c$) with elongation factor ($\Omega$) for different strut shapes at 95% porosity.

The variations in specific surface area of different strut shapes for various elongation factors are shown in Figure 3.20. The specific surface area increases with elongation. There is a gradual increase but no apparent order is yet found. As shown in the Table A.1 (see Appendix A), one could increase the specific surface area by a factor of 1.4 by keeping the same porosity in an *anisotropic* case compared to *isotropic* case. This fact is extremely useful for various geometrical and physical constraints. As expected, *anisotropic* foam structure of star strut cross section possesses highest value of specific surface area while the circular strut cross section possesses the lowest value for the same elongation factor.

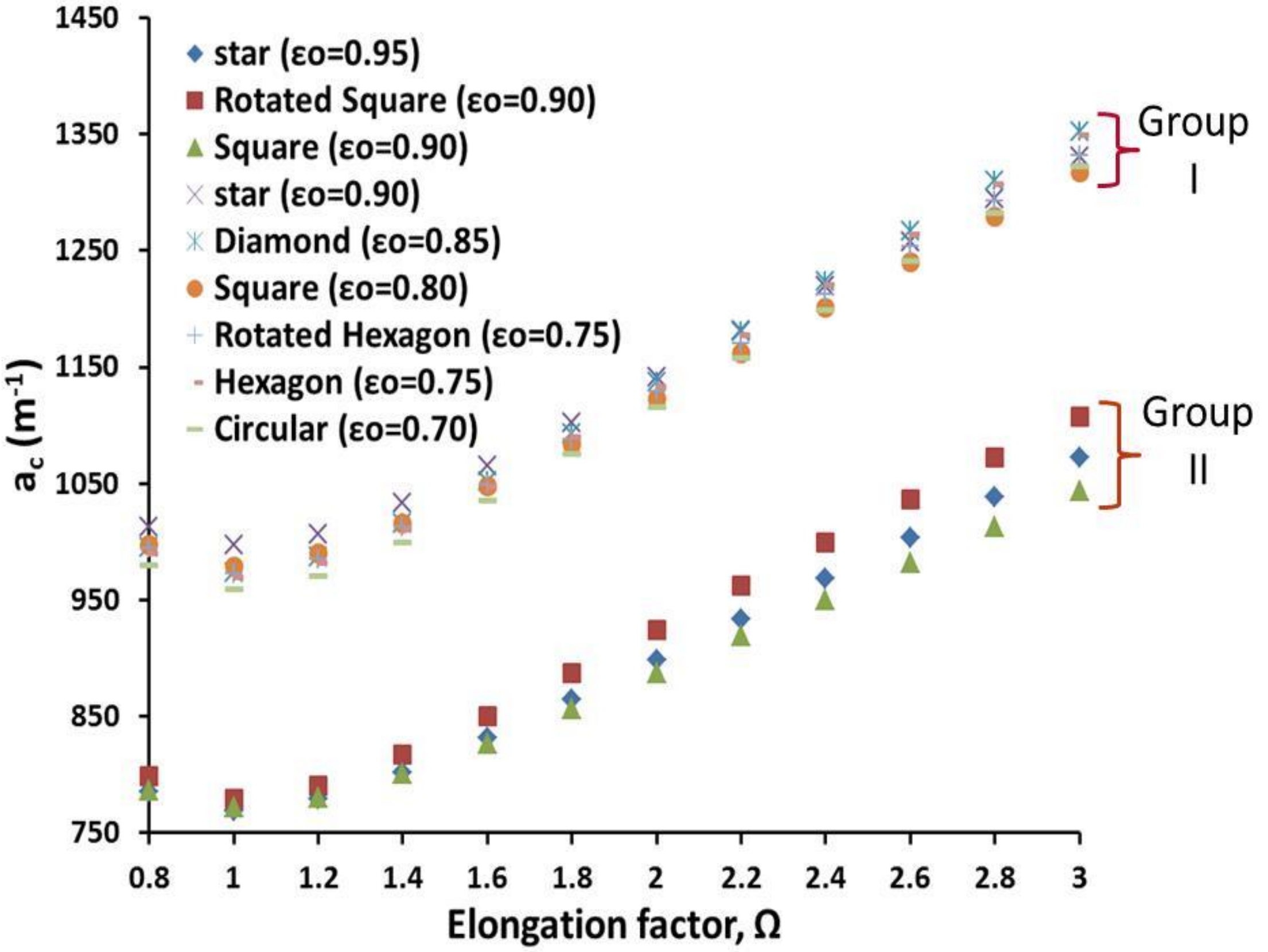

**Figure 3.21**. Classification of groups of same specific surface area with different strut shapes and porosity for all elongations ($\Omega$).

Anisotropy of different strut shapes helps in optimizing the specific surface area in order to achieve better thermo-hydraulic performances for a given geometrical constraint. In Figure 3.21, two groups, I and II of specific surface areas are clearly shown. In the group I, it is possible to obtain same specific surface area for different strut cross sections: circular at $\varepsilon_o$=0.70, hexagon and rotated hexagon at $\varepsilon_o$=0.75, square at $\varepsilon_o$=0.80, diamond at $\varepsilon_o$=0.85 and star at $\varepsilon_o$=0.90. These different strut shapes at different porosities for a given elongation factor provide the same specific surface area for different *anisotropic* foam samples.

Similarly, strut cross sections: square and rotated square at $\varepsilon_o$=0.90 and star at $\varepsilon_o$=0.95 of group II exhibit the same specific surface area for a given elongation factor. There is a very small variation in the specific surface area (about a factor of 1.02-1.06) between these strut shapes at extreme elongation.

It is thus, evident that different strut shapes at different porosity in the *anisotropic* case could exhibit same specific surface area which is not the case with *isotropic* foams. This fact could govern in optimizing thermo-hydraulic phenomena for various engineering applications where geometrical constraints are vital.

### 3.4.4 Correlations for predicting geometrical properties for *isotropic* ceramic foams

Depending upon the manufacturing processes (see section 2.2), most of the commercially available ceramic foams exhibit hollow strut. In this thesis, neither any virtual ceramic foam was modelled in CAD nor was any measurements performed on real geometry. Based on the measurements reported in the literature on circular or triangular strut shape with void in ceramic foams, the analytical correlation derived in section 3.4.2 for *isotropic* metal foams is extended to derive the correlation for ceramic foams (see Figure 3.22-left).

#### 3.4.4.1 Characterization of geometrical parameters

One of the typical hollow struts is shown in the Figure 3.22 (center). Based on tetrakaidecahedron geometry inside a cubic cell (see Figure 3.23) of circular strut cross section for total porosity range, $0.65 \leq \varepsilon_t \leq 0.90$, an analytical correlation that covers strut, open and total porosity was derived. Figure 3.22 (right) shows the dimension of circular strut radius, $R$ and approximated equilateral triangular void of side length, $N$.

In order to provide an approximate analytical solution, $L_s$ as strut length (without considering nodes) and $L$ as distance between two nodes (or length of solid truncated octahedron edge) were defined as shown in Figure 3.23.

Strut porosity due to void inside the strut is calculated as:

$$\varepsilon_{st} = \frac{V_{void}}{V_{strut}} = \frac{\sqrt{3}/4N^2L_s}{\pi R^2 L_s} \tag{3.78}$$

Equation 3.78 can be rewritten as:

$$N = \kappa R \qquad (3.79)$$

where, $\kappa = \sqrt{4\pi\varepsilon_{st}/\sqrt{3}}$

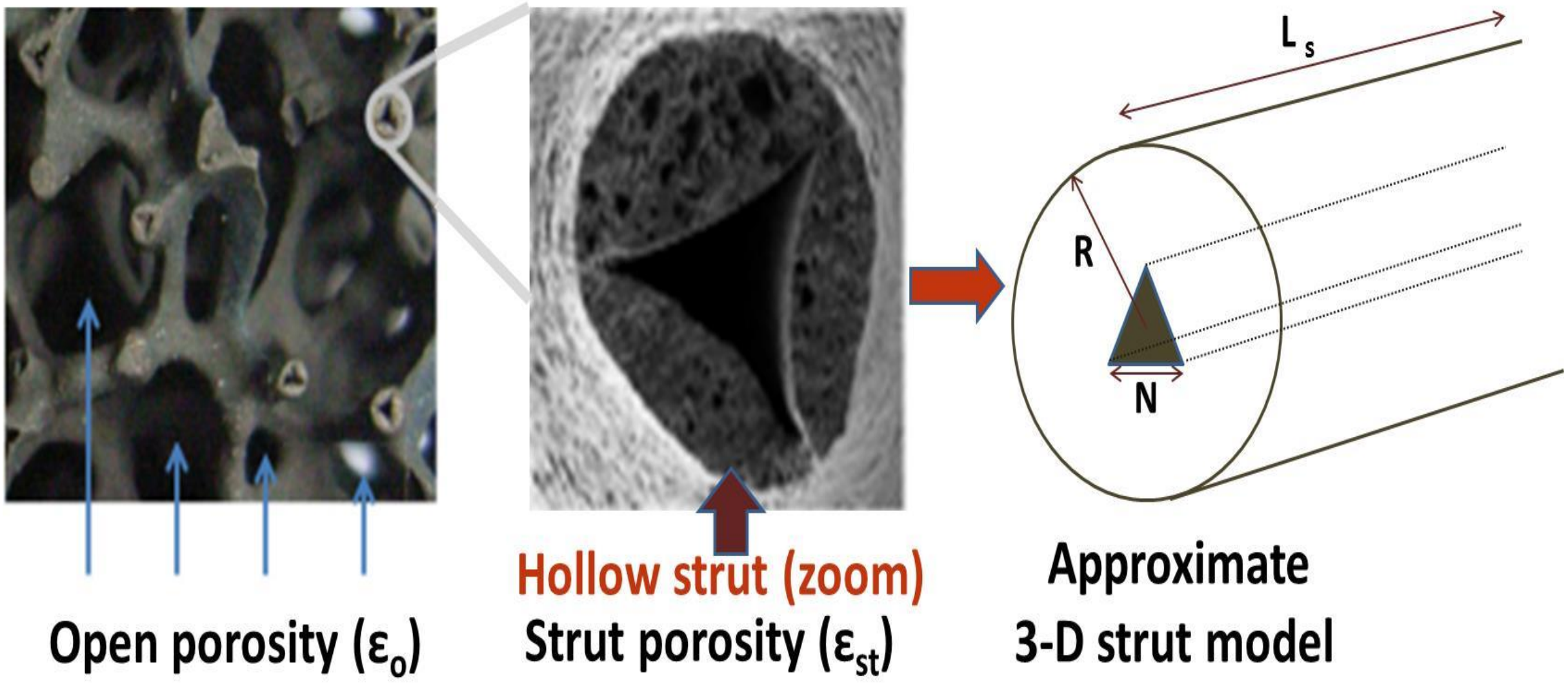

**Figure 3.22**. Left- Representation of ceramic foam where hollow struts are visible. Center- zoom view of hollow strut. Right: A detailed circular strut cross section with an equilateral triangular void inside the strut. The dimensions of strut (strut radius, $R$) and void cross section (void length, $N$) are clearly highlighted.

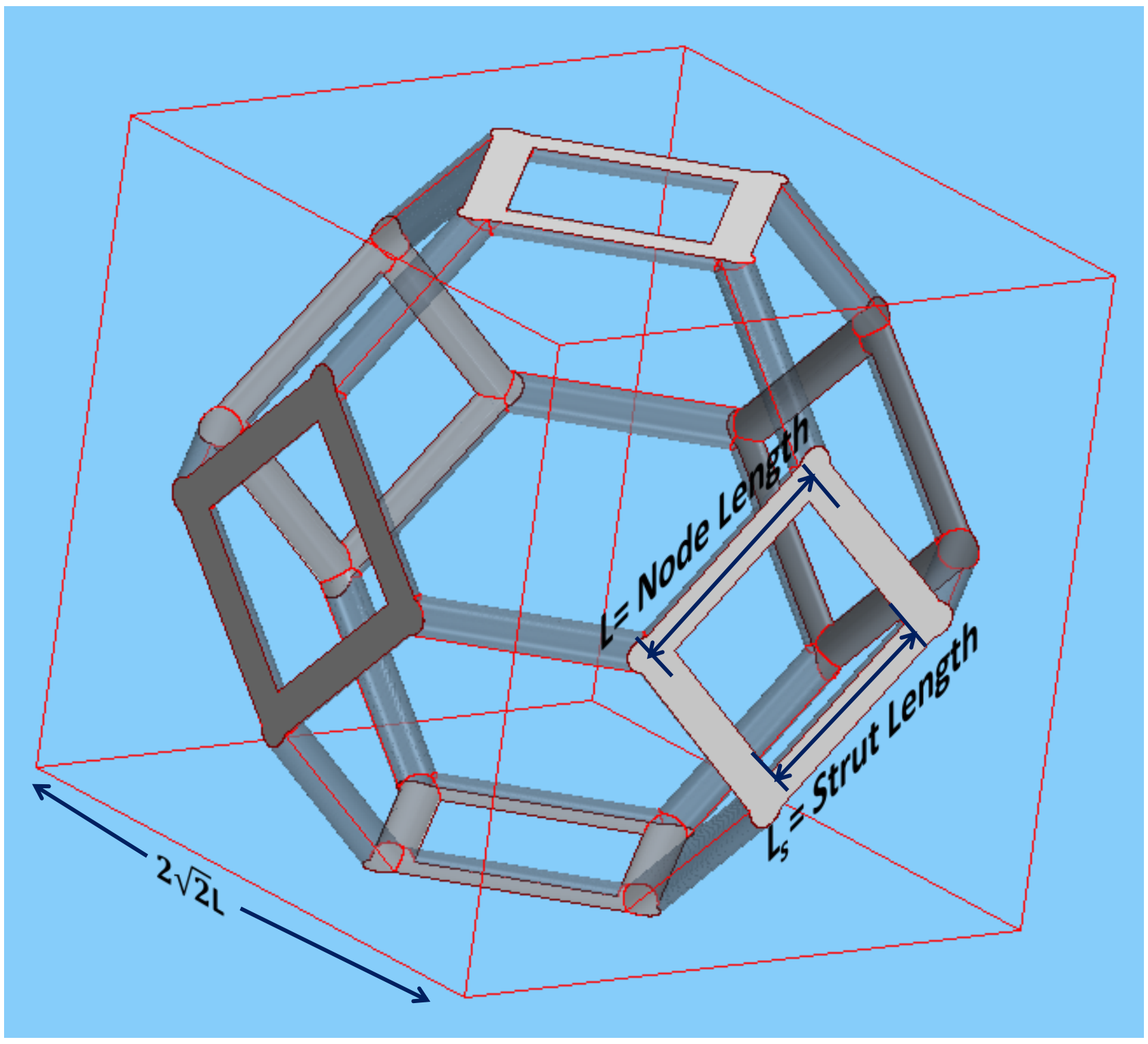

**Figure 3.23**. A tetrakaidecahedron foam model inside a cube. Node length $L$ (center to center distance of a node connection), strut length $L_s$ and cubic unit cell length $2\sqrt{2}L$ are clearly presented. The analytical correlation is based on the above unit cell having triangular void in struts.

The node volume calculation was chosen to base on the formulation given by Kanaun and Tkachenko (2008). Volume of node at the junction of four struts is given as (see Figure 3.24):

$$V_{node} = \frac{4}{3}\left(\pi R^3 - \frac{\sqrt{3}}{4}N^3\right) = \frac{4}{3}\pi R^3(1-\kappa\varepsilon_{st}) \quad (3.80)$$

Volume of the ligament is given as:

$$V_{ligament} = \pi R^2 L_s - \frac{\sqrt{3}}{4}N^2 L_s = \pi R^2 L_s(1-\varepsilon_{st}) \quad (3.81)$$

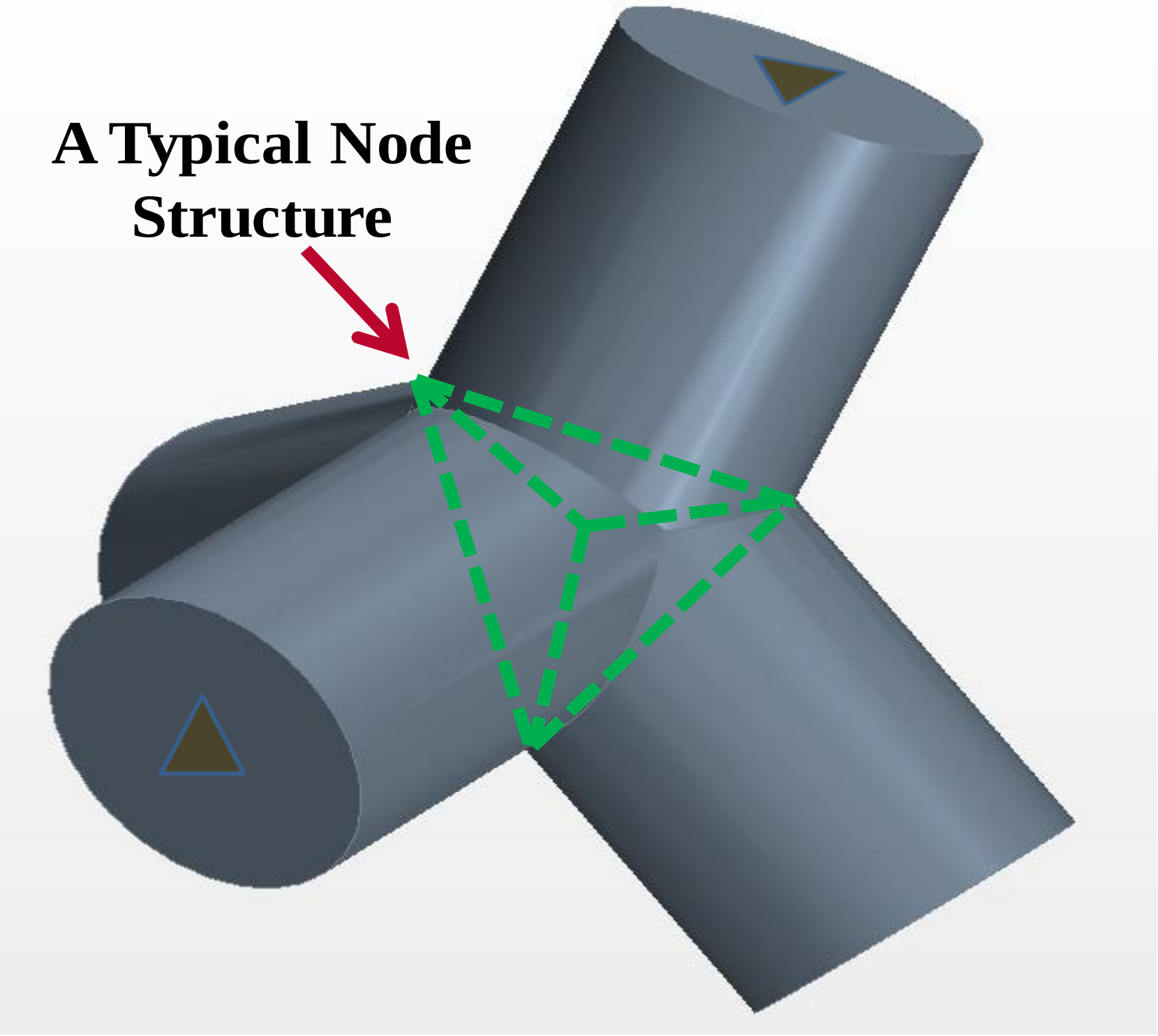

**Figure 3.24**. A typical node of ceramic foam structure. The 4 faces of a pyramid are shown which are taken into consideration in calculating volume of the node. The face of the nodes changes with the strut shape. Four struts of circular shape having equilateral triangular void at the node are shown which is approximated as triangular pyramid.

At the ligaments intersection, the node shape can be approximated by using geometrical interpretation as shown in Figure 3.24:

$$1.6R + L_s = L \quad (3.82)$$

In dimensionless form, we can rewrite Equation 3.82 as:

$$1.6\alpha + \beta = 1 \quad (3.83)$$

where, $\alpha = \frac{R}{L}$ and $\beta = \frac{L_s}{L}$

In a truncated octahedron structure (see Figure 3.23), there are 36 ligaments and 24 nodes but only 1/3[rd] of both, volume of ligament and node are included in the unit periodic cell. For periodic cellular foam in a unit cell, total solid volume ($V_s$), total truncated volume ( $V_T = 8\sqrt{2}L^3$) and total porosity ($\varepsilon_t$) are related as:

$$\varepsilon_t = \frac{1 - \frac{1}{3}\left(36V_{ligament} + 24V_{node}\right)}{V_T}$$
$$= \frac{1 - \frac{1}{3}\left(36\pi R^2 L_s(1-\varepsilon_{st}) + 24.\frac{4}{3}\pi R^3(1-\kappa\varepsilon_{st})\right)}{8\sqrt{2}L^3} \qquad (3.84)$$

In dimensionless form, we can rewrite Equation 3.84 as:

$$12\pi\alpha^2\beta(1-\varepsilon_{st}) + \frac{32}{3}\pi\alpha^3(1-\kappa\varepsilon_{st}) = 8\sqrt{2}(1-\varepsilon_t) \qquad (3.85)$$

Equation 3.85 gives a generic relation of total porosity as a function of geometrical parameters. Note that, $\alpha$ is ratio of strut radius to node length whereas $\beta$ is the ratio of strut length to node length. Equations 3.83 and 3.85 could be combined to get approximate values of $\alpha$ and $\beta$ as a function of $\varepsilon_t$. This approach can used to determine all the geometrical properties if full set of geometrical parameters are not known.

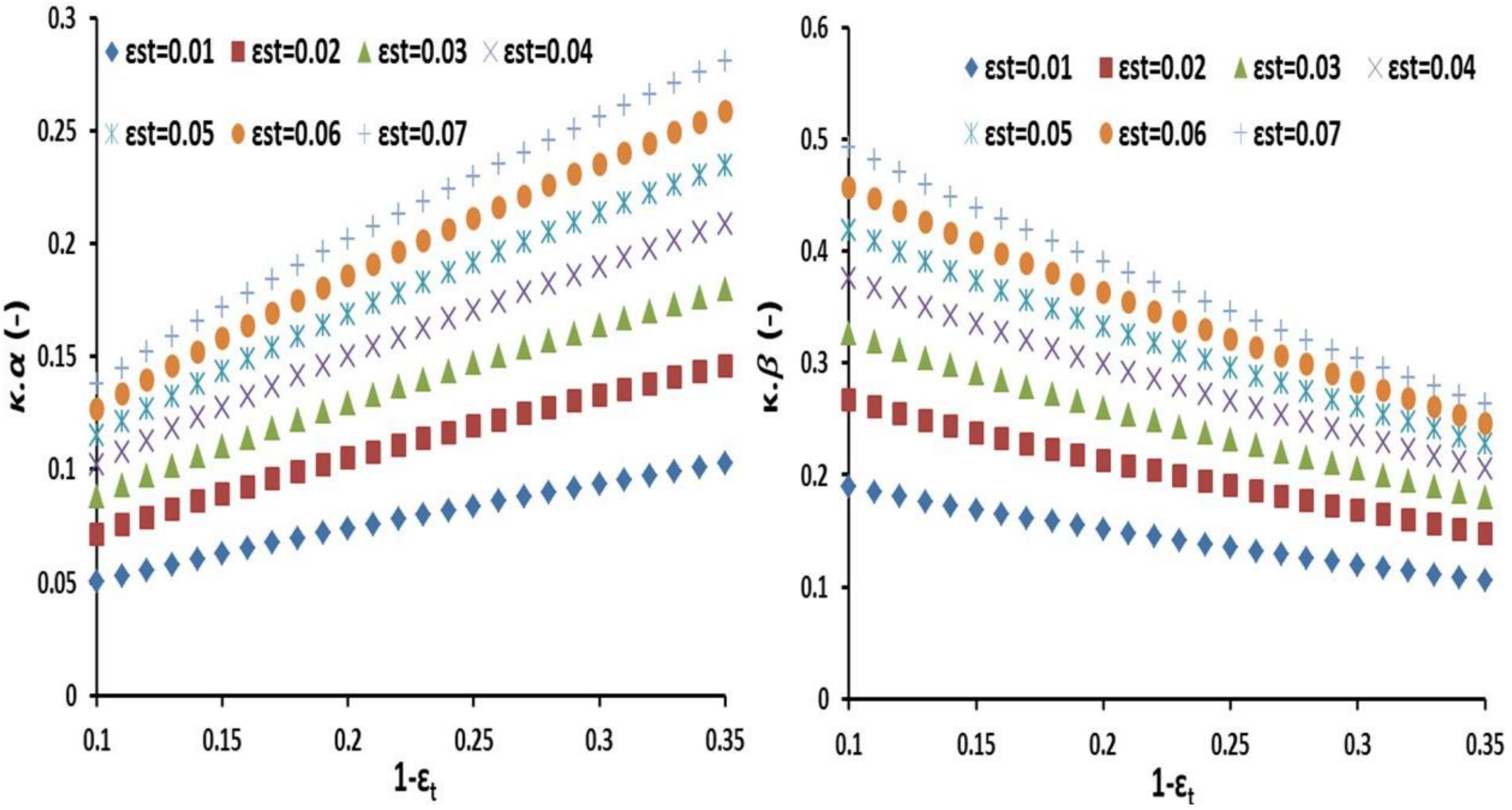

**Figure 3.25**. Plot of dimensionless geometrical parameters vs. $1 - \varepsilon_t$ for different strut porosities, $\varepsilon_s$. Left: $\kappa.\alpha$-function of hollow strut and strut diameter. Right: $\kappa.\beta$-function of hollow strut and strut length.

Strut porosity ($\varepsilon_{st}$), strut radius ($R$) and strut length ($L_s$) are inter-related. In order to see the influence of each parameter on geometrical characteristics, different values of $\kappa.\alpha$ and $\kappa.\beta$ for different porosities are plotted and they are found to follow power and exponential law respectively as shown in Figure 3.25 (left and right) and can be expressed as:

$$\kappa.\alpha = \Lambda_1(1-\varepsilon_t)^{\Lambda_2} \tag{3.86}$$

$$\kappa.\beta = \Lambda'_1 exp[-\Lambda'_2(1-\varepsilon_t)] \tag{3.87}$$

where, $\Lambda$ and $\Lambda'$ are the parameters that depend only on void shape and are functions of total porosity.

For instance, for a given hollow strut porosity and total porosity; one can quantify easily the strut radius and strut length. Moreover, depending upon the desired output quantity for industrial applications, one can tailor their own foam characteristics easily using Equations 3.86 and 3.87 and Figure 3.25.

### 3.4.4.2 Specific surface area

In the Figure 3.23, one can easily notice that there are 12 full ligaments and 24 half ligaments in a unit cell. Also, at the node junction, there are two half nodes and one one-fourth node. As it is far more convenient to calculate specific surface area using a cubic unit cell, the foam structure is considered in the cubic cell of volume $V_c$; specific surface area, $a_c$ can be written as:

$$a_c = \frac{\left(36\, S_{ligament} + 24\, S_{node}\right)}{V_c} = \frac{1}{\sqrt{2}L}\left(\frac{3}{2}\alpha\beta(2\pi - 3\kappa) + \frac{45}{32}\pi\alpha^2(1-\ \varepsilon_{st})\right) \tag{3.88}$$

where $S_{ligament}$ and $S_{node}$ are the surface area of one ligament and node contained in the cubic cell of volume, $V_c$ $(= 2V_T)$ is the volume of the cubic cell.

A dimensionless curve $\kappa.a_c.L$ against total porosity ( $\varepsilon_t$) is presented in Figure 3.26 (left). From this curve, one can identify either $a_c$ or $L$ for known total and strut porosity. The dimensionless parameter, $\kappa.a_c.L$ increases with increase in strut porosity, resulting in lower specific area for a given total porosity. A dimensionless curve relating $\kappa.\alpha.a_c.L$ with total porosity is also presented in Figure 3.26 (right). Using these curves, one can characterize all the geometrical parameters of any hollow strut. The derived correlations presented above can insight one to tailor their own foams accordingly. In this way, one can realize any number of foams depending upon the various engineering applications needs.

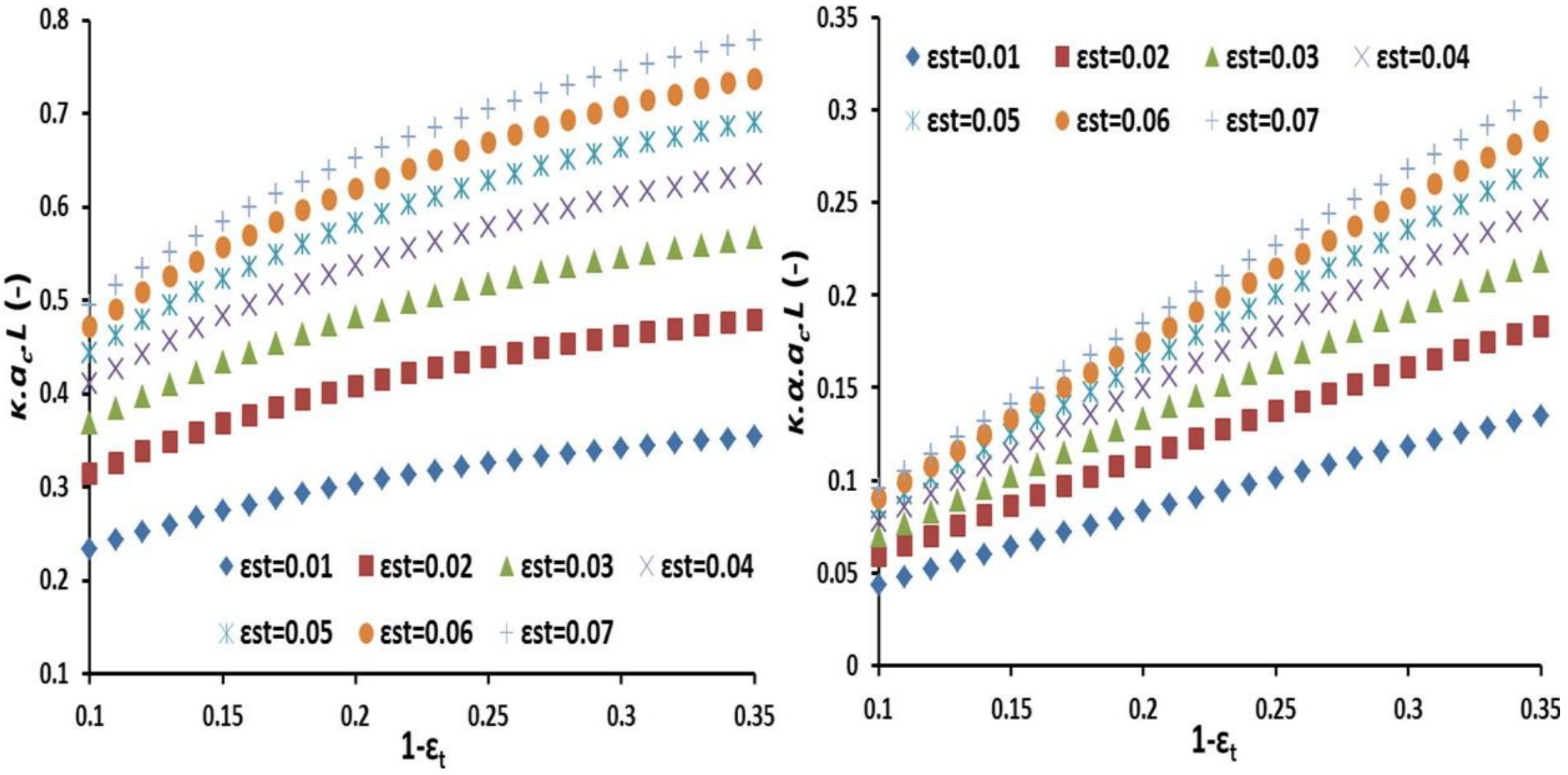

**Figure 3.26**. Plot of non-dimensional geometrical parameters vs. $1 - \varepsilon_t$ for different strut porosities, $\varepsilon_s$. Left: $\kappa.a_c.L$-function of hollow strut, specific surface area and node length. Right: $\kappa.\alpha.a_c.L$-function of hollow strut, strut diameter and specific surface area.

Note that the mathematical correlation is developed for porosity ranging from 65-90% for circular strut shape and triangular void. Richardson et al., (2000) proposed that for $\varepsilon_t$ >0.90, the strut shape changes from circular to triangular and possess a circular void. A mathematical formulation to determine geometrical properties and specific surface area for porosity, $\varepsilon_t$ >0.90 is presented in Appendix E.

## 3.5 Validation

In this section, the analytical correlations are validated against:

- Data measured on iMorph for cast metal foams.
- Data measured by classical CAD method and experimental data reported in the literature for various *isotropic* metal foams.
- Data measured by classical CAD method for virtual *anisotropic* foams.
- Experimental data reported in literature for ceramic foams.

### 3.5.1 Experimental validation of cast foam samples

Data measured on cast foams using iMorph and analytical approach are compared and validated for four foam samples. Specific surface area and node to node length (observed no ligament variation along its axis in cast foams) for different known strut diameters are compared and are presented in Table 3.7. From Table 3.7, is it clear that the hypothesis made while deriving analytical correlation works well for the cast CTIF foams. It shows that the CAD shape resembles well with the cast foam and precise value of geometrical parameters can be obtained analytically.

**Table 3.7**. Comparison and validation of morphological data and analytical approach for cast foam samples.

| Sample | $\varepsilon_o$ | Specific surface area, $a_c$ $(m^{-1})$ | | | Node to node length, $L$ $(mm)$ | | |
|---|---|---|---|---|---|---|---|
| | | iMorph Data | Analytical | Error | iMorph Data | Analytical | Error |
| 1 | 0.825 | 370.9 | 363 | 1.9% | 3.6 | 3.76 | 4.4% |
| 2 | 0.84 | 357.4 | 356 | 0.4% | 3.6 | 3.76 | 4.4% |
| 3 | 0.845 | 263.9 | 257 | 2.6% | 5.0 | 5.08 | 1.76% |
| 4 | 0.85 | 252.2 | 252 | 0.08% | 5.0 | 5.08 | 1.76% |

### 3.5.2 CAD and experimental validation of *isotropic* metal foams

In order to validate the proposed analytical correlation, data obtained on virtual CAD samples and experimental data are compared and found to be in excellent agreement. For CAD samples, strut and pore diameters and specific surface area of the different strut shapes for a given node length in the wide porosity range (low and high) are compared.

From the Figure 3.27, the bias in the strut diameters ($d_s$) and pore diameters ($d_p$) is observed up to 3% and 6% because of the geometrical approximations taken at the strut ligaments junction for complex strut shapes at low porosity only. The predicted values of the analytical $d_s$ and $d_p$ are underestimated and overestimated respectively.

The comparison of specific surface area obtained by classical CAD measurements and analytical approach is presented in Figure 3.28 (see also Table 2.3). The errors are within ±3% for different strut shapes but in the case of star strut shape, specific surface area is underestimated by 6% only at low porosities (0.75≤ $\varepsilon_o$ ≤0.80). The errors could be attributed to the approximated micro-structural node junction of complex strut shapes (e.g. star) at low porosities. The average deviation in the predicted specific surface area results is 0.88% and presented in Table 2.3. A maximum error of 4-6% is observed for circular, square, hexagon and rotated hexagon strut shape for porosity between 60-65% and can be attributed to the approximation of complex shape of node junction at low porosity.

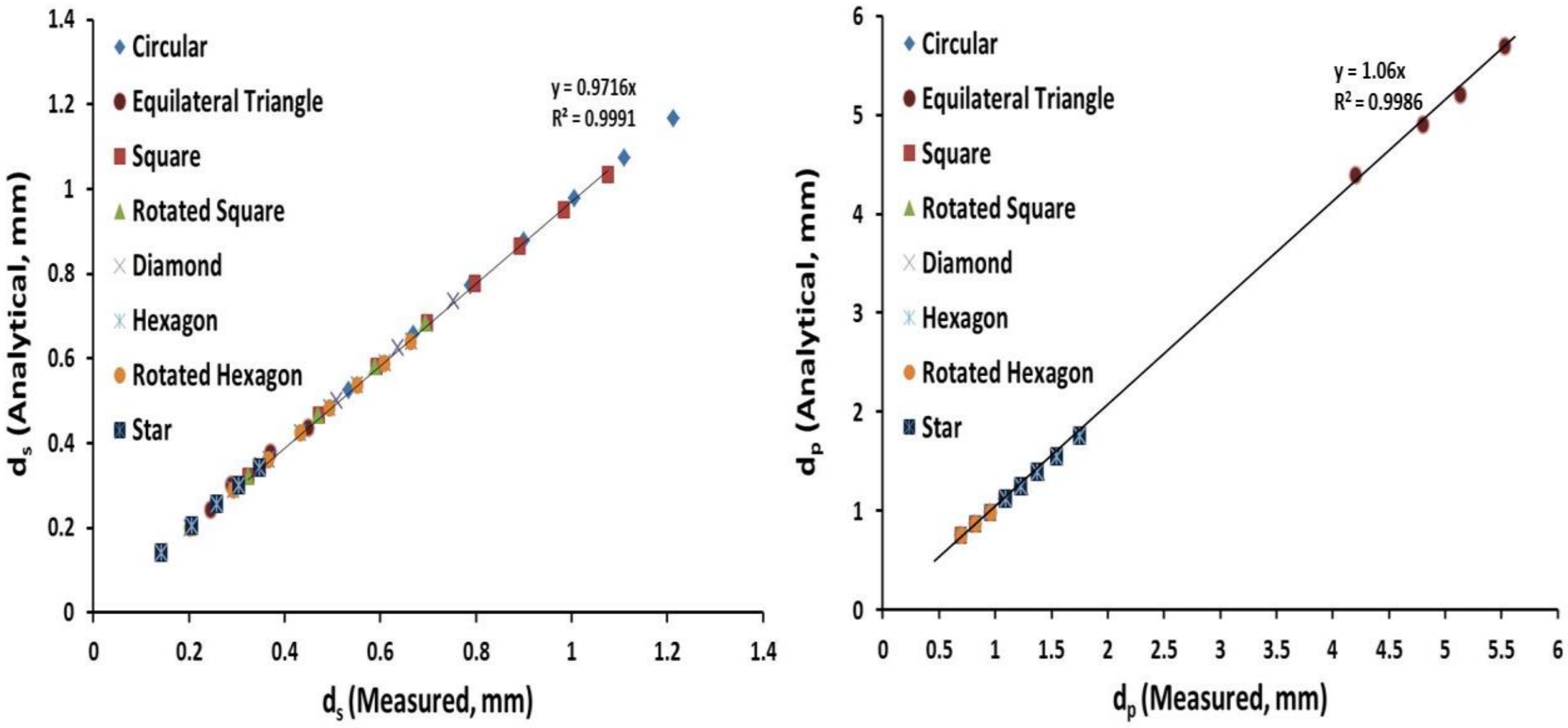

**Figure 3.27.** Validation of strut diameter (or side length) and pore diameter (left and right) of various shapes by classical CAD measurements and analytical approach.

Very few works exist in the literature with complete measurement of all geometrical properties (Perrot et al., 2007; Brun et al., 2009; De Jaeger et al., 2011). Moreover, there is always a bias existing in the literature over the tools to measure the geometrical parameters. In order to maintain a consistency in measuring the geometrical parameters, Brun et al., (2009) measured all the geometrical parameters of Recemat (Nickel-Chromium alloy), ERG (Al foams) and Fibernide (Ni foams) open cell foams. Their definition to measure pore diameter was based on equivalent included spherical diameter ($d_p{}^{eq}$) in the foam structure. Geometrical parameters of *isotropic* cast foam samples are already presented in Table 2.2. Perrot et al., (2007) and De Jaeger et al., (2011) measured geometrical parameters of their PU foams based on cell diameters that are orthotropic in nature and possess convex triangular strut shape. These yields 21 (Al/NC/Ni/PU) open cell foam samples and are presented in Table 3.8, allowing validation of the correlation. For validation, the foam samples of De Jaeger et al., (2011) and Perrot et al., (2007) were assumed as quasi-isotropic.

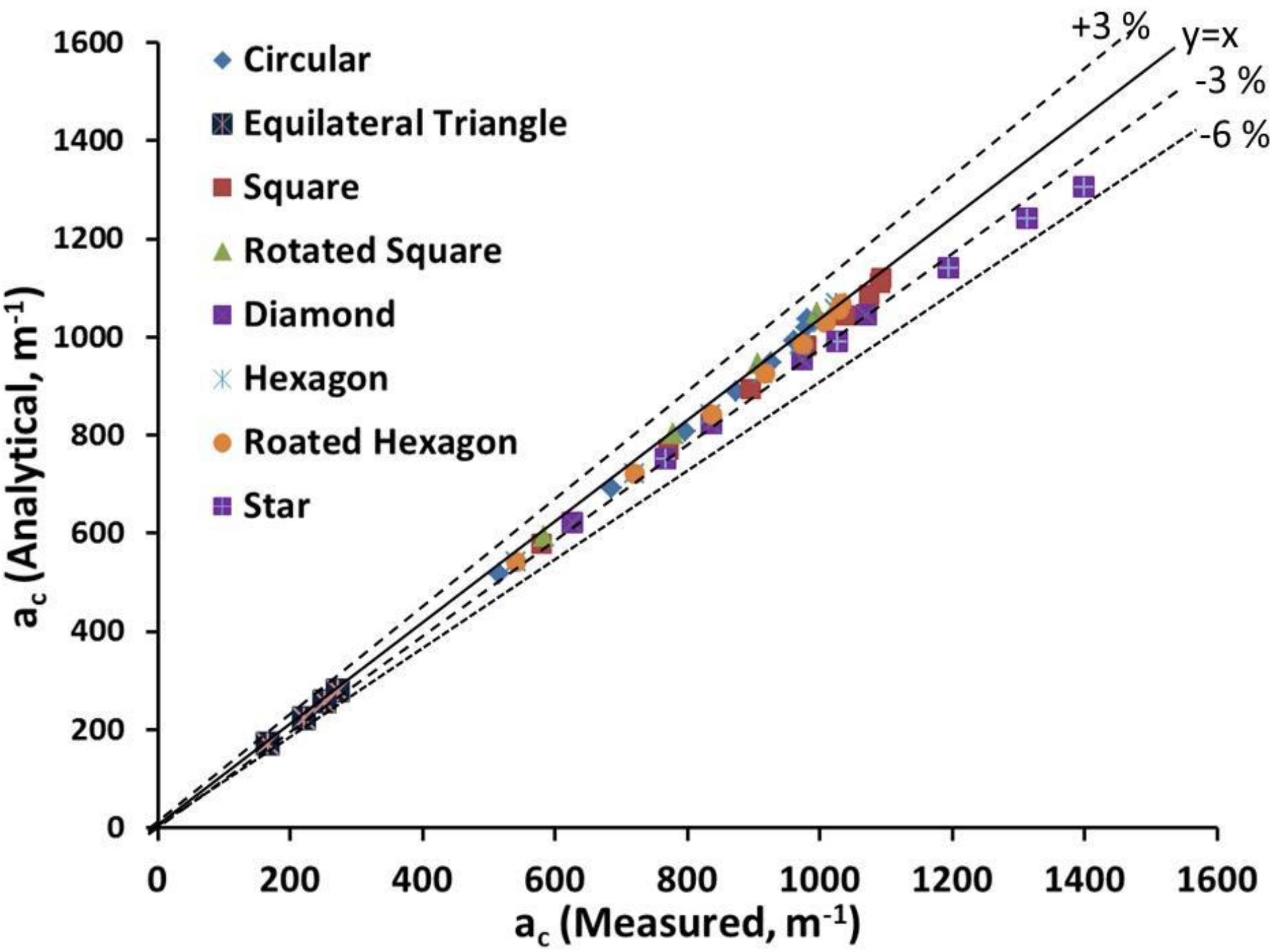

**Figure 3.28**. Validation of specific surface areas of various strut shapes by classical measurement and analytical approach.

For the 21 samples, the strut shape is approximated as equilateral triangle and using the methodology of equivalent radius, $R_{eq}$ presented in section 3.4.2 (see Table 3.6), specific surface area of 21 samples are compared and the errors associated with them are enlisted in Table 3.8. $\varepsilon_o$ was used to calculate $\alpha_t$ and $\beta$ followed by determining node length, $L$ using ${d_p}^{eq}$.

In Figure 3.29, it is obvious that the proposed analytical correlation is in good agreement with the measured and experimental specific surface areas within an error range of ±6%. The reason to compare different foams is to increase the scope of correlation validity over wide range of different strut shapes of different materials and different manufacturing techniques.

### 3.5.3 CAD validation of *anisotropic* metal foams

Specific surface area ($a_c$) obtained by classical CAD measurements and predicted values for virtual *anisotropic* open cell foams are compared and validated. In Figure 3.30, the analytical results of specific surface area are in excellent agreement for all elongation factors (error ±3%). However, the predicted results are underestimated (error -8%) only at low porosity (60-65%) which is due to the approximation of complex node junction and biased tendency of analytical $d_s$ (see also Figures 3.27 and 3.28). Specific surface areas of *anisotropic* open cell foams of different strut shapes are also compared at 90% porosity for entire range of elongation factor in Figure 3.31 and the predicted results are in excellent agreement. Furthermore, the specific surface area of 555 virtual *anisotropic* foam samples of all strut shapes are validated and presented in Table A.1 (see Appendix A). The maximum error observed is 9% for star strut shape at maximum elongation (*Ω*=3.0) at low porosity of 75%. The average error in analytical values of specific surface area for 555 *anisotropic* foam samples is 4%.

**Table 3.8**. Comparison of specific surface area by measured data and data reported in literature.

| | Data from direct measurement | | | | Analytical approach | | | | |
|---|---|---|---|---|---|---|---|---|---|
| Authors | Samples | $\varepsilon_o$ | $d_p{}^{eq}$ ($\mu m$) | $a_c$ ($m^{-1}$) | $\alpha_t$ | $\beta$ | $L$ ($\mu m$) | $a_c$ ($m^{-1}$) | Error (%) |
| Brun et al., (2009) | Recemat NC1116 | 89.6 | 2452 | 1300 | 0.4575 | 0.7283 | 913 | 1259 | -3.15 |
| | Recemat NC1723 | 87.3 | 1840 | 1740 | 0.5037 | 0.7008 | 691 | 1786 | 2.64 |
| | Recemat NC2733 | 90.9 | 831 | 4288 | 0.429 | 0.7452 | 270 | 4061 | -5.29 |
| | Recemat NC3743 | 87.3 | 569 | 5360 | 0.5037 | 0.7008 | 233 | 5299 | -1.14 |
| | ERG Al 10 | 89.2 | 4497 | 558 | 0.4721 | 0.7196 | 1780 | 626 | 12.19 |
| | ERG Al 20 | 88.9 | 3969 | 549 | 0.4659 | 0.7233 | 1920 | 607 | 10.56 |
| | ERG Al 40 | 88.5 | 3442 | 743 | 0.4802 | 0.7148 | 1480 | 805 | 8.34 |
| | Fibernide Ni10 | 89.5 | 4429 | 718 | 0.4597 | 0.727 | 1650 | 699 | -2.65 |
| Direct Mesaure -ment (iMorph) | Kelvin cell | 82.5 | 8700 | 371 | 0.5873 | 0.6512 | 3600 | 379 | 2.16 |
| | Kelvin cell | 84.0 | 8700 | 357 | 0.5627 | 0.6658 | 3600 | 368 | 3.08 |
| | Kelvin cell | 84.5 | 12400 | 263 | 0.5542 | 0.6708 | 5000 | 263 | 0.00 |
| | Kelvin cell | 85.0 | 12700 | 252 | 0.5456 | 0.6759 | 5000 | 260 | 3.17 |
| De Jaeger et al., (2011) | PPI 10 | 93.2 | 2540 | 440 | 0.3726 | 0.7787 | 2203 | 446 | 1.36 |
| | PPI 10 | 95.1 | 2540 | 380 | 0.3177 | 0.8113 | 2270 | 381 | 0.26 |
| | PPI 20 | 91.3 | 1270 | 860 | 0.4198 | 0.7506 | 1336 | 807 | -6.16 |
| | PPI 20 | 93.7 | 1270 | 720 | 0.3590 | 0.7867 | 1467 | 650 | -9.72 |
| | PPI 20 | 96.7 | 1270 | 580 | 0.262 | 0.8444 | 1297 | 565 | -2.59 |
| Perrot et al., (2007) | PPI 5 | 91.8 | 5080 | 431 | 0.408 | 0.7577 | 2312 | 456 | 5.80 |
| | PPI 10 | 91.8 | 2540 | 478 | 0.408 | 0.7577 | 2326 | 453 | -5.23 |
| | PPI 20 | 91.7 | 2540 | 624 | 0.4104 | 0.7563 | 1641 | 646 | 3.53 |
| | PPI 40 | 92.3 | 2540 | 700 | 0.3957 | 0.7649 | 1393 | 736 | 5.14 |
| Average Deviation | | | | | | | | | 1.06% |

*Pore diameter, $d_p{}^{eq}$ of grey blocks (samples in PPI) are estimated as $d_p{}^{eq} = 25.4/PPI$.

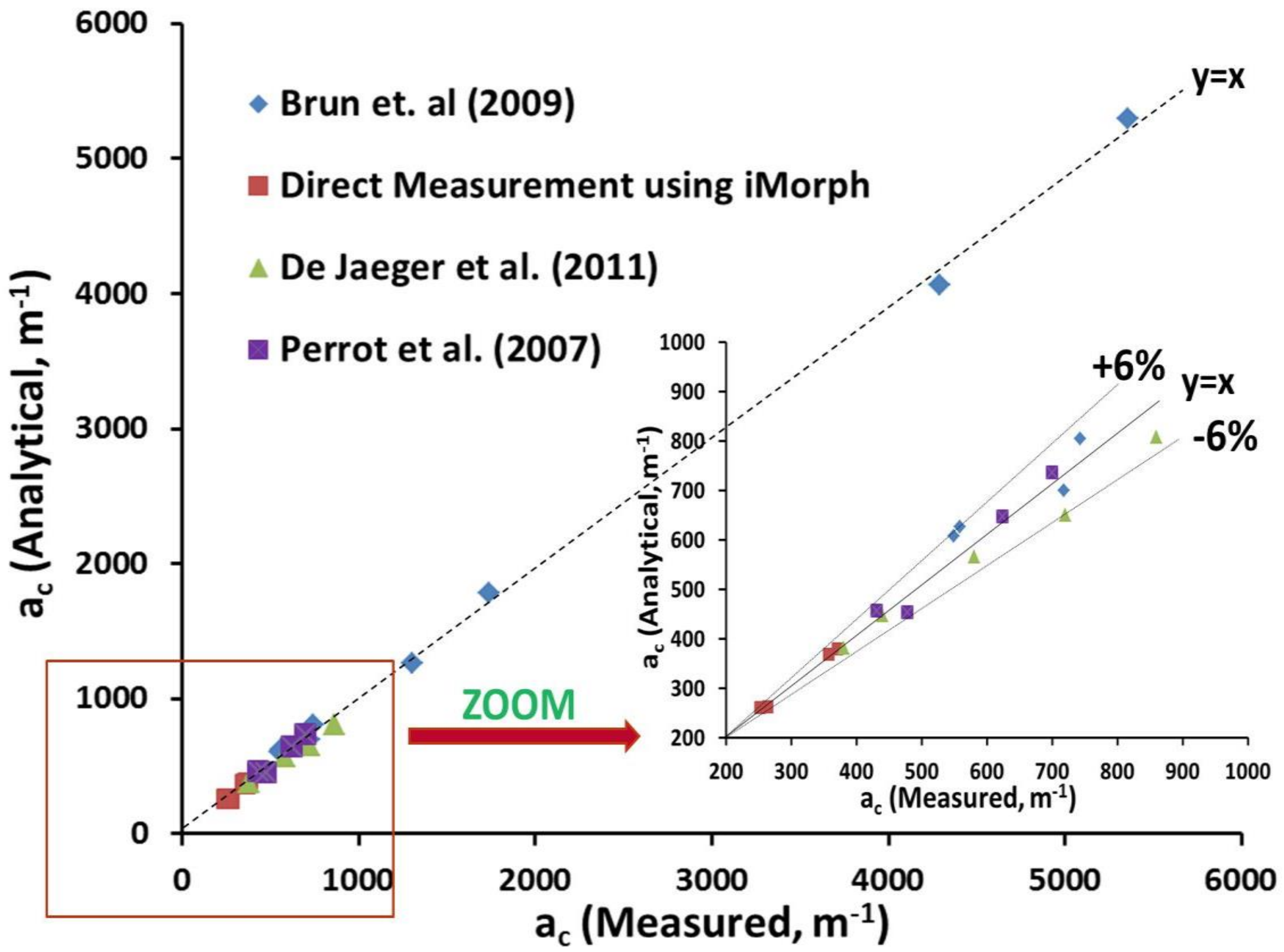

**Figure 3.29.** Comparison of analytical and measured values of specific surface area. Globally all data points lie within ±6%.

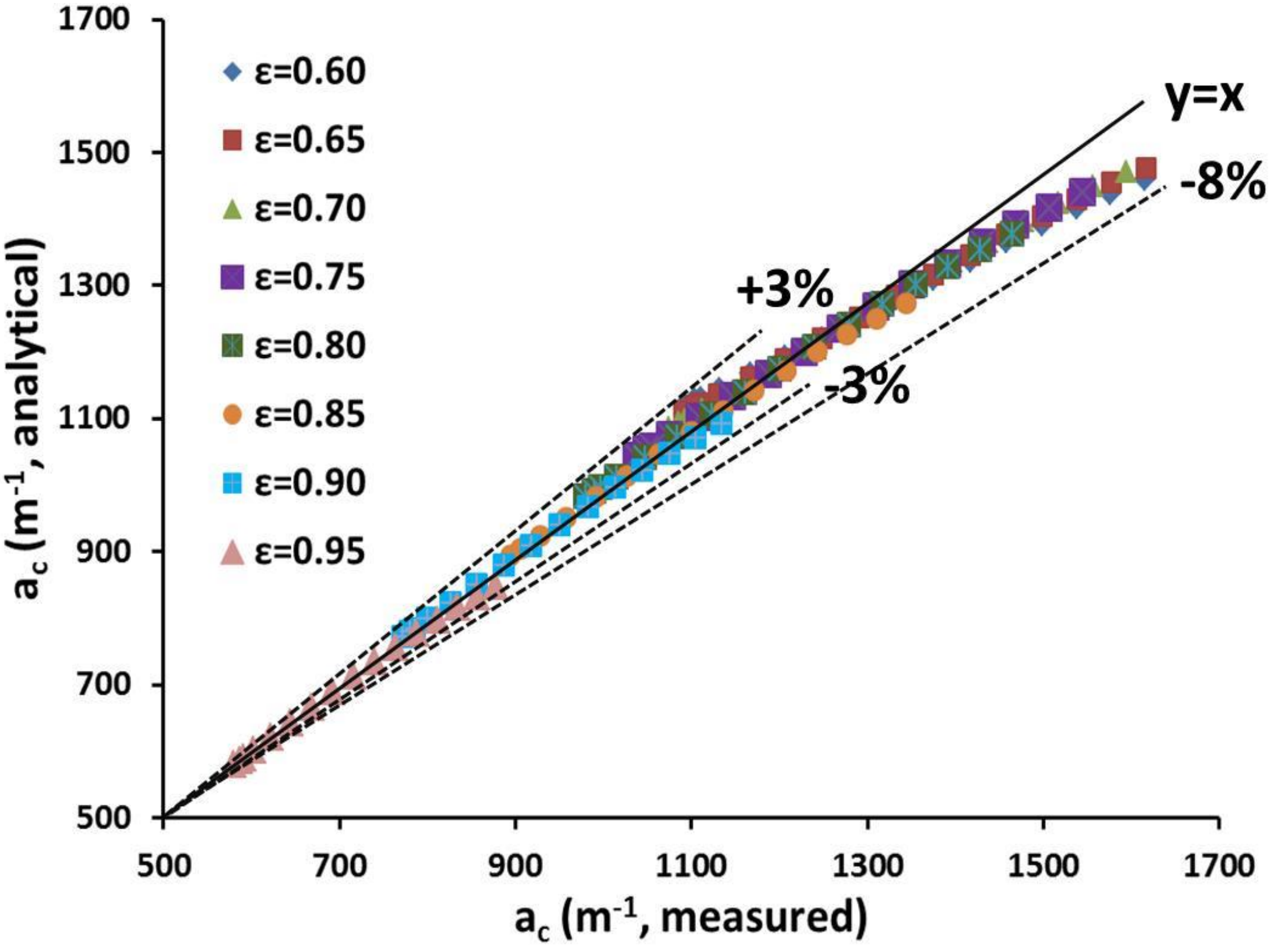

**Figure 3.30**. Comparison of specific surface area ($a_c$) obtained from CAD data with proposed analytical correlation: Square strut shape for entire range of porosity and elongation factors.

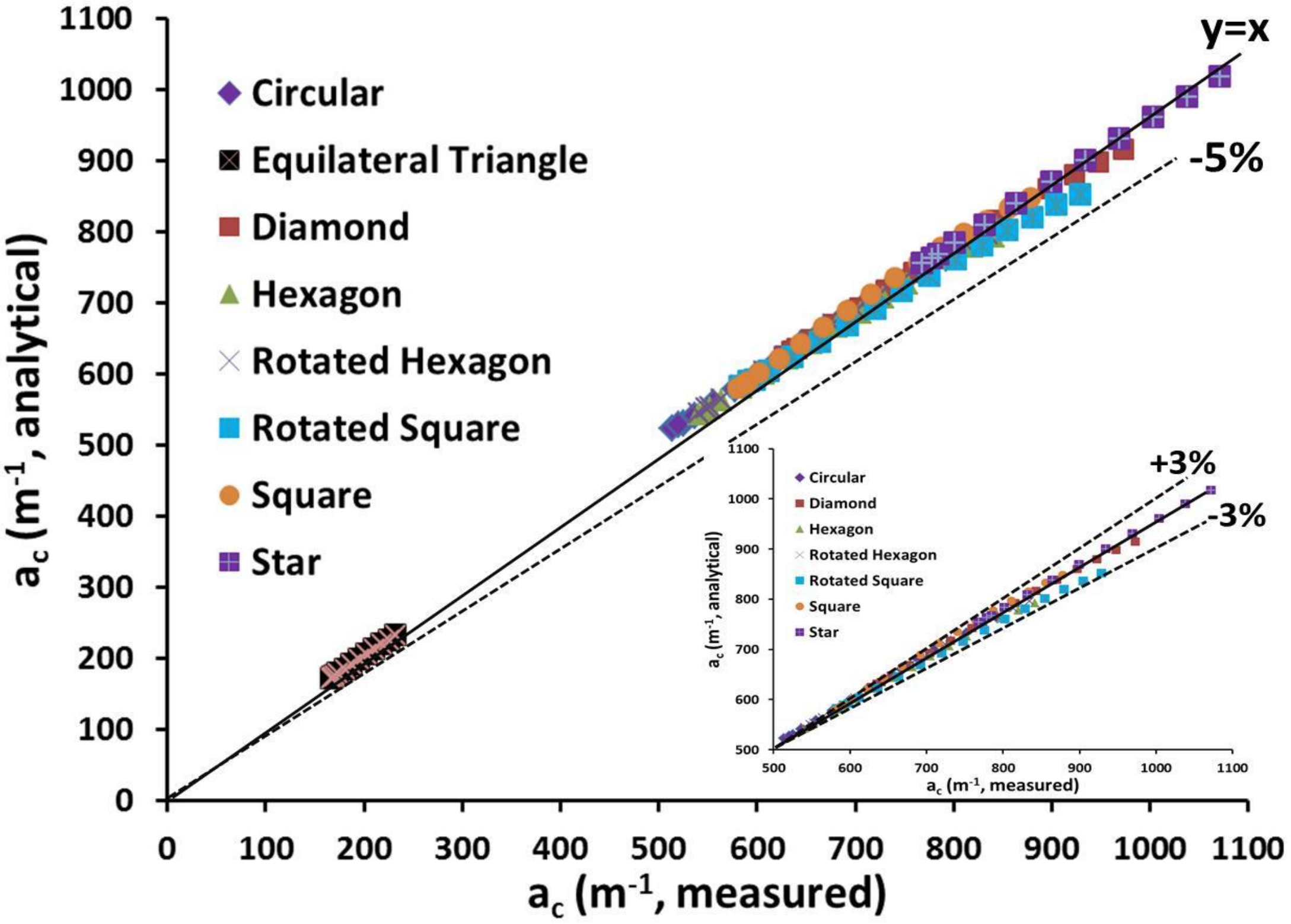

**Figure 3.31**. Comparison of specific surface area obtained from CAD data with proposed analytical correlation: Different strut shapes for all elongation factors at 90% porosity.

### 3.5.4 Experimental validation of *isotropic* ceramic foams

Data were gathered from experimental measurements performed on Alumina, Mullite, OBSiC, and SSiC ceramic foams as reported by several authors e.g. Garrido et al., 2008; Grosse et al., 2009; Dietrich et al., 2009; Inayat et al., 2011a. These are listed in Table 3.9. Theoretically, $\varepsilon_t$ and $\varepsilon_o$ are independent, but in the results shown in Figure 3.32, one observes that for existing material foams, there is a direct relation between $\varepsilon_t$ and $\varepsilon_o$ (see also Table 3.9). This point is used to derive an approximate value of $\varepsilon_o$ for a few Mullite and OBSiC ceramic samples that were not measured by Dietrich et al., (2009).

Following the general trend of void formation of ceramic foams of different materials, a fitting relation (see Figure 3.32) was determined between $\varepsilon_t$ and $\varepsilon_o$ which is given by Equation 3.89 as:

$$\varepsilon_t = 0.9942\varepsilon_o \tag{3.89}$$

This fitting relation (Equation 3.89) gives access to calculate the specific surface area analytically for the samples whose open porosity are unknown (see Table 3.9). A wide range of specific surface areas measured by different authors (Garrido et al., 2008; Grosse et al., 2009; Dietrich et al., 2009; Inayat et al., 2011a) was validated against the calculated results and are presented in Table 3.9. The analytical results are in good agreement and the average deviation of 28 different foam samples of different materials is 1.58%.

**Table 3.9**. Properties of $Al_2O_3$, Mullite, OBSiC and SSiC foams of different porosities and pore sizes. Experimental and analytical data are presented.

| | | Experimental Data | | | | | | | Analytical Data | | |
|---|---|---|---|---|---|---|---|---|---|---|---|
| Authors | Material | $d_w$ | $d_s$ | $\varepsilon_n$ | $\varepsilon_t$ | $\varepsilon_o$ | $a_c$ (MRI) | $a_c$ (*ref.) | $\varepsilon_o$ | $\varepsilon_s$ | $a_c$ |
| Garrido et al., (2008) | $Al_2O_3$ | 1.069 | 0.46 | 0.75 | 0.777 | 0.719 | 1290 | | | 0.058 | 1176 |
| | | 1.933 | 0.835 | 0.80 | 0.818 | 0.772 | 675 | | | 0.046 | 678 |
| | | 1.192 | 0.418 | 0.80 | 0.804 | 0.751 | 1187 | | | 0.053 | 1262 |
| | | 0.871 | 0.319 | 0.80 | 0.816 | 0.766 | 1438 | | | 0.05 | 1520 |
| | | 0.666 | 0.201 | 0.80 | 0.813 | 0.761 | 1884 | | | 0.052 | 2043 |
| | | 2.252 | 0.88 | 0.85 | 0.852 | 0.812 | 629 | | | 0.04 | 589 |
| | | 1.131 | 0.451 | 0.85 | 0.858 | 0.814 | 1109 | | | 0.044 | 1031 |
| | | 0.861 | 0.33 | 0.85 | 0.852 | 0.807 | 1422 | | | 0.045 | 1353 |
| | | 0.687 | 0.206 | 0.85 | 0.848 | 0.801 | 1816 | | | 0.047 | 2048 |
| Grosse et al., (2009) | | 1.974 | 1.001 | 0.75 | 0.75 | 0.688 | 639 | | | 0.062 | 646 |
| | | 1.070 | 0.651 | 0.75 | 0.736 | 0.719 | 1260 | | | 0.017 | 1176 |
| | | 1.796 | 0.944 | 0.80 | 0.794 | 0.773 | 664 | | | 0.021 | 641 |
| | | 0.955 | 0.509 | 0.80 | 0.814 | 0.745 | 1204 | | | 0.069 | 1107 |
| | | 0.847 | 0.391 | 0.80 | 0.816 | 0.754 | 1474 | | | 0.062 | 1246 |
| | | 0.781 | 0.276 | 0.80 | 0.801 | 0.763 | 1884 | | | 0.038 | 2012 |
| | | 1.952 | 0.809 | 0.85 | 0.848 | 0.812 | 629 | | | 0.036 | 593 |
| | | 1.137 | 0.544 | 0.85 | 0.853 | 0.813 | 1109 | | | 0.04 | 998 |
| | | 0.860 | 0.273 | 0.85 | 0.87 | 0.793 | 1520 | | | 0.077 | 1246 |
| | | 0.651 | 0.217 | 0.85 | 0.843 | 0.783 | 1816 | | | 0.06 | 1943 |
| Dietrich et al., (2009, *ref.) | $Al_2O_3$ | 1.529 | 0.651 | 0.75 | 0.754 | 0.69 | 1090 | | | 0.064 | 1155 |
| | | 2.253 | 0.967 | 0.80 | 0.808 | 0.765 | 664 | | | 0.043 | 612 |
| | | 1.091 | 0.476 | 0.80 | 0.802 | 0.748 | 1204 | | | 0.054 | 1290 |
| | | 0.884 | 0.391 | 0.80 | 0.806 | 0.752 | 1402 | | | 0.054 | 1542 |
| | | 0.625 | 0.195 | 0.80 | 0.809 | 0.757 | 1884 | | | 0.052 | 1801 |
| | | 1.464 | 0.544 | 0.85 | 0.854 | 0.811 | 1109 | | | 0.043 | 991 |
| | Mullite | 1.348 | 0.612 | 0.75 | 0.736 | - | - | 1035 | 0.695 | 0.041 | 1160 |
| | | 2.111 | 0.895 | 0.80 | 0.785 | - | - | 638 | 0.741 | 0.044 | 654 |
| | | 1.405 | 0.545 | 0.80 | 0.789 | 0.741 | 1291 | | | 0.048 | 1187 |
| | | 1.127 | 0.533 | 0.80 | 0.793 | 0.748 | 1395 | | | 0.045 | 1190 |
| | | 0.685 | 0.293 | 0.80 | 0.797 | 0.744 | 2126 | | | 0.053 | 2143 |
| | | 1.522 | 0.51 | 0.85 | 0.834 | - | - | 879 | 0.787 | 0.047 | 846 |
| | OBSiC | 1.361 | 0.896 | 0.75 | 0.742 | - | - | 899 | 0.701 | 0.041 | 858 |
| | | 2.257 | 1.063 | 0.80 | 0.791 | - | - | 578 | 0.747 | 0.044 | 601 |
| | | 1.489 | 0.719 | 0.80 | 0.791 | - | - | 869 | 0.747 | 0.044 | 889 |
| | | 1.107 | 0.544 | 0.80 | 0.791 | - | - | 1162 | 0.747 | 0.044 | 1175 |
| | | 0.715 | 0.275 | 0.80 | 0.786 | - | - | 1938 | 0.742 | 0.044 | 2374 |
| | | 1.467 | 0.622 | 0.85 | 0.845 | - | - | 855 | 0.798 | 0.047 | 785 |
| Inayat et al., (2011a) | SSiC | 1.800 | 0.701 | 0.88 | 0.878 | 0.853 | 732 | | | 0.025 | 683 |
| | | 1.297 | 0.480 | 0.90 | 0.896 | 0.873 | 858 | | | 0.023 | 784 |
| | | 1.030 | 0.399 | 0.90 | 0.885 | 0.862 | 1136 | | | 0.023 | 1042 |
| Average Deviation (%) | | | | | | | | | | | 1.58 |

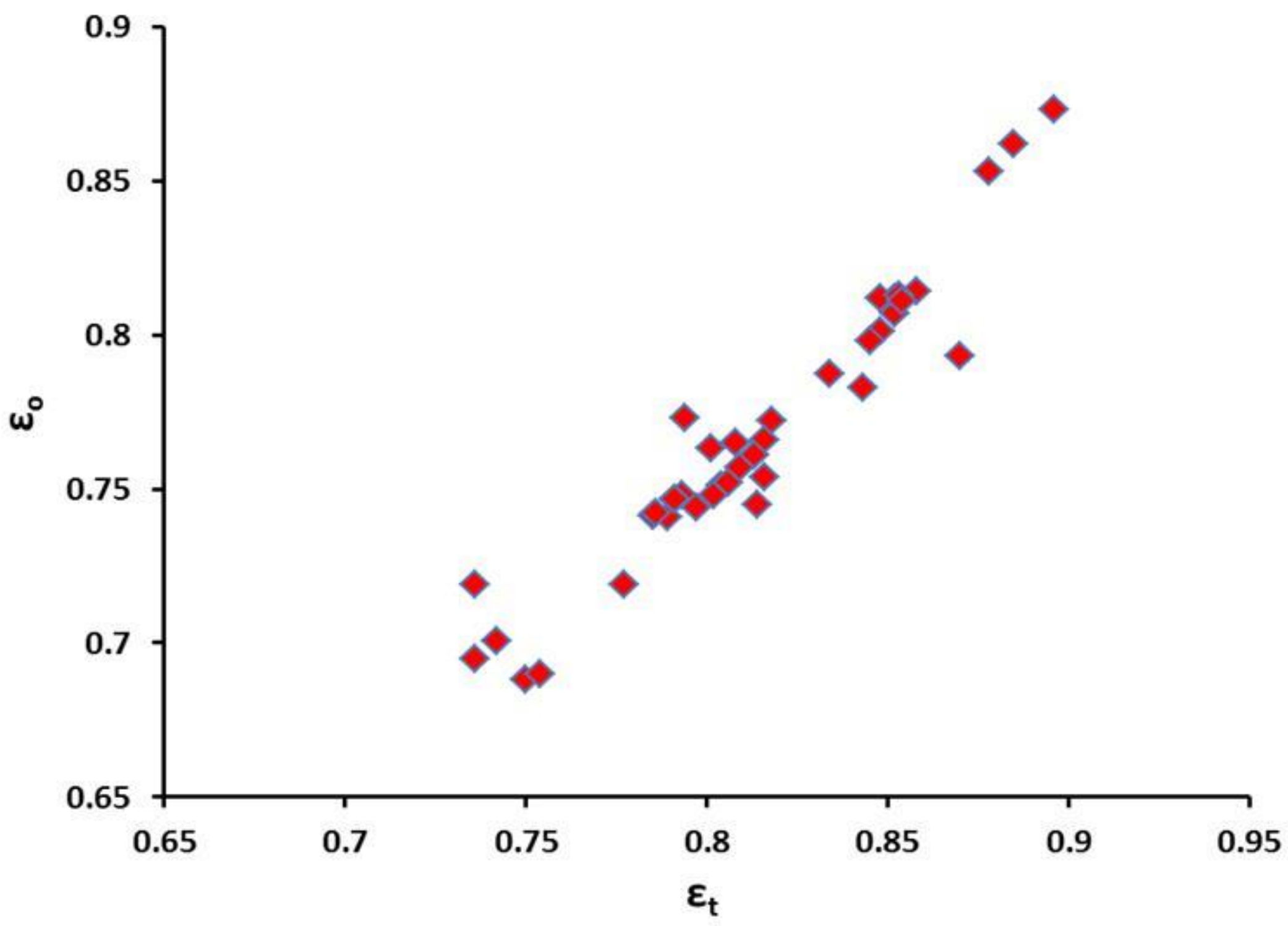

**Figure 3.32**. Relationship between total porosity, $\varepsilon_t$ and open porosity, $\varepsilon_o$ for different aterial ceramic foams. Experimental data are taken from Garrido et al., 2008; Grosse et al., 2009; Dietrich et al., 2009; Inayat et al., 2011a.

### 3.6 Summary and Conclusion

Open cell foams offer many advantages over known and established materials used for construction and engineering applications due to several attractive properties including acoustics, vibration absorption, heat and mass transfer enhancement etc. They inherit complex 3-D structure and their geometrical properties are very difficult to determine compared to the packed bed spheres. On the other hand, experimental characterization of open cell foams can be time-consuming and sometimes very expensive (e.g. Application of BET, MRI, µ-CT etc.). Alternatively, geometrical properties of foams can be predicted by using mathematical correlations (which can be derived using certain geometric model) that require some measured parameters for the calculation.

In the literature, several geometric models and correlations for predicting/estimating the foam properties have been proposed. In this chapter, an overview of important geometric models (cubic cell model, tetrakaidecahedron model, and pentagonal dodecahedron structure) and the resulting correlations to predict geometrical parameters and specific surface area of open cell foams has been presented. The validity of the correlations for predicting geometrical parameters and specific surface area has been evaluated by comparing the theoretical and experimental data of the foam structures from the literature as well as from the present work.

An important conclusion from the evaluation and comparison of theoretical and experimental data of foam properties from the state of the art is that, despite a number of publications on this subject during the last two decades, no generally applicable correlation for predicting geometrical properties and specific surface area has been proposed so far. It is interesting to note that in the literature; only a few authors have compared their predicted geometric specific surface area with the experimental data. The comparative study presented in this chapter shows that nearly all state of the art correlations tend to overestimate the experimental data in the low porosity range ($0.70< \varepsilon_o <0.90$). For high porosity range ($\varepsilon_o >0.90$), most of the experimental data and predicted results from correlations appear to be predicting the same results. These deviations of theoretical results from the experimental data can be ascribed to two main factors i.e. selection of geometric model and variation or change in strut cross section with porosity. Therefore, a single correlation developed for a certain strut shape cannot account for different strut cross sections.

It is worth noting that one could obtain the same geometrical properties between a CAD model and materialized cast foam. It is thus, of great significance to tailor the foams of desired output for numerous engineering applications. A generalized correlation to evaluate porosity has been established for different strut shapes to obtain geometrical parameters of *isotropic* metal foams and various relationships between them. Moreover, a certain strut shape cannot account for other strut cross sections and thus, different correlations were derived on the common basis to predict specific surface area for different strut shapes. Importantly, the methodology to predict either geometrical parameters or specific surface area of any strut shape is same.

To the best of our knowledge, anisotropy of open cell foams (metal foams) has not yet been studied to predict analytically the geometrical parameters and specific surface area of different strut shapes. *Anisotropic* foams have proven to be used as a replacement for strict restrictions in terms of porosity, strut shape and specific surface area. Strut shapes and porosities can be negotiated for a given application and structural constraint.

The correlation procedure of metal foams is extended to derive correlations for ceramic foams that can be easily applied to any void shape and strut shape. Thus, the analytical approach is of quite importance to derive tailored properties of ceramic foams and could optimize the strut shapes and foam matrices for different applications.

Moreover, the correlations (and curves) presented in section 3.4 for metal and ceramic foams are inter-dependent. If any of the two geometrical parameters are known, other geometrical properties can be easily determined. For different engineering problems and structural constraints, these correlations (and curves) would be very useful to predict the foam strut shape and geometry before end and also help in optimizing the foam structure.

New correlations derived in the present work have produced results with minimum deviation from the experimental and measured data of the foams of different materials in a wide range of porosities and complex strut shapes. It is to be noted that the correlations are based on the space filling tetrakaidecahedron geometry and is based on the theoretical aspects only as no empirical fitting of coefficients is involved. It is therefore, concluded that the tetrakaidecahedron model is the most suitable model to describe the geometrical configuration of the open cell foams.

The basics of the correlations derived for metal foams (either for family of convex and concave triangular shapes or different strut shapes) and ceramic foams lies on the same foundation. The algorithms that are based on only two inputs (or two measured geometrical properties) in case of metal foams and three inputs (due to internal cavity) in case of ceramic foams to characterize and predict other pertinent geometrical properties are provided. Figure 3.33 shows the algorithm of convex and concave triangular strut shapes of *isotropic* metal foam matrix. Similarly, an algorithm to predict all geometrical properties for both *isotropic* and *anisotropic* metal foam matrix of different strut shapes is presented in Figure 3.34. Lastly, an algorithm to predict geometrical properties of *isotropic* ceramic foams for different void sizes (internal cavity) and strut shapes is presented in Figure 3.35.

Internal cavity can also be found in metal (Ni, Cu, Ag...) foams produced by replication technique. In this work, for the sake of simplicity, solid foams are marked as metal foams while hollow foams as ceramic foams. Moreover, the correlations derived in this chapter are applicable to different types of solid and hollow foams irrespective of the technique used to produce them.

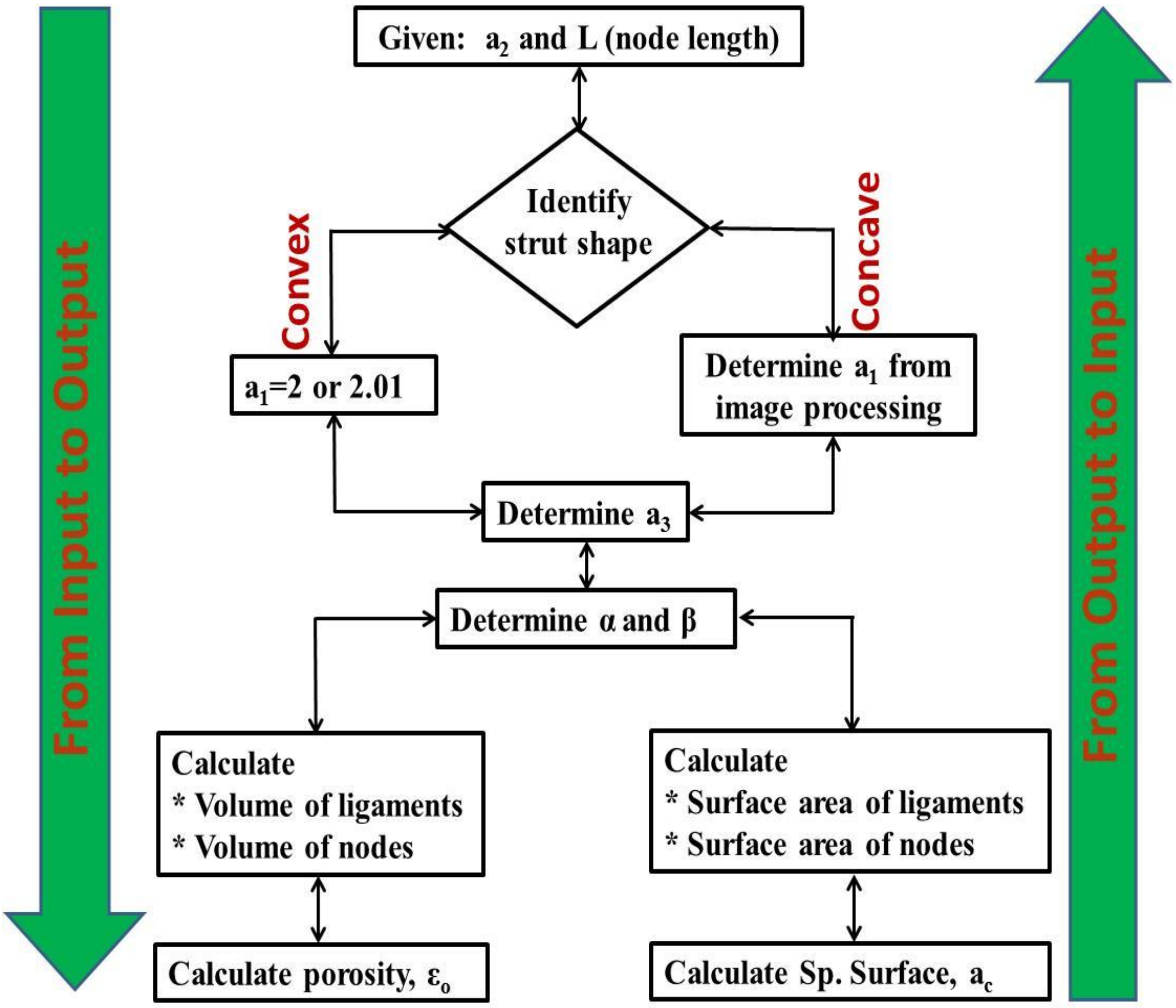

**Figure 3.33.** An algorithm to characterize all the geometrical characteristics of convex or concave triangular strut cross section *isotropic* open cell metal foams. A method to calculate porosity and specific surface area by knowing any two geometrical parameters. This algorithm can be used in reciprocal way- from input to output and vice versa.

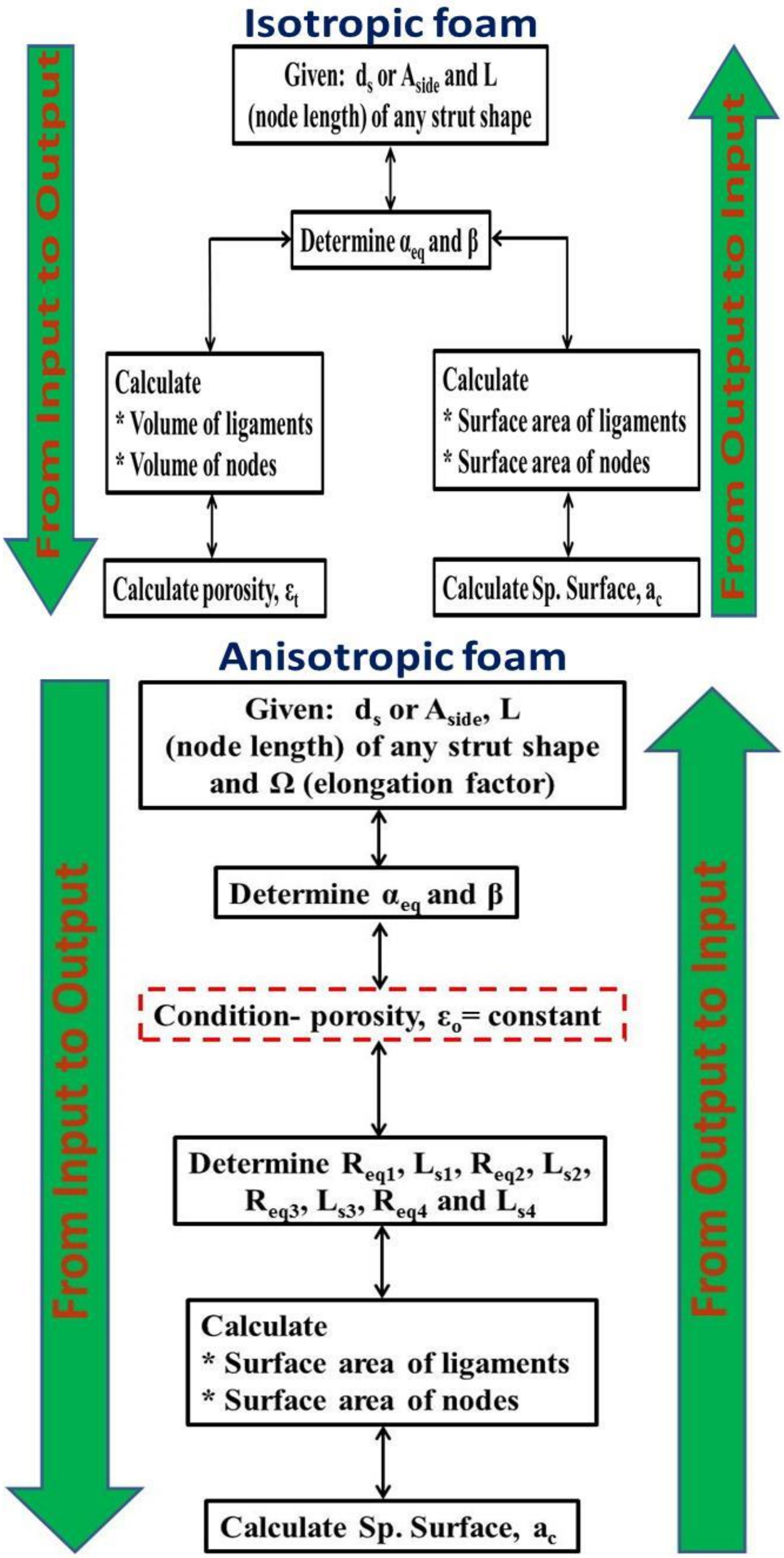

**Figure 3.34.** An algorithm to characterize all the geometrical characteristics of different strut cross section for *isotropic* and *anisotropic* open cell metal foams. A method to calculate porosity and specific surface area by knowing any two geometrical parameters. This algorithm can be used in reciprocal way- from input to output and vice versa.

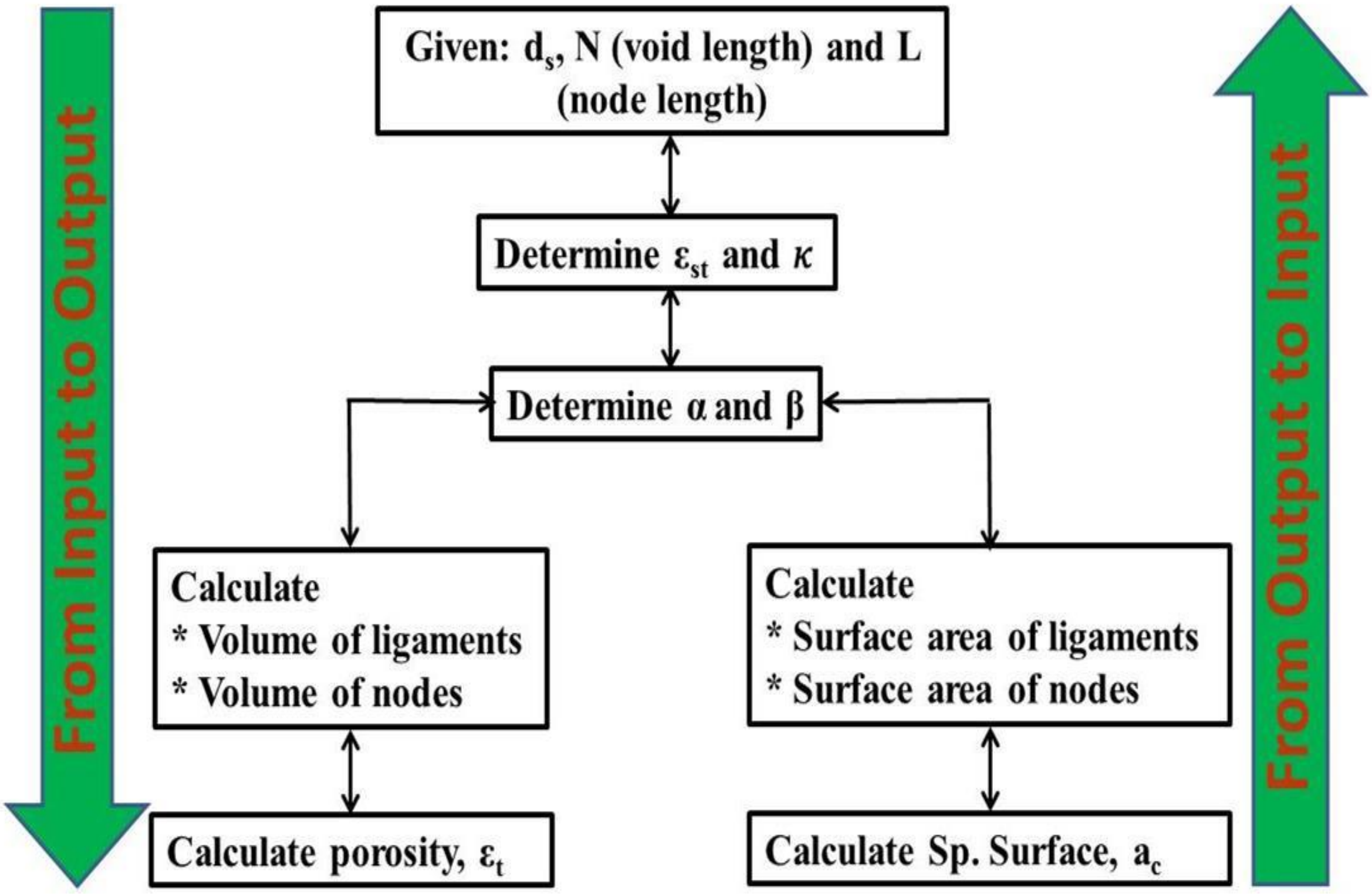

**Figure 3.35.** An algorithm to characterize all the geometrical characteristics of *isotropic* ceramic open cell foams. A method to calculate porosity and specific surface area by knowing three geometrical parameters. This algorithm can be used in reciprocal way- from input to output and vice versa.

# Chapter 4

# Pressure drop through open cell foams

Parts of this work were already published or submitted to Acta Materialia, Chemical Engineering Science, Transport in Porous Media and Defects and Diffusion Forum.

- **P. Kumar** & F. Topin, The geometric and thermo-hydraulic characterization of ceramic foams: An analytical approach, *Acta Materilia*, 75, pp. 273-286, 2014.

- **P. Kumar** & F. Topin, Investigation of fluid flow properties in open cell foams: Darcy and weak inertia regimes, *Chemical Engineering Science,* 116, pp. 793-805.

- **P. Kumar** & F. Topin, Micro-structural impact of different strut shapes and porosity on hydraulic properties of Kelvin like metal foams, *Transport in Porous Media*, 105 (1), pp. 57-81, 2014.

- **P. Kumar** & F. Topin, About thermo-hydraulic properties of open cell foams: Pore scale numerical analysis of strut shapes, *Defects and Diffusion Forum*, 354, pp. 195-200, 2014.

Parts of this work were already communicated in national/international conferences.

- **P. Kumar**, J.M. Hugo, F. Topin, J. Vicente, Influence of pore and strut shape on open cell metal foam bulk properties, *American Institute of Physics*, Conference proceedings, 1453, pp. 243-248, 2012.

- **P. Kumar** & F. Topin, Hydraulic properties of Kelvin like foam: Influence of porosity and strut shape, *INTERPORE-2013*, Prague, *Czech Republic*.

- **P. Kumar** & F. Topin, Propriétés Thermiques et Hydrauliques de Mousses Solides Régulières, *JEMP*-2012 (11$^{émé}$ Journées d'Etude sur les Milieux Poreux), Marseille, *France*.

- **P. Kumar**, J.M. Hugo & F. Topin, Experimental and pore scale numerical characterizations of thermo-physical properties of CTIF's Kelvin's cell foam, *METFOAM-2011*, Busan, *South Korea.*

## 4.1 Background

Although the transport phenomena in porous media have been studied for nearly two centuries, the work on highly porous materials are still relatively few and recent. Due to their specific structure in particular, and variability of texture related to the different routes of preparation, open cell foams are poorly characterized in term of transport properties.

For numerous industrial applications employing porous media e.g., filtration, heat exchange and chemical reaction etc., the pressure drop upon fluid flow in open cell foams has been extensively investigated. In case of heat exchangers or reactor designing, study of pressure drop is crucial to identify fan power requirement and a design goal is to minimize this power expenditure. Integration of open cell foams these days to various industrial systems has been increased tremendously due to their increasing compactness (higher heat transfer surface area per unit volume) that leads to an increase in pressure drop. This expectation is reinforced by the complex geometry of the foams which results in a high degree of boundary layer restarting and wake destruction by mixing.

The efficiency of these industrial processes depends greatly on the permeability of the porous medium employed therefore, it is very important to understand its fluid dynamics as well as the methods to evaluate its permeability or the pressure drop properties (Boomsma and Poulikakos, 2002; Richardson et al., 2002; Scheffler and Colombo, 2005; Innocentini et al., 2006; Garrido et al., 2008; Bonnet et al., 2008; Dietrich et al., 2009; Dukhan and Minjeur, 2011; Inayat et al., 2011b; Dietrich, 2012). Many researchers have performed fluid flow experiments through open cell foams (metal and ceramic) in order to pursue a complete flow characterization of these kinds of cellular materials. However, more efforts are needed in order to fully understand the behaviour of these materials in fluid flow applications.

Pressure drop in open cell foams are normally categorized in two ways: experiments and analytical correlations. Generally, extraction of flow properties i.e. permeability ($K$) and inertia coefficient ($C$) are the main concern as they do not depend on the fluid properties. In the literature, the overall results obtained for flow properties ($K$ and $C$) plotted against the morphological parameters (generally, PPI or porosity or pore diameter) of the foams are highly dispersed. Bonnet et al., (2008) showed that at fixed pore size, pressure drop generated by the foams disperse over three orders of magnitude (errors up to 300%) at low velocity, and two orders of magnitude at high velocity (errors up to 200%). These errors could be easily related to the choice of the flow law i.e. Darcy, Forcheimmer and Cubic law to extract flow properties and also pore size characterization introduces lot of discrepancies (Bonnet et al., 2008). Moreover, the discrepancies in both definition and measurement of pore size could also easily explain the dispersion of flow law results. In case of homothetic foams, one could conclude that pore diameter is sufficient enough to characterize flow law. However, in this work (foam samples based on controlled cell size), for almost the same pore size, it is difficult to characterise flow properties with pore diameter because it is not directly proportional to specific surface area but is proportional to the complex function of geometrical parameters (see Table 4.2). PPI is indeed not an adequate parameter to characterize the pore size (see chapter 3); different foam samples commercially available of same PPI are commonly different and exhibit different properties.

Based on the experimental results, various correlations have been developed to predict the pressure drop using Ergun-like approach based on the pore or particle diameter and porosity. Due to such phenomenal errors in the experimental flow properties, none of the correlations bears a general applicability. To the best of our knowledge, no pressure drop correlations in the literature are found to be integrated with the geometrical parameters of foam matrix.

In this chapter, a summary of the pressure drop research made with open cell foams is discussed. An overview of the state of the art of experimental findings and correlations for pressure drop prediction in open cell foams is presented in the following section 4.2. 3-D pore scale numerical simulations were performed in order to extract the flow properties is presented in the section 4.3. The numerically obtained pressure drop data are compared to explore the effect of foam geometry. A methodology to determine flow properties is presented. The applicability and validity of pressure drop correlations reported in the literature are discussed and examined with the present work in the section 4.4. As flow characteristics are strongly dependent on strut shapes, various correlations on a common basis are derived to predict the pressure drop accurately through metal and ceramic foams in the sections 4.5 and 4.6. Importantly, the flow characteristics are found to be strongly dependent on geometrical parameters of foam structure that are not generally discussed in the literature. A discussion about the constant numerical values of Ergun parameters is presented and a methodology to reduce the dispersion in friction factor using the numerical pressure drop data is presented. Lastly, all analytical results were compared and validated against the experimental and numerical pressure drop properties of metal and ceramic foams in the section 4.7.

## 4.2 Literature review of pressure drop experiments and correlations

This section is categorized in three parts: choice of flow laws, experimental flow properties and correlations. Depending upon the Reynolds number range, there is often an ambiguity to choose the flow law to extract the flow properties data. For a studied Reynolds number range, it is a common practice to extract the flow properties by fitting the pressure drop data either on polynomial curve or cubic one. Pressure drop correlations are generally derived by fitting the flow properties using Ergun-like approach. As discussed in the section 4.1 that permeability and inertia coefficient values are dispersed over three and two orders of magnitude; it is not surprising that one correlation could not be generalized and applied on variety of foam samples. Moreover, it is common in practice by various authors to provide the correlations only with porosity and pore diameter.

### 4.2.1 Choice of flow law

At the macroscopic scale, three forms of flow laws are mainly used to describe the pressure drop when a fluid flows in a porous medium. The laws of Darcy (Darcy, 1856) and Forchheimer (Forchheimer, 1901) are two most common laws that have been used in porous media. Historically, these laws have both been validated by experiments, for flows in media type "grain bed" where the porosity does not exceed 50 %.

However, several authors have shown by means of theoretical calculations that in certain flow conditions, the dependence of the pressure drop was cubic with the flow rate (Wodie and Levy, 1991; Mei and Auriault, 1991). These authors also validated their concept of flow law (called cubic law) with experimental data of Darcy (see also Firdaous et al., 1997).

#### 4.2.1.1 Darcy Law

The concept of permeability was first reported by Henry Darcy in 1856. The experiments were conducted on a local fountain through beds of sand for various thicknesses and it was showed that the velocity over the sand bed was directly proportional to the driving pressure and inversely proportional to the thickness of the bed. Darcy’s law (Equation 4.1) takes into account the viscous effects on the fluid pressure drop and establishes a linear relationship between pressure gradient and the fluid velocity through the porous media (Innocentini et al., 2006; Edouard et al., 2008a; Zeiser, 2008).

$$\frac{\Delta P}{\Delta x} = \frac{\mu}{K_D} V \tag{4.1}$$

The permeability $K_D$ characterizes the resistance to the flow of the porous medium. Note that, permeability is independent of the fluid nature. It is the function of the foam geometry of porous medium only.

Experimentally, the linearity of the relationship between the pressure drop $\Delta P$ and velocity $V$ is only checked at low velocity. This model is particularly useful in the field of hydrogeology where the flows take place in the environment with low porosity ($\varepsilon_o$ <0.5) and where flow rates are often below 0.1 $ms^{-1}$. However, as soon as the flow velocity increases, nonlinear effects start appearing and then this model has its limitations for the description and prediction of pressure drop.

#### 4.2.1.2 Forcheimmer Law

Forchheimer extended Darcy's equation by introducing the inertial effects in addition to the viscous effects on the fluid pressure drop in a porous medium. This gave a parabolic dependence of pressure gradient on the fluid velocity. Forchheimer equation can be expressed as follows:

$$\frac{\Delta P}{\Delta x} = \frac{\mu}{K_D} V + \rho C_{For} V^2 \tag{4.2}$$

#### 4.2.1.3 Cubic Law

The cubic law is proposed by some authors to describe the flow in the transition zone between viscous and inertia regimes. In this zone, the quadratic term of Forchheimer is replaced by a term that varies with the cubic form of velocity (Mei and Auriault 1991; Wodie and Levy, 1991; Firdaous et al., 1997).

$$\frac{\Delta P}{\Delta x} = \frac{\mu}{K_D} V + \frac{\rho^2}{\mu} C_L V^3 \tag{4.3}$$

In the above Equation 4.3, $C_L$ is a dimensionless parameter and is known as cubic coefficient. It is an intrinsic property of the porous medium through the fluid flow.

*Remarks about Cubic law and cubic coefficient*

The multiplicative factor $C_L.\rho^2/\mu$ involves the inverse of the dynamic viscosity. In this formulation, the viscous term of the pressure drop is proportional to $\mu$ while the transition term of the pressure drop is inversely proportional to $\mu$.

The cubic law, by analogy can be reduced to Forchheimer law in which expression of the inertial coefficient $C_{For}$ becomes a linear function of velocity.

$$\frac{\Delta P}{\Delta x} = \frac{\mu}{K_D} V + \rho C_{For}(V) V^2 \quad \Rightarrow C_{For}(V) = \frac{C_L.\rho}{\mu} V \tag{4.4}$$

The dependence of the inertial coefficient with velocity makes it more intrinsic to the porous medium through which it modifies in a fundamental way the properties of the Forchheimer model.

**Ergun Equation**

The most widely accepted interpretation of Forchheimer equation was presented by Ergun and Orning, 1949 (see also Ergun, 1952). These authors proposed the following equation to estimate the pressure drop in packed bed of spheres:

$$\frac{\Delta P}{\Delta x} = E_1 \frac{(1-\varepsilon_o)^2}{{\varepsilon_o}^3} {a_c}^2 \mu V + E_2 \frac{(1-\varepsilon_o)}{{\varepsilon_o}^3} a_c \rho V^2 \tag{4.5}$$

Note that the coefficients of viscous and inertial terms ($E_1$: 150 and $E_2$: 1.75) were empirically determined by Ergun (1952) by fitting a large amount of experimental pressure drop data on packed bed of spheres from the literature.

Many authors have further used Ergun equation to propose Ergun-like correlations for the pressure drop estimation/prediction in open cell foams (Macdonald et al., 1979; Du Plessis and Masliyah, 1991; Liu et al., 1994; Innocentini et al., 1999; Richardson et al., 2000; Tadrist et al., 2004; Moreira et al., 2004; Innocentini et al., 2006; Dukhan, 2006; Topin et al., 2006; Lacroix et al., 2007; Garrido et al., 2008; Dietrich et al., 2009; Huu et al., 2009; Mancin et al., 2010; Inayat et al., 2011b; Dietrich, 2012). Some authors (e.g. Lacrioix et al., 2007; Huu et al., 2009) have even used the same Ergun coefficients which were originally determined for the packed bed system. The permeability reported in all cases follows the Forchheimer law.

### 4.2.2 Monophasic flow experiments in open cell foams

The porous matrix of open cell foams consist of irregularly-shaped flow passages with a continuous disruption of any hydrodynamic boundary layers. The flow recirculates at the back of the solid fibers leading to turbulence and unsteady flows (Bastawros et al., 1998). The geometric complexity prevents the exact solutions of transport equations inside the pores (Lage et al., 1997; Bastawros 1998; Hwang et al., 2002). This led researchers to rely heavily on the experimentation and empirical models. (e.g. Seguin et al., 1998; Kim et al., 2000; Paek et al., 2000; Crosnier et al., 2003; Tadrist et al., 2004; Bonnet et al., 2008 etc.).

*It is worth noting that in this thesis, $K$(or $K_{For}$ or $K_{poly}$) and $C$ (or $C_{poly}$ ) are denoted to permeability and inertia coefficient respectively for the works where authors have extracted these values from second order polynomial curve fitting of pressure drop data.*

Lage et al., (1997) performed pressure drop experiments with 40 PPI samples in order verify what other authors reported in the literature as a decrease in pressure drop gradients beyond the Forchheimer regime. They found it contrary with experimental results with water and packed bed of spheres, that their results with air and aluminum porous medium layers yielded an increase in static pressure gradient as velocity was increased. They correlated their data and found that a cubic term model fitted best their experimental work.

Seguin et al., (1998) provided experimental characterization of flow regimes for various foam samples. These authors wanted to provide the limits to the laminar regime and found that the transition to turbulence for a pore diameter based Reynolds number ($Re_{d_p}$) was 470 which corresponds to the permeability based Reynolds number ($Re_{\sqrt{K}}$).

Bastawros et al., (1998) provided data for fluid flow experimental measurements of cellular metals subjected to transverse airflow in the Forchheimer regime. Experimental works on 30 PPI aluminum foams have been performed with porosities in the range of $\varepsilon_o$ >0.90. It was found that a power law was followed when pressure versus velocity plots were recorded.

Paek et al., (2000) performed experimental work with aluminum foams of different porosities in the range, 0.89< $\varepsilon_o$ <0.96 to determine the thermo-hydraulic properties. At a fixed porosity, increasing the cell size increased the specific surface area which therefore increased the flow resistance by lowering the permeability and increasing the pressure drop. So it was inferred that the permeability was influenced appreciably by both the porosity and the cell size. The friction factor was correlated with the permeability based Reynolds number.

Boomsma and Poulikakos (2002) performed experiments on aluminum foams of 40 PPI; compressed foam samples to various ratios and used water as the working fluid. The flow parameters for their samples were obtained with a fitting curve of the pressure drop against velocity data. It was found that the structural differences in the pre-compressed form between 92% and 95% porosity metal foams did not have a noticeable effect on the permeability and that similar compression factors had similar weighted effects. Increasing the compression factor decreased the permeability by regular incremental amounts and holding the porosity constant while decreasing the pore diameter decreased the permeability and increased the form coefficient. Finally, it was found that changing the velocity regime resulted in different values for the flow parameters.

Crosnier et al., (2003) performed experimental work with 20 and 40 PPI aluminum foam samples manufactured by ERG and a 20 PPI stainless steel sample with porosities greater than 90%. It was recorded that the transition to the turbulence regime was set to a Darican velocity of $1ms^{-1}$. It was revealed that with the increase in pore diameter of their foam samples, the value of the permeability increased with decreasing pressure drop. The passability was defined as the second order term in the Forchheimer Equation 5.2 which is the ratio of the inertia coefficient to the square root of the permeability. It was noted that as the pore size increased there was more variation in the permeability than the passability given to the fact that the permeability scales well with the square root of the pore size while the passability scaled well with the pore size. The permeability and the passability were functions of the porosity, pore size, surface area and solid structure of the foam.

Khayargoli et al., (2004) studied the effect of microstructure of the metal foams on flow parameters in the porosity range, 0.83< $\varepsilon_o$ <0.90. It was observed that the permeability increased and inertia coefficient decreased with increasing pore size but these properties did not show any clear relation with the porosity. These authors mentioned that there was a strong influence on the drag force exerted by the fluid due to which permeability increased with larger pores in the sample and finally concluded that although the flow phenomenon in the medium was complex, the flow parameters could be predicted using an Ergun-like equation.

Tadrist et al., (2004) investigated the aluminum metal foams for compact heat exchangers in high porosity range ($\varepsilon_o$ >0.90). Flow parameters (permeability and inertia coefficient) have been determined experimentally and used an Ergun-like approach to establish a correlation between the pressure drop and the fluid velocity.

Flow characteristics and pressure drop in open cell foams are crucial in many practical designs. There have always been inconsistencies in the reported data from the experiments. Generally, authors do not have full access to measure complete set of geometrical parameters of the foam matrix to relate them with pressure drop characteristics. Several authors (e.g. Madani et al., 2006; Bonnet et al., 2008, Edouard et al., 2008a) have reported that the standard deviation between experimental and theoretical values of the pressure drop in the literature can be as high as 3 orders of magnitude.

There is always a disparity with use of particle diameter ($D_p$) in Reynolds number. Most of the authors have replaced $D_p$ by $d_p$ to determine Reynolds number and subsequently, friction factor. For the random structure of porous media, the particle diameter is not trivial (see Dukhan and Patel, 2008).

The most widely used characteristic size is the pore diameter in determining Reynolds number (see Bonnet et al., 2008). This parameter is relatively easy to measure from 3-D tomography images. Nowadays, 3-D μ-CT scan are routinely used in research and industry and thus, both 3-D image and measurement software (open source or commercial) are easily accessible to measure geometrical properties of foam (see Vicente et al., 2006a). The strut size and the hydraulic diameter of the channel containing the porous medium are also often quoted. It has been proposed to use the value of $\sqrt{K}$ or $\sqrt{K}/\varepsilon_o$ (see Boomsma et al., 2003) which has the dimension of length and contains information about the viscous part of flow law. These formulations make it possible to evaluate the characteristic size from flow experiments, but cannot be used to quantify the structural influence of the solid matrix on the flow parameters as discussed by Bonnet et al., (2008).

During the flow experiments in open cell foams, it is very difficult to control the velocity of Darcy regime. All the studies presented in the literature review were performed at velocities which do not physically determine Darcy and inertia regimes separately. Madani et al., (2006) explained the feasibility to measure pressure drop precisely in Darcy regime as it is very small with the working fluids like water or air. Uncertainty analysis on experimental data was performed and it was shown that the viscous effects (or permeability) were not accurately measured. On the other hand, inertia coefficients were correctly measured.

Moreover, many authors have calculated universal inertial coefficient $c$ ($c = C\sqrt{K}$), that is frequently used to describe inertial effects in open cell foams, is meaningless due to great uncertainties concerning permeability. Lage and Antohe (2000) argued against the use of $c$.

In order to counter the dispersion in friction factor values, Dietrich et al., (2009) used hydraulic diameter ($d_h = 4\varepsilon_o/a_c$) that is one of the characteristic dimensions to determine correctly Reynolds number and should be related to the friction factor to improve the dispersion in friction factor (see Bonnet et al., 2008) as it contains porosity and specific surface area of the foam matrix.

Recently, Dukhan and Minjeur (2011) experimentally validated that fitting a quadratic curve for linear pressure drop ($\Delta P/L$) versus velocity ($V$) data to determine $K$ and $C$ can be misleading, as no flow regime change can be identified. Plotting data in this manner may encounter both the Darcy, transition and inertia regimes and can possibly mix the three, which leads to errors in $K$ and $C$. In the case of inertia regime, only inertia coefficient plays a significant role and extraction of permeability using polynomial curve is Forchheimer permeability ($K_{For}$ or $K_{poly}$) and that depends on the velocity range.

### 4.2.3 State of the art of pressure drop correlations

In the literature, several authors have presented their experimental work on pressure drop through the foam structures by proposing correlations for the pressure drop prediction (Du Plessis et al., 1994; Innocentini et al., 1999; Kim et al., 2000; Richardson et al., 2000; Fourie and Du Plessis, 2002; PhaniKumar and Mahajan, 2002; Moreira and Coury, 2004; Moreira et al., 2004; Giani et al., 2005; Liu et al., 2006; Dukhan, 2006; Lacroix et al., 2007; Garrido et al., 2008; Huu et al., 2009; Dietrich et al., 2009; Edouard et al., 2009; Eggenschwiler et al., 2009; Mancin et al., 2010; Bai and Chung, 2011; Inayat et al., 2011b).

Bonnet et al., (2008) and Edouard et al., (2008a) presented the synthesis of pressure drop correlations in the open cell foams. The correlations reported are the functions of several different geometric parameters such as pore size and porosity (e.g. Richardson et al., 2000; Moreira et al., 2004; Tadrist et al., 2004; Liu et al., 2006), tortuosity (e.g. Du Plessis et al., 1994; Fourie and Du Plessis, 2002) or even PPI grade foam (e.g. Dukhan 2006; Mancin et al., 2010). Various correlations by different authors are listed in Table 4.1.

It has been widely quoted by many authors (e.g. Innocentini et al., 1999; Richardson et al., 2000; Giani et al., 2005; Lacroix et al., 2007; Huu et al., 2009) to adapt the analogy between solid foam and spherical particles with the same specific surface area ($a_c$) and the same porosity ($\varepsilon_o$), leading to $D_p = 1.5d_s$ or $D_p = 6(1-\varepsilon_o)/a_c$. But this will not hold because pore size and pore density changes with manufacturing process (see Dukhan and Patel, 2008). Cellular materials do not exhibit the same geometry as packed beds of particles (e.g., void boundaries are concave rather than convex); yet, many authors (Innocentini et al., 1999; Richardson et al., 2000; Giani et al., 2005; Lacroix et al., 2007; Huu et al., 2009) followed the same procedures in developing pressure drop correlations for foams that inhibit significant errors in predicted pressure drop data. One of the serious limitations for these large spreads in proposed correlations is possibly to be a combination of measurement accuracy issues and over-simplification of the strut geometry and thus, cannot be applied directly to the open cell foams for pressure drop prediction.

The main limitation of these correlations (see Table 4.1) is that they cannot be applied to different sets of foam samples. For a given set of foam samples, one correlation allows achieving a good approximation of pressure drop only at very high porosity ($\varepsilon_o$ >0.90) but it does not take the geometrical parameters of foam structure into account. Few authors (e.g. Dukhan, 2006; Mancin et al., 2010) have correlated pressure drop with PPI. As discussed in chapters 2 and 3 that PPI represent nothing and is not at all suitable to derive or correlate flow properties.

Edouard et al., (2008a) gathered the pressure drop data from the literature and presented a review on the state of the art correlations of flow properties of foams. These authors concluded that it appears that *no model is perfect and that the standard deviation between experimental and theoretical values can be as high as 100 %*.

Mancin et al., (2010) presented a correlation to predict the pressure drop data with PPI in a high porosity range (0.90< $\varepsilon_o$ <0.93). These authors did not measure pore diameter and estimated pore diameter according to the simple relation of 2.54/PPI (in mm). PPI does not give any reliable information about the cell size and is quoted by manufacturer. These authors calculated Reynolds number based on square root of Forcheimmer permeability. All the relationships presented in their correlation were related to PPI and thus, may not be appropriate to other set of foam samples and could not be applied to the foams of low porosity ($\varepsilon_o$ <0.90).

Dietrich (2012) showed that the correlation derived by Dietrich et al., (2009) could predict the experimental pressure drop from open literature for both ceramic and metal foams with most of the data points lying between ± 40% in high porosity range (0.85< $\varepsilon_o$ <0.95). These authors obtained numerical constants as 110 and 1.45 for Ergun parameters ($E_1$ and $E_2$) respectively (see also Dietrich et al., 2009).

Commercially available foams of different materials presented in the literature lie in a small porosity range ($\varepsilon_o$~0.90±5%). Usually, at high porosity (0.85< $\varepsilon_o$ <0.95) and for a given pore size, the inertia coefficient does not vary much with respect to different strut shapes (circular, triangular, convex or concave triangular). For a known strut shape (e.g. triangular or circular strut shape) and small range of high porosity range (0.85< $\varepsilon_o$ <0.95) foam samples, constant numerical values of Ergun parameters could be possibly obtained.

On the other hand, in the case of low porosity (0.60< $\varepsilon_o$ <0.85) foam samples, inertia coefficient varies tremendously and depends strongly on strut shape. With respect to strut shapes of either metal or ceramic foams, Ergun parameters cannot have constant numerical values but they are the functions of foam geometry. In the literature, however, several authors agree that $E_1$ and $E_2$ are not constants but rather depend upon the properties of the porous medium (Richardson et al., 2000; Twigg and Richardson, 2002; Inayat et al., 2011b; Hugo and Topin, 2012).

Inayat et al., (2011b) developed a dimensionless correlation for Ergun parameters, $E_1$ and $E_2$; those depend upon the window diameter, strut diameter and open porosity of the foams. These authors argued that numerical values appearing in their correlations are geometric constants of foam geometry. Their correlations showed that Ergun-like equation with fixed parameters or unchanged coefficients ($E_1$ and $E_2$) cannot be applied to predict pressure drop for foam structures in the wide range of porosity 0.70< $\varepsilon_o$ <0.85.

In the literature, all the authors have compared and validated their experimental pressure drop against pressure drop predicted by the correlations for their set of foams where experiments are usually performed at moderate to high Reynolds number ($Re$ >10). Moreover, none of the authors have actually compared the flow properties (permeability and inertia coefficient) or pressure drop data separately in viscous and inertia regimes.

Usually, the experiments reported in the literature are performed at high velocity where the effect of viscous term is negligible and the pressure drop is mostly governed by inertia regime and thus, it is really difficult to stay in permeability regime as discussed by Madani et al., (2006), Bonnet et al. (2008) and Dukhan and Minjeur (2011).

The experimental data for the entire velocity range are mostly gathered in transition and inertia regimes and thus, supress the pressure drop values measured in viscous regime. One could obtain inertia coefficient ($C$) with sufficient accuracy from polynomial fitting but not permeability ($K$).

Literature review has suggested that pressure drop in open cell foams (metal or ceramic) is dependent not only on $d_p$ and $\varepsilon_o$ but also on the physical characteristics of the foam geometry and some of possible reasons of discrepancies in flow properties and correlations are highlighted below:

- Ill-measurement of geometrical parameters.
- Over-simplified geometry of unit cell (use of polyhedral, cubic lattice and pentagonal dodecahedron structure).
- Relation between pore diameter and strut diameter is inaccurate.
- Hypothesis of equivalent sphere particle and inaccurate models to predict specific surface area.
- Polynomial fitting of Forchheimer equation to extract of flow properties ($K$ and $C$).
- Characteristic dimension ($D_p$ or $d_p$) in Reynolds number is not defined on a common basis.

**Table 4.1.** State of the art correlation for predicting the pressure drop in open cell foams.

| Reference | Sample | PPI | $d_p \times 10^{-3}$ $(m)$ | $\varepsilon_o$ | $K \times 10^{-8}$ $(m^2)$ | $C$ $(m^{-1})$ | Correlation for pressure drop prediction |
|---|---|---|---|---|---|---|---|
| Giani et al., (2005) | Sample B | 5.4 | 4.7 | 0.927 | - | 161.58 | $f = 0.87 + \frac{13.56}{Re}; Re = \frac{\rho d_s V}{\mu}$ <br> $\frac{\Delta P}{\Delta x} = 13.56 \frac{{d_p}^3}{2(d_p - d_s)^4 . d_s} \mu V + 0.87 \frac{{d_p}^3}{2(d_p - d_s)^4 . d_s} \rho V^2$ |
| | Sample C | 11.5 | 2.2 | 0.938 | - | 531.93 | |
| | Sample D | 12.8 | 2.0 | 0.937 | - | 621.92 | |
| | Sample F | 5.6 | 4.6 | 0.911 | - | 285.22 | |
| Lu et al., (1998) | - | - | - | - | - | - | $f = \left[0.044 + \frac{0.008(d_p/d_s)}{(d_p/d_s - 1)^{0.43+1.13(d_s/d_p)}}\right] Re^{-0.15}$ <br> $Re = \frac{\rho d_s(\nu/1 - d_s/d_p)}{\mu}$ |
| Liu et al., (2006) | Sample 1 | 5 | 1.21 | 0.914 | 37.0 | 164.07 | $f = 0.22 + 22\frac{(1-\varepsilon_o)}{Re}; Re = \frac{\rho D_p V}{\mu}\ with\ D_p = 1.5 d_p \frac{(1-\varepsilon_o)}{\varepsilon_o}$ <br> $\frac{\Delta P}{\Delta x} = 22\frac{(1-\varepsilon_o)^2}{{\varepsilon_o}^3}\frac{\mu V}{{D_p}^2} + 0.22\frac{(1-\varepsilon_o)}{{\varepsilon_o}^3}\frac{\rho V^2}{D_p}$ |
| | Sample 2 | 10 | 1.19 | 0.918 | 62.3 | 230.58 | |
| | Sample 3 | 20 | 0.827 | 0.870 | 12.5 | 280.86 | |
| | Sample 4 | 20 | 0.0805 | 0.909 | 10.2 | 231.08 | |
| | Sample 5 | 20 | 0.0814 | 0.935 | 24.2 | 264.26 | |
| | Sample 6 | 20 | 0.08 | 0.958 | 142.0 | 285.32 | |
| | Sample 7 | 40 | 0.0685 | 0.935 | 13.3 | 282.43 | |
| Innocentini et al., (1999) | 30 PPI | 30 | 0.928 | 0.89 | 3.20 | 1587.3 | $f = 1.75 + 150\frac{(1-\varepsilon_o)}{Re}; Re = \frac{\rho D_p V}{\mu}\ with\ D_p = 1.5 d_p \frac{(1-\varepsilon_o)}{\varepsilon_o}$ <br> $\frac{\Delta P}{\Delta x} = 150\frac{(1-\varepsilon_o)^2}{{\varepsilon_o}^3}\frac{\mu V}{{D_p}^2} + 1.75\frac{(1-\varepsilon_o)}{{\varepsilon_o}^3}\frac{\rho V^2}{D_p}$ |
| | 45 PPI | 45 | 1.28 | 0.88 | 2.56 | 1176.4 | |
| | 60 PPI | 60 | 0.291 | 0.85 | 0.51 | 5555.5 | |
| | 75 PPI | 75 | 0.161 | 0.85 | 0.39 | 10000 | |
| Khayargoli et al., (2004) | NC 4753 | - | 0.4 | 0.86 | 0.162 | 2274.4 | $\frac{\Delta P}{\Delta x} = 100\frac{(1-\varepsilon_o)^2}{{\varepsilon_o}^3}\frac{\mu V}{{D_p}^2} + 1.0\frac{(1-\varepsilon_o)}{{\varepsilon_o}^3}\frac{\rho V^2}{D_p}; D_p = 1.5 d_p \frac{(1-\varepsilon_o)}{\varepsilon_o}$ |
| | NC3743 | - | 0.5 | 0.83 | 0.354 | 1748.1 | |
| | NC2733 | - | 0.6 | 0.9 | 0.501 | 1203.1 | |
| | NCX1723 | - | 0.9 | 0.885 | 1.53 | 488.72 | |
| | NCX1116 | - | 1.4 | 0.9 | 2.74 | 357.14 | |
| Richardson et al., (2000) | $Al_2O_3$ | 10 | 1.68 | 0.878 | 1.92 | 122.82 | $\frac{\Delta P}{\Delta x}=E_1\frac{(1-\varepsilon_o)^2}{{\varepsilon_o}^3}{a_c}^2\mu V+E_2\frac{(1-\varepsilon_o)}{{\varepsilon_o}^3}a_c\rho V^2$ |
| | $Al_2O_3$ | 30 | 0.826 | 0.874 | 0.482 | 431.03 | $E_1=973{d_p}^{0.743}(1-\varepsilon_o)^{-0.0982}$ |
| | $Al_2O_3$ | 45 | 0.619 | 0.802 | 0.396 | 876.55 | |
| | $Al_2O_3$ | 65 | 0.359 | 0.857 | 0.239 | 1948.2 | $E_2=368{d_p}^{-0.7523}(1-\varepsilon_o)^{0.07158}$ |
| Lacroix et al., (2007) | $SiC$ | - | 1.78 | 0.915 | 27.5 | 311.04 | $f=1.75+150\frac{(1-\varepsilon_o)}{Re};Re=\frac{\rho D_p V}{\mu}\ with\ D_p=1.5d_s$ |
| | $SiC$ | - | 2.68 | 0.91 | 58.9 | 214.43 | $\frac{\Delta P}{\Delta x}=150\frac{(1-\varepsilon_o)^2}{{\varepsilon_o}^3}\frac{\mu V}{{D_p}^2}+1.75\frac{(1-\varepsilon_o)}{{\varepsilon_o}^3}\frac{\rho V^2}{D_p}$ |
| | $SiC$ | - | 3.68 | 0.914 | 116 | 151.4 | |
| | $SiC$ | - | 1.78 | 0.89 | 20.9 | 372.1 | |
| | $SiC$ | - | 1.78 | 0.81 | 10.9 | 592.53 | |
| Moreira et al., (2004) | Sample 1 | 8 | 2.3 | 0.94 | 23.5 | 322 | $\frac{\Delta P}{\Delta x}=1.275\times10^9\frac{(1-\varepsilon_o)^2}{{\varepsilon_o}^3{d_p}^{-0.05}}\mu V+1.89\times10^4\frac{(1-\varepsilon_o)}{{\varepsilon_o}^3{d_p}^{-0.25}}\rho V^2$ |
| | Sample 2 | 20 | 0.8 | 0.88 | 8.07 | 580 | |
| | Sample 3 | 45 | 0.36 | 0.76 | 0.942 | 1708 | |
| Du Plessis et al., (1994); Fourie et Du Plessis (2002) | 45 PPI | 45 | 0.87 | 0.978 | 1.67 | 775 | $f=(3-\partial)(\partial-1)\frac{\rho_f\partial^2}{\mu{\varepsilon_o}^2(d_p+d_s)}\left[\frac{3A}{2}+\frac{B}{4}\right]V$ |
| | 60 PPI | 60 | 0.42 | 0.975 | 0.794 | 1014 | $A=\frac{24\varepsilon_o\mu}{\rho_f\partial(3-\partial)(d_p+d_s)V}$ |
| | 100 PPI | 100 | 0.25 | 0.973 | 0.234 | 2146 | $B=1+10\left[\frac{(d_p+d_s)(\partial-1)\rho_f V}{2\mu\varepsilon_o}\right]^{-0.667}$ |
| | 100 PPI | 100 | 0.25 | 0.973 | 0.181 | 3090 | $\frac{\Delta P}{\Delta x}=\frac{36\partial(\partial-1)(3-\partial)^2}{4{\varepsilon_o}^2{d_p}^2}\mu V+\frac{2.05\partial(\partial-1)}{2{\varepsilon_o}^2 d_p}\rho V^2$ |
| Tadrist et al., (2004) | 10 PPI | 10 | 3.97 | 0.917 | 13 | 128 | $\frac{\Delta P}{\Delta x}=c_1\frac{(1-\varepsilon_o)^2}{{\varepsilon_o}^3}\frac{\mu V}{{d_s}^2}+c_2\frac{(1-\varepsilon_o)}{{\varepsilon_o}^3}\frac{\rho V^2}{d_s}$ |
| | 20 PPI | 20 | 4.50 | 0.933 | 25 | 240 | $100\le c_1\le 8651\ \&\ 0.65\le c_2\le 2.6$ |
| | 40 PPI | 40 | 3.44 | 0.905 | 6.6 | 389 | |
| Topin et al., (2006) | NC3743 | - | 0.569 | 0.87 | 0.213 | 1330 | $\frac{\Delta P}{\Delta x}=\frac{1}{1.391\times10^{-4}}\frac{(1-\varepsilon_o)^2}{{\varepsilon_o}^3}\frac{\mu V}{{d_s}^2}+1.32a_c\rho V^2$ |
| | NC2733 | - | 0.831 | 0.91 | 0.444 | 1075 | |
| | NC1723 | - | 1.84 | 0.88 | 2.81 | 490 | |
| | NC1116 | - | 2.45 | 0.89 | 6.02 | 381 | |
| Dukhan (2006) | Sample 1 | 10 | - | 0.919 | 10 | 210 | $10\ PPI: K = 10^{-8}(0.0031e^{0.0955\varepsilon_o}), C = 100(-2.399\varepsilon_o + 222)$<br>$20\ PPI: K = 10^{-8}(0.0009e^{0.0946\varepsilon_o}), C = 100(-1.146\varepsilon_o + 108)$<br>$40\ PPI: K = 10^{-8}(8 \times 10^{-7}e^{0.0955\varepsilon_o}), C = 100(-0.613\varepsilon_o + 58)$ |
| | Sample 2 | 10 | - | 0.915 | 8.0 | 270 | |
| | Sample 3 | 20 | - | 0.919 | 6.3 | 290 | |
| | Sample 4 | 20 | - | 0.924 | 5.4 | 280 | |
| | Sample 5 | 40 | - | 0.923 | 4.7 | 380 | |
| Mancin et al., (2010) | Al-5-7.9 | 5 | 5.08 | 0.921 | 2.36 | 205 | $\frac{\Delta P}{\Delta x} = \frac{2FG^2}{\rho d_h} with\ G = \rho V\ \&\ F = \frac{1.765.\varepsilon_o{}^2.Re^{-0.1014}}{PPI^{0.6}}$<br>$Re = \frac{d_h.G}{\mu.\varepsilon_o}\ where, d_h = \frac{2\left(\frac{0.0254}{PPI} - d_s\right).L}{\left(\frac{0.0254}{PPI} - d_s + L\right)}$ |
| | Al-10-9.7 | 10 | 2.54 | 0.903 | 19.0 | 170 | |
| | Al-10-6.6 | 10 | 2.54 | 0.934 | 18.7 | 190 | |
| | Al-10-4.4 | 10 | 2.54 | 0.956 | 18.2 | 240 | |
| | Al-20-6.8 | 20 | 1.27 | 0.932 | 8.24 | 226 | |
| | Al-40-7.0 | 40 | 0.635 | 0.930 | 6.34 | 342 | |
| Inayat et al., (2011b) | Sample 1 | - | 3.085 | 0.871 | - | - | $\frac{\Delta P}{\Delta x} = E_1\frac{(1-\varepsilon_o)^2}{\varepsilon_o{}^3}a_c{}^2\mu V + E_2\frac{(1-\varepsilon_o)}{\varepsilon_o{}^3}a_c\rho V^2$<br>$E_1 = \left[\left(\frac{1-0.971(1-\varepsilon_o)^{0.5}}{0.6164(1-\varepsilon_o)^{0.5}}\right)\varepsilon_o\right]^{-1}$<br>$E_2 = \left[\left(\frac{1-0.971(1-\varepsilon_o)^{0.5}}{0.6164(1-\varepsilon_o)^{0.5}}\right)(1-\varepsilon_o)\right]$ |
| | Sample 2 | - | 2.397 | 0.846 | - | - | |
| | Sample 3 | - | 1.689 | 0.799 | - | - | |
| Dietrich et al., (2009) | $Al_2O_3$ | 20 | 1.529 | 0.69 | 13 | 1204.1 | $\frac{\Delta P}{\Delta x} = 110\frac{1}{\varepsilon_t}\frac{\mu V}{d_h{}^2} + 1.45\frac{1}{\varepsilon_t{}^2}\frac{\rho V^2}{d_h}; Hg = \frac{\Delta P}{\Delta x}\frac{d_h{}^3}{\rho V^2}$<br>$Hg = 110Re + 1.45Re^2\ with\ Re = \frac{\mu d_h}{\varepsilon_t \nu}$ |
| | $Al_2O_3$ | 10 | 2.253 | 0.765 | 7.7 | 492.88 | |
| | $Al_2O_3$ | 20 | 1.091 | 0.748 | 5.4 | 939.85 | |
| | $Al_2O_3$ | 30 | 0.884 | 0.752 | 3.2 | 1122.3 | |
| | $Al_2O_3$ | 45 | 0.625 | 0.757 | 2.0 | 1257.8 | |
| | $Al_2O_3$ | 20 | 1.464 | 0.811 | 14.4 | 496.44 | |
| | Mullite | 20 | 1.348 | - | 9.0 | 1179.7 | |
| | Mullite | 10 | 2.111 | - | 29.9 | 551.12 | |
| | Mullite | 20 | 1.405 | 0.741 | 8.8 | 753.64 | |
| | Mullite | 30 | 1.127 | 0.748 | 4.5 | 836.32 | |
| | Mullite | 45 | 0.685 | 0.744 | 2.9 | 1527.3 | |
| | Mullite | 20 | 1.522 | - | 12.0 | 506.55 | |
| | $OBSiC$ | 20 | 1.361 | - | 6.5 | 1004.6 | |
| | $OBSiC$ | 10 | 2.257 | - | 27.0 | 770.22 | |
| | $OBSiC$ | 20 | 1.489 | - | 5.6 | 831.72 | |
| | $OBSiC$ | 30 | 1.107 | - | 4.6 | 1203.8 | |

From the above paragraphs, it is very clear that no generally applicable correlation for the pressure drop prediction in the open cell foams has yet been proposed. Therefore, further work is definitely needed in this area. In the section 4.3, a new methodology is introduced to determine Darcian permeability ($K_D$) and Forcheimmer inertia coefficient ($C_{For}$) from pressure drop data. Further, new correlations are developed for different strut cross sections and validated against experimental and numerical pressure drop and flow properties data.

### 4.3 Pressure drop analysis

3-D pore scale numerical simulations were performed to determine the pressure gradient across the foam matrix in the section 4.3.1. Local analysis is performed to study the various flow regimes i.e. Darcy, transition and inertia regimes in the section 4.3.2. Further, an analysis is presented in the section 4.3.3 to determine flow law in the entire flow velocity range. Depending upon the flow regimes, a methodology is presented to determine $K_D$ and $C_{For}$ and their dependence on different pore sizes in the sections 4.3.4 and 4.3.5 respectively.

#### 4.3.1 Numerical simulations

Numerical simulations based on volume mesh generated from the actual solid surface were performed using CFD commercial code, StarCCM+ which is based on the finite volume method. The mesh is composed of core polyhedral meshes in the fluid phase. The sections in the X, Y and Z-directions and characteristic dimension of the unit periodic cell are presented in Figure 4.1. Navier-Stokes equations were solved using direct numerical simulations in the fluid phase with a segregated solver in order to determine flow laws parameters. These equations were solved for stationary and laminar conditions (no turbulence model) and only those results were taken into account to study flow properties where convergence was sufficient enough to consider that the flow is globally stationary. The symmetry conditions on four lateral faces and periodic velocity condition with prescribed pressure jump were applied between two faces normal to the main flow direction. The mesh independence study was performed for a 3-D periodic foam model.

The pressure and velocity fields were calculated for the entire fluid phase. From this data, pertinent values in order to determine flow parameters at macro-scale were extracted. The pressure gradients at each mesh point and pressure force exerted by the fluid on each solid-fluid interface point were calculated. From these local quantities, the averaged and integrated values were calculated in order to extract permeability and inertia coefficient. The calculations were performed until the values of Darcian permeability, $K_D$ and Forchheimer inertia coefficient, $C_{For}$ differed less than 1% between two consecutive mesh sizes. Mesh size (about 0.4 mm) was chosen in order to optimize the results reliability and computational time.

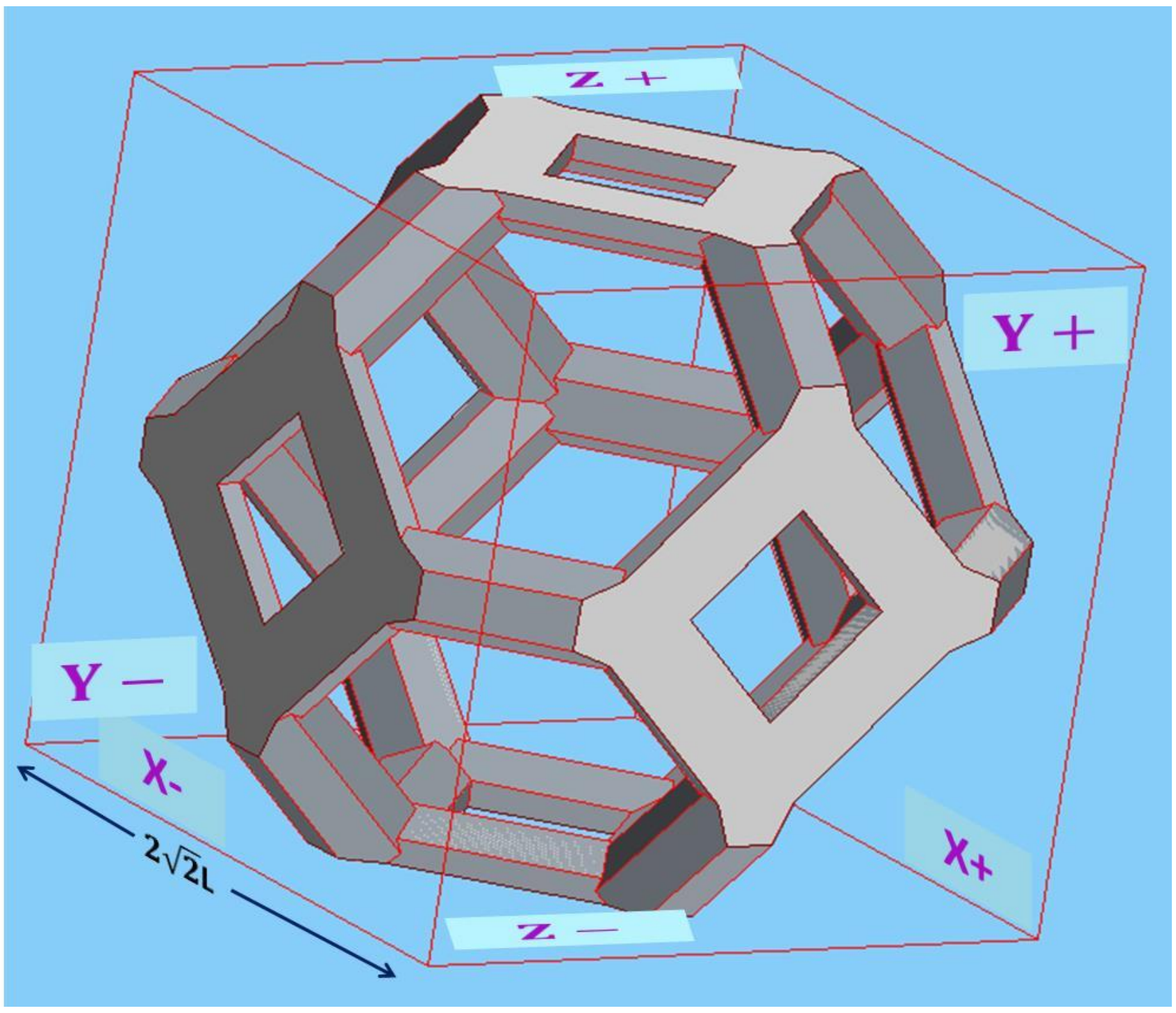

**Figure 4.1**. Presentation of tetrakaidecahedon model of square strut shape Kelvin like open-cell foam. Based on our construction method, the struts connecting the neighbouring cells are not included in the chosen cubic volume. Different sections in X, Y and Z directions are also marked for fluid flow calculations.

For each case, the convergence in terms of asymptotic behaviour of permeability $K_D$ was checked which was calculated at each iteration. The simulations were stopped when variations of $K_D$ were less than 0.1%. Moreover, the mass in-balance was also systematically checked and that there were no variations in the global flow.

Using a single unit cell model to simulate pressure drop in open cell foam takes advantage of the repeated cell structure of foams as well as the properties of flow through porous media. Here, periodicity was applied in only one direction (the X direction). In the other directions (the Y and Z directions), the remaining boundaries were set as symmetry planes as shown in the Figure 4.1. As permeability and inertia coefficient are independent of the fluid nature (see Bonnet et al., 2008), the fluid viscosity was varied from $10^{-7}$ to 1 $kgm^{-1}s^{-1}$ in order to carry out calculations for a wide range of Reynolds number. By doing so, a large amount of data points were generated to clearly identify Darcy, transition, and inertia regimes. The fluid medium used has a constant density of 998.5 $kgm^{-3}$. A pressure jump condition across the X direction (between sections X- and X+) was imposed and mass flow rate and subsequently, velocity was measured.

Below is the procedure used to determine macro-scale quantities from pore-scale numerically calculated information.

In the case of fluid flow in the Darcy regime, the macroscopic equation (see Whitaker, 1999) can be written as:

$$-\overline{\overline{K_D}}\nabla\langle P\rangle = \mu\langle V\rangle \tag{4.6}$$

$\nabla\langle P\rangle$ in Equation 4.6 can be determined by Equation 4.7:

$$\nabla\langle P\rangle = \langle\nabla P\rangle - \frac{1}{S}\int_S P.n_x\ dS \tag{4.7}$$

where, $S$ is the solid-fluid interface of the sample and $n_x$ is the unit vector normal to the elementary surface of integration $dS$.

For the flow at high Reynolds number, in which inertia effects are no longer negligible, the Forchheimer law is introduced. This law is rather empirical and many authors showed that it is well adapted to the fluid flow in foams (see Bonnet et al., 2008):

$$-\nabla\langle P\rangle = \mu\overline{\overline{K_D}^{-1}}\langle V\rangle + \rho\overline{\overline{C_{For}}}\|V\|\langle V\rangle \tag{4.8}$$

where, $\nabla\langle P\rangle$ is the average pressure gradient, $\overline{\overline{K_D}}$ is the permeability tensor of the foam, $\langle V\rangle$ is the average velocity over all the volume of the foam sample and $\overline{\overline{C_{For}}}$ is the tensor component of the inertia regime.

The flow properties relative to a given orientation of flow with respect to the foam were mainly studied. In case of *isotropic* foams, a check was made that the flow properties did not change when pressure jump was applied between sections Y+ and Y- (or Z+ and Z-) while making other sections symmetrical. All the calculations were carried out in X+ and X- sections for all the samples and thus, 1-D scale form of Equation 4.8 was used for which $\overline{\overline{K_D}}$ and $\overline{\overline{C_{For}}}$ reduce to scalars.

There are two ways to determine average pressure gradient. The first method consists of carrying out the measurements of the local pressure gradient over averaged volume ($\langle \nabla P \rangle$) and calculating the surface integral of the pressure exerted by the fluid on the solid foam surface ($\frac{1}{S}\int_S P.n_x \, ds$) as presented in Equation 4.7.

The second method (taking boundary conditions into account) is to measure the macroscopic pressure difference between inlet and outlet faces. In this case, one must take into account the surface porosity and is written as:

$$\nabla \langle P \rangle = \frac{\Delta \langle P \rangle_{fluid}^{surface}}{\Delta x} = \varepsilon_{sur} \cdot \frac{\Delta P}{\Delta x} \tag{4.9}$$

where, $\Delta \langle P \rangle_{fluid}^{surface}$ is the fluid pressure on foam surface and $\varepsilon_{sur}$ is the surface porosity at the faces of the inlet and outlet.

Similarly, the macroscopic velocity can be determined either by measuring velocity using mass flow rate or by volume averaging the local velocity over the entire cubic sample.

One case of 85% porosity (for equilateral triangular strut shape) is presented to verify the Equation 4.9 where $\varepsilon_{sur}$=0.76 was obtained. In the case of low Reynolds number ($Re$~0.5), the imposed macroscopic pressure difference between X+ and X- sections:

$$\frac{\Delta P}{\Delta x} = \begin{pmatrix} -7.042\text{E}-03 \\ 0 \\ 0 \end{pmatrix} \tag{4.10}$$

Measuring the volume average of local pressure and surface integral pressure at the solid surface, we get:

$$\langle \nabla P \rangle = \begin{pmatrix} -2.074\text{E}-03 \\ 7.172\text{E}-06 \\ 3.277\text{E}-07 \end{pmatrix} \text{and} \frac{1}{S}\int_S P.n_x \, dS = \begin{pmatrix} 3.510\text{E}-03 \\ 7.365\text{E}-06 \\ -2.150\text{E}-06 \end{pmatrix} \tag{4.11}$$

which gives,

$$\nabla \langle P \rangle = \begin{pmatrix} -5.345\text{E}-03 \\ -1.126\text{E}-06 \\ 2.435\text{E}-06 \end{pmatrix} = \varepsilon_{sur} \cdot \frac{\Delta P}{\Delta x} \tag{4.12}$$

The calculated components of $\nabla \langle P \rangle$ between Y+, Y-, Z+ and Z- sections are neglected. They are of the order of 0.1% of the main component in X+ and X- sections.

Similarly, the same procedure for high Reynolds number ($Re \sim 250$) was followed and imposed macroscopic pressure difference between X+ and X- sections:

$$\frac{\Delta P}{\Delta x} = \begin{pmatrix} -3.521\text{E}+01 \\ 0 \\ 0 \end{pmatrix} \tag{4.13}$$

Measuring the volume average of local pressure and surface integral pressure at the solid surface, we get:

$$\langle \nabla P \rangle = \begin{pmatrix} -1.802 \\ 1.232\text{E}-03 \\ 0.700\text{E}-03 \end{pmatrix} \text{ and } \frac{1}{S}\int_S P.n_x \; dS = \begin{pmatrix} 2.514\text{E}+01 \\ 0.796\text{E}-03 \\ 0.586\text{E}-03 \end{pmatrix} \tag{4.14}$$

which gives,

$$\nabla \langle P \rangle = \begin{pmatrix} -2.675\text{E}+01 \\ 0.436\text{E}-03 \\ -0.114\text{E}-03 \end{pmatrix} = \varepsilon_{sur} \cdot \frac{\Delta P}{\Delta x} \tag{4.15}$$

The two methods i.e. average pressure gradient method and imposed pressure drop are verified using numerical calculations in Darcy and inertia regimes using the periodic cubic volume of foam matrix; provide the same results (Equations 4.10-4.15). This also validates that using numerical simulations (see Equation 4.16); precise pressure fields can be easily determined:

$$\nabla \langle P \rangle = \langle \nabla P \rangle - \frac{1}{S}\int_S P.n_x \; ds = \varepsilon_{sur} \cdot \frac{\Delta P}{\Delta x} \tag{4.16}$$

Physically, there is a notable difference between the two calculations. At high Reynolds, the proportion of the pressure loss due to the fluid force on the solid ($\frac{1}{S}\int_S P.n_x \; ds$) is much larger than in the case of low Reynolds. All the flow properties were numerically calculated and listed in Table 4.2.

### 4.3.2 Local analysis of pressure drop

This section is divided in two parts: one part deals with only equilateral triangular strut shape while the other part deals with all other strut shapes. The foam matrix of equilateral triangular strut cross section has bigger pore size compared to other strut cross sections (see Table 4.2). The reason to separate them is to understand the flow fields which clearly depend on pore diameter or pore size.

*Case I: Impact of bigger pore size ($d_{cell}$=10 $\sqrt{2}$ mm)*

Using 3-D direct numerical calculations, the flow properties were studied at very low velocity in order to extract permeability in Darcy regime precisely and consequently, increased the velocity to enter in the inertia regime in both low and high porosity range (0.80< $\varepsilon_o$ <0.95). Equilateral triangular strut shape is studied in the case I. Hydraulic diameter ($d_h = 4\varepsilon_o/a_c$) is used to calculate Reynolds number in order to avoid discrepancies induced in Ergun-like approach and friction factor.

From the Figure 4.2 (top), it is clearly visible for $Re$ <1, flow characteristics are well settled in Darcy regime and upon increasing $Re$ ($Re$ >3.5), fluid characteristics starts to enter in the transition regime for a very limited range followed by inertia regime. This transition clearly depends on the porosity range. Steady flow conditions are obtained only up to $Re$=105 (see Figure 4.2-top) for low porosity (80%). On the other hand, for high porosity (95%), this condition holds true up to $Re$=700 (see Figure 4.2-bottom). The value of critical Reynolds number ($Re_c$) increases with increasing in porosity. In Figure 4.3, different critical Reynolds numbers at the start and finish of the transition regime are presented and vary between 3.5< $Re_c$ <25 in the porosity range, 0.80< $\varepsilon_o$ <0.95.

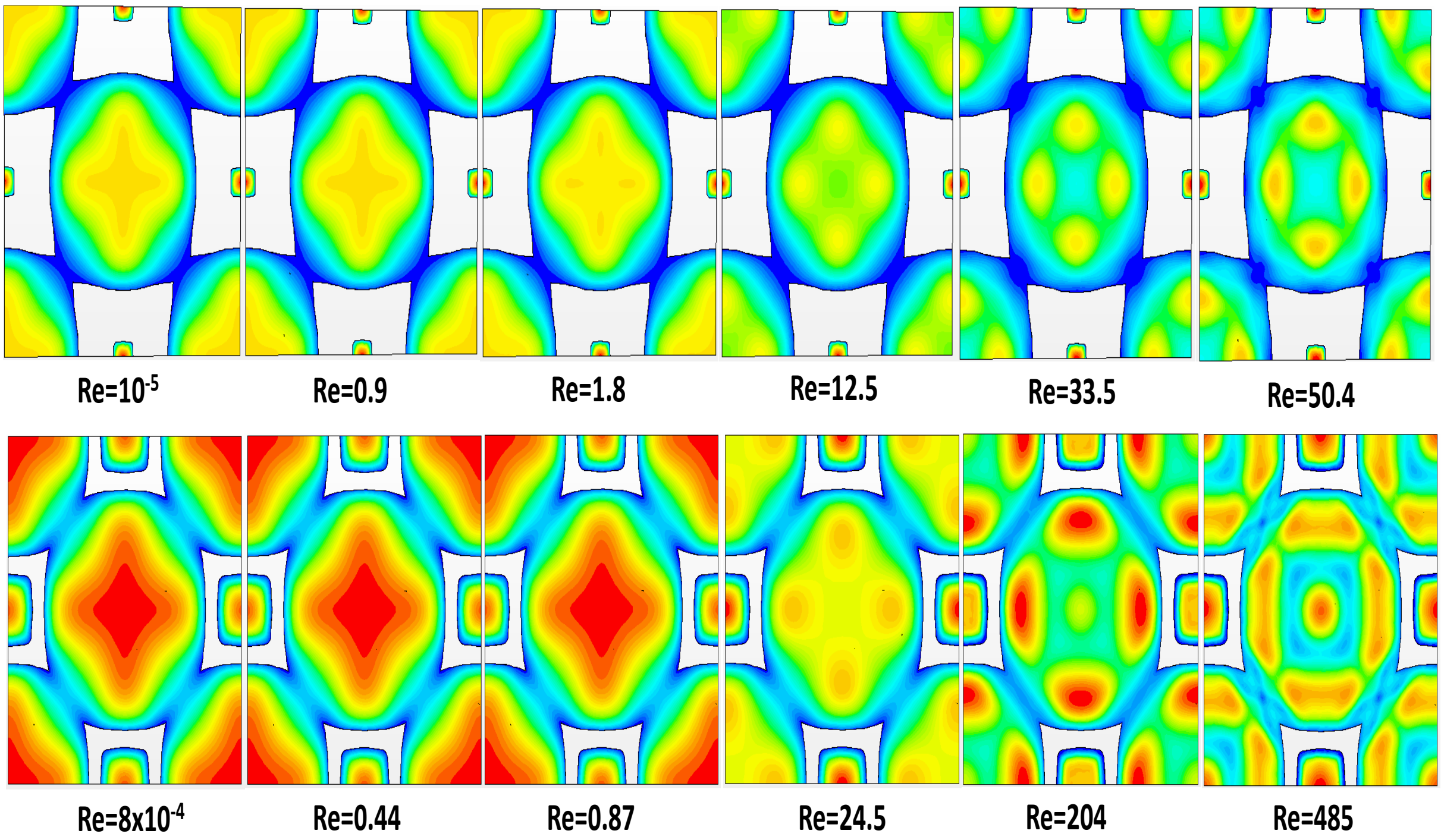

**Figure 4.2.** Top- Representation of average velocity flow field at different Reynolds number for 80% porosity. Bottom- Representation of average velocity flow field at different Reynolds number for 95% porosity. Flow in Darcy regime and its departure to inertial regime is also shown.

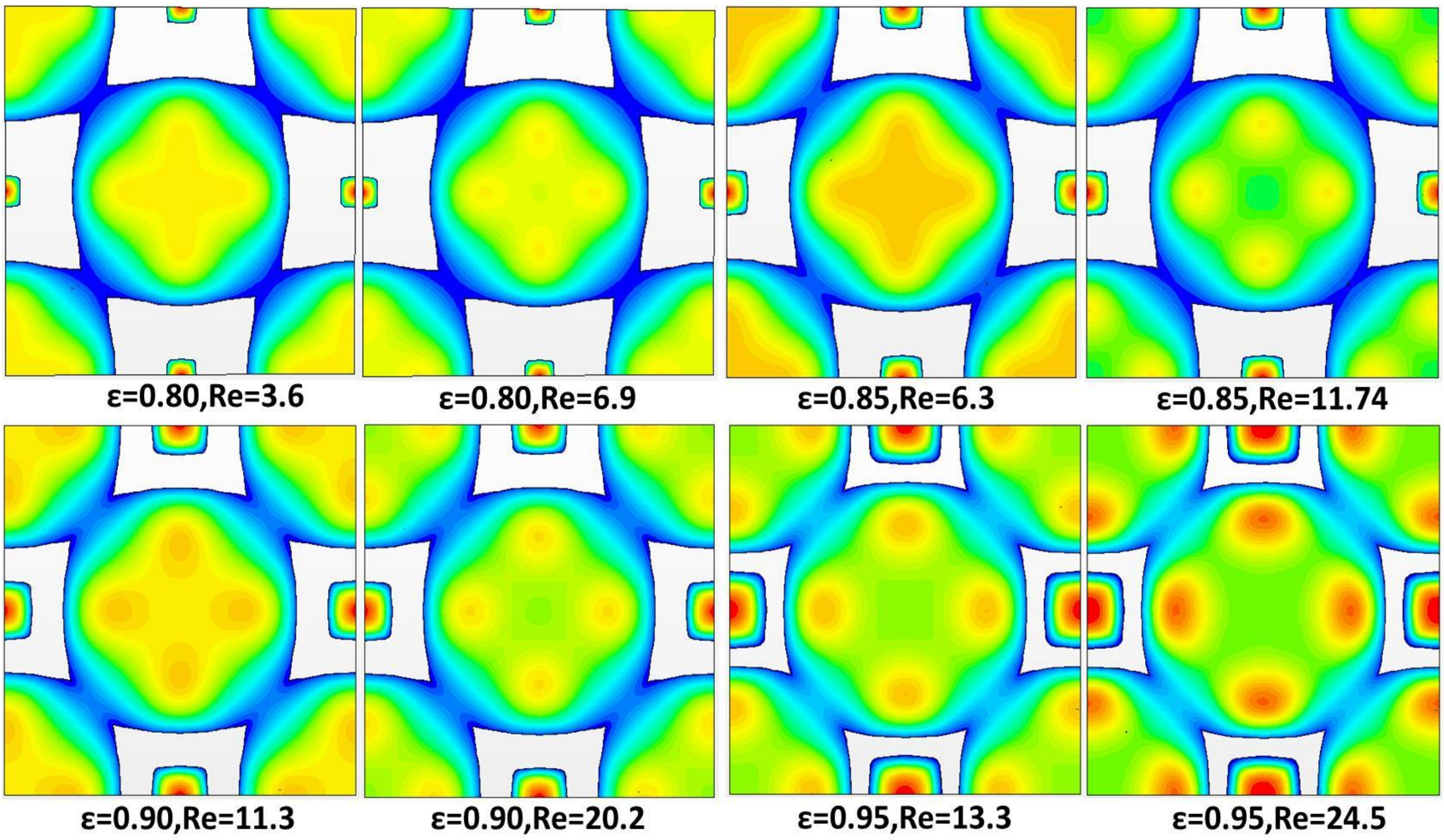

**Figure 4.3.** Representation of transition regimes for all studied range of porosities (80-95%). The critical $Re$ ($Re_c$) changes with the porosity and shift towards a higher value with increasing porosity. Flow field shows the start and finish of transition regime at different $Re$.

*Case II: Impact of smaller pore size ($d_{cell}$=4 mm)*

Same procedure was followed as in Case I to study the impact of smaller pore size of different strut shapes. Figure 4.4 and 4.5 show the velocity flow fields obtained in the case of square strut shape at 60% porosity and hexagon strut shape at 95% porosity.

From the Figure 4.4, it is clearly visible for $Re<1$, flow characteristics are well settled in Darcy regime and upon increasing $Re$ ($Re>5$), fluid flow starts to enter in transition regime for a very limited range followed by inertial regime. Steady flow conditions are obtained only up to $Re$=200 (see Figure 4.4) for low porosity (60%) while for high porosity (95%), this condition holds true up to $Re$=2000 (see Figure 4.5). Transition regime occurs almost at the same critical Reynolds number for different strut shapes for a given porosity. In the case of 80% porosity, $Re_c$=20 (see Figure 4.6) is observed for different strut shapes. The value of $Re_c$ changes with the porosity. In Figure 4.7, different critical Reynolds numbers are presented for circular strut shape in low and high porosity range ($0.60<\varepsilon_o<0.95$), varying from $5<Re_c<50$.

### *Analysis of critical Reynolds number ($Re_c$) for two different pore sizes*

Bonnet et al., (2008) showed that by plotting the reduced pressure drop ($\nabla\langle P\rangle/V$) against velocity ($V$), intersection of purely viscous and inertial regime gives critical Reynolds number (theoretically constant). These authors showed that for a close range of porosity, 0.89< $\varepsilon_o$ <0.92, critical Reynolds numbers ($Re_c$) are different for different pore sizes. For a bigger pore size, $Re_c$ is low ($Re$ ≈3) and increases with decreasing pore size.

As shown in Table 4.2, foam samples of equilateral triangular strut shape have bigger pore size than the other strut shapes. Same trend of critical Reynolds number is observed in case I and II for bigger and small pore sizes as shown by Bonnet et al., (2008). For instance, at $\varepsilon_o$=0.80, $Re_c$=3.6 for equilateral triangle strut shape (bigger pore size) while $Re_c$=20 for all other strut shapes (smaller pore size). It suggests that for the foam samples of same porosity, critical Reynolds number shifts close towards viscous regime for bigger pore size. Moreover, irrespective of strut shape, critical Reynolds number does not change much for a given pore size and porosity (see Figure 4.6). This analysis also suggests that the strut shape and strut length play an important role in fluid flow phenomena and should be introduced in correlations to predict accurate flow properties.

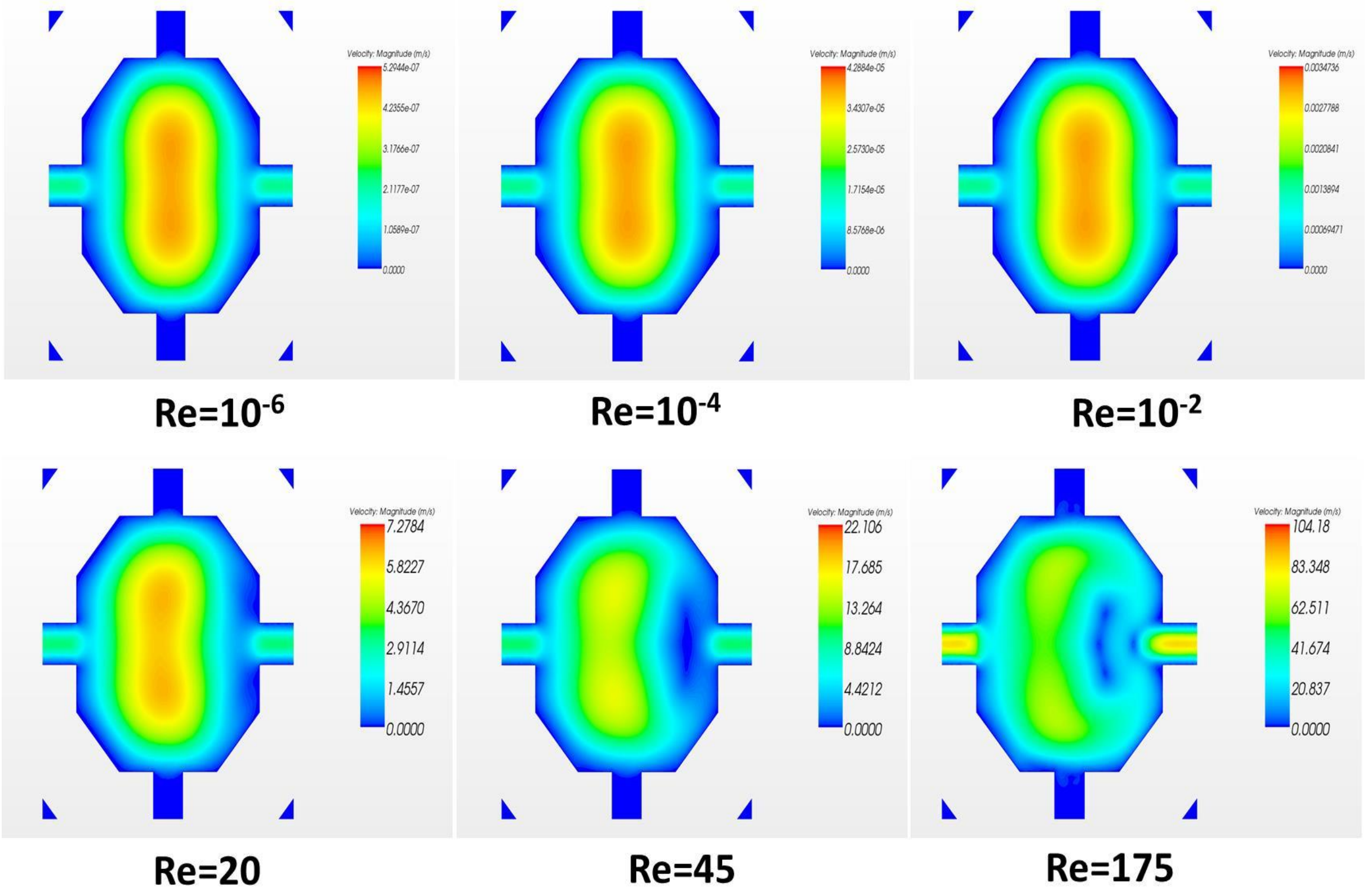

**Figure 4.4.** Representation of average velocity flow fields. Various regimes of square strut shape at 60% porosity are presented at different Reynolds number.

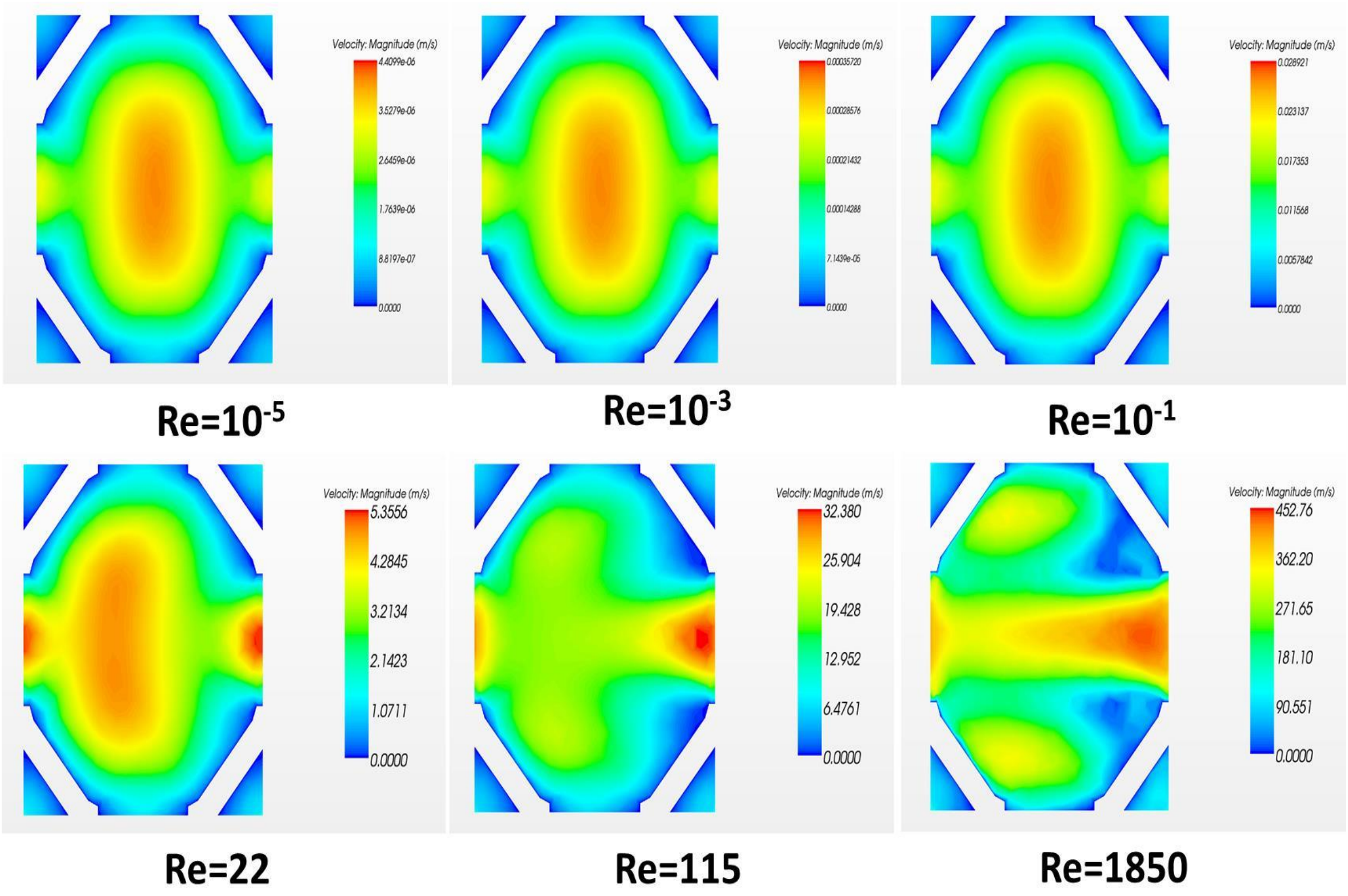

**Figure 4.5.** Representation of average velocity flow fields. Various regimes of hexagon strut shape at 95% porosity are presented at different Reynolds number.

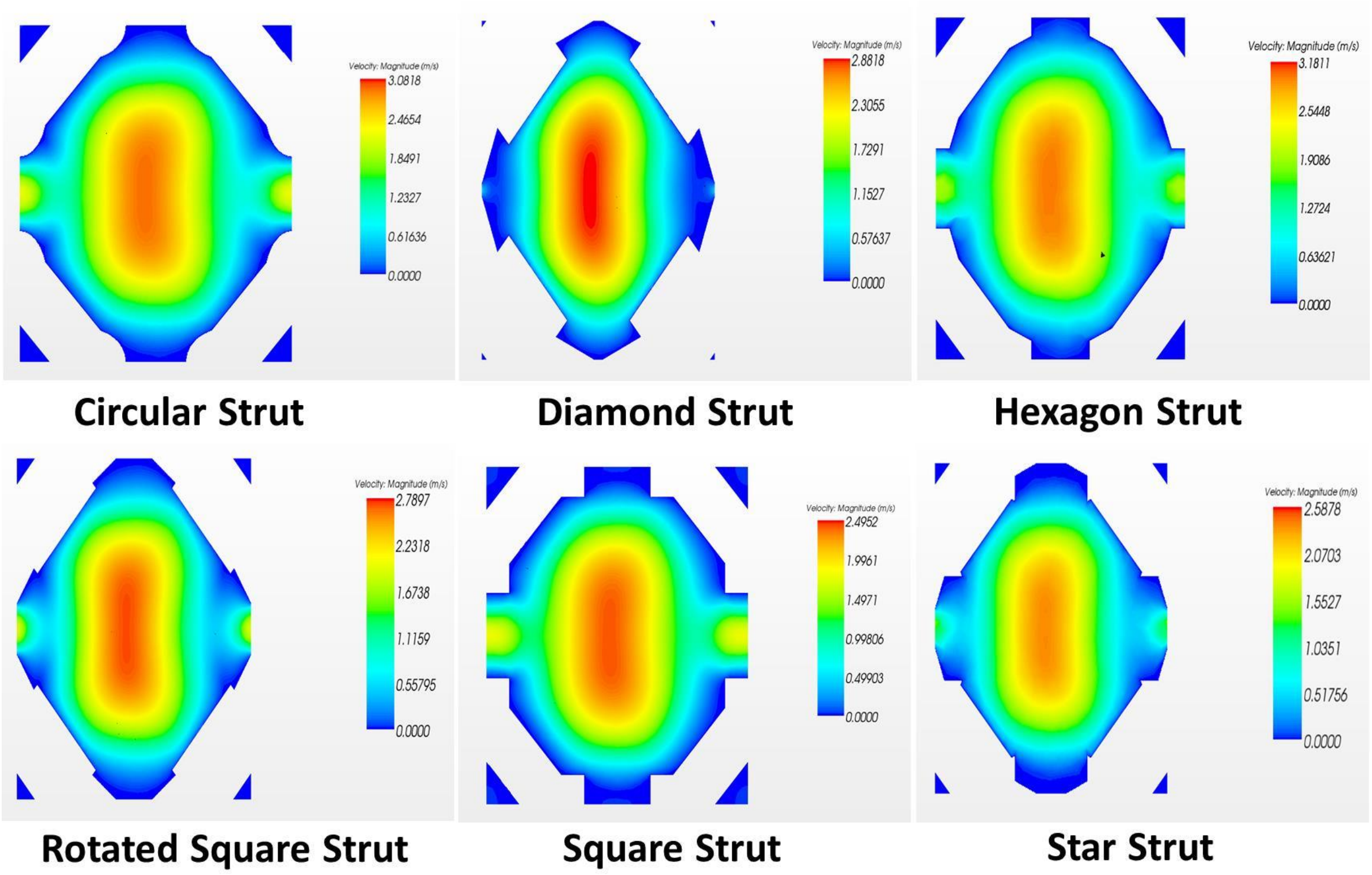

**Figure 4.6.** Representation of average velocity flow field for different strut shapes at the beginning of transient regime. The flow fields are shown for 80% porosity, observed at $Re_c$=20.

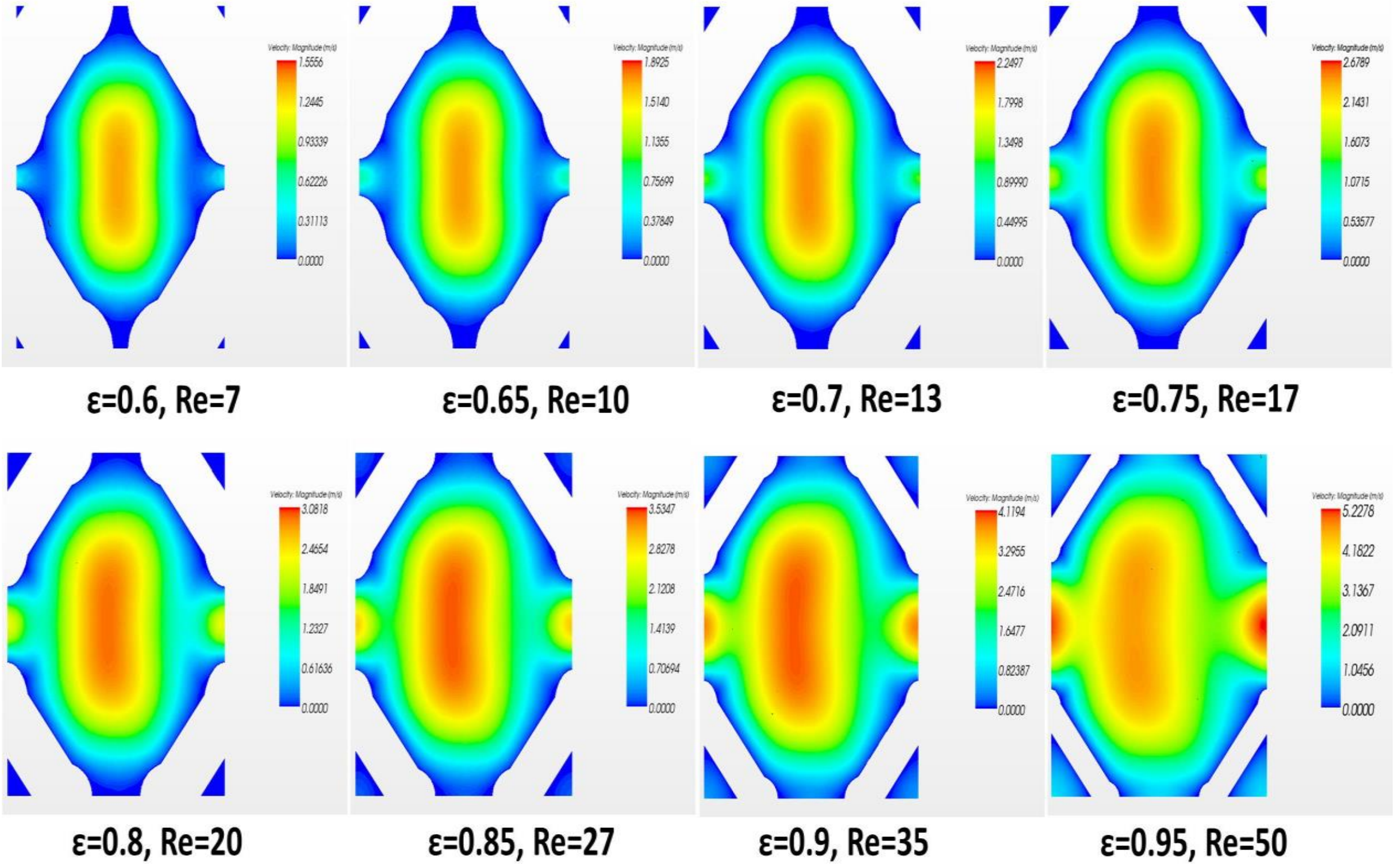

**Figure 4.7.** Representation of transient regimes for porosity range (60-95%) of circular strut shape. The critical Re ($Re_c$) changes with the porosity and shift towards a higher value with increasing porosity.

### 4.3.3 Methodology to choose the flow law

It is often quoted in the literature that choice of fluid flow law is a tricky problem (Firdaouss et al., 1997; Alder et al., 2013). Choice of flow law depends on the flow condition i.e. velocity. It was shown in the work of Bonnet et al., (2008) that fluid flow in open cell foams generally follows Forcheimmer law. These authors also analysed their experimental pressure drop data using Cubic law and found that the error is considerably higher compared to Forcheimmer law.

However, Firdaouss et al., (1997) proposed a methodology to identify the flow law. These authors proposed a normalization technique given by following Equation 4.17:

$$y_F = \frac{1 + (K_D.\langle \nabla P \rangle)/\mu V}{1 + (K_D.\langle \nabla P \rangle_{max})/\mu V_{max}} \qquad (4.17a)$$

and,

$$x_F = \frac{V}{V_{max}} \qquad (4.17b)$$

These authors demonstrated that if the experimental pressure drop data within some range of Reynolds numbers are on the line $y_F = x_F$, it means that the flow follows Forcheimmer law, whereas if the data collapse on parabola $y_F = {x_F}^2$, it follows Cubic law. If the data are on the line $y_F = 0$, it follows Darcy law.

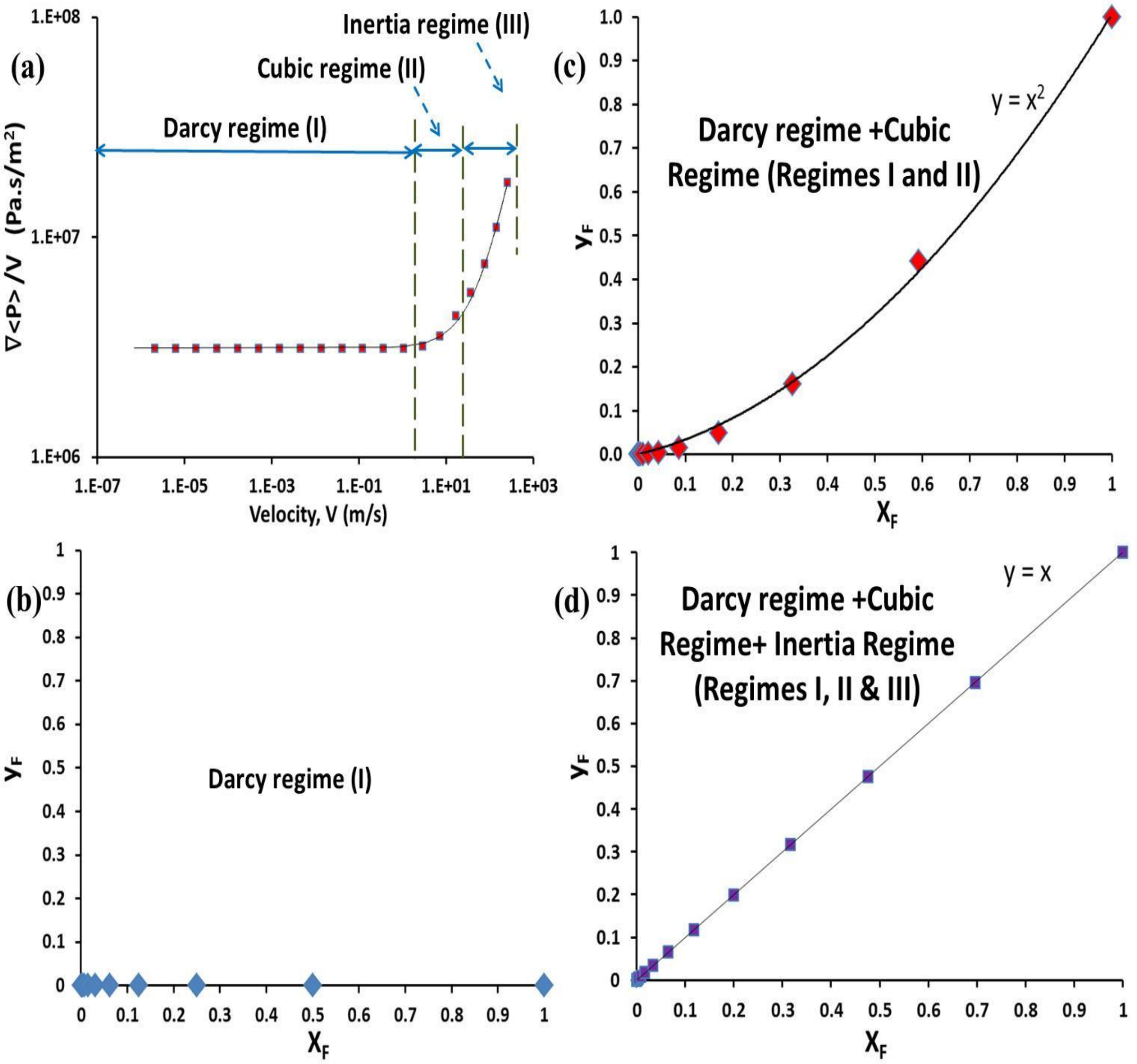

**Figure 4.8.** (a) Plot of $\nabla P/V$ vs. $V$. Darcy, transient and inertia regimes are shown. Plot of $y_F$ vs. $x_F$ to identify the flow law: (b) Darcy regime (c) Transition regime (d) Inertia regime (Circular strut shape of 60% porosity data is shown).

Typical pressure drop results are shown by two ways in Figure 4.8: by plotting $\langle\nabla P\rangle/V$ against $V$ and using Firdaouss normalized parameters (Equation 4.17). In Figure 4.8 (a), three regimes are clearly shown. It is clear that cubic law (transition regime, three-five data points in the numerical experiments) occurs between Darcy and inertia regimes for a very small range of Reynolds number. In Figure 4.8 (b), only Darcy regime (Regime I) is presented that clearly follows $y_F = 0$. Figure 4.8 (c) is shown for Darcy and transition regime (Regimes I and II) that follows cubic law i.e. $y_F = {x_F}^2$. Lastly, for the entire range of velocity (Regimes I, II and III), the pressure drop data follows Forcheimmer law i.e. $y_F = x_F$ as presented in Figure 4.8 (d). The impact of inertia regime is very significant compared to transition regime which suppresses its visibility in fluid flow.

With the fluid flow database generated in this work, we have two possibilities to study flow laws and flow characteristics: distinguish the three regimes and identify associated flow parameters that will be valid only for a given Reynolds number range and, choose a "global" flow law and identify associated flow parameters for wide range of Reynolds number. Generally, transition regime is not clearly identified as it occurs on a very limited Reynolds number range and thus, the latter method was chosen to obtain the flow characteristics.

### 4.3.4 Methodology to extract flow properties

A database of 1600 pressure gradient values was generated to determine permeability and inertia coefficients precisely in the low and high porosity range (0.60< $\varepsilon_o$ <0.95) for a wide range of Reynolds number ($10^{-7} < Re < 10^4$). Generally, plotting the pressure drop as a function of velocity as a polynomial function of Forchheimer equation directly gives $K$ and $C$ (or $K_{poly}$ and $C_{poly}$).

Based on rigorous analysis of flow properties reported in the literature and local analysis of Darcy, transition and inertia regimes, the global pressure drop curve for entire porosity range of equilateral triangular strut shape is presented in Figure 4.9. In Figure 4.9 (zoom-view), it is clear that the permeability obtained in the Darcy regime is significantly different than permeability obtained using polynomial fitting in the inertia regime.

Similarly, the global pressure drop of different strut shapes is presented in Figure 4.10. For a given porosity and same order of pore size, it is observed that the circular strut shape exhibits lower pressure drop while star strut shape exhibits higher pressure drop for the entire velocity range. In the high velocity range (inertia regime), star strut shape offers the maximum obstruction to fluid flow while circular strut shape offers the minimum. In addition, at low porosity range, specific surface area plays an important role in fluid flow behaviour in high velocity range.

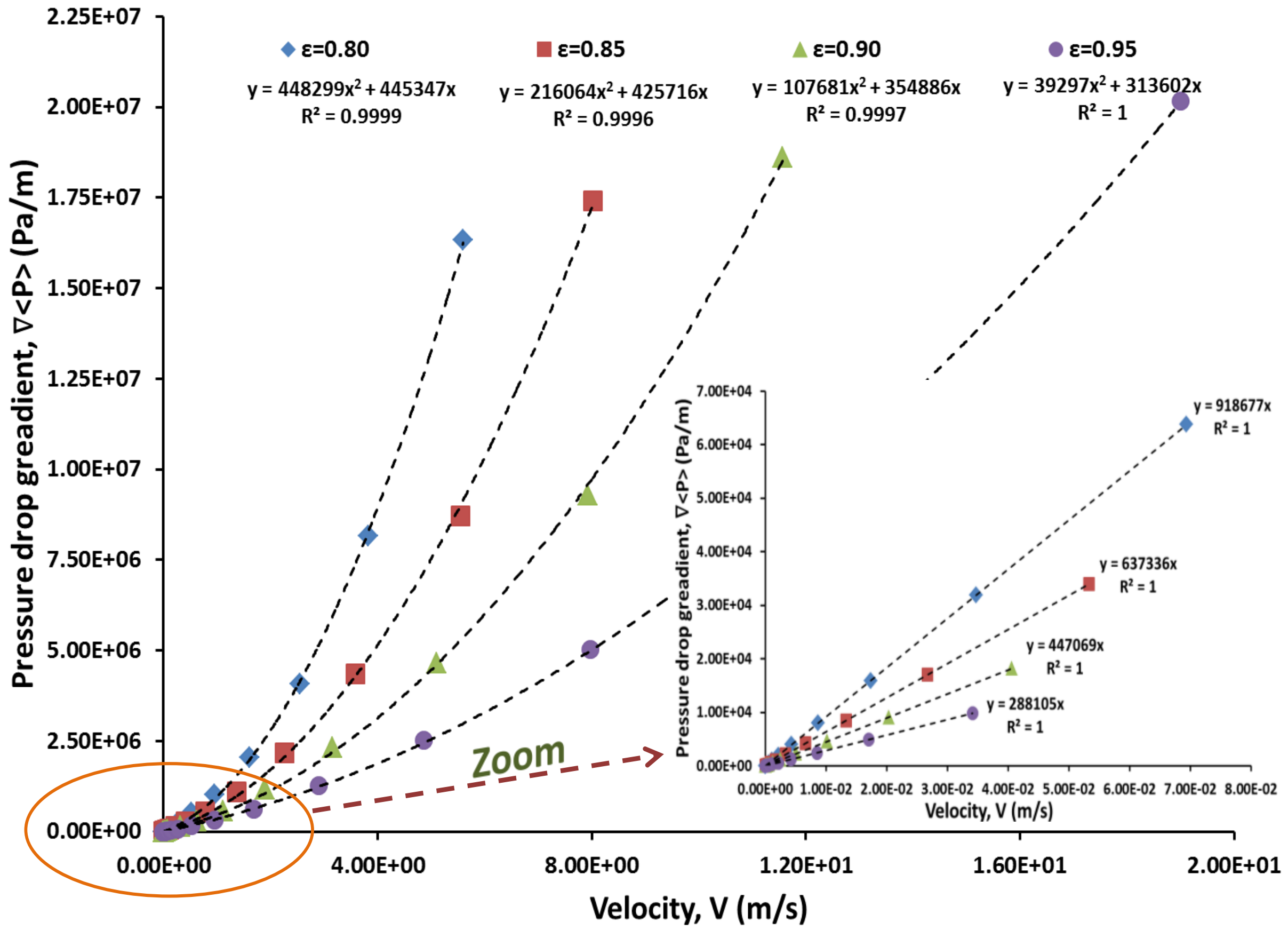

**Figure 4.9.** Global pressure drop, $\nabla\langle P\rangle$ Vs velocity, $V$ of different porosities for entire range of Reynolds number (Darcy, transition and inertia regime). Polynomial fitting of all porosity range (80-95%) is also presented. Zoom- Presentation of $\nabla\langle P\rangle$ Vs velocity, $V$ only in Darcy regime is presented. The polynomial fit in Darcy regime is clearly not same as obtained in inertia regime.

On the other hand, in case of low velocity range (pure viscous flow), it is also interesting to note that hexagon strut shape has the lowest slope while the square shape has the highest slope (see zoom view of Figure 4.10, presented only in Darcy regime). This implies that hexagon strut shape exhibits a higher value of $K_D$ while square strut shape exhibits lower value of $K_D$ ($K_D$ is inversely proportional to slope). However, there is no apparent order to identify this behaviour. No physical justification of this apparent order of permeability for different strut cross sections in Darcy regime is provided.

The polynomial curve in the entire range of velocity accounts for say, $K_{For}$ (or $K_{poly}$ obtained from the polynomial curve) but not Darcian permeability ($K_D$). One cannot trace back to same values of pressure drop by using $K$ ($K_{For}$ or $K_{poly}$) and $C$ (or $C_{poly}$) obtained from the polynomial curve fitting of pressure drop data (see illustration in Figure 4.11).

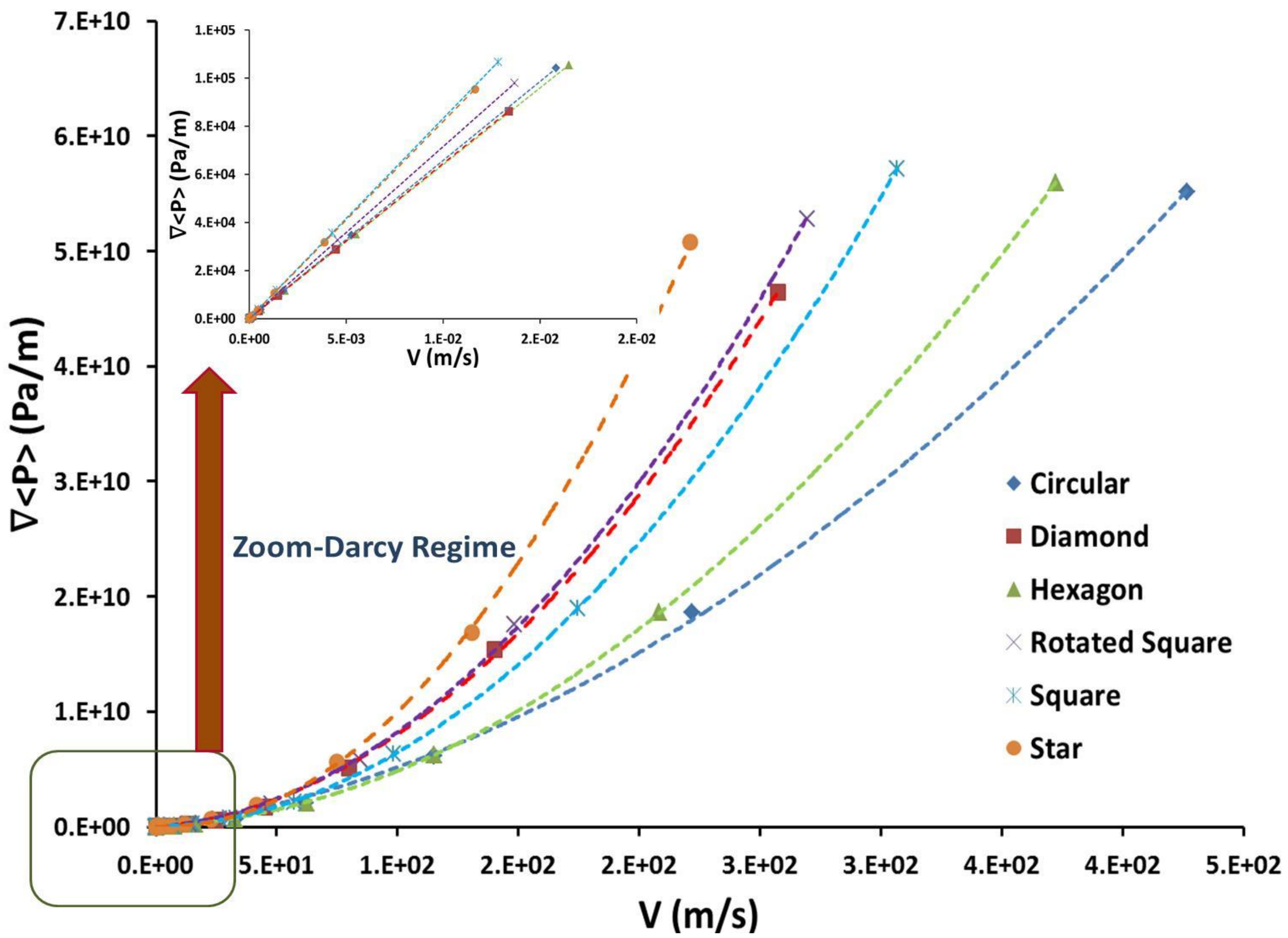

**Figure 4.10.** Global pressure drop, $\nabla\langle P\rangle$ Vs velocity, $V$ of different shapes for entire range of Reynolds number (Darcy, transition and inertia regime) of 80% porosity. Zoom- Presentation of $\nabla\langle P\rangle$ Vs velocity, $V$ only in Darcy regime.

In order to determine Darcian permeability ($K_D$) and Forchheimer inertia coeffcient ($C_{For}$); the pressure drop data was first analysed at very low velocities using Equation 4.6 to avoid discrepancies in permeability values. For high velocities, inertia coefficient is determined using Equation 4.18:

$$C_{For} = \frac{-\nabla\langle P\rangle - \frac{\mu}{K_D}V}{\rho V^2} \tag{4.18}$$

Using Equation 4.18, $C_{For}$ is calculated for the entire range of velocity. In Table 4.2, $K$ (obtained using polynomial curve), $C$ (obtained using polynomial curve), $K_D$ (obtained using Darcy Equation 4.6), $C_{For}$ (obtained using Forchheimer Equation 4.18) are presented. It is clear that permeability obtained using polynomial fitting presents an error of about 25-100% compared to permeability obtained using only Darcy regime data in the entire porosity range. As discussed in section 4.2.2, one can obtain inertia coefficient ($C$) with sufficient accuracy, the variations in $C_{poly}$ and $C_{For}$ are very close but significantly different (maximum difference up to 8% only at low porosity).

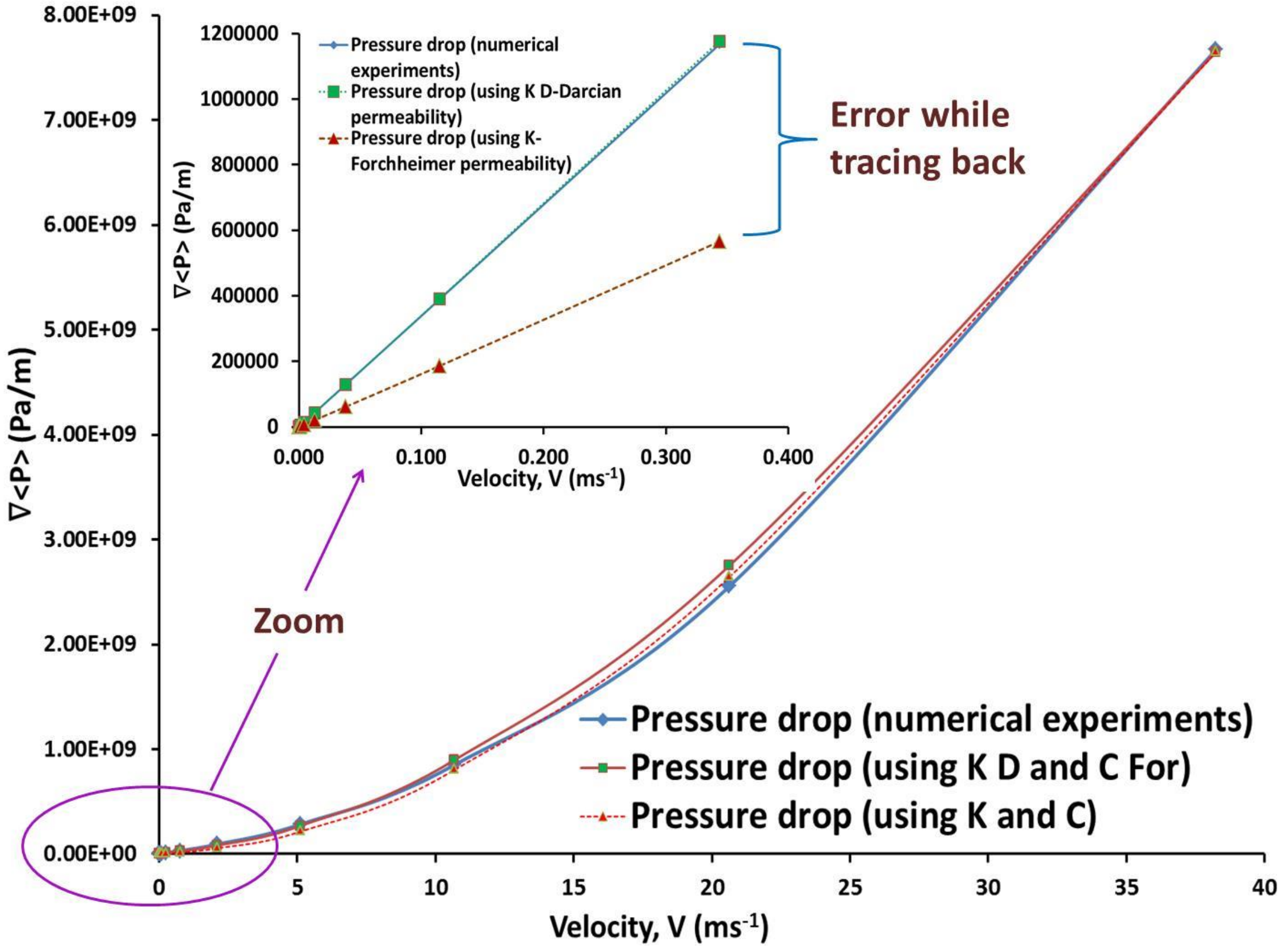

**Figure 4.11.** Comparison of $\nabla\langle P\rangle$in Darcy regime (zoom) and $\nabla\langle P\rangle$in inertial regime. The results are shown for 95% porosity for square strut shape. The zoom of Darcy regime clearly explains that $K$ (or $K_{For}$) obtained from the polynomial curve introduces discrepancy in flow properties and analytical solutions. Fluid properties: $\mu$-0.8887 ($kg.m^{-1}s^{-1}$) and $\rho$-998.5 ($kg.m^{-3}$) are used.

Using $K_D$ (obtained using Equation 4.6) and $C_{For}$ (obtained using equation 4.18), one can trace back the same values of experimental pressure drop as presented in Figure 4.11. Using $K$ and $C$ obtained from polynomial curve, one can track back only the inertial regime but significant error in the Darcy Regime at very low Reynolds number (in other words, velocity) is inevitable and presented in the Figure 4.11 (zoom view). From Figure 4.11 (zoom-view), it is also clear that plot of pressure drop against velocity only in pure Darcy regime using $K$ (or $K_{For}$) results in small slope value while it results in higher slope value when $K_D$ is used which in turn, indicates that $K_{For} > K_D$ (also see the comparison shown in Table 4.2). Thus, $K_{For}$ introduces significant discrepancies in pressure drop characteristics and in correlations which actually gets supressed due to inertia coefficient in high velocity regime.

**Table 4.2**. Representation of various strut shapes, porosities and pore diameters. Fluid flow properties, $K_{poly}$ and $C_{poly}$ (using polynomial curve), $K_D$ (using Darcy Equation 4.6) and $C_{For}$ (using Forchheimer equation 4.18) are also presented. Error (%) is also shown for permeability and inertia coefficient.

| | CAD data | | Direct Numerical Simulation Data | | | | Error (%) | |
|---|---|---|---|---|---|---|---|---|
| Strut Shape | $\varepsilon_o$ (%) | $d_p$ $(mm)$ | $K_{poly}$x$10^{-7}$ $(m^2)$ | $K_D$x$10^{-7}$ $(m^2)$ | $C_{poly}$ $(m^{-1})$ | $C_{For}$ $(m^{-1})$ | $\frac{K_{poly}-K_D}{K_D}$ | $\frac{C_{poly}-C_{For}}{C_{For}}$ |
| Circular | 60 | 0.694 | 0.982 | 0.6015 | 1229 | 1157 | 39 | 6 |
| | 65 | 0.822 | 1.079 | 0.7387 | 982 | 967 | 31 | 2 |
| | 70 | 0.951 | 1.238 | 0.9033 | 764 | 748 | 27 | 2 |
| | 75 | 1.083 | 1.436 | 1.1024 | 545 | 530 | 23 | 3 |
| | 80 | 1.223 | 1.851 | 1.3517 | 416 | 409 | 27 | 2 |
| | 85 | 1.372 | 2.466 | 1.6712 | 314 | 304 | 32 | 3 |
| | 90 | 1.541 | 2.927 | 2.1176 | 201 | 197 | 28 | 2 |
| | 95 | 1.749 | 3.457 | 2.8587 | 108 | 106 | 17 | 2 |
| Equilater-al Triangle | 80 | 3.646 | 19.96 | 9.6900 | 448.9 | 391.9 | 106 | 15 |
| | 85 | 4.370 | 20.88 | 14.0000 | 216.4 | 191.9 | 49 | 13 |
| | 90 | 5.132 | 25.04 | 19.9000 | 107.8 | 100.7 | 26 | 7 |
| | 95 | 5.531 | 28.34 | 30.9000 | 39.4 | 40.8 | -8 | 1.5 |
| Square | 60 | 0.692 | 0.629 | 0.3444 | 2733 | 2603 | 46 | 5 |
| | 65 | 0.822 | 0.887 | 0.4726 | 1714 | 1663 | 47 | 3 |
| | 70 | 0.951 | 1.091 | 0.6310 | 1157 | 1134 | 42 | 2 |
| | 75 | 1.085 | 1.256 | 0.8258 | 829 | 796 | 34 | 4 |
| | 80 | 1.224 | 1.603 | 1.0692 | 608 | 570 | 33 | 6 |
| | 85 | 1.374 | 2.542 | 1.3891 | 411 | 398 | 45 | 3 |
| | 90 | 1.543 | 2.825 | 1.8394 | 243 | 234 | 35 | 4 |
| | 95 | 1.750 | 4.536 | 2.6235 | 104 | 100 | 42 | 4 |
| Rotated Square | 80 | 1.243 | 1.974 | 1.2439 | 761 | 730 | 37 | 4 |
| | 85 | 1.387 | 2.889 | 1.5552 | 591 | 557 | 46 | 6 |
| | 90 | 1.550 | 3.016 | 2.0026 | 378 | 347 | 34 | 8 |
| | 95 | 1.753 | 4.271 | 2.7550 | 187 | 175 | 36 | 6 |
| Diamond | 80 | 1.238 | 2.252 | 1.3802 | 761 | 740 | 39 | 3 |
| | 85 | 1.382 | 2.766 | 1.6608 | 623 | 611 | 40 | 2 |
| | 90 | 1.546 | 3.456 | 2.0642 | 457 | 453 | 40 | 1 |
| | 95 | 1.750 | 4.183 | 2.7661 | 231 | 221 | 34 | 4 |
| Hexagon | 60 | 0.696 | 1.179 | 0.6010 | 1430 | 1362 | 49 | 5 |
| | 65 | 0.824 | 1.312 | 0.7419 | 1192 | 1139 | 44 | 4 |
| | 70 | 0.954 | 1.592 | 0.9116 | 795 | 784 | 43 | 1 |
| | 75 | 1.086 | 1.877 | 1.1250 | 582 | 561 | 40 | 4 |
| | 80 | 1.225 | 2.012 | 1.3899 | 431 | 416 | 31 | 3 |
| | 85 | 1.375 | 2.793 | 1.7525 | 282 | 277 | 37 | 2 |
| | 90 | 1.543 | 3.347 | 2.2561 | 188 | 175 | 32 | 7 |
| | 95 | 1.750 | 3.825 | 3.1054 | 96 | 95 | 19 | 1 |
| Star | 75 | 1.091 | 1.28 | 0.8050 | 1655 | 1502 | 37 | 9 |
| | 80 | 1.229 | 1.69 | 1.0858 | 913 | 896 | 36 | 2 |
| | 85 | 1.377 | 2.59 | 1.4108 | 611 | 603 | 46 | 1 |
| | 90 | 1.545 | 3.31 | 1.8704 | 382 | 366 | 44 | 4 |
| | 95 | 1.751 | 4.74 | 2.7363 | 152 | 152 | 42 | 0 |

Hence, to obtain flow characteristics with maximum accuracy in open cell foams, one should obtain the permeability only in Darcy regime (Equation 4.6) and then utilizing this permeability in Equation 4.18 to obtain inertia coefficient.

### 4.3.5 Influence of strut shapes on flow properties

Different strut shapes impact strongly on flow regimes and thus, flow properties. From Table 4.2, it is observed that for different strut shapes, $K_D$ increases by a factor of about 6 in the porosity range 0.60$< \varepsilon_o <$0.95. For a given porosity and same pore size, $K_D$ varies with different strut shapes which could be due to the sharp angles of strut shapes subjected to the fluid flow direction.

It is often reported in the literature that $K \propto {d_p}^2$ and $C \propto {d_p}^{-1}$ (see Madani et al., 2006; Bonnet et al., 2008). In Figure 4.12, it is observed that $K_D \propto {d_p}^2$. However, in the zoom view of Figure 4.12, it is clear that there are very small variations in $K_D$ for different strut shapes and these variations could be attributed to the pore diameter that was calculated as an equivalent diameter of hexagon and square faces of the foam matrix.

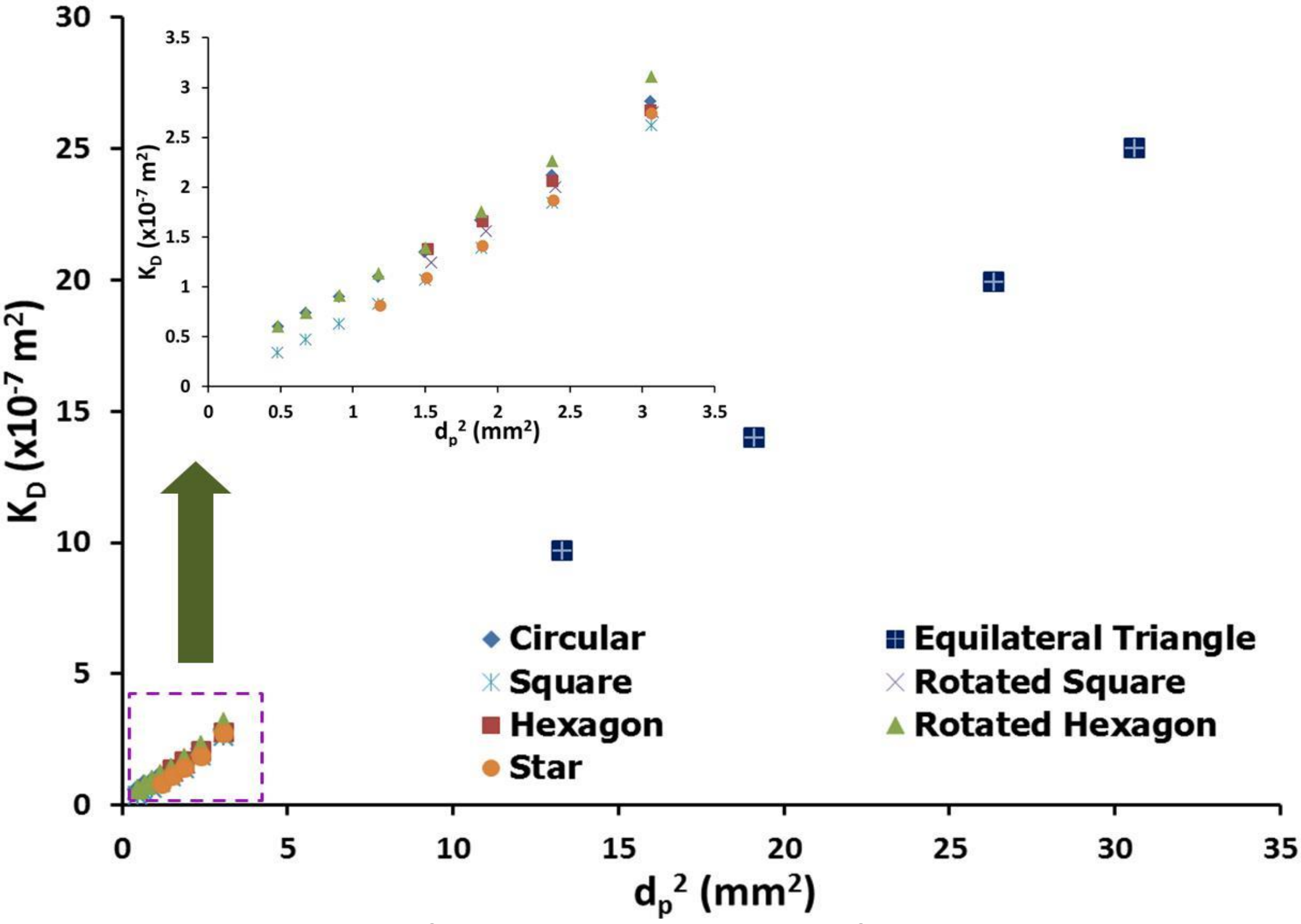

**Figure 4.12**. Plot of $K_D$ vs. ${d_p}^2$. It is clear that $K_D$ and ${d_p}^2$ are proportional to each other. Zoom view to present different strut shapes of same order of pore size.

It is thus important to know the impact of individual strut shape and porosity on Darcian permeability ($K_D$). In order to quantify the impact of strut shape on permeability for the performance of a system, ${K_D/d_p}^2$ (dimensionless) was plotted against different strut shapes and porosities in Figure 4.13. For different pore sizes, it is observed that the ratio ${K_D/d_p}^2$ is highest for hexagon strut shape and lowest for equilateral strut shape. This trend is similar for all the porosities. There is an increase in ${K_D/d_p}^2$ by a factor of about 1.25 when the strut shape changes from equilateral triangular shape to hexagon in the porosity range, 0.80$< \varepsilon_o <$0.95. For engineering applications in Darcy regime, low pressure drop can be easily obtained for equilateral triangular, star and square strut cross sections while for cases where high pressure drop is critical, hexagon strut shape could be used.

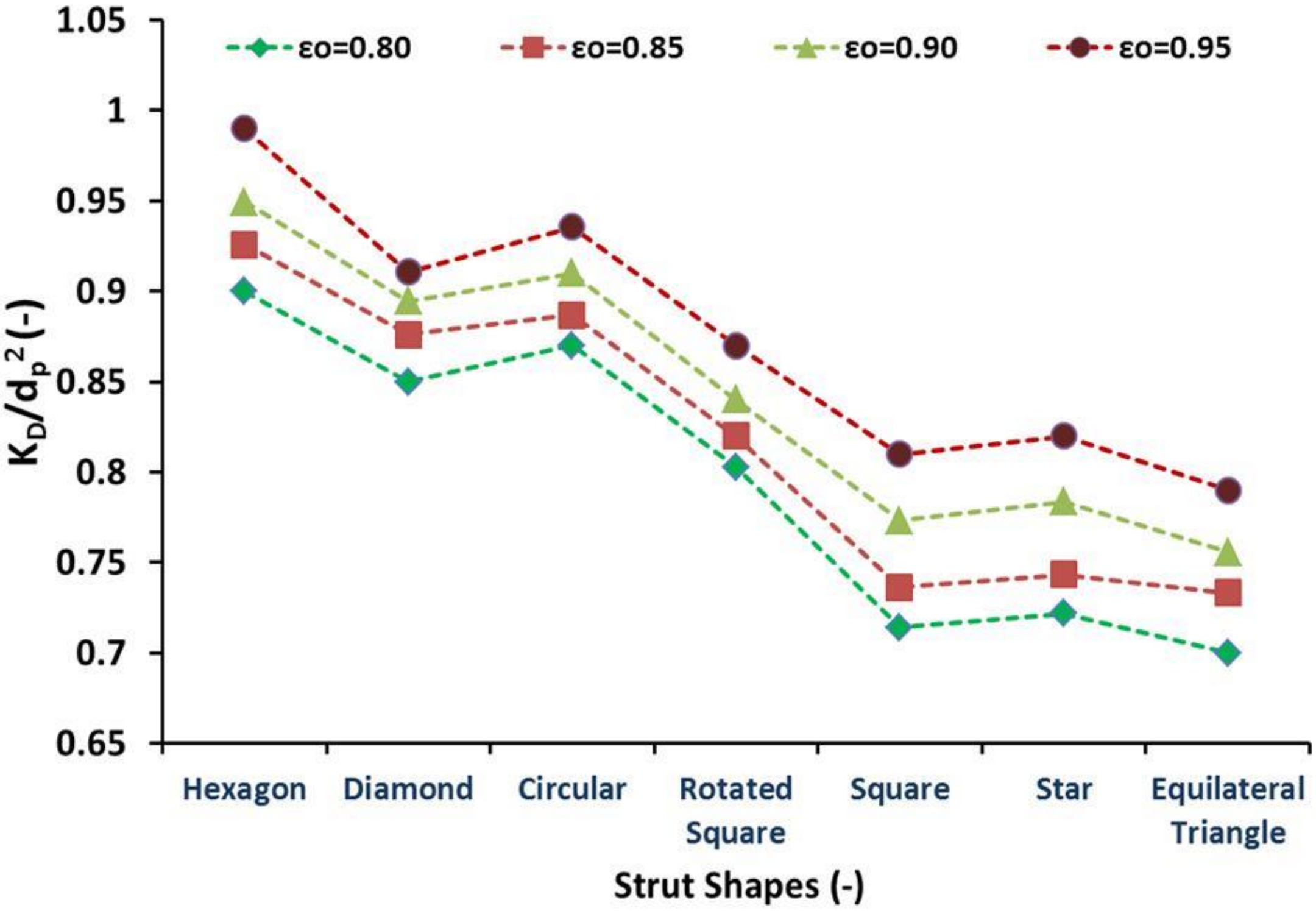

**Figure 4.13**. Plot of ${K_D/d_p}^2$ (dimensionless) vs. strut shapes at different porosities.

Unlike $K_D$, $C_{For}$ decreases with increase in porosity as presented in Table 4.2. At lower porosities, specific surface area (as well as sharp angles of strut shape with respect to fluid flow direction) contributes significantly in high inertia coefficient values. One can notice that the difference between $C_{poly}$ and $C_{For}$ at low porosities (0.60$< \varepsilon_o <$0.80) is around 6-8% and can be linked to the impact of high specific surface area in high velocity regimes. On the other hand, in the porosity range 0.80$< \varepsilon_o <$0.90, this difference starts to decrease (almost negligible) and these coefficients have quite similar values at very high porosity ($\varepsilon_o$=0.95).

$C_{For}$ is inversely proportional to $d_p$ as shown in Figure 4.14. It is clearly evident that the slopes are distinguishably different. The slopes of $C_{For}$ follow the same trend as already presented in Figure 4.10. $C_{For}.d_p$ (dimensionless) was plotted against different strut shapes for different porosities in Figure 4.14 (top-left: zoom view).

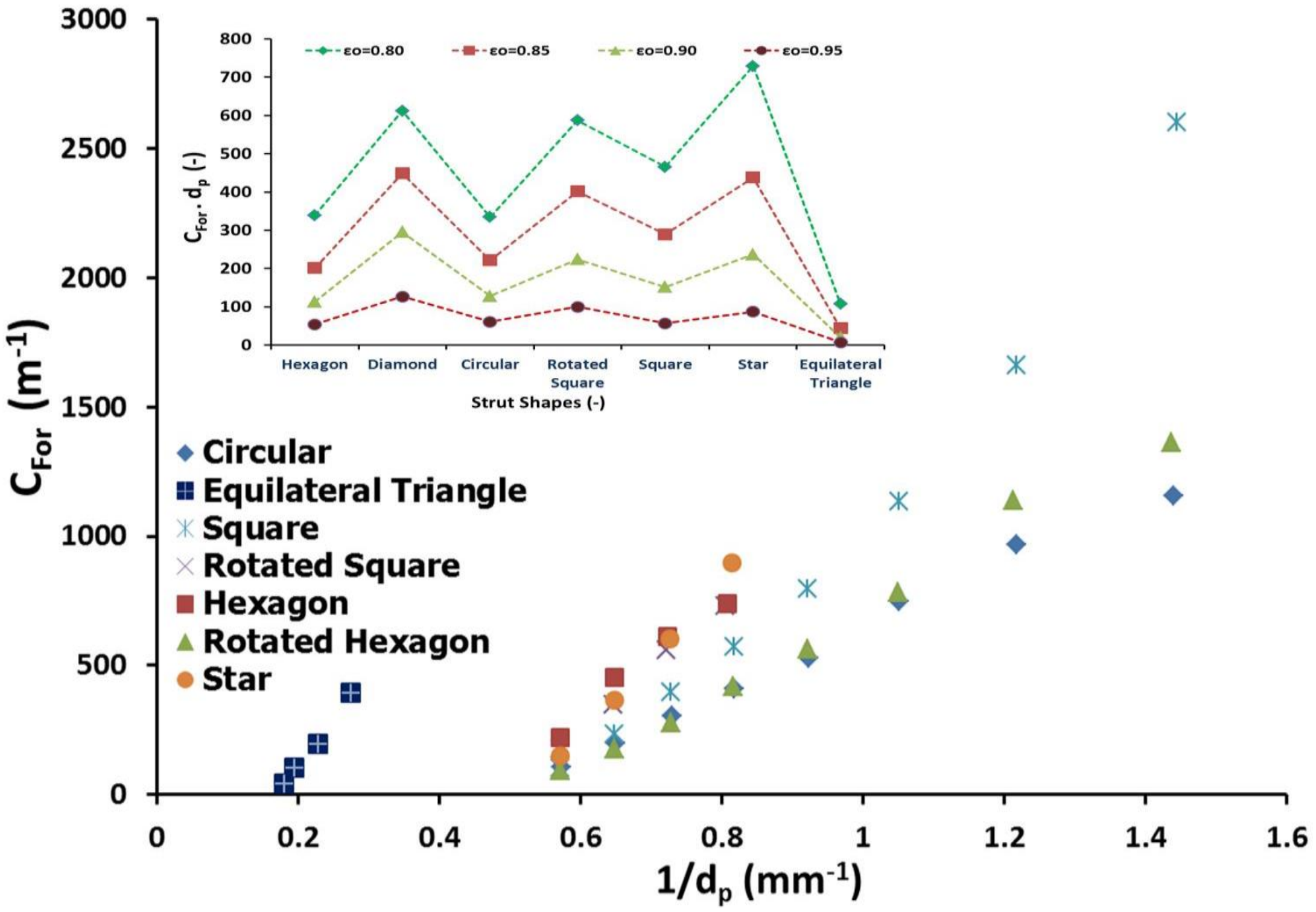

**Figure 4.14**. Plot of $C_{For}$ vs. ${d_p}^{-1}$. It is clear that $C_{For}$ and ${d_p}^{-1}$ are inversely proportional to each other. Zoom view is to present variation of $C_{For}$ for different struts of same order of pore size at different porosity.

The strut shapes impact very strongly on $C_{For}.d_p$ in the porosity range, 0.80< $\varepsilon_o$ <0.90 while in very high porosity range 0.90< $\varepsilon_o$ <0.95, the variation in $C_{For}.d_p$ is almost constant for different strut shapes. It is also worth noting that $C_{For}.d_p$ is minimum for equilateral triangular strut shape while it is maximum for star strut shape for all porosities. The high peaks are also obtained for diamond, rotated square and star strut shapes. These peaks are clearly attributed to their higher specific surface areas at low porosity (also see Figure 4.10).

## 4.4 Performance of state of the art correlations

Based on the numerical results and extracted flow properties presented in Table 4.2, flow characteristics obtained for equilateral triangle (bigger pore size) and circular (smaller pore size) strut cross sections are compared with the correlations presented in the literature. The reason to choose these two strut shapes for comparison is due to the fact that a few authors (e.g. Bhattacharya et al., 2002; Inayat et al., 2011 a, b) have reported a change in the strut shape from circular to convex or concave or equilateral triangular from low to high porosity.

The data measured in this work are mainly compared with the correlations presented by Moreira et al., (2004), Kharyagoli et al., (2004), Liu et al., (2006), Lacroix et al., (2007), Dietrich et al., (2009) and Inayat et al., (2011b) as presented in Figure 4.15 and 4.16 for equilateral triangular and circular strut cross sections respectively. In order to compare the measured pressure drop data investigated in present work with the calculated values from the correlations, pseudo experimental velocity values (very low velocity to determine viscous condition and high velocity to determine inertia condition) were chosen. The general observation suggests that none of the correlations from the literature could reproduce the experimental/numerical pressure drop values from the present work to a satisfactory level (see Figures 4.15 and 4.16). This is due to fact that the literature correlations were derived only for very high porosity foam samples where geometrical parameters do not play an important role (in most of the cases, pressure drop correlations were derived for equilateral triangular or circular strut shape). Also, none of the authors have validated the Darcian and inertia regimes separately. Impact of inertia regime suppresses strongly the error in the Darcy regime when deriving the correlations for whole velocity range.

Most of these correlations presented in Figures 4.15 and 4.16 underestimate the calculated pressure drop values because they have constant Ergun parameters and their numerical values are smaller than that were predicted either by Ergun (1952) or Dietrich et al., (2009). The correlations that overestimate the pressure drop values do not provide any constant numerical values of Ergun parameters but were based on simple geometrical model like cubic lattice or over-simplification of geometrical parameters. As discussed in the section 4.3.5, flow properties and thus, Ergun parameters are shape dependent and depend strongly on geometrical parameters of foam matrix which have not been taken into account in the correlations reported in the literature.

***In case of equilateral triangular strut shape (bigger pore size)***

The calculated values of pressure drop using the correlation of Dietrich et al., (2009) are underestimated and overestimated by a factor of 1.5-1.6 (error~150-160%) compared to the measured pressure drop data in the porosity range, 0.80< $\varepsilon_o$ <0.95. Moreover, for the porosity, $\varepsilon_o$=0.90, the error between calculated and measured pressure drop values is 10%.

When compared the calculated pressure drop values against the correlation established by Inayat et al., (2011b), the error is quite significant. The calculated values for all porosities are underestimated by a factor of 200-20000 in Darcy regime while it varies by a factor of 10-1400 in inertial regime.

The calculated values of pressure drop from the correlation of Lacroix et al., (2007) and Kharyagoli et al., (2004) are underestimated compared to measured values of pressure drop. The error is varying between 80-100% in Darcy and inertia regimes from low to high porosity. The Ergun parameters in the formulation of these authors are very close to each other and thus, their correlations provide almost the same order of results. The errors are more significant with increasing porosity.

In comparison to pressure drop values calculated using correlation of Liu et al., (2006), the calculated results are underestimated and the error is varying between 700-17000% in Darcy regime while 1000-5000% in inertia regime.

The calculated values of pressure drop from the correlation of Moreira et al., (2004) are overestimated in the entire range of porosity. The error estimated between calculated and measured values is 18-2500% in Darcy regime while it is 80-200% in inertia regime from low to high porosity. The overestimated results have less error for high porosity while the error increases with decreasing porosity.

***In case of circular strut shape (smaller pore size)***

The calculated values of pressure drop of circular strut shape are overestimated using the correlation of Dietrich et al., (2009). The error is 140-165% in both the regimes at low porosity ($\varepsilon_o$=0.60) only while it is significant when porosity increases.

The errors in the calculated pressure drop values in the Darcy and inertia regimes are more than 1000-25000% and 100-4000% respectively when compared to the correlations of Inayat et al., (2011b).

Correlation of Lacroix et al., (2007) provides overestimated and underestimated calculated values of pressure drop in the porosity range 0.60$< \varepsilon_o <$0.85 and 0.90$< \varepsilon_o <$0.95 respectively. The error ranges are 84-3000% and 60-400% in Darcy and inertia regimes. However, in the porosity range 0.85$< \varepsilon_o <$0.90, the correlation provides a good estimate in inertia regime (error range ∼10-30%) while the error in the Darcy regime is 70-300%.

Similar trend of underestimating and overestimating the calculated pressure drop values is observed for the correlation of Kharyagoli et al., (2004). However, the error is quite significant compared to correlation of Lacroix et al., (2007). The error lies in the range of 70-4500% and 40-600% respectively in the Darcy and inertia regimes.

The correlation of Liu et al., (2006) underestimates the calculated values of pressure drop by a factor of 2-200 (error ∼20-20000%) same as the case of equilateral triangular strut shape.

The correlation of Moreira et al., (2004) overestimates the pressure drop data. The errors in the porosity range, 0.60$< \varepsilon_o <$0.90 are significant. The error in pressure drop values is least, varying between 30-60% for 95% porosity only.

***Synopsis of performance of state of the art of pressure drop correlations***

Various correlations reported in the literature are investigated with the measured pressure drop data of equilateral triangular and circular strut cross sections in this work. From the comparison, it is evident that the correlations established by the authors are based on few critical parameters:

- Only high porosity range of foam samples where geometrical parameters do not play a significant role in fluid flow properties.
- Overestimation of specific surface area due to simplified geometry of foam structure.
- Extraction of $K_{For}$ instead of $K_D$ (error ∼50-90%, see Table 4.2).
- Validation of calculated against measured pressure drop data in the entire velocity range.
- No separate validation of flow characteristics i.e. permeability and inertia coefficient.

From Figures 4.15 and 4.16, the correlation proposed by Dietrich et al., (2009) exhibits the lowest error for the two strut cross sections while the other correlations exhibit enormous error and are incomparable to measured pressure drop values in this work. However, for other complex strut shapes, correlation of Dietrich et al., (2009) tends to overestimate the calculated pressure drop values to a great extent.

This comparison of two different strut shapes in the low and high porosity range suggests that Ergun parameters cannot have constant numerical values as strut shape does play an important role (see Figures 4.13 and 4.14; section 4.3.5). In case of high porosity, the error between calculated and measured pressure drop is less but in case of low porosity, specific surface area and nature of the strut shape start to impact very strongly on the pressure drop. Thus, a combination of geometrical parameters needs to be added to the classical Ergun-like equation in order to predict precise numerical values of $E_1$ and $E_2$.

Moreover, viscous term (or permeability) that accounts for Ergun parameter $E_1$ while establishing pressure drop correlation predicts wrong values of $E_1$ due to extraction of $K_{For}$ by polynomial fit. All the authors have tried to validate their correlations by comparing the calculated and measured pressure drop values but none of the authors compared the flow properties separately i.e. $K_D$ and $C_{For}$. When comparing the measured pressure drop values with experimental/numerical data in the wide velocity range, the pressure drop is mostly governed by inertia regime (where inertia coefficient is obtained with sufficient accuracy) and thus, suppresses the error in permeability of viscous regime that cannot be neglected. This is one of the reasons that correlations in the literature (e.g. Dietrich, 2012) are applicable to a very few foam samples of very high porosity and for a given strut shape.

It is thus, important to validate the pressure drop values in both the regimes i.e. viscous and inertia regimes separately. In other words, one must compare the measured and calculated permeability and inertia coefficients in order to improve the quality of the correlation that could help in reducing the dispersion in friction factor.

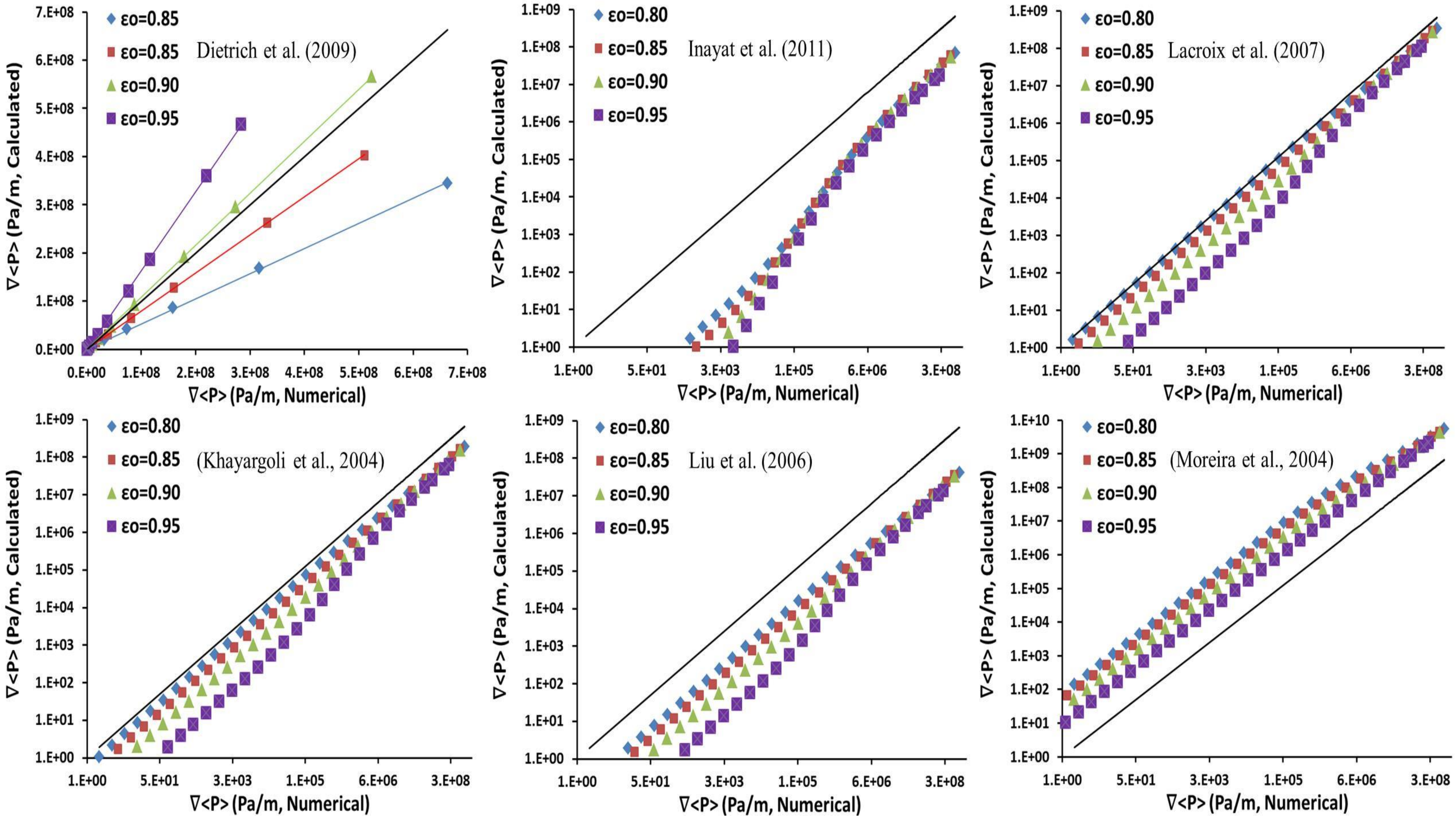

**Figure 4.15**. Performance of state of the art correlations (Black line corresponds to the measured numerical data: from present work). The comparison presented above is performed for equilateral triangular strut shape. (The calculated and experimental data are plotted in log-log scale in order to distinguish Darcy and inertia regimes clearly).

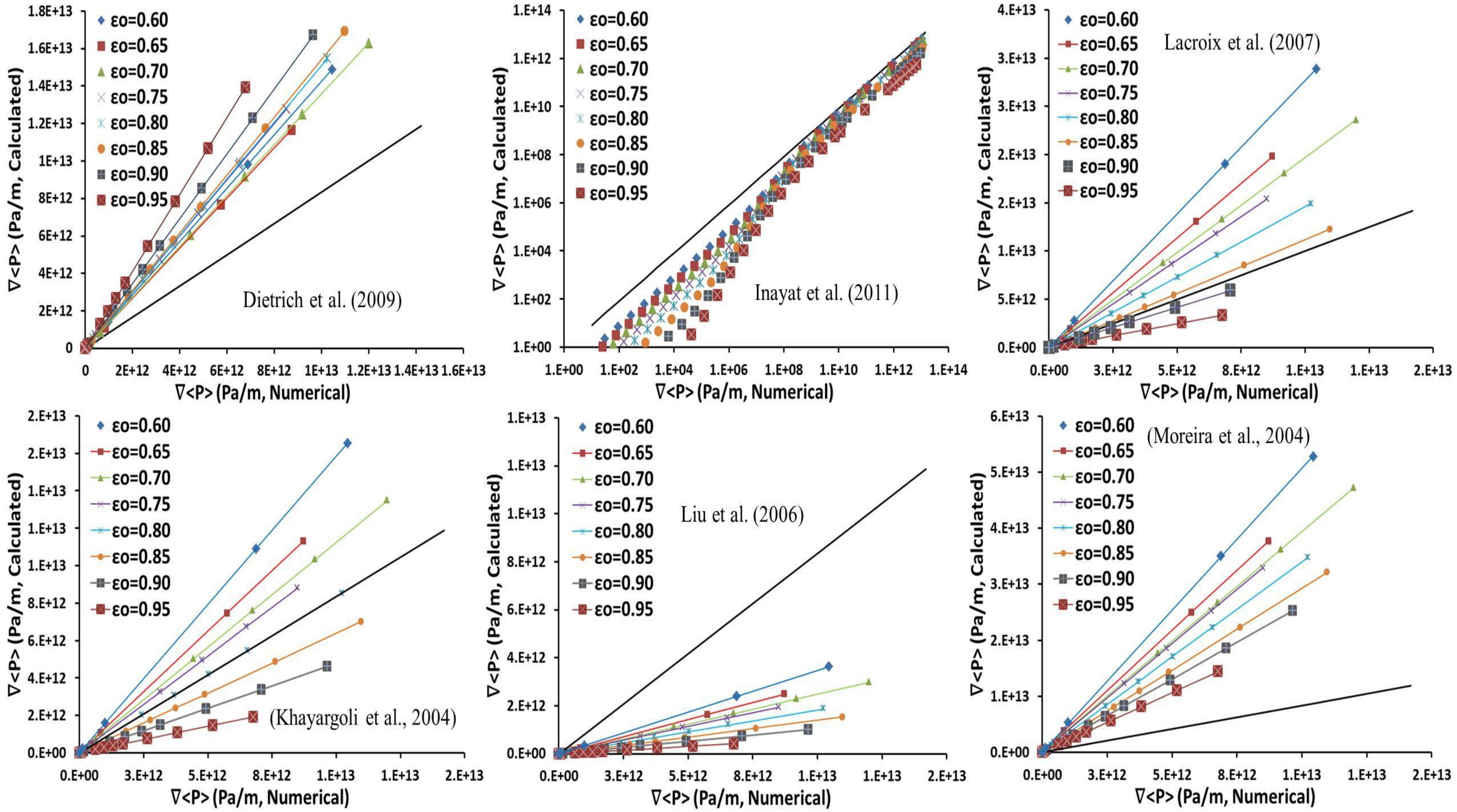

**Figure 4.16**. Performance of state of the art correlations (Black line corresponds to the measured numerical data: from present work). The comparison presented above is performed for circular strut shape. (The calculated and experimental data are plotted in log-log scale in order to distinguish Darcy and inertia regimes clearly).

## 4.5 Correlations for pressure drop in metal foams

In the literature, correlations of pressure drop are widely derived using Ergun-like approach (see Equation 4.5, section 4.2.1). Ergun (1952) obtained $E_1$=150 and $E_2$=1.75 that were actually determined for packed bed of spheres.

### 4.5.1. Ergun parameters have constant numerical values or not

There are several correlations proposed by various authors (see Table 4.1) in the literature. Literature review suggested that there are a few authors (e.g. Innocentini et al., 1999; Khayargoli et al, 2004; Liu et al., 2006; Lacroix et al., 2007; Dietrich et al., 2009) who predicted $E_1$ and $E_2$ to be constants. On the other hand, many authors (Richardson et al., 2000; Moreira et al., 2004; Tadrist et al., 2004; Topin et al., 2006; Dukhan et al., 2006; Inayat et al., 2011b) predicted that Ergun parameters cannot possess constant numerical values.

As $D_p$ is not a self-contained parameter to describe flow laws in open cell foams, a new form of Ergun approach is presented in the Equation 4.19 by replacing $a_c$ with the characteristic length of foam i.e. hydraulic diameter, $d_h$ (=$4\varepsilon_o/a_c$) and is given by the following expression:

$$\frac{\Delta P}{\Delta x} = 16E_1\frac{(1-\varepsilon_o)^2}{\varepsilon_o}\frac{\mu V}{{d_h}^2} + 4E_2\frac{(1-\varepsilon_o)}{{\varepsilon_o}^2}\frac{\rho V^2}{d_h} \tag{4.19}$$

Darcian permeability ($K_D$) and inertia coefficient ($C_{For}$) are related to the Ergun-like approach given in Equation 4.19 to evaluate precisely the parameters $E_1$ and $E_2$. Ergun parameters, $E_1$ and $E_2$ are then correlated with $\varepsilon_o$ and a combination of dimensionless geometrical parameters of foam matrix. A best fit was obtained in order to account any type of foam either based on homothetic transformation or based on constant cell diameter (our samples) and is shown in the Figure 4.17. Firstly, Ergun parameter, $E_1$ was determined so that it can be easily related to the flow property of Darcy regime (see figure 4.17-left). In order to account for both Darcy and inertia regimes, $E_1$ and $E_2$ were incorporated together (see figure 4.17-right). From Figure 4.17, it is evident that $E_1$ and $E_2$ are functions of geometrical parameters (ratio of strut radius or side length to strut length, see section 3.4) of foam matrix.

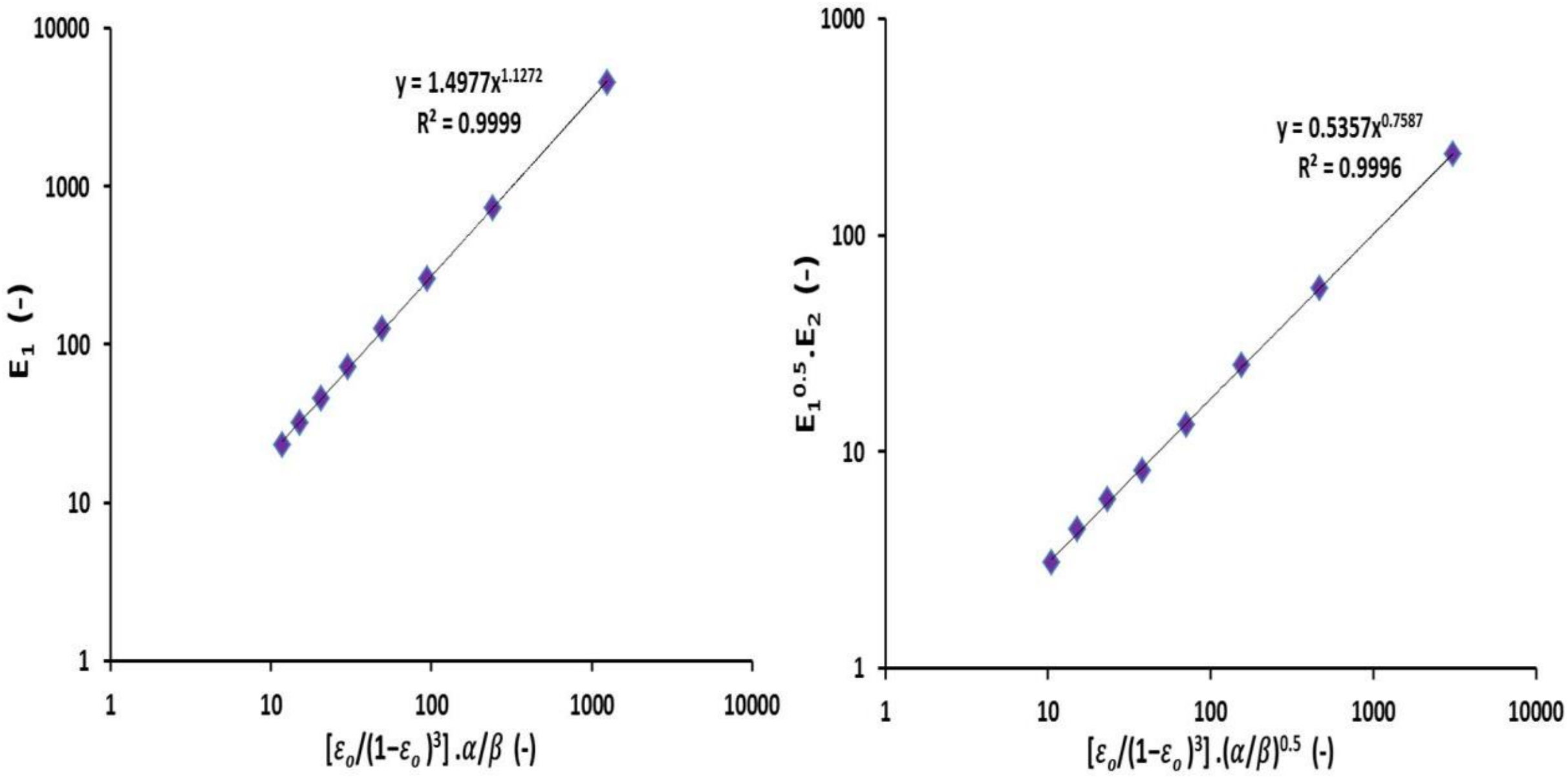

**Figure 4.17**. Plot of Ergun parameters $E_1$ (left) and $E_2$ (right) as combination of porosity and dimensionless geometrical parameter. The plots are shown for circular strut shape in the porosity range 0.60$<\varepsilon_o<$0.95.

In this way, one can obtain precise values of $E_1$ and $E_2$ just be measuring the correct geometrical parameters of foam matrix. From Figure 4.17, the pressure drop correlations of circular strut shape are given by Equation 4.20:

$$E_1 = 1.4977\left[\frac{\varepsilon_o}{(1-\varepsilon_o)^3}.\frac{\alpha_c}{\beta}\right]^{1.1272} \qquad (4.20a)$$

and,

$$E_2 = 0.4377\left[\frac{\varepsilon_o}{(1-\varepsilon_o)^3}.\frac{\alpha_t}{\beta}\right]^{0.1951}\left(\frac{\alpha_c}{\beta}\right)^{-0.3795} \qquad (4.20b)$$

No physical justification of the exponents obtained in the above expressions has been provided. The RMSD values of the parameters $E_1$ and $E_2$ are 1.96% and 2.82% respectively. The porosity of the commercially available foams is in the range ~0.90±0.05 where specific surface area and a known strut shape (generally equilateral triangle) do not impact strongly on flow characteristics and thus, simple correlations were reported in the literature.

As depicted from Figures 4.12, 4.13, 4.14, 4.15, 4.16 and 4.17, flow properties ($K_D$ and $C_{For}$) are strongly dependent on strut shapes and thus, different correlations are derived for different strut shapes on a common basis to predict parameters $E_1$ and $E_2$ and their RMSD values are presented in Table 4.3 (see Equations 4.21-4.26).

**Table 4.3**. Correlations and RMSD values for strut shapes to predict $E_1$ and $E_2$.

| Strut Shape | $E_1$ | RMSD | $E_2$ | RMSD | Equation no. |
|---|---|---|---|---|---|
| Equilateral Triangle | $2.5949\left[\frac{\varepsilon_o}{(1-\varepsilon_o)^3}.\frac{\alpha_t}{\beta}\right]^{0.9127}$ | 4.26% | $3.32423\left[\frac{\varepsilon_o}{(1-\varepsilon_o)^3}.\frac{\alpha_t}{\beta}\right]^{0.0045}\left(\frac{\alpha_t}{\beta}\right)^{-0.22545}$ | 5.8% | 4.21 |
| Diamond | $0.4706\left[\frac{\varepsilon_o}{(1-\varepsilon_o)^3}.\frac{\alpha_{det}}{\beta}\right]^{1.1351}$ | 1.26% | $0.6347\left[\frac{\varepsilon_o}{(1-\varepsilon_o)^3}.\frac{\alpha_{det}}{\beta}\right]^{0.2372}\left(\frac{\alpha_{det}}{\beta}\right)^{-0.40235}$ | 7.36% | 4.22 |
| Rotated Square | $0.7477\left[\frac{\varepsilon_o}{(1-\varepsilon_o)^3}.\frac{\alpha_{rs}}{\beta}\right]^{1.1045}$ | 0.25% | $0.8245\left[\frac{\varepsilon_o}{(1-\varepsilon_o)^3}.\frac{\alpha_{rs}}{\beta}\right]^{0.1805}\left(\frac{\alpha_{rs}}{\beta}\right)^{-0.36635}$ | 2.79% | 4.23 |
| Hexagon | $1.2958\left[\frac{\varepsilon_o}{(1-\varepsilon_o)^3}.\frac{\alpha_h}{\beta}\right]^{1.1074}$ | 1.55% | $0.5476\left[\frac{\varepsilon_o}{(1-\varepsilon_o)^3}.\frac{\alpha_h}{\beta}\right]^{0.1427}\left(\frac{\alpha_h}{\beta}\right)^{-0.3482}$ | 5.33% | 4.24 |
| Square | $1.1712\left[\frac{\varepsilon_o}{(1-\varepsilon_o)^3}.\frac{\alpha_s}{\beta}\right]^{1.0483}$ | 5.18% | $1.1192\left[\frac{\varepsilon_o}{(1-\varepsilon_o)^3}.\frac{\alpha_s}{\beta}\right]^{0.07245}\left(\frac{\alpha_s}{\beta}\right)^{-0.2983}$ | 8.37% | 4.25 |
| Star | $1.4753\left[\frac{\varepsilon_o}{(1-\varepsilon_o)^3}.\frac{\alpha_{st}}{\beta}\right]^{1.0573}$ | 2.12% | $1.0609\left[\frac{\varepsilon_o}{(1-\varepsilon_o)^3}.\frac{\alpha_{st}}{\beta}\right]^{0.0765}\left(\frac{\alpha_{st}}{\beta}\right)^{-0.30255}$ | 5.2% | 4.26 |

**Table 4.4**. Correlations between $K_D$ and $C_{For}$ for different strut shapes.

| Strut Shape | Relationship between $K_D$ and $C_{For}$ | $C_1$ | $C_2$ | RMSD (%) | Equation no. |
|---|---|---|---|---|---|
| Equilateral Triangle | $C_{For}.\sqrt{K_D} = e^{C_2}\left(\frac{\alpha_t}{\beta}.\sqrt{\varepsilon_o}\right)^{C_1}$ | 2.0211 | -0.724 | 10.22 | 4.28 |
| Diamond | $C_{For}.\sqrt{K_D} = e^{C_2}\left(\frac{\alpha_{det}}{\beta}.\sqrt{\varepsilon_o}\right)^{C_1}$ | 0.8477 | -1.0735 | 12.39 | 4.29 |
| Rotated Square | $C_{For}.\sqrt{K_D} = e^{C_2}\left(\frac{\alpha_{rs}}{\beta}.\sqrt{\varepsilon_o}\right)^{C_1}$ | 1.031 | -1.0557 | 7.51 | 4.30 |
| Hexagon | $C_{For}.\sqrt{K_D} = e^{C_2}\left(\frac{\alpha_h}{\beta}.\sqrt{\varepsilon_o}\right)^{C_1}$ | 1.0574 | -1.1138 | 7.38 | 4.31 |
| Square | $C_{For}.\sqrt{K_D} = e^{C_2}\left(\frac{\alpha_s}{\beta}.\sqrt{\varepsilon_o}\right)^{C_1}$ | 1.1356 | -1.4491 | 9.70 | 4.32 |
| Star | $C_{For}.\sqrt{K_D} = e^{C_2}\left(\frac{\alpha_{st}}{\beta}.\sqrt{\varepsilon_o}\right)^{C_1}$ | 1.3199 | 0.2529 | 7.38 | 4.33 |

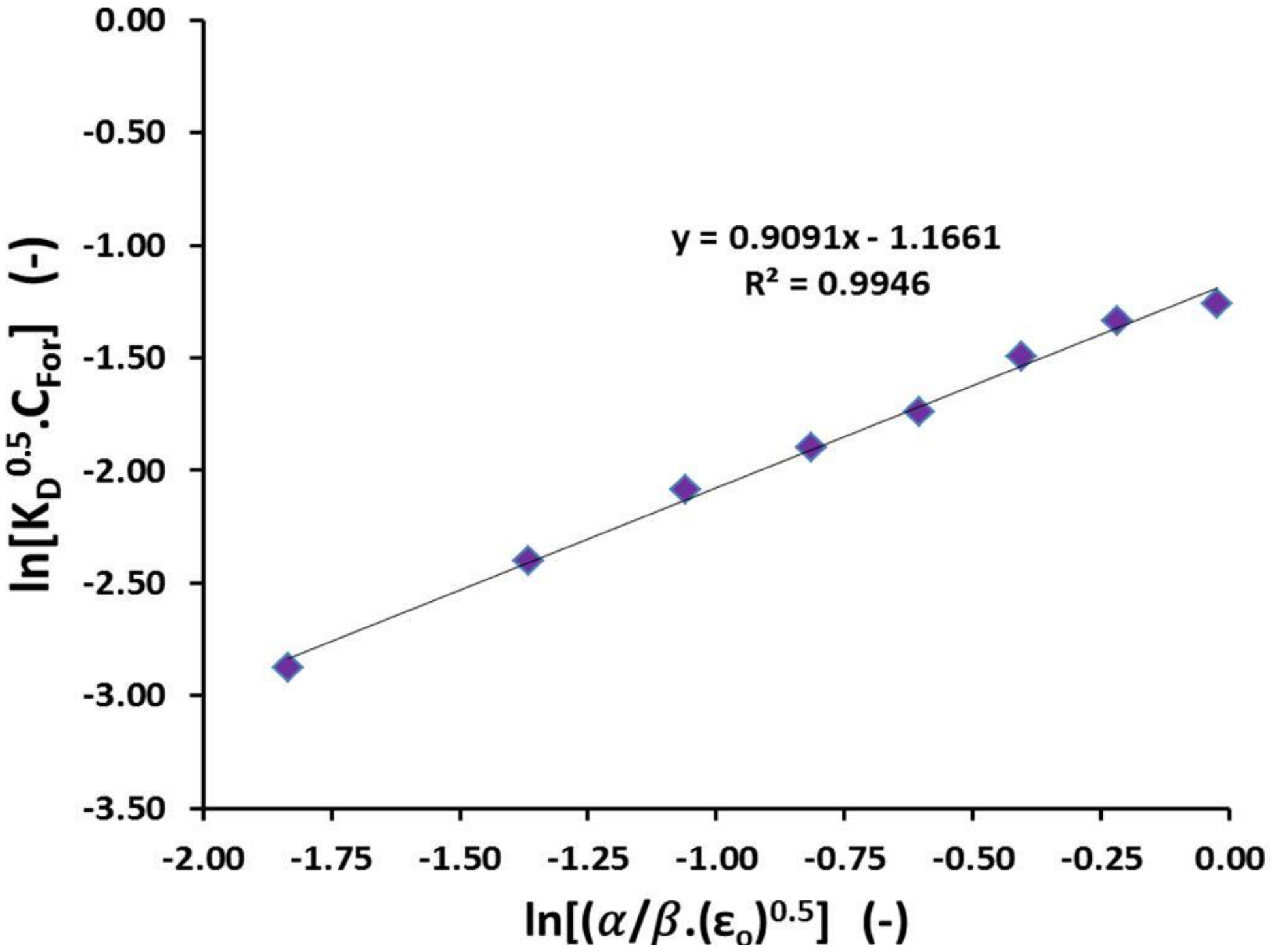

**Figure 4.18**. Relationship between $K_D$ and $C_{For}$. The plot is shown for circular strut shape in the porosity range 0.60< $\varepsilon_o$ <0.95.

Table 4.3 shows that numerical coefficients and exponents appear in the correlations are different for different strut cross sections which clearly suggests that Ergun parameters, $E_1$ and $E_2$ are strictly the functions of geometrical parameters and cannot possess constant numerical values in the wide range of porosity. However, there is a possibility that Ergun parameters could have constant numerical values only in the high porosity range ($\varepsilon_o$ >0.90) for a given strut shape.

It is also certain that it is highly difficult to control the experimental measurements in Darcy regime and thus, it is quite critical to predict $K_D$ if $C_{For}$ (or $\approx C_{poly}$) is correctly known. Several combinations have been attempted to relate $K_D$ and $C_{For}$ with the geometrical parameters of the foam matrix of different strut cross sections. For circular strut shape, the relationship between $K_D$ and $C_{For}$ is shown in Figure 4.18 and is given by the following expression:

$$C_{For}.\sqrt{K_D} = e^{C_2}\left(\frac{\alpha_c}{\beta}\ .\sqrt{\varepsilon_o}\right)^{C_1} \qquad (4.27)$$

where, $C_1$=0.9091 and $C_2$= -1.1661

An RMSD value of 7.92% was obtained for calculated values of $K_D$ in case of circular strut cross section in the entire range of porosity for known inertia coefficients ($C_{For}$). The correlations for other strut cross sections and RMSD values are presented in Table 4.4 (see Equations 4.28-4.33).

### 4.5.2 Friction factor

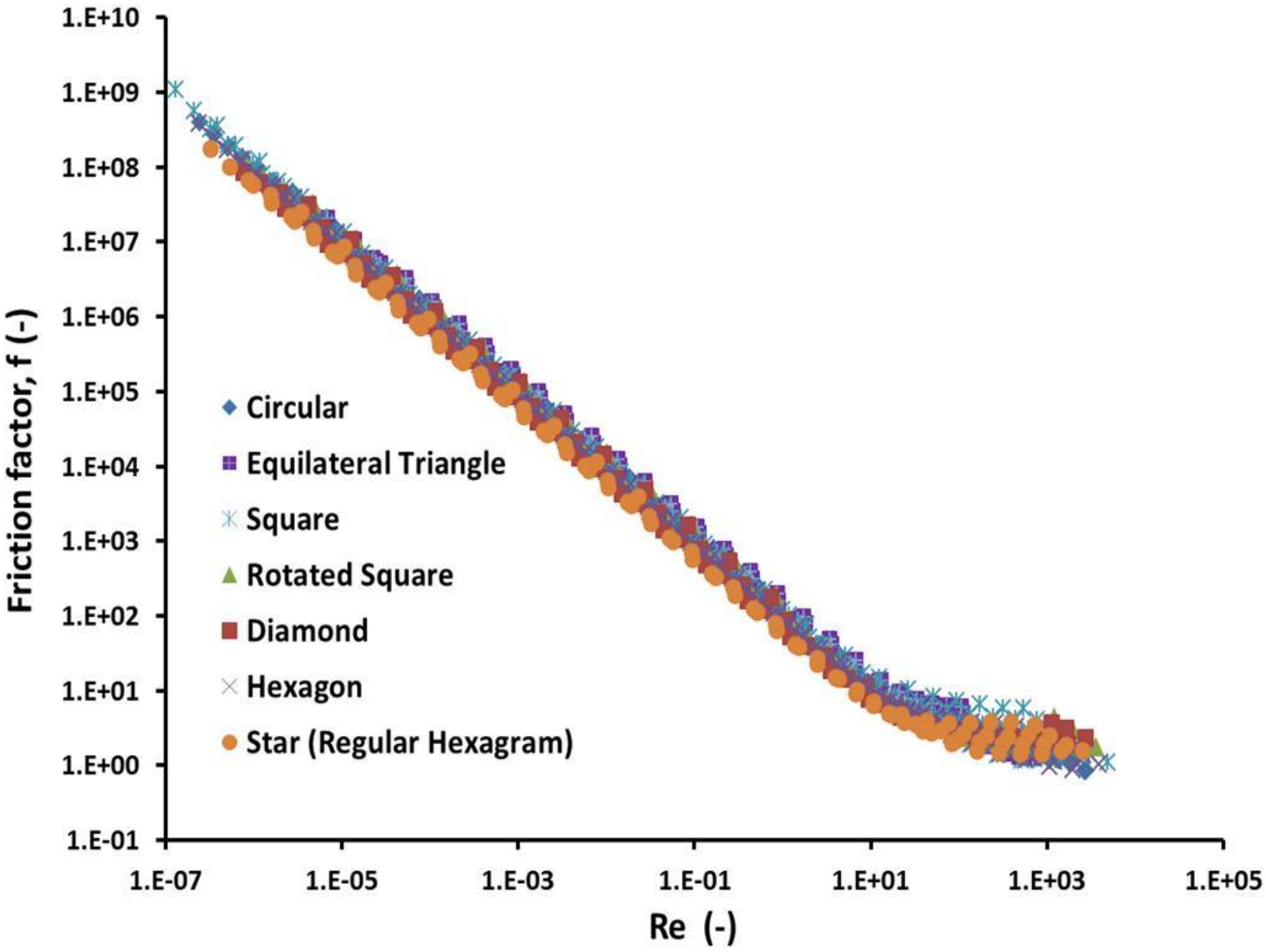

**Figure 4.19**. Plot of friction factor, $f$ and $Re$ for different strut cross sections.

In order to avoid ambiguities, characteristic length $d_h$ is used to calculate the friction factor and Reynolds number. Friction factor is calculated using numerically obtained pressure gradients across the foam sample and is given as:

$$f = \nabla\langle P\rangle . \frac{d_h}{\rho V^2} \tag{4.34}$$

Friction factor ($f$) is plotted against wide range of Reynolds number ($10^{-7} < Re < 10^4$) for different strut cross sections and is presented in Figure 4.19. It is clear that $f$ values are not dispersed and follows the same trend.

### 4.6 Correlations for pressure drop in ceramic foams

In this work, no fluid flow experiments were performed through the ceramic foams. For a given total porosity, the open porosity is different and thus, the flow properties are different for a given pore size as compared to metal foam (solid strut) of the same pore size. In case of ceramic foams, the ligament is more irregular as compared to metal ones and surface roughness is an important factor that influences the flow properties. Experimental data from the work of Dietrich et al., (2009) were gathered to develop new correlations to estimate pressure drop in case of ceramic forms. Moreover, it is highlighted in the present work (see sections 4.4 and 4.5) that a dimensionless geometrical parameter needs to be added in Ergun-like approach to account for strut shape and foam geometry in order to obtain precise correlation.

Dietrich et al., (2009) calculated the hydraulic diameter using total porosity ($\varepsilon_t$) but not open or hydrodynamic porosity ($\varepsilon_o$). These authors argued that the two values of porosities are close for their samples and that information about total porosity ($\varepsilon_t$) is more fully available. Permeability is very sensitive to flow conditions and porosity and in turn, will impact the global pressure drop quite significantly if correct open porosity is not considered (see Inayat et al., 2011b).

Dietrich et al., (2009) used a slightly different formula than the Ergun-like approach, given by following expression as:

$$\frac{\Delta P}{\Delta x} = E_1]_D \frac{1}{\varepsilon_t . d_h]_D^{\,2}} \mu V + E_2]_D \frac{1}{\varepsilon_t^{\,2} . d_h]_D} \rho V^2 \;\; with \;\; d_h]_D = \frac{4\varepsilon_t}{a_c} \tag{4.35}$$

Using the experimental values of flow properties $K$ and $C$ provided by Dietrich et al., 2009, parameters $E_1]_D$ and $E_2]_D$ (calculated using Equation 4.35) and parameters, $E_1$ and $E_2$ (calculated using Equation 4.5) are compared and the results are presented in Appendix F (Table F1 and Table F2). In Table F1, it is observed that the average deviations between parameters $E_1$ (Equation 4.5) and $E_1]_D$ (calculated either using hydraulic diameter from pressure drop measurements or specific surface area, Equation 4.35) are 35-40%. Similarly, the average deviations between parameters $E_2$ (Equation 4.5) and $E_2]_D$ (calculated either using hydraulic diameter from pressure drop measurements or specific surface area, Equation 4.35) are 26-28% as presented in Table F2.

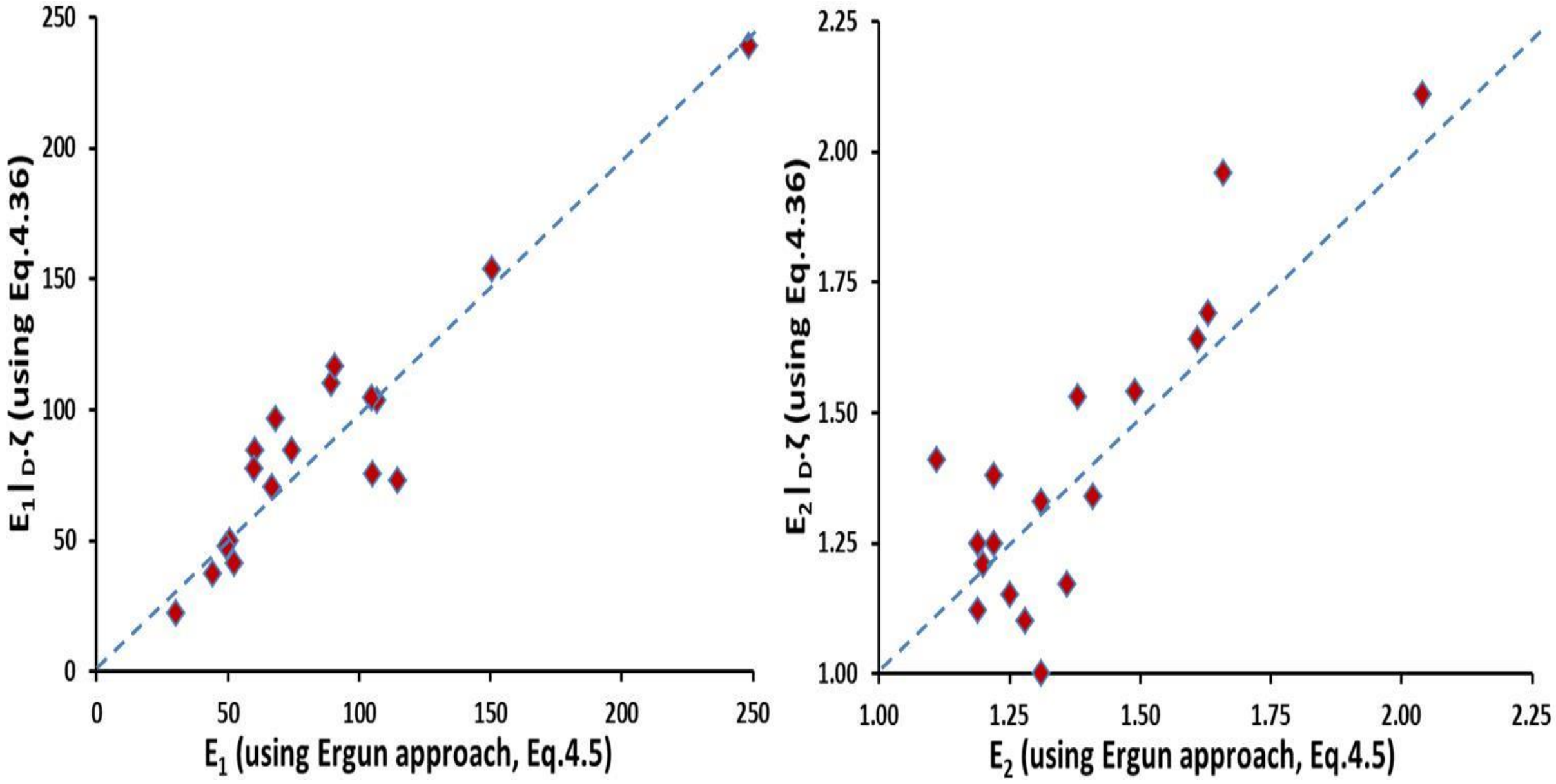

**Figure 4.20**. Left: Comparison and validation of Ergun parameter $E_1$ and $E_1]_D.\tau$. Right: Comparison and validation of Ergun parameter $E_2$ and $E_2]_D.\tau$.

These deviations in parameters $E_1]_D$ and $E_2]_D$ clearly suggest an inclusion of a correction factor that encompasses the characteristic dimensions of foam structure compared to window or strut diameter or open porosity even if the strut (or inner) porosity is unknown. A correction factor (a dimensionless geometrical parameter) $\tau = \varepsilon_o(\alpha/\beta)^k$ is proposed that needs to be multiplied with $E_1]_D$ and $E_2]_D$ to obtain comparable values of $E_1$ and $E_2$. The use of factor $\tau$ will improve the reliability of the correlation and help to reduce the dispersion of calculated values (For $\alpha$ and $\beta$, see the analytical approach presented in section 3.4.4).

The relationships between $E_1]_D$, $E_2]_D$, $E_1$ and $E_2$ are given as follows:

$E_1 = E_1]_D.\tau$ and $E_2 = E_2]_D.\tau$ with $d_h]_D$ from Equation (4.35) (4.36)

In Figure 4.20, the relationships presented in Equation 4.36 were plotted and the value of the exponent $k$ that appears in correction factor $\tau$ was identified. A value of $k = -0.1$ gives the best agreement between Ergun parameters $E_1$ and $E_2$ calculated using Equations 4.5 and 4.36. No physical interpretation of the empirical value of $k$ is yet provided. In order to provide a generic correlation, a systematic study needs to be done.

## 4.7 Validation of pressure drop correlations

All the correlations (corresponding to different strut shapes of metal and ceramic foams) developed in the present work are validated against measured numerical/experimental pressure drop data and flow properties. The validations are shown in the following subsections.

### 4.7.1 Pressure drop validation in metal foams

Flow characteristics ($K_D$ and $C_{For}$) and Ergun parameters ($E_1$ and $E_2$) are compared in the section 4.7.1.1. Based on the validation of flow properties, the calculated pressure drop data from correlations are validated against the measured values for different strut shapes in the entire porosity range in the section 4.7.1.2. The validation of friction factor is presented in the section 4.7.1.3 using the flow properties. Lastly, the correlations are validated against experimental flow properties reported in the literature in the section 4.7.1.4.

#### 4.7.1.1 Validation of flow characteristics

The calculated values of flow properties ($K_D$ and $C_{For}$) and Ergun parameters ($E_1$ and $E_2$) are presented in Table 4.5 and validated against numerical data of the present work. The RMSD of Ergun parameters are already provided in Table 4.3. The validation of Darcian permeability ($K_D$) and Forchheimer inertia coefficient ($C_{For}$) are also presented in the Figures 4.21 and 4.22 respectively and an RMSD value for each strut shape are provided. The predicted results are in excellent agreement. It is thus, evident that permeability and inertia coefficient are very sensitive to porosity range (mainly low porosity range) and strictly, are strut shape dependent.

#### 4.7.1.2 Validation of numerical pressure drop

The pressures drop variation in the entire range of velocity (viscous and inertia regimes) leads to undermine the error in Darcy regime due to the strong impact of inertia regime that is several orders of magnitude bigger than in Darcy regime (also illustrated in Figure 4.10). In most of the cases, it has been shown by authors (e.g. Dietrich, 2012) in the literature that the calculated pressure drop values are in the error range of ±40% which is mainly the error in inertia regime. Inertia regime, actually, supresses the error in Darcy regime to a great extent which is actually invisible for the entire pressure drop values range and it is one of the critical reasons that no correlation in the literature is universally accepted (see section 4.4).

The calculated pressure drop values are validated against measured pressure drop data using different correlations in the entire velocity range. Further, the comparison and validation in Darcy and inertia regimes are presented separately to quantity the impact of correlations and the errors associated with them in both the regimes.

Pressure drop values are validated for circular strut shape of 60% porosity in Figure 4.23. Three curves are presented: Darcy + Inertia regime (entire pressure drop values), Darcy regime and inertia regime. From Figure 4.23 (left), it is clear that the correlations predict excellent pressure drop values in the error range of ±4-6%. In Darcy regime (see Figure 4.23-right top), three distinct behaviours of pressure drop values are observed depending on the correlation used and the error is in the range of ±3-11%. Similarly, the error in the inertia regime (see Figure 4.23-right bottom) is exactly the same (±4-6%) like in case of Darcy + Inertia regimes which clearly highlights the suppression of significant errors in Darcy regime. Similarly, the calculated pressure drop for square strut shape of 90% porosity are validated and presented in Figure 4.24. Three distinct behaviours of pressure drop values are observed. The error in pressure drop values is in the range of ±3-9% for Darcy + inertia regimes (Figure 4.24-left). In Darcy regime, the error increases up to ±2-12% (see Figure 4.24-right top) while it is ±3-9% in inertia regime (see Figure 4.24-right bottom).

However, the error in calculated pressure drop values for square strut shape is higher than that of circular strut shape. This is due to the fact that it is very difficult to approximate the geometrical parameters at the node junction of such complex strut shape (see section 3.4.2) in the low porosity range.

In Figure 4.25, the calculated pressure drop of different strut shapes at 80% porosity is compared. It is clear that for the entire pressure drop values, the error is the range of ±1.5-5% (see Figure 4.25-left). However, the errors are ±1-3% in Darcy regime (see Figure 4.25-right top) and ±1.5-5% in inertia regime (see Figure 4.25-right bottom).

**Table 4.5**. Comparison and validation of flow properties and Ergun parameters from numerical measurements and correlations.

| | | Numerical Experiments | | | | Analytical/Calculated values | | | | | |
|---|---|---|---|---|---|---|---|---|---|---|---|
| Shape | $\varepsilon_o$ | $K_D$x$10^{-7}$ $(m^2)$ | $C_{For}$ $(m^{-1})$ | $E_1$ | $E_2$ | ${}_*K_D$x$10^{-7}$ $(m^2)$ | ${}_*C_{For}$ $(m^{-1})$ | $E_1$ | $E_2$ | ${}^*K_D$x$10^{-7}$ $(m^2)$ | ${}^*C_{For}$ $(m^{-1})$ |
| Circular | 0.60 | 0.6015 | 1157 | 23.3 | 0.637 | 0.5795 | 1180 | 24.179 | 0.649 | 0.6918 | 1241 |
| | 0.65 | 0.7387 | 967 | 31.7 | 0.775 | 0.7317 | 928 | 31.993 | 0.744 | 0.6991 | 940 |
| | 0.70 | 0.9033 | 748 | 45.8 | 0.891 | 0.9092 | 723 | 45.488 | 0.862 | 0.8298 | 717 |
| | 0.75 | 1.1024 | 530 | 71.4 | 0.966 | 1.1248 | 556 | 69.996 | 1.014 | 1.1514 | 542 |
| | 0.80 | 1.3517 | 409 | 124.2 | 1.198 | 1.3774 | 417 | 121.927 | 1.223 | 1.3242 | 404 |
| | 0.85 | 1.6712 | 304 | 257.4 | 1.564 | 1.7015 | 300 | 252.835 | 1.542 | 1.5261 | 291 |
| | 0.90 | 2.1176 | 197 | 731.3 | 2.089 | 2.1227 | 198 | 729.495 | 2.102 | 2.1003 | 196 |
| | 0.95 | 2.8587 | 106 | 4519.5 | 3.518 | 2.8011 | 105 | 4612.394 | 3.493 | 3.0886 | 110 |
| Equilateral Triangle | 0.80 | 9.69 | 391.9 | 177.2 | 3.675 | 9.0329 | 378 | 168.537 | 3.692 | 7.9550 | 355 |
| | 0.85 | 14.00 | 191.9 | 307.0 | 3.118 | 17.9390 | 217 | 317.122 | 3.995 | 16.5248 | 208 |
| | 0.90 | 19.90 | 100.7 | 750.0 | 3.322 | 24.9568 | 113 | 788.167 | 4.411 | 22.8088 | 108 |
| | 0.95 | 30.90 | 40.8 | 3979.6 | 4.189 | 33.4227 | 42 | 3860.807 | 5.130 | 27.8344 | 39 |
| Square | 0.60 | 0.3444 | 2603 | 32.9 | 1.287 | 0.2565 | 2246 | 29.945 | 1.086 | 0.3401 | 2586 |
| | 0.65 | 0.4726 | 1663 | 39.8 | 1.195 | 0.4727 | 1663 | 38.795 | 1.186 | 0.5058 | 1720 |
| | 0.70 | 0.6310 | 1134 | 52.4 | 1.208 | 0.7241 | 1214 | 53.381 | 1.299 | 0.6782 | 1175 |
| | 0.75 | 0.8258 | 796 | 76.0 | 1.295 | 0.9873 | 870 | 79.273 | 1.433 | 0.8562 | 810 |
| | 0.80 | 1.0692 | 570 | 124.8 | 1.489 | 1.1963 | 603 | 131.573 | 1.599 | 1.0104 | 554 |
| | 0.85 | 1.3891 | 398 | 245.4 | 1.823 | 1.3496 | 393 | 257.346 | 1.822 | 1.1689 | 366 |
| | 0.90 | 1.8394 | 234 | 665.4 | 2.208 | 1.7457 | 228 | 678.071 | 2.162 | 1.6667 | 223 |
| | 0.95 | 2.6235 | 100 | 3884.0 | 2.965 | 2.4836 | 98 | 3663.363 | 2.841 | 3.0920 | 109 |
| Rotated Square | 0.80 | 1.2439 | 730 | 103.7 | 1.875 | 1.2398 | 756 | 104.035 | 1.942 | 1.3624 | 764 |
| | 0.85 | 1.5552 | 557 | 213.7 | 2.516 | 1.5600 | 534 | 213.083 | 2.415 | 1.4128 | 531 |
| | 0.90 | 2.0026 | 347 | 600.5 | 3.249 | 2.0052 | 346 | 599.726 | 3.236 | 1.9265 | 340 |
| | 0.95 | 2.7550 | 175 | 3665.5 | 5.157 | 2.7510 | 177 | 3670.780 | 5.214 | 2.8796 | 179 |
| Diamond | 0.80 | 1.3802 | 740 | 81.0 | 1.772 | 1.3593 | 798 | 82.236 | 1.910 | 1.6022 | 798 |
| | 0.85 | 1.6608 | 611 | 173.1 | 2.568 | 1.6774 | 592 | 171.427 | 2.489 | 1.5498 | 590 |
| | 0.90 | 2.0642 | 453 | 503.2 | 3.939 | 2.0947 | 409 | 495.922 | 3.560 | 1.6731 | 408 |
| | 0.95 | 2.7661 | 221 | 3150.9 | 6.052 | 2.7410 | 235 | 3179.708 | 6.411 | 3.1520 | 236 |
| Hexagon | 0.60 | 0.6010 | 1362 | 21.4 | 0.718 | 0.5832 | 1339 | 22.041 | 0.706 | 0.6669 | 1435 |
| | 0.65 | 0.7419 | 1139 | 28.9 | 0.874 | 0.7356 | 1032 | 29.137 | 0.792 | 0.6401 | 1058 |
| | 0.70 | 0.9116 | 784 | 41.4 | 0.892 | 0.9215 | 787 | 40.973 | 0.896 | 0.9018 | 780 |
| | 0.75 | 1.1250 | 561 | 63.8 | 0.977 | 1.1445 | 589 | 62.747 | 1.026 | 1.1607 | 570 |
| | 0.80 | 1.3899 | 416 | 110.2 | 1.165 | 1.4145 | 428 | 108.298 | 1.198 | 1.3560 | 411 |
| | 0.85 | 1.7525 | 277 | 223.5 | 1.360 | 1.7672 | 295 | 221.657 | 1.446 | 1.8142 | 282 |
| | 0.90 | 2.2561 | 175 | 623.6 | 1.769 | 2.2488 | 183 | 625.638 | 1.858 | 2.3871 | 180 |
| | 0.95 | 3.1054 | 95 | 3780.6 | 3.011 | 3.0715 | 99 | 3822.377 | 3.156 | 2.9753 | 93 |
| Star | 0.75 | 0.8050 | 1502 | 42.8 | 1.811 | 0.8329 | 1391 | 41.411 | 1.678 | 0.7575 | 1457 |
| | 0.80 | 1.0858 | 896 | 68.2 | 1.745 | 1.0676 | 963 | 69.416 | 1.876 | 1.2121 | 947 |
| | 0.85 | 1.4108 | 603 | 135.4 | 2.064 | 1.3796 | 625 | 138.492 | 2.140 | 1.4216 | 605 |
| | 0.90 | 1.8704 | 366 | 369.8 | 2.598 | 1.8559 | 359 | 372.691 | 2.546 | 1.7077 | 350 |
| | 0.95 | 2.7363 | 152 | 2122.0 | 3.388 | 2.7747 | 150 | 2092.613 | 3.357 | 2.8321 | 154 |

${}_*K_D$ and ${}_*C_{For}$ are estimated from correlations derived for Ergun parameters ($E_1$ and $E_2$; see Table 4.3) .
${}^*K_D$ is estimated from a known $C_{For}$ while ${}^*C_{For}$ is estimated from a known $K_D$(see Table 4.4).

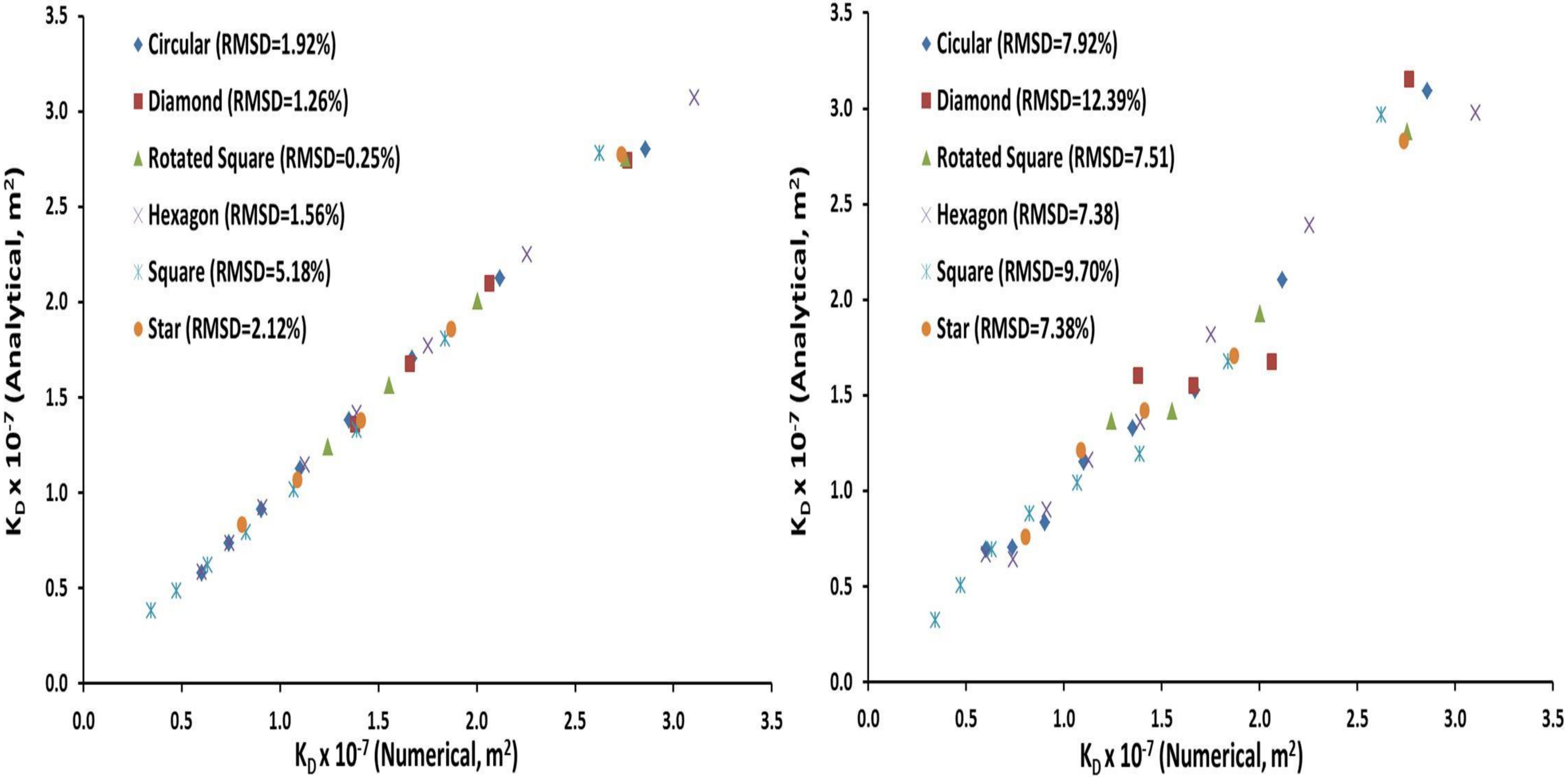

**Figure 4.21**. Comparison of analytical and numerically obtained Darcian permeability ($K_D$). Left- Analytically calculated using Ergun correlation presented in Table 4.3. Right- Analytically calculated using correlation presented in Table 4.4.

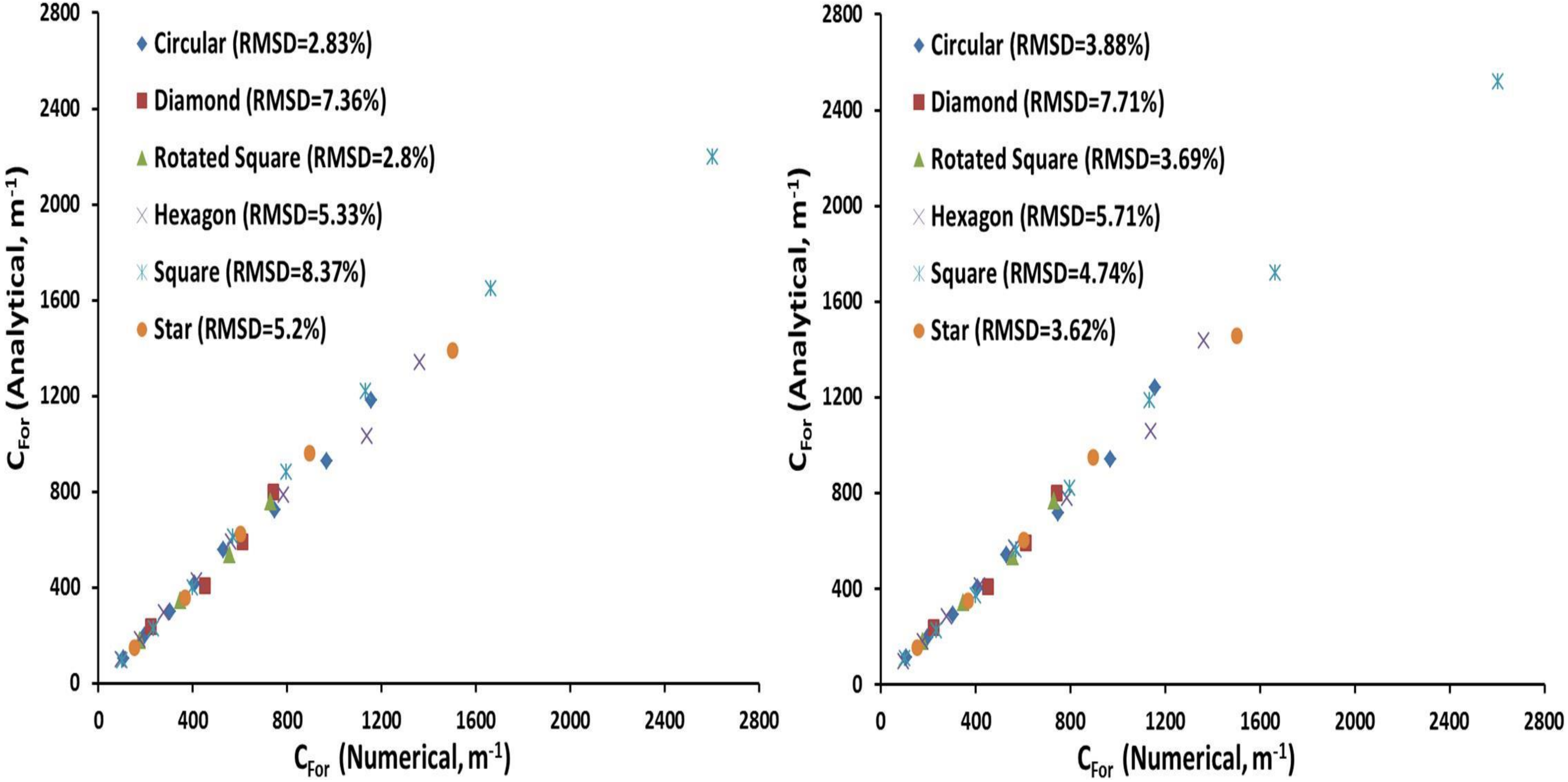

**Figure 4.22**. Comparison of analytical and numerically obtained Forchheimer inertia coefficient ($C_{For}$). Left- Analytically calculated using Ergun correlation presented in Table 4.3. Right- Analytically calculated using correlation presented in Table 4.4.

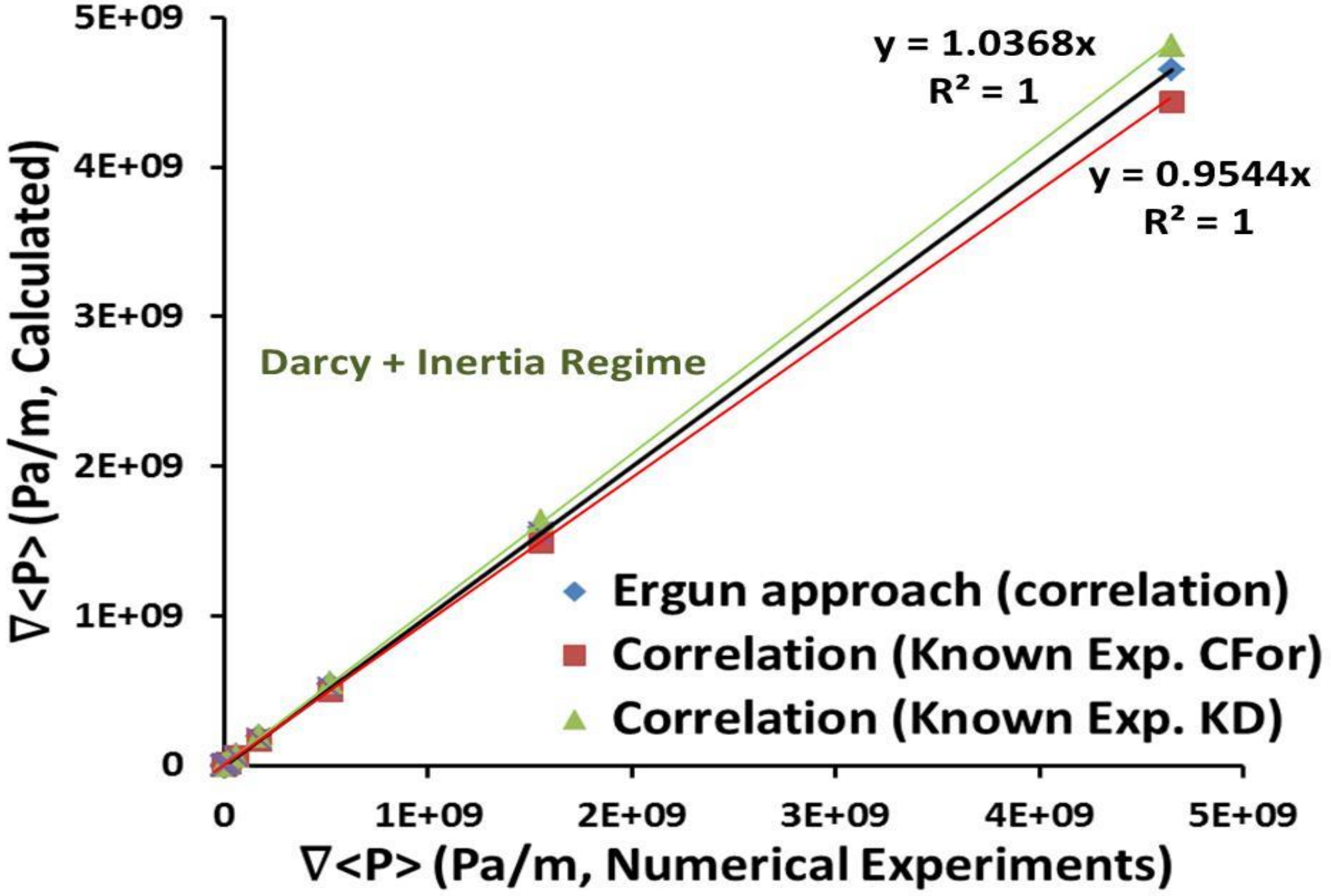

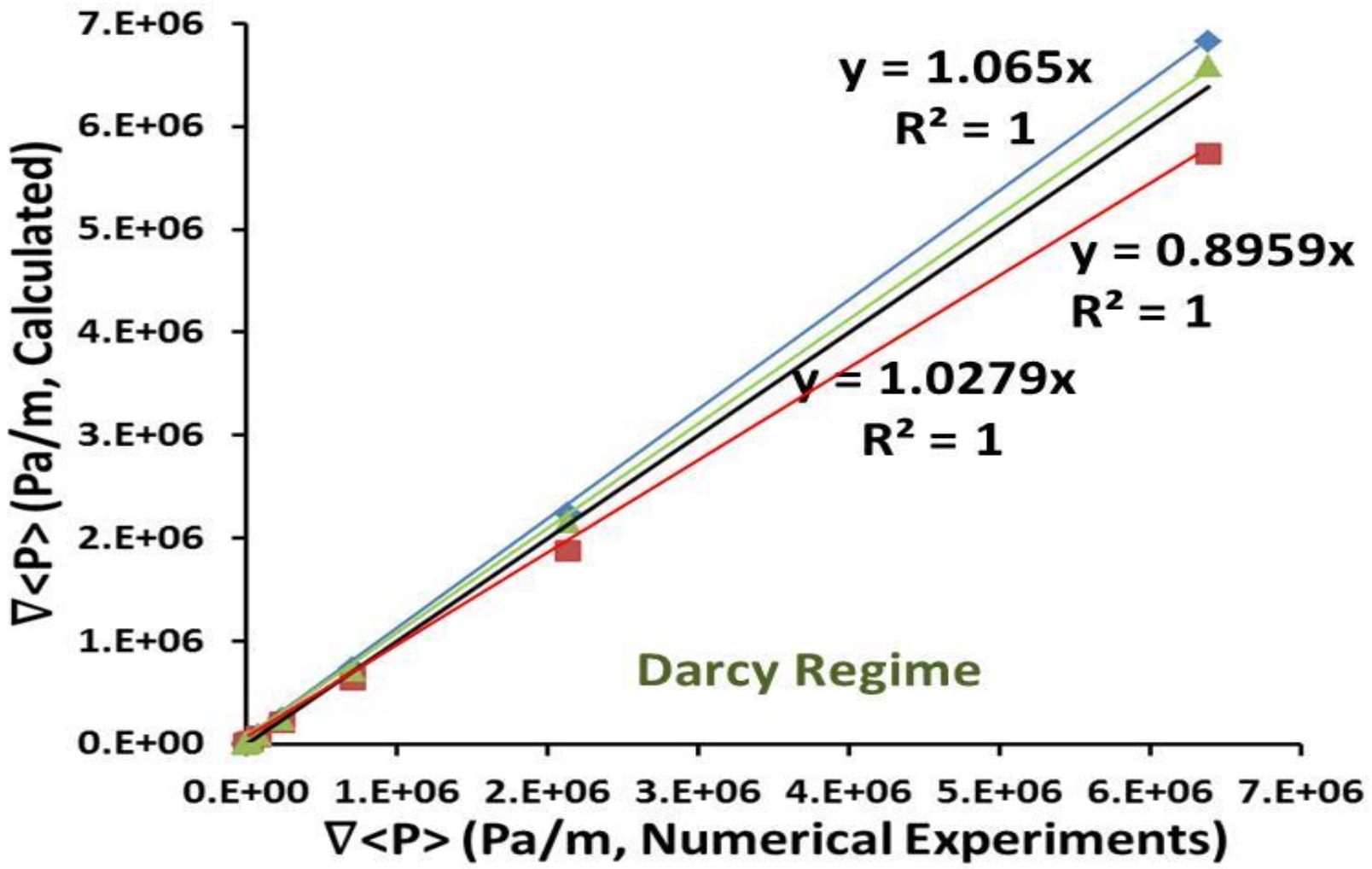

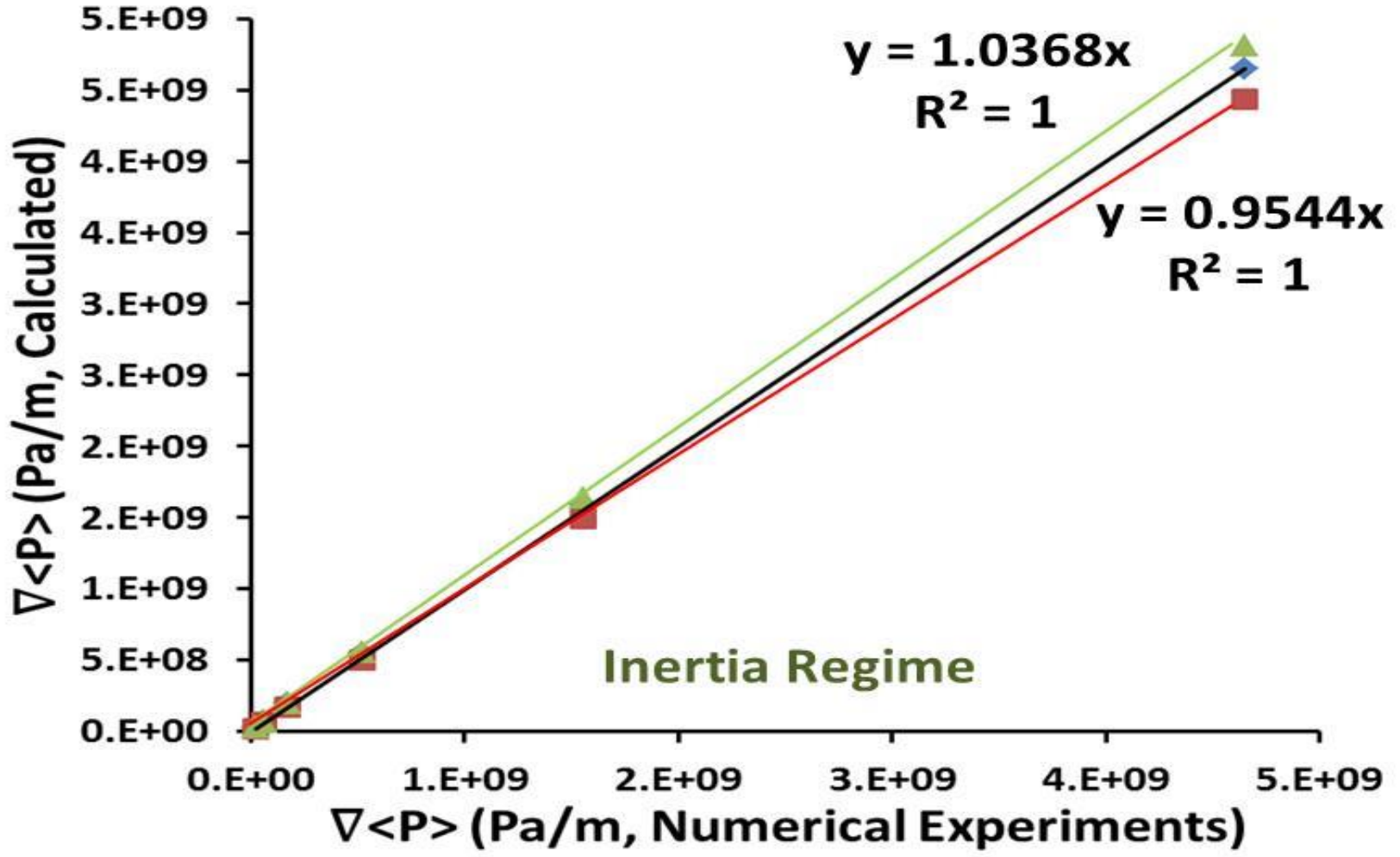

**Figure 4.23**. Comparison and validation of numerical experimental pressure drop against calculated values from different correlations. Left- For entire range of pressure drop (Darcy + Inertia regime). Right Top- Darcy regime, Right Bottom- Inertia Regime. The results presented here for circular strut shape of 60% porosity. (Black line corresponds to experimental pressure drop data).

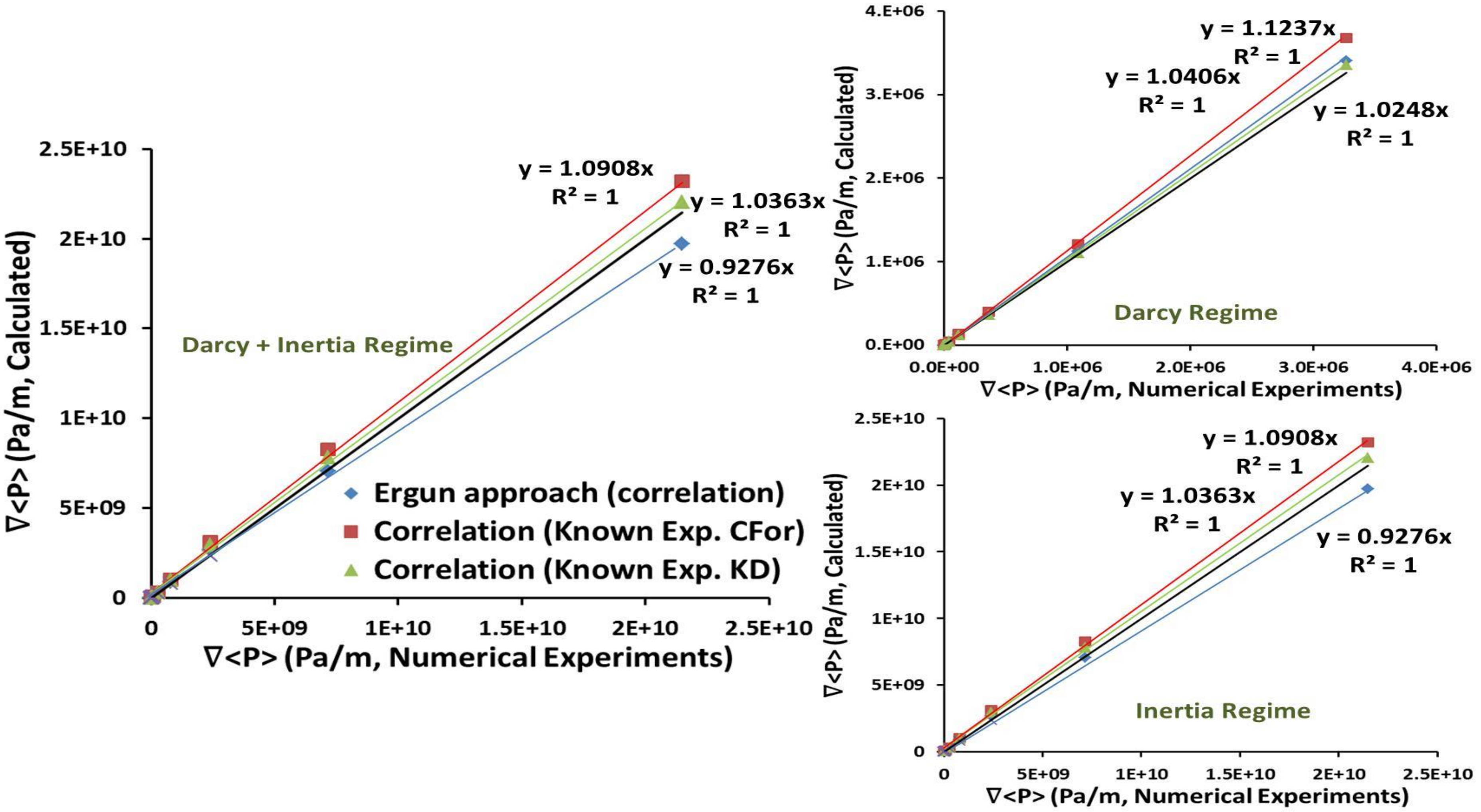

**Figure 4.24**. Comparison and validation of numerical experimental pressure drop against calculated values from different correlations. Left- For entire range of pressure drop (Darcy + Inertia regime). Right Top- Darcy regime, Right Bottom- Inertia Regime. The results presented here for square strut shape of 90% porosity. (Black line corresponds to experimental pressure drop data).

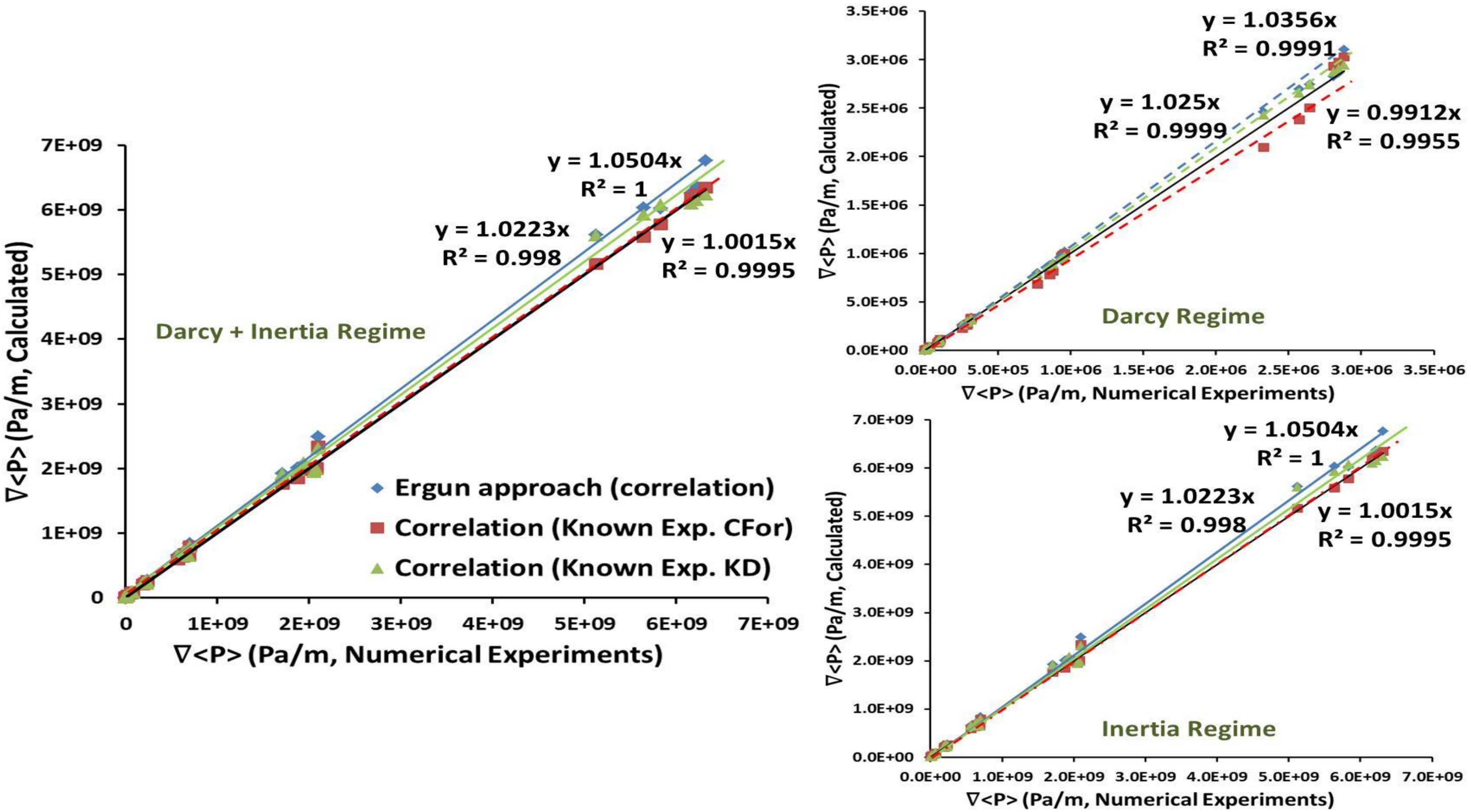

**Figure 4.25**. Comparison and validation of numerical experimental pressure drop against calculated values from different correlations. Left- For entire range of pressure drop (Darcy + Inertia regime). Right Top- Darcy regime, Right Bottom- Inertia Regime. The results presented here for circular, diamond, rotated square, square, hexagon and star strut shape of 80% porosity. (Black line corresponds to experimental pressure drop data).

After careful evaluation of Figures 4.23, 4.24 and 4.25, it is critical that few complex strut cross sections could suppress the apparent error even in Darcy regime over simple strut cross sections (e.g. circular strut cross section) which is really difficult to quantify and this fact could be easily attributed to impact of inertia regime on fluid flow in the foam structures. However, the correlations predict excellent pressure drop results for all strut shapes in both Darcy and inertia regimes.

### 4.7.1.3 Validation of numerical friction factor

In this section, the numerically obtained friction factor values studied in this work are validated. By applying the definition of Reynolds number using hydraulic diameter and combining with Equation 4.34 (1-D Equation), friction factor can be written as:

$$f = \frac{{d_h}^2}{Re.K_D} + C_{For}.d_h \tag{4.37}$$

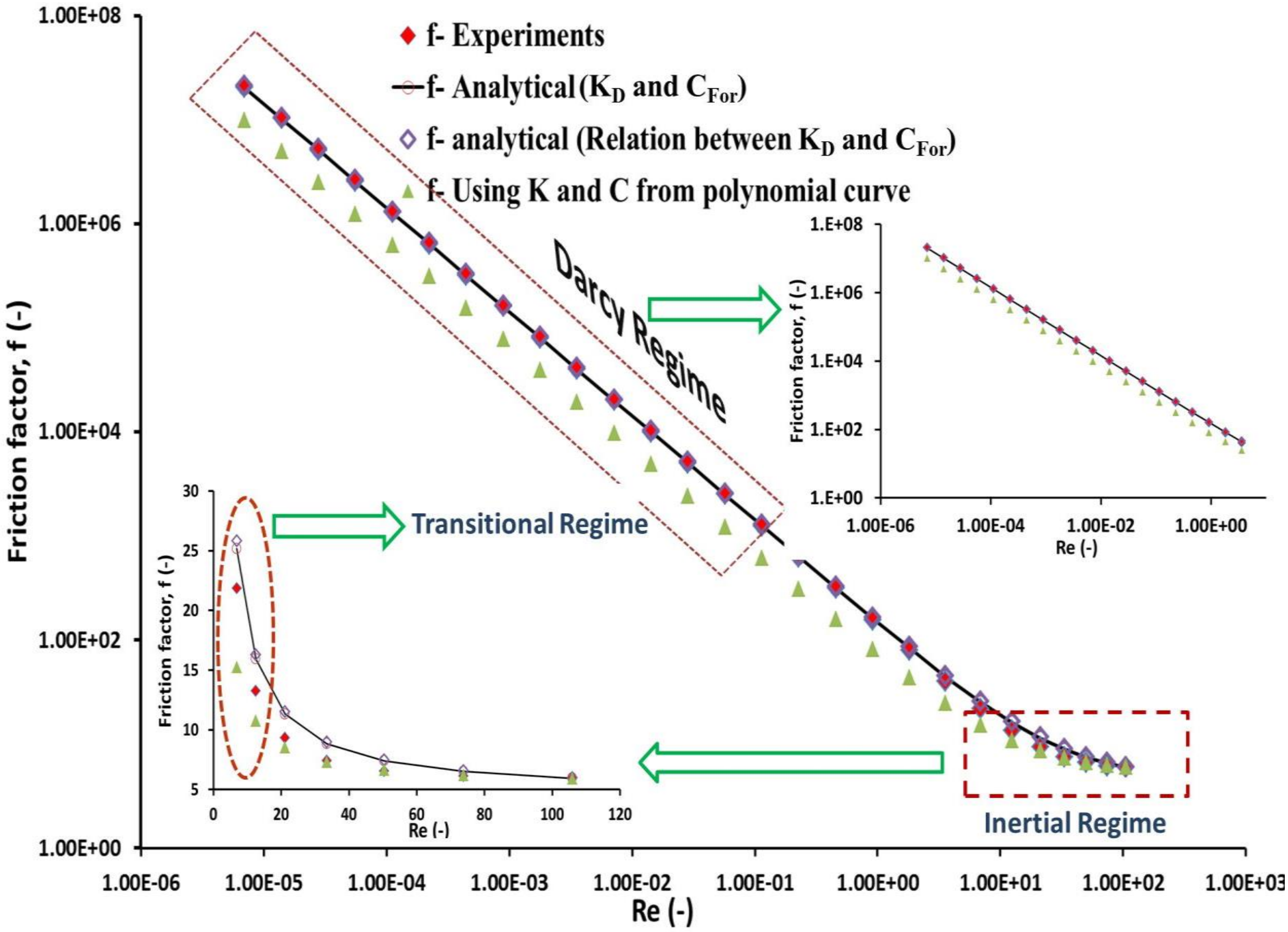

**Figure 4.26.** Plot of friction factor ($f$) Vs. Reynolds number ($Re$). Permeability ($K$) obtained using polynomial curve introduces significant amount of error in friction factor (zoom-right). Departure from transition to inertia regime, inertia coefficient ($C$) predicts numerical friction factor (zoom-left) as there is negligible effect of permeability ($K$). The error in friction factor using permeability $K$ is easily observed in transition regime. Results of equilateral triangular strut are presented.

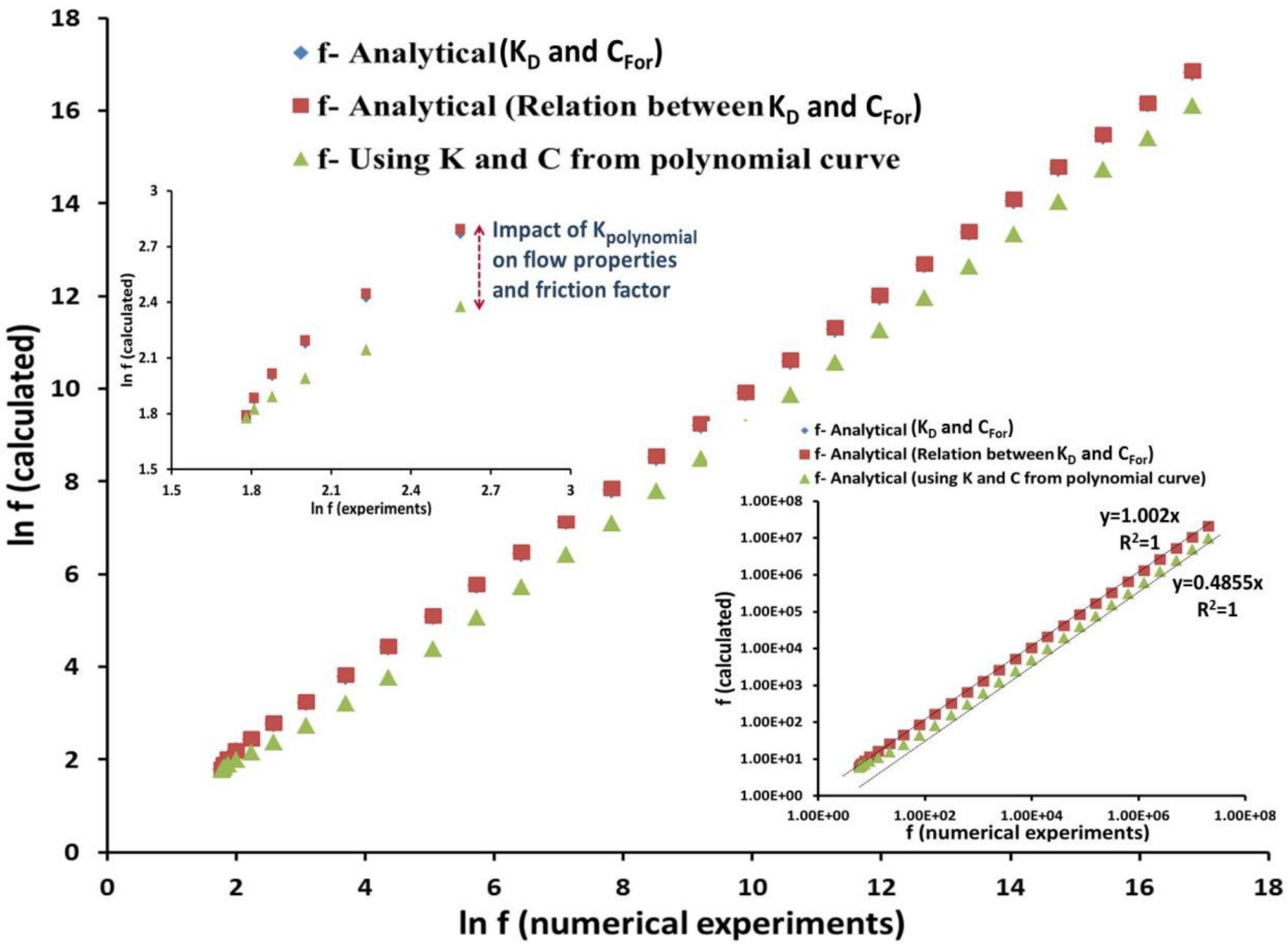

**Figure 4.27.** Comparison of experimental and analytical friction factors (in natural log function). Error increases enormously with permeability ($K$) obtained from polynomial curve (zoom-left). The error increases up to 50% in friction factor due to $K_{poly}$. Analytical correlations using $K_D$ and $C_{For}$ or their combination with geometrical parameters do not introduce errors and can easily be traced back to experimental values. Results of equilateral triangular strut are presented.

The above Equation 4.37 does not introduce any ambiguity according to parameters presented in the literature data (pore diameter, particle diameter, equivalent sphere diameter etc.). In Figure 4.26, friction factor obtained directly using simulations was compared against flow properties obtained using Equations 4.6 and 4.18, analytically determined flow properties using correlations (see Tables 4.3 and 4.4) and flow properties obtained directly using polynomial curve fitting (see Table 4.2). One can easily observe that using the new methodology discussed in the present work to determine $K_D$ and $C_{For}$, it is possible to trace back friction factor from inertia to Darcy regime. Similarly, the analytical formulations to determine the relationships between $K_D$ and $C_{For}$ presented in Tables 4.3 and 4.4 also lead to the same results. As it has been discussed in section 4.3.3 that $C_{For}$ and $C_{poly}$ do not vary to a great extent (see also Table 4.2) and thus, one can trace back friction factor only in inertial regime (see Figure 4.26-left & right bottom). As polynomial curve does not take into account the effects of Darcy and transition regimes, the possibility of tracing back starts to diminish in transition regime (see Figure 4.26-left bottom) which introduces significant error in friction factor in Darcy regime and cannot be neglected (see Figure 4.26-right top).

In the Figure 4.27, the simulated and calculated friction factors are compared in natural log terms. It is clearly evident that the direct polynomial extraction of flow properties ($K$ and $C$) does not provide precise flow characteristics (see figure 4.27-left top). The friction factor starts to collapse only in inertial regime with experimental or simulated values and introduces an error up to 50% in Darcy regime (see Figure 4.27-right bottom). The new methodology and analytical correlations are describing well the friction factors and do not introduce errors in flow properties. Figures 4.26 and 4.27 suggest that it is possible to trace the friction factor values accurately using $K_D$ and $C_{For}$.

#### 4.7.1.4 Validation of experimental pressure drop

The validity of the correlations presented in the present work is compared against experimental data of various authors (Bhattacharya et al., 2002; Boomsma et al., 2003; Dukhan, 2006; Mancin et al., 2010) and presented in Table 4.6. Ergun parameters $E_1$ and $E_2$ are determined analytically using correlations presented in Equation 4.21 (considering equilateral triangular strut shape) to calculate $K_D$ and $C_{For}$. The correlation was found to estimate the pressure drop data satisfactorily within an error range of ±8% for the entire range of pressure drop data (see Figure 4.28). Moreover, the pressure drop comparison is also shown separately in the Darcy regime (see Figure 4.28-zoom view). As discussed in section 4.3 (see also Table 4.2) that the permeability data reported in the literature is Forcheimmer permeability (see also Dukan and Minjeur, 2011), the error in Darcy regime (due to use of $K_{poly}$ and $K_D$) is ±15% as expected. However, the agreement is still fair taking into account that experimental error on $K$ is bigger than error on $C$.

### 4.7.2 Pressure drop validation in ceramic foams

The correlation derived by Dietrich et al., (2009) gives good estimate of pressure drop values for a given strut shape in the high porosity range. However, the correlations developed in the section 4.6 for ceramic foams by introducing a dimensionless geometrical parameter to the Ergun parameters ($E_1]_D$ and $E_2]_D$) derived by Dietrich et al., (2009), it is observed that the average deviations observed for the Ergun parameters $E_1$ and $E_2$ obtained from Equation 4.5 and Equation 4.36 are 7.18 % and 0.35% respectively (see Table F1 and F2).

The permeability and inertia coefficient $K$ and $C$ (see Table 4.7) are validated using the correction factor, $\tau$ (using Equation 4.2 and 4.36). The correlations tend to underestimate the experimental results, the average deviations in calculated flow properties, $K$ and $C$ are -2.89% and -1.6% respectively.

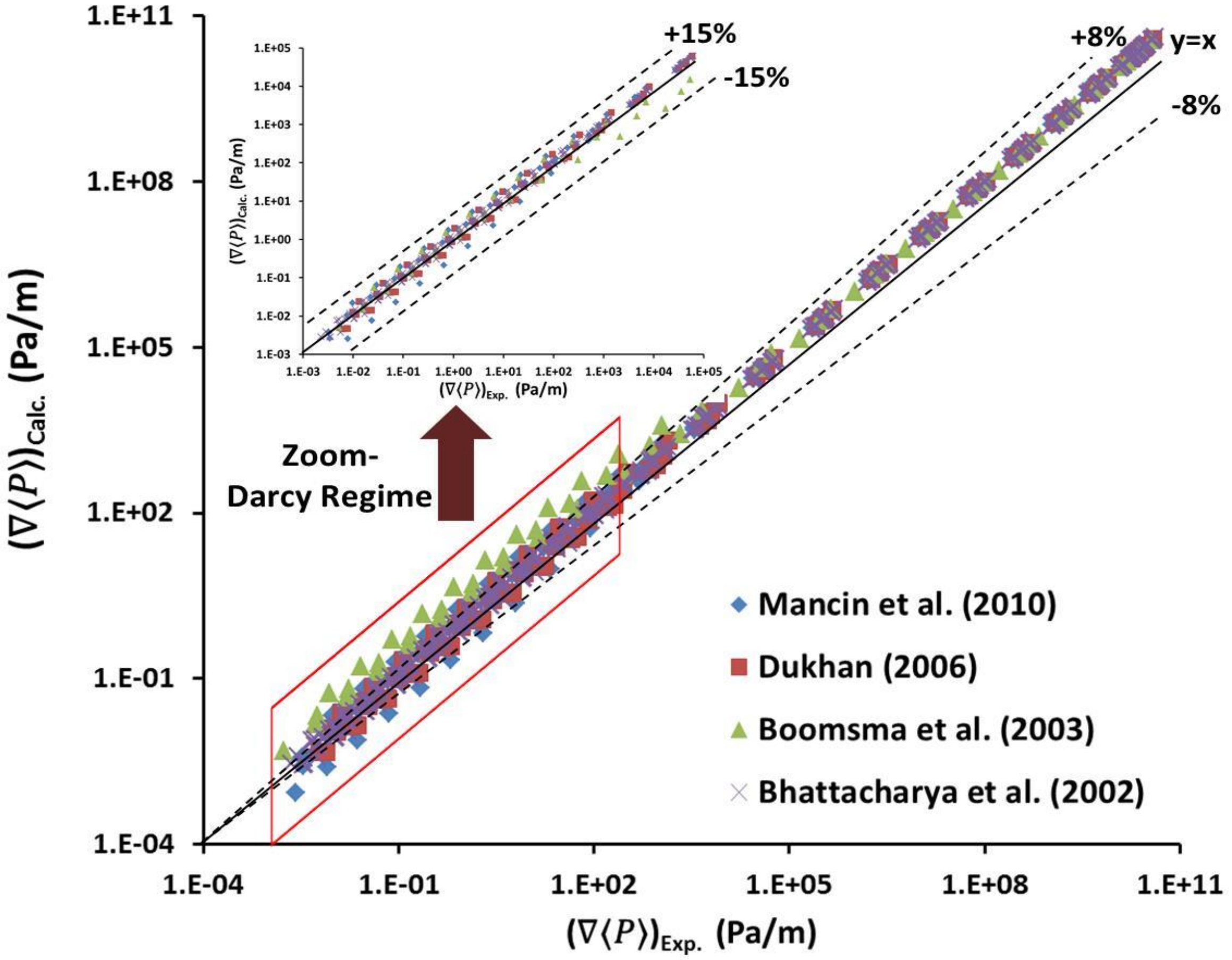

**Figure 4.28**. Calculated vs experimental pressure gradients: data from Table 4.6-present correlation. (The calculated and experimental data are plotted in log-log scale in order to distinguish Darcy and inertia regimes clearly).

One of the reasons for these deviations is the unavailability of experimental values for the specific surface area of a few Mullite foam samples and a complete set of $OBSiC$ foam samples (see Dietrich et al., 2009 and Table 3.9). Note that we extracted $a_c$ using the correlation presented in Equation 3.88 (see section 3.4.4) and used this value to carry out all analytical calculations.

## 4.8 Summary and Conclusion

Fluid flow and pressure drop in porous media are one of the key topics in many industries, especially for designing heat exchangers, filters, columns and reactors etc. that employ a certain porous medium (cellular materials or packed beds). For a similar geometric specific surface area, open cell foams offer a considerably lower pressure drop compared to the conventional randomly packed fixed-bed reactor or column configuration and thus present themselves as promising alternatives for reactor or column internals.

In this chapter, the problems of flow characteristics determination and their prediction in open cell foams have been addressed. For this purpose as a first step, pressure drop in periodic open cell foams of ideal tetrakaidecahedron geometry for different strut shapes was studied.

3-D numerical simulations at pore scale have been carried out to determine precisely the flow properties in the Darcy and inertia regimes. A methodology has been proposed to extract flow properties depending on the choice of flow regime (or flow law). Influence of strut shapes on Darcian permeability and Forchheimer inertia coefficient is analysed and discussed. It is found that different strut shapes impact strongly on flow properties. This tendency suggests us to look into detail the original form of Ergun-like equation and estimate whether Ergun parameters could possess constant numerical values or not.

Using the original form of the Ergun-like equation, different correlations (on a common basis for different strut shapes) for the pressure drop prediction were developed. The derived correlations are found to be dependent on the geometrical characteristics of foam matrix.

These correlations were validated against the numerical results of the pressure drop obtained for the periodic foam structures (based on tetrakaidecahedron geometry) investigated in this work. With the new correlations, it is possible to predict the pressure drop precisely by using any of the two measured geometrical parameters of foam matrix. It has been shown that the new correlations can predict a large amount of measured pressure drop data (of different foam sizes, porosities, and strut cross-sections) with most data points lying within the range of ±10%. It is thus, also possible to construct a foam matrix of the desired shape of any porosity (low and high) for a given pressure drop (or flow properties) using the correlations and methodology discussed in this chapter.

**Table 4.6**. Presentation of experimental data collected from the literature. Analytical values of Ergun parameters $E_1$ and $E_2$ are also presented.

| | | | Experimental data | | | Analytical data | |
|---|---|---|---|---|---|---|---|
| Authors | Sample | $\varepsilon_o$ | $a_c$ (m$^{-1}$) | $K_{poly}$ (x$10^{-7}$ $m^2$) | $C_{poly}$ $(m^{-1})$ | $E_1$ | $E_2$ |
| Mancin et al. (2010) | 5 PPI | 0.921 | 339 | 2.36 | 205 | 1498 | 5.70 |
| | 10 PPI | 0.934 | 692 | 1.87 | 190 | 2353 | 3.23 |
| | 10 PPI | 0.956 | 537 | 1.82 | 240 | 6489 | 8.45 |
| | 20 PPI | 0.932 | 1156 | 0.824 | 226 | 2183 | 2.22 |
| | 40 PPI | 0.93 | 1679 | 0.634 | 342 | 2030 | 2.23 |
| Dukhan (2006) | 10 PPI | 0.919 | 790 | 1 | 210 | 1407 | 2.43 |
| | 10 PPI | 0.915 | 810 | 0.8 | 270 | 1246 | 2.86 |
| | 20 PPI | 0.919 | 1300 | 0.63 | 290 | 1407 | 2.04 |
| | 20 PPI | 0.924 | 1200 | 0.54 | 280 | 1651 | 2.31 |
| | 40 PPI | 0.923 | 1800 | 0.47 | 380 | 1598 | 2.05 |
| Boomsma et al. (2003) | 10 PPI | 0.921 | 820 | 3.529 | 120 | 1498 | 1.38 |
| | 20 PPI | 0.92 | 1700 | 1.089 | 239 | 1451 | 1.30 |
| | 40 PPI | 0.928 | 2700 | 0.712 | 362 | 1891 | 1.42 |
| Bhattacharya et al. (2002) | 5 PPI | 0.973 | 516 | 2.7 | 186.68 | 21981 | 11.76 |
| | 5 PPI | 0.912 | 623 | 1.8 | 200.35 | 1142 | 2.64 |
| | 10 PPI | 0.949 | 843 | 1.2 | 280.01 | 4486 | 5.30 |
| | 10 PPI | 0.914 | 716 | 1.1 | 211.06 | 1210 | 2.49 |
| | 20 PPI | 0.955 | 934 | 1.3 | 257.94 | 6135 | 5.09 |
| | 20 PPI | 0.925 | 898 | 1.1 | 313.57 | 1707 | 3.51 |
| | 40 PPI | 0.927 | 1274 | 0.61 | 360.35 | 1827 | 2.94 |
| | 40 PPI | 0.913 | 1308 | 0.53 | 364.87 | 1175 | 2.32 |
| | 5 PPI | 0.946 | 689 | 2.17 | 212.52 | 3889 | 4.61 |
| | 5 PPI | 0.905 | 636 | 1.74 | 186.99 | 941 | 2.18 |
| | 20 PPI | 0.949 | 975 | 1.185 | 290.5 | 4486 | 4.76 |
| | 40 PPI | 0.952 | 1300 | 0.562 | 411.7 | 5221 | 5.42 |
| | 40 PPI | 0.937 | 1227 | 0.568 | 377.21 | 2643 | 3.82 |

**Table 4.7**. Experimentally and analytically determined flow parameters, $K$ and $C$ of $Al_2O_3$, Mullite and $ObSiC$ ceramic sponges of different pore sizes and porosities. Experimental data are taken from Dietrich et al., 2009.

| | | Experiments | Analytical | | Experiments | Analytical | |
|---|---|---|---|---|---|---|---|
| Material | $\varepsilon_n$ | $K$x $10^{-9}$ (m$^2$) | $K$x $10^{-9}$ (m$^2$) using Eq. (4.2 & 4.36) | $K$ x $10^{-9}$ (m$^2$) using Eq. (4.2 & 4.36) | $C$ x $10^{-5}$ (m) | $C$x $10^{-5}$ (m) using Eq. (4.2 & 4.36) | $C$ x $10^{-5}$ (m) using Eq. (4.2 & 4.36) |
| $Al_2O_3$ | 0.75 | 130 | 116 | 81 | 88 | 83 | 72 |
| | 0.80 | 77 | 90 | 108 | 187 | 202 | 225 |
| | | 54 | 49 | 47 | 114 | 107 | 107 |
| | | 32 | 29 | 28 | 98 | 89 | 91 |
| | | 20 | 22 | 24 | 76 | 80 | 83 |
| | 0.85 | 144 | 180 | 304 | 180 | 201 | 256 |
| Mullite | 0.75 | 90 | 71 | 54 | 95 | 85 | 79 |
| | 0.80 | 299 | 284 | 274 | 186 | 182 | 184 |
| | | 88 | 104 | 58 | 122 | 133 | 101 |
| | | 45 | 62 | 43 | 102 | 120 | 102 |
| | | 29 | 29 | 24 | 66 | 66 | 60 |
| | 0.85 | 120 | 129 | 169 | 190 | 197 | 223 |
| $OBSiC$ | 0.75 | 65 | 71 | 56 | 95 | 99 | 95 |
| | 0.80 | 276 | 255 | 257 | 135 | 130 | 134 |
| | | 56 | 54 | 54 | 123 | 120 | 124 |
| | | 46 | 45 | 45 | 84 | 83 | 86 |
| | | 17 | 11 | 11 | 50 | 41 | 42 |
| | 0.85 | 220 | 263 | 379 | 150 | 164 | 193 |
| *Average Deviation | | | 2.22 % | -2.89% | | 0.55% | -1.6% |

*Average deviation is calculated with respect to experimental values of $K$ and $C$ (Dietrich et al., 2009).

The extended applicability of the new correlations in the large range of porosities was examined by comparing the friction factors of different strut shapes for a wide range of Reynolds numbers. The new correlations presented in this work proved to be applicable for both, low and high porosity *isotropic* open cell foams.

On the other hand, the fluid flow properties for *anisotropic* foams are not presented in this work. It is extremely difficult to obtain permeability and inertia coefficient tensors. In such foams, velocity range is crucial in a given direction and no experimental data are reported in the literature to validate the numerical results.

An algorithm is presented in the Figure 4.29 to determine the flow properties (or Ergun parameters) by using any of the two geometrical parameters of the foam structure. This algorithm can be used in many ways. Some important points are highlighted below:

- Only two input parameters are required to derive all the flow characteristics.
- The combination of geometrical parameters is important to describe flow constants and properties.
- For any known output either geometrical or flow or combination of both geometrical and flow parameters, all other properties can be derived simultaneously.
- For a given constraint (geometrical or flow), one can tailor their own foams accordingly depending on industrial applications.
- One can modify the geometrical and flow properties for different processes.
- This algorithm can be used in reciprocal way (from input to output and vice versa).

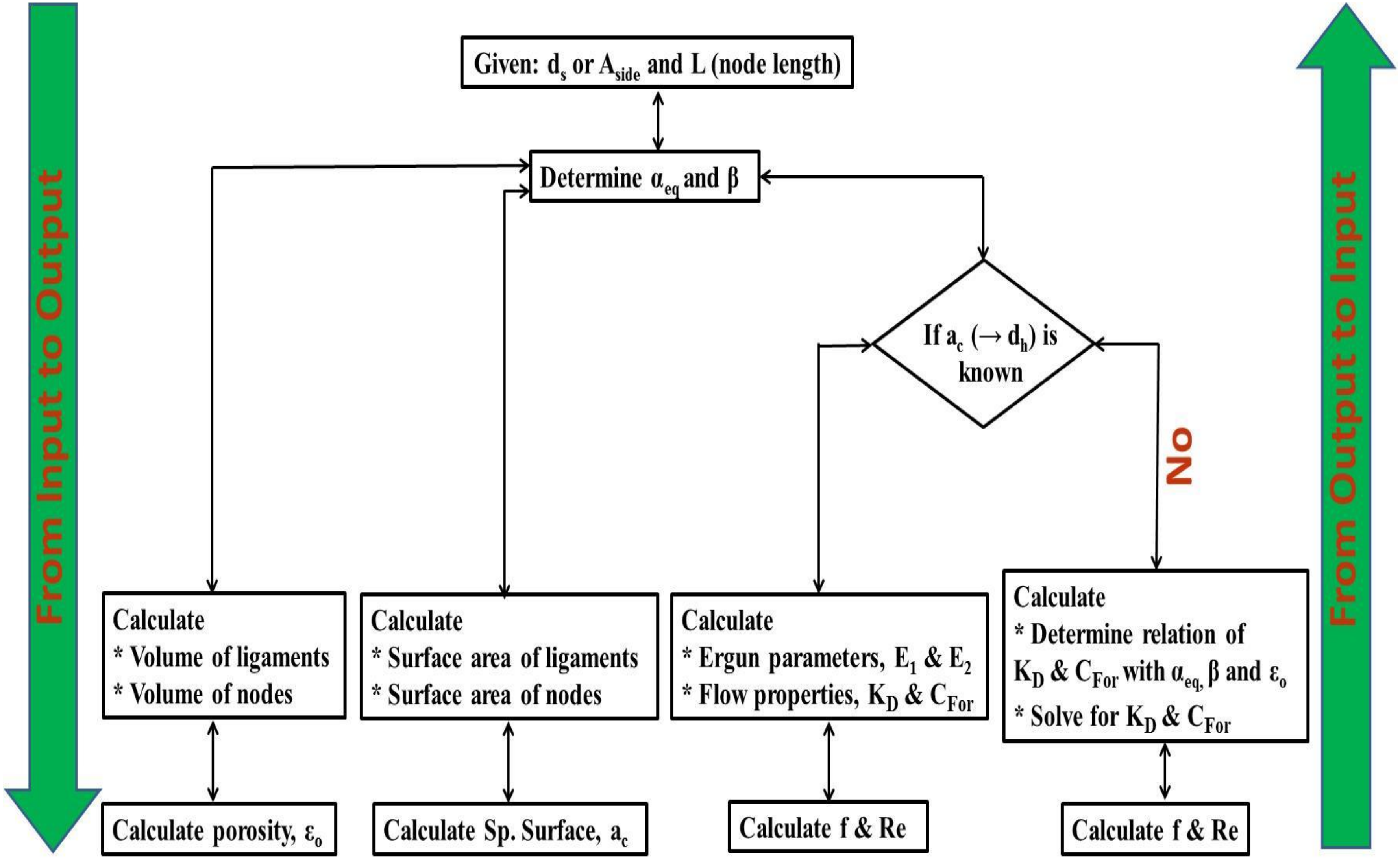

**Figure 4.29**. Algorithm to predict flow properties ($K_D$ and $C_{For}$) and Ergun parameters ($E_1$ and $E_2$) by geometrical characteristics of foam matrix.

# Chapter 5

# Effective thermal conductivity of open cell foams

Parts of this work were already published or submitted to Acta Materialia, Applied Thermal Engineering and Defects and Diffusion Forum.

- **P. Kumar** & F. Topin, The geometric and thermo-hydraulic characterization of ceramic foams: An analytical approach, *Acta Materilia*, 75, pp. 273-286, 2014.

- **P. Kumar** & F. Topin, Simultaneous determination of intrinsic solid phase conductivity and effective thermal conductivity of Kelvin like foams, *Applied Thermal Engineering*, 71, pp. 536-547, 2014.

- **P. Kumar** & F. Topin, About thermo-hydraulic properties of open cell foams: Pore scale numerical analysis of strut shapes, *Defects and Diffusion Forum*, 354, pp. 195-200, 2014.

Parts of this work were already communicated in national/international conferences.

- **P. Kumar** & F. Topin, Influence of strut shape and porosity on geometrical properties and effective thermal conductivity of Kelvin-like anisotropic metal foams, *The 15$^{th}$ Heat Transfer Conference (IHTC)-2014*, Kyoto, *Japan.*

- **P. Kumar** & F. Topin, Evaluation of thermal properties of metal and ceramic foams by geometrical characterization, *27$^{th}$ European Symposium on Applied Thermodynamics (ESAT)-2014*, Eindhoven, *The Netherlands.*

- About thermo-hydraulic properties of open cell foams: Pore scale numerical analysis of strut shapes, **P. Kumar** & F. Topin: *Diffusion in Solid and Liquids (DSL)-2013*, Madrid, *Spain*

- Propriétés Thermiques et Hydrauliques de Mousses Solides Régulières, **P. Kumar** & F. Topin: JEMP-2012 (11$^{émé}$ Journées d'Etude sur les Milieux Poreux), Marseille, *France*.

## 5.1 Background

Search for the materials of industrial importance and their characterization are always needed and that is a challenging task for engineers, mathematicians and physicists. Open cell foams have been used for a long time in the design of aircraft wing structures for aerospace industry, core structure for high-strength panels, and also in compact heat exchangers. In contrast to conventional packed beds formed by dense packing of granular material, the use of open cell foams has become very interesting because they offer to vary the geometry for solid-fluid contact. Solid foams present a high specific surface area with low pressure drop (e.g. Lacroix et al., 2007; Edouard et al., 2008a, Inayat et al., 2011b) over packed bed of spheres which can be advantageous in heat and mass transfer processes (e.g. Richardson et al., 2003; Giani et al., 2005; Garrido et al., 2008) and in multiphase reaction by intensification of hydrodynamic interactions between the fluid and solid phases (e.g. Edouard et al., 2008b; Stemmet et al., 2007; Stemmet et al., 2008; Stemmet et al., 2010; Tschentscher et al., 2010).

The effective thermal conductivity of open cell foams is a complex affair that depends on the various characteristics of the foam structure. The complexity of geometry encountered in the open cell foams, along with the large difference in thermal conductivities of the constituent phases make it difficult to predict the effective thermal conductivity. In this context, the knowledge of effective thermal conductivity is often a key characteristic for the planning and designing chemical engineering processes (Tronconi and Groppi, 2002). Therefore, there is a need of experimental/numerical and theoretical works concerning effective thermal conductivity in solid foams.

Various works exist in the literature where many authors have mainly focused to measure effective thermal conductivity of different foam samples (Paek et al., 2000; Bhattacharya et al., 2002; Druma et al., 2006; Solorzano et al., 2008). During experiments, effective thermal conductivity has been obtained by measuring the diffusivity of the foam samples. In general, intrinsic solid phase thermal conductivity of strut or foam is unknown when performing experiments to determine effective thermal conductivity. Dietrich et al., (2010) measured intrinsic solid phase conductivity of their ceramic foam samples. Recently, using numerical simulations, Randrianalisoa et al., (2013) also measured solid phase conductivity of different foams.

There are different empirical correlations proposed in the literature to estimate the effective thermal conductivity of open cell foams that depends on the foam morphology and the conductivities of the solid and fluid phases. These correlations can be classified into two parts: either the correlations are based on the combination of resistances or they are based on ideal unit cells. The effective thermal conductivity correlations reported by several authors were derived either on very high porosity (metal foams) or low porosity (ceramic foams) foam samples but it has not been yet correlated with the geometrical properties of foam matrix. One or two fitting parameters were obtained from the effective thermal conductivity experiments to derive a correlation but one of the correlations reported by a author could not be applied to predict effective thermal conductivity of other set of foam samples with variable porosities. The effective thermal conductivity experiments were performed on the samples where the fluid phase conductivity is almost negligible compared to solid phase conductivity. Moreover, very few works reported in the literature where authors have measured intrinsic solid phase thermal conductivity of the foam sample. Manufacturing process impacts greatly the solid conductivity of foam matrix that depends on the grain size of the parent material.

Despite the critical importance of effective thermal conductivity in various industrial systems, the knowledge of geometrical properties of foam matrix and conductivities of constituent phases on effective thermal conductivity is poorly understood. Moreover, the effect of constituent phase conductivities of the same order of magnitude on effective thermal conductivity is still unknown.

In this chapter, an overview of the state of the art of correlations for effective thermal conductivity prediction in open cell foams is presented in the following section 5.2. 3-D pore scale numerical simulations were performed in order to extract the effective thermal conductivities of *isotropic* and *anisotropic* metal foams that are presented in the section 5.3. Local analysis of effective thermal conductivity in local thermal equilibrium condition is studied for different strut shapes, variable porosities (0.60< $\varepsilon_o$ <0.95) and wide range of solid to fluid phase conductivity ratio ($\lambda_s/\lambda_f$=10-30000). The validity and applicability of effective thermal conductivity models and correlations presented in the literature for open cell foams (metal and ceramic) are examined in the section 5.4. Three different correlations are developed and any of the two correlations could be used simultaneously to predict precisely the intrinsic solid phase conductivity and effective thermal conductivity of metal (*isotropic* and *anisotropic* foams) and ceramic foams in the sections 5.5 and 5.6. The validation of correlations with numerical and experimental data of metal and ceramic foams is presented in the section 5.7.

## 5.2 State of the art of effective thermal conductivity correlations

The widespread range of applications of open cell foams has led to increase in the interest of modelling the heat transfer phenomena in such porous media. It is pointed out that, the precise calculation of effective thermal conductivity ($\lambda_{eff}$) is required for an accurate modelling of thermal transport through open cell foams as well as foam heat exchangers when there are large differences in the thermal conductivities of the solid and fluid phases (2-3 orders of magnitude) as well as the high porosity of the medium.

There are several kinds of models in the literature to determine effective thermal conductivity using analytical approach. One group of studies focuses on asymptotic bound approach while the other deals with micro-structural approach. Asymptotic bound approach studies are performed in the case of spherical cells or packed bed spheres and micro-structural approach studies show the importance of various geometrical parameters of the foam matrix in determining effective thermal conductivity.

In asymptotic approach, the very simple existing models are parallel and series models that assume fluid and solid phases in parallel or perpendicular to the heat flow direction provide the highest and lowest bounds of the effective thermal conductivity of a porous medium. They are given by the Equations 5.1 and 5.2:

$$\lambda_{parallel} = (1 - \varepsilon_o)\lambda_s + \varepsilon_o \lambda_f \tag{5.1}$$

$$\lambda_{series} = \frac{\lambda_s . \lambda_f}{(1 - \varepsilon_o)\lambda_f + \varepsilon_o \lambda_s} \tag{5.2}$$

Maxwell-Eucken (1954) developed a model for discontinuous phase in a medium to determine effective thermal conductivity by using Equations 5.1 and 5.2 (Table 5.1, Equations 5.3 and 5.4). Maxwell-Eucken upper and lower models provide us more tight limits resulting more close values to the true thermal conductivity (see Zimmermann, 1989). These models assume that the inclusions of the dispersed phase (fluid phase) do not encounter with the similar neighbouring inclusions of continuous solid phase (see Rocha and Cruz, 2001; Awad and Muzychka, 2008).

Since the pore structure of most of the stones (especially Sander sandstone) consists of large and small pores interconnected through capillary tubes, therefore, Maxwell-Eucken upper and lower models do not provide satisfactory results. In the case of open cell foams, both solid and fluid phases are fully connected. Thus, these upper and lower bound approaches of Maxwell-Eucken are not applicable in case of open cell foams.

The Landauer's effective medium theory model (EMT) (see Landauer, 1952; Kirkpatrick, 1973) uses a similar approach to the Maxwell-Eucken models to establish a relationship for the effective thermal conductivity of the medium (Table 5.1, Equation 5.5). However, this model assumes a completely random distribution of each phase. EMT is a statistical approach that is often used to model thermal conductivity of random mixtures of component materials, particularly when one component has higher thermal conductivity than the other component. EMT is also applicable for the estimation of electrical resistances for a network of resistors. Unlike the Maxwell-Eucken models, EMT does not have any continuous and dispersed phases. According to this theory, the effective thermal conductivity of a two-phase system can be estimated (see Nait-Ali et al., 2006).

By combining these structural models or from heuristic approach, several other models have been developed. Krischer and Kast (1978) proposed a weighted harmonic mean of the series and parallel models to predict effective thermal conductivity of heterogeneous materials (Table 5.1, Equation 5.6). Their model is often used in drying studies and provides a rough estimate of the effective thermal conductivity. It is sensitive to the weighting parameter $\sigma$ which must be set for each material and porosity range.

Lemlich (1978) developed an analogy to predict the electrical conductivity of polyhedral liquid foam of high porosity and is given by the Equation 5.7 (Table 5.1). The limitation of using Lemlich model (1978) is that it does not predict an approximate value of effective thermal conductivity when water is used as fluid medium but works well with air as a fluid medium. In fact, this model takes into account only heat conduction in the solid phase. When fluid phase conductivity is of the same order of magnitude as solid phase conductivity, this model is no more valid because of significant heat exchange between foam ligament and interstitial fluid and therefore, Equation 5.7 is not appropriate for determining effective thermal conductivity.

In the case of micro-structural approach to predict effective thermal conductivity of foam structures, a unit cell approach has been generally taken to represent the foam microstructure (e.g. Calmidi and Mahajan, 2000; Bhattacharya et al., 2002; Fourie and Du Plessis, 2004), and it is assumed that this unit cell can be repeated throughout the medium by virtue of periodicity. The unit cell approach breaks the problem into distinct conduction paths in solid and fluid phases; and calculates the conductivity of the medium as a combination of the individual resistances from series and parallel models for those paths. Applying the energy equation to the suggested unit cell, the effective thermal conductivity can be found analytically or numerically depending on the complexity of the unit cell. Proposed unit cells studied in the literature include 2-D hexagon, 3-D dodecahedron, 3-D tetrakaidecahedron and polyhedral structures.

There are various one, two and three dimensional conduction models of open cell foams in the literature. Hsu et al., (1995) introduced a one-dimensional conduction model based on in-line touching cubes, and carried out an elegant analysis to show good agreement of calculated results with the experimental data in the case of packed beds (Table 5.1, Equation 5.8). The majority of the empirical correlations in the literature were verified for samples made of metal or reticulated vitreous carbon (RVC) with porosities larger than 0.89. The model of Abramenko et al., (1999) correlates the experimental effective thermal conductivity results of ceramic foam samples with porosities in the range, $0.69 < \varepsilon_o < 0.79$ (Table 5.1, Equation 5.9). Paek et al., (2000) considered 1-D heat conduction to determine effective thermal conductivity. Their results indicate that effective thermal conductivity increases when porosity decreases. However, no noticeable changes in effective thermal conductivity were detected from variations of the cell size of the foam samples at a fixed porosity (Table 5.1, Equation 5.10).

Ashby (2006) proposed an empirical correlation for cellular structures by adding two terms to the Lemlich model (1978). This model considers conduction in both solid and fluid phases and is suitable for a medium with a small solid to fluid thermal conductivity ratio (e.g., RVC foam-air). The heat transfer through the structure was described by the sum of heat conducted through the struts and that through the still fluid contained in the cells. Equation 5.11 (Table 5.1) is an adequate approximation of effective thermal conductivity for very low-density foams, but it obviously breaks down when the ratio of foam and solid densities ($\tilde{\rho}/\rho_s$) approaches unity. This is because joints are shared by the struts, and as $\tilde{\rho}/\rho_s$ rises; the joints occupy a larger and larger fraction of the volume. This accounts to introduce $(\tilde{\rho}/\rho_s)^{1.5}$ as an additional term in conduction through the solid cell (Table 5.1, Equation 5.12). This model provides good estimate of effective thermal conductivity when $\lambda_f/\lambda_s \approx 0$ (e.g. RVC foam-water), but highly overestimate the effective thermal conductivity when $\lambda_f/\lambda_s \approx 10^{-2}$.

Calmidi and Mahajan (2000) proposed 1-D conductivity model considering the porous medium to be formed by a 2-D array of hexagonal cells with square lumps at nodes of the high porosity metal foams. These authors described a dimensionless parameter of value 0.09 which was obtained through the fitting of experimental data. Bhattacharya et al., (2002) extended the model of Calmidi and Mahajan (2000) with square and circular lumps at nodes and obtained a dimensionless parameter of value 0.19 to fit analytical model with experimental data. Both the models can accurately predict the thermal conductivity of Al foams, but they overestimate the effective thermal conductivity for other foam materials.

Ozmat et al., (2004) proposed a compact analytical model using regular dodecahedron structure having 12 pentagon-shaped faces with triangular cross sectional ligaments. These authors considered no lumped materials at the intersections of ligaments and found close agreement with their experimental data for low thermal conductivity ratios (Table 5.1, Equation 5.13). Their analytical model does not include heat conduction in the fluid phase. Edouard (2011) predicted effective thermal conductivity of open cell foams using the cubic lattice structure. The use of “slim” and “fat” description of the structure according to strut variation i.e. cylindrical struts for porosities below 90% and triangular struts above 90% was proposed. The proposed model (using “slim” and “fat” foams) predicts effective thermal conductivity like lower and upper bound of asymptotic approach (Table 5.1, Equation 5.14).

Various effective thermal conductivity correlations based on 3-D tetrakaidecahedron structure proposed in the literature by different authors. Boomsma and Poulikakos (2001) proposed a 3-D model with cubic nodes at the intersection of cylindrical ligaments (Table 5.1, Equation 5.15). Using the fitting of experimental data, a dimensionless geometrical parameter of numerical value 0.339 was proposed but the model becomes unrealistic when $\varepsilon_o$ <0.90. Schmierer and Razani (2006) also presented 3-D model with spherical nodes at the intersection of ligaments. These authors performed image and geometrical analyses of the microstructure to find the node size. Numerical finite element analysis was performed to calculate the effective thermal conductivity.

The unit cell composed of cylinders as strut ligaments and spheres or cubes as lumps at nodes assumed by many authors (e.g. Calmidi and Mahajan, 2000; Bhattacharya et al., 2002; Hsu et al., 1995; Boomsma and Poulikakos, 2001; Schmierer and Razani, 2006) is not applicable to the majority of foams that possess triangular or any other strut cross section.

3-D numerical simulation tool has also been used to determine effective thermal conductivity on the actual foam structure. The 3-D foam geometry is usually obtained using X-ray μCT and calculations for effective thermal conductivity are performed either on simplified resistor network model (Vicente et al., 2006b; Bodla et al., 2010) or on full geometry (Krishnan et al., 2008; Hugo, 2012). Vicente et al., (2006b) measured directional tortuosity of the solid matrix and correlate it to the cell shape and orientation. These authors quantified the dependence of the effective thermal conductivity with tortuosity. Bodla et al., (2010) performed numerical simulations using resistor network model on three samples of grades 10, 20 and 40 PPI of very close porosities. The effective thermal conductivity is estimated through a 1-D conduction model, representing individual ligament as an effective thermal resistance using the topological information from the scanned data.

Body-Centered-Cubic (BCC) structure used by Krishnan et al., (2010) proposed a numerical model to determine the effective thermal conductivity. Their results were in agreement with experimental data only when the porosity $\varepsilon_o$ >0.94 because of geometry limitations. Hugo (2012) showed that the porosity is not the only parameter to relate with effective thermal conductivity. It was shown that the change in the ratio of thermal conductivities of solid to fluid phase impacts strongly on local heat conduction. Moreover, effective thermal conductivity changes for slightly elongated foams by keeping the same porosity. Thus, porosity alone cannot be identified as a function of effective thermal conductivity.

One group of the studies also exists in the literature that deals with the asymptotic approach method and direct measurements of effective thermal conductivity on foams for analytical modelling.

Singh and Kasana (2004) established the effective thermal conductivity correlation only for higher porosities ($\varepsilon_o$ >0.90) where the curve fitting followed a straight line. Their correlation has involved the solid and fluid phase conductivities where fluid phase is the order of 0.02~0.7 and solid phase constitutes Al and RVC (Table 5.1, Equation 5.16).

The empirical models proposed by several authors e.g. Hsu et al., 1995; Calmidi and Mahajan, 2000; Paek et al., 2000; Boomsma and Poulikakos, 2001; Bhattacharya et al., 2002; Fourie and Du Plessis, 2004; Singh and Kasana, 2004; Ashby, 2006 used the parent material thermal conductivity of solid phase (for e.g. foams of pure Al/Al 6101 T alloy; $\lambda_s$=218-240 W/mK) as its intrinsic thermal conductivity value in their analytical correlations in order to fit their experimental data as fluid phase like air or water does not play a significant role in predicting an approximate value of effective thermal conductivity. On contrary, $\lambda_s$=218-240 W/mK is not the true intrinsic solid phase thermal conductivity of fabricated foams and is far from actual intrinsic solid phase conductivity of foam and should not be considered in determining effective thermal conductivity either analytically or numerically. Manufacturing processes greatly impact the solid phase thermal conductivity of parent material when transformed into foams. As different commercially available foams employ different manufacturing techniques, and that lead to significant changes in intrinsic solid phase thermal conductivity of foams compared to the same parent material one, and none of these authors have measured the intrinsic value of solid phase thermal conductivity of the foam materials.

In general, intrinsic solid phase thermal conductivity of strut or foam structure is unknown when performing experiments to determine effective thermal conductivity. Dietrich et al., (2010) measured intrinsic solid phase conductivities of their ceramic foam samples. These authors derived effective thermal conductivity correlation using porosity and a fitting parameter from the experimental data (Table 5.1, Equation 5.17).

In case of cast foams studied in this work, the contact surface between cast material (here e.g. pure Al) and mould (often sand) that contains Si or contamination is important where some alloying may occur. Due to rapid solidification and casting conditions, grain size are often different from the one observed in the massive object. For several cases, a relatively large amount of micro-bubbles were also observed. All these effects may impact thermal conductivity and could possibly be a cause of decrease in the solid phase thermal conductivity of foams.

**Table 5.1**. A synthesis of pertinent correlations of effective thermal conductivity of foams.

| Authors | Effective thermal conductivity ( $\lambda_{eff}$), W/mK | Eq. No. |
|---|---|---|
| Maxwell-Eucken (1954) | $\lambda_{eff_{M-E}}{}^{UL} = \lambda_f \left[\frac{2\lambda_f + \lambda_s - 2(\lambda_f - \lambda_s)(1-\varepsilon_o)}{2\lambda_f + \lambda_s + (\lambda_f - \lambda_s)(1-\varepsilon_o)}\right]$ | 5.3 |
| | $\lambda_{eff_{M-E}}{}^{LL} = \lambda_s \left[\frac{2\lambda_s + \lambda_f - 2(\lambda_s - \lambda_f)(1-\varepsilon_o)}{2\lambda_s + \lambda_f + (\lambda_s - \lambda_f)(1-\varepsilon_o)}\right]$ | 5.4 |
| Landauer (1952) | $\lambda_{eff}{}^{EMT} = \frac{1}{4}\left[\lambda_f(3V_f - 1) + \lambda_s(3V_s - 1) + \sqrt{\left\{\lambda_f(3V_f - 1) + \lambda_s(3V_s - 1)\right\}^2 + 8\lambda_f\lambda_s}\right]$ | 5.5 |
| Krischer and Kast (1978) | $\lambda_{eff}{}^{K} = \frac{1}{\sigma/\lambda_{series} + (1-\sigma)/\lambda_{parallel}}$ | 5.6 |
| Lemlich (1978) | $\lambda_{eff} = \frac{1}{3}\lambda_s(1-\varepsilon_o)$ | 5.7 |
| Hsu et al. (1995) | $\frac{\lambda_{eff}}{\lambda_f} = \left[\left[1-\sqrt{1-\varepsilon_o}\right] + \frac{\sqrt{1-\varepsilon_o}}{\lambda_f/\lambda_s} + \left[\sqrt{\varepsilon_o} + \sqrt{1-\varepsilon_o} - 1\right]\right].$ $\left\{\frac{\beta\left(1-\frac{\lambda_f}{\lambda_s}\right)}{\left(1-\beta\frac{\lambda_f}{\lambda_s}\right)^2} ln\frac{1}{\beta\frac{\lambda_f}{\lambda_s}} - \frac{\beta - 1}{1-\beta\frac{\lambda_f}{\lambda_s}}\right\};\ \beta = \left(\frac{1-\varepsilon_o}{\varepsilon_o}\right)^{0.9676}$ | 5.8 |
| Abramenko et al. (1999) | $\lambda_{eff} = 0.13\frac{\lambda_s}{1-\varepsilon_o\left(1-\frac{\lambda_f}{\lambda_s}\right)} + 0.87\frac{\lambda_f}{1+\varepsilon_o\left(\frac{\lambda_s}{\lambda_f} - 1\right)}$ | 5.9 |
| Paek et al. (2000) | $\lambda_{eff} = \lambda_s t^2 + \lambda_f(1-t)^2 + \frac{2t(1-t)\lambda_s\lambda_f}{\lambda_s(1-t) + \lambda_f t};\ t = \frac{1}{2} + cos\left[\frac{4\pi}{3} + \frac{1}{3}cos^{-1}(2\varepsilon_o - 1)\right]$ | 5.10 |
| *Ashby (2006) | $\lambda_{eff} = \frac{1}{3}\left(\frac{\tilde{\rho}}{\rho_s}\right)\lambda_s + \left(1-\frac{\tilde{\rho}}{\rho_s}\right)\lambda_f$ | 5.11 |
| | $\lambda_{eff} = \frac{1}{3}\left[\frac{\tilde{\rho}}{\rho_s} + 2\left(\frac{\tilde{\rho}}{\rho_s}\right)^{1.5}\right]\lambda_s + \left(1-\frac{\tilde{\rho}}{\rho_s}\right)\lambda_f$ | 5.12 |
| Ozmat et al. (2004) | $\lambda_{eff} = 0.346\lambda_s\rho_s$ | 5.13 |
| Edouard (2011) | $\frac{1}{\lambda_{eff}} = \frac{2d}{\{d^2(4y^2 - 2\pi y) + \pi d\}\,\lambda_s + \{d^2(2\pi y - 4y^2) - \pi d + 1\}\,\lambda_s} + \frac{2d(y-1)}{4y^2d^2\,\lambda_s + (1 - 4y^2d^2)\,\lambda_f} + \frac{(1-2dy)}{\pi d^2\,\lambda_s + (1 - \pi d^2)\,\lambda_f}$ | 5.14 |
| Boomsma and Poulikakos (2001) | $\frac{\lambda_{eff}}{\lambda_s} = \frac{1}{\sqrt{2}}\left\{\frac{4p}{2e^2 + \pi p(1-e)} + \frac{3e - 2p}{e^2} + \frac{\left(\sqrt{2} - 2e\right)^2}{2\pi p^2(1 - 2e\sqrt{2})}\right\}$ <br> $p = \sqrt{\frac{\sqrt{2}(2 - (5/8)e^3\sqrt{2} - 2\varepsilon_o)}{\pi(3 - 4e\sqrt{2} - e)}}\ ; e = 0.339$ | 5.15 |
| Singh and Kasana (2004) | $\lambda_{eff} = \left[(1-\varepsilon_o)\lambda_s + \varepsilon_o\lambda_f\right]^F \left[\frac{\lambda_s.\lambda_f}{(1-\varepsilon_o)\lambda_f + \varepsilon_o\lambda_s}\right]^{1-F}\ ;\ F = 0.3031 + 0.0623ln\left(\varepsilon_o\frac{\lambda_s}{\lambda_f}\right)$ | 5.16 |
| Dietrich et al. (2010) | $\lambda_{eff} = 0.51\left\{\frac{\lambda_s.\lambda_f}{(1-\varepsilon_o)\lambda_f + \varepsilon_o\lambda_s}\right\} + 0.49\left\{(1-\varepsilon_o)\lambda_s + \varepsilon_o\lambda_f\right\}$ | 5.17 |

* $\tilde{\rho}$ is the density of the foam, $\rho_s$ is the density of the solid, $d$and $y$ are dimensions of cubic lattice and $p$ is constant.

Literature survey shows that all the 3-D empirical correlations of effective thermal conductivity are simply derived either on the basis of asymptotic approach (Hsu et al., 1995; Abramenko et al., 1999; Paek et al., 2000; Ashby, 2006; Singh and Kasana, 2004; Dietrich et al., 2010) or on micro-structural approach (Boomsma and Poulikakos, 2001; Edouard, 2011). Asymptotic approach was mainly used for packed bed of spheres and correlated only with porosity. Geometrical characteristics of foams have been ignored while deriving the empirical correlations of effective thermal conductivity. In micro-structural approach, using the analogy with the simplest periodic structure (modified cubic lattice), Edouard (2011) simply calculated the thermal conductivities of the individual layers in series of the cubic lattice. The mean error is small compared to other models (due to consideration of geometrical parameters) but models of Edouard (2011) do not take into account the non-linear flow of heat flux lines as effective thermal conductivity obviously lies between series and parallel bounds.

## 5.3 Effective thermal conductivity analysis

In this section, 3-D pore scale numerical simulations were performed to determine the heat flux and temperature gradient across the foam matrix as schematically presented in the Figure 5.1. Volume averaging technique was further used to calculate effective thermal conductivity. Local analysis of temperature gradient and constituent phase conductivities are studied on *isotropic* and *anisotropic* metal foams in the sections 5.3.2 and 5.3.3 respectively.

### 5.3.1 Numerical simulations

Only pure conduction in both the phases is considered in an established stagnant steady state. Neither radiative nor mass transfer at the interface nor chemical reaction nor heat neither source nor phase change is studied. The energy conservation equations are solved numerically over the entire volume of the test sample and allow us to obtain the temperature fields and their gradients at any point of the two phases. The boundary conditions on the open cell foam are presented in Figure 5.1.

The methodology proposed by Topin (2006) was used to determine effective properties and was already validated in the thesis of Hugo (2012). The local thermal equilibrium condition (LTE) was checked for all calculations. It can be written as: $\langle T\rangle^s = \langle T\rangle^f$. Thus, one-temperature model can be used to determine $\nabla\langle T\rangle$ and $\langle \Phi\rangle$. Whitaker (1999) showed that $\Phi = \overline{\overline{\lambda_{eff}}}.\nabla\langle T\rangle$ when local thermal equilibrium is reached and can be written using an effective thermal conductivity tensor defined symmetric and positive:

$$\Phi = \overline{\overline{\lambda_{eff}}}.\nabla\langle T\rangle \ with \ \overline{\overline{\lambda_{eff}}} = \begin{bmatrix} \lambda_{xx} & \lambda_{xy} & \lambda_{xz} \\ \lambda_{xy} & \lambda_{yy} & \lambda_{yz} \\ \lambda_{xz} & \lambda_{yz} & \lambda_{zz} \end{bmatrix} \tag{5.18}$$

where, $\Phi$ is the macroscopic heat flux, $\overline{\overline{\lambda_{eff}}}$ is 2nd order symmetric tensor of effective thermal conductivity and $\nabla\langle T\rangle$ is macroscopic temperature gradient.

The definition of effective thermal conductivity requires knowledge of the entire distribution of temperature and heat flux inside the sample. However, the volume integral can be replaced by surface integral (Sánchez-Vila et al., 1995). For example, the average temperature gradient in the X direction is the scalar product of the averaged gradient by the unit normal vector $i$, in the X direction (Equation 5.19). Integrating by parts allows replacement of the volume integral by a surface integral in Equation 5.20:

$$\nabla < T_x >= \frac{1}{V}\int_V \nabla T(x).n_x dV \tag{5.19}$$

and,

$$< \nabla T_x >= \frac{1}{V}\int_S T(x)\, i.n_x \, dS = \frac{\Delta < T_x >}{\Delta x} \tag{5.20}$$

where, $S$ is the boundary of the sample and $n_x$ is the unit vector normal to the elementary surface of integration $dS$.

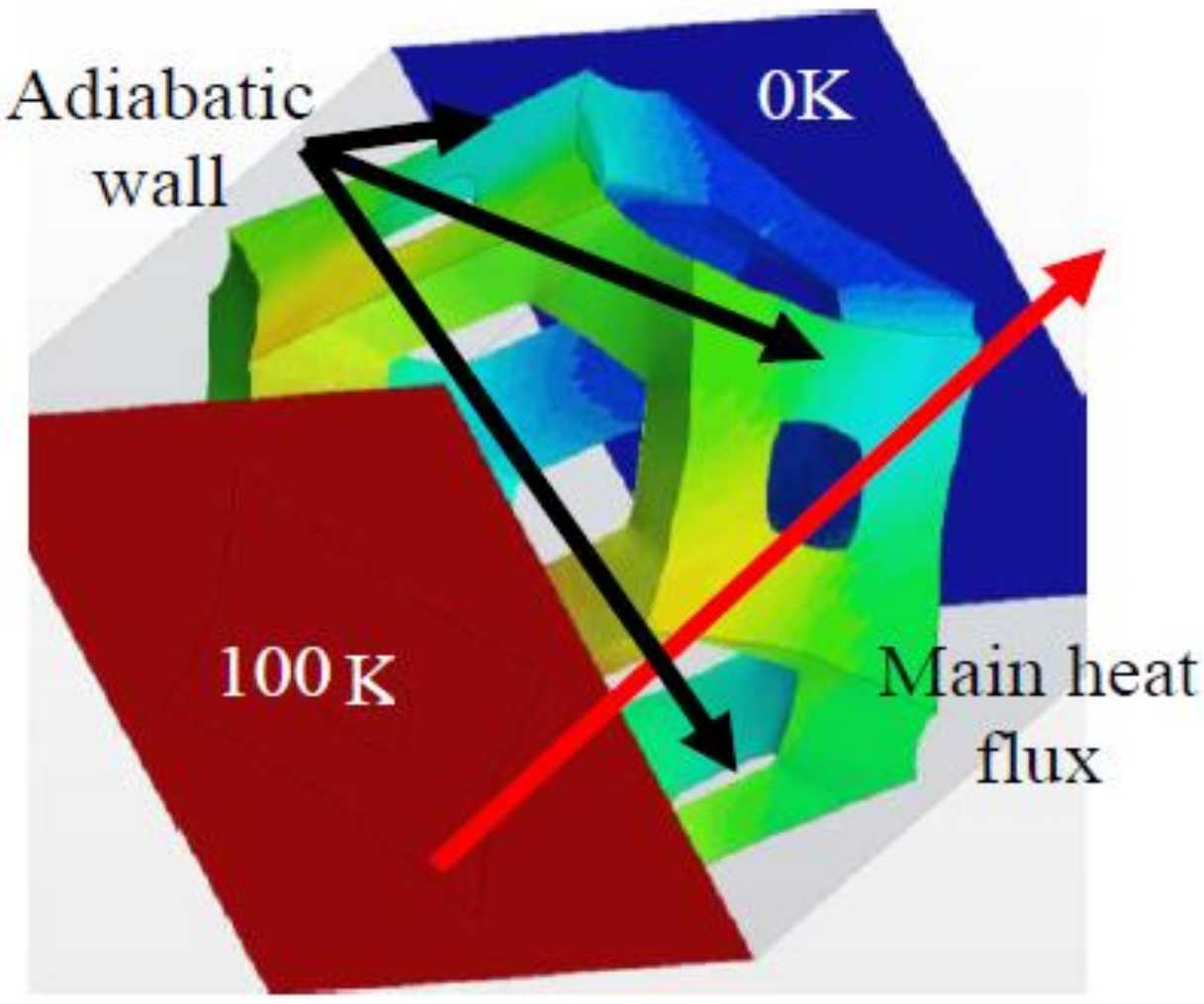

**Figure 5.1**. Boundary conditions and unit cell model: Heat flux in the main direction, no fluxes in other direction (adiabatic walls). Temperature difference is imposed in heat flux direction.

Knowing the geometry of the sample and the distribution of temperature on its surface are therefore sufficient to calculate the average temperature gradient inside the sample. Similarly, the averaged specific heat flux in the X-direction is defined by a volume integral. However, integration by parts shows that it can be replaced by a surface integral under steady state conditions.

$$\varphi_x = \frac{1}{V}\int_S x\, \varphi . n_x \; dS = \frac{P_x}{S_x} \tag{5.21}$$

where, $P_x$ is the heat flux in Watts

Again, this integral can be easily evaluated by knowing the geometry of the sample and measuring the distribution of the heat fluxes on the boundary of the sample. Numerically, volume averaged and surface averaged quantities could be obtained. It has been previously checked that both the methods lead to the same values (see Hugo, 2012). Equation 5.19 provides a system of three equations with six unknowns: the components of the tensor. This linear system is underdetermined.

The following boundary conditions are used: Prescribe temperature difference between the two opposite faces and null fluxes across the other faces as shown in Figure 5.1. Temperature difference along each direction on the two opposite faces and null flux on the four other faces was successively imposed. The boundary conditions were repeated three times to perform the experiments in all the directions. The components of the averaged specific heat flux vector and the components of the average temperature gradient for each flux experiment can then be easily determined by using Equations 5.20 and 5.21.

Thermal conductivity tensors (for *anisotropic* open cell foams) are then calculated which verify at best, to the least square criteria, Equation 5.21 for all flux directions. 6 unknowns are obtained and 9 equations underdetermined system to solve at the least square sense. In case of *isotropic* open cell foams, the diagonal components of the matrix presented in Equation 5.18 are equal while the non-diagonal terms are zero i.e. Equation 5.18 will be reduced to 1-D scalar.

### 5.3.2 Local analysis in *isotropic* metal foams

A database of more than 2000 values was created in the entire solid to fluid thermal conductivity ratio ($\lambda_s/\lambda_f$=10-30,000) for different strut shapes in the porosity range, 0.60$<$ $\varepsilon_o$ <0.95. The arbitrary chosen fluid phase conductivity ($\lambda_f$) used in the present work is 10 W/mK. Using 3-D direct numerical simulations, the effective thermal conductivity can be precisely calculated for known input properties: geometrical parameters of foams and intrinsic solid to fluid thermal conductivity ratio. Some of the effective thermal conductivity values of different strut shapes are presented in Table 5.2. In the Figures 5.2 and 5.3, the temperature fields of various porosities and different strut shapes in LTE condition are shown. According to strut shape, there is a slight difference in the values of effective thermal conductivity. This difference is linked to the local difference in temperature field contours.

From the Figure 5.4 (left), it is clear that the temperature field and hence effective thermal conductivity depends strongly on the ratio of solid to fluid phase thermal conductivity. Moreover, from the Figure 5.4 (right), it is evident that effective thermal conductivity cannot be related only with porosity. The difference in $\lambda_{eff}/\lambda_f$ starts to decrease when porosities increase. This difference is less at higher porosity because the fluid medium plays an important role in a homogenous medium. Similarly, this difference at lower porosity is high because the solid medium plays an important role compared to fluid medium. Moreover, when $\varepsilon_o$=0, $\lambda_{eff}/\lambda_f$ will attain their maximum value while for $\varepsilon_o$=1, $\lambda_{eff}/\lambda_f$ will coincide at one point.

### 5.3.3 Local analysis in *anisotropic* metal foams

For *anisotropic* foams, a database of more than 14000 $\overline{\overline{\lambda_{eff}}}$ values is generated in the entire range solid to fluid thermal conductivity ratio for all strut shapes and different porosities. Some of the $\overline{\overline{\lambda_{eff}}}$ values of different strut shapes are presented in Table 5.3. In the Figure 5.5, the temperature fields of square strut shape at a given porosity (80%) in LTE condition are presented for various anisotropies. There is a slight difference in the values of $\overline{\overline{\lambda_{eff}}}$ according to strut shape. Similarly, in Figure 5.6, the temperature fields of solid and fluid phases (top and bottom) for the elongation factor of $\Omega$=1.4 of various strut shapes are presented.

This difference is linked to the local difference in temperature field contours. As compressions of the foam structure along Y and Z directions are same, the effective thermal conductivity tensor components, $\lambda_{yy}$ and $\lambda_{zz}$ are equal. Note that $\lambda_{xx}$, $\lambda_{yy}$ and $\lambda_{zz}$ are effective thermal conductivities in X, Y and Z-directions respectively.

**Table 5.2**. Representation of numerical $\lambda_{eff}$ (W/mK) of different strut shapes for $\lambda_s/\lambda_f$ ($\lambda_f$=10 W/mK).

| Shape | $\varepsilon_o$ (%) | $\lambda_s/\lambda_f$ 25 | 100 | 500 | 1000 | 2000 | 5000 | 10000 | 20000 | 30000 |
|---|---|---|---|---|---|---|---|---|---|---|
| | | $\lambda_{eff}$ | $\lambda_{eff}$ | $\lambda_{eff}$ | $\lambda_{eff}$ | $\lambda_{eff}$ | $\lambda_{eff}$ | $\lambda_{eff}$ | $\lambda_{eff}$ | $\lambda_{eff}$ |
| Circular | 60 | 73 | 262 | 1268 | 2526 | 5041 | 12588 | 25166 | 50321 | 75476 |
| | 65 | 63 | 219 | 1055 | 2099 | 4187 | 10451 | 20891 | 41772 | 62652 |
| | 70 | 53 | 180 | 857 | 1703 | 3396 | 8473 | 16935 | 33858 | 50782 |
| | 75 | 44 | 144 | 676 | 1342 | 2672 | 6665 | 13319 | 26627 | 39934 |
| | 80 | 36 | 111 | 511 | 1011 | 2011 | 5011 | 10011 | 20011 | 30011 |
| | 85 | 28 | 81 | 361 | 712 | 1413 | 3517 | 7023 | 14035 | 21048 |
| | 90 | 21 | 54 | 228 | 445 | 879 | 2183 | 4355 | 8699 | 13043 |
| | 95 | 15 | 30 | 110 | 209 | 408 | 1005 | 1999 | 3988 | 5977 |
| Equilateral Triangle | 80 | 37 | 117 | 543 | 1076 | 2041 | 5036 | 10061 | 20005 | 30061 |
| | 85 | 29 | 85 | 381 | 751 | 1492 | 3514 | 7016 | 14082 | 21028 |
| | 90 | 22 | 56 | 238 | 466 | 921 | 2186 | 4361 | 8711 | 13061 |
| | 95 | 15 | 31 | 112 | 215 | 419 | 1031 | 2051 | 4092 | 6033 |
| Square | 60 | 74 | 265 | 1285 | 2559 | 5107 | 12753 | 25496 | 50981 | 76446 |
| | 65 | 63 | 222 | 1067 | 2124 | 4237 | 10576 | 21142 | 42273 | 63404 |
| | 70 | 53 | 182 | 866 | 1720 | 3430 | 8558 | 17105 | 34200 | 51294 |
| | 75 | 44 | 145 | 682 | 1352 | 2693 | 6717 | 13423 | 26835 | 40248 |
| | 80 | 36 | 112 | 515 | 1018 | 2025 | 5046 | 10081 | 20150 | 30220 |
| | 85 | 28 | 81 | 364 | 716 | 1421 | 3537 | 7063 | 14115 | 21167 |
| | 90 | 22 | 54 | 229 | 445 | 883 | 2191 | 4371 | 8731 | 13090 |
| | 95 | 15 | 30 | 110 | 209 | 408 | 1005 | 2001 | 3991 | 5981 |
| Rotated Square | 80 | 36 | 111 | 513 | 1016 | 2021 | 5036 | 10061 | 20110 | 30159 |
| | 85 | 28 | 81 | 363 | 716 | 1420 | 3535 | 7059 | 14107 | 21154 |
| | 90 | 21 | 54 | 229 | 446 | 882 | 2189 | 4367 | 8723 | 13078 |
| | 95 | 15 | 30 | 110 | 209 | 409 | 1006 | 2001 | 3992 | 5983 |
| Diamond | 80 | 36 | 112 | 518 | 1024 | 2037 | 5075 | 10139 | 20267 | 30395 |
| | 85 | 29 | 82 | 368 | 724 | 1437 | 3576 | 7141 | 14271 | 21400 |
| | 90 | 22 | 55 | 231 | 451 | 892 | 2213 | 4416 | 8820 | 13225 |
| | 95 | 15 | 31 | 111 | 211 | 412 | 1013 | 2017 | 4023 | 6028 |
| Hexagon | 60 | 73 | 261 | 1264 | 2518 | 5025 | 12547 | 25084 | 50158 | 75232 |
| | 65 | 63 | 219 | 1054 | 2097 | 4182 | 10440 | 20869 | 41727 | 62586 |
| | 70 | 53 | 180 | 855 | 1699 | 3386 | 8449 | 16887 | 33764 | 50640 |
| | 75 | 44 | 144 | 675 | 1339 | 2667 | 6650 | 13290 | 26568 | 39846 |
| | 80 | 36 | 111 | 509 | 1008 | 2005 | 4996 | 9982 | 19953 | 29923 |
| | 85 | 28 | 81 | 361 | 711 | 1411 | 3511 | 7011 | 14012 | 21012 |
| | 90 | 21 | 54 | 227 | 444 | 877 | 2177 | 4343 | 8676 | 13008 |
| | 95 | 15 | 30 | 109 | 208 | 406 | 1000 | 1990 | 3970 | 5951 |
| Rotated Hexagon | 60 | 73 | 260 | 1258 | 2506 | 5001 | 12487 | 24963 | 49915 | 74867 |
| | 65 | 62 | 218 | 1047 | 2083 | 4155 | 10373 | 20734 | 41458 | 62182 |
| | 70 | 53 | 179 | 852 | 1694 | 3376 | 8425 | 16839 | 33667 | 50495 |
| | 75 | 44 | 143 | 670 | 1330 | 2649 | 6606 | 13201 | 26392 | 39581 |
| | 80 | 36 | 110 | 507 | 1004 | 1996 | 4974 | 9938 | 19865 | 29792 |
| | 85 | 28 | 80 | 359 | 708 | 1405 | 3495 | 6979 | 13948 | 20917 |
| | 90 | 21 | 54 | 226 | 442 | 873 | 2167 | 4324 | 8637 | 12950 |
| | 95 | 15 | 30 | 109 | 208 | 405 | 998 | 1986 | 3961 | 5936 |
| Star | 75 | 45 | 146 | 685 | 1360 | 2708 | 6753 | 13495 | 26979 | 40463 |
| | 80 | 36 | 113 | 518 | 1025 | 2039 | 5080 | 10149 | 20286 | 30424 |
| | 85 | 29 | 82 | 366 | 720 | 1429 | 3556 | 7100 | 14190 | 21279 |
| | 90 | 22 | 55 | 230 | 448 | 886 | 2199 | 4388 | 8765 | 13142 |
| | 95 | 15 | 30 | 110 | 210 | 410 | 1009 | 2008 | 4007 | 6004 |

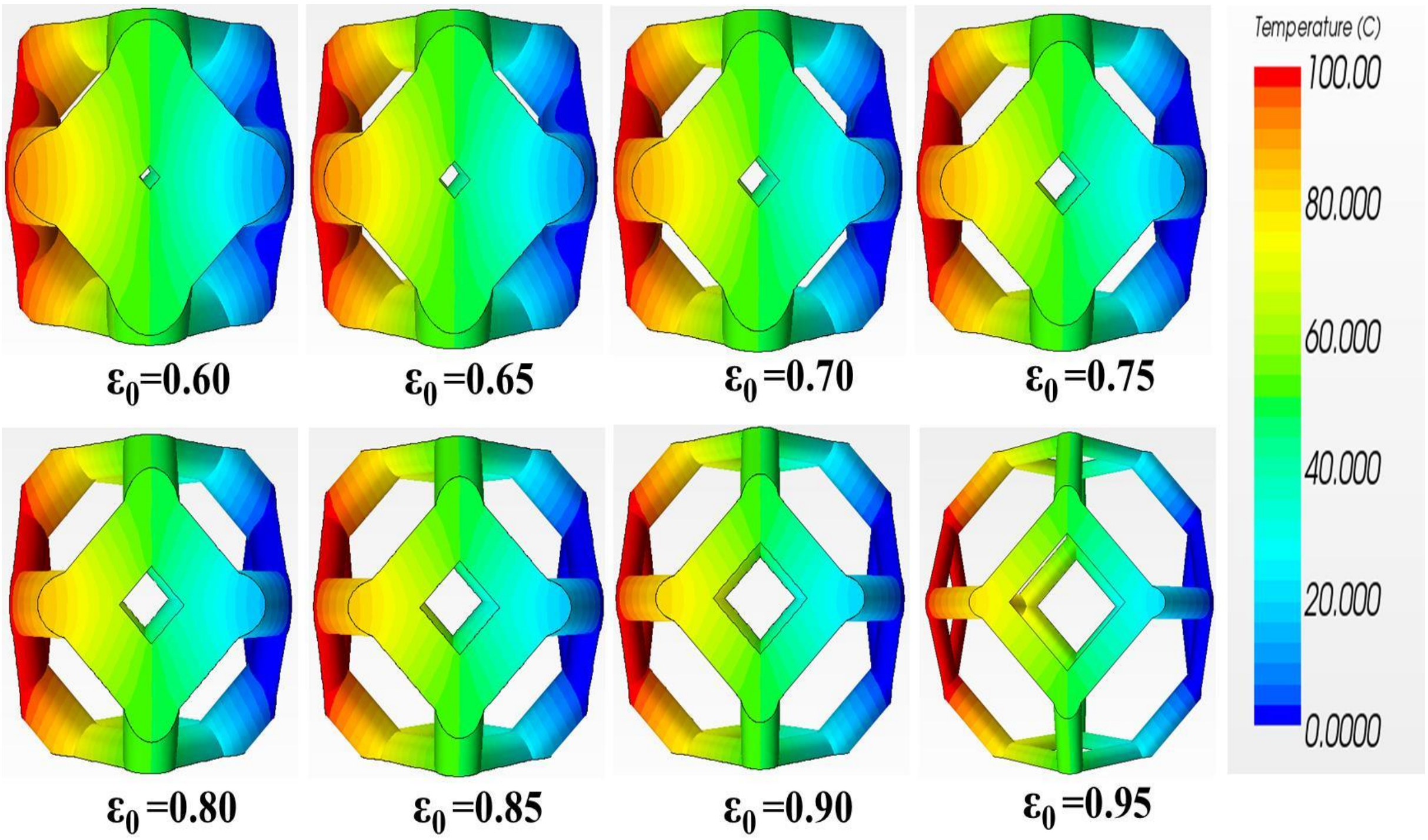

**Figure 5.2**. Temperature field in LTE at different porosities of circular strut shape.

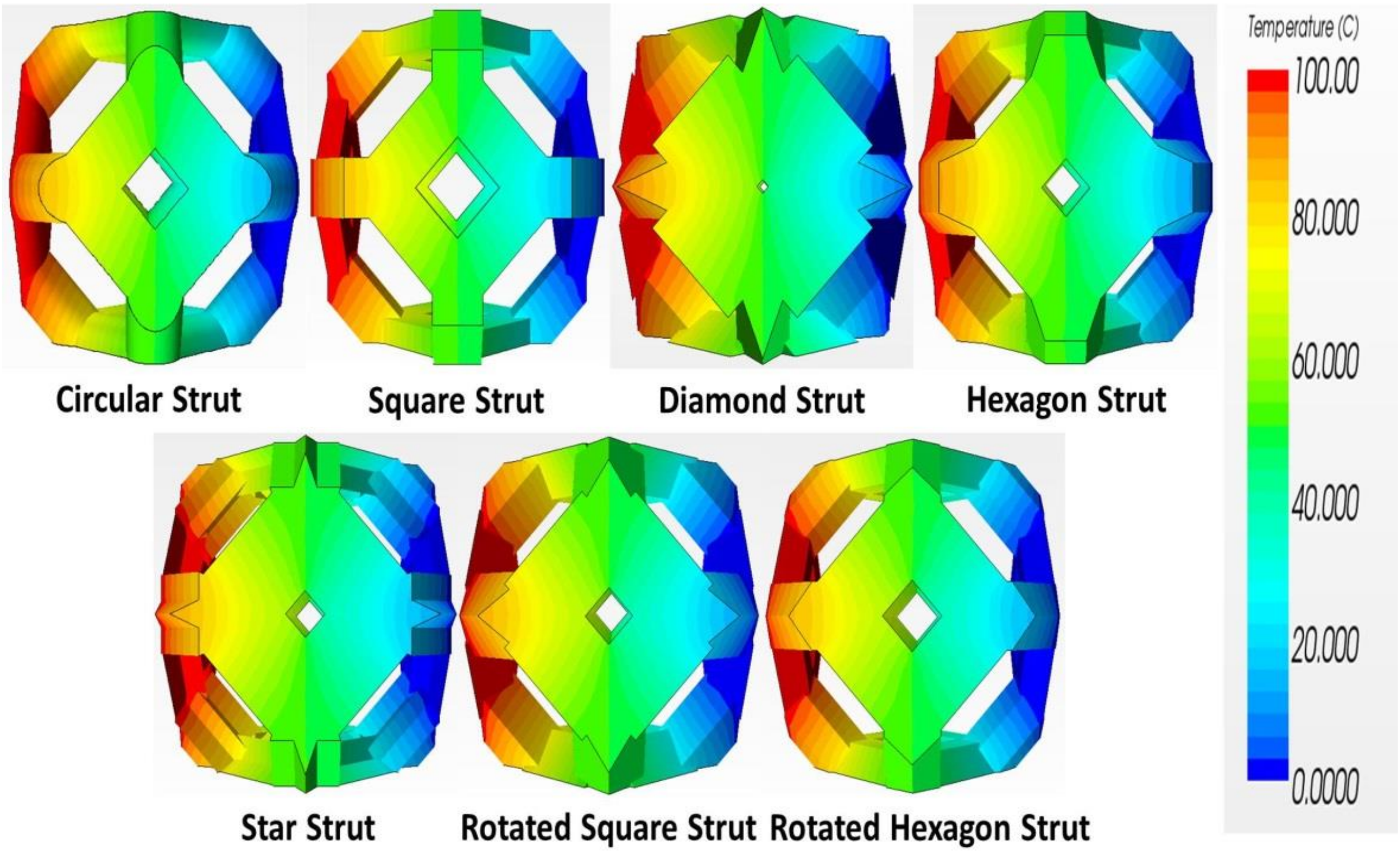

**Figure 5.3**. Temperature field in LTE of different strut shapes at 80% porosity. This figure shows that there is no strong impact of strut shape on temperature field.

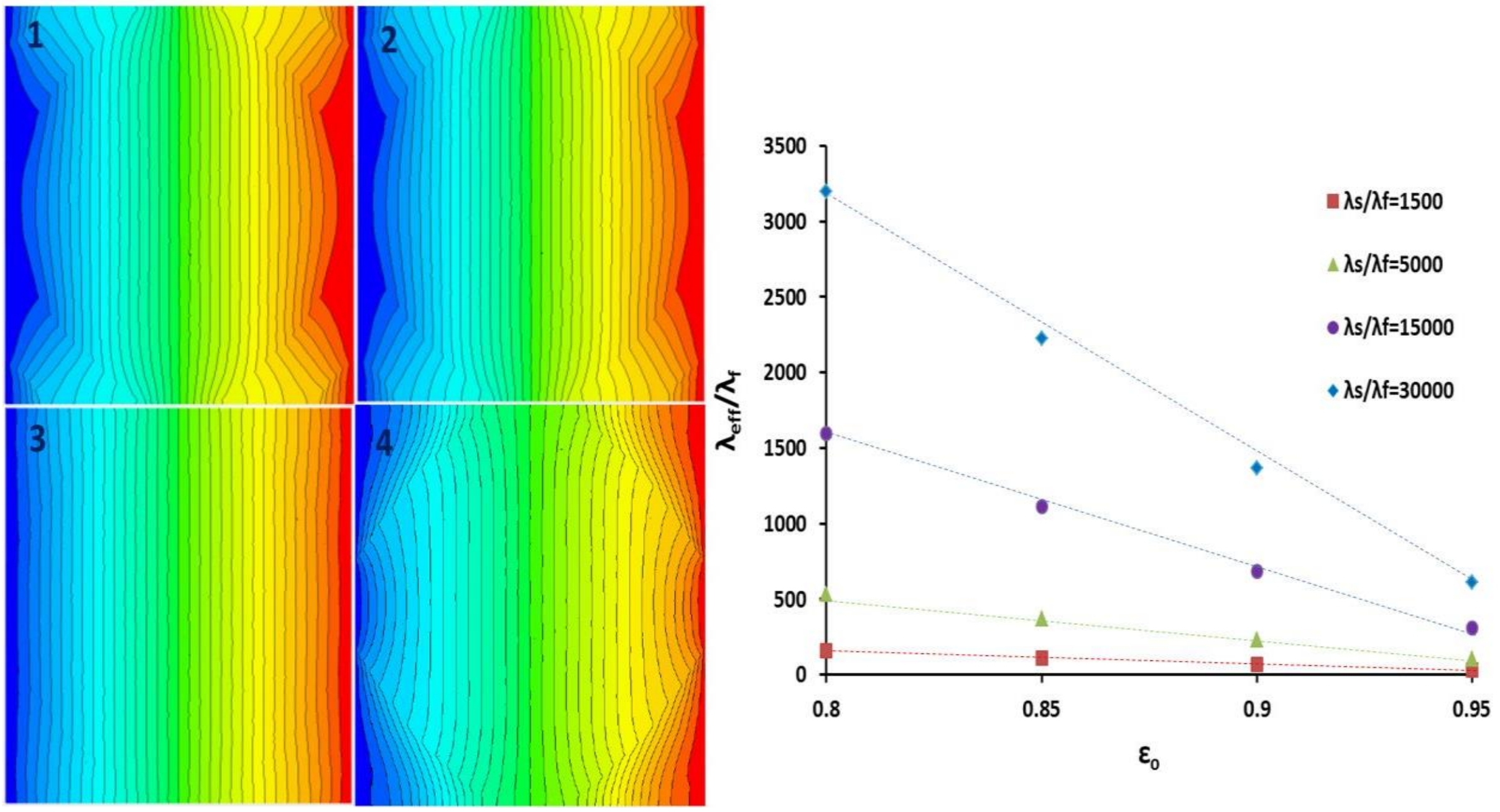

**Figure 5.4**. Left- Temperature field (1) $\lambda_s/\lambda_f$ =10000 (2) $\lambda_s/\lambda_f$=10 (3) $\lambda_s/\lambda_f$=1 (4) $\lambda_s/\lambda_f$=0.1 at $\varepsilon_o$=0.85. Right- Plot of $\lambda_{eff}/\lambda_f$ versus porosity for different $\lambda_s/\lambda_f$ (Visual evidence of effect of different ratios of $\lambda_s/\lambda_f$ in figure 6.3(left) and clear dependence of $\lambda_s/\lambda_f$ and $\varepsilon_o$). The effective thermal conductivity data of equilateral triangular strut shape is presented.

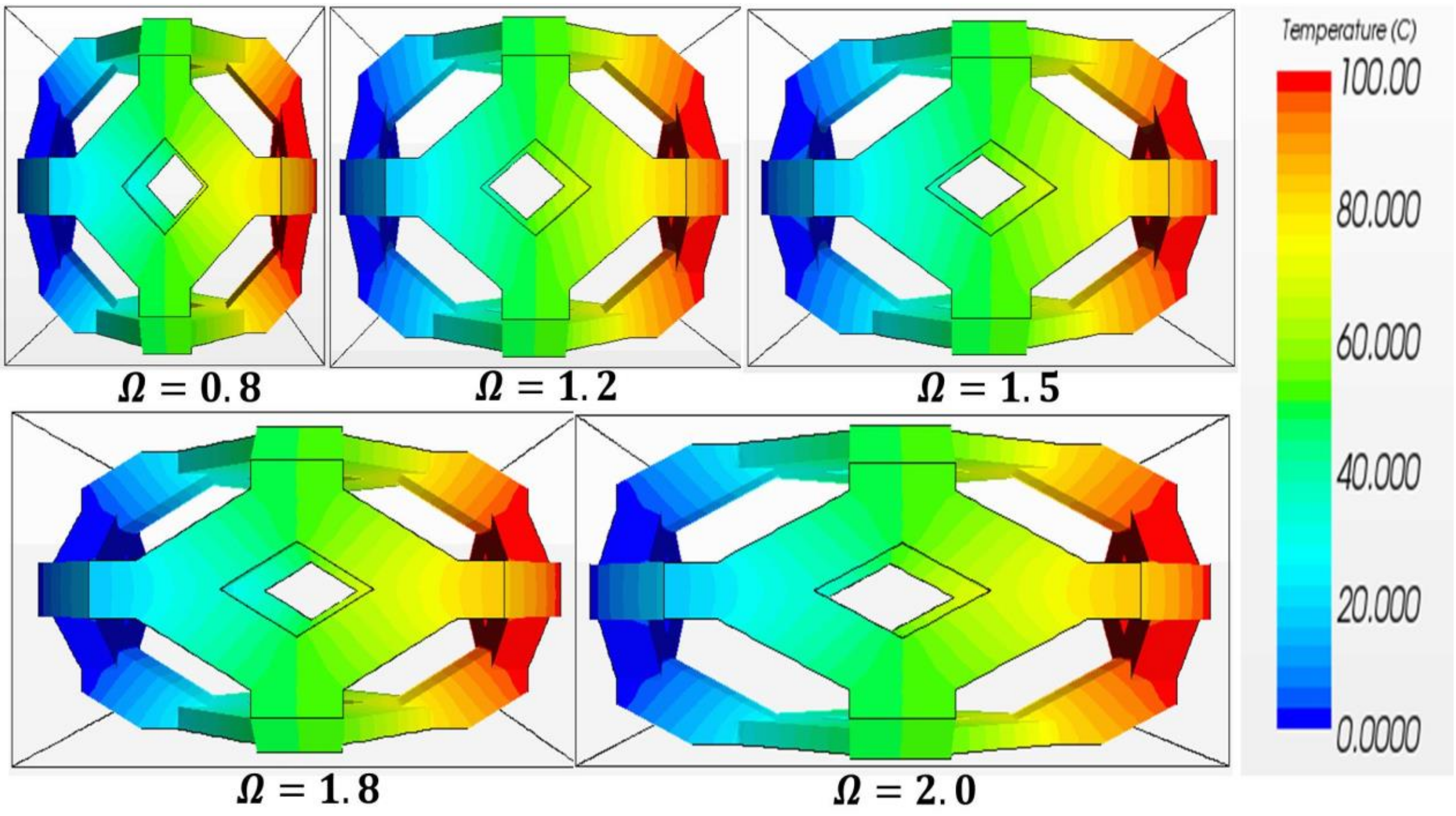

**Figure 5.5**. Left- Temperature field in LTE at different elongations of square strut shape at 80% porosity.

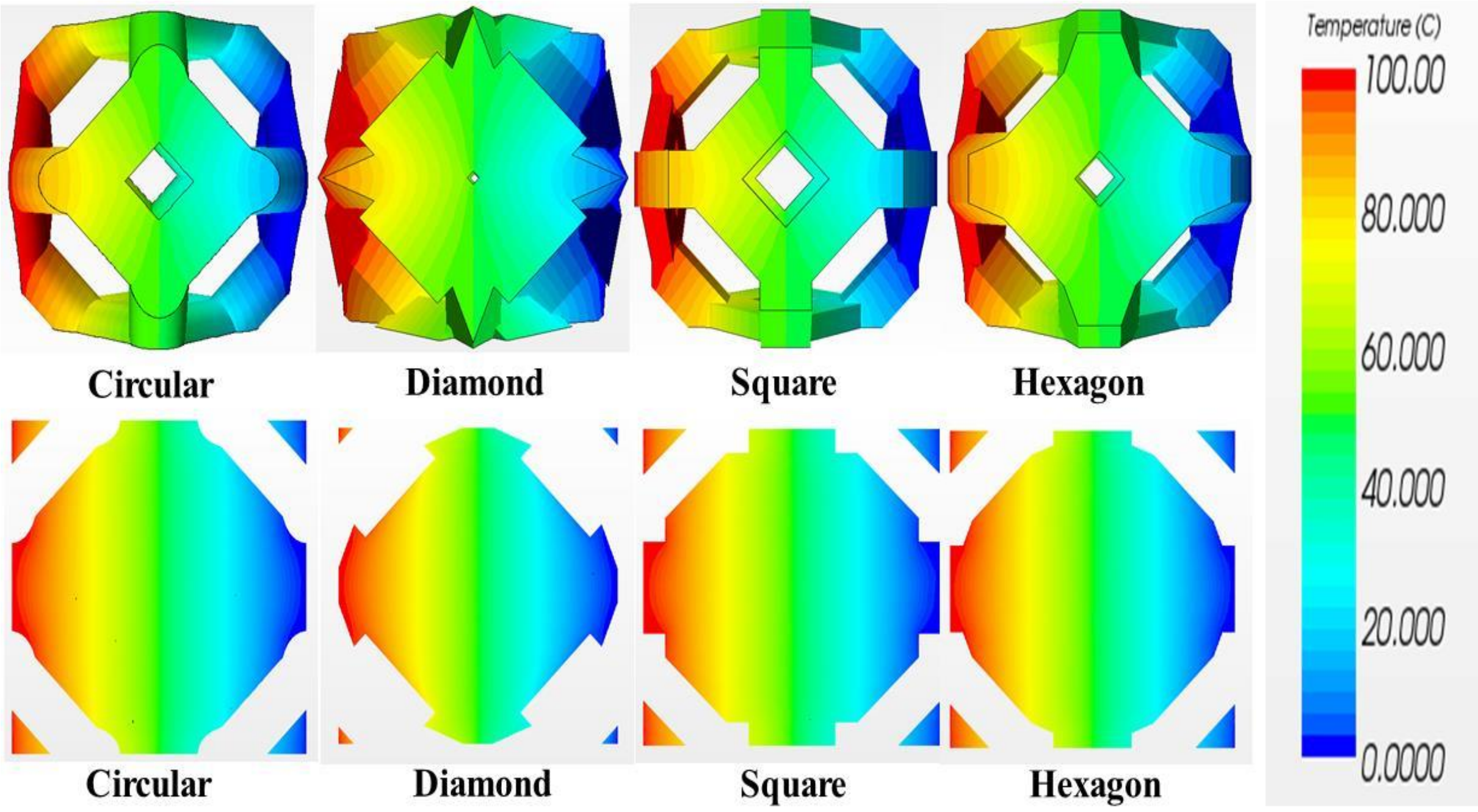

**Figure 5.6**. Temperature field in: Right (Top) - solid phase and Right (bottom) - fluid phase of different strut shapes at $\Omega$=1.4 (80% porosity). This figure shows that there is no strong impact of strut shape on temperature field.

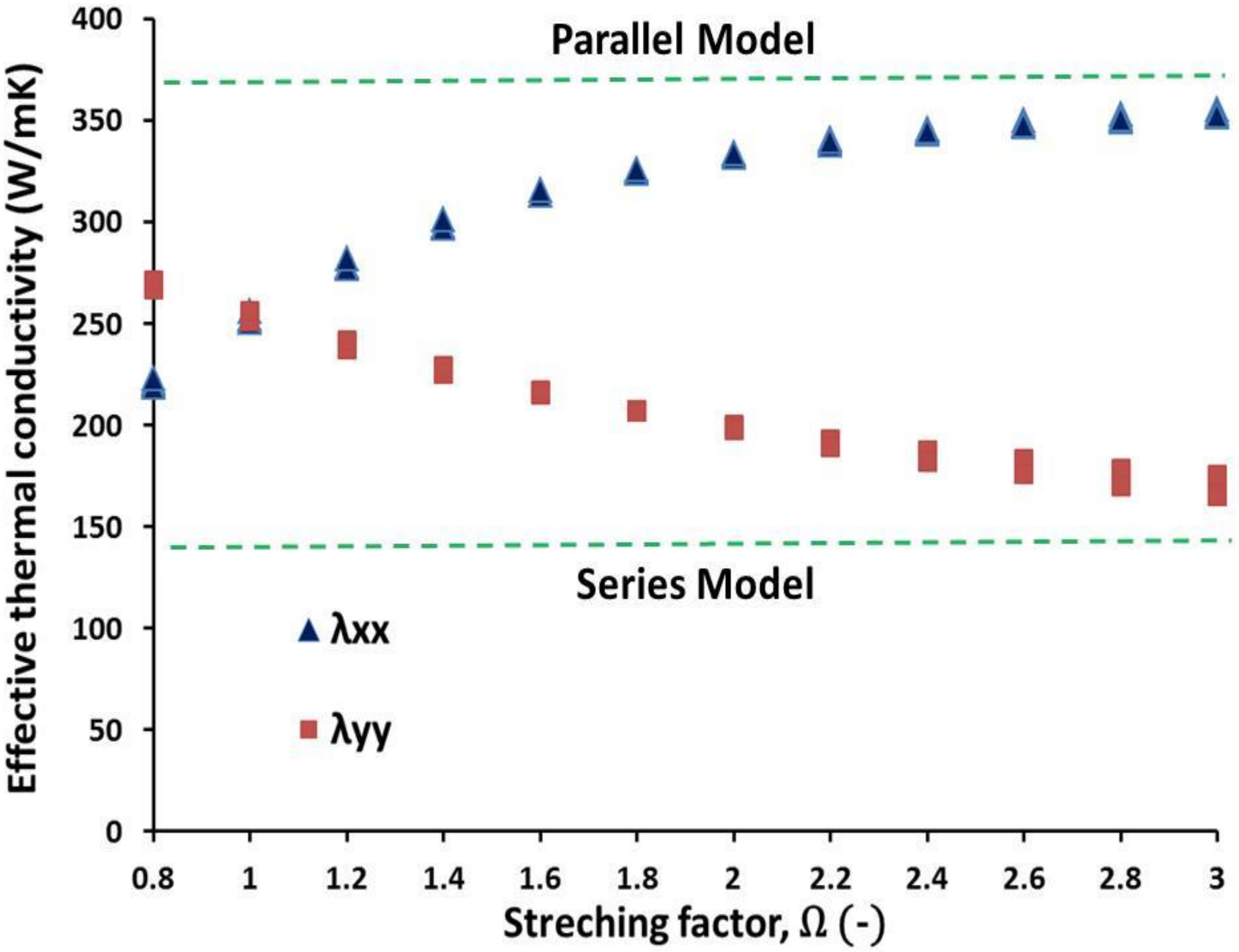

**Figure 5.7**. Plot of effective thermal conductivity, $\lambda_{xx}$ and $\lambda_{yy}$ for different strut shapes at various elongation/stretching factor for $\lambda_s/\lambda_f$=1000 and $\varepsilon_o$=0.60. At $\Omega$=1, $\lambda_{xx}$=$\lambda_{yy}$=$\lambda_{zz}$.

**Table 5.3**. Representation of specific surface area, $a_c$ (CAD data) and numerical $\overline{\overline{\lambda_{eff}}}$ of different strut shapes for $\lambda_s/\lambda_f$ of various anisotropic foams ($\lambda_f$=10 W/mK).

| | | $\varepsilon_o$ = 0.85 | | | | | $\varepsilon_o$ = 0.95 | | | | |
|---|---|---|---|---|---|---|---|---|---|---|---|
| | | | $\lambda_s/\lambda_f$=500 | | $\lambda_s/\lambda_f$=5000 | | | $\lambda_s/\lambda_f$=1000 | | $\lambda_s/\lambda_f$=10000 | |
| | $\Omega$ | $a_c$ | $\lambda_{xx}$ | $\lambda_{yy}$ | $\lambda_{xx}$ | $\lambda_{yy}$ | $a_c$ | $\lambda_{xx}$ | $\lambda_{yy}$ | $\lambda_{xx}$ | $\lambda_{yy}$ |
| Circular | 0.8 | 812 | 282 | 394 | 2717 | 3844 | 525 | 154 | 231 | 1449 | 2220 |
| | 1.2 | 805 | 421 | 325 | 4110 | 3153 | 521 | 251 | 184 | 2418 | 1743 |
| | 1.6 | 859 | 501 | 256 | 4920 | 2457 | 556 | 306 | 136 | 2969 | 1260 |
| | 2.2 | 962 | 564 | 173 | 5544 | 1608 | 622 | 346 | 84 | 3366 | 734 |
| Square | 1.8 | 992 | 525 | 226 | 5156 | 2148 | 644 | 319 | 119 | 3093 | 1091 |
| | 2.2 | 1063 | 557 | 173 | 5474 | 1607 | 692 | 339 | 87 | 3293 | 771 |
| | 2.6 | 1136 | 575 | 133 | 5658 | 1205 | 740 | 350 | 66 | 3403 | 552 |
| | 3.0 | 1207 | 586 | 105 | 5767 | 915 | 788 | 356 | 51 | 3467 | 404 |
| Rotated Square | 1.4 | 951 | 470 | 294 | 4609 | 2839 | 612 | 285 | 160 | 2762 | 1501 |
| | 1.8 | 1032 | 532 | 233 | 5229 | 2219 | 664 | 327 | 117 | 3178 | 1065 |
| | 2.4 | 1163 | 582 | 164 | 5730 | 1522 | 748 | 360 | 73 | 3505 | 627 |
| | 2.8 | 1248 | 600 | 133 | 5912 | 1201 | 803 | 372 | 56 | 3624 | 451 |
| Diamond | 0.8 | 995 | 294 | 400 | 2839 | 3899 | 641 | 158 | 234 | 1485 | 2247 |
| | 1.6 | 1052 | 501 | 275 | 4917 | 2643 | 678 | 307 | 140 | 2977 | 1305 |
| | 2.2 | 1180 | 562 | 205 | 5523 | 1932 | 760 | 346 | 89 | 3367 | 787 |
| | 2.8 | 1310 | 592 | 158 | 5827 | 1453 | 843 | 364 | 59 | 3551 | 486 |
| Hexagon | 1.2 | 846 | 422 | 326 | 4129 | 3160 | 548 | 252 | 184 | 2427 | 1750 |
| | 1.6 | 903 | 503 | 255 | 4936 | 2447 | 585 | 306 | 136 | 2972 | 1258 |
| | 2.2 | 1014 | 565 | 173 | 5561 | 1610 | 656 | 347 | 84 | 3381 | 738 |
| | 3.0 | 1163 | 602 | 106 | 5927 | 926 | 752 | 370 | 49 | 3604 | 378 |
| Rotated Hexagon | 0.8 | 855 | 281 | 393 | 2710 | 3831 | 553 | 156 | 230 | 1464 | 2209 |
| | 1.4 | 872 | 465 | 290 | 4555 | 2802 | 564 | 280 | 161 | 2712 | 1519 |
| | 2.0 | 973 | 542 | 200 | 5324 | 1883 | 630 | 330 | 102 | 3205 | 923 |
| | 2.6 | 1081 | 575 | 138 | 5659 | 1256 | 701 | 350 | 67 | 3402 | 561 |
| Star | 1.4 | 1247 | 473 | 297 | 4635 | 2868 | 802 | 287 | 161 | 2772 | 1510 |
| | 2.2 | 1453 | 570 | 187 | 5612 | 1751 | 934 | 352 | 86 | 3429 | 753 |
| | 2.4 | 1508 | 583 | 167 | 5734 | 1550 | 969 | 360 | 74 | 3510 | 635 |
| | 2.8 | 1614 | 600 | 137 | 5909 | 1241 | 1038 | 372 | 57 | 3628 | 456 |

The variations in $\lambda_{xx}$ and $\lambda_{yy}$ based on the elongation factors are presented in Figure 5.7. It is found that $\lambda_{xx}$ increases with elongation factor in the principal heat flux direction while $\lambda_{yy}$ (or $\lambda_{zz}$) decreases in the other two directions. No apparent order is yet found about the increase and decrease of effective thermal conductivity tensors. Moreover, when the elongation in X-direction and compression in Y and Z-directions are applied simultaneously on the foam structure (to maintain the same porosity as of *isotropic* foam structure), $\lambda_{xx}$ approaches to attain the conductivity of parallel model while $\lambda_{yy}$ approaches to attain the conductivity of series model.

### 5.4 Performance of state of the art correlations

Based on the discussions presented in the section 5.2, it is worth noting that most of the correlations were based on the solid phase thermal conductivity of parent material and porosity. In the present work, the validity and performance of some of the correlations presented by few authors e.g. Hsu et al., 1995; Abramenko et al., 1999; Calmidi and Mahajan, 2000; Paek et al., 2000; Boomsma and Poulikakos, 2001; Bhattacharya et al., 2002; Fourie and Du Plessis, 2004; Singh and Kasana, 2004; Ashby, 2006 are not presented due to very high errors in predicted effective thermal conductivity values that were already shown in the works of Dietrich et al., (2010) and Edouard (2011).

In this section, the validity and performance of the correlations derived by Dietrich et al., (2010) and Edouard (2011) are mainly examined with the numerical effective thermal conductivity data of *isotropic* open cell foams of the present work. In the Figure 5.8, the validity of effective thermal conductivity correlations of Dietrich et al., (2010) and Edouard (2011) are presented for different strut cross sections in the porosity range, $0.60< \varepsilon_o <0.95$ from a low to high intrinsic solid to fluid phase conductivity ratios ($\lambda_s/\lambda_f$=25, 100, 1000, 5000, 30000). Two ranges of porosity i.e. $0.60< \varepsilon_o \leq 0.80$ and $0.80\leq \varepsilon_o <0.95$ are distinguished for the comparison in Figure 5.8 (I). It is clear that the correlation of Dietrich et al., (2010) predicts good $\lambda_{eff}$ results in the high porosity range, $0.80\leq \varepsilon_o <0.95$ with an average error of ±3%. On the other hand, the correlation of Dietrich et al., (2010) underestimates the numerical data in the low porosity range, $0.60< \varepsilon_o \leq 0.80$ and the error increases with decreasing porosity (average error: 9% - 15%). However, the correlation of Edouard (2011) underestimates the predicted results. The average error lies in the range 23%-35% for the two (low and high) porosity ranges. In the Figure 5.8 (II, III, IV and V), the correlation of Dietrich et al., (2010) follows the same trend of estimating good results only in the high porosity range, $0.80\leq \varepsilon_o <0.95$ but underestimates the calculated results in low porosity range $0.60< \varepsilon_o \leq 0.80$ for different values of $\lambda_s/\lambda_f$. However, it is seen that when $\lambda_s/\lambda_f$ increases, the correlation of Edouard (2011) underestimates the results with an average error of 14% in the porosity range, $0.90< \varepsilon_o <0.95$ and it increases up to 28% in the porosity range, $0.60\leq \varepsilon_o \leq 0.90$.

The correlation of Dietrich et al., (2010) was derived using porosity in the range, $0.75< \varepsilon_o <0.95$ where geometrical parameters of foam matrix do not play an important role in determining effective thermal conductivity. However, in the low porosity range, geometrical parameters play a significant role. This could be a possible reason that is attributed to the underestimation of predicted results by the correlation of Dietrich et al., (2010). On the other hand, Edouard (2011) used geometrical parameters of a foam matrix to derive effective thermal conductivity correlation based on cubic lattice unit cell structure which does not correspond to the real foam structure and thus, the predicted results are always underestimated with high errors.

From the above comparisons, it is evident that the correlations established by authors in the literature are based on a few critical parameters that induce high errors:

- Relationship with porosity only.
- Few set of foam samples.
- Availability of experimental values of effective thermal conductivity only.
- No prior information/experimental measurements of intrinsic solid phase thermal conductivity.
- Derived only in the porosity range, 0.75< $\varepsilon_o$ <0.95.

After careful evaluation, few critical remarks have been suggested in order to derive precisely effective thermal conductivity:

- Prior knowledge/experimental determination of intrinsic solid phase conductivity.
- Development of foam samples of low porosity.
- Understanding the heat conduction when fluid phase conductivity is not negligible.
- Inclusion of geometrical parameters over porosity.
- Non-linear heat flux in foam geometry in order to incorporate geometrical parameters (like the model of Singh and Kasana, 2004).

In the present work, solid conductivity of parent material is assigned to the solid conductivity of material by nature (e.g. Al) while the material that constitutes phase or form of the foam (e.g. thermal conductivity of strut of Al) is assigned as intrinsic thermal conductivity which usually differs from the parent material one.

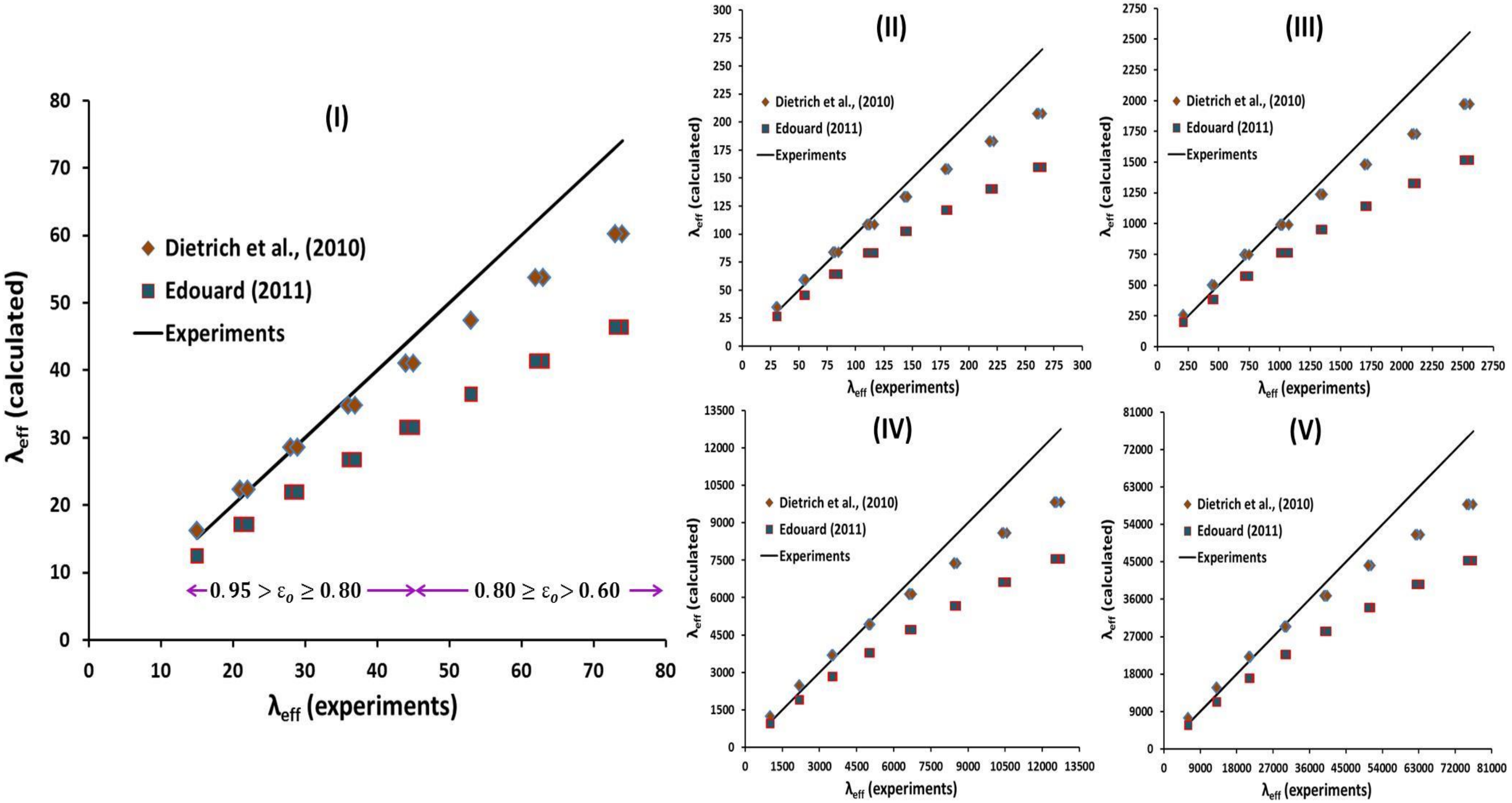

**Figure 5.8**. Comparison of the correlations (Dietrich et al., 2010; Edouard, 2011) with numerically obtained effective thermal conductivity in the present work. The comparison is shown in the porosity range 0.60< $\varepsilon_o$ <0.95 for a wide range of solid to fluid phase conductivity ratios ($\lambda_s/\lambda_f$= (I) 25, (II) 100, (III) 1000, (IV) 5000, (V) 30000). Black line shows the numerical data of effective thermal conductivity. The comparison is performed on various strut cross sections.

## 5.5 Effective thermal conductivity correlations for metal foams

In this section, three correlations are developed, namely, resistor model, modified Lemlich model and PF model. The purpose of these models is to predict precisely effective thermal conductivity depending on the known input parameters for both *isotropic* and *anisotropic* metal foams. Note that PF model is not applicable on *anisotropic* nature of foams that will be discussed in section 5.5.1.3.

### 5.5.1 Correlations for predicting effective thermal conductivity in *isotropic* metal foams

Three different correlations are developed in this section that can be used simultaneously to predict intrinsic solid phase thermal conductivity ($\lambda_s$) and effective thermal conductivity ($\lambda_{eff}$) if fluid phase conductivity ($\lambda_f$) is known.

#### 5.5.1.1 Resistor model

The resistor model approach (see Singh and Kasana, 2004) is applied on the unit cell in order to incorporate varying individual geometries and non-linear flow of heat flux lines generated by the difference in the thermal conductivity of the constituent phases. The effective thermal conductivity lies between the parallel model and series model of a two phase system and can be found by incorporating a correction factor $F$. This relationship is given by Equation 5.22 as:

$$\lambda_{eff} = \lambda_{parallel}{}^{F} . \lambda_{series}{}^{1-F} \qquad F \geq 0, 0 \leq F \leq 1 \tag{5.22}$$

where, $F^{th}$ fraction of the material is oriented in the direction of heat flow and remaining $(1-F)^{th}$ fraction is oriented in the perpendicular direction.

On substitution of Equations 3.56, 3.57 and 3.60 (see chapter3, section 3.4.2) of any strut shape in Equations 5.1 and 5.2 of parallel and series conductivity models, we get:

$$\lambda_{parallel} = \frac{\frac{1}{3}\left(36V_{ligament} + 24V_{node}\right)\left(\lambda_s - \lambda_f\right)}{V_T} + \lambda_f \tag{5.23}$$

$$\lambda_{parallel} = \frac{\left(12\pi {\alpha_{eq}}^2 \beta + \frac{32}{3}\pi {\alpha_{eq}}^3\right)\left(\lambda_s - \lambda_f\right)}{8\sqrt{2}} + \lambda_f \tag{5.23a}$$

Similarly,

$$\frac{1}{\lambda_{series}} = \left[\frac{1}{3}\left(\frac{36V_{ligament} + 24V_{node}}{V_T}\right)\left(\frac{1}{\lambda_s} - \frac{1}{\lambda_f}\right)\right] + \frac{1}{\lambda_f} \tag{5.24}$$

$$\frac{1}{\lambda_{series}} = \left[\left(\frac{12\pi{\alpha_{eq}}^2\beta + \frac{32}{3}\pi{\alpha_{eq}}^3}{8\sqrt{2}}\right)\left(\frac{1}{\lambda_s} - \frac{1}{\lambda_f}\right)\right] + \frac{1}{\lambda_f} \tag{5.24a}$$

Knowing precisely the geometrical parameters of metal foams and strut shape, one can obtain the parallel and series combination of thermal conductivity given by Equations 5.25 and 5.26:

$$\lambda_{parallel} = \left(\zeta{\alpha_{eq}}^2\beta + \chi{\alpha_{eq}}^3\right)\left(\lambda_s - \lambda_f\right) + \lambda_f \tag{5.25}$$

$$\frac{1}{\lambda_{series}} = \left\{\left(\zeta{\alpha_{eq}}^2\beta + \chi{\alpha_{eq}}^3\right)\left(\frac{1}{\lambda_s} - \frac{1}{\lambda_f}\right) + \frac{1}{\lambda_f}\right\} \tag{5.26}$$

where, $\zeta = 3.3322$ and $\chi = 2.96192$ are numerical values from Equations 5.23a and 5.24a .

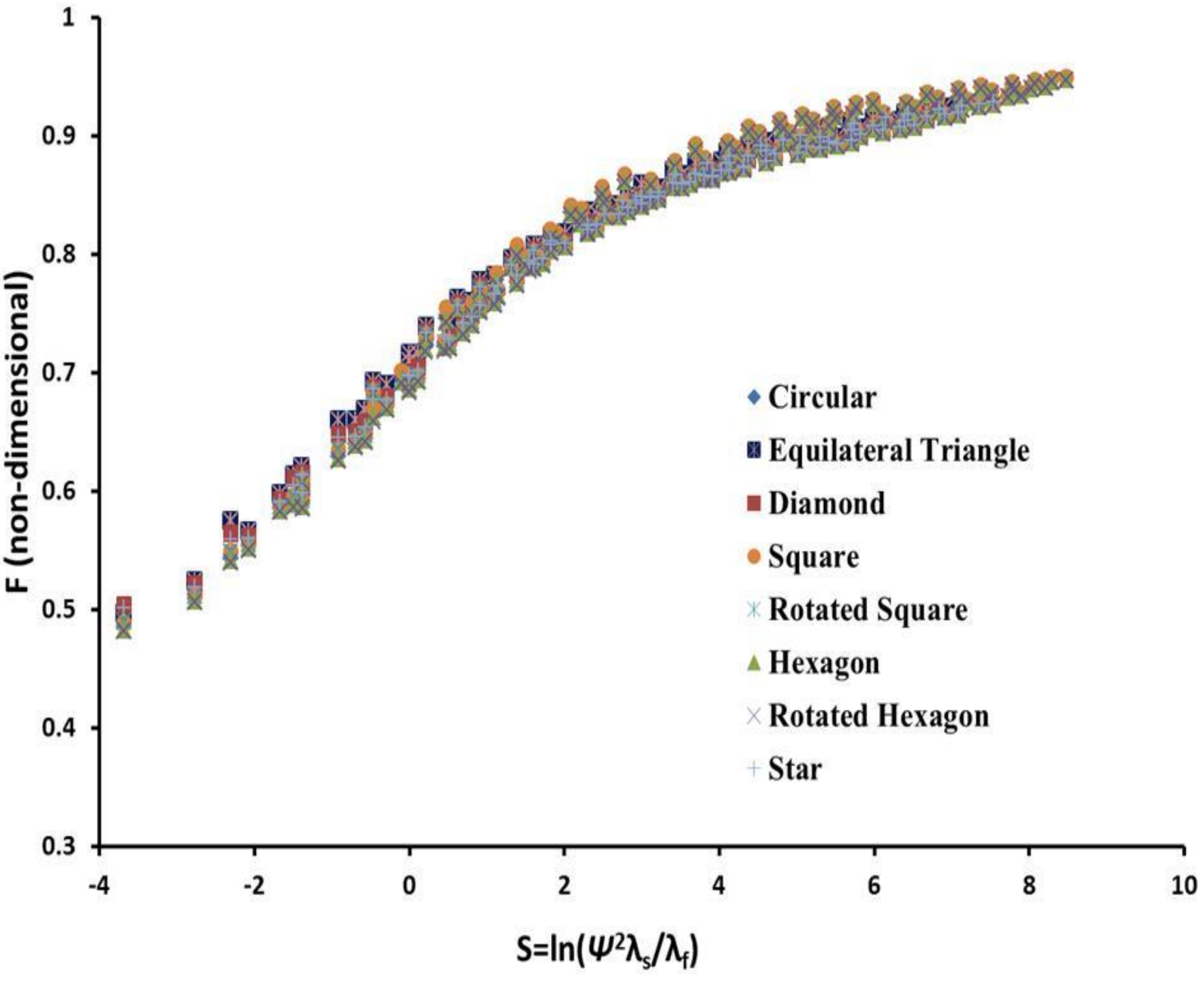

**Figure 5.9**. Plot of $F$ (dimensionless) and $S$ (dimensionless) to determine $\lambda_{eff}$. The uniqueness of this curve is due to the similarity of local temperature field at LTE.

In simple form, we can rewrite the Equations 5.25 and 5.26 as:

$$\lambda_{parallel} = \psi(\lambda_s - \lambda_f) + \lambda_f \tag{5.27}$$

$$\frac{1}{\lambda_{series}} = \left\{\psi\left(\frac{1}{\lambda_s} - \frac{1}{\lambda_f}\right) + \frac{1}{\lambda_f}\right\} \tag{5.28}$$

where, $\psi = \zeta {\alpha_{eq}}^2 \beta + \chi {\alpha_{eq}}^3$

Equation 5.22 is solved for $F$ in terms of $\lambda_{parallel}$, $\lambda_{series}$ and $\lambda_{eff}$ that contains geometric function, $\psi$ (see Singh and Kasana, 2004). The solution is:

$$F = \frac{\ln\left[(1-\psi)\frac{\lambda_{eff}}{\lambda_f} + \psi\frac{\lambda_{eff}}{\lambda_s}\right]}{ln\left[1 + \psi(1-\psi)\left(\frac{\lambda_s}{\lambda_f} + \frac{\lambda_f}{\lambda_s} - 2\right)\right]} \tag{5.29}$$

For a known $\varepsilon_o$, one can have only one $\lambda_{parallel}$ and $\lambda_{series}$ which leads to only one value of $\lambda_{eff}$ for a given value of $F$. Thus, using this approach implies that the influence of solid matrix geometry (as well as thermal properties) will be taken into account only through $F$ value.

For the same porosity, one can obtain different effective thermal conductivity values in case of *anisotropic* foams as shown in the Table 5.3 which implies the necessity of additional geometrical parameters to be correlated in determining $\lambda_{eff}$. Moreover, for variable strut cross section along the ligament axis by keeping the same porosity in case of *isotropic* foams, effective thermal conductivity decreases when the node accumulates more mass compared to the centre of the ligament.

The correction factor, $F$ proposed by Singh and Kasana (2004) will not hold for the cases where the solid to fluid conductivity ratio are of the same order and its relation as a function of porosity only. In order to determine a more precise correlation compared to Singh and Kasana (2004) which is valid for wider porosity range and solid to fluid thermal conductivity ratios, numerical experiments have been performed to better support the analytical model (see also section 5.3.1).

From these 2000 numerically obtained values of $\lambda_{eff}$ from the database, values of $F$ using Equation 5.29 are extracted and the best correlation that includes geometrical parameters of foam of different strut shapes ($\psi$) and ratio of constituent phases ($\lambda_s/\lambda_f$) has been derived. The plot of $F$ as a function of $S = ln(\psi^2 . \lambda_s/\lambda_f)$ is shown in the Figure 5.9. It is observed that $F$ increases and follows roughly a quadratic polynomial function with increase in $S$ where all the values of correction factor $F$ for different porosities collapsed on a single curve.

There is no physical reason to choose this quadratic polynomial function (Equation 5.30) and no physical meaning to the curve fitting is claimed. The quadratic polynomial function is the simplest function that gives a good approximation of effective thermal conductivity data. From the Figure 5.9, it is found that:

$$F = -0.0039S^2 + 0.0593S + 0.704971 \quad (5.30)$$

The values of $F$ obtained for $\lambda_{eff}$ are independent of strut shape. This behavior was expected due to similar temperature fields in LTE condition (see Figure 5.2 and 5.3). The parameter $\psi$ is a function of strut shape and size which is an indirect function of porosity.

Due to the scattering of numerical data presented in Figure 5.9, the root mean square deviation (RMSD) of the fitting relationship was calculated using Equation 5.31:

$$RMSD = 10^{RMS(ELOG)} \quad with\ ELOG = log\left(\frac{\lambda_{eff}}{\lambda_f}\right)_{calc} - log\left(\frac{\lambda_{eff}}{\lambda_f}\right)_{exp} \quad (5.31)$$

An RMSD value of 0.0512 (or 5.12%) was obtained for calculated values of the effective thermal conductivity. From the Equation 5.30 and Figure 5.9, it is evident that $F$ is a function of geometrical parameters and ratio of thermal conductivities of constituent phases and is applicable for wide range of thermal conductivity ratios ($\lambda_s/\lambda_f$).

The expression of analytical resistor model is rather complex but validity range is explicit. All hypotheses could be easily checked by the knowledge of geometry and solid to fluid phase thermal conductivity ratio in the form of $F$ function. The importance of this analytical model is that it can be extended and generalized to *anisotropic* nature of foams. For instance, in the case of an *anisotropic* foams (elongating in one direction and compressing in other two directions to maintain the constant porosity), one has to determine only $\alpha_{eq}$ and $\beta$ (see section 3.4.3) by measuring geometrical parameters of foams. This analytical model is useful in the cases when two foams acquire same porosity and are independent of the foam nature.

### 5.5.1.2 Modified Lemlich model

Equation 5.7 (see Table 5.1) with exponent of 1 on solid porosity $(1-\varepsilon_o)$ has been directly used as a check by several authors to predict their effective thermal conductivity values (e.g. Calmidi and Mahajan, 2000; Boomsma and Poulikakos, 2001; Bhattacharya et al., 2002; Singh and Kasana, 2004). These authors used the same value of parent material thermal conductivity (for foams of pure Al/Al 6101 T alloy; $\lambda_s$=218-240 W/mK) as its intrinsic value of thermal conductivity in their analytical correlations in order to fit their experimental data as fluid phase like air or water does not play a significant role in predicting an approximate value of $\lambda_{eff}$.

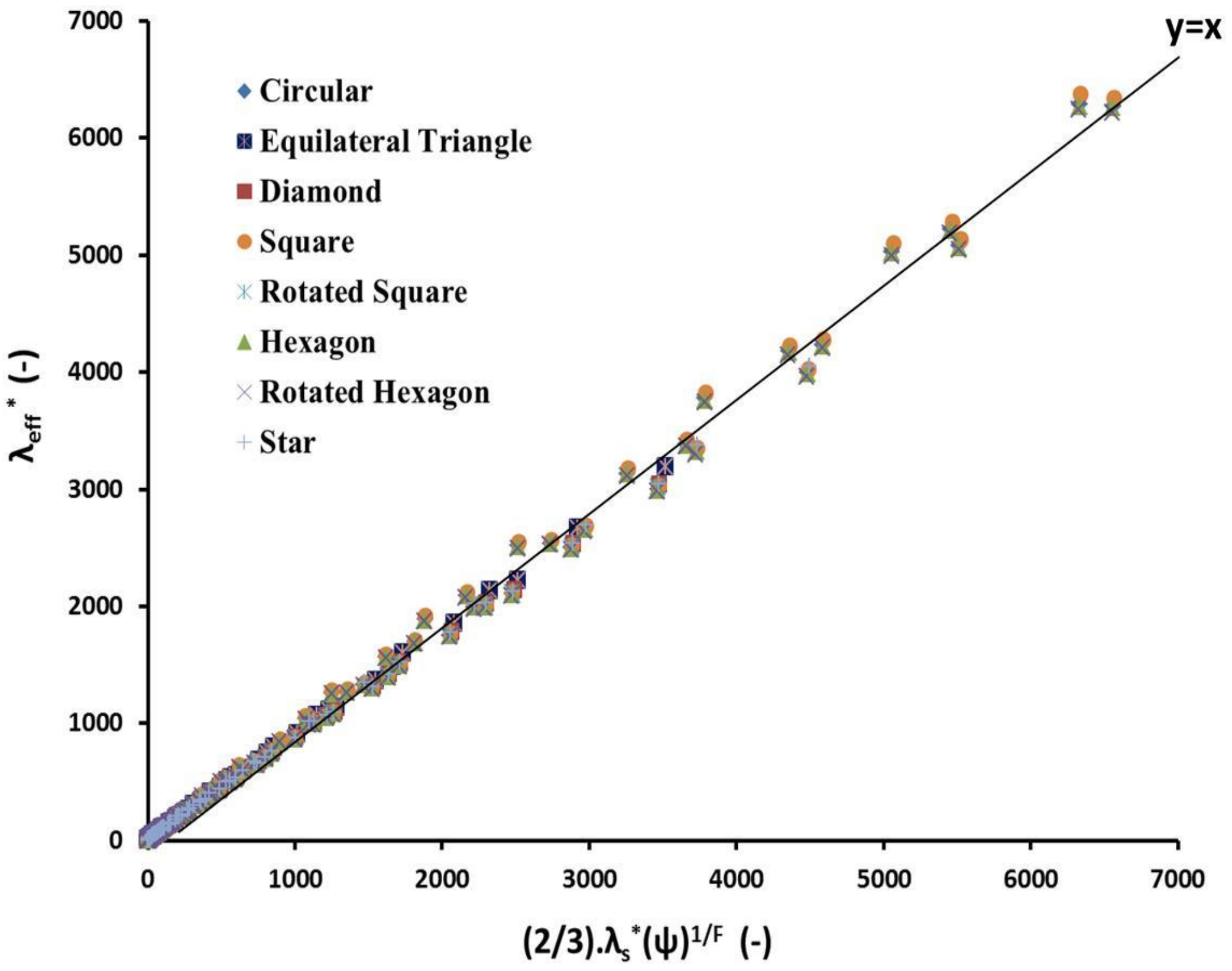

**Figure 5.10**. Plot of $\lambda_{eff}{}^{*}$ vs. $\frac{2}{3}\lambda_s{}^{*}(\psi)^{1/F}$ (dimensionless)

Thus, there is a need to find an empirical correlation which incorporates intrinsic solid phase thermal conductivity of foams and fluid phase to replace Lemlich model (1978). In comparison to the resistor model discussed in section 5.5.1.1, another model is developed based on scattering of $\lambda_{eff}$ values obtained using direct numerical simulations. A similar model like Lemlich (1978) with an exponent $n$ on the solid porosity $(1-\varepsilon_o)$ and that is replaced by a function of geometrical parameters, $\psi$ is proposed. This exponent takes into account the structural impact on effective thermal conductivity.

Several combinations between $\lambda_{eff}$, $\lambda_s$, $\lambda_f$, $\psi$ and $F$ (from section 5.5.1.1) have been tested. The modified Lemlich model for *isotropic* foams is presented in Equation 5.32. Using the database, the best fit for all porosities and all strut shapes in the studied range of $\lambda_s/\lambda_f$ (10 - 30000) collapsed on a single curve and is presented in the Figure 5.10. From Figure 5.10, the relation is observed as a straight line and is given by:

$$\lambda_{eff}{}^* = \frac{2}{3}\lambda_s{}^*(\psi)^{1/F} \tag{5.32}$$

where, $\lambda_{eff}{}^* = \frac{\lambda_{eff}}{\lambda_f}$ and $\lambda_s{}^* = \frac{\lambda_s}{\lambda_f}$

There are very small deviations for the complex strut cross sections (square, hexagon and rotated hexagon) at low porosity. Due to the scattering of the data shown in Figure 5.10, the RMSD of all calculated effective thermal conductivity values for different strut shapes in the wide porosity range is 2.19%.

Like resistor model, the importance of modified Lemlich analytical model is that it can be easily extended and generalized on *anisotropic* foams. Equation 5.32 contains correction factor $F$ which is dependent on strut shape and thermal conductivity ratio of constituent phases.

### 5.5.1.3 PF model

Another correlation, PF model, which is a direct curve fitting of effective thermal conductivity values that is applicable only on *isotropic* foams is developed and is very similar to Lemlich model (1978). As discussed in the section 5.5.1.2, the exponent $n$ is replaced by a constant numerical value while solid porosity $(1-\varepsilon_o)$ by a function of geometrical parameters, $\psi$ of the Lemlich model.

Several combinations between $\lambda_{eff}$, $\lambda_s$, $\lambda_f$ and $\psi$ have been tested using the effective thermal conductivity database. The best fit for all porosities and strut shapes in the studied $\lambda_s/\lambda_f$range, 10 – 30000 is found to collapse on a single curve and is presented in the Figure 5.11. From Figure 5.11, the relation is observed as a straight line and is given by:

$$\lambda_{eff}{}^{*} = \eta.\lambda_s{}^{*}(\psi)^{1.3} \tag{5.33}$$

where, $\lambda_{eff}{}^{*} = \frac{\lambda_{eff}}{\lambda_f}$, $\lambda_s{}^{*} = \frac{\lambda_s}{\lambda_f}$ and $\eta$ =0.89

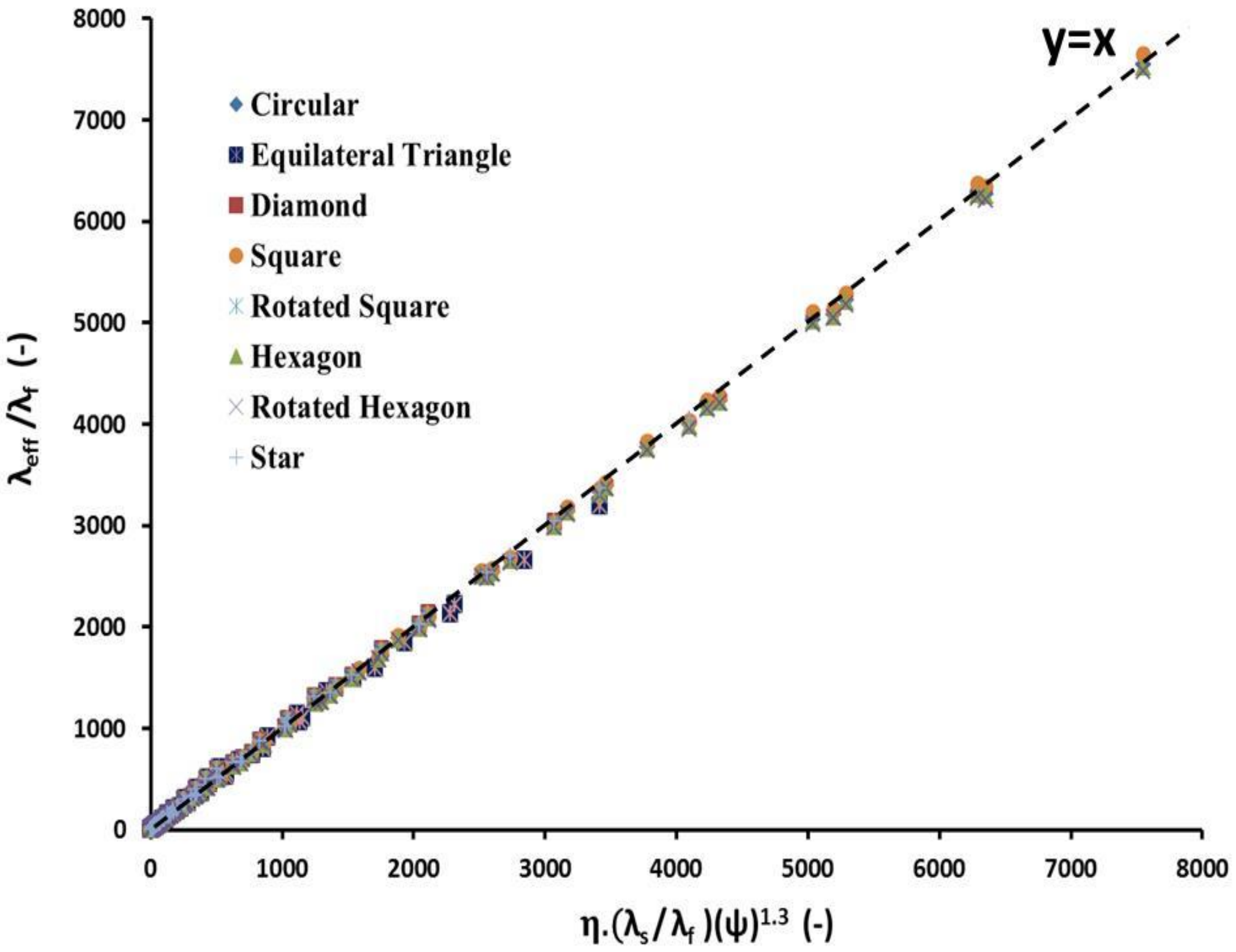

**Figure 5.11**. Presentation of PF model and its relation is found as $\lambda_{eff}{}^{*} = \eta.\lambda_s{}^{*}(\psi)^{1.3}$.

Equation 5.33 is obtained as best fit for $n$ =1.3. No physical interpretation of exponent has been yet found but other geometrical properties of foams obviously control this exponent value. PF model is the simplest model and accommodates the influence of correction factor $F$ in the exponent $n$ and parameter $\eta$. The RMSD of PF model is 4.19% on the calculated effective thermal conductivities for all strut shapes and is valid for wide porosity range.

This model can be used in two ways: to determine effective thermal conductivity if intrinsic solid phase thermal conductivity is known or vice versa. However, in most of the experimental data, intrinsic solid phase thermal conductivity of open cell foams is not measured and is not reported. Thus, this model can be used to predict the intrinsic solid phase conductivity to perform various numerical and analytical calculations of a given problem.

PF model has rather simple expression than resistor model but the validity range is not explicit as the combination of geometrical parameters impact the exponent in an unknown way. The effect of *anisotropic* foams (keeping the same porosity same as *isotropic* foams) is not taken into account in the expression 5.33.

### 5.5.2 Correlations for effective thermal conductivity in *anisotropic* metal foams

In this section, two models i.e. resistor and modified Lemlich models are developed. Note that PF model does not contain any geometrical parameters of foam matrix explicitly and thus, is not developed in the case of *anisotropic* foams.

#### 5.5.2.1 Resistor model

The resistor model is derived the same way like in the case of *isotropic* metal foams (see section 5.5.1.1). The parallel and series models to determine conductivities are given by Equations 5.34 and 5.35:

$$\lambda_{parallel} = (1-\varepsilon_o)\lambda_s + \varepsilon_o\lambda_f = \frac{\frac{1}{3}(V'_L + V'_N)(\lambda_s - \lambda_f)}{V_T} + \lambda_f \tag{5.34}$$

and

$$\frac{1}{\lambda_{series}} = \left\{\frac{\lambda_s.\lambda_f}{(1-\varepsilon_o)\lambda_f + \varepsilon_o\lambda_s}\right\} = \left[\frac{1}{3}\left(\frac{V'_L + V'_N}{V_T}\right)\left(\frac{1}{\lambda_s} - \frac{1}{\lambda_f}\right)\right] + \frac{1}{\lambda_f} \tag{5.35}$$

where, $V'_L$ and $V'_N$ are the total volume of ligaments and nodes of any strut shape.

An *anisotropic* foam structure in a cubic cell (see Figure 2.18-bottom and section 3.4.3), there are 16 strut lengths $L_{s1}$of square face in horizontal direction, 8 strut lengths $L_{s2}$ of square face in vertical direction, 8 strut lengths $L_{s3}$ of hexagon face in horizontal direction and 4 struts lengths $L_{s4}$ of hexagon face in vertical direction.

Total ligament and node volumes ($V'_L$ and $V'_N$) of any strut shape are given as:

$$V'_L = \left[16\left(\pi {R_{eq1}}^2 L_{s1}\right) + 8\left(\pi {R_{eq2}}^2 L_{s2}\right) + 8\left(\pi {R_{eq3}}^2 L_{s3}\right) + 4\left(\pi {R_{eq4}}^2 L_{s4}\right)\right] \tag{5.36}$$

$$V'_N = 8\left[\frac{2}{3}\pi\left({R_{eq1}}^3 + {R_{eq4}}^3\right) + \frac{2}{3}\pi\left({R_{eq1}}^3 + {R_{eq3}}^3\right) + \frac{2}{3}\pi\left({R_{eq2}}^3 + {R_{eq3}}^3\right)\right] \tag{5.37}$$

We can rewrite Equations 5.34 and 5.35 using Equations 5.36 and 5.37 as:

$$\lambda_{parallel} = \left[\left\{\frac{\pi {\alpha_{eq}}^2 \beta}{6\sqrt{2}}\left(6+\frac{2}{\Pi}(\delta)^{\frac{2}{3}}+\frac{1}{\zeta}(\omega)^{\frac{2}{3}}\right)+\frac{\sqrt{2}\pi {\alpha_{eq}}^3}{9}\left(2(\Pi)^{\frac{3}{2}}+(\zeta)^{\frac{3}{2}}+2\delta+\omega\right)\right\}\right]\left(\lambda_s - \lambda_f\right)+\lambda_f \quad (5.38)$$

$$\frac{1}{\lambda_{series}} = \left[\left\{\frac{\pi {\alpha_{eq}}^2 \beta}{6\sqrt{2}}\left(6+\frac{2}{\Pi}(\delta)^{\frac{2}{3}}+\frac{1}{\zeta}(\omega)^{\frac{2}{3}}\right)+\frac{\sqrt{2}\pi {\alpha_{eq}}^3}{9}\left(2(\Pi)^{\frac{3}{2}}+(\zeta)^{\frac{3}{2}}+2\delta+\omega\right)\right\}\left(\frac{1}{\lambda_s}-\frac{1}{\lambda_f}\right)\right]+\frac{1}{\lambda_f} \quad (5.39)$$

In simpler words, Equations 5.38 and 5.39 can be written as:

$$\lambda_{parallel} = \psi''\left(\lambda_s - \lambda_f\right)+\lambda_f \quad (5.40)$$

$$\frac{1}{\lambda_{series}} = \psi''\left(\frac{1}{\lambda_s}-\frac{1}{\lambda_f}\right)+\frac{1}{\lambda_f} \quad (5.41)$$

where, $\psi''$ is a function of strut shape and its geometrical parameters (see also section 5.5.1.1).

Equation 5.22 (see section 5.5.1.1) is solved for $F$ in terms of $\lambda_{parallel}$, $\lambda_{series}$ and $\overline{\overline{\lambda_{eff}}}$ that contains geometric function, $\psi''$ (see Singh and Kasana, 2004). The solution is:

$$F = \frac{ln\left[(1-\psi'')\frac{\overline{\overline{\lambda_{eff}}}}{\lambda_f}+\psi''\frac{\overline{\overline{\lambda_{eff}}}}{\lambda_s}\right]}{ln\left[1+\psi''(1-\psi'')\left(\frac{\lambda_s}{\lambda_f}+\frac{\lambda_f}{\lambda_s}-2\right)\right]} \quad (5.42)$$

From the database of 14000 values of $\overline{\overline{\lambda_{eff}}}$ obtained numerically, values of $F_{xx}$ and $F_{yy}$ using Equation 5.42 are extracted and the best correlation (see Equation 5.43) that includes geometrical properties of foam of different shapes at various elongation factors and ratio of constituent phases is found. Note that $F_{xx}$ and $F_{yy}$ are the correction factors in X and Y (or in Z) directions respectively and are calculated from Equation 5.42 (using $\lambda_{xx}$ and $\lambda_{yy}$) for each direction.

The plots: $F_{xx}$ as a function of $S_{xx} = ln(\psi''^3.(\Omega.\zeta)^4.\lambda_s/\lambda_f)$ and $F_{yy}$ as a function of $S_{yy} = ln(\psi''^3.(\Omega)^{-4}.\lambda_s/\lambda_f)$ are presented in the Figure 5.12. One can notice that all the values of $F$ in relation with $S$ (in each direction) are collapsed on a single curve for different strut shapes. The $F$ values obtained for $\overline{\overline{\lambda_{eff}}}$ are independent of strut shape. This behaviour was expected due to similar temperature field in LTE condition (see Figures 5.5 and 5.6).

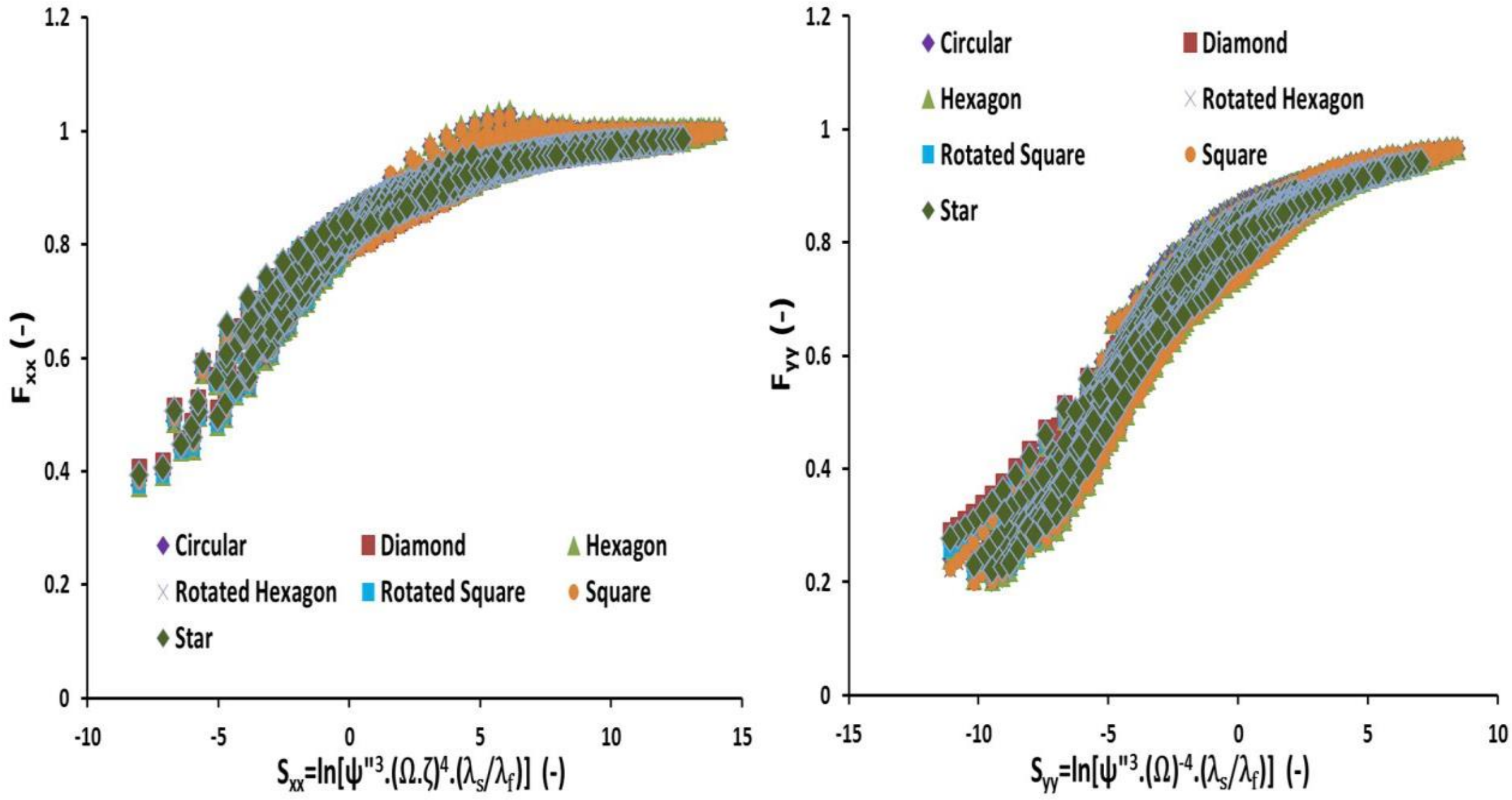

**Figure 5.12**. Plot of $F$ (dimensionless) and $S$ (dimensionless) to determine $\overline{\overline{\lambda_{eff}}}$. The uniqueness of these curves is due to the similarity of local temperature field at LTE.

The geometrical function $\psi''$ (see Equations 5.40 and 5.41) is a function of strut shape and size which is an indirect function of porosity. It is observed that $F$ increases roughly parabolically with increasing $S$. From the Figure 5.12 (left and right), simple relations of $F_{xx}$ and $F_{yy}$ are estimated and are given by Equation 5.43:

$$F_{xx} = -0.0017{S_{xx}}^2 + 0.0351S_{xx} + 0.79 \quad (5.43a)$$

$$F_{yy} = -0.0025{S_{yy}}^2 + 0.0396S_{yy} + 0.811 \quad (5.43b)$$

For 14000 calculated values of $\overline{\overline{\lambda_{eff}}}$, RMSD of Equations 5.43a and 5.43b are 2.87% and 8.86%.

### 5.5.2.2 Modified Lemlich model

In comparison to the resistor model discussed in section 5.5.2.1, modified Lemlich model was developed based on scattering of $\overline{\overline{\lambda_{eff}}}$ values. Several combinations between $\overline{\overline{\lambda_{eff}}}$ , $\lambda_s$, $\lambda_f$, $\psi''$, $\Omega$ and $F$ (from section 5.5.2.1) were tested and the best fit for all porosities of different materials in the wide range of intrinsic solid to fluid thermal conductivity ratio ($\lambda_s/\lambda_f$=10-30000) collapsed on a single curve and is presented in the Figure 5.13. From Figure 5.13, the relation is observed as a straight line and is given by:

$$\lambda_{xx}^* = \frac{2}{3}\lambda_s^*(\Omega)^{1/4}(\psi")^{1/F_{xx}} \quad (5.44a)$$

and,

$$\lambda_{yy}^* = \frac{2}{3}\lambda_s^*(\Pi.\Omega)^{-1/4}(\psi")^{1/F_{yy}} \quad (5.44b)$$

where, $\lambda_{xx}^* = \frac{\lambda_{xx}}{\lambda_f}$, $\lambda_{yy}^* = \frac{\lambda_{yy}}{\lambda_f}$ and $\lambda_s^* = \frac{\lambda_s}{\lambda_f}$

For 14000 calculated values of $\overline{\overline{\lambda_{eff}}}$ using Equations 5.44a and 5.44b, the average errors are ±5% and ±7% respectively.

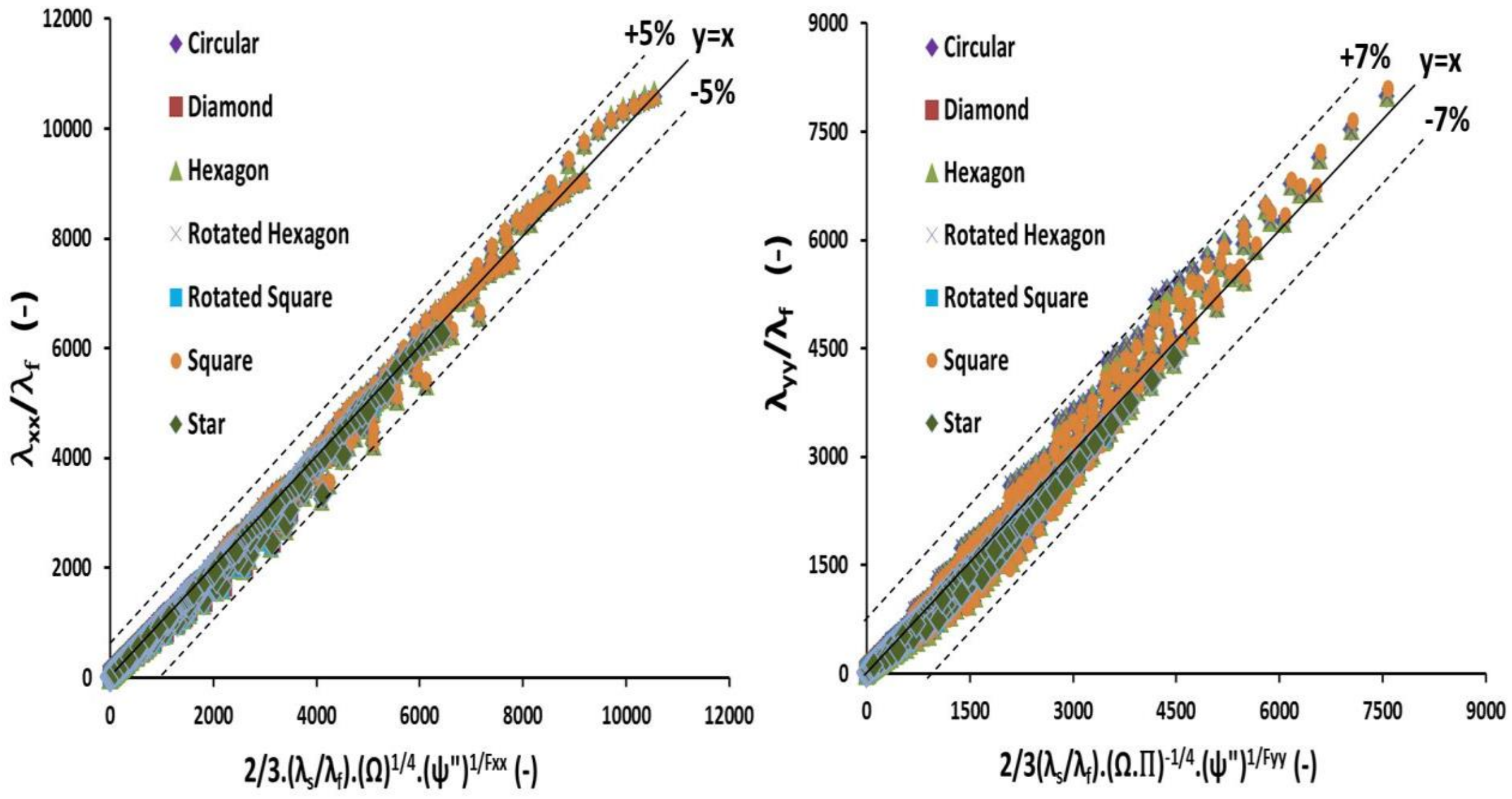

**Figure 5.13**. Plot of effective thermal conductivity tensors ($\overline{\overline{\lambda_{eff}}}$) as a function of modified Lemlich model.

### 5.6 Effective thermal conductivity correlations for ceramic foams

Like in section 5.5, three correlations, namely, resistor model, modified Lemlich model and PF model are derived for ceramic foams. The experimental effective thermal conductivity data from the literature (Dietrich et al., 2010) were gathered to derive various correlations.

### 5.6.1 Resistor model

The resistor model (see section 5.5.1.1) on the ceramic foam structure inside a cubic unit cell was applied considering *isotropic* nature of ceramic foams.

The parallel and series models to determine conductivities are given by Equations 5.45 and 5.46:

$$\lambda_{parallel} = \lambda_s(1-\varepsilon_t) + \varepsilon_t\lambda_f = \psi'(\lambda_s - \lambda_f) + \lambda_f \quad (5.45)$$

$$\frac{1}{\lambda_{series}} = \left\{\frac{1-\varepsilon_t}{\lambda_s} + \frac{\varepsilon_t}{\lambda_f}\right\} = \left\{\psi'\left(\frac{1}{\lambda_s} - \frac{1}{\lambda_f}\right) + \frac{1}{\lambda_f}\right\} \quad (5.46)$$

where, $\psi' = \varrho_1\alpha^2\beta(1-\varepsilon_{st}) + \varrho_2\alpha^3(1-\Omega\varepsilon_{st})$, $\varrho_1$ and $\varrho_2$ are numerical values from Equation 3.85 (see section 3.4.4 of ceramic foam in chapter 3).

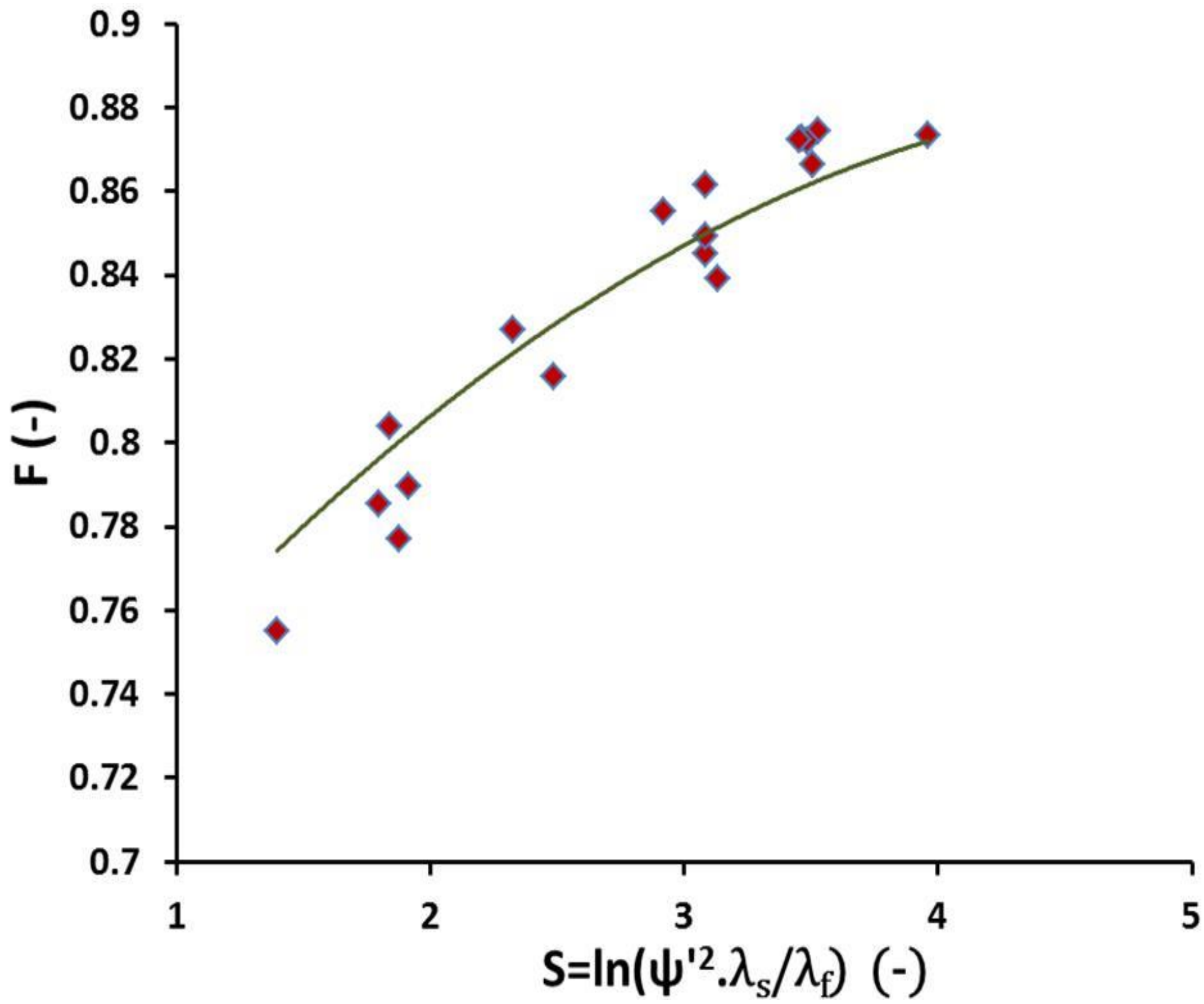

**Figure 5.14**. Plot of $F$ (dimensionless) and $S$ (dimensionless).

Equation 5.22 (see section 5.5.1.1) is solved for $F$ in terms of $\lambda_{parallel}$, $\lambda_{series}$ and $\lambda_{eff}$ that contains geometric function, $\psi'$ (see Singh and Kasana, 2004). The solution is:

$$F = n\left[(1-\psi')\frac{\lambda_{eff}}{\lambda_f} + \psi'\frac{\lambda_{eff}}{\lambda_s}\right] / ln\left[1 + \psi'(1-\psi')\left(\frac{\lambda_s}{\lambda_f} + \frac{\lambda_f}{\lambda_s} - 2\right)\right] \quad (5.47)$$

Dietrich et al., (2010) performed effective thermal conductivity measurements on Alumina, Mullite and OBSiC foams for intrinsic solid to fluid thermal conductivity ratio ranging from 140- 900.

From $\lambda_{eff}$ results of Dietrich et al. (2010), values of $F$ using Equation 5.47 are extracted and a best fit that includes geometrical parameters $(\psi')$ of foam structure and ratio of constituent phases $(\lambda_s/\lambda_f)$ is obtained. The plot of $F$ as function of $S = ln(\psi'^2 . \lambda_s/\lambda_f)$ is shown in the Figure 5.14. It is observed that $F$ increases roughly quadratic polynomial with increase in $S$ and all the values of factor, $F$ for different porosities collapsed on a single curve. From the Figure 5.14, it is found that:

$$F = -0.004S^2 + 0.0593\, S + 0.7144 \quad (5.48)$$

Due to scattering of experimental data presented in Figure 5.14, an RMSD value of 0.1088 (or 10.88%) on calculated values of effective thermal conductivity is obtained (see also Equation 5.31). From the Equation 5.48 and Figure 5.14, it is evident that $F$ is a function of geometrical parameters and ratios of thermal conductivities of constituent phases and is applicable for wide range of thermal conductivity ratios $(\lambda_s/\lambda_f)$ of ceramic foams.

### 5.6.2 Modified Lemlich Model

Modified Lemlich model for ceramic foams is derived the same way as in case of *isotropic* metal foams. In case of ceramic foams (see chapter 3 and section 3.4.4), it is important to distinguish between total porosity $(\varepsilon_t)$ and open porosity $(\varepsilon_o)$. The modified Lemlich model was derived by trying the several combinations between $\lambda_{eff}$, $\lambda_s$, $\lambda_f$, $\psi'$ and $F$ (from section 5.6.1). The best fit for all porosities of different materials and intrinsic solid to fluid thermal conductivity ratios ranging from 140-900 (see Dietrich et al. 2010) collapsed on a single curve and is presented in the Figure 5.15. From Figure 5.15, the relation is observed as a straight line and is given by Equation 5.49. An RMSD value of 0.1074 (or 10.74%) was obtained on calculated values of effective thermal conductivity.

$$\lambda_{eff}^* = \frac{2}{3}\lambda_s^*(\psi')^{1/F} \quad (5.49)$$

where, $\lambda_{eff}^* = \frac{\lambda_{eff}}{\lambda_f}$ and $\lambda_s^* = \frac{\lambda_s}{\lambda_f}$

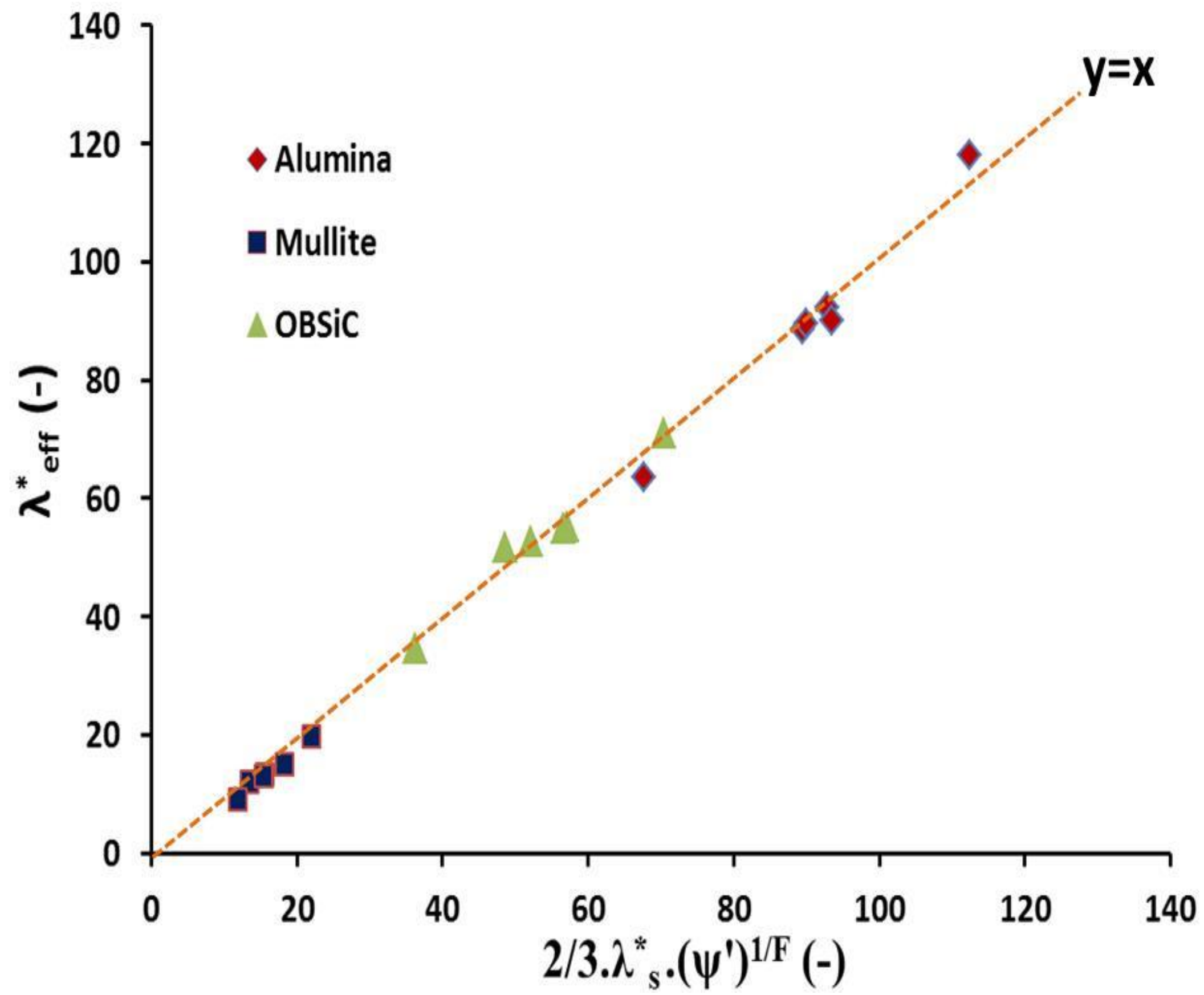

**Figure 5.15**. Plot of $\lambda_{eff}{}^{*}$ vs. $\frac{2}{3}\lambda_s{}^{*}(\psi')^{1/F}$ (dimensionless)

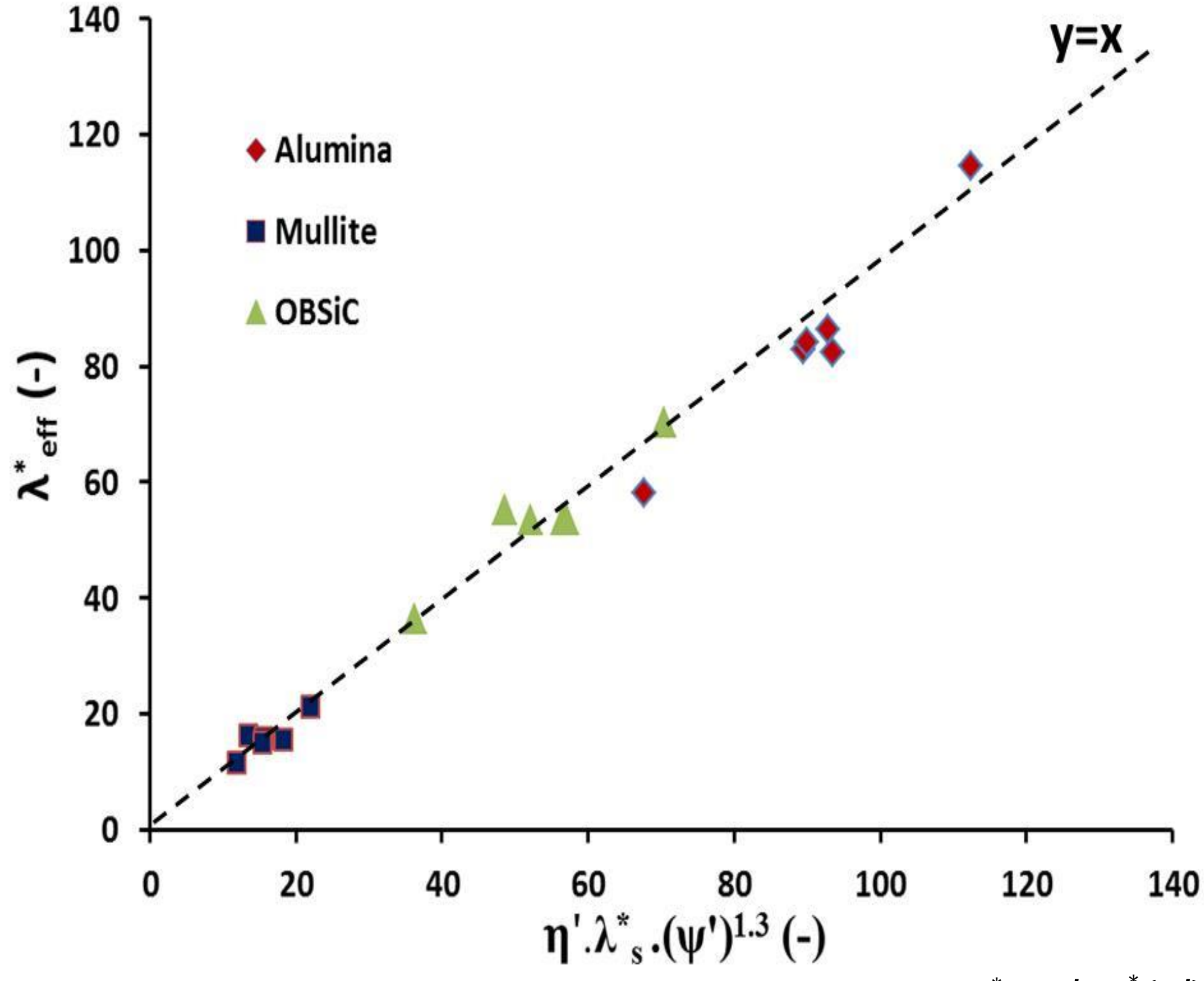

**Figure 5.16**. Presentation of PF model and its relation is found as $\lambda_{eff}{}^{*} = \eta'.\lambda_s{}^{*}(\psi')^{1.3}$.

### 5.6.3 PF model

PF model in case of ceramic foams was derived in the same way like *isotropic* metal foams. Several combinations between $\lambda_{eff}$, $\lambda_s$, $\lambda_f$ and $\psi'$ from the data presented in the work of Dietrich et al. (2010) have been tested. The best fit for all porosities of different materials and intrinsic solid to fluid thermal conductivity ratios ranging from 140-900 collapsed on a single curve and is presented in the Figure 5.16. From Figure 5.16, the relation is observed as a straight line and is given by:

$$\lambda_{eff}^{\ *} = \eta' \lambda_s^{\ *} (\psi')^{1.3} \qquad (5.50)$$

where, $\lambda_{eff}^{\ *} = \frac{\lambda_{eff}}{\lambda_f}$, $\lambda_s^{\ *} = \frac{\lambda_s}{\lambda_f}$ and $\eta'$ =0.89

An RMSD value of 0.0914 (or 9.14%) was obtained for calculated values of effective thermal conductivity and no physical interpretation of exponent $n$ =1.3 is yet provided.

## 5.7 Validation of effective thermal conductivity correlations

In this section, the analytical correlations are validated against:

- Numerically measured effective thermal conductivity data on virtual *isotropic* metal foam samples.
- Numerically measured effective thermal conductivity data on virtual *anisotropic* metal foam samples.
- Experimental effective thermal conductivity data reported in the literature for metal foams.
- Experimental effective thermal conductivity data reported in the literature for ceramic foams.

### 5.7.1 Numerical validation of *isotropic* metal foams

In this section, the three correlations i.e. resistor model, modified Lemlich model and PF model are validated against numerical data of effective thermal conductivity.

#### 5.7.1.1 Validation of resistor model

From the Figure 5.17, it is found that the analytical effective thermal conductivity values calculated using by Equation 5.22 and 5.30 underestimates and overestimates the measured thermal conductivity but the error lies within ±6% which clarifies that the effective thermal conductivity can be precisely calculated using resistor model approach if the geometrical parameters or the relations between them are known. The errors are observed only at low porosities where the node exhibits very complex shape.

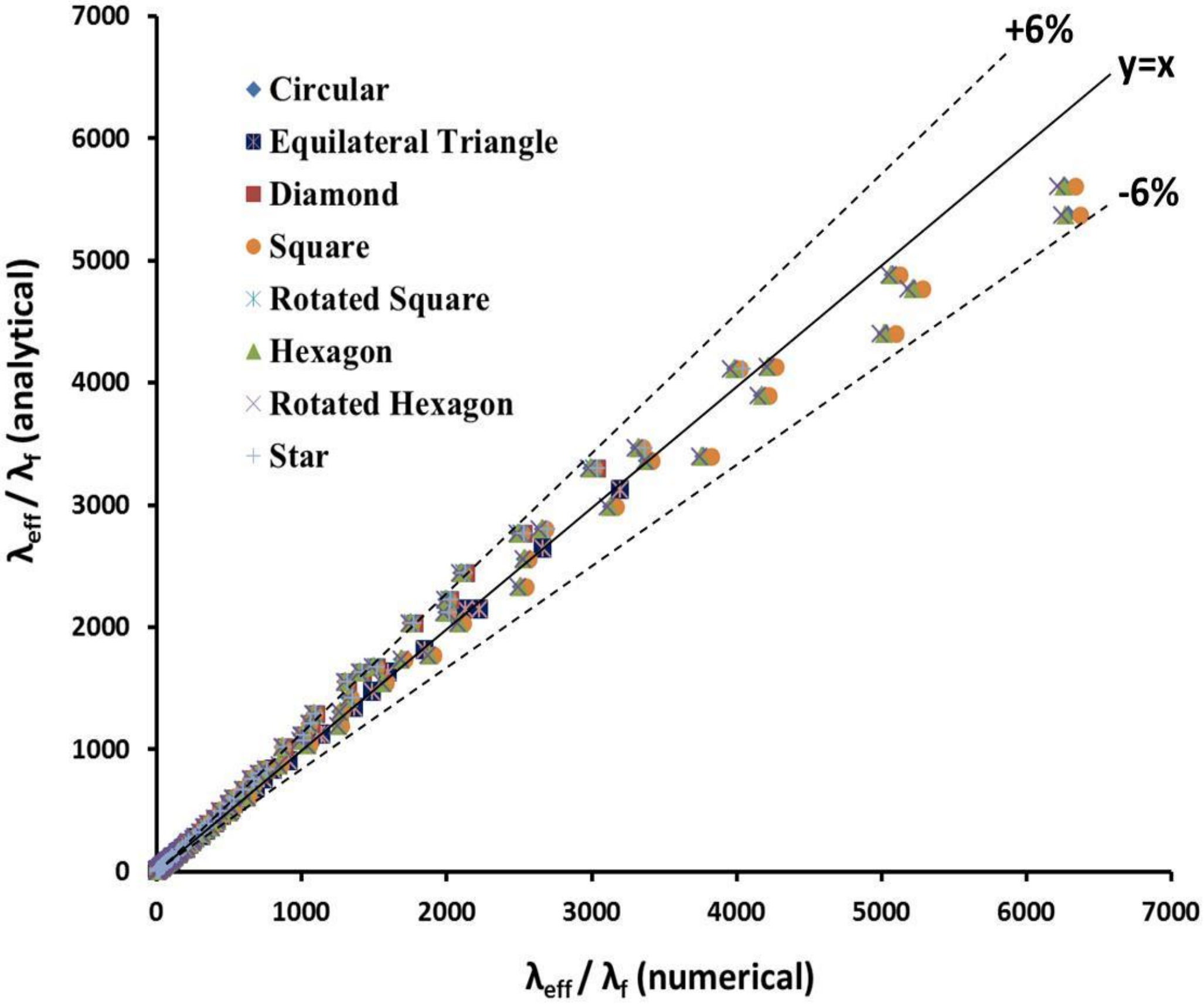

**Figure 5.17**. Comparison and validation of numerical and analytical $\lambda_{eff}$. The error is mainly due to node approximation for strut shapes at low porosity ($\varepsilon_o$<0.70).

For the porosities $\varepsilon_o > 0.70$, one group is identified for which the node approximation of any strut shape does not introduce significant error and an accuracy of ±3% is obtained. Another group of complex node shapes is found in the case of hexagon, rotated hexagon and square strut shapes for porosities $\varepsilon_o < 0.70$. In this group, the node volume approximation could not be geometrically accurate as for low porosity, node volume is significant and thus accuracy decreases. This node approximation introduces a loss in accuracy only at low porosity ($0.60 < \varepsilon_o < 0.70$) where the error as high as ±6% is obtained.

The validation states that the resistor model gives precise results for the whole range of porosity ($0.60 < \varepsilon_o < 0.95$) as it can encompass all the strut shapes and even complex ones depicts model's robustness.

### 5.7.1.2 Validation of modified Lemlich model

The validation of the modified Lemlich model for different strut shapes in the wide range of porosity is presented in the Figure 5.18. The effective thermal conductivity results are compared using correction factor $F$ obtained experimentally in Equation 5.29 and analytically using correlation presented in Equation 5.30. The total error for all strut shapes in the studied porosity range for $\lambda_s/\lambda_f$ ≈10-30000 is ±7%. This suggests that the correction factor $F$ calculated from Equation 5.30 predicts excellent effective thermal conductivity results for known intrinsic solid phase thermal conductivities. The error is mainly due to the node approximation at low porosity ($\varepsilon_o$ <0.70) of complex strut shapes.

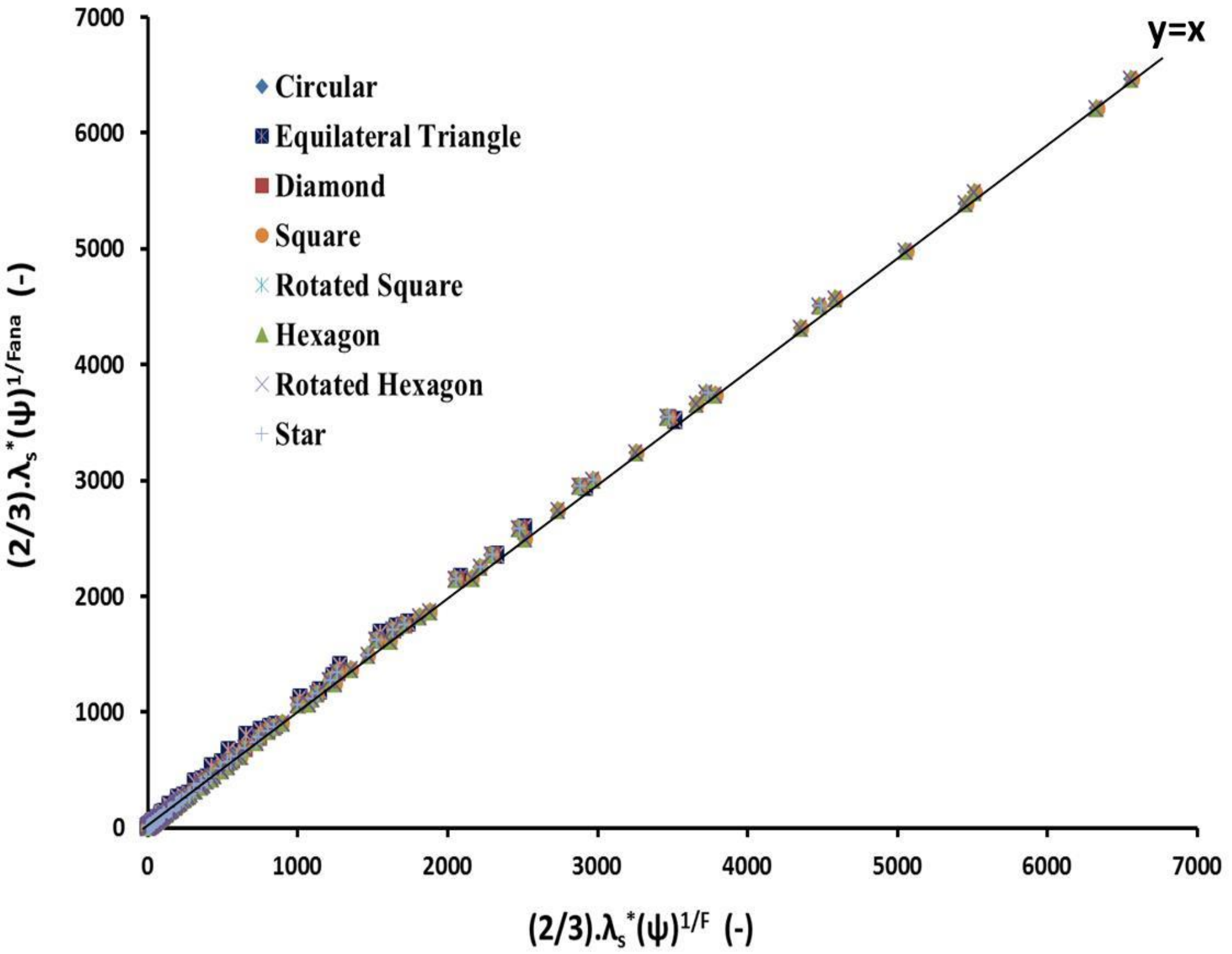

**Figure 5.18**. Validation of modified Lemlich model using correction factor $F$ obtained experimentally and analytically. ($F_{ana}$ is obtained from the analytical correlation presented in Equation 5.30).

### 5.7.1.3 Validation of PF model

The validation of PF model is presented in Figure 5.11. The fit works very well for all known intrinsic $\lambda_s$. An error of ±4% is observed for $\lambda_{eff}$ values that are associated to slight difference in temperature field contours. PF model could also be used a check to identify either effective thermal conductivity or intrinsic solid phase conductivity if any one of these properties is known. The more detailed validation of PF model is shown for ceramic foams in section 5.7.4.2.

### 5.7.2 Numerical validation of *anisotropic* metal foams

In this section, the two correlations i.e. resistor model and modified Lemlich model against numerical data of effective thermal conductivity tensors ($\overline{\overline{\lambda_{eff}}}$) are validated.

#### 5.7.2.1 Validation of resistor model

The numerically obtained effective thermal conductivities in X and Y directions i.e. $\lambda_{xx}$ and $\lambda_{yy}$ against the correlations presented in Equation 5.43a and 5.43b are compared and validated. The correlations predict excellent results in the porosity range 0.70< $\varepsilon_o$ <0.95 for different strut cross sections as presented in Figure 5.19 (left) for $\lambda_{xx}$.

In the Figure 5.19 (left), a "high error region" is presented which is actually a low porosity region (0.60< $\varepsilon_o$ <0.70) where node junction of complex strut shapes (e.g. diamond, star) is difficult to approximate. Moreover, at very high elongation factor of the foam sample, the node junction is extremely difficult to visualize and due to its complex nature (see section 3.5.3, CAD validation of *anisotropic* metal foams: Table A.1, Appendix A), it introduces biased errors in geometrical characteristics and therefore, in $\lambda_{xx}$ but the error is limited to 14% in low porosity range (0.60< $\varepsilon_o$ <0.70).

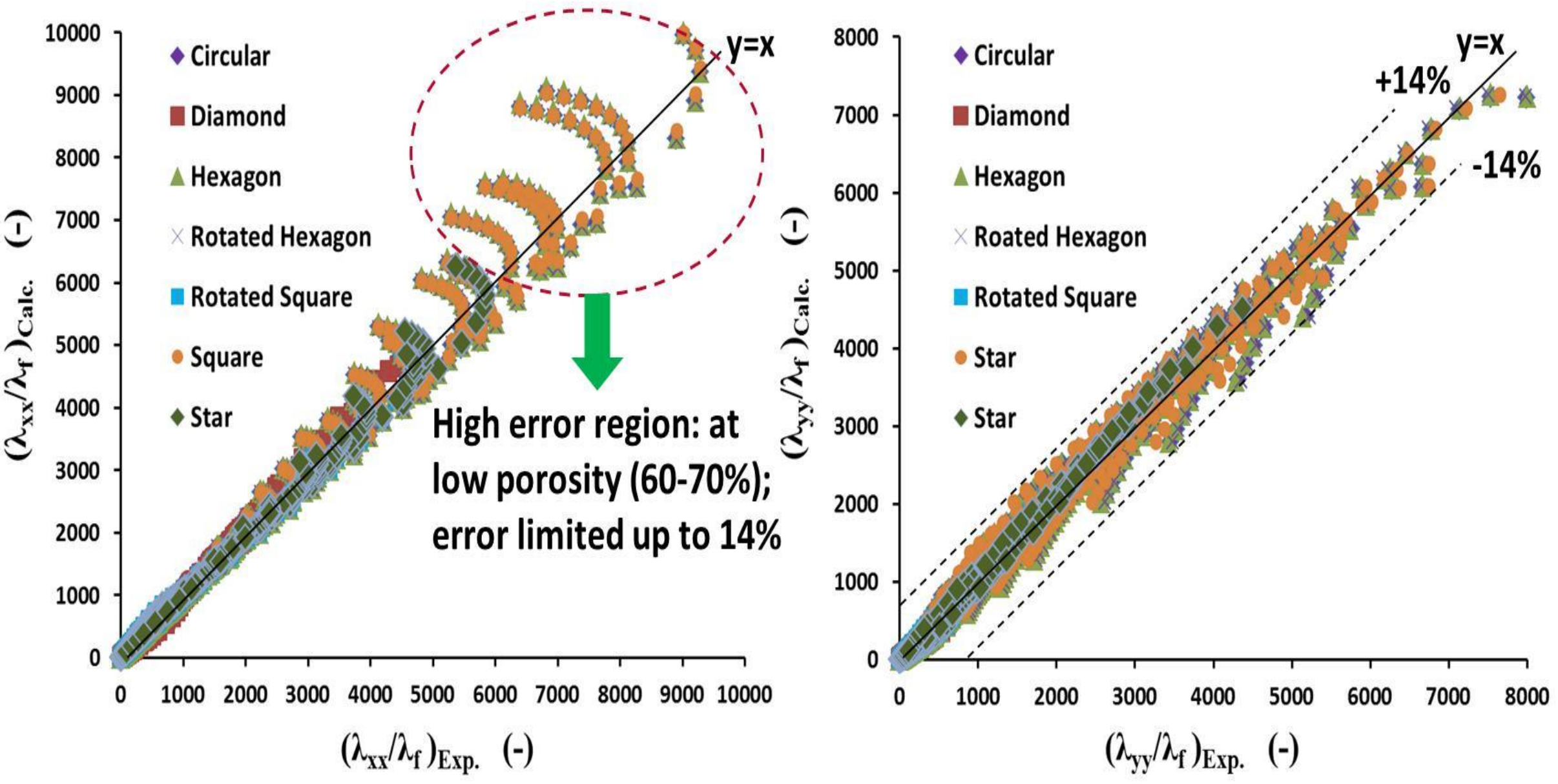

**Figure 5.19**. Validation of numerical and calculated effective thermal conductivity tensors using resistor model. Left: $\lambda_{xx}$. Right: $\lambda_{yy}$. The validation is made for the entire porosity and solid to fluid thermal conductivity ratio range for all elongation of foam samples.

In the Figure 5.19 (right), the validation of numerical and calculated values of $\lambda_{yy}$ is presented. The correlation predicts excellent results within the error range of ±14%. The error is mainly due to approximation of node junction of complex strut shapes at low porosity and at very high elongation factor.

### 5.7.2.2 Validation of modified Lemlich model

The effective thermal conductivities in X and Y directions i.e. $\lambda_{xx}$ and $\lambda_{yy}$ are validated against the correlations derived in Equations 5.44a and 5.44b (see section 5.5.2.2) and are presented in Figure 5.20. In the Figure 5.20, the numerically obtained $F_{xx}$ and $F_{yy}$ are mainly compared and validated against correlations presented in Equation 5.43a and 5.43b. Note that $\lambda_{xx}^*(=\lambda_{xx}/\lambda_f)$ and $\lambda_{yy}^*(=\lambda_{yy}/\lambda_f)$ are functions of $F_{xx}$ and $F_{yy}$. It is evident that the correction factors $F_{xx}$ and $F_{yy}$ are in excellent agreement and predict excellent results of $\lambda_{xx}^*$ and $\lambda_{yy}^*$. The cumulative errors in calculated $\lambda_{xx}^*$ and $\lambda_{yy}^*$ are ±14% respectively.

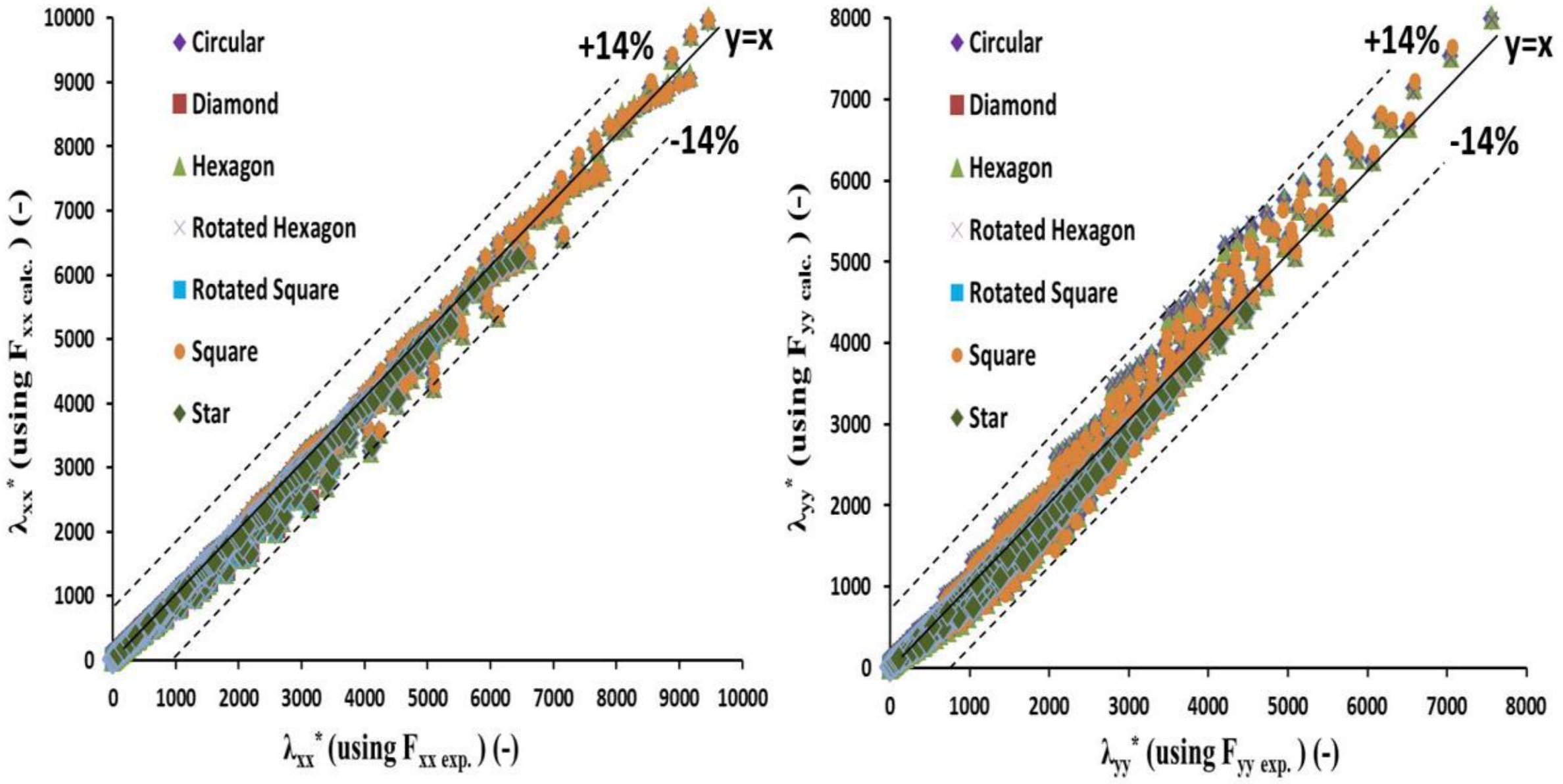

**Figure 5.20**. Validation of experimental and calculated effective thermal conductivity tensors using modified Lemlich model. Left:$\lambda_{xx}^*$. Right:$\lambda_{yy}^*$. The validation is made for the entire porosity and solid to fluid thermal conductivity ratio range.

### 5.7.3 Experimental validation of metal foams

***Intrinsic solid phase thermal conductivity determination***

In order to validate the analytical models developed in the present work with the experimental data where no prior information of intrinsic $\lambda_s$ is known, various $\lambda_{eff}$ values reported by different authors: Takegoshi et al., 1992; Paek et al., 2000; Bhattacharya et al., 2002 are listed in Table 5.4.

To increase the scope and validity of the correlations over a wide range of different strut shapes of different materials and different manufacturing techniques, intrinsic $\lambda_s$ was calculated from the experimental effective thermal conductivity values of Takegoshi et al., 1992; Paek et al., 2000; Bhattacharya et al., 2002 using (and solving simultaneously) Equation 5.29 and 5.30 of the resistor model and also from Equation 5.33 of PF model (see Table 5.4) considering equilateral triangular strut of *isotropic* metal foams (as illustrated by these authors).

Intrinsic $\lambda_s$ obtained from resistor model (Equation 5.29 and 5.30) is used in PF model (Equation 5.33) to determine analytically $\lambda_{eff}$. Similarly, intrinsic $\lambda_s$ obtained from PF model (Equation 5.33) is used in resistor model (Equation 5.30 followed by Equation 5.29) to determine analytically $\lambda_{eff}$ (see Table 5.4).

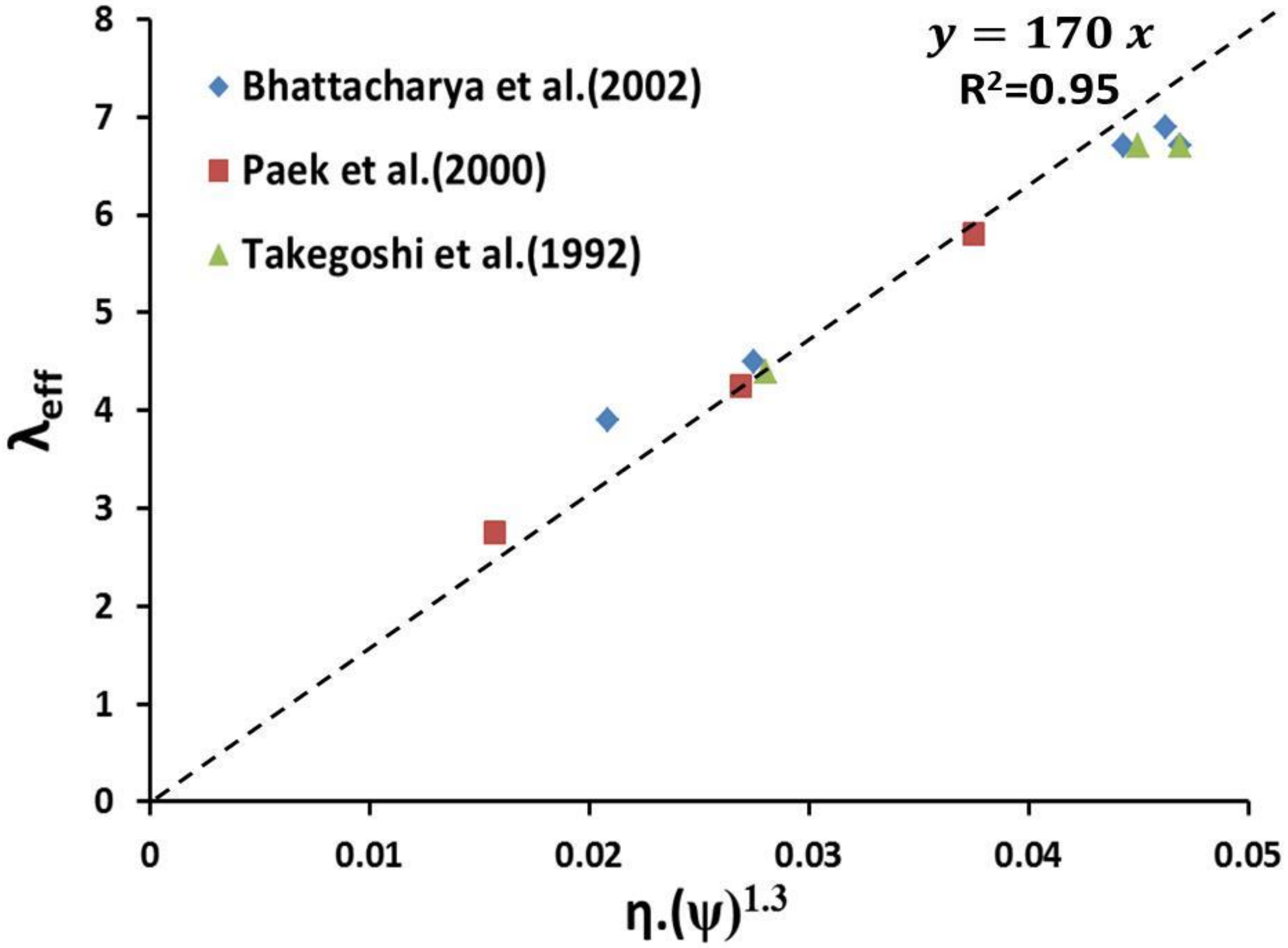

**Figure 5.21**. Plot of experimental values of $\lambda_{eff}$ from various authors (Bhattacharya et al., 2002; Paek et al., 2000; Takegoshi et al., 1992) against $\eta.(\psi)^{1.3}$ to determine an approximate value of $\lambda_s$ for all ERG (Al 6101 T alloy) foams. An approximate value of $\lambda_s$=170 W/mK is obtained using PF model.

**Table 5.4**. Comparison of experimental and calculated $\lambda_{eff}$ and associated errors (in %).

| | | Exp. | Analytical using Eqs. 5.29 and 5.30 | Calculated using Eq. 5.33 | Error (%) | Analytical using Eq. 5.33 | Calculated using Eqs. 5.29 and 5.30 | Error (%) | $\lambda_s$=170 (from Fig. 5.21) | Error (%) |
|---|---|---|---|---|---|---|---|---|---|---|
| Authors | $\varepsilon_o$ | $\lambda_{eff}$ | $\lambda_s$ | $\lambda_{eff}$ (calc.) | | $\lambda_s$ | $\lambda_{eff}$ (calc.) | | $\lambda_{eff}$ (calc.) | |
| Bhattacharya et al., (2000) | 0.905 | 6.7 | 166 | 6.93 | 3.50 | 161 | 7.11 | 6.13 | 7.09 | 5.9 |
| | 0.949 | 3.9 | 222 | 4.13 | 5.88 | 210 | 4.31 | 10.45 | 3.16 | -19.0 |
| | 0.909 | 6.7 | 178 | 7.01 | 4.64 | 170 | 7.15 | 6.73 | 6.71 | 0.1 |
| | 0.906 | 6.9 | 176 | 7.22 | 4.70 | 168 | 7.34 | 6.45 | 7.00 | 1.4 |
| | 0.937 | 4.5 | 191 | 4.66 | 3.66 | 184 | 4.90 | 8.98 | 4.16 | -7.6 |
| Paek et al., (2000) | 0.938 | 4.25 | 183 | 4.39 | 3.21 | 177 | 4.62 | 8.79 | 4.07 | -4.2 |
| | 0.92 | 5.8 | 181 | 6.02 | 3.87 | 174 | 6.24 | 7.60 | 5.67 | -2.2 |
| | 0.959 | 2.75 | 205 | 2.88 | 4.57 | 196 | 3.03 | 10.35 | 2.38 | -13.5 |
| Takegoshi et al., (1994) | 0.936 | 4.4 | 182 | 4.54 | 3.18 | 176 | 4.78 | 8.64 | 4.24 | -3.5 |
| | 0.908 | 6.7 | 174 | 6.97 | 4.03 | 167 | 7.14 | 6.58 | 6.80 | 1.6 |
| | 0.905 | 6.7 | 166 | 6.93 | 3.50 | 161 | 7.11 | 6.13 | 7.09 | 5.9 |
| Average Deviation (%) | | | | | 4.07 | | | 7.89 | | -3.18 |

Both models (resister and PF models) have predicted an approximate decrease in 25% in intrinsic $\lambda_s$ than the conductivity of parent material. Depending upon the manufacturing process employed and constituent materials, an intrinsic $\lambda_s$~150-180 W/mK was observed for most of porosities in the case of pure Al/Al 6101 T alloy foams.

In Table 5.4, intrinsic values of $\lambda_s$ from two methodologies (resister and PF models) are varying by 4% which confirms the validity of two models over wide range of porosity and material properties. Average deviations in $\lambda_{eff}$ are found to be 4.07% (for PF model) and 7.89% (for resistor model) respectively.

Experimental $\lambda_{eff}$ values of foam samples reported by Takegoshi et al., 1992; Paek et al., 2000; Bhattacharya et al., 2002 are plotted against their $\eta.(\psi)^{1.3}$ (using PF model) to determine only one value of intrinsic $\lambda_s$ as traced in Figure 5.21 and a value of $\lambda_s \approx 170$ $W/mK$ is obtained. Using $\lambda_s$=170 $W/mK$ from Figure 5.21, $\lambda_{eff}$ are calculated using resistor model and the average deviation for calculated values is -3.18%. Deviation is least for the empirical models to predict effective thermal conductivity analytically. Both methodologies are equivalent in terms of precision.

In determining effective thermal conductivity, prior knowledge of intrinsic $\lambda_s$ is crucial. By developing analytical models i.e. resistor and PF models, one does not need to measure intrinsic $\lambda_s$. Equations 5.29 (and 5.30) and 5.33 form a system of two linear equations as intrinsic $\lambda_s$ and $\lambda_{eff}$ are unknown parameters for a known $\lambda_f$. This approach expands the potential use of the analytical effective thermal conductivity correlations.

### 5.7.4 Experimental validation of ceramic foams

#### 5.7.4.1 Validation of resistor and modified Lemlich models

The effective thermal conductivity results of ceramic foam samples are validated using resistor model and modified Lemlich model approaches and presented in Table 5.5. The correction factor $F$ that is obtained from experimental effective thermal conductivity values of Dietrich et al., (2010) is compared first using Equation 5.48. The average deviation observed is 1.66% and suggests that the results of analytical correlation (Equation 5.48) is a good prediction of $F$ obtained from experiments.

Further, the analytical results of $\lambda_{eff}$ obtained from resistor model using Equation 5.48 (followed by Equation 5.47) and experimental $F$ obtained from measured effective thermal conductivity values are compared. The average deviations are 5.04% and 5.51% respectively which clarifies that the effective thermal conductivity can be precisely calculated using resistor model approach if the geometrical parameters or relationship between parameters are known. Lastly, the $\lambda_{eff}$ results are calculated using modified Lemlich model (Equation 5.49) using analytically obtained $F$ (from Equation 5.48) and compared. The average deviation is 2.36 %. All the three approaches lead to same values. This signifies the importance of geometrical parameters of foams on thermal properties.

**Table 5.5**. Experimentally and analytically determined effective thermal conductivity, $\lambda_{eff}$ of $Al_2O_3$, Mullite and OBSiC ceramic sponges of different pore sizes and porosities.

| Experiments | | | | | Analytical | | |
|---|---|---|---|---|---|---|---|
| | | | | | Resistor model | | Modified Lemlich Model |
| Material | $\varepsilon_n$ | $\lambda_{eff}/\lambda_f$ | $F$ (*Exp.) | $F$ (*Ana.) | $\lambda_{eff}/\lambda_f$ (Using $F$ Ana.) | $\lambda_{eff}/\lambda_f$ (Using $F$ Exp.) | $\lambda_{eff}/\lambda_f$ (Using $F$ Ana.) |
| $Al_2O_3$ | 0.75 | 112.39 | 0.873 | 0.886 | 120.15 | 117.91 | 120.75 |
| | 0.80 | 89.49 | 0.873 | 0.872 | 89.15 | 88.63 | 88.48 |
| | | 92.79 | 0.874 | 0.874 | 92.53 | 92.15 | 92.04 |
| | | 90.01 | 0.872 | 0.872 | 90.28 | 89.55 | 89.67 |
| | | 93.42 | 0.882 | 0.871 | 88.59 | 89.96 | 87.89 |
| | 0.85 | 67.67 | 0.865 | 0.853 | 64.03 | 63.53 | 61.61 |
| Mullite | 0.75 | 22.01 | 0.827 | 0.831 | 22.29 | 19.86 | 20.00 |
| | 0.80 | 13.47 | 0.730 | 0.813 | 17.66 | 12.09 | 15.02 |
| | | 15.48 | 0.777 | 0.812 | 17.29 | 13.43 | 14.62 |
| | | 18.28 | 0.834 | 0.810 | 16.93 | 15.03 | 14.22 |
| | | 15.4 | 0.785 | 0.808 | 16.56 | 13.05 | 13.82 |
| | 0.85 | 11.94 | 0.755 | 0.789 | 13.26 | 9.21 | 10.22 |
| OBSiC | 0.75 | 70.43 | 0.866 | 0.873 | 72.67 | 70.93 | 71.81 |
| | 0.80 | 52.16 | 0.841 | 0.859 | 56.49 | 52.70 | 54.80 |
| | | 56.53 | 0.859 | 0.859 | 56.49 | 54.82 | 54.80 |
| | | 57.09 | 0.862 | 0.859 | 56.49 | 55.08 | 54.80 |
| | | 48.52 | 0.820 | 0.861 | 58.11 | 51.73 | 56.52 |
| | 0.85 | 36.2 | 0.816 | 0.837 | 39.57 | 34.49 | 36.54 |
| Average Deviation | | | | 1.66 % | 5.04 % | 5.51 % | 2.36 % |

*Exp. -Experiments, *Ana. – Analytical

The modified Lemlich model validation is already presented in Figure 5.15 using correction factor $F$ obtained experimentally from measured $\lambda_{eff}$ values. The fit works very well for all known intrinsic $\lambda_s$ of ceramic foams. An error of ±4% is observed in entire range of $\lambda_{eff}$ values which is comparable to uncertainties on intrinsic solid phase conductivity values of Dietrich et al., (2010).

### 5.7.4.2 Validation of PF model

In the case of metal foams, intrinsic solid phase thermal conductivity is not reported in the literature. However, using the experimental data of Dietrich et al., (2010), PF model for ceramic foams is validated by comparing the intrinsic solid phase conductivity presented in Table 5.6. The intrinsic $\lambda_s$ values for each effective thermal conductivity value and its averaged value for same type of material is calculated and presented. The comparison shows an excellent validation of PF model's prediction of intrinsic value of solid phase conductivity from $\lambda_{eff}$ values. Note that Dietrich et al., (2010) used $\lambda_f$=0.03 for air as fluid medium at 50C.

**Table 5.6**. Comparison and validation of intrinsic solid phase conductivity ($\lambda_s$) using PF model with experimental data of Dietrich et al., (2010).

| Material | $\varepsilon_t$ | $\lambda_{eff}/\lambda_f$ (*exp.) | $\lambda_s$ (*exp.) | $\lambda_s$ (calculated using Eq. 5.50) | | $\lambda_s$ (calculated using Figure 5.22) |
|---|---|---|---|---|---|---|
| | | | | $\lambda_s$ | Average $\lambda_s$ | |
| Alumina | 0.754 | 112.39 | 26 | 23.46 | 25.9 | 25.4 |
| | 0.808 | 89.49 | | 25.78 | | |
| | 0.802 | 92.79 | | 25.68 | | |
| | 0.806 | 90.01 | | 25.58 | | |
| | 0.809 | 93.42 | | 27.09 | | |
| | 0.854 | 67.67 | | 27.83 | | |
| Mullite | 0.736 | 22.01 | 4.4 | 4.47 | 4.36 | 4.28 |
| | 0.785 | 13.47 | | 3.57 | | |
| | 0.789 | 15.48 | | 4.21 | | |
| | 0.793 | 18.28 | | 5.09 | | |
| | 0.797 | 15.4 | | 4.40 | | |
| | 0.834 | 11.94 | | 4.43 | | |
| OBSiC | 0.742 | 70.43 | 15 | 14.74 | 14.66 | 14.62 |
| | 0.791 | 52.16 | | 14.35 | | |
| | 0.791 | 56.53 | | 15.55 | | |
| | 0.791 | 57.09 | | 15.71 | | |
| | 0.786 | 48.52 | | 12.95 | | |
| | 0.845 | 36.2 | | 14.69 | | |

*Exp values are taken from Dietrich et al., (2010).

Moreover, $\lambda_{eff}{}^{*}$ is plotted against $\eta'.(\psi')^{1.3}$ in Figure 5.22 to identify the intrinsic $\lambda_s$ of the same material in the entire porosity range. By linear fit, the $\lambda_s$ values are calculated and presented in Table 5.6. The predictions of intrinsic $\lambda_s$ are in excellent agreement with the experimental data.

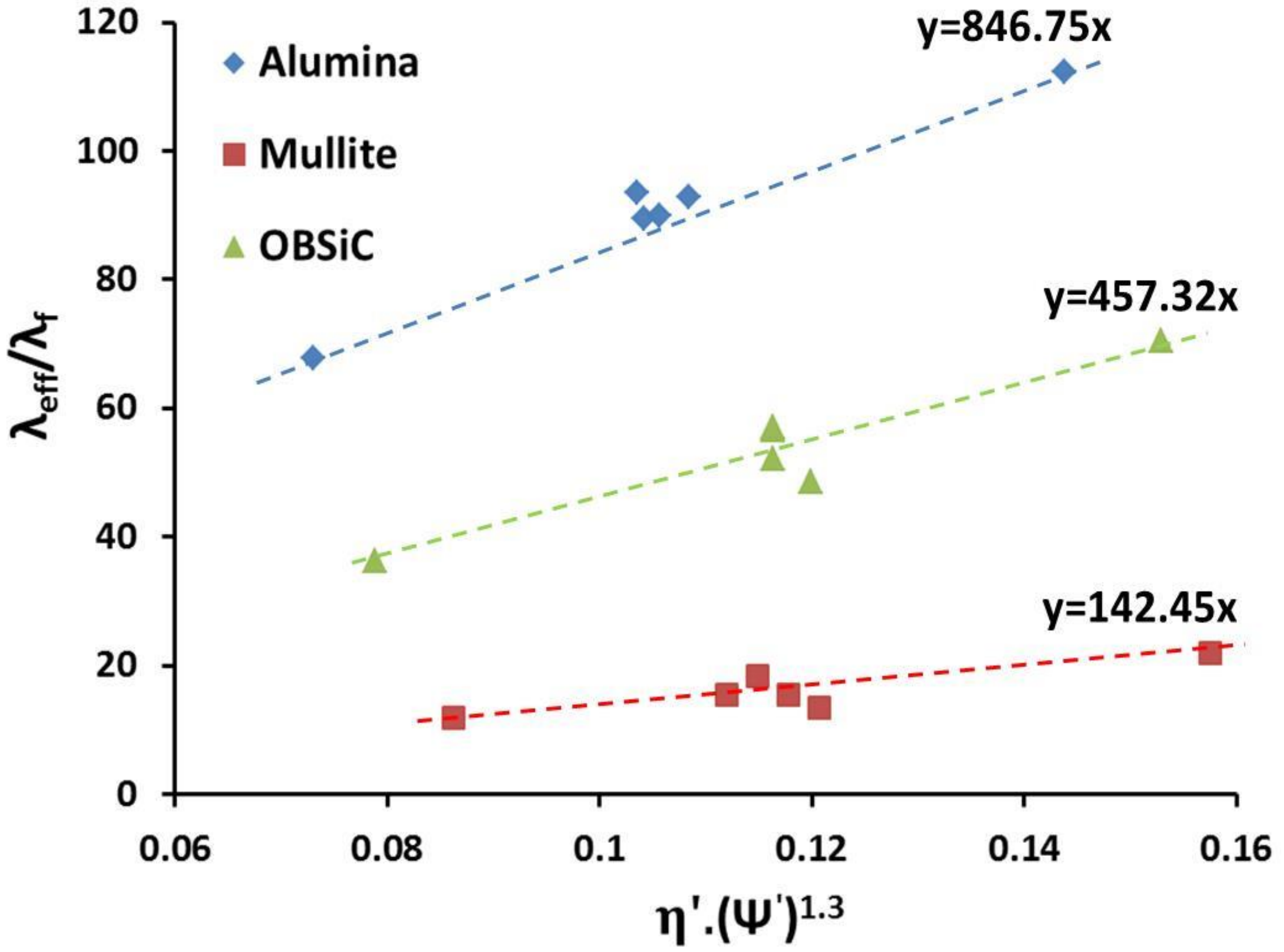

**Figure 5.22**. Plot of $\lambda_{eff}/\lambda_f$ vs. $\eta'.(\psi')^{1.3}$.

### 5.8 Summary and Conclusion

Open cell foams are commonly used for a wide range of heat and mass transfer applications, such as, combustion systems, evaporators, heat exchangers, thermal insulators, filtration systems, etc. Most research efforts on thermal modelling are based on homogenization approach, where the transport processes are analysed using macroscopic models, dealing with physical quantities averaged on a large number of pores and thereby neglecting the detailed micro-scale effects. In order to obtain accurate predictions, these macroscopic models require reliable information on the effective properties of foam structures. These properties do depend on the micro-structure of the porous media (e.g. porosity, pore size distribution, construction of the struts, etc.) and also on the properties of the constituting materials. In this regard, the effective thermal conductivity of open cell foams is an essential parameter whenever conductive heat transfer is present.

The physicochemical structures of open cell foams are very complex, which virtually rules out the possibility of analytical approaches. Most research studies have therefore been directed on experimental measurements; however, published accounts on this subject are scarce, and the scope and depth of prior experiments have been limited. As it is not often possible to conduct experiments to study the effective thermal conductivity, a theoretical expression is needed to predict its values. In the literature, several correlations for predicting/estimating the effective thermal conductivity based on either asymptotic or micro-structural have been proposed. In this chapter, an overview of various effective thermal conductivity correlations of open cell foams has been presented. The validity of the correlations for predicting effective thermal conductivity has been evaluated by comparing the numerical data from the present work.

The correlations in the literature are derived for very small range of high porosity samples ($0.85 < \varepsilon_o < 0.95$) and very high solid to fluid conductivity ratio ($\lambda_s/\lambda_f > 10{,}000$) and often contain a fitting parameter that changes with the assumed unit cell considered by the authors. The correlations are usually related only with the porosity of the foam sample. There is, thus, a need for a correlation that would be valid on wide range of porosities, wide range of solid to fluid conductivity ratios and could relate geometrical characteristics of the foam matrix. In this context, three different correlations namely resistor model, modified Lemlich model and PF model based on an ideal periodic structure (tetrakaidecahedra) of the solid foam are established for metal and ceramic foams. Critical importance of intrinsic solid phase thermal conductivity in predicting effective thermal conductivity has been discussed. New correlations derived in the present work have produced results with minimum deviation from the experimental and numerical data of the foams of different materials for a wide range of porosities and complex strut shapes.

The correlations can be used in either ways: when $\lambda_s$ is known to determine $\lambda_{eff}$ or vice versa. These correlations can also be used to solve a problem simultaneously as a system of two linear equations where both $\lambda_s$ and $\lambda_{eff}$ are unknowns for a known $\lambda_f$. Since intrinsic solid phase conductivity is usually unknown, this is extremely useful as it allows the tailoring of foams for many different engineering applications.

An algorithm is presented in Figure 5.23 to predict effective thermal conductivity just by knowing micro-structural foam characteristics and solid to fluid phase thermal conductivity ratios of open cell foams.

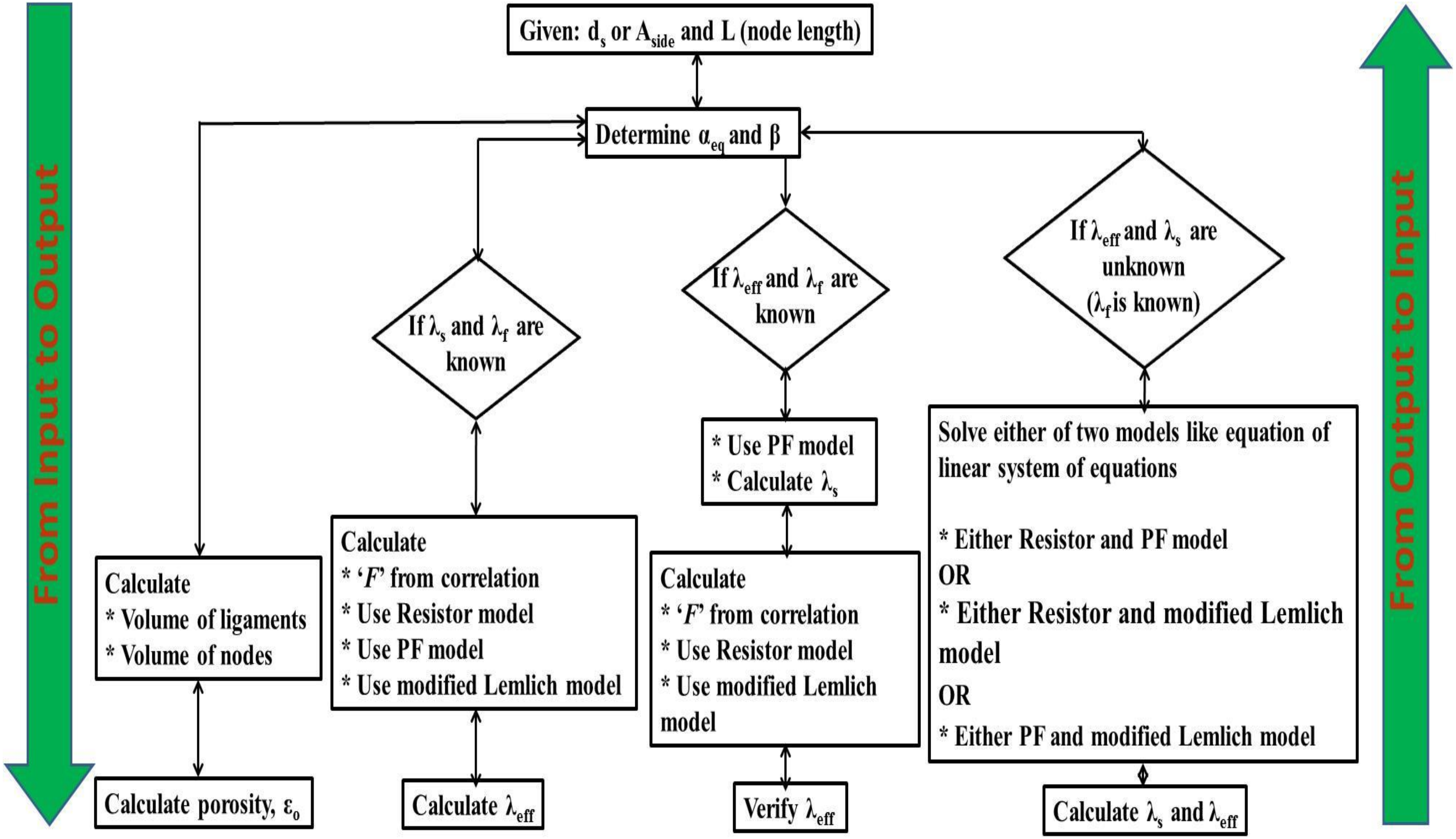

**Figure 5.23**. Algorithm to predict effective thermal conductivity by geometrical characteristics of a foam matrix.

# Chapter 6

# Conclusion and recommendations

## 6.1 Conclusion

This thesis is devoted to the investigation of morphological and thermo-hydraulic properties of open cell foams as they have diverse industrial applications such as light weight structures, filters, energy absorptions, heat exchangers and even medical applications. The purpose of this work was to characterize the morphological properties of foam matrix and study their influence on fluid and effective thermal conductivity properties. Further, the purpose was to develop generally applicable correlations for estimating the geometrical characteristics, pressure drop and effective thermal conductivity of foam structures having different strut cross sections that are key parameters for designing heat exchangers, solar receivers, columns and reactors with foams as internals.

In the following paragraphs, a complete summary of the present work along with the concluding remarks is given:

- In the second chapter, the morphological characterization of cast open cell foams of periodic cellular structures of tetrakaidecahedron geometry (an efficiently space filling and widely accepted representative geometry of foams) has been accomplished. Virtual foam samples were modelled in CAD and further materialized using casting method through foundry route. The morphological characterization was carried out by using in-house code *iMorph* on tomography samples.

  The conventional way of representing the cell size i.e. PPI does not give reliable information about the cell size. The PPI value should be treated merely as a nominal value. Its use as a modelling parameter or in developing correlations is not recommended.

  Depending upon the porosity, commercially available open cell foams exhibit different strut cross sections, namely circular, convex and concave triangular. The shape of the strut cross section affects greatly the geometrical as well as thermo-hydraulic properties of foams. Therefore, it is recommended to take the shape of the strut cross section into account for geometric or thermo-hydraulic modelling on open cell foams.

  Depending on the manufacturing route, most of the replicated open cell foams have hollow struts and hence exhibit certain strut porosity. The strut porosity is hardly accessible by the fluid that flows through the foam structure. Therefore, in order to determine the open or so-called hydrodynamic related porosity, the strut porosity must be subtracted from the total porosity of the foam. In such foams, volume image analysis e.g. X-ray µ-CT or MRI can be applied to determine the geometrical characteristics (e.g. specific surface area) of open cell foams.

- A comprehensive experimental characterization of open cell foams can be time consuming and expensive. Therefore, it is important to derive mathematical correlations that allow the prediction of the important geometrical properties of foams by using some easily measureable parameters.

  In this regard, in the third chapter, geometrical modelling of open cell foams was presented where an overview of the state of the art of geometric models (polyhedral, tetrakaidecahedron and pentagonal dodecahedron structure) as well as correlations for predicting the geometrical properties of foam matrix has been presented and extensively discussed. The performance and validity of state of the art of correlations were examined by comparing the experimental data (from the present work as well as the literature) with the predictions of the correlations.

  It was observed that the correlations reported in the literature for geometrical properties of foams are not predictive and a large deviation between the predicted and the measured values is seen. The deviation of predicted values from experimental data can be attributed to the selection of geometric model used to derive the correlation and the variation in the strut cross sections with/without porosity change.

  With the advancement in technology to fabricate the controlled morphological foam structures for numerous applications, virtual *isotropic* foam samples of different strut shapes were developed in CAD. A new generalized correlation for the theoretical estimation of the geometrical characteristics of open cell foams was proposed. The correlation was derived by using tetrakaidecahedron as representative geometry and taking the different strut shapes of foam structure into account without any fitting parameter. The generalized correlation to predict specific surface area has different forms a common basis depending on the strut shape and its characteristic dimension.

  These virtual *isotropic* foam structures have been further transformed into *anisotropic* ones. Impact of anisotropy on geometrical parameters and specific surface area is studied and a generalized correlation to predict specific surface area (different forms according to the strut shape) has been developed.

  The correlations derived in the present work were validated against the measured values of the pore diameter, strut diameter and specific surface area from the present work as well as from the literature for foams of different materials in a wide range of pore size and porosity. It has been demonstrated that the proposed correlation can be extended to other complex strut shapes and can predict the geometrical characteristics of foams with least error than any other state of the art correlation either empirical or theoretical.

- Pressure drop through open cell foams is one of the key parameters for designing a heat exchanger or chemical reactor or column. It plays an important economic role in an industrial operation as it is directly related to the energy consumption. For similar exchange surface area (geometric specific surface area), open cell foams offer much lower pressure drop compared to conventional randomly packed bed spheres.

  Pressure drop measurement and modelling on open cell foams along with state of the art correlations for pressure drop estimation have been discussed in the fourth chapter. Most of the studies performed on pressure drop through open cell foams were based on experiments. 3-D numerical simulations at pore scale using commercial software based on finite volume method were performed in a very wide range of Reynolds number to distinguish Darcy and inertia regimes. It has been shown that it is possible to obtain the flow characteristics precisely using numerical simulations. A methodology is given to extract the flow characteristics of different regimes in order to avoid dispersion in friction factor. Further, the state of the art correlations were compared and validated against the numerical data of the present work and it has been found that the correlations reported in the literature are not predictive and induce very high errors.

  As a first step, a correlation (different forms on a common basis according to the strut shape) for pressure drop in periodic open cell foams was developed. Impact of geometrical parameters of foam matrix on flow characteristics is discussed and these parameters are included in the pressure drop correlations that have not been yet reported. Based on the correlation, it has been discussed that whether Ergun parameters could possess constant numerical values or not. The validity of the correlations was tested for foam structures of low and high porosities. For the entire range of porosity (low and high) studied in the present work, the correlations predicted the numerically obtained flow properties precisely. It has been recommended that flow characteristics should be compared and validated while deriving the correlations instead of pressure drop values for the entire range of velocity.

- Open cell metal foams have emerged as a result of recent developments in processing technology and are one of the most promising emerging materials for thermal management applications where a large amount of heat needs to be transported over a small volume. Compared to conventional packed bed of spheres, open cell foams offer much higher effective thermal conductivity (and thus, heat transfer) due to continuous connection of the foam structure. The widespread range of applications of open cell foams has led to an increase in the interest of modelling the heat transfer phenomena in porous media.

  It is pointed out that the precise calculation of effective thermal conductivity is required for accurate modelling of thermal transport through open cell foams. The state of the art of effective thermal conductivity correlations has been discussed in the fifth chapter. Like pressure drop through open cell foams, most of the studies on effective thermal conductivity were based on experiments. 3-D numerical simulations at pore scale in local thermal equilibrium condition (LTE) were performed on *isotropic* and *anisotropic* foam samples for a very wide range of solid to fluid phase thermal conductivity ratios. The validity and performance of state of the art correlations were examined by comparing with the numerical effective thermal conductivity data of the present work.

  Large deviations have been observed between the predicted and measured values of effective thermal conductivity which suggests that the correlations proposed in the literature are not predictive. The deviation of correlations from the numerical data can be attributed to the selection of geometric model used to derive the correlation, absence of experimental data in a wide porosity range, variation in the strut cross sections with/without porosity change and no prior knowledge of intrinsic solid phase thermal conductivity.

  A database of effective thermal conductivity is generated for both *isotropic* and *anisotropic* open cell foams. Three different dependent/independent correlations have been derived based on tetrakaidecahedron as representative geometry and taking into account the different strut shapes of the foam structure. Emphasis on prior knowledge of intrinsic solid phase thermal conductivity of foam is discussed. Any of the two correlations could be easily used a system of linear equations to predict simultaneously intrinsic solid phase thermal conductivity and effective thermal conductivity of the foam structure.

The correlations were validated against the experimental data as well as numerical data from the present work for foams of different materials in a wide range of porosity and different strut shapes. It has been demonstrated that the proposed correlations can predict the effective thermal conductivity of open cell foams with more precision than any other state of the art correlation either empirical or theoretical.

## 6.2 Recommendations

Integration of open cell foams in numerous industrial applications is essential and for the better performance of the systems, modifications in geometrical structure (strut shape and size) of open cell foams and their thermo-hydraulic properties can be varied for the desired outputs. Based on the work done in this thesis and the literature survey, some of the interesting points are recommended for future research:

- *Isotropic* foam modelling of variable strut cross sections along the ligament axis for a known cell size. This development will increase the specific surface area due to accumulation of mass at the node junction than in the centre of the ligament. Thus, for a given porosity, one could obtain different specific surface areas as a function of the parameter that controls the variation of strut cross sections. Creation of *anisotropic* foams with variable strut cross sections will also allow us a better understanding of morphological properties. The correlation to determine geometrical characteristics of foam matrix presented in this thesis can be easily adopted and extended to develop new correlations that encompass the parameter which governs the variable strut cross sections. This development will help in realizing the foams manufactured from any processing technology and thus, will help in optimizing the foam shape and its structure for a given application.
- It was demonstrated that using numerical simulations, it is possible to obtain the thermo-hydraulic properties precisely. A database can be easily created with abundant geometrical and thermo-hydraulic properties for variable strut cross sections. This will expand various dimensions to understand the influence of individual geometrical parameter of foam structure on physical properties.
- Pressure drop (or flow properties) will change while varying the strut cross section along the ligament axis for a given porosity and strut shape. This additional function needs to be added in the pressure drop correlations that were presented in this thesis. For a given structural and hydraulic constraint, foam structure can be easily optimized to meet the demand of desired outputs. Permeability and inertia coefficient need to be understood in case of *anisotropic* foam samples. Depending on the various outputs and constraints of the system, *isotropic* and *anisotropic* foams can be used as an alternative to each other.

- Creation of database of effective thermal conductivity for variable strut cross sections for *isotropic* and *anisotropic* foams is equally indispensable. As presented in this thesis that porosity is not the sole parameter to be correlated with effective thermal conductivity. Variable strut cross sections will greatly impact the effective thermal conductivity due to more mass accumulation at the node junction. Thus, the correlation presented in this thesis can be easily adopted and extended by adding the governing parameter of variable strut cross section.
- Finally, the determination of properties such as dispersion, volume and parietal heat exchange coefficients, radiative heat transfer based morphological parameters remain a major issue in the design of thermal devices such as heat exchangers or reactor column that need to be systematically addressed and studied.

# Chapter 7

**Contd.**

**Appendix A**: Presentation of specific surface area of virtual *anisotropic* foam samples.

**Table A.1**. Presentation and validation of measured and analytically calculated specific surface area of different strut shapes for each elongation of the virtual foam samples.

| Shapes | | Circular | | Equilateral Triangle | | Diamond | | Square | | Rotated Square | | Hexagon | | Rotated Hexagon | | Star | |
|---|---|---|---|---|---|---|---|---|---|---|---|---|---|---|---|---|---|
| $\varepsilon_o$ | $\Omega$ | Exp. $a_c$ | Cal. $a_c$ | Exp. $a_c$ | Cal. $a_c$ | Exp. $a_c$ | Cal. $a_c$ | Exp. $a_c$ | Cal. $a_c$ | Exp. $a_c$ | Cal. $a_c$ | Exp. $a_c$ | Cal. $a_c$ | Exp. $a_c$ | Cal. $a_c$ | Exp. $a_c$ | Cal. $a_c$ |
| 0.60 | 0.8 | 1001 | 1045 | | | | | 1111 | 1132 | | | 1047 | 1079 | 1054 | 1080 | | |
| | 1 | 980 | 1037 | | | | | 1092 | 1122 | | | 1025 | 1070 | 1033 | 1072 | | |
| | 1.2 | 993 | 1042 | | | | | 1104 | 1128 | | | 1038 | 1076 | 1046 | 1078 | | |
| | 1.4 | 1021 | 1056 | | | | | 1131 | 1145 | | | 1069 | 1091 | 1075 | 1092 | | |
| | 1.6 | 1057 | 1076 | | | | | 1167 | 1167 | | | 1108 | 1111 | 1112 | 1111 | | |
| | 1.8 | 1097 | 1098 | | | | | 1206 | 1193 | | | 1152 | 1135 | 1152 | 1133 | | |
| | 2 | 1139 | 1122 | | | | | 1248 | 1221 | | | 1198 | 1161 | 1195 | 1157 | | |
| | 2.2 | 1181 | 1147 | | | | | 1290 | 1250 | | | 1244 | 1187 | 1238 | 1183 | | |
| | 2.4 | 1223 | 1173 | | | | | 1332 | 1279 | | | 1291 | 1214 | 1281 | 1208 | | |
| | 2.6 | 1265 | 1199 | | | | | 1375 | 1308 | | | 1337 | 1241 | 1323 | 1234 | | |
| | 2.8 | 1307 | 1224 | | | | | 1416 | 1337 | | | 1382 | 1268 | 1365 | 1260 | | |
| | 3 | 1348 | 1248 | | | | | 1457 | 1364 | | | 1426 | 1293 | 1406 | 1284 | | |
| Avg. Dev. | | | -0.6 | | | | | | -1.6 | | | | -2.2 | | -2.27 | | |
| 0.65 | 0.8 | 998 | 1038 | | | | | 1111 | 1122 | | | 1045 | 1066 | 1051 | 1072 | | |
| | 1 | 977 | 1028 | | | | | 1091 | 1111 | | | 1023 | 1056 | 1029 | 1062 | | |
| | 1.2 | 990 | 1035 | | | | | 1103 | 1118 | | | 1036 | 1063 | 1042 | 1069 | | |
| | 1.4 | 1018 | 1050 | | | | | 1131 | 1137 | | | 1067 | 1080 | 1071 | 1085 | | |
| | 1.6 | 1054 | 1070 | | | | | 1166 | 1162 | | | 1106 | 1102 | 1108 | 1107 | | |
| | 1.8 | 1094 | 1094 | | | | | 1206 | 1190 | | | 1150 | 1129 | 1149 | 1132 | | |
| | 2 | 1136 | 1120 | | | | | 1248 | 1221 | | | 1195 | 1157 | 1192 | 1160 | | |
| | 2.2 | 1178 | 1147 | | | | | 1290 | 1252 | | | 1242 | 1186 | 1235 | 1188 | | |
| | 2.4 | 1221 | 1174 | | | | | 1333 | 1284 | | | 1288 | 1215 | 1277 | 1217 | | |

**Contd.**

| | | | | | | | | | | | | | | | | | |
|---|---|---|---|---|---|---|---|---|---|---|---|---|---|---|---|---|---|
| | 2.6 | 1263 | 1201 | | | | | 1375 | 1316 | | | 1334 | 1245 | 1320 | 1245 | | |
| | 2.8 | 1304 | 1227 | | | | | 1417 | 1346 | | | 1379 | 1273 | 1362 | 1273 | | |
| | 3 | 1345 | 1253 | | | | | 1458 | 1376 | | | 1423 | 1301 | 1403 | 1300 | | |
| Avg. Dev. | | | -0.59 | | | | | | -1.72 | | | | -2.45 | | -1.93 | | |
| 0.70 | 0.8 | 980 | 1012 | | | | | 1093 | 1098 | | | 1026 | 1040 | 1031 | 1048 | | |
| | 1 | 959 | 1002 | | | | | 1074 | 1085 | | | 1005 | 1028 | 1009 | 1037 | | |
| | 1.2 | 971 | 1009 | | | | | 1085 | 1094 | | | 1018 | 1036 | 1022 | 1044 | | |
| | 1.4 | 999 | 1026 | | | | | 1113 | 1114 | | | 1048 | 1054 | 1051 | 1062 | | |
| | 1.6 | 1035 | 1049 | | | | | 1148 | 1140 | | | 1086 | 1079 | 1088 | 1086 | | |
| | 1.8 | 1075 | 1075 | | | | | 1187 | 1171 | | | 1129 | 1107 | 1128 | 1114 | | |
| | 2 | 1116 | 1103 | | | | | 1228 | 1203 | | | 1174 | 1137 | 1170 | 1144 | | |
| | 2.2 | 1158 | 1131 | | | | | 1270 | 1237 | | | 1220 | 1168 | 1213 | 1174 | | |
| | 2.4 | 1199 | 1161 | | | | | 1312 | 1270 | | | 1265 | 1199 | 1255 | 1205 | | |
| | 2.6 | 1241 | 1189 | | | | | 1354 | 1303 | | | 1310 | 1229 | 1297 | 1235 | | |
| | 2.8 | 1282 | 1217 | | | | | 1396 | 1335 | | | 1355 | 1259 | 1338 | 1265 | | |
| | 3 | 1322 | 1244 | | | | | 1437 | 1366 | | | 1398 | 1288 | 1379 | 1293 | | |
| Avg. Dev. | | | -0.5 | | | | | | -1.7 | | | | -2.5 | | -1.64 | | |
| 0.75 | 0.8 | 945 | 971 | | | | | 1056 | 1057 | | | 990 | 998 | 994 | 1007 | 1429 | 1332 |
| | 1 | 925 | 960 | | | | | 1037 | 1043 | | | 970 | 986 | 973 | 995 | 1399 | 1314 |
| | 1.2 | 937 | 967 | | | | | 1049 | 1052 | | | 982 | 994 | 985 | 1003 | 1417 | 1326 |
| | 1.4 | 964 | 985 | | | | | 1075 | 1073 | | | 1011 | 1013 | 1013 | 1022 | 1460 | 1354 |
| | 1.6 | 998 | 1009 | | | | | 1109 | 1101 | | | 1048 | 1039 | 1049 | 1048 | 1514 | 1391 |
| | 1.8 | 1037 | 1037 | | | | | 1147 | 1132 | | | 1090 | 1068 | 1088 | 1077 | 1574 | 1434 |
| | 2 | 1077 | 1066 | | | | | 1188 | 1166 | | | 1133 | 1099 | 1129 | 1107 | 1637 | 1478 |
| | 2.2 | 1117 | 1096 | | | | | 1229 | 1200 | | | 1177 | 1131 | 1170 | 1139 | 1701 | 1524 |
| | 2.4 | 1158 | 1126 | | | | | 1270 | 1234 | | | 1221 | 1162 | 1212 | 1171 | 1764 | 1570 |
| | 2.6 | 1198 | 1156 | | | | | 1311 | 1268 | | | 1264 | 1194 | 1253 | 1202 | 1827 | 1615 |
| | 2.8 | 1237 | 1185 | | | | | 1351 | 1301 | | | 1307 | 1224 | 1293 | 1232 | 1889 | 1658 |
| | 3 | 1276 | 1213 | | | | | 1391 | 1332 | | | 1349 | 1253 | 1332 | 1262 | 1950 | 1700 |
| Avg. Dev. | | | -0.4 | | | | | | -1.6 | | | | -2.5 | | -1.4 | | -9.28 |
| 0.80 | 0.8 | 891 | 911 | 283 | 279 | 1093 | 1070 | 997 | 995 | 1022 | 1003 | 934 | 938 | 937 | 946 | 1343 | 1267 |
| | 1 | 871 | 899 | 279 | 273 | 1070 | 1055 | 979 | 982 | 996 | 989 | 915 | 925 | 917 | 934 | 1314 | 1249 |
| | 1.2 | 883 | 907 | 282 | 277 | 1084 | 1065 | 990 | 991 | 1011 | 998 | 926 | 934 | 929 | 942 | 1332 | 1261 |
| | 1.4 | 909 | 926 | 288 | 285 | 1115 | 1088 | 1016 | 1012 | 1045 | 1020 | 954 | 953 | 955 | 962 | 1371 | 1290 |
| | 1.6 | 941 | 950 | 297 | 296 | 1156 | 1118 | 1048 | 1040 | 1088 | 1047 | 989 | 979 | 989 | 988 | 1422 | 1327 |
| | 1.8 | 978 | 978 | 307 | 308 | 1201 | 1152 | 1084 | 1072 | 1134 | 1079 | 1028 | 1008 | 1026 | 1017 | 1479 | 1369 |
| | 2 | 1016 | 1007 | 317 | 320 | 1248 | 1188 | 1123 | 1105 | 1182 | 1113 | 1069 | 1039 | 1065 | 1048 | 1538 | 1414 |
| | 2.2 | 1054 | 1038 | 328 | 332 | 1296 | 1225 | 1162 | 1139 | 1231 | 1147 | 1111 | 1071 | 1105 | 1080 | 1598 | 1460 |
| | 2.4 | 1092 | 1068 | 338 | 345 | 1345 | 1262 | 1201 | 1173 | 1278 | 1181 | 1152 | 1102 | 1144 | 1111 | 1658 | 1505 |
| | 2.6 | 1130 | 1097 | 348 | 357 | 1392 | 1298 | 1240 | 1207 | 1325 | 1215 | 1193 | 1134 | 1183 | 1143 | 1717 | 1550 |
| | 2.8 | 1167 | 1126 | 358 | 369 | 1439 | 1333 | 1279 | 1239 | 1371 | 1247 | 1234 | 1164 | 1221 | 1173 | 1775 | 1593 |
| | 3 | 1204 | 1154 | 368 | 381 | 1485 | 1367 | 1317 | 1270 | 1417 | 1278 | 1273 | 1193 | 1259 | 1202 | 1831 | 1634 |
| Avg. Dev. | | | -0.4 | | 0.58 | | -4.5 | | -1.4 | | -5.2 | | -2.4 | | -1.22 | | -7.72 |
| | 0.8 | 812 | 828 | 260 | 257 | 995 | 978 | 911 | 908 | 930 | 915 | 853 | 853 | 855 | 861 | 1221 | 1166 |
| | 1 | 795 | 816 | 256 | 252 | 974 | 964 | 895 | 895 | 906 | 901 | 835 | 841 | 837 | 849 | 1195 | 1148 |
| | 1.2 | 805 | 824 | 258 | 255 | 987 | 973 | 905 | 904 | 920 | 910 | 846 | 849 | 847 | 857 | 1211 | 1160 |
| | 1.4 | 829 | 842 | 265 | 263 | 1016 | 995 | 928 | 924 | 951 | 931 | 871 | 868 | 872 | 876 | 1247 | 1188 |
| | 1.6 | 859 | 866 | 273 | 272 | 1052 | 1025 | 958 | 951 | 990 | 958 | 903 | 893 | 903 | 901 | 1293 | 1224 |
| 0.85 | 1.8 | 892 | 893 | 282 | 283 | 1093 | 1058 | 992 | 982 | 1032 | 989 | 939 | 922 | 937 | 929 | 1345 | 1265 |
| | 2 | 927 | 921 | 292 | 294 | 1137 | 1093 | 1027 | 1014 | 1076 | 1021 | 976 | 952 | 973 | 959 | 1399 | 1308 |
| | 2.2 | 962 | 950 | 302 | 306 | 1180 | 1128 | 1063 | 1047 | 1120 | 1054 | 1014 | 982 | 1009 | 990 | 1453 | 1351 |
| | 2.4 | 997 | 979 | 312 | 317 | 1224 | 1164 | 1100 | 1080 | 1163 | 1087 | 1052 | 1012 | 1045 | 1020 | 1508 | 1395 |
| | 2.6 | 1031 | 1008 | 321 | 329 | 1267 | 1198 | 1136 | 1112 | 1206 | 1119 | 1089 | 1042 | 1081 | 1050 | 1561 | 1437 |
| | 2.8 | 1065 | 1035 | 331 | 340 | 1310 | 1232 | 1172 | 1143 | 1248 | 1150 | 1126 | 1071 | 1116 | 1079 | 1614 | 1479 |
| | 3 | 1099 | 1062 | 339 | 351 | 1352 | 1264 | 1207 | 1172 | 1289 | 1180 | 1163 | 1098 | 1151 | 1107 | 1666 | 1518 |
| Avg. Dev. | | | -0.2 | | 0.64 | | -3.6 | | -1.2 | | -4.5 | | -2.2 | | -1.09 | | -6.24 |
| | 0.8 | 700 | 710 | 227 | 226 | 856 | 844 | 786 | 783 | 799 | 788 | 735 | 734 | 736 | 739 | 1049 | 1013 |
| | 1 | 685 | 699 | 223 | 221 | 838 | 831 | 772 | 771 | 779 | 776 | 720 | 723 | 721 | 728 | 1027 | 997 |
| | 1.2 | 694 | 706 | 226 | 224 | 848 | 839 | 780 | 779 | 791 | 784 | 729 | 730 | 730 | 736 | 1040 | 1007 |
| 0.90 | 1.4 | 714 | 723 | 231 | 230 | 873 | 860 | 801 | 798 | 817 | 803 | 751 | 748 | 751 | 753 | 1071 | 1033 |
| | 1.6 | 740 | 745 | 239 | 239 | 905 | 887 | 827 | 822 | 850 | 827 | 778 | 771 | 778 | 776 | 1111 | 1065 |
| | 1.8 | 769 | 769 | 247 | 248 | 940 | 917 | 856 | 850 | 887 | 855 | 809 | 796 | 808 | 802 | 1155 | 1102 |
| | 2 | 799 | 795 | 256 | 258 | 977 | 948 | 887 | 879 | 924 | 884 | 841 | 823 | 839 | 829 | 1201 | 1141 |
| | 2.2 | 829 | 822 | 265 | 268 | 1015 | 980 | 919 | 909 | 962 | 914 | 874 | 851 | 870 | 857 | 1248 | 1181 |

**Contd.**

| | | | | | | | | | | | | | | | | | |
|---|---|---|---|---|---|---|---|---|---|---|---|---|---|---|---|---|---|
| | 2.4 | 859 | 848 | 273 | 278 | 1052 | 1012 | 950 | 938 | 999 | 944 | 907 | 878 | 901 | 884 | 1295 | 1220 |
| | 2.6 | 889 | 874 | 282 | 288 | 1089 | 1044 | 982 | 967 | 1036 | 973 | 939 | 905 | 933 | 911 | 1341 | 1258 |
| | 2.8 | 918 | 899 | 290 | 298 | 1126 | 1074 | 1013 | 995 | 1072 | 1001 | 971 | 931 | 963 | 938 | 1387 | 1295 |
| | 3 | 947 | 922 | 298 | 308 | 1162 | 1103 | 1044 | 1022 | 1107 | 1028 | 1002 | 956 | 994 | 963 | 1431 | 1331 |
| Avg. Dev. | | | -0.2 | | 0.81 | | -2.8 | | -0.9 | | -3.8 | | -1.9 | | -0.93 | | -4.81 |
| | 0.8 | 525 | 531 | 175 | 171 | 641 | 635 | 591 | 589 | 598 | 591 | 552 | 551 | 553 | 554 | 785 | 767 |
| | 1 | 514 | 522 | 172 | 167 | 627 | 624 | 580 | 579 | 583 | 582 | 540 | 542 | 541 | 545 | 769 | 754 |
| | 1.2 | 521 | 528 | 174 | 169 | 635 | 631 | 587 | 586 | 592 | 588 | 548 | 548 | 548 | 551 | 779 | 763 |
| | 1.4 | 536 | 541 | 178 | 174 | 654 | 647 | 602 | 601 | 612 | 603 | 564 | 561 | 564 | 565 | 802 | 783 |
| | 1.6 | 556 | 558 | 184 | 181 | 678 | 669 | 622 | 620 | 637 | 623 | 585 | 580 | 584 | 583 | 832 | 809 |
| 0.95 | 1.8 | 577 | 578 | 191 | 188 | 704 | 692 | 644 | 642 | 664 | 645 | 608 | 600 | 607 | 603 | 865 | 838 |
| | 2 | 600 | 599 | 198 | 195 | 732 | 717 | 668 | 665 | 692 | 668 | 632 | 622 | 630 | 625 | 899 | 869 |
| | 2.2 | 622 | 619 | 205 | 203 | 760 | 743 | 692 | 689 | 720 | 692 | 656 | 643 | 654 | 647 | 934 | 900 |
| | 2.4 | 645 | 640 | 212 | 211 | 788 | 768 | 716 | 712 | 748 | 715 | 681 | 665 | 678 | 668 | 969 | 930 |
| | 2.6 | 667 | 660 | 219 | 218 | 816 | 792 | 740 | 735 | 776 | 738 | 705 | 686 | 701 | 690 | 1004 | 961 |
| | 2.8 | 689 | 680 | 225 | 226 | 843 | 816 | 764 | 757 | 803 | 760 | 729 | 706 | 725 | 710 | 1038 | 990 |
| | 3 | 711 | 699 | 232 | 233 | 870 | 839 | 788 | 778 | 829 | 781 | 752 | 726 | 747 | 730 | 1072 | 1017 |
| Avg. Dev. | | | 0.01 | | -1.4 | | -1.9 | | -0.5 | | -3.0 | | -1.5 | | -0.7 | | -3.3 |

*Avg. Dev. is the average deviation presented in % and is calculated between the error in calculated and measured values of specific surface areas for a given porosity in the entire range of elongation factor.

In case of anisotropic foam samples, the formulae to predict specific surface area of different strut cross sections are given below.

- Surface area of equilateral triangular shape:

$$a_c = \frac{\left[12\alpha_t\beta\left[\frac{2}{\sqrt{\Pi}}+\frac{1}{\sqrt{\zeta}}+\frac{2}{\Pi}.(\delta)^{\frac{1}{3}}+\frac{1}{\zeta}.(\omega)^{\frac{1}{3}}\right]+\frac{3\sqrt{3}}{2}\alpha_t{}^2\left[2\Pi+\zeta+\frac{1}{2}(\delta)^{\frac{2}{3}}+\frac{(\omega)^{\frac{2}{3}}}{4}\right]\right]}{16\sqrt{2}L} \quad (A.1)$$

where, $\alpha_t = \frac{A_t}{L}$ and $\beta = \frac{L_s}{L}$

- Surface area of diamond shape:

$$a_c = \frac{\left[16\alpha_{det}\beta\left[\frac{2}{\sqrt{\Pi}}+\frac{1}{\sqrt{\zeta}}+\frac{2}{\Pi}.(\delta)^{\frac{1}{3}}+\frac{1}{\zeta}.(\omega)^{\frac{1}{3}}\right]+3\sqrt{3}\alpha_{det}{}^2\left[2\Pi+\zeta+\frac{1}{2}(\delta)^{\frac{2}{3}}+\frac{(\omega)^{\frac{2}{3}}}{4}\right]\right]}{16\sqrt{2}L} \quad (A.2)$$

where, $\alpha_{det} = \frac{A_{det}}{L}$ and $\beta = \frac{L_s}{L}$

- Surface area of square shape:

$$a_c = \frac{\left[16\alpha_s\beta\left[\frac{2}{\sqrt{\Pi}}+\frac{1}{\sqrt{\zeta}}+\frac{2}{\Pi}.(\delta)^{\frac{1}{3}}+\frac{1}{\zeta}.(\omega)^{\frac{1}{3}}\right]+6\alpha_s{}^2\left[2\Pi+\zeta+\frac{1}{2}(\delta)^{\frac{2}{3}}+\frac{(\omega)^{\frac{2}{3}}}{4}\right]\right]}{16\sqrt{2}L} \quad (A.3)$$

where, $\alpha_s = \frac{A_s}{L}$ and $\beta = \frac{L_s}{L}$

- Surface area of rotated square shape:

$$a_c = \frac{\left[16\alpha_{rs}\beta\left[\frac{2}{\sqrt{\Pi}}+\frac{1}{\sqrt{\zeta}}+\frac{2}{\Pi}.(\delta)^{\frac{1}{3}}+\frac{1}{\zeta}.(\omega)^{\frac{1}{3}}\right]+6\alpha_{rs}{}^2\left[2\Pi+\zeta+\frac{1}{2}(\delta)^{\frac{2}{3}}+\frac{(\omega)^{\frac{2}{3}}}{4}\right]\right]}{16\sqrt{2}L} \quad (A.4)$$

where, $\alpha_{rs} = \frac{A_{rs}}{L}$ and $\beta = \frac{L_s}{L}$

Surface area of hexagon shape:

$$a_c = \frac{\left[24\alpha_h\beta\left[\frac{2}{\sqrt{\Pi}}+\frac{1}{\sqrt{\zeta}}+\frac{2}{\Pi}.(\delta)^{\frac{1}{3}}+\frac{1}{\zeta}.(\omega)^{\frac{1}{3}}\right]+9\sqrt{3}\alpha_h{}^2\left[2\Pi+\zeta+\frac{1}{2}(\delta)^{\frac{2}{3}}+\frac{(\omega)^{\frac{2}{3}}}{4}\right]\right]}{16\sqrt{2}L} \quad (A.5)$$

**Contd.**

where, $\alpha_h = \frac{A_h}{L}$ and $\beta = \frac{L_s}{L}$

- Surface area of rotated hexagon shape:

$$a_c = \frac{\left[24\alpha_{rh}\beta\left[\frac{2}{\sqrt{\Pi}} + \frac{1}{\sqrt{\zeta}} + \frac{2}{\Pi}.(\delta)^{\frac{1}{3}} + \frac{1}{\zeta}.(\omega)^{\frac{1}{3}}\right] + 9\sqrt{3}\alpha_{rh}{}^{2}\left[2\Pi + \zeta + \frac{1}{2}(\delta)^{\frac{2}{3}} + \frac{(\omega)^{\frac{2}{3}}}{4}\right]\right]}{16\sqrt{2}L} \quad (A.6)$$

where, $\alpha_h = \frac{A_{rh}}{L}$ and $\beta = \frac{L_s}{L}$

- Surface area of star shape:

$$a_c = \frac{\left[48\alpha_{st}\beta\left[\frac{2}{\sqrt{\Pi}} + \frac{1}{\sqrt{\zeta}} + \frac{2}{\Pi}.(\delta)^{\frac{1}{3}} + \frac{1}{\zeta}.(\omega)^{\frac{1}{3}}\right] + 18\sqrt{3}\alpha_{st}{}^{2}\left[2\Pi + \zeta + \frac{1}{2}(\delta)^{\frac{2}{3}} + \frac{(\omega)^{\frac{2}{3}}}{4}\right]\right]}{16\sqrt{2}L} \quad (A.7)$$

where, $\alpha_{st} = \frac{A_{st}}{L}$ and $\beta = \frac{L_s}{L}$

**Appendix B:** Presentation of measured data of geometrical properties of foam structure**.**

**Table B.1**. Presentation of geometrical parameters of foam matrix from the literature.

| Reference | Sample | $\varepsilon_o$ | $d_p$ (mm) | $d_s$ (mm) | $a_c$ (m$^{-1}$) |
|---|---|---|---|---|---|
| Du Plessis (1994) | G100 | 0.973 | 0.254 | 0.047 | - |
| | G60 | 0.975 | 0.423 | 0.054 | - |
| | G45 | 0.978 | 0.564 | 0.054 | - |
| Lu et al. (1998) | 10 PPI | 0.96 | 0.5 | 0.092 | - |
| | 10 PPI | 0.96 | 1.0 | 0.19 | - |
| | 10 PPI | 0.96 | 2.0 | 0.36 | - |
| | 20 PPI | 0.92 | 0.5 | 0.11 | - |
| | 20 PPI | 0.92 | 1.0 | 0.215 | - |
| | 20 PPI | 0.92 | 2.0 | 0.44 | - |
| | 40 PPI | 0.88 | 0.5 | 0.13 | - |
| | 40 PPI | 0.88 | 1.0 | 0.25 | - |
| | 40 PPI | 0.88 | 2.0 | 0.49 | - |
| Zhao et al. (2001) | 10 PPI | 0.954 | 3.131 | 0.287 | - |
| | 10 PPI | 0.875 | 3.109 | 0.351 | - |
| | 30 PPI | 0.959 | 1.999 | 0.215 | - |
| | 30 PPI | 0.907 | 2.089 | 0.267 | - |
| | 30 PPI | 0.946 | 1.998 | 0.241 | - |
| | 60 PPI | 0.945 | 0.975 | 0.124 | - |
| | 10 PPI | 0.926 | 2.645 | 0.263 | - |
| | 10 PPI | 0.885 | 2.697 | 0.270 | - |
| | 30 PPI | 0.940 | 1.284 | 0.122 | - |
| | 30 PPI | 0.881 | 1.431 | 0.127 | - |
| | 60 PPI | 0.927 | 0.554 | 0.0888 | - |
| | 60 PPI | 0.915 | 0.657 | 0.0932 | - |
| Bhattarchya et al. (2002) | 5 PPI | 0.973 | 4.02 | 0.5 | 516 |
| | 5 PPI | 0.912 | 3.8 | 0.55 | 623 |
| | 10 PPI | 0.949 | 3.13 | 0.4 | 843 |
| | 10 PPI | 0.914 | 3.28 | 0.45 | 716 |
| | 20 PPI | 0.955 | 2.7 | 0.3 | 934 |
| | 20 PPI | 0.925 | 2.9 | 0.35 | 898 |
| | 20 PPI | 0.901 | 2.58 | 0.35 | 949 |
| | 40 PPI | 0.927 | 2.02 | 0.25 | 1274 |
| | 40 PPI | 0.913 | 1.8 | 0.2 | 1308 |
| | 5 PPI | 0.946 | 3.9 | 0.47 | 689 |
| | 5 PPI | 0.905 | 3.8 | 0.49 | 636 |
| | 10 PPI | 0.909 | 2.96 | 0.38 | 807 |
| | 20 PPI | 0.949 | 2.7 | 0.32 | 975 |
| | 40 PPI | 0.952 | 1.98 | 0.24 | 1300 |
| | 40 PPI | 0.937 | 2.0 | 0.24 | 1227 |
| (Boomsma and Poulikakos, 2002) | 10 | 0.921 | 6.9 | - | 820 |
| | 20 | 0.920 | 3.6 | - | 1700 |
| | 40 | 0.928 | 2.3 | - | 2700 |
| Kharyagoli et al. (2004) | NCX 2733 | 0.9 | 0.6 | - | 2500 |
| | NC 3743 | 0.83 | 0.5 | - | 3700 |
| | NC 4753 | 0.86 | 0.4 | - | 5600 |
| Moreira and Coury | 8 PPI | 0.94 | 2.3 | - | 1830 |

**Contd.**

| | | | | | |
|---|---|---|---|---|---|
| (2004) | 20 PPI | 0.88 | 0.8 | - | 1920 |
| | 45 PPI | 0.76 | 0.36 | - | 2340 |
| Giani et al. (2005) | Sample A | 0.945 | 4.3 | 0.66 | 333 |
| | Sample B | 0.927 | 4.7 | 0.82 | 352 |
| | Sample C | 0.938 | 2.2 | 0.37 | 696 |
| | Sample D | 0.937 | 2.0 | 0.33 | 767 |
| | Sample E | 0.932 | 1.7 | 0.28 | 942 |
| | Sample F | 0.911 | 4.6 | 0.80 | 449 |
| Topin et al. (2006) | NC 3743 | 0.870 | 0.569 | - | 5303 |
| | NC 2733 | 0.910 | 0.831 | - | 3614 |
| | NC 1723 | 0.880 | 1.84 | - | 1658 |
| | NC 1116 | 0.890 | 2.452 | - | 1295 |
| Dukhan et al. (2006) | 10 | 0.919 | - | - | 790 |
| | 10 | 0.915 | - | - | 810 |
| | 20 | 0.919 | - | - | 1300 |
| | 20 | 0.924 | - | - | 1200 |
| | 40 | 0.923 | - | - | 1800 |
| Liu et al. (2006) | 5 | 0.914 | 1.208 | - | - |
| | 10 | 0.918 | 1.190 | - | - |
| | 20 | 0.909 | 0.805 | - | - |
| | 40 | 0.935 | 0.685 | - | - |
| Ozmat et al. (2006) | 30 PPI | 0.97 | - | - | 1447 |
| | 20 PPI | 0.96 | - | - | 1266 |
| | 10 PPI | 0.94 | - | - | 899 |
| Stemmet et al. (2006) | 5 PPI | 0.931 | 2.45 | 0.553 | - |
| | 10 PPI | 0.932 | 0.612 | 0.138 | - |
| | 40 PPI | 0.936 | 0.314 | 0.066 | - |
| Perrot et al. (2007) | PPI 5 | 0.918 | - | - | 431 |
| | PPI 10 | 0.918 | - | - | 478 |
| | PPI 20 | 0.917 | - | - | 624 |
| | PPI 40 | 0.923 | - | - | 700 |
| Huu et al. (2009) | Sample A | 0.91 | 1.326 | 0.405 | - |
| | Sample B | 0.90 | 1.2 | 0.456 | - |
| | Sample C | 0.915 | 0.392 | 0.14 | - |
| | Sample D | 0.91 | 1.053 | 0.225 | - |
| | Sample E | 0.88 | 0.75 | 0.226 | - |
| | Sample F | 0.96 | 1.259 | 0.303 | - |
| | Sample G | 0.955 | 0.893 | 0.284 | - |
| | Sample H | 0.98 | 0.591 | 0.12 | - |
| Brun et al. (2009) | NC1116 | 0.896 | 2.452 | 0.337 | 1300 |
| | NC1723 | 0.873 | 1.840 | 0.255 | 1740 |
| | NC2733 | 0.909 | 0.831 | 0.120 | 4288 |
| | NC3743 | 0.873 | 0.569 | 0.088 | 5360 |
| | ERG Al 10 | 0.892 | 4.497 | 0.366 | 558 |
| | ERG Al 20 | 0.889 | 3.969 | 0.232 | 549 |
| | ERG Al 40 | 0.885 | 3.442 | 0.189 | 743 |
| Mancin et al. (2010) | 5 | 0.921 | 5.08 | 0.540 | 339 |
| | 10 | 0.903 | 2.54 | 0.529 | 839 |
| | 10 | 0.934 | 2.54 | 0.450 | 692 |
| | 10 | 0.956 | 2.54 | 0.445 | 537 |
| | 20 | 0.932 | 1.27 | 0.367 | 1156 |
| | 40 | 0.93 | 0.635 | 0.324 | 1679 |
| Mancin et al. (2011) | 5 | 0.92 | 5.08 | 0.490 | 341 |
| | 10 | 0.926 | 2.54 | 0.553 | 736 |
| | 20 | 0.93 | 1.27 | 0.315 | 1169 |
| | 40 | 0.926 | 0.635 | 0.282 | 1721 |
| De Jaeger et al. (2011) | PPI 10 | 0.932 | - | - | 440 |
| | PPI 10 | 0.951 | - | - | 380 |
| | PPI 20 | 0.913 | - | - | 860 |
| | PPI 20 | 0.937 | - | - | 720 |
| | PPI 20 | 0.967 | - | - | 580 |

**Appendix C:** Correlations to relate geometrical parameters of the foam structure to porosity.

- For an equilateral triangular strut shape, $R_{eq} = A_t.\sqrt{\sqrt{3}/4\pi}$

On substitution, we get:

$$\varepsilon_o = \frac{1-\frac{1}{3}\left(36\frac{\sqrt{3}}{4}{A_t}^2L_s + 24.\frac{4}{3}.\frac{\sqrt{3}}{4}.\sqrt{\frac{\sqrt{3}}{4\pi}}{A_t}^3\right)}{8\sqrt{2}L^3} \Rightarrow 3\sqrt{3}{\alpha_t}^2\beta + \frac{8}{\sqrt{3}}\sqrt{\frac{\sqrt{3}}{4\pi}}{\alpha_t}^3$$

$$= 8\sqrt{2}(1-\varepsilon_o) \qquad (C.1)$$

where, $\alpha_{det} = \frac{A_{det}}{L}$ and $\beta = \frac{L_s}{L}$

- For a square strut shape, $R_{eq} = A_s/\sqrt{\pi}$

On substitution, we get:

$$\varepsilon_o = \frac{1-\frac{1}{3}\left(36{A_s}^2L_s + 24.\frac{4}{3}{A_s}^3/\sqrt{\pi}\right)}{8\sqrt{2}L^3} \Rightarrow 12{\alpha_s}^2\beta + \frac{32}{3\sqrt{\pi}}{\alpha_s}^3 = 8\sqrt{2}(1-\varepsilon_o) \qquad (C.2)$$

where, $\alpha_s = \frac{A_s}{L}$ and $\beta = \frac{L_s}{L}$

- For a rotated square strut shape, $R_{eq} = A_{rs}/\sqrt{\pi}$

On substitution, we get:

$$\varepsilon_o = \frac{1-\frac{1}{3}\left(36{A_{rs}}^2L_s + 24.\frac{4}{3}{A_{rs}}^3/\sqrt{\pi}\right)}{8\sqrt{2}L^3} \Rightarrow 12{\alpha_{rs}}^2\beta + \frac{32}{3\sqrt{\pi}}{\alpha_{rs}}^3 = 8\sqrt{2}(1-\varepsilon_o) \qquad (C.3)$$

where, $\alpha_{rs} = \frac{A_{rs}}{L}$ and $\beta = \frac{L_s}{L}$

- For a diamond strut shape, $R_{eq} = A_{det}.\sqrt{\sqrt{3}/2\pi}$

On substitution, we get:

$$\varepsilon_o = \frac{1-\frac{1}{3}\left(36\frac{\sqrt{3}}{2}{A_{det}}^2L_s + 24.\frac{4}{3}.\frac{\sqrt{3}}{2}.\sqrt{\frac{\sqrt{3}}{2\pi}}{A_{det}}^3\right)}{8\sqrt{2}L^3} \Rightarrow 6\sqrt{3}{\alpha_{det}}^2\beta + \frac{16}{\sqrt{3}}\sqrt{\frac{\sqrt{3}}{2\pi}}{\alpha_{det}}^3$$

$$= 8\sqrt{2}(1-\varepsilon_o) \qquad (C.4)$$

where, $\alpha_{det} = \frac{A_{det}}{L}$ and $\beta = \frac{L_s}{L}$

- For a hexagon strut shape, $R_{eq} = A_h.\sqrt{3\sqrt{3}/2\pi}$

On substitution, we get:

**Contd.**

$$\varepsilon_o = \frac{1 - \frac{1}{3}\left(36\frac{3\sqrt{3}}{2}{A_h}^2 L_s + 24.\frac{4}{3}.\frac{3\sqrt{3}}{2}.\sqrt{\frac{3\sqrt{3}}{2\pi}}{A_h}^3\right)}{8\sqrt{2}L^3} \Longrightarrow 18\sqrt{3}{\alpha_h}^2\beta + 16\sqrt{3}\sqrt{\frac{3\sqrt{3}}{2\pi}}{\alpha_h}^3$$

$$= 8\sqrt{2}(1 - \varepsilon_o) \qquad (C.5)$$

where, $\alpha_h = \frac{A_h}{L}$ and $\beta = \frac{L_s}{L}$

- For a rotated hexagon strut shape, $R_{eq} = A_{rh}.\sqrt{3\sqrt{3}/2\pi}$

On substitution, we get,

$$\varepsilon_o = \frac{1 - \frac{1}{3}\left(36\frac{3\sqrt{3}}{2}{A_{rh}}^2 L_s + 24.\frac{4}{3}.\frac{3\sqrt{3}}{2}.\sqrt{\frac{3\sqrt{3}}{2\pi}}{A_{rh}}^3\right)}{8\sqrt{2}L^3} \Longrightarrow 18\sqrt{3}{\alpha_{rh}}^2\beta + 4\sqrt{3}\sqrt{\frac{3\sqrt{3}}{2\pi}}{\alpha_{rh}}^3$$

$$= 8\sqrt{2}(1 - \varepsilon_o) \qquad (C.6)$$

where, $\alpha_h = \frac{A_{rh}}{L}$ and $\beta = \frac{L_s}{L}$

- For a star (regular hexagram) strut shape, $R_{eq} = A_{st}.\sqrt{3\sqrt{3}/\pi}$

On substitution, we get,

$$\varepsilon_o = \frac{1 - \frac{1}{3}\left(36\sqrt{3}{A_{st}}^2 L_s + 24.\frac{4}{3}.3\sqrt{3}.\sqrt{\frac{3\sqrt{3}}{\pi}}{A_{st}}^3\right)}{8\sqrt{2}L^3} \Longrightarrow 36\sqrt{3}{\alpha_{st}}^2\beta + 32\sqrt{3}\sqrt{\frac{3\sqrt{3}}{\pi}}{\alpha_{st}}^3$$

$$= 8\sqrt{2}(1 - \varepsilon_o) \qquad (C.7)$$

where, $\alpha_{st} = \frac{A_{st}}{L}$ and $\beta = \frac{L_s}{L}$

**Appendix D:** Correlations to predict specific surface area from geometrical parameters of the foam structure.

- Specific surface area of an equilateral triangular strut shape is given as

$$a_c = \frac{\left\{72A_tL_s + 24.\frac{3}{4}\left(\frac{5}{4}\left(\frac{\sqrt{3}}{4}A_t{}^2\right)\right)\right\}}{2\left(8\sqrt{2}L^3\right)} = \frac{1}{\sqrt{2}L}\left(4.5\alpha_t\beta + \frac{45\sqrt{3}}{128}\alpha_t{}^2\right) \qquad (D.1)$$

- Specific surface area of a square strut shape is given as

$$a_c = \frac{\left\{96A_sL_s + 24.\frac{3}{4}\left(\frac{5}{4}A_s{}^2\right)\right\}}{2\left(8\sqrt{2}L^3\right)} = \frac{1}{\sqrt{2}L}\left(6\alpha_s\beta + \frac{45}{32}\alpha_s{}^2\right) \qquad (D.2)$$

- Specific surface area of a rotated square strut shape is given as

$$a_c = \frac{\left\{96A_{rs}L_s + 24.\frac{3}{4}\left(\frac{5}{4}A_{rs}{}^2\right)\right\}}{2\left(8\sqrt{2}L^3\right)} = \frac{1}{\sqrt{2}L}\left(6\alpha_{rs}\beta + \frac{45}{32}\alpha_{rs}{}^2\right) \qquad (D.3)$$

- Specific surface area of a diamond strut shape is given as

$$a_c = \frac{\left\{96A_{det}L_s + 24.\frac{3}{4}\left(\frac{5}{4}\left(\frac{\sqrt{3}}{2}A_{det}{}^2\right)\right)\right\}}{2\left(8\sqrt{2}L^3\right)} = \frac{1}{\sqrt{2}L}\left(6\alpha_{det}\beta + \frac{45\sqrt{3}}{64}\alpha_{det}{}^2\right) \qquad (D.4)$$

- Specific surface area of a hexagon strut shape is given as

$$a_c = \frac{\left\{144A_hL_s + 24.\frac{3}{4}\left(\frac{5}{4}\left(\frac{3\sqrt{3}}{2}A_h{}^2\right)\right)\right\}}{2\left(8\sqrt{2}L^3\right)} = \frac{1}{\sqrt{2}L}\left(9\alpha_h\beta + \frac{135\sqrt{3}}{64}\alpha_h{}^2\right) \qquad (D.5)$$

- Specific surface area of a rotated hexagon strut shape is given as

$$a_c = \frac{\left\{144A_{rh}L_s + 24.\frac{3}{4}\left(\frac{5}{4}\left(\frac{3\sqrt{3}}{2}A_{rh}{}^2\right)\right)\right\}}{2\left(8\sqrt{2}L^3\right)} = \frac{1}{\sqrt{2}L}\left(9\alpha_{rh}\beta + \frac{135\sqrt{3}}{64}\alpha_{rh}{}^2\right) \qquad (D.6)$$

Specific surface area of a star strut shape is given as

$$a_c = \frac{\left\{288A_{st}L_s + 24.\frac{3}{4}\left(\frac{5}{4}\left(3\sqrt{3}A_{st}{}^2\right)\right)\right\}}{2\left(8\sqrt{2}L^3\right)} = \frac{1}{\sqrt{2}L}\left(18\alpha_{st}\beta + \frac{135\sqrt{3}}{32}\alpha_{st}{}^2\right) \qquad (D.7)$$

**Appendix E:** Analytical modelling for porosity, $\varepsilon_t$>0.90 in case of ceramic foams.

We have followed the same methodology as presented in section 3.4.4. We have shown a circular void inside an equilateral triangular strut in Figure E1.

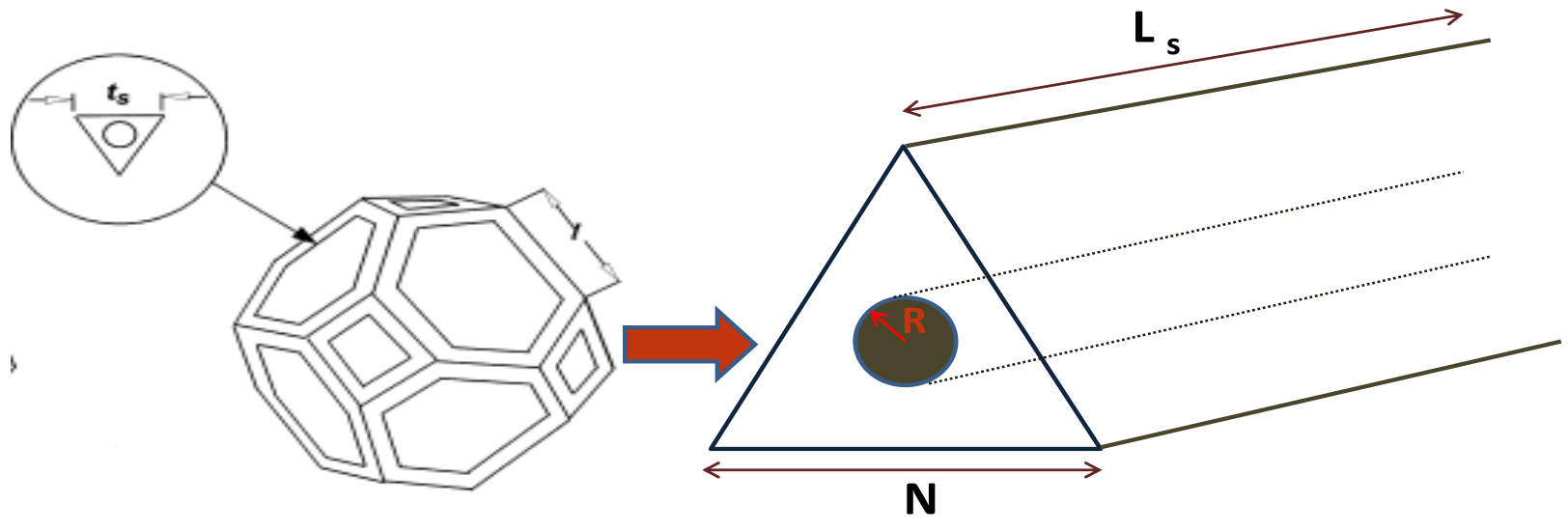

**Figure E1**. Left: Presentation of a typical hollow ceramic strut with circular void and triangular strut shape. The image is taken from the work of Richardson et al., (2000). Right: Equilateral triangular strut cross-section with circular void inside the strut is detailed. The dimensions of strut (side length, $N$) and void cross section (void radius, $R$) are clearly highlighted.

Strut porosity due to void inside the strut is calculated as:

$$\varepsilon_{st} = \frac{V_{void}}{V_{strut}} = \frac{\pi R^2 L_s}{\sqrt{3}/4N^2 L_s} \qquad (E.1)$$

Equation E.1 can be rewritten as

$$R = \kappa' N \qquad (E.2)$$

where, $\kappa' = \sqrt{\varepsilon_{st}\sqrt{3}/4\pi}$

Note that approximation at the node junction for a triangular strut will be different than circular strut cross section (see Kanaun and Tkachenko, 2008) and is given as:

$$0.594N + L_s = L \qquad (E.3)$$

Equation A.3 in non-dimension form can be rewritten as

$$0.594\alpha' + \beta' = 1 \qquad (E.4)$$

where, $\alpha' = \frac{N}{L}$ and $\beta' = \frac{L_s}{L}$

Total porosity, $\varepsilon_t$ as a function of geometrical parameters is given by Equation E.5:

$$3\sqrt{3}\alpha'^2\beta'(1-\varepsilon_{st}) + \frac{8}{\sqrt{3}}\alpha'^3(1-\kappa'\varepsilon_{st}) = 8\sqrt{2}(1-\varepsilon_t) \qquad (E.5)$$

Specific surface area is calculated by the same procedure as derived in section 3.4.4 and is given by the Equation E.6:

$$a_c = \frac{1}{\sqrt{2}L}\left(\frac{3}{2}\alpha'\beta'(3-2\pi\kappa') + \frac{45\sqrt{3}}{128}\alpha'^2(1-\ \varepsilon_{st})\right) \qquad (E.6)$$

**Appendix F:** Comparison of Ergun parameters $E_1$ and $E_1]_D$.

**Table F1**. Comparison of Ergun parameter, $E_1]_D$ (pressure drop and specific surface area approach) and $E_1$ (Ergun approach) of $Al_2O_3$, Mullite and OBSiC ceramic foams.

| | | $\varepsilon_t$ | | $\varepsilon_t$ | | $\varepsilon_o$ | |
|---|---|---|---|---|---|---|---|
| | | Experiments | | Calculated | | Analytical | |
| Material | $\varepsilon_n$ | $d_{h,\Delta P}$ (mm) | $d_{h,a_c}$ (mm) | $E_1]_D$ [ref.] Using $d_{h,\Delta P}$ | $E_1]_D$ [ref.] Using $d_{h,a_c}$ | $E_1$ Eq. (4.5) Using Ergun | $E_1$ Eq. (4.36) Using $d_{h,a_c}$ |
| $Al_2O_3$ | 0.75 | 3.46 | 2.75 | 69.44 | 43.86 | 22.13 | 30.31 |
| | 0.80 | 4.1 | 4.82 | 176.40 | 243.79 | 238.79 | 248.38 |
| | | 2.86 | 2.66 | 121.48 | 105.09 | 84.19 | 74.52 |
| | | 2.34 | 2.28 | 137.92 | 130.93 | 109.93 | 89.42 |
| | | 1.84 | 1.7 | 136.95 | 116.90 | 103.49 | 107 |
| | 0.85 | 4.24 | 3.07 | 106.62 | 55.89 | 84.32 | 60.35 |
| Mullite | 0.75 | 2.94 | 2.9 | 70.69 | 68.78 | 37.40 | 44.34 |
| | 0.80 | 4.78 | 5.01 | 59.99 | 65.90 | 49.95 | 50.94 |
| | | 3.01 | 3.25 | 81.23 | 94.70 | 41.35 | 52.51 |
| | | 2.34 | 2.77 | 96.49 | 135.21 | 75.26 | 105.08 |
| | | 1.4 | 1.63 | 53.87 | 73.02 | 47.94 | 49.48 |
| | 0.85 | 4.36 | 3.87 | 132.12 | 104.09 | 116.59 | 90.97 |
| OBSiC | 0.75 | 3.22 | 3.34 | 118.36 | 127.35 | 73.02 | 114.73 |
| | 0.80 | 5.6 | 5.54 | 89.88 | 87.96 | 70.51 | 66.83 |
| | | 3.33 | 3.68 | 156.63 | 191.29 | 153.74 | 150.61 |
| | | 2.58 | 2.75 | 114.46 | 130.04 | 104.67 | 104.92 |
| | | 1.86 | 1.65 | 159.96 | 125.88 | 96.28 | 68.26 |
| | 0.85 | 4.18 | 3.98 | 67.11 | 60.84 | 77.28 | 59.94 |
| *Average Deviation | | | | 40.52% | 35.94% | | 7.18% |

*Average deviation is calculated with respect to $E_1$ of Ergun approach using Equation 4.5 considering open porosity.

[ref.] - the data are taken from the work of Dietrich et al., (2009).

**Contd.**

**Table F2**. Comparison of Ergun parameter, $E_2]_D$ (pressure drop and specific surface area approach) and $E_2$ (Ergun approach) of $Al_2O_3$, Mullite and OBSiC ceramic foams.

| | | $\varepsilon_t$ | | $\varepsilon_t$ | | $\varepsilon_o$ | |
|---|---|---|---|---|---|---|---|
| | | Experiments | | Calculated | | Analytical | |
| Material | $\varepsilon_n$ | $d_{h,\Delta P}$ (mm) | $d_{h,a_c}$ (mm) | $E_2]_D$ [ref.] Using $d_{h,\Delta P}$ | $E_2]_D$ [ref.] Using $d_{h,a_c}$ | $E_2$ Eq. (4.5) Using Ergun | $E_2$ Eq. (4.36) Using $d_{h,a_c}$ |
| $Al_2O_3$ | 0.75 | 3.46 | 2.75 | 2.24 | 1.78 | 1.10 | 1.28 |
| | 0.80 | 4.1 | 4.82 | 1.43 | 1.68 | 1.53 | 1.38 |
| | | 2.86 | 2.66 | 1.61 | 1.50 | 1.21 | 1.20 |
| | | 2.34 | 2.28 | 1.55 | 1.51 | 1.25 | 1.22 |
| | | 1.84 | 1.7 | 1.58 | 1.46 | 1.25 | 1.19 |
| | 0.85 | 4.24 | 3.07 | 1.72 | 1.24 | 1.41 | 1.11 |
| Mullite | 0.75 | 2.94 | 2.9 | 1.68 | 1.65 | 1.12 | 1.19 |
| | 0.80 | 4.78 | 5.01 | 1.58 | 1.66 | 1.33 | 1.31 |
| | | 3.01 | 3.25 | 1.54 | 1.66 | 1.00 | 1.31 |
| | | 2.34 | 2.77 | 1.44 | 1.71 | 1.17 | 1.36 |
| | | 1.4 | 1.63 | 1.35 | 1.57 | 1.15 | 1.25 |
| | 0.85 | 4.36 | 3.87 | 1.60 | 1.42 | 1.38 | 1.22 |
| OBSiC | 0.75 | 3.22 | 3.34 | 1.87 | 1.94 | 1.34 | 1.41 |
| | 0.80 | 5.6 | 5.54 | 2.60 | 2.57 | 2.11 | 2.04 |
| | | 3.33 | 3.68 | 1.69 | 1.87 | 1.54 | 1.49 |
| | | 2.58 | 2.75 | 1.92 | 2.05 | 1.69 | 1.63 |
| | | 1.86 | 1.65 | 2.30 | 2.04 | 1.64 | 1.61 |
| | 0.85 | 4.18 | 3.98 | 1.99 | 1.89 | 1.96 | 1.66 |
| *Average Deviation | | | | 28.38% | 26.47% | | 0.35% |

*Average deviation is calculated with respect to $E_2$ of Ergun approach using Equation 4.5 considering open porosity.

[ref.] - the data are taken from the work of Dietrich et al., (2009).

# Curriculum Vitae

## Prashant KUMAR

IUSTI, CNRS UMR 7343, Technopole de Chateau Gombert, 5 Rue Enrico Fermi, 13453 Marseille Cedex 13, France
Email: prashant.kumar@etu.univ-amu.fr, Mobile : +33 624 042 771

## ACADEMIC POSITIONS

***Doctor of Philosophy*** Jun. 2011- Sep. 2014

Title: Investigation of Kelvin-like solid foams for potential engineering applications: Attractive set of geometrical and thermo-hydraulic properties.

Department Mécanique-Energétique, IUSTI, CNRS (UMR 7343), Aix-Marseille Université, France.

Supervisor: Frederic TOPIN, Assistant Professor.

***Master Research degree* with Merit** Sep. 2009- Jun. 2010

Title: Numerical study of buoyancy effects in a channel flow generated by asymmetric wall injection.

Department de Génie Mécanique, Speciality: Heat Transfer.

INSA-Lyon, France.

***Master of Science degree* with Distinction** Oct. 2008- May. 2009

Title: Numerical simulation of unsteady flow in heat exchanger tube bundle.

Department of Mechanical and Manufacturing Engineering, Specialty: Computational Fluid Dynamics.

Trinity College Dublin, Ireland.

***Bachelor of Technology degree* (GPA: 8.50/10)** Aug. 2003-May. 2007

Department of Mechanical Engineering, Specialty: Fluid Mechanics.

SVNIT-Surat, India.

## HONORS

- **Young Scientist Award** at International *DSL* Conference-2011, *Portugal*.
- **Erasmus Mundus Fellowship** by European Union for *EMMME* program (2008-2010).

## LANGUAGES

- French (Fluent), English (Fluent).
- Hindi (Native), Russian (Basic).

## COMPETENCES

- Programming Languages: C, C++.
- Operating System: Windows 7/8, LINUX (Ubuntu).
- Software Skills: Ansys, Fluent, StarCCM+, CAO, Thermoptim, Abaqus, SAP R6.4, Primavera Ver.5, MS office.

## JOURNAL PUBLICATIONS

*Published articles (Peer-reviewed)*

[1] **P. Kumar** & F. Topin, Micro-structural impact of different strut shapes and porosity on hydraulic properties of Kelvin like metal foams, *Transport in Porous Media*, 105 (1), pp. 57-81, 2014.

[2] **P. Kumar** & F. Topin, Simultaneous determination of intrinsic solid phase conductivity and effective thermal conductivity of Kelvin like foams, *Applied Thermal Engineering*, 71, pp. 536-547, 2014.

[3] **P. Kumar** & F. Topin, Investigation of fluid flow properties in open cell foams: Darcy and weak inertia regimes, *Chemical Engineering Science,* 116, pp. 793-805.

[4] **P. Kumar** & F. Topin, The geometric and thermo-hydraulic characterization of ceramic foams: An analytical approach, *Acta Materilia*, 75, pp. 273-286, 2014.

[5] **P. Kumar**, F. Topin & J. Vicente, Determination of effective thermal conductivity from geometrical properties: Application to open cell foams, *International Journal of Thermal Sciences*, 81, pp. 13-28, 2014 (top ten downloaded papers in July 2014).

[6] **P. Kumar** & F. Topin, About thermo-hydraulic properties of open cell foams: Pore scale numerical analysis of strut shapes, *Defects and Diffusion Forum*, 354, pp. 195-200, 2014.

[7] **P. Kumar**, F. Topin & L. Tadrist, Enhancement of Heat Transfer over Spatial Stationary and Moving Sinusoidal Wavy Wall: A Numerical Analysis, *Defects and Diffusion Forum*, 326-328, pp. 341-347, 2012.

*Book Chapter*

[8] **P. Kumar**, F. Topin, M. Miscevic, P. Lavieille & L. Tadrist, Heat Transfer enhancement in short corrugated mini-tubes, *Numerical Analysis of Heat and Mass Transfer in Porous Media, Advanced Structured Materials,* 27, pp 181-208, 2012.

*Under Review/Submitted*

[9] **P. Kumar**, F. Topin & L. Tadrist, Geometrical characterization of Kelvin like metal foams for different strut shapes and porosity, *Journal of Porous Media* (under review).

[10] **P. Kumar** & F. Topin, A review of geometrical characterization and effective thermal conductivity of open-celled foams, *AIChE* (under review).

[11] **P. Kumar** & F. Topin, Pressure drop correlations of metallic foams for volumetric solar receiver applications, *Applied Energy* (under review).

## CONFERENCE PROCEEDINGS

*Peer-reviewed*

[1] **P. Kumar** & F. Topin, Influence of strut shape and porosity on geometrical properties and effective thermal conductivity of Kelvin-like anisotropic metal foams, *The 15$^{th}$ Heat Transfer Conference*-2014, Japan (accepted).

[2] **P. Kumar**, J.M. Hugo, F. Topin, J. Vicente, Influence of pore and strut shape on open cell metal foam bulk properties, *American Institute of Physics*, Conference proceedings, 1453, pp. 243-248, 2012.

## COMMUNICATIONS

*Abstracts and Oral presentations in conferences*

Aug. 2014 — Influence of strut shape and porosity on geometrical properties and effective thermal conductivity of Kelvin-like anisotropic metal foams, **P. Kumar** & F. Topin: *The 15$^{th}$ Heat Transfer Conference (IHTC)-2014*, Kyoto, *Japan*.

Jun. 2013 — About thermo-hydraulic properties of open cell foams: Pore scale numerical analysis of strut shapes, **P. Kumar** & F. Topin: *Diffusion in Solid and Liquids (DSL)-2013*, Madrid, *Spain*.

Nov. 2012 — Propriétés Thermiques et Hydrauliques de Mousses Solides Régulières, **P. Kumar** & F. Topin: JEMP-2012 (11$^{émé}$ Journées d'Etude sur les Milieux Poreux), Marseille, *France*.

Sep. 2011 — Experimental and pore scale numerical characterizations of thermo-physical properties of CTIF's Kelvin's cell foam, **P. Kumar**, J.M. Hugo & F. Topin: *METFOAM-2011*, Busan, *South Korea.*

Jun. 2011 — Enhancement of Heat Transfer over Spatial Stationary and Moving Sinusoidal Wavy Wall: A Numerical Analysis, **P. Kumar**, F. Topin & L. Tadrist*: Diffusion in Solid and Liquids (DSL)-2011*, Algrave, *Portugal.*

*Poster presentations in conferences*

Jul. 2014 Evaluation of thermal properties of metal and ceramic foams by geometrical characterization, **P. Kumar** & F. Topin: *27th European Symposium on Applied Thermodynamics (ESAT)-2014*, Eindhoven, *The Netherlands.*

May. 2013 Hydraulic properties of Kelvin like foam: Influence of porosity and strut shape, **P. Kumar** & F. Topin: *INTERPORE-2013*, Prague, *Czech Republic*.

## INDUSTRIAL EXPERIENCE

*Research Engineer* Nov. 2010- May. 2011

**IUSTI, CNRS (UMR 7343)**

*Projects undertaken*

Concept of mini-channel heat exchanger: Dynamic deformation of channel wall by piezoelectric actuation.

*Responsibilities held*

- Design and analysis of pressure drop and heat transfer in static and dynamic corrugated mini-tubes.
- Optimal designs of deformable corrugated mini-tubes for improved performance of heat exchangers.

*Project Planner* Jul. 2007- Aug. 2008

**ALSTOM** Projects India Ltd.

*Projects undertaken*

BUJAGALI (HEPP) Country-Uganda - 5x51 MW vertical Kaplan turbine project and manufacturing of 4680 generator bars for 5 units.

*Responsibilities held*

- Involved in monitoring of progress compared to schedule commitments, management of the project float, proactive identification of risks and opportunities, and recommendation of actions to drive on-time performance with the goal to complete the project on time.
- Established MTS (Master Time Schedule) for stator bars and created Project in PRIMAVERA and uploaded WBS (Work Breakdown Structure).

## MEMBERSHIPS

- International Society for Porous Media, *INTERPORE* since 2013
- French Thermal Society (*Société Française de Thermique - SFT*) since 2011.

## MISCELLANEOUS

- Sports: Badminton, Volley-ball, Cricket, Trekking.
- Hobbies: Traveling, Writing, Cooking.